\documentclass[twoside,12pt]{report}

\usepackage{latexsym}
\usepackage{amsmath,amssymb,amsfonts}
\usepackage{graphicx}
\usepackage{feynmp}
\usepackage{fancyhdr}

\begin{document}   


\newbox\SlashedBox  
\def\fs#1{\setbox\SlashedBox=\hbox{#1} 
\hbox to 0pt{\hbox to 1\wd\SlashedBox{\hfil/\hfil}\hss}{#1}} 
\def\hboxtosizeof#1#2{\setbox\SlashedBox=\hbox{#1} 
\hbox to 1\wd\SlashedBox{#2}} 
\def\littleFraction#1#2{\hbox{$#1\over#2$}} 
\def\ms#1{\setbox\SlashedBox=\hbox{$#1$}
\hbox to 0pt{\hbox to 1\wd\SlashedBox{\hfil/\hfil}\hss}#1}
\def\partialslash{\mathslashed{\partial}}
\newcommand{\dd}{\raisebox{10pt}
           {\tiny$\scriptscriptstyle\longleftrightarrow$}\hspace{-11.7pt}}
\newcommand{\db}{\raisebox{10pt}
           {\tiny$\scriptscriptstyle\longleftarrow$}\hspace{-11pt}}
\newcommand{\dbo}{\raisebox{11.5pt}
           {\tiny$\scriptscriptstyle\longleftarrow$}\hspace{-11pt}}           
\newcommand{\Dsm}{\,{\raisebox{1pt}{$/$} \hspace{-8.7pt} D}}
\newcommand{\Ds}{\,{\raisebox{1pt}{$/$} \hspace{-15.7pt} D}}
\newcommand{\desm}{\,{\raisebox{1pt}{$/$} \hspace{-7pt} \partial}}
\newcommand{\des}{\,{\raisebox{1pt}{$/$} \hspace{-14.5pt} $\partial$}}
\newcommand{\napp}{\raisebox{0.5pt}{/} \hspace{-10.5pt} \in}
\def\noblackbox{\overfullrule=0pt}
\def\I {1 \hspace{-1.1mm} {\rm I}}
\def\R {{\rm I}\! {\rm R}}
\def\C {\rule{0.6pt}{8pt} \hspace{-1.35mm} {\rm C}}
\def\sC {\mbox{\rule{0.3pt}{4pt}} \hspace{-0.75mm} \mbox{C}} 
\def\Z {\mbox{\sf{Z}} \hspace{-1.6mm} \mbox{\sf{Z}} \hspace{0.4mm}}
\newcommand{\tr}{{\rm tr}}
\newcommand{\ie}{{\em i.e.~}}
\newcommand{\eg}{{\em e.g.~}}
\newcommand{\RR}{$R\otimes R$ }
\newcommand{\NSNS}{$NS\otimes NS$ }
\newcommand{\RNS}{$R\otimes NS$ }
\newcommand{\NSR}{$NS\otimes R$ }
\newcommand{\alpr}{{\alpha^{\prime}}}
\newcommand{\gym}{g_{_{{\rm YM}}}}
\newcommand{\tym}{\theta_{_{{\rm YM}}}}



\thispagestyle{empty}

\vspace*{-2cm}
\begin{flushright}
ROM2F/99/19
\end{flushright}

\vspace{1cm}

\begin{center}

{\LARGE {\bf ${\cal N}$=4 supersymmetric Yang--Mills theory and 
\rule{0pt}{22pt}the\rule{0pt}{22pt} AdS/SCFT\rule{0pt}{22pt} 
correspondence \rule{0pt}{22pt}}}

\vspace{1cm}
{\large Stefano Kovacs} \\ 
\vspace{0.6cm} 
{\large {\it Dipartimento di Fisica, \ Universit{\`a} di Roma \  
``Tor Vergata''}} \\  
{\large {\it I.N.F.N.\ -\ Sezione di Roma \ ``Tor Vergata''}} \\ 
{\large {\it Via della Ricerca  Scientifica, 1}} \\ 
{\large {\it 00173 \ Roma, \ ITALY}} 
\end{center}

\vspace*{1cm}
{\small 
\begin{center} {\bf Abstract} \end{center}

\noindent
This dissertation reviews various aspects of the ${\cal N}$=4 
supersymmetric Yang--Mills theory in particular in relation with the 
AdS/CFT correspondence. 

The first two chapters are introductory. The first one contains a 
description of the general properties of rigid supersymmetric theories 
in four dimensions both at the classical and at the quantum level. 
The second chapter is a review of the main properties of the ${\cal N}$=4 
SYM theory under consideration. 

Original results are reported in chapters 3, 4 and 5. A systematic 
re-analysis of the perturbative properties of the theory is presented 
in the third chapter. Two-, three- and four-point Green functions of 
elementary fields are computed using the component formulation and/or 
the superfield approach and subtleties related to the gauge-fixing are 
pointed out. In the fourth chapter, after an introduction to instanton 
calculus in supersymmetric gauge theories, the computation of 
the one-instanton contributions to Green functions of gauge invariant 
composite operators in the semiclassical approximation is reported. 
The calculations of four-, eight- and sixteen-point Green functions of 
operators in the supercurrent multiplet are reviewed in detail. 
The final chapter is devoted to the AdS/SCFT correspondence. Some 
general aspects are discussed. Then the attention is focused on the 
relation between instantons in ${\cal N}$=4 SYM and D-instanton 
effects in type IIB string theory. The comparison between instanton 
contributions to Green functions of composite operators in the boundary 
field theory and D-instanton generated terms in the amplitudes 
computed in type IIB string theory is performed and agreement between 
these two sources of non-perturbative effects is shown. 
}

\newpage

\baselineskip=18pt                 %


\thispagestyle{empty}

\vspace*{-1.5truecm}

\hspace*{-0.6cm}\makebox[\textwidth]{
	\includegraphics[height=1.45truecm]{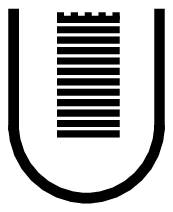}
	\raisebox{0.68cm}{
	\begin{minipage}[h]{13truecm}
        \hspace*{0.2cm}{\LARGE UNIVERSIT\`A DEGLI STUDI DI ROMA 
        \hspace*{3.76cm}``TOR \rule{0pt}{24pt}VERGATA''}
    \end{minipage}}}

\vspace{0.8cm}

\begin{center}
FACOLT\`A DI SCIENZE MATEMATICHE, FISICHE E NATURALI \\
Dipartimento di Fisica \\
\end{center}

\vspace{2cm}

\begin{center}
{\LARGE {\bf ${\cal N}$=4 supersymmetric Yang--Mills theory and 
\rule{0pt}{22pt}the\rule{0pt}{22pt} AdS/SCFT\rule{0pt}{22pt} 
correspondence \rule{0pt}{22pt}}}
\end{center}
   
\vspace{1.5cm}
                  
\begin{center}
{\large Tesi di dottorato di ricerca in Fisica} \\
\rule{0pt}{22pt} presentata da \rule{0pt}{22pt} \\ 
\rule{0pt}{22pt}{\large {\em Stefano Kovacs}} \rule{0pt}{22pt}
\end{center}

\vspace{1.5cm}                  
\noindent
Relatori \\
\rule{0pt}{20pt}{\large Professor {\em Giancarlo Rossi} \\ 
\rule{0pt}{20pt}Professor {\em Massimo Bianchi}} \\ 
Coordinatore del dottorato \rule{0pt}{30pt} \\ 
\rule{0pt}{20pt}{\large Professor {\em Piergiorgio Picozza}} \\ 

\vspace{1cm}
\begin{center}
\underline{Ciclo XI} \\
\vspace{0.5cm}
Anno Accademico 1997-1998
\end{center}



%
%


\pagestyle{fancy}
\renewcommand{\headrulewidth}{0.4pt}
\addtolength{\headwidth}{1cm}
\fancyhead{}
\fancyfoot{}

 

\pagenumbering{roman}

\setcounter{page}{1}         

\renewcommand{\headrulewidth}{0.4pt}


\fancyhead[RO,LE]{\thepage}
\fancyhead[RE,LO]{{\footnotesize {\rm Table of contents}}}

\tableofcontents

\newpage
\fancyhead{}
\fancyfoot{}
\renewcommand{\headrulewidth}{0pt}

\addcontentsline{toc}{chapter}{Acknowledgments}
\addcontentsline{toc}{chapter}{Note added}
\thispagestyle{plain}

{\Large \textit{\textbf{Acknowledgments}}}

\vspace*{0.5cm}
\noindent
{\em 
This dissertation is based on research done at the Department of 
Physics of the {\rm Universit\`a di Roma ``Tor Vergata''}, during the period 
November 1995 - October 1998, under the supervision of Prof. Massimo Bianchi 
and Prof. Giancarlo Rossi. I would like to thank them for introducing me to 
the arguments covered in this thesis, for all what they have taught me and 
for their willingness and constant encouragement. I also wish to thank Prof. 
Michael B. Green for the very enjoyable and fruitful collaboration on 
the subjects reported in the last two chapters. I am grateful to Yassen 
Stanev for many useful and interesting discussions. Finally I thank 
the whole theoretical physics group at the Department of Physics of the 
{\rm Universit\`a di Roma ``Tor Vergata''} for the very stimulating 
environment I have found there.
}

\vspace*{0.5cm}
\begin{center}\rule{3cm}{0.4pt}\end{center}

\vspace*{0.5cm}
\noindent
\textbf{Note added}

\vspace*{0.3cm}
\noindent
This thesis was presented to the ``Universit\`a di Roma {\it Tor Vergata}'' 
in candidacy for the degree of ``Dottore di Ricerca in Fisica'' and 
it was defended in April 1999. Hence more recent developments are not 
discussed here and moreover the references are not updated. However I 
cannot help mentioning at least the very detailed and comprehensive 
review by O. Arhony, S. Gubser, J. Maldacena, H. Ooguri and Y. Oz on 
``Large $N$ Field Theories, String Theory and Gravity'' 
(hep-th/9905111), that is not cited in the bibliography. 
Actually this paper would have made my writing of the last chapter 
much simpler if it came out earlier!

There are some direct developments of this thesis that have been 
carried on during the last few months. In particular the analysis of 
logarithmic singularities in four-point Green functions and their relation 
to anomalous dimensions of un-protected operators in ${\cal N}$=4 SYM has 
been finalised in collaboration with Massimo Bianchi, Giancarlo Rossi and 
Yassen Stanev. This issue is only briefly addressed in the concluding 
remarks. The interested reader is referred to the paper ``On the 
logarithmic behaviour in ${\cal N}$=4 SYM theory'', hep-th/9906188.


\fancyhead{}
\fancyfoot{}
\renewcommand{\headrulewidth}{0.4pt}




\chapter{Rigid supersymmetric theories in four dimensions}
\label{cap1}
\pagenumbering{arabic}
\setcounter{page}{1}
\vspace*{2cm}
\fancyhead[RO,LE]{\thepage}
\fancyhead[RE]{{\footnotesize {\rm Chapter 1.}~~{\it Rigid 4$d$ 
supersymmetry}}} 

\noindent
Supersymmetry was introduced in \cite{golfand} as a generalization of 
Poincar\'e invariance and then implemented in the context of four 
dimensional quantum field theory in \cite{wesszumino}. The 
idea of a symmetry transformation exchanging bosonic and fermionic 
degrees of freedom has proved extremely rich of consequences. Although 
up to now there exists no experimental evidence for supersymmetry, it 
plays a central r\^ole in theoretical physics.

At the present day the interest in the study of supersymmetric 
theories relies on one side on the r\^ole it plays in the 
construction of a more fundamental model beyond the standard model of 
particle physics (SM) and on the other on the possibility of deriving 
results that can hopefully be extended to phenomenologically more relevant 
models.

The standard model describes the strong and electro-weak interactions 
within the framework of a SU(3)$\times$SU(2)$\times$U(1) non-Abelian 
gauge theory; despite the full agreement with experimental data there 
are various motivations for believing that it cannot be a truly 
fundamental theory. A first problem arises in relation with the grand 
unification of particle interactions. The gauge couplings are 
believed to be unified at an energy scale of the order of $10^{16}$ GeV. 
The standard model contains many free parameters (masses and mixing 
angles) that must be ``fine-tuned'' in order to achieve the grand 
unification ({\em hierarchy problem}): the problem follows from the 
existence of ultra-violet quadratic divergences and is solved in 
supersymmetric models in which the divergences, as will be discussed 
later, are milder.

Moreover the standard model does not include gravitational 
interactions: at present the only consistent description of 
gravity at the quantum level is provided by (super)string theories 
and supersymmetry appears to be a fundamental ingredient of such 
theories.

In a more immediate perspective supersymmetric theories can be of help 
in understanding aspects of phenomenologically interesting models. 
For example a description of confinement in asymptotically free 
theories as dual superconductivity has been proposed and various 
realizations of this mechanism have been observed in supersymmetric 
models.

More recently a correspondence between four dimensional quantum 
field theories and supergravity has been suggested; this particular 
subject appears very promising and will be extensively studied in 
chapter 5.

This first chapter contains a review of rigid supersymmetry in 
four dimensions. The material presented is by no means original, the 
aim is to establish the notations and to recall ideas and results, 
mainly about the quantization of supersymmetric theories, that will 
be extensively used in the following.

In the first part of the chapter the supersymmetry 
algebra and its realizations are described. Then the formalism of 
superspace is introduced in somewhat more detail both at the 
classical and at the quantum level. The last section reviews general 
properties of supersymmetric models.

An introduction to supersymmetry is provided by the books by P. West 
\cite{west} and J. Wess and J. Bagger \cite{wessbagger} (whose 
notations will be followed in this thesis) and by the review papers 
\cite{sohnius,lykken}; superspace techniques are treated in more 
detail in \cite{onethousandone}; for phenomenological aspects of 
supersymmetry see \cite{nilles}; an extensive introduction to superstring 
theory is given in the book by M.B. Green, J.H. Schwarz and E. Witten 
\cite{greenschwarzwitten}. 

\fancyhead[LO]{{\footnotesize{\it Supersymmetry algebra}}}
\section{General supersymmetry algebra in four dimensions}
\label{susyalgebra}
\fancyhead[LO]{{\footnotesize 1.1~~{\it Supersymmetry algebra}}}

Quantum field theory models of interest for the description of 
fundamental interactions combine the Poincar\'e invariance with the 
invariance under a global internal symmetry group G. In \cite{colmandu} 
Coleman and Mandula proved, under general assumptions following from 
the axioms of quantum field theory, a ``no-go'' theorem that, states 
that the most general symmetry group for a theory with non trivial 
$S$-matrix is the direct product of the Poincar\'e group and 
an internal group G, which must be of the form of a semisimple Lie 
group times possible U(1) factors.

Historically supersymmetry was discovered as an attempt to evade the 
constraints of the Coleman--Mandula no-go theorem .

The Poincar\'e group P=ISO(3,1) is the semidirect product of the group 
T$_{4}$ of four dimensional translations and the Lorentz group SO(3,1); 
it is generated by $P_{\mu}$, $M_{\mu \nu}$ satisfying:
\begin{eqnarray}
    &&[ P_{\mu},P_{\nu}] = 0  \nonumber \\
    &&[ P_{\lambda}, M_{\mu \nu}] = (\eta_{\lambda \mu} P_{\nu} - 
    \eta_{\lambda \nu} P_{\mu}) 
    \label{poincare} \\
    &&[ M_{\mu \nu}, M_{\rho \sigma}] = - (\eta_{\mu \rho} M_{\nu \sigma} 
    + \eta_{\nu \sigma} M_{\mu \rho} - \eta_{\mu \sigma} M_{\nu \rho} 
    - \eta_{\nu \rho} M_{\mu \sigma}) \, . \nonumber 
\end{eqnarray}
(See appendix \ref{appa} for the conventions and a collection of useful 
formulas).

The generic internal symmetry group is a Lie group defined by the 
commutation relations:
\begin{equation}
	[T^{a},T^{b}] = i f^{ab}{}_{c} T^{c} \,.
\label{internal}
\end{equation}
Internal symmetries are supplemented by the discrete ones C, P and T.

The supersymmetry algebra is a graded extension of (\ref{poincare}): 
the theorem of Coleman and Mandula is bypassed by allowing 
anticommuting as well as commuting generators. In \cite{hagsohlop} 
the most general algebra allowed by this weakening of the hypothesis 
of \cite{colmandu} was constructed: it contains, beyond (\ref{poincare}) 
and (\ref{internal}), anticommuting generators $Q_{\alpha}^{i}$, 
$\overline{Q}_{{\dot \alpha}j}$ that are Weyl spinors transforming in 
the $(\frac{1}{2},0)$ and $(0,\frac{1}{2})$ representations of the 
Lorentz group respectively, so that they do not commute with the 
generators $P_{\mu}$ and $M_{\mu \nu}$. 
Here $\alpha, {\dot \alpha} = 1,2$ are 
spinor indices and $i,j$ ranging from 1 to ${\cal N} \geq 1$ 
label the various of supersymmetries.

The most general symmetry algebra of a supersymmetric quantum field 
theory is therefore given by (\ref{poincare}) and (\ref{internal}) 
supplemented by 
\begin{eqnarray}
	&& \{ Q_{\alpha}^{i},{\overline Q}_{\beta}^{j} \} = 
	2 \sigma^{\mu}_{\alpha {\dot \alpha}}  
	P_{\mu} \delta^{ij} \nonumber \\
	&&[P_{\mu},Q_{\alpha}^{i}] = [P_{\mu}, {\overline Q}_{{\dot 
	\alpha}j}] = 0 \nonumber \\
	&&[Q_{\alpha}^{i}, M_{\mu \nu}] = (\sigma_{\mu \nu})_{\alpha}{}^{\beta} 
	Q_{\beta}^{i} \nonumber 
\end{eqnarray}
\begin{eqnarray}
	&&[{\overline Q}_{{\dot \alpha} j}, M_{\mu \nu}] = 
	({\overline \sigma}_{\mu \nu})^{\dot \alpha}{}_{\dot \beta} 
	{\overline Q}^{\dot \beta}_{j} \nonumber  \\
	&&\{ Q_{\alpha}^{i}, Q_{\beta}^{j} \} = 
	\varepsilon_{\alpha \beta} Z^{ij} \nonumber \\
	&&\{  {\overline Q}_{{\dot \alpha}i},{\overline Q}_{{\dot \beta}j} \} = 
	\varepsilon_{{\dot \alpha} {\dot \beta}} Z^{\dagger}_{ij} 
	\label{susyalg} \\
	&&[Q_{\alpha}^{i}, T^{a}]  = B^{ai}{}_{j} Q_{\alpha}^{j}  \nonumber  \\
	&&[T^{a},{\overline Q}_{{\dot \alpha}i}] = B^{\dagger a}_{j}{}^{i} 
	{\overline Q}_{{\dot \alpha}i} \nonumber \\
	&&[Z^{ij}, X] = [Z^{\dagger}_{lk}, X] = 0 \, ,\nonumber
\end{eqnarray}
where $B^{\dagger a}_{i}{}^{j} = B^{a j}{}_{i}$. In (\ref{susyalg}) $X$ 
denotes an arbitrary generator in the algebra so that $Z^{ij}=-Z^{ji}$ 
(central charges) generate the center of the algebra~\footnote{In principle 
the antisymmetry of $Z^{ij}$ is expected to imply that no central 
extension can exist in non extended (${\cal N}$=1) theories. It 
can be proved \cite{shifmanasi} that a central extension 
can actually be present in ${\cal N}$=1 supersymmetric Yang--Mills 
theory. See \cite{ferpor} for a recent discussion in the context of 
the AdS/CFT correspondence.}. 
Central charges play a major r\^ole in the quantization of 
${\cal N}$-extended supersymmetric theories. They generate an Abelian 
subalgebra so that 
\begin{displaymath}
	Z^{ij} = a_{b}^{ij} T^{b} \, , 
\end{displaymath}
with $B_{a}^{i}{}_{j} a^{b \, jk} = - a^{b \, ij}B_{a \, j}{}^{k}$.

The supersymmetry algebra (\ref{susyalg}) considered as a graded Lie 
algebra possesses a group of automorphisms which is U(${\cal N}$) in 
the general case . The U(${\cal N}$) group of automorphisms acts on 
the charges $Q_{\alpha}^{i}$ and ${\overline Q}_{{\dot \alpha}j}$ as
\begin{displaymath}
	Q_{\alpha}^{i} \longrightarrow U^{i}{}_{k}Q_{\alpha}^{k} \, 
	\hspace{1cm} {\overline Q}_{{\dot \alpha}j} \longrightarrow 
	{\overline Q}_{{\dot \alpha}k} U^{\dagger k}_{j} \, .
\end{displaymath}
In the particular case of ${\cal N}$=1 supersymmetry the U(1) group of 
automorphisms is generated by $R$ such that 
\begin{displaymath}
	[Q_{\alpha},R]=Q_{\alpha} \, , \hspace{1cm} [{\overline Q}_{\dot 
	\alpha},R]=-{\overline Q}_{\dot \alpha} \, .  
\end{displaymath}

The supersymmetry charges $Q_{\alpha}^{i}$ and ${\overline Q}_{{\dot 
\alpha}j}$ transform in spin $\frac{1}{2}$ representations of the 
Lorentz group; this implies that acting with $Q_{\alpha}^{i}$ or 
${\overline Q}_{{\dot \alpha}j}$ on a state of spin $j$ produces a 
state of spin $j\pm \frac{1}{2}$: supersymmetry generators exchange 
bosonic and fermionic states. Reversing the line of reasoning and 
starting with a symmetry transforming the fields among themselves in 
such a way as to mix bosons and fermions, one concludes that the different 
physical dimension of bosonic and fermionic fields implies that the 
transformation must involve derivatives, {\em i.e.} space-time 
translations. This gives an intuitive explanation of how the 
requirement of a non trivial mixing of internal and space-time 
symmetries naturally leads to supersymmetry.

An immediate consequence of the supersymmetry algebra is that in 
supersymmetric theories the energy is positive definite and vanishes 
only on supersymmetric (ground) states. The first equation in 
(\ref{susyalg}) implies 
\begin{displaymath}
	0 \leq \sum_{i} \left( \{ Q_{1}^{i},(Q_{1 i})^{\dagger} \} + 
	\{ Q_{2}^{i},(Q_{2 i})^{\dagger} \} \right) = -4 {\cal N} P_{0} =
	4{\cal N} H \, ,
\end{displaymath}
so that 
\begin{displaymath}
	\langle \psi | H | \psi \rangle \geq 0
\end{displaymath}
for every state $|\psi\rangle$ and $H|\Psi\rangle = 0$ if and only if 
$Q|\psi\rangle =0$, {\em i.e.} if $|\psi\rangle$ is supersymmetric.

\fancyhead[LO]{{\footnotesize{\it Representations of supersymmetry}}}
\section{Representations of the supersymmetry algebra}
\label{susyreps}
\fancyhead[LO]{{\footnotesize 1.2~~{\it Representations of supersymmetry}}}

The irreducible representations of the general supersymmetry algebra 
can be constructed starting from the (anti)commutation relations of 
the preceding section and the definition of the Casimir operators. 
For the Poincar\'e group the quadratic Casimir operators are 
$P^{2}=P_{\mu}P^{\mu}$ and $W^{2}=W_{\mu}W^{\mu}$, where $W_{\mu}$ is the 
Pauli--Lubanski vector
\begin{displaymath}
	W_{\mu} = \frac{1}{2} \varepsilon_{\mu \nu \rho \sigma} P^{\nu} 
	M^{\rho \sigma} \, .
\end{displaymath}
Here $W^{2}$ is not a Casimir anymore, because $M_{\mu \nu}$ does not 
commute with the supersymmetry generators $Q_{\alpha}^{i}$ and 
${\overline Q}_{{\dot \alpha}j}$, and is substituted by $C^{2}$ 
defined (for the case ${\cal N}$=1) by
\begin{eqnarray*}
	C^{2} & = & C_{\mu \nu} C^{\mu \nu}  \\
	C_{\mu \nu}  & = & B_{\mu} P_{\nu} - B_{\nu} P_{\mu}  \\
	B_{\mu} & = & W_{\mu} - \frac{1}{4} {\overline Q}_{\dot \alpha} 
	{\overline \sigma}_{\mu}^{{\dot \alpha}\alpha} Q_{\alpha} \, .
\end{eqnarray*}
The irreducible representations can be constructed by Wigner's 
technique of induced representations. 

A general result following from the supersymmetry algebra is that 
every irreducible representation contains an equal number of bosonic 
and fermionic states. Defining the fermion number operator 
$(-)^{N_{f}}$ such that 
\begin{displaymath}
  (-)^{N_{f}} | B \rangle = + | B \rangle \, , \hspace{1cm} 
  (-)^{N_{f}} | F \rangle = - | F \rangle \, ,
\end{displaymath}
where $| B \rangle$ denotes a bosonic state and $| F \rangle$ a 
fermionic one, it follows that
\begin{displaymath}
	(-)^{N_{f}} Q_{\alpha} = - Q_{\alpha} (-)^{N_{f}} \, , \hspace{1cm}
	(-)^{N_{f}} {\overline Q}_{\dot \alpha} = 
	- {\overline Q}_{\dot \alpha} (-)^{N_{f}} \, .
\end{displaymath}
These relations imply
\begin{displaymath}
	0 = \tr \left[ (-)^{N_{f}} \{Q_{\alpha},{\overline Q}_{\dot \alpha} \}
	\right] = 2 \sigma^{\mu}_{\alpha {\dot \alpha}} \delta^{ij} P_{\mu}
	\tr (-)^{N_{f}} \, ,
\end{displaymath}
so that in conclusion for non vanishing $P_{\mu}$ 
\begin{displaymath}
	\tr (-)^{N_{f}} = 0
\end{displaymath}
from which the statement follows.

\vspace{0.7cm}
{\sl Massive representations without central charges.} 
\vspace{0.3cm}

\noindent
For particles of mass $M$ in the rest frame $P_{\mu}=(-M,0,0,0)$ the 
supersymmetry algebra reduces to 
\begin{eqnarray}
	&&\{Q_{\alpha}^{i},{\overline Q}_{{\dot \beta}j} \} = 2M 
	\delta_{\alpha {\dot \alpha}} \delta^{i}_{j} \nonumber  \\
	 &&\{Q_{\alpha},Q_{\beta}\} = \{ {\overline Q}_{{\dot \alpha}i}
	 {\overline Q}_{{\dot \beta}j} \} = 0 \, ,
\end{eqnarray}
and creation and annihilation operators can be defined as follows
\begin{eqnarray}
	a_{\alpha}^{i} & = & \frac{1}{\sqrt{2M}} Q_{\alpha}^{i} 
	\nonumber \\
	\left( a_{\alpha}^{i} \right)^{\dagger} & = & \frac{1}{\sqrt{2M}}
	{\overline Q}_{{\dot \alpha}i} \, .
	\label{masscreation}
\end{eqnarray}
$a_{\alpha}^{i}$ and $\left( a_{\alpha}^{i} \right)^{\dagger}$ 
satisfy the algebra of fermionic creation and annihilation operators. 
In fact by direct calculation one finds
\begin{eqnarray}
	\{ a_{\alpha}^{i}, \left( a_{\beta}^{j} \right)^{\dagger} \} 
	& = & \delta_{\alpha}^{\beta} \delta^{i}_{j} 
	\nonumber  \\
	\{a_{\alpha}^{i}, a_{\beta}^{j} \} & = & 
	\{ \left( a_{\alpha}^{i} \right)^{\dagger}, 
	\left( a_{\beta}^{j} \right)^{\dagger} \} = 0 \, .
	\label{cliffmass}
\end{eqnarray}
Given a Clifford vacuum $|\Omega \rangle$ defined by 
\begin{displaymath}
	a_{\alpha}^{i} |\Omega \rangle = 0 \hspace{1.5cm} \forall \, i,
	\alpha \, ,
\end{displaymath}
with $P^{2} |\Omega \rangle = - M^{2} |\Omega \rangle$, the states 
are constructed by acting with the operators $\left( a_{\alpha}^{i} 
\right)^{\dagger}$ on $|\Omega \rangle$:
\begin{equation}
	|\Omega^{(n)\, \alpha_{1} \ldots \alpha_{n}}_{i_{1} \ldots i_{n}} 
	\rangle = \frac{1}{\sqrt{n!}} \left( a_{\alpha_{n}}^{i_{n}} 
    \right)^{\dagger} \ldots \left( a_{\alpha_{1}}^{i_{1}} 
    \right)^{\dagger} |\Omega \rangle \, .
	\label{massivestate}
\end{equation}
Antisymmetry under the exchange of pairs $(\alpha_{k},i_{k})$, 
$(\alpha_{l},i_{l})$ implies $n \leq 2{\cal N}$. For each $n$ the number 
of states (degeneracy) is 
$\left( \begin{array}{c} 2{\cal N} \\ n \end{array} \right)$, so that the 
total number of states (\ie the dimension of the representation) is 
\begin{displaymath}
	d = \sum_{n=0}^{2{\cal N}} \left( \begin{array}{c} 2{\cal N} \\ 
	n \end{array} \right) = 2^{2{\cal N}} \, .
\end{displaymath}
The representation contains $2^{2{\cal N}-1}$ bosonic and 
$2^{2{\cal N}-1}$ fermionic states. 
The highest spin state is obtained symmetrizing the 
maximum number of spinor indices, namely ${\cal N}$, leading to the 
value $\frac{1}{2} {\cal N}$ for the highest spin in the multiplet.

\vspace{0.7cm}
{\sl Massless representations without central charges.} 
\vspace{0.3cm}

\noindent
In the ``light-like'' reference frame $P_{\mu}=(-E,0,0,E)$, with 
$P^{2}=0$, the supersymmetry algebra reads 
\begin{eqnarray*}
	\{ Q_{\alpha}^{i}, {\overline Q}_{{\dot \beta} j} \} & = & 2\left( 
	\begin{array}{cc} 2E & 0 \\ 0 & 0 \end{array} \right) 
	\delta^{i}_{j}  \\
	\{Q_{\alpha}^{i} , Q_{\beta }^{j} \} & = & \{ {\overline Q}_{{\dot 
	\alpha}i} , {\overline Q}_{{\dot \beta}j} \} = 0 \, .
\end{eqnarray*}
The natural definition of creation and annihilation operators is then
\begin{eqnarray}
	a^{i} & = & \frac{1}{2\sqrt{E}} Q_{1}^{i} \nonumber \\
	a_{i}^{\dagger} & = & (a^{i})^{\dagger} = \frac{1}{2\sqrt{E}} 
	{\overline Q}_{\dot 1}^{i} \, .
	\label{masslesscreation}
\end{eqnarray}
They generate the Clifford algebra
\begin{eqnarray}
	\{a^{i} , a_{j}^{\dagger} \} & = & \delta^{i}_{j} \nonumber \\
	\{a^{i},a^{j} \} & = & \{a_{i}^{\dagger}, a_{j}^{\dagger} \} = 0
	\label{cliffmassless}
\end{eqnarray}
Notice that in this case only half of the generators $Q_{\alpha}^{i}$ 
can be used to construct the fermionic oscillators because $Q_{2}^{i}$ 
and ${\overline Q}_{{\dot 2}j}$ anticommute. Starting with a state of 
lowest helicity $\lambda$, $| \Omega_{\lambda}\rangle$, playing the 
r\^ole of a Clifford vacuum, defined by 
\begin{displaymath}
  a^{i} |\Omega_{\lambda} \rangle = 0
\end{displaymath}
the states which constitute the multiplet are constructed as 
\begin{equation}
	|\Omega^{(n)}_{\lambda+\frac{n}{2};i_{1},\ldots,i_{n}} \rangle =
	\frac{1}{\sqrt{n!}} a_{i_{n}}^{\dagger}\ldots a_{i_{1}}^{\dagger} 
	|\Omega_{\lambda} \rangle \, .
	\label{masslesstate}
\end{equation}
The state $|\Omega^{(n)}_{\lambda+\frac{n}{2};i_{1},\ldots,i_{n}} 
\rangle$ has helicity $\lambda+\frac{n}{2}$ and because of 
antisymmetry in the indices $i_{1},\ldots,i_{n}$ is 
$\left( \begin{array}{c} {\cal N} \\ n \end{array} \right)$ times 
degenerate. The highest helicity in the multiplet is ${\overline 
\lambda} = \lambda +\frac{{\cal N}}{2} $. The dimension of the 
representation given by the total number of states is 
\begin{displaymath}
d = \sum_{i=0}^{{\cal N}} \left( \begin{array}{c} {\cal N} \\ 
n \end{array} \right) = 2^{{\cal N}} \, .
\end{displaymath}
CPT invariance implies in general a doubling of the number of states 
because it reverses the helicity. So for example the ${\cal N}$=3 
multiplet in four dimensions coincides with the ${\cal N}$=4 
multiplet if CPT invariance is required. Multiplets of particular 
relevance such as ${\cal N}$=2 with $\lambda=-\frac{1}{2}$, ${\cal N}$=4 
with $\lambda=-1$ and ${\cal N}$=8 with $\lambda=-2$ are 
automatically CPT invariant.

\vspace{0.7cm}
{\sl Supersymmetry representations in the presence of central charges.} 
\vspace{0.3cm}

\noindent
For $P^{2}=-M^{2}$ in the rest frame $P_{\mu}=(-M,0,0,0)$ the 
supersymmetry algebra with non vanishing central charges is given by
\begin{eqnarray}
	\{Q_{\alpha}^{i},{\overline Q}_{{\dot \alpha}j} \} & = & 
	2M \delta^{i}_{j} \delta_{\alpha{\dot \alpha}} \nonumber  \\
	\{ Q_{\alpha}^{i},Q_{\beta}^{j}\} & = & \varepsilon_{\alpha \beta} 
	Z^{ij}
	\label{centralsusy}  \\
	\{ {\overline Q}_{{\dot \alpha}i}, {\overline Q}_{{\dot \beta}j} 
	\} & = & -\varepsilon_{{\dot \alpha}{\dot \beta}} 
	Z_{ij}^{\dagger} \, , \nonumber
\end{eqnarray}
with $Z^{ij}=-Z^{ji}$, $Z_{ij}=-Z^{ij}$. Since $Z^{ij}$ commute with 
all the generators, by invoking Schuur's lemma there exists a basis in 
which they are diagonal with eigenvalues $Z^{ij}$. The antisymmetric 
${\cal N}\times {\cal N}$ matrix with elements $Z^{ij}$ can be put in 
the following form 
\begin{displaymath}
	Z^{ij} = U^{i}{}_{k}\tilde{Z}^{kl}(U^{T})_{l}{}^{j} \, ,
\end{displaymath}
with $U$ is a unitary matrix and 
\begin{eqnarray*}
	\tilde{Z} & = & \varepsilon \otimes D \hspace{3.6cm} ({\rm 
	for}~~{\cal N}~~{\rm even})  \\
	\tilde{Z} & = & \left( 
	   \begin{array}{cc} \varepsilon \otimes D & 0 \\
	                     0 & 0 
	   \end{array}                  
	\right) \hspace{2cm}({\rm for}~~{\cal N}~~{\rm odd}) \, ,
\end{eqnarray*}
where $\varepsilon = \left( \begin{array}{cc} 0 & 1 \\ -1 & 0 
\end{array} \right)$ and $D$ is a diagonal $\frac{{\cal N}}{2} \times 
\frac{{\cal N}}{2}$ matrix with eigenvalues $Z_{R}$, such that 
$Z_{R}^{*} = Z_{R}$, $Z_{R}>0$. 
Considering the case with ${\cal N}$ even the algebra can be rewritten 
in a more suitable form by decomposing the indices $i$ and $j$ as 
\begin{displaymath}
	i=(a,R) \, , \hspace{1.5cm} j=(b,S) \, ,
\end{displaymath}
with $a,b=1,2$ and $R,S=1,\ldots,\frac{{\cal N}}{2}$. Now the algebra 
can be put in the form
\begin{eqnarray}
	\{ Q_{\alpha}^{aR}, {\overline Q}_{{\dot \beta}}^{bS} \} & = & 
	2M\delta_{\alpha{\dot \alpha}} \delta^{ab}\delta^{RS} \nonumber \\
	\{Q_{\alpha}^{aR},Q_{\beta}^{bS} \} & = & \varepsilon_{\alpha \beta} 
	\varepsilon^{ab} \delta^{RS} Z_{S}
	\label{susycentral}  \\
	\{ {\overline Q}_{\dot \alpha}^{aR},{\overline Q}_{\dot \beta}^{bS} 
	\} & = & \varepsilon_{{\dot \alpha}{\dot \beta}}\varepsilon^{ab} 
	\delta^{RS} Z_{S} \, , \nonumber
\end{eqnarray}
so that creation and annihilation operators can be defined in the 
following way:
\begin{eqnarray*}
	a_{\alpha}^{R} & = & \frac{1}{\sqrt{2}} \left[ 
	Q_{\alpha}^{1R}+\varepsilon_{\alpha 
	\beta}(Q_{\beta}^{2R})^{\dagger} \right] \, , \hspace{1.5cm} 
	(a_{\alpha}^{R})^{\dagger} = a^{\dagger}_{\alpha}{}^{R}  \\
	b_{\alpha}^{R} & = & \frac{1}{\sqrt{2}} \left[ 
	Q_{\alpha}^{1R}-\varepsilon_{\alpha 
	\beta}(Q_{\beta}^{2R})^{\dagger} \right] \, , \hspace{1.5cm} 
	(b_{\alpha}^{R})^{\dagger} = b^{\dagger}_{\alpha}{}^{R}  
\end{eqnarray*}
These operators satisfy the algebra of fermionic creation and 
annihilation operators
\begin{eqnarray}
	\{a_{\alpha}^{R},a_{\beta}^{S} \} & = & \{b_{\alpha}^{R},
	b_{\beta}^{S} \} = \{ a_{\alpha}^{R},b_{\beta}^{S} \} = 0
	\nonumber  \\
	\{a_{\alpha}^{R},(a_{\beta}^{S})^{\dagger} \} & = & 
	\delta_{\alpha\beta}\delta^{RS} (2M+Z_{S})
	\label{centralcliff}  \\
	\{b_{\alpha}^{R},(b_{\beta}^{S})^{\dagger} \} & = & 
	\delta_{\alpha\beta}\delta^{RS} (2M-Z_{S}) \, .
	\nonumber
\end{eqnarray}
Since $\{ A,A^{\dagger} \}$ is a positive definite operator $\forall 
A$, it immediately follows that $Z_{R}\leq 2M$. For $Z_{R}<2M$ $\forall R$ 
the structure of the multiplet is the same as in the case of no 
central charges. For $Z_{R} = 2M$ with $R=1,\ldots,{\overline R}$ a 
subset of operators $b$ cancels out and the algebra reduces to a 
$2({\cal N}-{\overline R})$ dimensional Clifford algebra giving rise to 
so called {\em short multiplets}. States belonging to short multiplets, 
also referred to as {\em Bogomol'nyi Prasad Sommerfield (BPS) saturated 
states}, play a crucial r\^ole in the non perturbative dynamics of 
supersymmetric gauge theories as will be discussed later.

As an example of the previous construction consider the simplest case, 
namely ${\cal N}$=1 massive multiplet with a spin zero Clifford 
vacuum $|\Omega \rangle$: the states are 
\begin{eqnarray*}
	&|\Omega \rangle  & ({\rm spin}~~0) \\
	& (a_{\alpha})^{\dagger} |\Omega \rangle 
	&({\rm spin}~~\frac{1}{2}) \\
	& \frac{1}{\sqrt{2}} (a_{\alpha})^{\dagger} (a_{\beta})^{\dagger}
	|\Omega \rangle = - \frac{1}{2 \sqrt{2}}\varepsilon_{\alpha \beta}
	(a^{\gamma})^{\dagger} (a_{\gamma})^{\dagger} |\Omega \rangle 
	\hspace{1cm} & ({\rm spin}~~0)
\end{eqnarray*}

In the following the discussion will focus mainly on the ${\cal N}$=4 
massless case. For a rigid supersymmetric theory with fields of spin 
not larger than one the multiplet contains the following states with 
relative degeneracies
\vspace{0.5cm}

\begin{center}
\begin{tabular}{|l|c|c|c|c|c|}
	\hline
	~~{\rm helicity}~~\rule[-7pt]{0pt}{21pt} & $~~-1~~~$ & 
	$~~-\frac{1}{2} ~~~$ & $~~~0~~~$ & $~~~\frac{1}{2}~~~$ & $~~~1~~~$  \\
	\hline
	~~{\rm degeneracy}~~\rule[-7pt]{0pt}{21pt} & $~~~1~~~$ & $~~~4~~~$ & 
	$~~~6~~~$ & $~~~4~~~$ & $~~~1~~~$  \\
	\hline
\end{tabular}
\end{center}
\vspace{0.5cm}
Given the irreducible representations of the supersymmetry algebra 
the corresponding realizations in terms of fields can be immediately 
deduced. Although this was the historical pattern in the construction of 
supersymmetric quantum field theory models, it appears more natural 
to derive the structure of supermultiplets using the superfield formalism, 
so this will be the approach followed here. 

Notice that Wigner's method here employed gives representations of 
``on-shell'' states. It is always possible to construct sets of fields 
realizing such irreducible representations, but on the contrary it is 
usually not possible, for ${\cal N} > 1$, to achieve the 
closure of the supersymmetry algebra on a set of ``off-shell'' 
fields, {\em i.e.} without use of the equations of motion.

The irreducible representations of supersymmetry have automatically a 
well defined transformation under the group of automorphisms of the 
algebra (\ref{susyalg}). For massless representations the largest 
subgroup of the automorphisms group that respects 
helicity is the whole U(${\cal N}$), whereas for massive states it 
depends on the presence and number of central charges. In the absence 
of central charges, or if none of them saturates the BPS bound, the spin 
preserving automorphisms group is USp($2{\cal N}$). It reduces to 
USp(${\cal N}$) for ${\cal N}$ even and to USp(${\cal N}+1$) for 
${\cal N}$ odd if one central charge satisfies      the condition $Z=2M$.

\fancyhead[LO]{{\footnotesize{\it Superspace}}}
\section{${\cal N}$=1 superspace}
\label{superspace1}
\fancyhead[LO]{{\footnotesize 1.3~~{\it Superspace}}}

Ordinary quantum field theories are constructed in terms of field 
operators that are functions (actually distributions) defined on 
Minkowskian space-time, parametrized by the coordinates $x_{\mu}$. 
The action is required to be invariant under Poincar\'e 
space-time symmetry as well as under transformations of an internal 
group G. The transformation of a generic field $f(x)$ is of the form 
\begin{equation}
	R(g) f(x) = f(\tau(g) x) \, ,
	\label{fieldtrans}
\end{equation}
with $R(g)$ and $\tau(g)$ suitable representations of the relevant 
symmetry group.

In supersymmetric models the space-time symmetry is extended by the 
introduction of the transformations generated by anticommuting charges 
$Q_{\alpha}^{i}$, ${\overline Q}_{{\dot \alpha}j}$. The natural way 
to implement the extended symmetry algebra is to construct the theory 
in terms of generalized fields, called {\em superfields} 
\cite{superspace1,superspace2}, depending 
on the coordinates $x_{\mu}$ as well as on ``fermionic'' coordinates 
$\theta_{\alpha}$ and ${\overline \theta}_{\dot \alpha}$
\begin{equation}
	F = F(x_{\mu}, \theta_{\alpha}, {\overline \theta}_{\dot \alpha})
	\, .
	\label{eq:superfield}
\end{equation}
Starting from the supersymmetry algebra a generic element of the 
corresponding group of transformations can be written in the form
\begin{displaymath}
	g(x,\theta,{\overline \theta}) = 
	e^{i(x^{\mu}P_{\mu}+\theta^{\alpha}Q_{\alpha}+{\overline 
	\theta}_{\dot \alpha}{\overline Q}^{\dot \alpha})} \, .
\end{displaymath}
The action on superfields is then defined through a differential 
representation of the operators $P_{\mu}$, $Q_{\alpha}$ and 
${\overline Q}^{\dot \alpha}$.

This construction can be carried out in a more general fashion, 
suitable for generalization to the case of ${\cal N}$-extended 
supersymmetry, defining the superspace parametrized by 
\begin{displaymath}
	z_{m} = (x_{\mu}, \theta_{\alpha}, {\overline \theta}_{\dot \alpha})
\end{displaymath}
as a coset space. 
${\cal N}$=1 superspace under consideration can be obtained as the 
coset SP/L, where SP is the super Poincar\'e group generated by 
(\ref{susyalg}) and L is the Lorentz group  SO(3,1). 

A coset group K=G/H can be parametrized by coordinates $\xi_{m}$, 
$m=1,\ldots,({\rm dim~G}-{\rm dim~H})$. By writing an element of G 
through  the exponential map 
\begin{displaymath}
	g = e^{i \xi^{m}L_{m}}e^{i \zeta^{k}H_{k}} \, ,
\end{displaymath}
where the generators of G split into the generators of H ($H_{k}$) 
and the remaining $L_{n}$, elements of the coset are obtained by 
taking $\zeta^{k}=0$.

For the present case K=SP/L this implies that one should write 
\begin{equation}
	g = e^{i(a^{\mu}P_{\mu}+\eta^{\alpha}Q_{\alpha}+{\overline 
    \eta}_{\dot \alpha}{\overline Q}^{\dot \alpha})} 
	e^{\frac{1}{2} w^{\mu \nu}M_{\mu \nu}}\, 
	\label{eq:cosetel}
\end{equation}
so that $w^{\mu \nu}=0$ gives the elements of K  
\begin{displaymath}
	k = e^{i(x^{\mu}P_{\mu}+\theta^{\alpha}Q_{\alpha}+{\overline 
    \theta}_{\dot \alpha}{\overline Q}^{\dot \alpha})} = 
	e^{z^{m}K_{m}} 
\end{displaymath}
which are parametrized by $z_{m} = (x_{\mu}, \theta_{\alpha}, 
{\overline \theta}_{\dot \alpha})$; as a result ${\cal N}$=1 
superspace is an eight dimensional manifold. Under the action of a 
group element $g_{_{0}} = (a_{\mu},\eta_{\alpha},
{\overline \eta}_{\dot \alpha})$
\begin{displaymath}
	g_{_{0}} \circ e^{z^{m}K_{m}} = 
	e^{{z^{\prime}}^{m}K_{m}}
	e^{\frac{1}{2} {w^{\prime}}^{\mu \nu} M_{\mu \nu}} 
	\equiv g^{\prime} \, .
\end{displaymath}
An explicit calculation using Becker--Hausdorff's formula and 
anticommutation of $Q$ and $P$ gives the following transformations 
of the coordinates $z_{m}$
\begin{eqnarray}
	{x^{\prime}}^{\mu}  & = & x^{\mu}+a^{\mu}+i\eta^{\alpha} 
	\sigma^{\mu}_{\alpha{\dot \alpha}}{\overline \theta}^{\dot \alpha}
	-i\theta^{\alpha}\sigma^{\mu}_{\alpha{\dot \alpha}}{\overline 
	\eta}^{\dot \alpha}+w^{\mu \nu}x_{\nu} \nonumber  \\
	{\theta^{\prime}}^{\alpha} & = & 
	\theta^{\alpha}+\eta^{\alpha}+\frac{1}{4}w_{\mu \nu}
	\sigma^{\mu \nu \, \alpha}{}_{\beta} \theta_{\beta}
	\label{scoordtrans}  \\
	{{\overline \theta}^{\prime}}^{\dot \alpha} & = & {\overline 
	\theta}^{\dot \alpha}+{\overline \eta}^{\dot \alpha}-\frac{1}{4}
	{\overline \theta}^{\dot \beta}{\overline \sigma}^{\mu \nu}_{\dot 
	\beta}{}^{\dot \alpha}w_{\mu \nu} \nonumber \, .
\end{eqnarray}
Now comparison of $(g_{1}\circ g_{2}) \circ e^{z^{m}K_{m}}~$ 
with $~g_{1}\circ (g_{2}\circ e^{z^{m}K_{m}})~$ 
allows one to prove that the generators $Q_{\alpha}$ and ${\overline 
Q}_{\dot \alpha}$ satisfy the supersymmetry algebra (\ref{susyalg}) 
for ${\cal N}$=1.

As already mentioned a superfield is a function of $z_{m}$
\begin{displaymath}
	F(z)=F(x,\theta,{\overline \theta}) \, .
\end{displaymath}
The transformation under the group action is the generalization of 
(\ref{fieldtrans}) that implements $z^{\prime}=\tau(g)z$ with 
$z^{\prime}$ given by (\ref{scoordtrans}). For infinitesimal 
transformations 
\begin{displaymath}
	\delta_{g}F=F(z+\delta z)-F(z)=\delta g_{m}X^{m}F(z) \, ,
\end{displaymath}
which is achieved realizing the super Poincar\'e generators through 
the differential operators  
$X_{m}=(\ell_{m},\ell_{\alpha},{\overline \ell}_{\dot \alpha},\ell_{\mu 
\nu})$
\begin{eqnarray}
	\ell_{\mu} & = & \partial_{\mu} \nonumber  \\
	\ell_{\alpha} & = & \frac{\partial}{\partial \theta^{\alpha}}
	+i\sigma^{\mu}_{\alpha{\dot \alpha}}{\overline 
	\theta}^{\dot \alpha} \partial_{\mu} \nonumber  \\
	{\overline \ell}_{\dot \alpha} & = & 
	\frac{\partial}{\partial{\overline \theta}^{\dot \alpha}} 
	+i\theta^{\alpha}\sigma^{\mu}_{\alpha{\dot \alpha}} \partial_{\mu}
	\label{diffgener}  \\
	\ell_{\mu \nu} & = & 
	-(x_{\mu}\partial_{\nu}-x_{\nu}\partial_{\mu})-\frac{1}{2}
	\theta^{\beta}\sigma^{\mu 
	\nu}_{\beta}{}^{\alpha}\frac{\partial}{\partial \theta^{\alpha}} 
	+\frac{1}{2} {\overline \theta}^{\dot \beta}{\overline \sigma}^{\mu 
	\nu}_{\dot \beta}{}^{\dot \alpha}\frac{\partial}{\partial 
	{\overline \theta}^{\dot \alpha}} \nonumber \: .
\end{eqnarray}
A calculation shows that they satisfy the super Poincar\'e 
algebra
\begin{eqnarray}
	& \{ \ell_{\alpha},\ell_{\beta} \} = 0 \, ,\hspace{1cm}  
	& \{ \ell_{\alpha} ,{\overline \ell}_{\dot \alpha} \} = 
	2i \sigma^{\mu}_{\alpha{\dot \alpha}}\ell_{\mu} \, , 
	\nonumber  \\
	 &  & 
	\label{killalg}  \\
	& [\ell_{\mu},\ell_{\alpha}] = 0  \, , 
	\hspace{1cm} & [\ell_{\alpha},\ell_{\mu \nu}] = 
	\frac{1}{2} \sigma_{\mu \nu \alpha}{}^{\beta}\ell_{\beta} \, .
	\nonumber
\end{eqnarray}
Superfields defined in this way form linear representations of 
supersymmetry. Since supersymmetry acts on superfields through linear 
differential operators it follows that the product of superfields is a 
new superfield.

The objects considered up to now are scalar 
superfields~\footnote{From now on scalar superfields will be referred 
to simply as superfields.}; more general 
superfields transforming in a non trivial way under Lorentz 
transformations can be considered as well. In general a superfield 
carrying a set of Lorentz indices encoded in a single label $p$ 
transforms as 
\begin{displaymath}
	R(g) f_{p}(z) = D_{p}{}^{q} \left( e^{-\frac{1}{2}w^{\mu \nu}
	M_{\mu \nu}} \right) F_{q}(\tau(g)z) \, ,
\end{displaymath}
where $D_{p}^{q}$ is a suitable representation of the Lorentz group.

Physical fields directly related to the states considered in section 
\ref{susyreps} can be obtained from the superfields by expanding in 
the $\theta$ and ${\overline \theta}$ variables. Since $\theta_{\alpha}$  
and ${\overline \theta}_{\dot \alpha}$ are Weyl spinors the  expansion 
reduces to a polynomial. For a generic superfield
\begin{eqnarray}
	F(x,\theta,{\overline \theta}) &=& 
	f(x)+\theta^{\alpha}\chi_{\alpha}(x)+{\overline \theta}_{\dot \alpha}
	{\overline \psi}^{\dot \alpha}(x)+\theta^{\alpha}\theta_{\alpha} 
	g(x) +{\overline \theta}_{\dot \alpha} {\overline \theta}^{\dot \alpha} 
	h(x)+ \nonumber \\
	&& \hspace{-2.7cm}+\theta^{\alpha}\sigma^{\mu}_{\alpha {\dot \alpha}} 
	{\overline \theta}^{\dot \alpha}r_{\mu}(x) 
	+\theta^{\alpha}\theta_{\alpha}{\overline \theta}_{\dot \alpha}
	{\overline \lambda}^{\dot \alpha}(x)+{\overline \theta}_{\dot \alpha}
	{\overline \theta}^{\dot \alpha}\theta^{\alpha}\xi_{\alpha}(x) +
	\theta^{\alpha}\theta_{\alpha}{\overline \theta}_{\dot \alpha}
	{\overline \theta}^{\dot \alpha}s(x) \, .
	\label{eq:sfieldexp}
\end{eqnarray}
The fields $f(x),~\chi(x),~{\overline 
\psi}(x),~g(x),~h(x),~r_{\mu}(x),~{\overline \lambda}(x),~\xi(x)~{\rm 
and}~s(x)$ constitute a supermultiplet and their transformation law is 
derived from the transformation of the superfield by defining 
\begin{eqnarray}
	\delta_{\eta} F(x,\theta,{\overline \theta}) = && \hspace{-0.6cm}
	\delta_{\eta}f(x)+\theta^{\alpha}\delta_{\eta}\chi_{\alpha}(x)+
	{\overline \theta}_{\dot \alpha}
	\delta_{\eta}{\overline \psi}^{\dot \alpha}(x)+
	\theta^{\alpha}\theta_{\alpha} \delta_{\eta}g(x) +
	{\overline \theta}_{\dot \alpha} {\overline \theta}^{\dot \alpha} 
	\delta_{\eta}h(x)+ \nonumber \\ 
	&& \hspace{-3.2cm}+\theta^{\alpha}\sigma^{\mu}_{\alpha {\dot \alpha}} 
	{\overline \theta}^{\dot \alpha}\delta_{\eta}r^{\mu}(x) + 
	\theta^{\alpha}\theta_{\alpha}{\overline \theta}_{\dot \alpha}
	\delta_{\eta}{\overline \lambda}^{\dot \alpha}(x)+
	{\overline \theta}_{\dot \alpha}{\overline \theta}^{\dot \alpha}
	\theta^{\alpha}\delta_{\eta}\xi_{\alpha}(x) +
	\theta^{\alpha}\theta_{\alpha}{\overline \theta}_{\dot \alpha}
	{\overline \theta}^{\dot \alpha}\delta_{\eta}s(x)
	\label{transfcomp}
\end{eqnarray}
and matching terms with the same powers of $\theta$ and ${\overline 
\theta}$.

Linear representations of supersymmetry constructed in this way are in 
general reducible; the irreducible representations studied in section 
\ref{susyreps} are in correspondence with constrained superfields. 
All the representations of rigid ${\cal N}$=1 supersymmetry in four 
dimensions are realized in terms of two basic superfields, namely {\em 
chiral superfields} and {\em vector superfields}. 

From now on the differential operators $\ell_{\alpha}$ and 
${\overline \ell}_{\dot \alpha}$ will be denoted as $Q_{\alpha}$ and 
${\overline Q}_{\dot \alpha}$ like the abstract generators since they 
satisfy the same algebra. The operators $Q_{\alpha}$ and 
${\overline Q}_{\dot \alpha}$ realize the left regular representation 
of supersymmetry on superfields, it is useful to analogously define 
the right regular representation realized by the differential 
operators
\begin{equation}
	D_{\alpha}=\frac{\partial}{\partial 
	\theta^{\alpha}}+i\sigma^{\mu}_{\alpha {\dot \alpha}}{\overline 
	\theta}^{\dot \alpha} \partial_{\mu} \, ,\hspace{1cm}
	{\overline D}_{\dot \alpha}=-\frac{\partial}{\partial 
	{\overline \theta}^{\dot \alpha}}-
	i \theta^{\alpha}\sigma^{\mu}_{\alpha {\dot \alpha}} \partial_{\mu}
	\label{eq:superderiv}
\end{equation}
satisfying
\begin{eqnarray*}
	\{D_{\alpha},{\overline D}_{\dot \alpha}\} & = & 
	-2i\sigma^{\mu}_{\alpha {\dot \alpha}}\partial_{\mu}  \\
	\{D_{\alpha},D_{\beta}\} & = & \{{\overline D}_{\dot \alpha},
	{\overline D}_{\dot \beta} \}  = 0 \, .
\end{eqnarray*}
$D_{\alpha}$ and ${\overline D}_{\dot \alpha}$ are usually referred 
to as superspace covariant derivatives, actually they can be  obtained 
from the general expression for the covariant derivative in the coset 
space SP/L: the fact that $D_{\alpha}$ and ${\overline D}_{\dot \alpha}$ 
do not reduce to $\partial_{\alpha}$ and ${\overline \partial}_{\dot 
\alpha}$ despite the flatness of the ${\cal N}$=1 superspace is a 
consequence of the presence of a non vanishing torsion.

\vspace{0.7cm}
{\sl Chiral superfields.} 
\vspace{0.3cm}

\noindent
Chiral superfields are characterized by the constraint
\begin{displaymath}
	{\overline D}_{\dot \alpha} \Phi(x,\theta,{\overline \theta}) = 0 \, ,
\end{displaymath}
which takes a particularly simple form expressing 
$\Phi(x,\theta,{\overline \theta})$, $D$ and ${\overline D}$ in terms 
of the variable $y^{\mu}=x^{\mu}+i\theta\sigma^{\mu}{\overline \theta}$. 
As functions of $y$, $\theta$ and ${\overline \theta}$ one gets
\begin{eqnarray*}
	D_{\alpha} & = & \frac{\partial}{\partial 
	\theta^{\alpha}} + 2i 
	\sigma^{\mu}_{\alpha{\dot \alpha}}{\overline \theta}^{\dot \alpha}
	\frac{\partial}{\partial y^{\mu}} \\
	{\overline D}_{\dot \alpha} & = & - 
	\frac{\partial}{\partial{\overline \theta}^{\dot \alpha}}  \, ,
\end{eqnarray*}
so that the general chiral superfield assumes the form
\begin{displaymath}
	\Phi(y,\theta) = \phi(y) + \sqrt{2} \theta \psi(y)+
	\theta \theta F(y) \, .
\end{displaymath}
Substituting $y^{\mu}=x^{\mu}+i\theta\sigma^{\mu}{\overline \theta}$ 
and expanding in $\theta$ and ${\overline \theta}$ yields
\begin{eqnarray}
	\Phi(x,\theta,{\overline \theta}) & = & \phi(x) +\sqrt{2}\theta 
	\psi(x)+i\theta\sigma^{\mu}{\overline \theta}
	\partial_{\mu}\phi(x) + \nonumber \\
	&& +\theta\theta F(x)
	-\frac{i}{\sqrt{2}}\theta\theta\partial_{\mu}\psi(x)\sigma^{\mu}
	{\overline \theta}+\frac{1}{4}\theta\theta {\overline \theta}
	{\overline \theta} \Box \phi(x) \, .
\label{chiral}
\end{eqnarray}

Analogously antichiral superfields are defined by
\begin{displaymath}
	D_{\alpha}\Phi^{\dagger}(x,\theta,{\overline \theta}) = 0 \, .
\end{displaymath}
The natural variable is $y^{\dagger\,\mu}=x^{\mu}-i\theta 
\sigma^{\mu}{\overline \theta}$; in terms of $y^{\dagger\,\mu}$
\begin{displaymath}
	\Phi^{\dagger} = \Phi^{\dagger}(y^{\dagger}, 
	{\overline \theta}) = \phi^{*}(y^{\dagger}) + \sqrt{2}{\overline 
	\theta}{\overline \psi}(y^{\dagger}) + {\overline \theta}{\overline 
	\theta}F^{*}(y^{\dagger}) \, .
\end{displaymath}

The field components associated to chiral and antichiral 
multiplets are of two complex scalars ($\phi$, $F$) and one Weyl 
spinor ($\psi$) corresponding to twice the number of states in the 
massive on-shell ${\cal N}$=1 multiplet of section \ref{susyreps}.

Products of (anti) chiral superfields are still (anti) chiral 
superfields. For example
\begin{eqnarray*}
	\Phi_{i}\Phi_{j} & = & \phi_{i}(y)\phi_{j}(y) + \sqrt{2} \theta 
	[\psi_{i}(y)\phi_{j}(y) + \phi_{i}(y) \psi_{j}(y)] + \\
	&& + \theta\theta[\phi_{i}(y)F_{j}(y) + \phi_{j}(y)F_{i}(y) 
	- \psi_{i}(y)\psi_{j}(y)] 
\end{eqnarray*}
The product $\Phi^{\dagger}\Phi$ is neither chiral nor antichiral. 
Moreover since $D^{3}={\overline D}^{3}=0$ it immediately follows 
that for a generic superfield $F$
\begin{eqnarray*}
	&{\overline D}^{2} F = \Phi & \hspace{2cm}  {\rm is~~chiral}  \\
	& D^{2} F = \Phi^{\dagger} & \hspace{2cm} {\rm is~~antichiral} \, .
\end{eqnarray*}
Notice that the double constraint $D_{\alpha}\Phi={\overline D}_{\dot 
\alpha}\Phi=0$ implies that $\Phi$ is constant.

\vspace{0.7cm}
{\sl Vector superfields.} 
\vspace{0.3cm}

\noindent
Vector superfields obey the constraint 
\begin{displaymath}
	V^{\dagger}(x,\theta,{\overline \theta}) = 
	V(x,\theta,{\overline \theta}) \, , 
\end{displaymath}
which gives rise to the component  expansion
\begin{eqnarray}
	&& V(x,\theta,{\overline \theta}) = C(x) + i \theta \chi(x) - i 
	{\overline \theta} {\overline \chi}(x) + \frac{i}{\sqrt{2}}\theta 
	\theta S(x) + \nonumber \\ 
	&& -\frac{i}{\sqrt{2}} {\overline \theta}{\overline \theta} 
	S^{\dagger}(x) - \theta \sigma^{\mu}{\overline \theta} A_{\mu}(x) 
	+i\theta \theta {\overline \theta} \left[ {\overline \lambda}(x) + 
	\frac{i}{2}{\overline \sigma}^{\mu}\partial_{\mu}\chi(x) \right] + 
	\nonumber \\
	&& -i{\overline \theta}{\overline \theta} \theta \left[ 
	\lambda(x) +\frac{i}{2}\sigma^{\mu}\partial_{\mu}{\overline \chi}(x) 
	\right]+\frac{1}{2} \theta \theta {\overline \theta}{\overline \theta} 
	\left[ D(x) + \frac{1}{2} \Box C(x) \right] .
\label{vector}
\end{eqnarray}
The supermultiplet contains two real scalars ($C$, $D$), 
one complex scalar ($S$), four Weyl spinors ($\chi$, ${\overline 
\chi}$, $\lambda$, ${\overline \lambda}$) and one real vector 
($A_{\mu}$). This is twice the content of the massive ${\cal N}$=1 
multiplet obtained starting with a Clifford vacuum of spin 
$\frac{1}{2}$ and also of the massless CPT invariant multiplet with 
minimum helicity $\lambda=-1$.

Since the vector multiplet contains a vector field $A_{\mu}(x)$ it is 
the basic ingredient for the construction of supersymmetric gauge 
theories. The supersymmetric generalization of an Abelian gauge 
transformation \cite{superqed1,superqed2} is
\begin{displaymath}
	V \longrightarrow V + \Phi + \Phi^{\dagger} \, ,
\end{displaymath}
with
\begin{eqnarray*}
	\Phi+\Phi^{\dagger} & = & \phi+\phi^{*}+\sqrt{2}(\theta \psi + 
	{\overline \theta} {\overline \psi}) + \theta \theta F + 
	{\overline \theta}{\overline \theta} F^{*} + 
    i\theta \sigma^{\mu}{\overline \theta}\partial_{\mu} 
	(\phi - \phi^{*}) + \nonumber \\
	&& + \frac{1}{\sqrt{2}}\theta \theta {\overline \theta} {\overline 
	\sigma}^{\mu}\partial_{\mu}\psi+\frac{1}{\sqrt{2}}
	{\overline \theta} {\overline \theta} \sigma^{\mu}\theta
	\partial_{\mu}{\overline \psi}+\frac{1}{4} \theta \theta 
	{\overline \theta} {\overline \theta} \Box (\phi+\phi^{*}) \, .
\end{eqnarray*}
For the components one gets
\begin{eqnarray*}
	C & \longrightarrow & C+\phi+\phi^{*}  \\
	\chi & \longrightarrow & \chi - i \sqrt{2} \psi  \\
	S & \longrightarrow & S - i\sqrt{2} F \\
	A_{\mu} & \longrightarrow & A_{\mu} - i \partial_{\mu} 
	(\phi - \phi^{*}) \\
	\lambda & \longrightarrow & \lambda  \\
	D & \longrightarrow & D \: .
\end{eqnarray*}
Notice that the transformations for the lowest components $C$, $\chi$ 
and $S$ are purely algebraic (no derivatives involved), so they can 
always be used to put these fields to zero. This particular choice of 
gauge is known as {\em Wess--Zumino} (WZ) {\em gauge}. Subtleties related 
to the choice of the Wess--Zumino gauge will be discussed in chapter 3.
It is important to note that in this gauge one has 
\begin{eqnarray*}
	V(x,\theta,{\overline \theta}) & = & -\theta\sigma^{\mu}
	{\overline \theta} A_{\mu}(x) +i \theta \theta 
	{\overline \theta} {\overline \lambda}(x) - i 
	{\overline \theta}{\overline \theta}  \theta \lambda(x) +\frac{1}{2}
	\theta\theta {\overline \theta}{\overline \theta} D(x) \\
	V^{2}(x,\theta,{\overline \theta}) & = & -\frac{1}{2} 
	\theta\theta {\overline \theta}{\overline \theta} A_{\mu}(x)
	A^{\mu}(x)\\
	V^{n}(x,\theta,{\overline \theta}) & = & 0 \, , \hspace{1cm}
	\forall n \geq 3  \: .
\end{eqnarray*}
Fixing the Wess--Zumino gauge still leaves an arbitrary Abelian 
gauge freedom on $A_{\mu}$.

From the vector superfield it is straightforward to construct a 
superfield containing the field strength for the field $A_{\mu}$: it 
is defined as
\begin{equation}
	W_{\alpha}(x,\theta,{\overline \theta}) = -\frac{1}{4}{\overline D}
	{\overline D} D_{\alpha} V(x,\theta,{\overline \theta})
	\label{eq:fieldstr}
\end{equation}
and is a spinor superfield. Besides $W_{\alpha}$ it is useful to 
introduce
\begin{equation}
	{\overline W}_{\alpha}(x,\theta,{\overline \theta}) = 
	-\frac{1}{4}DD {\overline D}_{\dot \alpha} V(x,\theta,
	{\overline \theta}) \, .
	\label{eq:afieldstr}
\end{equation}
They are respectively chiral and antichiral:
\begin{displaymath}
	{\overline D}_{\dot \alpha} W_{\alpha} = 0 \, , \hspace{2cm}
	D_{\alpha}{\overline W}_{\dot \alpha} = 0
\end{displaymath}
and gauge invariant 
\begin{displaymath}
	W[V+\Phi+\Phi^{\dagger}] = W[V] \, .
\end{displaymath}
Gauge invariance of $W_{\alpha}$ allows to compute its component 
expansion in the Wess--Zumino gauge; in terms of $y^{\mu}$ one gets
\begin{eqnarray}
	W_{\alpha}(y,\theta) &=& -i  \lambda(y) +\left[ 
	\delta_{\alpha}{}^{\beta} D - \frac{i}{2} 
	(\sigma^{\mu}\sigma^{\nu})_{\alpha}{}^{\beta} 
	(\partial_{\mu}A_{\nu}(y)-\partial_{\nu}A_{\mu}(y)) \right] 
	\theta_{\beta} + \nonumber \\ 
	&& + \theta \theta \sigma^{\mu}_{\alpha {\dot \alpha}}
	\partial_{\mu} {\overline \lambda}^{\dot \alpha}(y) = 
\label{fieldstrength} \\
	&& = -i  \lambda(y) +\left[ \delta_{\alpha}{}^{\beta} D - 
	\frac{i}{2} (\sigma^{\mu}\sigma^{\nu})_{\alpha}{}^{\beta} 
	F_{\mu \nu}(y) \right] 
	\theta_{\beta} + \theta \theta \sigma^{\mu}_{\alpha {\dot \alpha}}
	\partial_{\mu} {\overline \lambda}^{\dot \alpha}(y) 
	\nonumber 
\end{eqnarray}
and analogously for ${\overline W}_{\alpha}$.

\fancyhead[LO]{{\footnotesize{\it Supersymmetric models}}}
\section{${\cal N}$=1 supersymmetric models}
\label{susymod}
\fancyhead[LO]{{\footnotesize 1.4~~{\it Supersymmetric models}}}

The superfield formalism allows a straightforward construction of 
supersymmetric actions. The supersymmetry transformation of the 
(${\overline \theta}{\overline \theta}$) $\theta \theta$ component 
($F$-component) of a (anti) chiral superfield is a total derivative, 
as can be deduced by specializing (\ref{transfcomp}) to the case of a 
chiral superfield, so that an action of the form~\footnote{The properties 
of Grassmannian integration and the conventions here employed are 
summarized in appendix \ref{appa}.} 
\begin{displaymath}
	S = \int d^{4}x \left[ \int d^{2}\theta \, \Phi + 
	\int d^{2}{\overline \theta} \, \Phi^{\dagger} \right]
\end{displaymath}
is automatically supersymmetric for an arbitrary chiral $\Phi$.

The same situation holds for the $\theta \theta {\overline 
\theta}{\overline \theta}$ component ($D$-component) of a vector 
superfield yielding the supersymmetric expression
\begin{displaymath}
	S = \int d^{4}x \int d^{4} \theta \, V \,  ,
\end{displaymath}
for all $V$'s such that $V^{\dagger}=V$.

Requirement of invariance under the group of automorphisms allows to 
further restrict the possible couplings in the action for 
supersymmetric models; in this context the symmetry under 
transformations of the group of automorphisms is referred to as 
R-{\em symmetry}.

\subsection{Supersymmetric chiral theories: Wess--Zumino model}

The action for the Wess--Zumino model \cite{wesszuminomod} 
is constructed in terms of only chiral superfields and reads
\begin{eqnarray}
	S &=& \int d^{4}x \left\{ d^{4}\theta \, \Phi^{\dagger}_{i} \Phi_{i}
	+ \left[ \int d^{2}\theta \left( \frac{1}{2} m_{ij}\Phi_{i}\Phi_{j}
	+ \right. \right. \right. \nonumber \\ 
	&& \hspace{1cm} \left. \left. \left. \rule{0pt}{18pt} 
	+ \frac{1}{3} g_{ijk} \Phi_{i} \Phi_{j} \Phi_{k} + 
	\lambda_{i}\Phi_{i}\right) + {\rm h.c.} \right] \right\} .
\label{wzsup}
\end{eqnarray}
In terms of component fields it becomes
\begin{eqnarray}
	S &=& \int d^{4}x \left\{ -\partial_{\mu}\phi_{i}^{*}
	\partial^{\mu}\phi_{i} - i{\overline \psi}_{i}{\overline 
	\sigma}^{\mu}\partial_{\mu}\psi_{i}+F^{*}_{i}F_{i}+\left[ 
	m_{ij}(\phi_{i}F_{j}-\frac{1}{2} \psi_{i} \psi_{j}) +  
	\nonumber \right. \right. \\
	&& \hspace{1cm} \left. \left. 
	+ g_{ijk}(\phi_{i}\phi_{j}F_{k}-\psi_{i}
	\psi_{j}\phi_{k})+ \lambda_{i}F_{i} + 
	{\rm h.c.}\rule{0pt}{18pt} \right] \right\}  \, .
	\label{wzcomp}
\end{eqnarray}
The action of the Wess--Zumino model is invariant under the 
supersymmetry transformations
\begin{eqnarray*}
	\delta_{\eta}\phi_{i} & = & \sqrt{2} \eta \psi_{i}  \\
	\delta_{\eta}\psi_{i} & = & i\sqrt{2} \sigma^{\mu}{\overline \eta}
	\partial_{\mu}\phi_{i} +\sqrt{2} \eta F_{i}  \\
	\delta_{\eta}F_{i} & = & i\sqrt{2}{\overline \eta}{\overline 
	\sigma}^{\mu}\partial_{\mu} \psi_{i}
\end{eqnarray*}
The action (\ref{wzcomp}) does not contain derivatives of the fields 
$F_{i}$. They are {\em auxiliary fields}: their equations of motion are 
algebraic and can be used to eliminate them from the action. 
Solving the equations for $F_{i}$
\begin{eqnarray*}
	\frac{\partial {\cal L}}{\partial F^{*}_{k}} & = & 
	F_{k}+\lambda^{*}_{k}+m^{*}_{ik} \phi_{i}^{*}+g^{*}_{ijk}
	\phi^{*}_{i} \phi^{*}_{j} = 0 \\
	\frac{\partial {\cal L}}{\partial F_{k}} & = & 
	F^{*}_{k}+\lambda_{k}+m_{ik} \phi_{i}+g_{ijk}\phi_{i} 
	\phi_{j} = 0 \, ,
\end{eqnarray*}
yields the Lagrangian
\begin{eqnarray*}
	{\cal L} & = & -i {\overline \psi}_{i} {\overline 
	\sigma}^{\mu}\partial_{\mu}\psi_{i} -\partial^{\mu} \phi^{*} 
	\partial_{\mu} \phi -\frac{1}{2} m_{ij} \psi_{i}\psi_{j} 
	-\frac{1}{2}m_{ij}^{*}{\overline \psi}_{i}{\overline \psi}_{j}+ \\
	 && - g_{ijk}\psi_{i}\psi_{j}\phi_{k} - g_{ijk}^{*}{\overline 
	 \psi}_{i}{\overline \psi}_{j}\phi_{k}^{*} - 
	 {\cal V}(\phi_{i},\phi^{*}_{i}) \, ,
\end{eqnarray*}
where the potential ${\cal V}$ is given (for $\lambda_{i}=0$) by the 
expression
\begin{equation}
{\cal V} = (F^{*}_{k} F_{k}) = 
(m_{ik}\phi_{i}+g_{ijk}\phi_{i}\phi_{j}) (m^{*}_{mk}\phi^{*}_{m}+
g^{*}_{mnk}\phi^{*}_{m}\phi^{*}_{n}) \, .
\label{fterm}
\end{equation}
This scalar potential in particular contains a $\phi^{4}$ interaction 
and the mass terms for the scalars $\phi_{k}$; 
actually the Wess--Zumino model is the most general renormalizable 
supersymmetric theory involving only chiral superfields.

The action (\ref{wzcomp}) is the most general renormalizable 
supersymmetric action that can be constructed in terms of only chiral 
superfields. More generally one can consider an invariant action of 
the form 
\begin{displaymath}
	S=\int d^{4}x \left\{ \int d^{4}\theta \, 
	\Phi_{i}^{\dagger}\Phi_{i} + \left[ \int d^{2}\theta \, {\rm 
	W}(\Phi) + {\rm h.c.} \right] \right\} \, ,
\end{displaymath}
where W($\Phi$) is a holomorphic function of $\Phi$ called {\em 
superpotential}. In this case the scalar potential becomes 
\begin{displaymath}
	{\cal V}(\Phi) = \left| \frac{\partial{\rm W}(\Phi)}{\partial \Phi} 
	\right|^{2} \, .
\end{displaymath}

The Wess--Zumino model can be constructed in terms of unconstrained 
superfields using the property $D^{3}={\overline D}^{3}=0$ and 
defining
\begin{displaymath}
	\Phi = {\overline D}^{2}\Psi \, \hspace{1cm} 
	\Phi^{\dagger} = D^{2} \Psi^{\dagger} \, ,
\end{displaymath}
The unconstrained action that reduces to (\ref{wzsup}), for the 
case of one single field $\Phi$, is
\begin{eqnarray}
	S &=& \int d^{4}x d^{4}\theta \, \left\{ (D^{2}\Psi^{\dagger})
	({\overline D}^{2}\Psi) - 2m (\Psi{\overline D}^{2}\Psi 
	+\Psi^{\dagger}D^{2}\Psi + \right. \nonumber \\ 
	&& \hspace{2cm} \left. \rule{0pt}{18pt} - \frac{4g}{3!} \left[ 
	\Psi({\overline D}^{2}\Psi)^{2} +\Psi^{\dagger}
	(D^{2}\Psi^{\dagger})^{2} \right] \right\} \, . 
\end{eqnarray}
The equivalence follows using the identity
\begin{displaymath}
	\int d^{4}x \int d^{2}\theta \, \Psi = 
	\int d^{4}x \int d^{4}\theta \, \left(-\frac{1}{4} 
	{\overline D}^{2}\right)\Psi \, .
\end{displaymath}

\subsection{Supersymmetric gauge theories}

Gauge invariant interactions can be constructed in terms of vector 
superfields and more precisely in terms of the superfield $W_{\alpha}$ 
defined in section \ref{superspace1}, which contains the gauge field 
strength among its components.

Given a chiral superfield $\Phi$ and a vector one $V$ an Abelian 
gauge theory is obtained as follows. The gauge transformation for the 
chiral superfield is 
\begin{eqnarray*}
	&\Phi_{k}^{\prime} = e^{-i q_{k} \Lambda} \Phi_{k}\, , \hspace{0.8cm} 
	{\rm with} \hspace{0.8cm} & {\overline D}_{\dot \alpha} \Lambda = 0 \\
	&\Phi^{\dagger \, \prime}_{k} = e^{i q_{k} \Lambda^{\dagger}} 
	\Phi^{\dagger}_{k} \, , \hspace{0.8cm} {\rm with} \hspace{0.8cm}  
	& D_{\alpha} \Lambda^{\dagger} =  0  \, , 
\end{eqnarray*}
where $q_{k}$ denotes the charge of the superfield $\Phi_{k}$. 
The gauge parameter must be a chiral superfield $\Lambda$. Supplementing 
these transformations with the following one for the vector 
superfield $V$
\begin{displaymath}
	V^{\prime} = V + i (\Lambda - \Lambda^{\dagger}) \, ,
\end{displaymath}
an invariant action can be written in the form 
\begin{eqnarray}
	S & = & \int d^{4}x \left\{ \int d^{2}\theta \,
	\frac{1}{4} W^{\alpha} W_{\alpha} + \int d^{2}{\overline \theta} \,
	\frac{1}{4} {\overline W}_{\dot \alpha} {\overline W}^{\dot \alpha} +
	\int d^{4}\theta \, \Phi_{i}^{\dagger} e^{q_{i}V} \Phi_{i} +  
	\right. \nonumber \\ 
	&&\left. + \left[ \int d^{2}\theta \, \left( \frac{1}{2} m_{ij} 
	\Phi_{i}\Phi_{j} + \frac{1}{3} g_{ijk}\Phi_{i} \Phi_{j} \Phi_{k} 
	\right) + ~{\rm h.c.} \right] \right\} 
	\label{susyabelian}
\end{eqnarray}
Invariance under the above defined transformations is immediately verified. 
Notice that the action (\ref{susyabelian}) is non-polynomial: this a 
general feature of supersymmetric gauge theories and is a consequence 
of the adimensionality of the vector superfield $V$. Choosing the 
Wess--Zumino gauge reduces (\ref{susyabelian}) to a polynomial form 
since it implies $V^{3} = 0$. This choice allows one to prove the
renormalizability of the model. 

The vector superfield $V$ contains the 
gauge dependent components $C$, $\chi$ and $S$ that can be put  to 
zero by choosing the Wess--Zumino gauge as well as the real scalar $D$. 
The latter is an auxiliary field since the corresponding 
equations of motion obtained from (\ref{susyabelian}) are purely 
algebraic. 

This construction can be generalized to the non Abelian case
\cite{sym1,sym2}. 
For a generic non Abelian compact gauge group G the fields $\Phi$, 
$\Phi^{\dagger}$ and $V$ are matrices. 
\begin{displaymath}
	\left( \Phi \right)_{ij} = T^{a}_{ij} \Phi_{a} \, , \hspace{1cm}
	\left( \Phi^{\dagger} \right)_{ij} = T^{a}_{ij} 
	\Phi_{a}^{\dagger} \, , \hspace{1cm} V_{ij} = T^{a}_{ij} V_{a} \, ,
\end{displaymath}
with generators $T^{a}$ of G satisfying
\begin{displaymath}
	\tr (T^{a}T^{b}) = C({\bf r}) \delta^{ab} \hspace{1.5cm} 
	[T^{a},T^{b}] = i f^{ab}{}_{c} T^{c} \, ,
\end{displaymath}
$C({\bf r}) = d_{{\bf r}}$ being the Dynkin index of the representation 
{\bf r}.

The gauge transformation takes the form 
\begin{displaymath}
	\Phi^{\prime} = e^{-i \Lambda} \Phi \, , \hspace{1.5cm} 
	\Phi^{\dagger \, \prime} = \Phi^{\dagger} e^{i \Lambda^{\dagger}}
\end{displaymath}
and for the vector superfield
\begin{equation}
	V \longrightarrow V^{\prime} \, : \hspace{0.5cm} e^{V^{\prime}} = 
	e^{-i\Lambda^{\dagger}} e^{V} e^{i\Lambda} \, .
	\label{vtransf}
\end{equation}
It can be proved that in this case as well it is possible to choose the 
Wess--Zumino gauge in which $V^{3}=0$. 

For infinitesimal gauge transformations use of the Becker--Hausdorff's 
formula allows to write 
\begin{equation}
	\delta V = V^{\prime} - V = i {\cal L}_{V/2} \left[ (\Lambda + 
	\Lambda^{\dagger} + \coth({\cal L}_{V/2}) (\Lambda - \Lambda^{\dagger})
	\right] \, ,
	\label{vvariaz}
\end{equation}
where ${\cal L}_{A}(B) = [A,B]$ is the Lie derivative and 
(\ref{vvariaz}) is a compact form to be understood in terms of the 
power expansion of $\coth({\cal L})$.

The non Abelian generalization of the field strength superfield 
$W_{\alpha}$ is 
\begin{displaymath}
W_{\alpha} = - \frac{1}{4} {\overline D}{\overline D} e^{-V} 
D_{\alpha} e^{V} \, .
\end{displaymath}
In the Wess--Zumino gauge the non Abelian $W_{\alpha}$ has the same 
form as in (\ref{fieldstrength}) with 
\begin{displaymath}
	F_{\mu \nu} = \partial_{\mu}A_{\nu}-\partial_{\nu}A_{\mu} +i g 
	[A_{\mu},A_{\nu}] \, .
\end{displaymath}
The transformation law of $W_{\alpha}$ induced by (\ref{vtransf}) is 
easily obtained and reads
\begin{displaymath}
	W_{\alpha} \longrightarrow W_{\alpha}^{\prime} = e^{-i \Lambda} 
	W_{\alpha} e^{i \Lambda} \, ,
\end{displaymath}
so that while in the Abelian case $W_{\alpha}$ was gauge invariant 
for a non Abelian gauge group G it turns out to be covariant.

The most general renormalizable action for interacting chiral and 
vector superfields is therefore 
\begin{eqnarray}
	S &=& \int d^{4}x \: \frac{1}{d_{{\bf r}}}\tr \left\{ 
	\frac{1}{16g^{2}} \left[ \int d^{2}\theta \, 
	W^{\alpha} W_{\alpha} + \int d^{2}{\overline \theta} \, 
	{\overline W}_{\dot \alpha} {\overline W}^{\dot \alpha} \right] 
	+ \int d^{4}\theta \, e^{-V} \Phi^{\dagger} e^{V} \Phi + \right. 
	\nonumber \\
	&& \left. + \left[ \int d^{2}\theta \, \left( \frac{1}{2} m_{ij} 
	\Phi_{i} \Phi_{j} + \frac{1}{3} g_{ijk} \Phi_{i}\Phi_{j}\Phi_{k}
	\right) + {\rm h.c.} \right] \right\} \, .
\label{susyyangmills}
\end{eqnarray}
The corresponding component field expansion in the Wess--Zumino gauge 
for the case $m_{ij}=0,~g_{ijk}=0$ is 
\begin{eqnarray}
	&& \hspace{-1.6cm} S = \int d^{4}x \, \left\{ - \frac{1}{4} 
	F^{a}_{\mu \nu} F^{a \mu \nu} - i {\overline \lambda}^{a} 
	{\overline \sigma}^{\mu} {\cal D}_{\mu} \lambda^{a} +
	\frac{1}{2} D^{a}D^{a} - {\cal D}_{\mu} 
	\phi^{\dagger}{\cal D}^{\mu} \phi + \right. \nonumber  \\
	&&\hspace{-1.6cm} \left. -i{\overline \psi}{\overline \sigma}^{\mu} 
	{\cal D}_{\mu} \psi + F^{\dagger} F + i \sqrt{2} g 
	\left( \phi^{\dagger} T^{a} \psi \lambda^{a} - 
	{\overline \lambda}^{a} T^{a} \phi {\overline \psi} \right) + 
	g D^{a} \phi^{\dagger} T^{a} \phi \rule{0pt}{18pt} \right\} \, ,
	\label{nonabcomp}
\end{eqnarray}
where the correct factors of the coupling constant $g$ have been 
recovered substituting in (\ref{susyyangmills}) $V \longrightarrow 2gV$.
In (\ref{nonabcomp}) ${\cal D}_{\mu}$ denotes the ordinary covariant 
derivative
\begin{eqnarray*}
	{\cal D}_{\mu}\phi & = & \partial_{\mu} \phi + 
	i g A_{\mu}^{a} T^{a} \phi  \\
	{\cal D}_{\mu}\psi & = & \partial_{\mu} \psi + 
	i g A_{\mu}^{a} T^{a} \psi  \\
	{\cal D}_{\mu}\lambda^{a} & = & \partial_{\mu} \lambda^{a} +  
	i g f^{a}{}_{bc} A_{\mu}^{b} \lambda^{c}  \, .
\end{eqnarray*}
Eliminating the auxiliary fields $F$ and $D$ through the equations of 
motion produces the scalar potential given by (\ref{fterm}) plus the 
so called ``$D$-term''
\begin{displaymath}
	{\cal V}_{D} = \frac{1}{8} g^{2}\left( 
	[\phi^{\dagger},\phi] \right)^{2} \, .
\end{displaymath}

The action (\ref{nonabcomp}) can be proved to be invariant under the 
transformations
\begin{eqnarray}
	\delta_{\xi}\phi & = & \sqrt{2} \xi \psi
	\nonumber  \\
	\delta_{\xi}\psi & = & i \sqrt{2}\sigma^{\mu}{\overline \xi} 
	{\cal D}_{\mu} \phi + \sqrt{2} \xi F
	\nonumber  \\
	\delta_{\xi}F & = & i \sqrt{2}\sigma^{\mu}{\overline \xi} 
	{\cal D}_{\mu} \psi + 2i g T^{a}\phi {\overline \xi}
	{\overline \lambda}^{a} 
	\nonumber  \\
	\delta_{\xi}A_{\mu}^{a} & = & -i {\overline \lambda}^{a}{\overline 
	\sigma}_{\mu}\xi + i {\overline \xi} {\overline \sigma}_{\mu} 
	\lambda^{a} 
\label{symtrans}  \\
	\delta_{\xi} \lambda^{a} & = & \sigma^{\mu \nu} \xi F_{\mu \nu}^{a} 
	+i \xi D^{a}
	\nonumber  \\
	\delta_{\xi}D^{a} & = & -\xi \sigma^{\mu}{\cal D}_{\mu}{\overline 
	\lambda}^{a} - {\cal D}_{\mu}\lambda^{a} \sigma^{\mu} 
	{\overline \xi} \, .
	\nonumber
\end{eqnarray}
The Wess--Zumino condition is not supersymmetric so that it not only 
(partially) breaks gauge invariance, but it also breaks 
supersymmetry, that is recovered through a non linear realization 
given by equations (\ref{symtrans}). These transformations are more 
precisely a mixture of supersymmetry and the residual non Abelian 
gauge symmetry. The choice of the Wess--Zumino gauge leads to subtleties 
at the quantum level, as will be discussed in chapter \ref{cap3}.

The super-gauge invariant action (\ref{susyyangmills}) can be 
generalized by introducing a complexified coupling constant
\begin{displaymath}
\tau = \frac{\theta}{2\pi}+\frac{4\pi i}{g^{2}}
\end{displaymath} 
obtaining, in the pure super Yang--Mills case, the analytic expression
\begin{eqnarray*}
	S &=& \frac{1}{8\pi} {\rm Im}\left[ \tau \int d^{4}x d^{2}\theta 
	\, \tr W^{\alpha}W_{\alpha} \right] = \\
    &=& \frac{1}{g^{2}} \int d^{4}x \, \tr \left[ -\frac{1}{4} 
    F_{\mu \nu}F^{\mu \nu} - i \lambda \sigma^{\mu}{\cal 
    D}_{\mu}{\overline \lambda} + \frac{1}{2} D^{2} \right] + \\
    && - \frac{\theta}{32 \pi^{2}} \int d^{4}x \, \tr F_{\mu \nu} 
    \tilde{F}^{\mu \nu} \, ,
\end{eqnarray*}
where $\tilde{F}^{\mu \nu}=\frac{1}{2} 
\varepsilon^{\mu\nu\rho\sigma}F_{\rho\sigma}$.

\section{Theories of extended supersymmetry}
\label{extendedsusy}

The most straightforward way to construct multiplets for theories 
with extended (${\cal N}>1$) supersymmetry is to assemble together 
${\cal N}$=1 multiplets in such a way as to obtain a field content 
corresponding to the representations of section \ref{susyreps}. It 
turns out that in general this program can be achieved only 
on-shell; in other words it is not possible to realize ${\cal 
N}$-extended supersymmetry in a closed form through an off-shell 
multiplet of fields. With the exception of ${\cal N}$=2 
supersymmetric Yang--Mills theories closure of the algebra  without 
imposing the equations of motion would require an infinite set of 
auxiliary fields. This problem is related to the presence of central 
charges in the supersymmetry algebra for ${\cal N}>1$.

As in the ${\cal N}$=1 case the notion of superspace will be 
introduced to start with and then explicit realizations in specific 
models will be considered.

\fancyhead[LO]{{\footnotesize 1.5~~{\it Extended supersymmetry}}}

\subsection{Extended superspace: a brief survey}

The aim of this section is to give a brief account of ideas beyond 
the superspace formulation for theories with extended supersymmetry,  
no claim of completeness is made. In the following extended superspace 
will not be used in explicit calculations, but it will rather be invoked 
to justify some statements. A detailed and self-contained review of 
extended superspace is given in \cite{extsuperspace}.

The general construction of superspace as a coset space introduced in 
section \ref{superspace1} can be generalized to define extended 
superspace as the coset K=SP$^{*}$/L, where SP$^{*}$ is the extended 
super Poincar\'e group generated by the general algebra 
(\ref{poincare}), (\ref{susyalg}) and L is the Lorentz group SO(3,1).
The generic element of SP$^{*}$ can be written  
\begin{displaymath}
	g = e^{i(a^{\mu}P_{\mu}+\eta_{i}^{\alpha}Q_{\alpha}^{i}+{\overline 
	\eta}_{\dot \alpha}^{i}{\overline Q}_{i}^{\dot 
	\alpha}+ b^{r}Z_{r}+\frac{1}{2} w_{\mu \nu}M^{\mu \nu})} \, ,
\end{displaymath}
where $Z_{ij} = \Omega_{ij}^{r}Z_{r}$ are the central charges 
of the supersymmetry algebra and in this case there are ${\cal N}$ 
anticommuting generators $Q_{\alpha}^{i}$. Elements of the coset are 
obtained setting $w_{\mu \nu}=0$ and can be expressed in the form
\begin{equation}
    k = e^{i(x^{\mu}P_{\mu}+\theta_{i}^{\alpha}Q_{\alpha}^{i}+{\overline 
	\theta}_{\dot \alpha}^{i}{\overline Q}_{i}^{\dot 
	\alpha}+\zeta^{r}Z_{r})} \, .	
	\label{extcosetele}
\end{equation}
Consequently they are parametrized by the enlarged set of coordinates 
\begin{equation}
    z_{m} = (x_{\mu},\theta_{\alpha}^{i}, {\overline 
    \theta}_{\dot \alpha}^{j}, \zeta_{r}) \, .
	\label{superspcoord}
\end{equation}
Repeating the steps of section \ref{superspace1} allows to obtain the 
following supersymmetric coordinate transformations  
\begin{eqnarray}
	x^{\prime \, \mu} & = & 
	x^{\mu}+i\eta^{\alpha}_{i}\sigma^{\mu}_{\alpha{\dot 
	\alpha}}{\overline \theta}^{{\dot \alpha} i} -i \theta^{\alpha}_{i}
	\sigma^{\mu}_{\alpha{\dot \alpha}}{\overline \eta}^{{\dot \alpha} i}
	\nonumber \\
	\theta^{\prime\, \alpha}_{i} & = & \theta^{\alpha}_{i} + 
	\eta^{\alpha}_{i}
	\nonumber  \\
    {\overline \theta}^{\prime\,{\dot \alpha}j} & = & {\overline 
    \theta}^{{\dot \alpha}j} + {\overline \eta}^{{\dot \alpha}j}
    \label{extcoordtransf} \\
	\zeta^{\prime \, r} & = & \zeta^{r}+i\eta^{\alpha}_{i}\theta_{\alpha 
	j}(\Omega^{r})^{ij}+i{\overline \eta}_{\dot \alpha}^{i}{\overline 
	\theta}^{{\dot \alpha}j}(\Omega^{r})_{ij} \, .
	\nonumber
\end{eqnarray}
Under the action of the central charges $x^{\mu}$, 
$\theta_{\alpha}^{i}$ and ${\overline \theta}_{{\dot \alpha}j}$ do 
not transform while
 \begin{equation}
 	\zeta^{\prime \, r} = \zeta^{r} + b^{r} \, .
 	\label{centraltransf}
 \end{equation} 

Superfields are functions defined on the manifold parametrized by 
$z_{m}$
\begin{equation}
	F = F(x_{\mu}, \theta_{\alpha}^{i}, 
	{\overline \theta}_{\dot \alpha}^{j}, \zeta_{r}) \, .
 \label{extsuperfield}
\end{equation}
The action of the super Poincar\'e group on superfields is defined 
through 
\begin{displaymath}
	R(g) \, F(z) = F(z^{\prime}) 
\end{displaymath}
where $z^{\prime}$ is given in (\ref{extcoordtransf}) and $R(g)$ acts 
as the left regular representation.

Differential operators generating the infinitesimal transformations 
now read
\begin{eqnarray*}
	\ell_{\mu} & = & \partial_{\mu}  \\
	\ell_{\alpha}^{i} & = & \frac{\partial}{\partial 
	\theta^{\alpha}_{i}} + i \sigma^{\mu}_{\alpha{\dot \alpha}} 
	{\overline \theta}^{{\dot \alpha}i} \partial_{\mu} + 
	i(\Omega^{r})^{ij}\theta_{\alpha j} 
	\frac{\partial}{\partial \zeta^{r}} \\
	{\overline \ell}_{{\dot \alpha}i} & = & \frac{\partial}{\partial 
	{\overline \theta}^{{\dot \alpha}i}} + i \theta^{\alpha}_{i}
	\sigma^{\mu}_{\alpha{\dot \alpha}} \partial_{\mu} + 
	i {\overline \theta}_{\alpha}^{j} (\Omega^{r})_{ij}
	\frac{\partial}{\partial \zeta^{r}} \\
	\ell^{r} & = &  \frac{\partial}{\partial \zeta_{r}} \\
	\ell_{\mu \nu} & = & 
	-(x_{\mu}\partial_{\nu}-x_{\nu}\partial_{\mu})-\frac{1}{2}
	\theta^{\beta}\sigma^{\mu 
	\nu}_{\beta}{}^{\alpha}\frac{\partial}{\partial \theta^{\alpha}} 
	+\frac{1}{2} {\overline \theta}^{\dot \beta}{\overline \sigma}^{\mu 
	\nu}_{\dot \beta}{}^{\dot \alpha}\frac{\partial}{\partial 
	{\overline \theta}^{\dot \alpha}} \nonumber \\
	\ell_{a} & = & 
	-\theta^{\alpha}_{i}B_{a}^{i}{}_{j}\frac{\partial}{\partial 
	\theta^{\alpha}_{j}} - {\overline \theta}^{{\dot \alpha}i} 
	\left( B_{a}^{i}{}_{j} \right)^{*} \frac{\partial}{\partial 
	{\overline \theta}^{\dot \alpha}_{j}} \: .
\end{eqnarray*}

Relation to component field multiplets is established by Taylor expanding 
in the $\theta$, ${\overline \theta}$ as well as in 
the $\zeta$ coordinates. Notice that the $\zeta$'s are ``bosonic'' 
coordinates, so that the expansion is a truly infinite series
\begin{equation}
	F(x,\theta,{\overline \theta}, \zeta) = F^{(0)}
	(x,\theta,{\overline \theta}) + F_{(r)}^{(1)}
	(x,\theta,{\overline \theta}) \, \zeta^{(r)} + \ldots \, ,
	\label{superfieldexp}
\end{equation}
where 
\begin{eqnarray*}
	F^{(0)}(x,\theta,{\overline \theta}) & = & F(x,\theta, 
	{\overline \theta}, 0) \\
	F^{(1)}_{(r)}(x,\theta,{\overline \theta}) & = & \frac{\partial}
	{\partial \zeta^{r}} \left. F(x,\theta, {\overline \theta}, \zeta)
	\rule[-8pt]{0pt}{12pt}\right|_{\zeta =0} \, .
\end{eqnarray*}

``Super-covariant'' derivatives generalizing those defined in 
section \ref{superspace1} are given by
\begin{eqnarray}
     D_{\mu} & = & \partial_{\mu} \nonumber \\ 
	 D_{\alpha}^{i} & = & \frac{\partial}{\partial 
	\theta^{\alpha}_{i}}+i\sigma^{\mu}_{\alpha {\dot \alpha}}{\overline 
	\theta}^{{\dot \alpha}i} \partial_{\mu} - i (\Omega^{r})^{ij}
	\theta_{\alpha j} \frac{\partial}{\partial \zeta^{r}} \nonumber \\
	{\overline D}_{{\dot \alpha}i} & = & -\frac{\partial}{\partial 
	{\overline \theta}^{{\dot \alpha}i}}-
	i \theta^{\alpha}_{i}\sigma^{\mu}_{\alpha {\dot \alpha}} 
	\partial_{\mu}-i (\Omega^{r})_{ij}{\overline \theta}_{\dot 
	\alpha}^{j} \frac{\partial}{\partial \zeta^{r}}
	\label{extcovariant}  \\
	D_{r} & = & \frac{\partial}{\partial \zeta^{r}}
	\nonumber \: .
\end{eqnarray}
Just as in the case of ${\cal N}$=1 superspace constraints defining 
superfields corresponding to irreducible representations of 
supersymmetry are implemented through the covariant derivatives 
(\ref{extcovariant}). In general constrained superfields still depend 
on the coordinates $\zeta$, so that the component supermultiplets 
contain an infinite number of fields; the only exception is the 
${\cal N}$=2 {\em vector superfield} defined by the condition 
\begin{equation}
	D_{r} V(x,\theta,{\overline \theta}, \zeta) = 
	\frac{\partial}{\partial \zeta^{r}} 
	V(x,\theta,{\overline \theta}, \zeta) = 0 \, . 
	\label{vectorext}
\end{equation}
As will be discussed in the next subsection the component fields 
associated to this superfield are those defining the ${\cal N}$=2 
supersymmetric Yang--Mills model.

${\cal N}>1$ irreducible on-shell representations constructed in 
section \ref{susyreps} contain a finite number of states 
($2^{2{\cal N}}$ in the massive case and $2^{\cal N}$ in the massless 
case). The reason for the appearance here of an infinite number of 
component fields is that closure of the supersymmetry algebra 
off-shell in the presence of central charges requires infinitely many 
auxiliary fields, the only exception being the ${\cal N}$=2 super 
Yang--Mills multiplet related to the superfield in equation 
(\ref{vectorext}).

The ${\cal N}$=2 supersymmetry algebra contains in principle two 
central charges, but one of them can always be eliminated through a 
chiral rotation, so that the only relevant anticommuting relation is 
\begin{displaymath}
	\{ D_{\alpha}^{i},D_{\beta}^{j} \} = 2\varepsilon_{\alpha \beta}
	\varepsilon^{ij} D_{\zeta} \, .
\end{displaymath}
One can consider a superfield $W$ satisfying 
\begin{equation}
	{\overline D}^{i}_{\dot \alpha} W = 0 \, , \hspace{2cm} 
	D_{\zeta}W = 0 \, .
	\label{chiralext}
\end{equation}
A multiplet of fields can be obtained taking the $\theta = 0$ 
component of the fields
\begin{eqnarray}
	&& W\, , ~~ D_{\alpha}^{i}W\, , ~~ D^{\alpha i}D_{\alpha}^{j}W \, ,
	~~D^{i}_{\alpha}D_{\beta i}W \, , \nonumber \\ 
	&& D_{\alpha}^{i}D_{\beta}^{j}D_{\gamma}^{k}W\, ,
	~~ D_{\alpha}^{i}D_{\beta}^{j}D_{\gamma}^{k} D_{\delta}^{l} W \, .
	\label{2symmultiplet}
\end{eqnarray}
Exploiting the conditions (\ref{chiralext}) in (\ref{2symmultiplet}) 
does not produce an irreducible multiplet. A further reduction is 
necessary which is achieved by imposing the reality condition 
\begin{displaymath}
	D^{\alpha i}D^{j}_{\alpha} W = {\overline D}^{i}_{\dot 
	\alpha}{\overline D}^{{\dot \alpha}j}{\overline W} \, .
\end{displaymath}
This gives rise to a multiplet containing the fields 
\begin{equation}
	D\, , ~~ \chi_{\alpha}^{i}\, , ~~ C^{ij}\, , ~~ F_{\mu \nu} \, ,
	\label{2symcomp}
\end{equation}
where $D$ and $C^{ij}=C^{ji}$ are real scalars, $\chi^{i}_{\alpha}$ Weyl 
spinors and $F_{\mu \nu} = \partial_{\mu}A_{\nu} - \partial_{\nu}A_{\mu}$ 
is the field strength for a real vector field $A_{\mu}$. These are the 
fields of the ${\cal N}$=2 super Yang--Mills model: they sum up a 
chiral and a vector ${\cal N}$=1 multiplet.

An example of a ${\cal N}$=2 multiplet possessing central charge 
arises if one considers a superfield satisfying 
\begin{eqnarray}
	&& D_{\alpha}^{i} \Phi_{j} = \frac{1}{2} \delta^{i}_{j} 
	D^{k}_{\alpha} \Phi_{k} \nonumber  \\
	&& {\overline D}_{{\dot \alpha}i}\Phi_{j} + 
	{\overline D}_{{\dot \alpha}j}\Phi_{i} = 0 \, .
	\label{hypersup}
\end{eqnarray}
Using these constraints a multiplet of fields can be obtained taking 
the $\theta=0$ component of the fields
\begin{displaymath}
	\Phi\, , ~~ D_{\alpha}^{k}\Phi_{k}\, , ~~ 
	{\overline D}_{\dot \alpha}^{k}\Phi_{k}\, , ~~ D_{\zeta} \Phi \, .
\end{displaymath}
Some $D$-algebra manipulations allow to construct a supermultiplet 
containing two Weyl spinors and four complex scalars
\begin{equation}
	\psi_{\alpha} \, , ~~ \chi_{\alpha}\, , ~~
	\phi_{i}\, , ~~ F_{i} \, .
	\label{eq:hypermult}
\end{equation}
This multiplet is known as {\em hypermultiplet} and can be 
viewed as the combination of two ${\cal N}$=1 chiral multiplets. It 
will be shown to describe ${\cal N}$=2 matter in supersymmetric gauge 
theories.

In general the constraints that need to be imposed on ${\cal N}>1$ 
superfields cannot be explicitly solved for the whole superfield; in 
other words considering all the $\theta$-components yields a set of 
non independent fields. The lack of an explicit solution of the constraints 
does not allow to exploit the potentiality of the superspace formalism 
for extended supersymmetry, so that it does not prove so powerful a 
tool as in the ${\cal N}$=1 case. 

These problems were partially solved in \cite{gikos} by the 
introduction of {\em harmonic superspace}. It is based on a 
parametrization of ${\cal N}$=2 superspace as a homogeneous space 
that leads to different bosonic coordinates associated to the central 
charges. Use of such coordinates allows to construct ${\cal N}$=2 
hypermultiplets in terms of unconstrained superfields. A 
generalization of this approach leading to the so called {\em analytic 
superspace} has been proposed in order to give a manifestly 
supersymmetric description of ${\cal N}$=4 theories (see the review 
\cite{extsuperspace} for the details), however it only works on-shell.

In the following all the explicit calculations using superfield 
formalism will be carried out within the framework of 
${\cal N}$=1 superspace.

\subsection{${\cal N}>1$ supersymmetric models}

In this section only examples of ${\cal N}$=2 theories will be 
discussed, the general properties of ${\cal N}$=4 models, that are 
the main subject of this work, will be studied in detail in next 
chapter.

\vspace{0.7cm}
{\sl ${\cal N}$=2  supersymmetric Yang--Mills model.} 
\vspace{0.3cm}

\noindent
The superfield formulation of ${\cal N}$=2 pure supersymmetric 
Yang--Mills models is a straightforward generalization of the ${\cal 
N}=1$ case because it is constructed in terms of a ${\cal N}$=2 
superfield satisfying the constraint (\ref{vectorext}), which implies no 
dependence on the central charges. Since no central charges are 
present a full off-shell formulation is possible. Superspace in this 
case is parametrized by 
\begin{displaymath}
	z_{m} = (x_{\mu},\theta^{i}_{\alpha}, {\overline \theta}_{{\dot 
	\alpha} j}) = (x_{\mu}, \theta_{\alpha},{\overline \theta}_{\dot 
	\alpha}, \tilde{\theta}_{\alpha}, \overline{\tilde{\theta}}_{\dot 
	\alpha})
\end{displaymath}
The starting point is a chiral superfield $\Psi$ defined by
\begin{equation}
	{\overline D}_{\dot \alpha} \Psi = 0 \, , \hspace{1.5cm} 
	\overline{\tilde{D}}_{\dot \alpha} \Psi = 0 
\label{chiraln2}
\end{equation}
There is a natural choice of variables to describe $\Psi$, namely
\begin{displaymath}
	\tilde{y}_{\mu} = x_{\mu}+i\theta \sigma_{\mu} {\overline \theta} 
	+i\tilde{\theta}\sigma_{\mu}\overline{\tilde{\theta}}
\end{displaymath}
$\Psi$ can be expressed in terms of $\tilde{y}$ as
\begin{displaymath}
	\Psi(\tilde{y},\theta) = \Phi(\tilde{y},\theta) +i\sqrt{2} 
	\tilde{\theta}^{\alpha}W_{\alpha}(\tilde{y},\theta) +
	\tilde{\theta}\tilde{\theta} G(\tilde{y},\theta) \, ,
\end{displaymath}
where $\Phi$ and $G$ are ${\cal N}$=1 chiral superfields and 
$W_{\alpha}$ is a chiral spinor superfield.

To construct a non Abelian gauge theory $\Psi$ is taken to be a 
matrix
\begin{displaymath}
	\Psi_{ij} = T^{a}_{ij} \Psi_{a} \, .
\end{displaymath}
A ${\cal N}$=2 supersymmetric action can then be written as
\begin{equation}
	S = \frac{1}{4\pi} {\rm Im} \left\{ \tau \int d^{4}x \int d^{2}\theta
	d^{2}\tilde{\theta} \, \tr \left( \frac{1}{2} \Psi^{2} \right) 
	\right\} \, ,
\label{sym2sup}
\end{equation}
where 
\begin{displaymath}
	\left. \Psi^{2}\right|_{{\theta}^{2}\tilde{\theta}^{2}} = 
	\left. W^{\alpha}W_{\alpha}\right|_{\theta^{2}} +  
	\left. 2 G \Phi \right|_{\theta^{2}} \, .
\end{displaymath}
${\cal N}$=2 super Yang--Mills is obtained by further constraining 
the chiral superfield $G$. This is achieved by requiring $\Psi$ to satisfy
\begin{displaymath}
	D^{\alpha i}D^{j}_{\alpha} \Psi = {\overline D}_{\dot \alpha}^{i} 
	{\overline D}^{{\dot \alpha} j} \Psi^{\dagger} \, .
\end{displaymath}
Solving the constraint provides for $G$ the following expression
\begin{displaymath}
	G = \int d^{2}\theta \, 
	\Phi^{\dagger}(\tilde{y}-i\theta\sigma{\overline \theta}, 
	{\overline \theta}) e^{V(\tilde{y}-i\theta\sigma{\overline \theta}, 
	{\overline \theta})} \, .
\end{displaymath}
Substituting into the general form (\ref{sym2sup}) gives the action 
of ${\cal N}$=2 super Yang--Mills theory. It can be viewed as a particular 
${\cal N}$=1 supersymmetric gauge theory coupled to chiral matter, 
namely it corresponds to the case of one single flavor in the adjoint 
representation of the gauge group. Supersymmetry does not allow a 
superpotential for the model. Integration over the $\theta$ variables 
gives the component expansion. The superfield $\Psi$ is a singlet 
under the SU(2) R-symmetry group. The resulting component multiplet 
obtained eliminating the auxiliary fields in the Wess--Zumino gauge 
contains a vector $A_{\mu}$ and a complex scalar $\phi$ that are 
SU(2) singlets and two Weyl spinors $(\psi,\lambda)$ that transform 
in the {\bf 2} of SU(2). As already noticed, this field content is what 
would be obtained putting together a vector and a chiral ${\cal N}$=1 
superfield both in the adjoint of the gauge group.

The most general ${\cal N}$=2 action that can be constructed in terms 
of the superfield $\Psi$ of equation (\ref{chiraln2}) is 
\begin{displaymath}
	S = \frac{1}{4\pi} {\rm Im} \left\{ \int d^{4}x \int d^{2}\theta 
	d^{2}\tilde{\theta} \, {\cal F}(\Psi) \right\} \, ,
\end{displaymath}
where ${\cal F}$ is a holomorphic function of $\Psi$ called {\em 
prepotential}.

\vspace{0.7cm}
{\sl ${\cal N}$=2  matter.} 
\vspace{0.3cm}

\noindent
The ${\cal N}$=2 super Yang--Mills model can be generalized 
introducing the coupling to matter 
described by hypermultiplets. Hypermultiplets are associated to 
superfields satisfying (\ref{hypersup}). A kinetic term for these 
superfields can be written in the form
\begin{displaymath}
	S_{K} = \int d^{4}x \, D^{\alpha i}D^{j}_{\alpha} \left[ 
	\Phi^{i \, \dagger} D^{\beta k}D^{k}_{\beta}\Phi^{j} \right] \, .
\end{displaymath}
Explicit dependence on the central charge does not allow a manifestly 
${\cal N}$=2 formulation without resorting to harmonic superspace. 
Anyway an on-shell action can be constructed in terms of ${\cal 
N}=1$ superfields. Denoting the two chiral superfields in the 
hypermultiplet by $Q$ and $\tilde{Q}$ the action in ${\cal N}$=1 
language is given by
\begin{eqnarray}
	&& \hspace{-1.6cm}
	S=\int d^{4}x \left\{ \int d^{4}\theta \, \left[ Q^{\dagger}e^{2V}Q 
	+ \tilde{Q}^{\dagger} e^{2V}\tilde{Q} + \Phi^{\dagger} e^{2V} \Phi
	\right] \right. + \nonumber \\
	&& \hspace{-1.6cm}
	+ \left. \frac{1}{4\pi} {\rm Im} \left[ \tau \int d^{2} \theta \,
	W^{\alpha} W_{\alpha} \right] + \left[ \int d^{2}\theta \, \left( 
	\sqrt{2} \tilde{Q}\Phi Q + m_{i} \tilde{Q}_{i} Q_{i} \right) +
	{\rm h.c.} \right] \right\} .
\label{sqcd2}
\end{eqnarray}
The action (\ref{sqcd2}) contains the ordinary kinetic terms, a mass 
term for chiral fields in the hypermultiplet and a superpotential that 
couples $Q$, $\tilde{Q}$ and $\Phi$. Notice that no potential term 
coming from a self interaction of the fields $Q$ and $\tilde{Q}$ is 
allowed by ${\cal N}$=2 supersymmetry. This follows from the 
observation that $Q$ and $\tilde{Q}$ are assembled into a superfield 
$\Phi^{i}$ transforming in the {\bf 2} of SU(2) R-symmetry and there 
exists no trilinear invariant in SU(2).

\section{Quantization of supersymmetric theories}
\label{quantization}
\fancyhead[LO]{{\footnotesize 1.6~~{\it Quantization of SUSY 
theories}}}

The functional approach to the quantization of field theories, based 
on the construction of a generating functional for the time-ordered 
products of fields, can be applied with no further difficulty to 
supersymmetric theories in their component field formulation. 
However, although this formulation allows to extract all the peculiar 
properties of supersymmetric theories at the quantum level, it makes 
rather obscure how such properties are a direct consequence of 
supersymmetry. On the contrary a quantization procedure can be 
carried out directly in superspace in a step by step manifestly 
supersymmetric fashion.
This approach, that allows to construct directly Green functions for 
the superfields, has proved extremely powerful and it is perhaps the most 
useful application of superspace techniques.

Super Feynman rules for the computation of Green functions of the 
superfields were first proposed in \cite{superspace1} and then 
reformulated in a more powerful way in \cite{grisaruroceksiegel}. This 
latter formulation will be reviewed here. Once the Green functions for 
the superfields are known contact with the ``physical'' component 
field formulation is established by further integrating the fermionic 
coordinates ($\theta,{\overline \theta}$) to extract correlators for 
the various components.

The discussion will focus on the quantization of ${\cal N}$=1 
superfields; it is also possible to generalize the method to ${\cal 
N}$-extended superspace, {\em e.g.} to harmonic superspace, but 
the ${\cal N}$=1 formulation appears to be the most suitable for 
application to explicit calculations, even in the extended case.

\subsection{General formalism}

In ordinary quantum field theories the path integral quantization is 
based on the construction of a generating functional for the Green 
functions. Denoting by $\phi_{i}$ the generic fields of the model and 
by $S[\phi]$ the classical action the generating functional is 
defined as
\begin{displaymath}
	Z[J] = N \int [{\cal D} \phi] \, e^{iS[\phi]+i \int dx \phi_{i}(x) 
	J_{i}(x)} \, ,
\end{displaymath}
where $N$ is a normalization such that $Z[0]=1$ and $J_{i}(x)$ are 
external sources associated to the fields $\phi_{i}(x)$. Green functions 
are obtained taking functional derivatives with respect to the 
sources $J_{i}$. 

This approach has a natural extension that leads to a generating 
functional for the Green functions of the superfields in a 
supersymmetric theory. 

The general form of the action of a supersymmetric theory involves, 
in ${\cal N}$=1 language, chiral and vector superfields and can be 
written
\begin{equation}
	S = \int d^{4}x d^{4}\theta \, K(\Phi,\Phi^{\dagger},V) +
	\left[ \int d^{4}x d^{2}\theta \, {\rm W}[\Phi] + {\rm h.c.} 
	\right] \, .
\label{generalaction}
\end{equation}
To construct a generating functional classical sources for the 
different superfields must be introduced, that are in turn superfields.
Since functional derivatives are to be taken with respect to the 
classical sources, such sources must be arbitrary. A problem arises 
with chiral superfields as sources coupled to them must themselves be
chiral. This difficulty can be solved by using a suitable definition of 
functional derivatives with respect to chiral superfields.

For a generic superfield $F(x,\theta,{\overline \theta})$ functional 
derivatives are defined according to the rule
\begin{displaymath}
	\frac{\delta F(x^{\prime},\theta^{\prime},
	{\overline \theta}^{\prime})}{\delta F(x,\theta,{\overline \theta})}
	= \delta_{4}(x-x^{\prime}) \delta_{4}(\theta-\theta^{\prime}) \, ,
\end{displaymath}
where the $\delta$ function of fermionic variables is to be 
interpreted as
\begin{displaymath}
	\delta_{4}(\theta-\theta^{\prime}) = \delta_{2}
	(\theta-\theta^{\prime}) \delta_{2}({\overline \theta}-{\overline 
	\theta}^{\prime}) = (\theta-\theta^{\prime})^{2} 
	({\overline \theta}-{\overline \theta}^{\prime})^{2} \, ,
\end{displaymath}
so that
\begin{displaymath}
	\int d^{4}\theta \, \delta_{4}(\theta-\theta^{\prime}) F(x,\theta, 
	{\overline \theta}) = F(x,\theta^{\prime}, {\overline 
	\theta}^{\prime}) 
\end{displaymath}
and 
\begin{equation}
	\frac{\delta}{\delta F(x,\theta, {\overline \theta})} \int 
	d^{4}x^{\prime} d^{4}\theta^{\prime} \, 
	F(x^{\prime},\theta^{\prime}, {\overline \theta}^{\prime}) 
	G(x^{\prime},\theta^{\prime}, {\overline \theta}^{\prime}) = 
	G(x, \theta, {\overline \theta}) \, .
\label{functder}
\end{equation}
For chiral superfields $\Phi$ one defines in turn
\begin{displaymath}
	\frac{\delta \Phi(x^{\prime},\theta^{\prime},
	{\overline \theta}^{\prime})}{\delta \Phi(x,\theta, {\overline \theta})}
	= -\frac{1}{4} {\overline D}^{2} \delta_{4}(x-x^{\prime}) 
	\delta_{4}(\theta-\theta^{\prime}) \, .
\end{displaymath}
In this way the same result as in equation (\ref{functder}) is reproduced 
\begin{eqnarray*}
	&& \frac{\delta}{\delta  \Phi(x,\theta, {\overline \theta})} \int 
	d^{4}x^{\prime} d^{2}\theta^{\prime} \, \Phi(x^{\prime},\theta^{\prime},
	{\overline \theta}^{\prime}) \tilde{\Phi}(x^{\prime},\theta^{\prime},
	{\overline \theta}^{\prime}) = \\
	&& = \int d^{4}x^{\prime} d^{2}\theta^{\prime} \left( 
	-\frac{1}{4} {\overline D}^{2} \right) \delta_{4}(x-x^{\prime})
	\delta_{4}(\theta-\theta^{\prime}) \tilde{\Phi}(x^{\prime},
	\theta^{\prime},{\overline \theta}^{\prime}) = \\
	&& = \int d^{4}x^{\prime} d^{4}\theta^{\prime} \delta_{4}(x-x^{\prime})
	\delta_{4}(\theta-\theta^{\prime}) \tilde{\Phi}(x^{\prime},
	\theta^{\prime},{\overline \theta}^{\prime}) = 
	\tilde{\Phi}(x,\theta,{\overline \theta}) \, , 
\end{eqnarray*}
where in the last line it was used the property that 
$d^{2}{\overline \theta}$ and $-\frac{1}{4} {\overline D}^{2}$ are 
equivalent under space-time integration, as can be proved by direct 
calculation by an integration by parts.

Given these definitions of functional derivatives the generating 
functional can be written as
 \begin{equation}
	Z[J,J^{\dagger},J_{V}] = N \int [{\cal D}\Phi {\cal D}\Phi^{\dagger}
	{\cal D}V] \, e^{iS[\Phi, \Phi^{\dagger},V] + i(\Phi,J) + 
	i(\Phi^{\dagger},J^{\dagger}) + i(V,J_{V})} \, ,
\label{generfunct}
\end{equation}
where $S$ is the action (\ref{generalaction}), $J$ ($J^{\dagger}$) is 
a chiral (antichiral) source, $J_{V}$ is real and 
\begin{eqnarray*}
	(\Phi,J) & = & \int d^{4}x d^{2}\theta \, \Phi(x,\theta,{\overline 
	\theta}) J(x,\theta,{\overline \theta}) \, ,  \\
	(\Phi^{\dagger},J^{\dagger}) & = & \int d^{4}x d^{2}{\overline \theta}
	\, \Phi^{\dagger}(x,\theta,{\overline \theta}) 
	J^{\dagger}(x,\theta,{\overline \theta}) \, ,  \\
	(V,J_{V}) & = & \int d^{4}x d^{4}\theta \, V(x,\theta,{\overline 
	\theta}) J_{V}(x,\theta,{\overline \theta}) \, .
\end{eqnarray*}
Green functions can then be calculated by the formula
\begin{eqnarray}
	&& \hspace{-.5cm}{\cal G}_{n}(z_{1},\ldots,z_{i},z_{i+1},
	\ldots,z_{j},z_{j+1},\ldots,z_{n}) = \nonumber \\
	&& \hspace{-.5cm}= \langle 0|T\{ \Phi(z_{1})\ldots\Phi(z_{i})
	\Phi^{\dagger}(z_{i+1})
	\ldots \Phi^{\dagger}(z_{j})V(z_{j+1})\ldots V(z_{n}) \}
	|0\rangle =  
	\label{greensup}  \\
	&& \hspace{-.5cm}= \left. (-i)^{n}\frac{\delta^{n} 
	Z[J,J^{\dagger},J_{V}]}{\delta J(z_{1})
	\ldots\delta J(z_{i})\delta J^{\dagger}(z_{i+1})\ldots
	\delta J^{\dagger}(z_{j})\delta J_{V}(z_{j+1})
	\ldots\delta J_{V}(z_{n})} \right|_{J=J^{\dagger}=J_{V}=0} \, .
	\nonumber
\end{eqnarray}

A generating functional for the connected Green functions can be 
obtained by taking the logarithm of $Z$
\begin{displaymath}
	W[J,J^{\dagger},J_{V}] = (-i) \, 
	\log \left( Z[J,J^{\dagger},J_{V}] \right) \, .
\end{displaymath}
The quantum effective action $\Gamma$ that acts as a generating 
functional for the one particle irreducible Green functions is 
obtained through the Legendre transform. Defining the ``classical'' 
fields
\begin{equation}
   \tilde{\Phi} = \frac{\delta W}{\delta J} \, , \hspace{1cm} 
   \tilde{\Phi}^{\dagger} = \frac{\delta W}{\delta J^{\dagger}} \, ,
   \hspace{1cm} \tilde{V} = \frac{\delta W}{\delta J_{V}} \: ,
\label{classfields}
\end{equation}
$\Gamma$ is given by
\begin{displaymath}
	\Gamma[\tilde{\Phi},\tilde{\Phi}^{\dagger},\tilde{V}] = 
	\left. \left( W[J,J^{\dagger},J_{V}] - \left[ 
	(J,\Phi)+(J^{\dagger},\Phi^{\dagger})+(J_{V},V) \right] \right) 
	\right|_{(\tilde{\Phi},\tilde{\Phi}^{\dagger},\tilde{J}_{V})} \, ,
\end{displaymath}
where on the right hand side $J=J[\tilde{\Phi},\tilde{\Phi}^{\dagger},
\tilde{V}]$, $J^{\dagger}=J^{\dagger}[\tilde{\Phi},\tilde{\Phi}^{\dagger},
\tilde{V}]$ and $J_{V}=J_{V}[\tilde{\Phi},\tilde{\Phi}^{\dagger},
\tilde{V}]$ come from the inverting  (\ref{classfields}). 

The generating functional $Z$ can be expressed in terms of the 
corresponding functional in the free theory, $Z_{0}$. The general 
action (\ref{generalaction}) can be divided up into a free part and an 
interacting part
\begin{displaymath}
	S[\Phi,\Phi^{\dagger},V] = S_{0}[\Phi,\Phi^{\dagger},V] + 
	S_{_{{\rm int}}}[\Phi,\Phi^{\dagger},V] \, ,
\end{displaymath}
where $S_{0}$ is quadratic in the fields and contains the kinetic and 
mass terms. Then one can write
\begin{equation}
	Z[J,J^{\dagger},J_{V}] = e^{iS_{_{{\rm int}}}\left[ 
	\frac{\delta}{\delta J}, \frac{\delta}{\delta J^{\dagger}}, 
	\frac{\delta}{\delta J_{V}} \right]} Z_{0}[J,J^{\dagger},J_{V}] \, ,
\label{zinter}
\end{equation}
where 
\begin{displaymath}
	Z_{0}[J,J^{\dagger}J_{V}] = N \int [{\cal D}\Phi {\cal 
	D}\Phi^{\dagger}{\cal D}V] \, e^{iS_{0}[\Phi,\Phi^{\dagger},V] +
	i(\Phi,J)+i(\Phi^{\dagger},J^{\dagger})+i(V,J_{V})}
\end{displaymath}
and $e^{iS_{_{{\rm int}}}[\frac{\delta}{\delta J},
\frac{\delta}{\delta {J^{\dagger}}},\frac{\delta}{\delta J_{V}}]}$ 
is to be understood as a power expansion. 
Since $S_{0}$ is quadratic $Z_{0}$ is a Gaussian integral that can be 
explicitly computed. Writing 
\begin{eqnarray}
	&& Z_{0}[J,J^{\dagger},J_{V}] = {\rm exp}\left\{ 
	-\frac{i}{2} \int d^{4}x d^{4}\theta \, (\Phi, \Phi^{\dagger}) 
	{\cal M}_{\Phi} \left( 
	\begin{array}{c} \Phi \\ \Phi^{\dagger} \end{array} 
	\right)  + \right. \nonumber \\
 	&& \left. + \frac{i}{2} \int d^{4}x d^{4}\theta \, 
 	V {\cal M}_{V} V + i(\Phi,J) + i(\Phi^{\dagger},J^{\dagger}) + 
 	i(V,J_{V}) \rule{0pt}{18pt} \right\}
	\label{zgauss}
\end{eqnarray}
one gets 
\begin{eqnarray}
	Z_{0}[J,J^{\dagger},J_{V}] &=& {\rm exp} \left\{  
	\frac{i}{2} \int d^{8}z \, d^{8}z^{\prime} \, (J(z),J^{\dagger}(z)) 
	\Delta_{\Phi}(z,z^{\prime}) \left( \begin{array}{c} 
	J(z^{\prime}) \\ J^{\dagger}(z^{\prime}) \end{array} \right) 
	\right. + \nonumber \\ 
	&+& \left. \frac{i}{2} \int d^{8}z \, d^{8}z^{\prime} \, 
	J_{V}(z) \Delta_{V}(z,z^{\prime}) J_{V}(z^{\prime}) \right\} \, ,
\label{z0prop}
\end{eqnarray}
where superfield propagators, $\Delta_{\Phi}$ and $\Delta_{V}$, have 
been introduced.

\subsection{Superfield propagators}

To compute the propagators the free action must be put in a 
canonical quadratic form. Since the computation leads to different
kinds of  difficulties for chiral and vector fields the two cases will be 
considered separately.

\vspace{0.7cm}
{\sl Chiral superfield propagator.} 
\vspace{0.3cm}

\noindent
The general form of the free action is 
\begin{eqnarray}
	S_{0} &=& \int d^{4}x \left\{ \int d^{4}\theta \, \Phi^{\dagger}_{i}
	\Phi_{i} + \left[ \int d^{2}\theta \, \frac{1}{2} m_{ij} 
	\Phi_{i}\Phi_{j} + \int d^{2}{\overline \theta} \, \frac{1}{2} m_{ij}
	\Phi^{\dagger}_{i}\Phi^{\dagger}_{j} \right] \right\} = \nonumber \\
	&=& \int d^{4}x d^{4}\theta \left[ \Phi^{\dagger}_{i}\Phi_{i} -
	\frac{1}{8} m_{ij} \left( \Phi_{i} \frac{DD}{\Box} \Phi_{j} +
	\Phi^{\dagger}_{i}\frac{{\overline D}{\overline D}}{\Box} 
	\Phi^{\dagger}_{j} \right) \right] \, ,
\label{chiralfreeact}
\end{eqnarray}
where the last line follows from the equivalence $d^{2}\theta 
\leftrightarrow -\frac{1}{4} D^{2}$, 
$d^{2}{\overline \theta} \leftrightarrow  
-\frac{1}{4} {\overline D}^{2}$, valid under a space-time integration, 
and the action of the projection operators
\begin{eqnarray}
	& P_{1} = \displaystyle{\frac{1}{16} 
	\frac{D^{2}{\overline D}^{2}}{\Box}} \, , 
	\hspace{1.5cm} & P_{1}\Phi^{\dagger} = \Phi^{\dagger} \, , \hspace{1cm}
	P_{1}\Phi = 0 \nonumber \\
	& P_{2} = \displaystyle{\frac{1}{16} 
	\frac{{\overline D}^{2} D^{2}}{\Box}} \, ,
	\hspace{1.5cm} & P_{2}\Phi = \Phi \, , \hspace{1cm}
	P_{2}\Phi^{\dagger} = 0 \, .
	\label{p1p2project}
\end{eqnarray}
Hence the action can be rewritten as
\begin{equation}
	S_{0} = \frac{1}{2} \int d^{4}x d^{4}\theta \, 
	\left(\Phi_{i},\Phi^{\dagger}_{i} \right) {\cal M}_{ij} \left(
	\begin{array}{c} \Phi_{j} \\ \Phi_{j}^{\dagger} \end{array} \right) 
	\, , 
\label{chiralgauss}
\end{equation}
where 
\begin{eqnarray*}
   {\cal M}_{ij} = \left( \begin{array}{cc} 
                   -\frac{1}{4}\frac{m_{ij}}{\Box} DD & \delta_{ij}
				   \rule[-6pt]{0pt}{10pt}\\
				   \delta_{ij} \rule{0pt}{16pt}
				   & -\frac{1}{4}\frac{m_{ij}}{\Box} 
				   {\overline D}{\overline D} \end{array} \right) \: .
\end{eqnarray*}
Considering for simplicity the case of one single flavor and taking 
into account the source terms, the exponent of the Gaussian integral 
that yields $Z_{0}$ turns out to be
\begin{eqnarray*}
	i \int d^{4}x d^{4}\theta \, \left[ \frac{1}{2} 
	\left(\Phi,\Phi^{\dagger}\right) {\cal M} \left( 
	\begin{array}{c} \Phi \\ \Phi^{\dagger} \end{array} \right) 
    +\left(\Phi,\Phi^{\dagger}\right) \left(
	\begin{array}{cc} -\frac{1}{4}\frac{D^{2}}{\Box} & 0 \\
		0 & -\frac{1}{4}\frac{{\overline D}^{2}}{\Box} \end{array}
	\right) \left( 
	\begin{array}{c} J \\ J^{\dagger} \end{array} \right)
		\right] \: .
\end{eqnarray*}
A simple but lengthy calculation using $D$-algebra and properties 
of the projection operators allows to write $Z_{0}$ in the form of 
equation (\ref{z0prop}), where the chiral superfield propagator, known 
as {\em Grisaru--Ro\v{c}ek--Siegel propagator} \cite{grisaruroceksiegel}, 
reads
\begin{eqnarray}
	\Delta_{\Phi}(x,\theta,{\overline \theta};x^{\prime},\theta^{\prime},
	{\overline \theta}^{\prime}) &=& \frac{1}{\Box - m^{2}} \left(
	\begin{array}{cc} \frac{m}{4} \frac{D^{2}}{\Box} & 1 \\
		1 & \frac{m}{4} \frac{{\overline D}^{2}}{\Box} \end{array}
	\right) \, \delta_{8}(z-z^{\prime}) 	
\end{eqnarray}
and $\delta_{8}(z-z^{\prime})=\delta_{4}(x-x^{\prime}) 
\delta_{4}(\theta -\theta^{\prime})$.

\vspace{0.7cm}
{\sl Vector superfield propagator.} 
\vspace{0.3cm}

\noindent
The free field action for a vector superfield is
\begin{equation}
	S_{0}  = \int d^{4}x \left\{ \int d^{2}\theta \, \frac{1}{4} W^{\alpha} 
	W_{\alpha} + \int d^{2}{\overline \theta} \, \frac{1}{4} {\overline 
	W}_{\dot \alpha} {\overline W}^{\dot \alpha} + \int d^{4}\theta \, 
	m^{2} V^{2} \right\} \, .
\label{freevect}
\end{equation}
No massive vector field will enter the models considered in the 
following so $m$ will be put to zero from the beginning.

The kinetic operator in (\ref{freevect}) has zero-modes so that just 
like in ordinary gauge theories a gauge fixing procedure {\em \`a la} 
Faddeev--Popov is necessary.
This requires the introduction of a gauge fixing term in 
(\ref{freevect}) and a further integration over the 
Faddeev--Popov ghost fields will appear. 
The gauge-fixed action thus becomes 
\begin{eqnarray}
	&& S = S_{0} + S_{_{{\rm GF}}} + 
	S_{_{{\rm FP}}} =  \nonumber \\
	&& = \int d^{4}x d^{4}\theta \, \left[ \frac{1}{4} 
	W^{\alpha}W_{\alpha} \delta_{2}({\overline \theta}) +
	\frac{1}{4}{\overline W}_{\dot \alpha}{\overline W}^{\dot \alpha}
	\delta_{2}({\theta})\right] + \nonumber \\
	&& + \int d^{4}x d^{4}\theta \, \left[ 
	-\frac{\xi}{8}({\overline D}^{2}V)(D^{2}V) \right] + \nonumber \\
	&& +i \int d^{4}x d^{4}\theta \,(C^{\prime}+{\overline C}^{\prime})
	\, {\cal L}_{\frac{V}{2}}\left[ (C+{\overline C}) +
	\coth {\cal L}_{\frac{V}{2}}(C-{\overline C}) \right] \, , 
\label{vectoraction}
\end{eqnarray}
where $C$, $C^{\prime}$ are chiral superfields for the ghosts and no 
source term has been introduced for such fields. 
The gauge fixing term here introduced corresponds to a class of 
gauges that for the component field $A_{\mu}$ interpolate between the 
Lorentz gauge ($\xi=\infty$) and the Fermi--Feynman gauge ($\xi=1$). 
 
The free propagators for the ghosts are exactly the same as for 
ordinary chiral superfields. Furthermore there is a non polynomial 
interaction with the vector superfield coming from the expansion of 
$\coth({\cal L}_{V/2})$.

The vector superfield propagator is calculated from the quadratic 
part of the action
\begin{eqnarray}
	S &=& S_{0} + S_{_{{\rm GF}}} = \nonumber \\
	&=& \int d^{4}x d^{4}\theta \, \left\{ V\left[ -\Box P_{T} - \xi
	(P_{1}+P_{2})\Box \right] V \right\} = \nonumber \\
	&=& \int d^{4}x d^{4}\theta \, V {\cal M}_{V} V \, . 
\label{vectgauss}
\end{eqnarray}
Introducing the source term for $V$ the generating functional becomes 
\begin{displaymath}
	Z[J_{V}] = \int [{\cal D}V] \, e^{i\int V {\cal M}_{V}V +i(V,J_{V})} 
	\, .
\end{displaymath}
The Gaussian integral can now be performed yielding a result of the 
form (\ref{z0prop}), where the vector superfield propagator is
\begin{eqnarray}
	\Delta_{V}(x,\theta,{\overline \theta};x^{\prime},\theta^{\prime},
	{\overline \theta}^{\prime}) &=& \left[ -\frac{1}{\Box} P_{T} - 
	\frac{\alpha}{\Box} (P_{1}+P_{2}) \right] \delta_{8}(z-z^{\prime}) 
	= \nonumber \\
	&=& -\frac{1}{\Box} \left[ 1+(\alpha -1) (P_{1}+P_{2}) \right]
	\delta_{8}(z-z^{\prime}) \, ,
\label{propvect}
\end{eqnarray}
with $\alpha = \frac{1}{\xi}$.

\subsection{Feynman rules}

Once the free field propagators are known the perturbative expansion 
for the Green functions of the superfields can be derived from the 
generating functional in the form (\ref{zinter}) that yields
\begin{eqnarray}
	&& {\cal G}_{n}(z_{1},\ldots,z_{i},z_{i+1},
	\ldots,z_{j},z_{j+1},\ldots,z_{n}) = \nonumber \\
	&& = \langle 0|T\{ \Phi(z_{1})\ldots\Phi(z_{i})\Phi^{\dagger}(z_{i+1})
	\ldots \Phi^{\dagger}(z_{j})V(z_{j+1})\ldots V(z_{n}) \}|0\rangle =  
	\nonumber \\
	&& = \frac{\delta}{\delta J(z_{1})}\ldots\frac{\delta}
	{\delta J(z_{i})}\frac{\delta}{\delta J^{\dagger}(z_{i+1})}\ldots
	\frac{\delta}{\delta J^{\dagger}(z_{j})}
	\frac{\delta}{\delta J_{V}(z_{j+1})}\ldots\frac{\delta}
	{\delta J_{V}(z_{n})} \cdot \nonumber \\
	&& \left. \cdot \sum_{k=0}^{\infty} \frac{(-i)^{k}}{k!} 
    \left( S_{_{{\rm int}}}\left[ \frac{\delta}{\delta J},
    \frac{\delta}{\delta J^{\dagger}},\frac{\delta}{\delta J_{V}}
    \right] \right)^{k} \, 
    Z_{0}[J,J^{\dagger},J_{V}]\right|_{J=J^{\dagger}=J_{V}=0} 
    \label{greensup2} \: .
\end{eqnarray}
A super Feynman graph technique can be developed which is a 
straightforward generalization of what is done in ordinary field 
theories. Super Feynman rules for the evaluation of one particle 
irreducible Green functions can be derived given the above generating 
functional. The rules can be summarized as follows.
\begin{itemize}
	\item  To each external line is associated the corresponding 
	superfield.

	\item  At each vertex there is an integration over the whole 
	superspace $d^{4}x d^{4}\theta$; factors, as for example powers of the 
	coupling constant, to be associated to the vertices, are read directly 
	from the action.

	\item  A factor of $-\frac{1}{4} {\overline D}^{2}$ or $-\frac{1}{4} 
	D^{2}$ acting on the internal chiral or antichiral lines respectively 
	must be included at each vertex. Such factors come from the 
	definition of the functional derivatives. At purely chiral 
	(antichiral) vertices one factor of $-\frac{1}{4} {\overline D}^{2}$ 
	($-\frac{1}{4} D^{2}$) must be suppressed, because it is used to 
	reconstruct the measure over the whole superspace.

	\item  For internal lines the propagators previously derived are used.

	\item  The specific form of the interaction at each vertex as well 
	as the combinatoric analysis are worked out exactly like in ordinary 
	perturbation theory.
	
	\item  The expressions for the one particle irreducible Green 
	functions in momentum space are obtained by Fourier transforming in 
	the $x$ variables, but not in the fermionic $\theta$ coordinates.
\end{itemize}

\section{General properties of supersymmetric theories}
\label{generalprop}
\fancyhead[LO]{{\footnotesize 1.7~~{\it General properties}}}

Supersymmetry has far reaching consequences in the quantum theory both 
at the perturbative and at the non-perturbative level. Some general 
results will be briefly recorded in this section. More can be 
found in the books and reviews quoted in the references.

Some dramatic effects on the divergences encountered in perturbation 
theory can be easily derived from the Feynman rules. 
Explicit calculation shows that in Feynman diagrams four covariant derivatives
are associated at each vertex, with the exception of vertices 
involving external (anti) chiral lines that have two derivatives less. 
Moreover $\Phi \Phi$ ($\Phi^{\dagger}\Phi^{\dagger}$) propagators with 
momentum $p$ have an additional factor of $\frac{D^{2}}{p^{2}}$  
($\frac{{\overline D}^{2}}{p^{2}}$). Simple dimensional analysis 
allows to compute the superficial degree of divergence $d_{s}$ of a 
diagram which is
\begin{displaymath}
	d_{s} = 4L-2P+2V-C-E-2L \, ,
\end{displaymath}
where $L$ is the number of loops, $P$ the number of internal 
propagators, $V$ the number of vertices, $C$ the number of $\Phi\Phi$ 
or $\Phi^{\dagger}\Phi^{\dagger}$ propagators and $E$ the number of 
external chiral or antichiral lines. Use of the topological constraint 
$L-P+V=1$ yields \cite{divdegree}
\begin{displaymath}
	d_{s} = 2-C-E \, .
\end{displaymath}
For graphs with only external $V$ lines gauge invariance requires the 
presence of four $D$ or ${\overline D}$ factors on the external lines 
so that the result is
\begin{displaymath}
	d_{s} = -C
\end{displaymath}
For diagrams with only lines of a given chirality there are further 
factors of $D^{2}$ or ${\overline D}^{2}$ associated to the external 
lines so that one obtains 
\begin{displaymath}
	d_{s} = 1-C-n \, ,
\end{displaymath}
where $n$ is the number of external chiral or antichiral lines.

This calculation allows to determine all the potentially divergent 
diagrams. The only possible divergences are logarithmic and can arise 
in diagrams with only external $V$ lines with $C=0$ ({\em e.g.} 
corrections to the vector superfield propagator) and in 
$\Phi^{\dagger}\Phi$ propagators with $C=0$. This shows that in 
supersymmetric models only wave function renormalizations may be 
required; there is no direct mass or coupling constant 
renormalization, \ie with ${\rm W}=m\phi^{2}$ one has $Z_{{\rm W}}=1$, 
but $Z_{\phi}\neq 1$ implies $Z_{m}=Z_{\phi}^{-1/2}\neq 1$. 

This and other results can be obtained from the general non 
renormalization theorem for ${\cal N}$=1 theories \cite{n1nonrenormal}:
{\em Any perturbative quantum contribution to the effective action 
$\Gamma$ can be expressed as an integral over the whole superspace 
and is local in the $\theta$ variables}, namely
\begin{eqnarray}
	\Gamma & = & \sum_{n} \int d^{4}x_{1} \ldots d^{4}x_{n} d^{4}\theta 
	\, G(x_{1},\ldots,x_{n}) \cdot 
	\nonumber  \\
	&& \hspace{1.1cm} \cdot 
	f_{n}[\Phi,\Phi^{\dagger},V,D_{\alpha}\Phi,
	{\overline D}_{\dot \alpha}\Phi^{\dagger},D_{\alpha}V,
	{\overline D}_{\dot \alpha}V,\ldots] \, .
\label{gamma}
\end{eqnarray}
The theorem can be easily proved using Feynman rules and successively 
integrating by parts the $\theta$ variables exploiting the properties 
of Grassmannian integration. Explicit calculations in the ${\cal 
N}$=4 supersymmetric Yang--Mills theory showing the mechanism leading 
to this result will be presented in chapter 3, so the proof of the 
above statement will not be discussed here.

Notice that the non renormalization theorem does not forbid in 
principle quantum corrections to the superpotential. It is easy to see 
that a contribution of the form of (\ref{gamma}) can produce a 
correction to the superpotential \cite{nononrenorm1,nononrenorm2}. 
As an example consider a term of the form
\begin{displaymath}
	\int d^{4}x d^{2}\theta d^{2}{\overline \theta} \frac{D^{2}}{16 
	\Box} \Phi^{n} = \int d^{4}x d^{2}\theta \frac{{\overline D}^{2}
	D^{2}}{16 \Box} \Phi^{n} = \int d^{4}x d^{2}\theta \Phi^{n} \, ,
\end{displaymath}
where $\Phi$ is chiral. The right hand side, which follows since 
$P_{1}=\frac{{\overline D}^{2}D^{2}}{16 \Box}$ is the projector on 
chiral superfields, has the form of a correction to the superpotential. 
This example shows that in principle an arbitrary 
correction to the superpotential can be generated perturbatively. 
This effect is related to an infrared singular behaviour that can be 
present only in theories containing massless fields. The example shows 
that the correction, which is non-local when written as an integral 
over the whole superspace, can be rewritten as a local term integrated 
over a subspace. The same construction could not be repeated if the 
theory does not contain massless particles, since in that case one 
would obtain a factor of $\frac{1}{\Box+m^{2}}$, which cannot be 
re-expressed as a subintegral. In \cite{nononrenorm2} it was shown 
that such a correction as the one considered above actually occurs at 
the two-loop level in the Wess--Zumino model. In a ${\cal N}$=1 super 
Yang--Mills theory coupled to the Wess--Zumino model the correction 
appears at one loop \cite{nononrenorm3}. An alternative approach to 
the study of this kind of effects has been given in 
\cite{nononrenorm4,nononrenorm5}.

Other non renormalization theorems hold for theories of extended 
supersymmetry. In particular considerations based on the so called 
{\em anomalies argument} allow to prove the finiteness of ${\cal N}$=2 
theories beyond one loop. More precisely it can be proved that the 
only perturbative contributions to the $\beta$ function of extended 
supersymmetric theories can come from one loop diagrams. This result 
allows to single out a class of finite supersymmetric models and 
among them the ${\cal N}$=4 super Yang--Mills theory that will be 
studied in detail in the following chapters. 

Supersymmetry puts severe restrictions on the non-perturbative 
properties as well. A review of the general results on the 
non-perturbative dynamics of supersymmetric theories is provided by 
\cite{peskin,shifman}. Many astonishing results in this context can 
be traced back to the properties of {\em holomorphy} of supersymmetric 
theories. In supersymmetric models for example the superpotential is a 
holomorphic function of the chiral superfields. Analogously, since the 
parameters of the theories (such as masses and coupling constants) 
can be viewed as expectation values of auxiliary fields, various 
quantities depend analytically on such parameters. Using holomorphy 
properties many exact result have been derived. In \cite{nsvz} an 
expression for the $\beta$ function of supersymmetric gauge theories 
was proposed that is exact at the perturbative as well as at the 
non-perturbative level for a class of models. Also on holomorphy 
relies the construction of the exact solution for the low energy dynamics 
of a class of ${\cal N}$=2 theories proposed by Seiberg and Witten 
\cite{sw}.


\chapter{${\cal N}=4$ supersymmetric Yang--Mills theory}
\label{cap2}
\vspace*{2cm}
\fancyhead[RO,LE]{\thepage}
\fancyhead[RE]{{\footnotesize {\rm Chapter 2.}~~{\it ${\cal N}$=4 
                super Yang--Mills theory}}} 
\fancyhead[LO]{}

\noindent 
Four dimensional ${\cal N}$=4 supersymmetric Yang--Mills theory is a 
very special quantum field theory. It possesses several peculiar 
properties that will be reviewed in this chapter. The general 
properties of the model are discussed here, original results in the 
study of ${\cal N}$=4 Yang--Mills will be presented in the following 
chapters.

The action of the theory was given for the first time in 
\cite{brinkscherkschwarz,gso} within the framework of string theory 
toroidal compactifications. The theory has the maximal amount of 
supersymmetry for a rigid supersymmetric theory in four dimensions, 
namely sixteen real supercharges; a larger number of 
supercharges would require fields of spin larger than one and 
therefore the inclusion of gravity.

Historically the interest in this model was raised by its property of 
finiteness; the $\beta$ function of Gell-Mann and Low has been proved 
to be vanishing in perturbation theory and the same is supposed to be 
true at the non-perturbative level. As a consequence the 
superconformal invariance displayed by the classical theory is 
believed to be preserved after quantization.

Moreover ${\cal N}$=4 super Yang--Mills theory possesses exact 
electric-magnetic duality. It contains beyond the elementary fields 
an infinite set of dyonic states which are believed to realize exactly 
the duality symmetry proposed by Montonen and Olive in 
\cite{olivemont}.

More recently there has been a renewal of interest in ${\cal N}$=4 
Yang--Mills theory following the proposal made by Maldacena in 
\cite{maldacena} of a new duality relating type IIB supergravity in 
$d+1$ dimensional anti-de Sitter space and $d$ dimensional 
(super)conformal theories. This subject will be discussed in detail 
in chapter 5.

The present chapter is organized as follows. In section \ref{formulations} 
the various formulations of the model, which will be employed 
at different stages in the following, are reviewed. Section 
\ref{symmandcurr} presents the classical symmetries and conserved currents 
of the theory. The properties of finiteness are discussed in section 2.3. 
The final section is then devoted to electric-magnetic duality in 
${\cal N}$=4 supersymmetric Yang--Mills theory. The general aspects of 
electric-magnetic duality are reported separately in appendix 
\ref{appb}.

\section{Diverse formulations of ${\cal N}$=4 supersymmetric 
Yang--Mills theory}
\label{formulations}
\fancyhead[LO]{{\footnotesize 2.1~~{\it Formulations of ${\cal N}$=4 SYM}}}

${\cal N}$=4 supersymmetric Yang--Mills theory in four dimensions was 
obtained  for the first time in \cite{brinkscherkschwarz,gso} by 
applying the method of dimensional reduction to ${\cal N}$=1 super 
Yang--Mills in ten dimensions. The latter is the low energy effective 
theory coming from type I superstring theory and describes a ${\cal 
N}=1$ vector multiplet in ten dimensions consisting of one real vector 
and one Majorana--Weyl spinor. The action is~\footnote{Capital Greek 
indices refer to ten dimensional space-time.}
\begin{equation}
	S = \int d^{10} x \, \tr \left\{ -\frac{1}{4} F_{\Gamma \Lambda}
	F^{\Gamma \Lambda} +\frac{i}{2}{\overline \lambda} 
	\Gamma^{\Lambda} D_{\Lambda} \lambda \right\} \, ,
\label{tendimaction}
\end{equation}
where $F_{\Gamma \Lambda}=\partial_{\Gamma} A_{\Lambda} - 
\partial_{\Lambda} A_{\Gamma} + i g [A_{\Gamma},A_{\Lambda}]$ and 
$\lambda$ satisfies the Majorana and Weyl conditions
\begin{displaymath}
	\lambda = C_{(10)} {\overline \lambda}^{T} \, , \hspace{1.5cm} 
	\lambda = \pm \Gamma_{5} \lambda \, ,
\end{displaymath}
where $C_{(10)}$ is the charge conjugation operator. The action 
(\ref{tendimaction}) is invariant under the supersymmetry 
transformations
\begin{eqnarray}
	\delta_{\eta} A_{\Lambda} & = & i{\overline \eta} \Gamma_{\Lambda} 
	\lambda  \nonumber \\
	\delta_{\eta}\lambda & = & \Sigma_{\Gamma\Lambda}
	F^{\Gamma \Lambda} \eta \, .
	\label{tendimsusy}
\end{eqnarray}
In the previous equations capital Greek letters $\Gamma$ and $\Sigma$ 
denote ten dimensional matrices.

Dimensional reduction {\em \'a la} Kaluza--Klein on a six dimensional 
torus $T^{6}$ leads to a multiplet of fields in four dimensions 
possessing an additional SU(4)$\sim$SO(6) global symmetry, which is 
a direct consequence of the ten dimensional Lorentz invariance.
More precisely it is a consequence of the fact that the torus $T^{6}$ 
has a trivial holonomy group. In general in the dimensional reduction 
on a Riemannian manifold of dimension $k$ the spin connection is a 
SO($k$) gauge field and consequently the spinors transform, upon 
parallel transport around a closed contractible curve, under a 
subgroup of SO($k$), which is the holonomy group. In the case at hand 
of $T^{6}$ every spinor is covariantly constant, so that the whole 
SO(6)$\sim$SU(4) becomes a global symmetry of the model. 
From the point of view of the ${\cal N}$=4 theory this 
SU(4) global symmetry is identified with the R-symmetry 
group of the ${\cal N}$=4 supersymmetry algebra, as will be discussed 
in the next section.

In the bosonic sector the compactification gives rise to a four 
dimensional real vector and to six real scalars from the internal 
components of the ten dimensional vector. More precisely one defines 
\begin{eqnarray}
	& A_{\mu}^{a} = A_{\mu}^{a} &\mu=0,1,2,3 \nonumber \\
	& \varphi_{B4}^{a}=\frac{1}{\sqrt{2}} (A^{a}_{B+3}+iA^{a}_{B+6}) 
	\hspace{1cm} & B=1,2,3 \nonumber \\
	& \varphi^{a\, AB} = \frac{1}{2} \varepsilon^{ABC4}\varphi^{a}_{C4}
	= \left(\varphi^{a\,AB}\right)^{*} & 
	A,B,C=1,2,3 \label{bosoncomp} \, .
\end{eqnarray}
In the fermionic sector a suitable choice of the ten-dimensional 
$\Gamma$ matrices allows to write, in Majorana notation, the 
32-component Majorana--Weyl spinor $\lambda$ as
\begin{displaymath}
	\lambda = \left( \begin{array}{c} L\chi^{1} \\ \vdots \\ L\chi^{4} \\
	                  R \widetilde{\chi}_{1} \\ \vdots \\ 
					  R \widetilde{\chi}_{4} \end{array} \right) \, ,
	\hspace{1.5cm} \tilde{\chi}_{A} = C ({\overline \chi}^{A})^{T} 
	\, ,				  
\end{displaymath}
where $L$ and $R$ denote the left and right chirality projection 
operators and $C$ is the four-dimensional charge conjugation operator 
related to the ten-dimensional one by $C_{(10)}=C \otimes \left( 
\begin{array}{cc} 0 & \I_{4} \\ \I_{4} & 0 \end{array} \right)$.  
In four dimensions one obtains four Majorana (or equivalently Weyl) 
spinors 
\begin{equation}
	\lambda^{A} = \left( \begin{array}{c} L\chi^{A} \\ 
	\rule{0pt}{20pt} R \widetilde{\chi}_{A}
	\end{array} \right) \, \hspace{0,5cm} A=1,2,3,4 \, .
\label{fermioncomp}	
\end{equation}
At the end of the day the result of the compactification is the ${\cal 
N}=4$ multiplet which contains six real scalars $\varphi^{a\,AB}$ 
satisfying (\ref{bosoncomp}), one real vector $A^{a}_{\mu}$ 
and four Weyl spinors $\lambda^{a A}_{\alpha}$, all in 
the adjoint representation of the gauge group. The construction here 
described from the ten dimensional theory allows to derive 
immediately the transformation of the fields under the SU(4) global 
symmetry. Because of the constraint (\ref{bosoncomp}) the scalars 
$\varphi^{a\,AB}$ are in the second rank complex self dual {\bf 6} of 
SU(4); the spinors $\lambda^{aA}$ are in the {\bf 4} while the 
${\overline \lambda}_{A}^{a}$ transform in the ${\overline {\bf 4}}$; 
the gauge field $A_{\mu}^{a}$ is a singlet. In the following a 
different notation for the scalar fields will also be used, namely one 
can define 
\begin{displaymath}
	\varphi^{i}=\frac{1}{2} {\overline{t}^{i}}_{AB} 
    \varphi^{AB} \, , \hspace{1.5cm} i=1,2,\ldots,6 \, ,
\end{displaymath}
where $(t_{i})^{AB}$ are Clebsch--Gordan coefficients that  couple 
two {\bf 4}'s to a {\bf 6} (these are six-dimensional generalizations 
of the four-dimensional ${\sigma^{\mu}}_{\alpha {\dot \alpha}}$ 
matrices).

In Weyl notation the action of the four dimensional ${\cal N}$=4 
theory turns out to be 
\begin{eqnarray} 
	&& \hspace{-0.8cm} S = \int d^{4}x \: \tr \Big\{ 
    (D_{\mu}\varphi^{AB})(D^{\mu}{\overline \varphi}_{AB}) 
	-\frac{1}{2}i({\lambda^{\alpha}}^{A} 
	{\dd \ms{D}}_{\alpha{\dot \alpha}} 
	{{\overline{\lambda}}^{{\dot \alpha}}}_{A}) - 
	\frac{1}{4}F_{\mu \nu}F^{\mu \nu} + 
	\nonumber \\ 
 	&& \hspace{-0.7cm} - g {\lambda^{\alpha}}^{A}
 	[{\lambda_{\alpha}}^{B}, {\overline{\varphi}}_{AB}] 
	-g {\overline{\lambda}}_{{\dot \alpha}A}[{{\overline 
    {\lambda}} ^{{\dot \alpha}}}_{B}, \varphi^{AB}] + 2g^{2} 
	[\varphi^{AB},\varphi^{CD}][{\overline{\varphi}}_{AB}, 
	{\overline{\varphi}}_{CD}]\Big\} \, , 
	\label{action} ~~~
\end{eqnarray} 
where $\tr$ denotes a trace over the colour indices and 
$D_{\mu}^{ab} = \delta^{ab} \partial_{\mu} + i g f^{abc} A_{\mu c}$ 
is the covariant derivative in the adjoint representation. 

The action (\ref{action}) is invariant under the supersymmetry 
transformations
\begin{eqnarray} 
	\delta \varphi^{AB} &=& \frac{1}{2} ({\lambda^{\alpha}}^{A}  
    {\eta_{\alpha}}^{B}- 
	{\lambda^{\alpha}}^{B} {\eta_{\alpha}}^{A}) + \frac{1}{2}  
        \varepsilon^{ABCD} 
	{{\overline{\eta}}_{\dot \alpha}}_{C}  
	{\overline \lambda}^{\dot \alpha}_{D} \nonumber \\ 
	\delta {\lambda_{\alpha}}^{A} &=& -\frac{1}{2} F^-_{\mu \nu}  
	{{\sigma^{\mu \nu}}_{\alpha}}^{\beta} {\eta_{\beta}}^{A} + 4i 
        (\ms{D}_{\alpha {\dot \alpha}} \varphi^{AB}) 
        {\overline{\eta}}^{{\dot \alpha}}_{B} - 
	8g[{\overline{\varphi}}_{BC}, \varphi^{CA}] 
        {\eta_{\alpha}}^{B}  \nonumber \\ 
	\delta A^{\mu} &=& -i {\lambda^{\alpha}}^{A}  
	{\sigma^{\mu}}_{\alpha{\dot \alpha}} 
	{{\overline \eta}^{\dot \alpha}}_{A} 
	-i {\eta^{\alpha}}^{A} {\sigma^{\mu}}_{\alpha {\dot \alpha}} 
	{{\overline \lambda}^{\dot \alpha}}_{A} \, ,  
	\label{transs} 
\end{eqnarray} 
which can be easily obtained from the ten dimensional transformations 
(\ref{tendimsusy}) using (\ref{bosoncomp}) and (\ref{fermioncomp}).

A manifestly off-shell formulation of ${\cal N}$=4 supersymmetric 
Yang--Mills theory would require the introduction of an infinite set 
of auxiliary fields as a consequence of the presence of bosonic 
coordinates associated to the twelve central charges in the ${\cal N}$=4 
superspace, as noticed in \cite{siegelrocek,rivelles}. 
A superfield formulation can be constructed using 
analytic superspace, however such an approach is rather involved 
and cannot be applied to explicit calculations because it is 
intrinsically on-shell. An on-shell superfield 
description can be given in terms of a superfield 
$W^{AB}=W^{AB}(x,\theta^{A}_{\alpha},{\overline \theta}_{{\dot 
\alpha}A})$ satisfying \cite{townsenda} the reality condition 
\begin{equation}
	{\overline W}_{A B} =\frac{1}{2} \varepsilon_{A B C D} W^{C D} 
	\label{wrealcond} \, ,  
\end{equation}
together with the constraint, 
\begin{equation}
    {\cal D}_{\alpha}^{A}  W^{BC} = {\cal D}_{\alpha}^{[A} W^{BC]} \, , 
\label{wchiralcond} 
\end{equation}  
where ${\cal D}_\alpha$ is the super-covariant derivative. The 
fields entering the action (\ref{action}) are combined into the lowest 
components of $W^{AB}$.

In the next chapter perturbative calculations in ${\cal N}$=4 Yang--Mills 
will be carried out using a formulation of the model in terms of ${\cal 
N}=1$ superfields. The field content of the theory can be 
obtained by taking one ${\cal N}$=1 vector superfield $V$ and three 
${\cal N}$=1 chiral superfields $\Phi^{I}$ all in the adjoint 
representation of the gauge group. In this approach the six real 
scalars are combined into three complex fields that are the scalar 
components of the chiral superfields $\Phi^{I}$, three of the Weyl 
fermions are the spinors of $\Phi^{I}$ and the fourth fermion is the 
gaugino that together with the real vector constitute the vector superfield 
$V$. As a result only a SU(3)$\times$U(1) subgroup of the original SU(4) 
symmetry is manifest in this formulation. More precisely the 
representations of SU(4) decompose according to {\bf 6} $\rightarrow$ 
{\bf 3}+${\overline {\bf 3}}$, {\bf 4} $\rightarrow$ {\bf 3}+{\bf 1}, 
so that the chiral superfields $\Phi^{I}$ transform in the {\bf 3} of 
SU(3), the antichiral $\Phi^{\dagger}_{I}$ in the ${\overline {\bf 
3}}$ and the vector $V$ is a singlet under SU(3). 

The action in the ${\cal N}$=1 superfield formulation reads
\begin{eqnarray}
	S & = & \frac{1}{d_{{\bf r}}} \tr \left\{ \int d^{4}x \left[ 
	d^{4}\theta \, e^{-gV}\Phi^{\dagger}_{I}e^{gV}\Phi^{I} + 
	\frac{1}{4g^{2}} \left( \int d^{2}\theta \frac{1}{4} W^{\alpha}
	W_{\alpha} + {\rm h.c.} \right) \right. \right. + \nonumber \\
	& + & \left. \left. i g \frac{\sqrt{2}}{3!} \left( \int d^{2}\theta \,  
	\varepsilon_{IJK} \Phi^{I}[\Phi^{J},\Phi^{K}] + \int d^{2}{\overline 
	\theta} \, \varepsilon^{IJK} 
	\Phi^{\dagger}_{I}[\Phi^{\dagger}_{J},\Phi^{\dagger}_{K}] 
	\rule{0pt}{18pt} \right) \right] \right\} \, ,
	\label{n1superfield}
\end{eqnarray}
where $d_{{\bf r}}$ denotes the Dynkin index of the representation to 
which the fields belong. To recover the correct dependance on the 
coupling constant $g$ one must substitute $v\to 2gV$ in the second 
term in (\ref{n1superfield}). 

The action (\ref{n1superfield}) is non-polynomial as is always the 
case in supersymmetric gauge theories in ${\cal N}$=1 superspace. In 
particular the scalar potential of (\ref{action}) comes from the 
superpotential term
\begin{eqnarray*}
	&& \int d^{4}x \left\{ \int d^{2}\theta \, {\rm W}(x,\theta, {\overline 
	\theta}) + {\rm h.c.} \right\} = \\ 
	&& = \int d^{4}x \left\{ \int d^{2}\theta \,\left[ -i\frac{\sqrt{2}}{3!} 
	\varepsilon_{IJK} f^{abc} (\Phi^{I}_{a}\Phi^{J}_{b}\Phi^{K}_{c}) 
	(x,\theta,{\overline \theta}) \right] +{\rm h.c.} \right\}
\end{eqnarray*}

By power expanding $e^{V}$ one obtains 
\begin{eqnarray}
    && \hspace{-0.65cm} S  =  \int d^{4}x\:d^{4}\theta \; 
	\left\{ \rule{0pt}{16pt}  
	V^{a}(x,\theta,{\overline \theta}) \left[ - \Box P_{T} - \xi 
	(P_{1}+P_{2}) \Box \right] V_{a}(x,\theta,{\overline \theta}) + 
	\right. \nonumber \\ 
	&& \hspace{-0.65cm} + {\Phi^{\dagger}}^{a}_{I}(x,\theta,
	{\overline \theta}) 
	\Phi_{a}^{I}(x,\theta,{\overline \theta}) 
	+ i g f_{abc}{\Phi^{\dagger}}^{a}_{I}
	(x,\theta,{\overline \theta})
	V^{b}(x,\theta,{\overline \theta}) \Phi^{Ic}(x,\theta,
	{\overline \theta}) + \nonumber \\
	&& \hspace{-0.65cm}  
	- \frac{1}{2} g^{2}f_{ab}{}^{e}f_{ecd} {\Phi^{\dagger}}^{a}_{I}
	(x,\theta,{\overline \theta})
	V^{b}(x,\theta,{\overline \theta}) V^{c}(x,\theta,{\overline \theta})
	\Phi^{Id}(x,\theta,{\overline \theta}) + \ldots \nonumber \\
	&& \hspace{-0.65cm} - \frac{i}{4} g f_{abc} 
	\left[ {\overline D}^{2} 
	\left( D^{\alpha}V^{a}(x,\theta,{\overline \theta}) \right) 
	\right] V^{b}(x,\theta,{\overline \theta}) 
	\left( D_{\alpha}V^{c}(x,\theta,
	{\overline \theta}) \right) + \nonumber \\
	&& \hspace{-0.65cm} -\frac{1}{8} g^{2} f_{ab}{}^{e}f_{ecd} 
    V^{a}(x,\theta,{\overline \theta}) \left( D^{\alpha}
	V^{b}(x,\theta,{\overline 
	\theta}) \right) \left[ \left({\overline D}^{2} 
	V^{c}(x,\theta,{\overline \theta}) 
	\right) \left( D_{\alpha}V^{d}(x,\theta,{\overline \theta}) \right) 
	\right] \hspace{-0.15cm} + \nonumber \\
	&& \hspace{-0.65cm} +\ldots - \frac{\sqrt{2}}{3!} g f^{abc} 
	\left[ \varepsilon_{IJK}
	\Phi_{a}^{I}(x,\theta,{\overline \theta}) 
	\Phi_{b}^{J}(x,\theta,{\overline \theta}) 
	\Phi_{c}^{K}(x,\theta,{\overline \theta}) \delta({\overline 
	\theta}) + \right. \nonumber \\
	&& \hspace{-0.65cm} + \left. \varepsilon^{IJK} 
	{\Phi^{\dagger}}_{Ia}(x,\theta,{\overline \theta}) 
	{\Phi^{\dagger}}_{Jb}(x,\theta,{\overline \theta}) 
	{\Phi^{\dagger}}_{Kc}(x,\theta,{\overline \theta}) 
	\delta(\theta) \right] + 
	\left( {{\overline C}^{\prime}}_{a}(x,\theta,{\overline \theta})
	C^{a}(x,\theta,{\overline \theta}) + \right. \nonumber \\
	&& \hspace{-0.65cm} \left. - {C^{\prime}}_{a}
	(x,\theta,{\overline \theta})
	{\overline C}^{a}(x,\theta,{\overline \theta}) \right)  
    +\frac{i}{2\sqrt{2}} g f_{abc} \left( {C^{\prime}}^{a}
	(x,\theta,{\overline \theta}) +
	{{\overline C}^{\prime}}^{a}(x,\theta,{\overline \theta}) \right) 
	\cdot \nonumber \\ 
	&& \hspace{-0.65cm} \cdot V^{b}(x,\theta,{\overline \theta}) 
	\left( C^{c}(x,\theta,{\overline \theta}) +
	{\overline C}^{c}(x,\theta,{\overline \theta}) \right) - 
	\frac{1}{8} g^{2} {f_{ab}}^{e}f_{ecd}  \left( 
	{C^{\prime}}^{a}(x,\theta,{\overline \theta}) +
	{{\overline C}^{\prime}}^{a}(x,\theta,{\overline \theta}) 
	\right) \cdot \nonumber \\
	&& \hspace{-0.65cm} \left. \cdot  
	V^{b}(x,\theta,{\overline \theta}) 
	V^{c}(x,\theta,{\overline \theta}) 
	\left( C^{d}(x,\theta,{\overline \theta}) +
	{\overline C}^{d}(x,\theta,{\overline \theta}) \right)  
    + \ldots \rule{0pt}{18pt} \right\} \, ,
\label{actionsuperfields}	
\end{eqnarray}
which of course can be made polynomial by choosing the Wess--Zumino 
gauge which implies $V^{3}=0$. In (\ref{actionsuperfields}) the gauge 
fixing term as well as the corresponding ghost fields have been 
included following the prescription of (\ref{vectoraction}) of chapter 
1 and the expansion has been truncated to the terms that will be 
relevant in future calculations.

Integrating over the $\theta$ variables and eliminating the auxiliary 
fields through the equations of motion results in a different component 
field formulation, which again has a manifest SU(3)$\times$U(1) 
global symmetry. Such a formulation will be employed in the next chapter 
in the discussion of subtleties related to the problem of gauge fixing 
in this model. Choosing the Wess--Zumino gauge one obtains
\begin{eqnarray}
	S & = & \int d^{4}x \, \left\{ -\frac{1}{4} F^{a}_{\mu\nu}
	F^{a\,\mu\nu} - D_{\mu}\varphi^{\dagger}_{I}D^{\mu}\varphi^{I} +
	\frac{i}{2} {\overline \lambda}^{a} \ms{D} \lambda^{a} + \frac{i}{2}
	{\overline \psi}^{a}_{I}\ms{D}\psi^{aI} + \nonumber \right. \\
	&& \hspace{-1.5cm} \left. + ig \sqrt{2} f_{abc} \left(
	{\overline \lambda}^{a}{\overline \varphi}^{b}_{I}\psi^{cI} -
	{\overline \psi}^{a}_{I}\varphi^{bI}\lambda^{c} \right) 
    -\frac{i\sqrt{2}}{2} f_{abc} \left( \varepsilon^{I}_{JK}
	{\overline \psi}_{I}^{a}\varphi^{bJ}\psi^{cK}  
	\right. \right. +  
	\label{n1sym4} \\
	&& \hspace{-1.5cm} \left. \left. - \varepsilon^{IJ}_{K}
	{\overline \psi}_{I}^{a}\varphi^{\dagger \,b}_{J}\psi^{cK}
	\right) -\frac{1}{2} g^{2} (f_{abc}\varphi^{\dagger a}_{I} 
	\varphi^{b}_{I})^{2}+\frac{1}{2}g^{2}f_{abc}f^{a}_{de}
	\varepsilon_{IJK}\varepsilon^{LMK}
	\varphi^{bI}\varphi^{cJ}\varphi^{\dagger d}_{L}
	\varphi^{\dagger e}_{M} \rule{0pt}{18pt} \right\} \, .
	\nonumber
\end{eqnarray} 
This formulation can be easily seen to be perfectly equivalent to 
(\ref{action}) apart from the the manifest R-symmetry, that is only 
SU(3)$\times$U(1).

It is also useful to give a third component field formulation of 
${\cal N}$=4 supersymmetric Yang--Mills theory in terms of ${\cal 
N}=2$ multiplets. This approach will actually be used in instanton 
calculations in chapter 4. The ${\cal N}$=4 super Yang--Mills 
multiplet decomposes in terms of ${\cal N}$=2 multiplets into a 
vector plus a hypermultiplet. In this 
description the subgroup of the original SU(4) R-symmetry that is 
realized explicitly is 
SU(2)$_{{\cal V}}\times$SU(2)$_{{\cal H}}\times$U(1), 
where the subscripts ${\cal V}$ and ${\cal H}$ refer to the vector 
and hypermultiplet respectively: fields in the two multiplets are 
charged only with respect to one of the two SU(2) factors. 
The representations of  SU(2)$_{\cal V}\times$SU(2)$_{\cal H}
\times$U(1) will be denoted by  ({\bf r}$_{{\cal V}}$, 
{\bf r}$_{{\cal H}}$)$_{q}$  with the subscript $q$ referring to 
the $U(1)$ charge  and the following notation for the component 
fields will be used
\begin{eqnarray} 
	{\cal V} ~~~ && \rightarrow ~~~ \lambda^{u} \in {({\bf 2}, 
       {\bf  1})}_{+1}, 
	~~~\varphi \in {({\bf 1}, {\bf 1})}_{+2}, ~~~A_\mu \in {({\bf 1}, 
       {\bf 1})}_{0} \nonumber \\ 
	{\cal H} ~~~ && \rightarrow ~~~ \psi^{{\dot u}} \in {({\bf 1},  
	{\bf 2})}_{-1}, ~~~ q_{u{\dot u}} \in {({\bf 2}, {\bf 2})}_{0} ~,  
\label{entworeps}
\end{eqnarray}  
where $u,{\dot u} = 1,2$. In particular the form of the Yukawa 
couplings in this formulation will be of relevance in the 
following calculations. In ${\cal N}$=2 notation 
the Yukawa couplings read
\begin{eqnarray} 
	{\cal L}^{({\cal N}=2)}_{Y} = \frac{\sqrt{2}}{2} g f_{abc} 
	&& \hspace{-0.8cm} \left\{ 
	q^a_{_S}\left(\lambda^{b \alpha \,u}  
	{\sigma^{S}}_{u {\dot u}} {\psi_{\alpha}}^{{c\dot u}} + 
	{\overline{\psi}}^b_{{\dot \alpha}\,{\dot u}}  
	{\overline{\sigma}}^{S{\dot u} u}  
	{\overline{\lambda}}^{c{\dot \alpha}}_{u} \right) \right. + 
	\nonumber \\ 
	&+& \left. \! \varphi^a \left( \varepsilon^{uv}  
	{\overline{\lambda}}^b_{{\dot \alpha}\,u} 
    {\overline{\lambda}}^{c{\dot \alpha}}_{v} 
	+ \varepsilon_{{\dot u}{\dot v}} \psi^{b\alpha\,{\dot u}}  
	\psi_{\alpha}^{c\,{\dot v}} \right)  \right. + \nonumber \\ 
	&+&  \left. \! {\overline \varphi}^a \left( \varepsilon_{u v} 
	\lambda^{b\alpha \,u} \lambda_{\alpha}^{c\,v} +  
	\varepsilon^{{\dot u}{\dot v}}  
	{\overline{\psi}}^b_{{\dot \alpha}\,{\dot u}} 
	{\overline{\psi}}^{c{\dot \alpha}}_{{\dot v}} \right) \right\}\, ,
\label{yukawa} 
\end{eqnarray} 

In conclusion of this review of the diverse formulations of ${\cal 
N}=4$ super Yang--Mills it is worth recalling that a truly 
off-shell analysis of the theory, in components and in the 
Wess--Zumino gauge, was given in \cite{white} within the framework of 
algebraic renormalization. The problems related to the necessity of 
including an infinite set of auxiliary fields are dealt with using the 
method of Batalin and Vilkoviski. This approach, though technically 
rather involved, allows to prove that all the classical 
symmetries of the theory survive quantization.

\section{Classical symmetries and conserved currents}
\label{symmandcurr}
\fancyhead[LO]{{\footnotesize 2.2~~{\it Symmetries and currents}}}

The field content of ${\cal N}$=4 supersymmetric Yang--Mills theory 
described in the preceding section is unique apart from the choice of 
the gauge group G. In the Abelian case the theory is free as follows 
immediately by observing that all the fields belong to the adjoint 
representation of the gauge group, so that all the covariant 
derivatives reduce to ordinary ones if G is Abelian, and all the 
potential terms are proportional to the structure constants $f_{abc}$. 
In the non Abelian case the theory has a moduli space of vacua 
parameterized by the vacuum expectation values (vev's) of the six real 
scalars, corresponding to a Coulomb phase. The manifold of vacua is 
determined by the condition of vanishing scalar potential ($F$-flatness 
plus $D$-flatness) which reads
\begin{displaymath}
	\tr\left( [\varphi^{AB},\varphi^{CD}][{\overline \varphi}_{AB},
	{\overline \varphi}_{CD}]\right) = 0 \quad \Longleftrightarrow \quad 
	\tr\left([\varphi^{i},\varphi^{j}][\varphi_{i},\varphi_{j}]\right) 
	= 0 \, .
\end{displaymath}
By writing $\varphi^{i}$ explicitly in terms of generators of the 
adjoint representation one obtains
\begin{equation}
	\sum_{i,j}^{1,6} \sum_{a=1}^{{\rm dim\,G}}\varphi^{2}_{ia}
	\varphi_{jb}^{2} \, \tr\left([T^{a},T^{b}]\right)^{2} = 0 \, ,
	\label{vacuum}
\end{equation}
so that for a rank $r$ gauge group G at most $r$ components 
$\varphi_{ia}$ for each field can have non vanishing vacuum expectation 
value. As a result the space of vacua is parameterized by $6r$ real 
coordinates. The moduli space is 
\begin{equation}
	{\cal M} = {\mathbb R}^{6r}/{\cal S}_{r} \, ,
	\label{moduli}
\end{equation}
where ${\cal S}_{r}$ is the group of permutations of $r$ elements, 
because configurations in which two or more components of 
$\varphi^{i}$ are exchanged are physically indistinguishable. In 
conclusion the manifold ${\cal M}$ is a $6r$-dimensional real space 
with {\em orbifold type singularities}. 

In a generic point of the Coulomb phase with nonvanishing vev's for the 
scalar fields the gauge group G is broken to U(1)$^{r}$, so that 
the theory is again Abelian and thus free. On the contrary the origin 
of the moduli space, where the vev's of all the scalars vanish, 
corresponds to a phase that is expected to describe a highly non 
trivial superconformal field theory. Note that the superconformal 
phase corresponds to an orbifold singularity of the manifold  ${\cal M}$.
In this phase the classical action 
(\ref{action}) is actually invariant under a larger group of global 
transformations than that generated by the super Poincar\'e algebra 
(\ref{susyalg}), namely the ${\cal N}$=4 superconformal group.

The ${\cal N}$-extended superconformal algebra in four dimensions is 
an extension of (\ref{susyalg}) which includes additional generators 
$D$, $K_{\mu}$, $S^{i}_{\alpha}$ and ${\overline S}_{{\dot \alpha}i}$ 
corresponding respectively to dilatations, special conformal 
transformations and the associated special supersymmetry 
transformations. 
The complete algebra further includes the following (anti)commutation 
relations, supplementing (\ref{susyalg})
\begin{equation}
\begin{array}{ll}
    \begin{displaystyle}
	\mbox{}\rule{0pt}{22pt}[M_{\mu\nu},K_{\lambda}] = 
	\eta_{\nu \lambda}K_{\mu} - 
	\eta_{\mu\lambda}K_{\nu} \end{displaystyle}  & \begin{displaystyle}  
    \end{displaystyle} \\
	\begin{displaystyle}
	\mbox{}\rule{0pt}{22pt}[D,P_{\mu}] = -P_{\mu} 
    \end{displaystyle}  & \hspace{-2.5cm} \begin{displaystyle}  
	[D,K_{\mu}] = K_{\mu} \end{displaystyle} \\
    \begin{displaystyle}
	\mbox{}\rule{0pt}{22pt}[P_{\mu},K_{\nu}] = 
	-2M_{\mu\nu}+2\eta_{\mu\nu}D 
	\end{displaystyle}  & \hspace{-2.5cm} \begin{displaystyle}  
	[K_{\mu},K_{\nu}] = 0 \end{displaystyle} \\
	\begin{displaystyle}
	\mbox{}\rule{0pt}{22pt}
	[S_{\alpha i},M^{\mu\nu}]=(\sigma^{\mu\nu})_{\alpha}{}^{\beta}
	S_{\beta i} \end{displaystyle}  & \hspace{-2.5cm} 
	\begin{displaystyle}  [{\overline S}_{\dot 
	\alpha}^{i},M^{\mu\nu}]=
	({\overline \sigma})^{\dot \alpha}{}_{\dot \beta}
	{\overline S}^{{\dot \beta}i} \end{displaystyle} \\
	\begin{displaystyle}
	\mbox{}\rule{0pt}{22pt}\{S_{\alpha i},
	{\overline S}_{\dot \alpha}^{j}\} = 
	2\sigma^{\mu}_{\alpha{\dot \alpha}}K_{\mu}\delta_{i}^{j} 
    \end{displaystyle} & \hspace{-2.5cm} \begin{displaystyle}   
    \end{displaystyle} \\
	\begin{displaystyle}
	\mbox{}\rule{0pt}{22pt}\{S_{\alpha i},
	S_{\beta j}\} = 0 \end{displaystyle}  & \hspace{-2.5cm} 
	\begin{displaystyle}  \{{\overline 
	S}_{\dot \alpha}^{i},{\overline S}_{\dot \beta}^{j}\} = 0 
    \end{displaystyle} \\
	\begin{displaystyle}
	\mbox{}\rule{0pt}{22pt}[Q_{\alpha}^{i},D]=
	\frac{1}{2}Q_{\alpha}^{i} \end{displaystyle}  & \hspace{-2.5cm} 
	\begin{displaystyle}  
	[{\overline Q}_{{\dot \alpha}i},D]=\frac{1}{2}
	{\overline Q}_{{\dot \alpha}i} \end{displaystyle} \\
	\begin{displaystyle}
	\mbox{}\rule{0pt}{22pt}[S_{\alpha i},D]=
	-\frac{1}{2}S_{\alpha i} \end{displaystyle}  & \hspace{-2.5cm}
	\begin{displaystyle}  
	[{\overline S}_{\dot \alpha}^{i},D]=\frac{1}{2}
	{\overline S}_{\dot \alpha}^{i} \end{displaystyle} \\ 
	\begin{displaystyle}
	\mbox{}\rule{0pt}{22pt}[Q_{\alpha}^{i},
	K^{\mu}]=i\sigma^{\mu}_{\alpha{\dot 
	\alpha}}{\overline S}^{{\dot \alpha}i} \end{displaystyle} 
	& \hspace{-2.5cm} \begin{displaystyle}  [{\overline 
	Q}^{\dot \alpha}_{i},K^{\mu}]=
	-i{\overline \sigma}^{\mu\,{\dot \alpha}\alpha}S_{\alpha i} 
    \end{displaystyle} \\
	\begin{displaystyle}
	\mbox{}\rule{0pt}{22pt}[S_{\alpha i},
	K^{\mu}]=0 \end{displaystyle}  & \hspace{-2.5cm} \begin{displaystyle}  
	[{\overline 
	S}^{{\dot \alpha}i},K^{\mu}]=0 \end{displaystyle} \\
	\begin{displaystyle}
	\mbox{}\rule{0pt}{22pt}[S_{\alpha i},P^{\mu}]=
	-i\sigma^{\mu}_{\alpha{\dot 
	\alpha}}{\overline Q}^{\dot \alpha}_{i} \end{displaystyle}  & 
	\hspace{-2.5cm} \begin{displaystyle}  [{\overline 
	S}^{{\dot \alpha}i},P^{\mu}]=i{\overline \sigma}^{\mu\,{\dot 
	\alpha}\alpha}Q_{\alpha}^{i} \end{displaystyle} \\
	\begin{displaystyle}
	\mbox{}\rule{0pt}{22pt}[S_{\alpha i}, 
	T^{a}]  = - B^{a}_{i}{}^{j} S_{\alpha j}  
	\end{displaystyle}  & \hspace{-2.5cm} \begin{displaystyle}  
	[T^{a},{\overline S}_{\dot \alpha}^{i}] = - B^{\dagger a i}{}_{j} 
	{\overline S}_{\dot \alpha}^{j} \end{displaystyle} \\
	\begin{displaystyle}
	\mbox{}\rule{0pt}{22pt}\{Q_{\alpha}^{i},
	S_{\beta j}\} = 2\varepsilon_{\alpha\beta}\delta^{i}_{j}D-
    i(\sigma^{\mu\nu})_{\alpha}{}^{\gamma}\varepsilon_{\gamma\beta}
    M_{\mu\nu}\delta^{i}_{j} - 
    4i\varepsilon_{\alpha\beta}\delta^{i}_{j}A
    +B^{a i}{}_{j}T_{a}\end{displaystyle} \hspace{-2cm} &  \\
	\begin{displaystyle}
	\mbox{}\rule{0pt}{22pt}
	\{{\overline Q}_{{\dot \alpha}i}, 
	{\overline S}_{\dot \beta}^{j}\} = 2\varepsilon_{{\dot \alpha}{\dot 
	\beta}}\delta_{i}^{j}D-i\varepsilon_{{\dot \alpha}{\dot \gamma}}
    ({\overline \sigma}^{\mu\nu})^{\dot \gamma}{}_{\dot \beta} 
    M_{\mu\nu}\delta_{i}^{j}+4i\varepsilon_{{\dot \alpha}{\dot \beta}}
    \delta_{i}^{j}A+
	B^{\dagger \,a}_{i}{}^{j}T_{a}
    \end{displaystyle} \hspace{-2cm} & \\ 
	\begin{displaystyle}
	\mbox{}\rule{0pt}{22pt}[Q_{\alpha}^{i},A]=
	-i\left(\begin{displaystyle}
	\frac{4-{\cal N}}{4{\cal N}} \end{displaystyle} \right)
	Q_{\alpha}^{i} \end{displaystyle}  & \hspace*{-2.5cm}
	\begin{displaystyle}  [{\overline Q}_{{\dot \alpha}i},A]=
	i\left(\frac{4-{\cal N}}{4{\cal N}} \right) {\overline Q}_{{\dot 
	\alpha}i} \end{displaystyle} \\
	\begin{displaystyle}
	\mbox{}\rule{0pt}{22pt}[S_{\alpha i},A]=
	i\left(\begin{displaystyle}
	\frac{4-{\cal N}}{4{\cal N}}\end{displaystyle} \right)
	S_{\alpha i} \end{displaystyle}  & \hspace*{-2.5cm}
	\begin{displaystyle}  [{\overline S}_{\dot \alpha}^{i},A]=
	-i\left(\frac{4-{\cal N}}{4{\cal N}} \right) 
	{\overline S}_{\dot \alpha}^{i}\end{displaystyle} \, . 
\end{array}
\label{supconfalg}
\end{equation}
The operator $A$ in these relations is the generator of the U(1) 
factor of the group U(${\cal N}$)=SU(${\cal N}$)$\times$U(1) of the 
automorphisms of the algebra. The ${\cal N}$=4 case is singular since 
it implies 
\begin{displaymath}
	[Q_{\alpha}^{i},A]=[{\overline Q}_{{\dot \alpha}i},A]=0 \, .
\end{displaymath}
 
The chiral rotation operator $A$ must therefore have the same action 
on all the states in the multiplet. This implies in particular that it 
must act in the same way on states of opposite helicities 
$-\frac{1}{2}$ and $\frac{1}{2}$, which is possible only if $A$ is 
actually zero. For this reason the R-symmetry group in the ${\cal 
N}=4$ case is SU(4) and not U(4) as would be expected from general 
arguments. 

As already noticed in section \ref{formulations} a SU(4)$\sim$SO(6) 
global symmetry, that is identified with the R-symmetry, naturally 
arises in the compactification on $T^{6}$ that leads to ${\cal N}$=4 
super Yang--Mills in $D=4$ from the ten dimensional ${\cal N}$=1 
theory. In this context the reduction of the global symmetry from 
U(4) to SU(4) is understood as a consequence of the lack of a real six 
dimensional representation in U(4). More precisely it follows from 
the form of the Yukawa coupling ${\lambda^{\alpha}}^{A}
[{\lambda_{\alpha}}^{B}, {\overline{\varphi}}_{AB}]$ in the action 
(\ref{action}). In fact a transformation, under the U(1) factor 
contained in U(4), of the type $\lambda \rightarrow 
e^{i\alpha}\lambda$ for the spinor would require 
$\varphi^{AB}\rightarrow e^{-2i\alpha}\varphi^{AB}$ for the scalars 
which is prohibited by the reality condition (\ref{bosoncomp}). 

The Noether currents associated with the superconformal 
transformations together with those corresponding to chiral SU(4) 
R-transformations are components of a unique on-shell multiplet 
\cite{fzsupercurr}. The complete multiplet of 
currents was given for the Abelian case in \cite{currents}. The 
conserved currents are
\begin{eqnarray} 
	{\cal T}^{\mu \nu} &=& \frac{1}{2} [\delta^{\mu \nu} 
	(F^{-}_{\rho \sigma})^{2}-4{{F^{-}}^{\mu}}_{ \rho} {F^{-}}^{\nu  
    \rho} +  
	\mbox{ h.c.}] - \frac{1}{2} \lambda^{\alpha A} 
	{\sigma^{( \mu}}_{\alpha {\dot \alpha}} 
	\dd \partial^{\nu )}{\overline{\lambda}}^{{\dot \alpha}}_{A} 
	\nonumber \\ 
	&+& \delta^{\mu \nu} (\partial_{\rho} 
	{\overline{\varphi}}_{AB})(\partial^{\rho} \varphi^{AB})  
	-2(\partial^{\mu}{\overline{\varphi}}_{AB}) 
	(\partial^{\nu} \varphi^{AB})  \nonumber \\ 
	&-& \frac{1}{3} (\delta^{\mu \nu} \Box 
	- \partial^{\mu }\partial^{\nu})({\overline{\varphi}}_{AB}  
	\varphi^{AB})  \nonumber \\ 
	{\Sigma^{\mu}}_{\alpha A} &=& -\sigma^{\kappa \nu} 
    F^{-}_{\kappa \nu}  
	{\sigma^{\mu}}_{\alpha {\dot \alpha}}  
	{{\overline \lambda}^{{\dot \alpha}}}_{A} + 
	2i {\overline \varphi}_{AB} \dd \partial^{\mu} 
	{\lambda_{\alpha}}^{B} + \frac{4}{3}i 
	{{\sigma^{\mu \nu}}_{\alpha}}^{\beta}  
	\partial_{\nu}({\overline \varphi}_{AB} {\lambda_{\beta}}^{B})  
    \nonumber \\ 
	{{{\cal J}^{\mu}}_{A}}^{B} &=& {\overline{\varphi}}_{AC} 
    \dd \partial^{\mu}  
	\varphi^{CB} +  {\overline \lambda}_{{\dot \alpha} A} 
	{\overline \sigma}^{\mu{\dot \alpha}\alpha}  
	{{\lambda}_{\alpha}}^{B} 
	-\frac{1}{4} {\delta_{A}}^{B} \lambda^{\alpha C} 
	{\sigma^{\mu}}_{\alpha{\dot \alpha}}  
	{{\overline{\lambda}}^{\dot \alpha}}_{C} \label{currentdef1} \, .
\end{eqnarray}
The remaining components of the supermultiplet are obtained by 
supersymmetry using the equations of motion and read
\begin{eqnarray}	 
	{\cal C} &=& ({F^{-}}_{\mu \nu})^{2} \nonumber \\ 
	{{\hat \Lambda}_{\alpha}}^{A} &=& 
	-{{\sigma^{\mu \nu}}_{\alpha}}^{\beta}  
	{F^{-}}_{\mu \nu} {\lambda_{\beta}}^{A}  \nonumber \\ 
	{\cal E}^{AB} &=& {\lambda^{\alpha}}^{A}  
	{\lambda_{\alpha}}^{B} \nonumber \\ 
	{{\cal B}_{\mu \nu}}^{AB} &=& \lambda^{\alpha A}  
	\sigma_{\mu \nu\alpha}{}^{\beta} 
	{\lambda_{\beta }}^{B} + 2i {\varphi}^{AB} {F^{-}}_{\mu \nu} 
     \label{currentdef2} \\ 
	{\hat \chi}_{\alpha\, AB}^C &=& \frac{1}{2}  
     \varepsilon_{ABDE}(\varphi^{DE}  
	{\lambda_{\alpha}}^{C}+\varphi^{CE} {\lambda_{\alpha}}^{D})  
     \nonumber \\ 
	{{\cal Q}^{AB}}_{CD} &=& \varphi^{AB} {\overline{\varphi}}_{CD} -  
    \frac{1}{12}{\delta^{A}}_{[C} {\delta^{B}}_{D]} \varphi^{EF}  
	{\overline{\varphi}}_{EF} \nonumber \;  . 
\end{eqnarray}  
In (\ref{currentdef1}) ${\cal T}_{\mu\nu}$ is the (improved) energy 
momentum tensor, $\Sigma^{\mu}_{\alpha A}$ are the spin $\frac{1}{2}$ 
supersymmetry currents and ${\cal J}^{\mu}_{A}{}^{B}$ are the SU(4) 
currents. ${\cal T}_{\mu\nu}$ is a singlet under SU(4) whereas 
$\Sigma^{\mu}_{\alpha A}$ transforms in the ${\overline {\bf 4}}$ and 
${\cal J}^{\mu}_{A}{}^{B}$ in the {\bf 4}$\times {\overline {\bf 4}}$. 
The multiplet further includes three scalars ${\cal C}$, 
${\cal E}^{(AB)}$ and ${\cal Q}^{ij}$, two fermionic spin 
$\frac{1}{2}$ components ${\hat \chi}_{\alpha\, AB}^C$ and 
${\hat \Lambda}_{\alpha}^{A}$ and one antisymmetric tensor 
${\cal B}_{\mu \nu}^{[AB]}$. Their transformation under SU(4) is as 
follows: the scalars ${\cal C}$, ${\cal E}^{(AB)}$ and ${\cal 
Q}^{ij}$ are respectively in the {\bf 1}, {\bf 10} and {\bf 20}, the 
two fermions ${\hat \chi}_{\alpha\, AB}^C$ and 
${\hat \Lambda}_{\alpha}^{A}$ belong to the 
{\bf 4}$\times {\bf 6}$ and {\bf 4} respectively 
and ${\cal B}_{\mu \nu}^{[AB]}$ is in the {\bf 6}.

All of these fields can be obtained as lowest components of an 
on-shell superfield ${\cal W}_{(2)}^{ij}$ which is a bilinear in 
$W^i \equiv \frac{1}{2}t_{AB}^i W^{AB}$
\begin{displaymath}
	{\cal W}_{(2)}^{ij} = \tr \left( W^i W^j - \frac{\delta^{ij}}{6} 
	W_k W^k \right) \, .
\end{displaymath}
In the non Abelian case in (\ref{currentdef1}) and (\ref{currentdef2}) 
there is a trace over the colour indices and additional terms are present. 
Such terms comprise beyond those required for the covariantization of all 
the derivatives also `potential' terms. For example the complete 
non Abelian expression for ${\cal E}^{(AB)}$ is 
\begin{equation} 
	{\cal E}^{(AB)} = \tr \left( \lambda^{\alpha A} 
	{\lambda_{\alpha}}^{B} \right) + g \, t_{[ijk]}^{(AB)_+} 
	\tr \left( \varphi^{i}\varphi^{j}\varphi^{k} \right) \, ,
	\label{enonabelian}
\end{equation}
where $t_{[ijk]}^{(AB)_+}$ is the completely antisymmetrized product 
of three $t_{i}^{AB}$ matrices.

\section{Finiteness of ${\cal N}$=4 supersymmetric Yang--Mills theory}
\label{finiteness}

The first argument suggesting the possibility that the ${\cal N}$=4 
supersymmetric Yang--Mills theory could be a finite theory at the 
quantum level, free of ultraviolet as well as infrared divergences, 
was given in \cite{sym1}. In that paper within the study of the general 
problem of the coupling of scalar (chiral) multiplets to vector 
multiplets it was observed that a number of three chiral multiplets 
in a supersymmetric Yang--Mills theory would lead to a Gell-Mann and 
Low $\beta$ function vanishing in the one-loop approximation. In fact 
for a gauge group G=SU($N$) the one loop $\beta$ function is 
given by 
\begin{displaymath}
	\beta = -\frac{g^{2}}{16 \pi^{2}} (3-n) N \, ,
\end{displaymath}
where $n$ is the number of chiral multiplets coupled to the vector 
multiplet in the ${\cal N}$=1 language.

Successively this result was improved in \cite{2loop}, where the same 
was shown to hold at the two loop level by an explicit 
computation in the component field formulation. Further improvement of 
this result beyond two loops proves extremely complicated using 
ordinary techniques based on Feynman rules for the component theory.
There exist a computer calculation \cite{tarasov} of the 
three-loop $\beta$ function performed by using ordinary Feynman rules 
which gives a vanishing result as well.
\fancyhead[LO]{{\footnotesize 2.3~~{\it Finiteness of ${\cal N}$=4 SYM}}}

Superspace techniques allow to derive this result in a much simpler 
way as was shown in \cite{grisaruroceksiegel,grirocsieg2,caswzanon}. 
The basic ingredient is the observation that a vertex coupling three 
chiral superfields is finite as a direct consequence of dimensional 
analysis. By applying the rules of section \ref{generalprop} the 
superficial degree of divergence of a Green function with three 
external (anti) chiral superfields is negative, $D=2-3=-1$, 
{\em i.e.} the diagram is finite. 
This means that the $\beta$ function is completely determined by the 
wave function renormalization  constant calculated from the divergent 
part of the propagator of, say, the chiral superfield. The equation for 
the $\beta$ function \cite{caswzanon} is 
\begin{displaymath}
	\beta(g) = \frac{3}{2} g^{2} \frac{\partial Z^{(1)}_{g}}{\partial 
	g} \, ,
\end{displaymath}
where $Z^{(1)}_{g}$ is the coefficient of the simple pole term in the 
Fourier transform of the chiral superfield propagator. Calculations 
presented in the papers of references \cite{grirocsieg2,caswzanon} 
show that the propagator of the chiral superfield is finite, in the 
supersymmetric generalization of the Fermi--Feynman gauge ($\alpha$=1 
in the notation of section \ref{quantization}), up to three loops, 
implying the vanishing of the $\beta$ function at the same order in 
perturbation theory. 

The lack of manifest ${\cal N}$=2 supersymmetry in the calculations 
described up to now prevents from using the non 
renormalization theorem for ${\cal N}$=2 in order to extend the 
result to every order. A formulation based on unconstrained ${\cal 
N}=2$ superfields (which is basically a version of harmonic 
superspace) was developed in \cite{n2finite}; such an approach allows 
a description of ${\cal N}$=4 super Yang--Mills with manifest 
${\cal N}$=2 supersymmetry. The finiteness of the theory at all orders 
in perturbation theory is then a consequence of the ${\cal N}$=2 non 
renormalization theorem once the absence of divergences is proved 
at one loop. 

Another proof of the finiteness of the theory was given in 
\cite{mandelstam} using a light-cone superspace formulation. The 
advantage of this formulation relies on the fact that only physical 
fields are involved so that no auxiliary fields are necessary, 
however the lack of Lorentz invariance introduces other 
difficulties and makes the interpretation of the results less 
transparent.

A different approach based on anomaly arguments has been presented 
in \cite{sohniuswest}. The vanishing of the $\beta$ function is 
obtained as a consequence the superconformal invariance since (see 
\cite{adler}) 
\begin{displaymath}
	\frac{\beta(g)}{g} F_{\mu\nu}^{a}F^{\mu\nu}_{a} = 
	{\cal T}_{\mu}^{\mu} \, ,
\end{displaymath}
so that invariance under dilatations, implying the vanishing of ${\cal 
T}_{\mu}{}^{\mu}$, immediately gives $\beta(g)=0$. The problem is 
therefore the proof of superconformal invariance, which is obtained as 
a consequence of the SU(4) symmetry being preserved at the quantum 
level. As already remarked a different demonstration of superconformal 
symmetry was given in \cite{white} using somewhat more general 
arguments.

All of the results here discussed either apply to gauge invariant 
quantities or are obtained avoiding possible subtleties related to the 
choice of the gauge. As will be discussed in the next chapter subtleties 
are actually present in the calculation of non gauge invariant quantities 
such as Green functions both in the component field formulation and in 
the description in terms of ${\cal N}$=1 superfields.

\section{S-duality}
\label{sduality}
\fancyhead[LO]{{\footnotesize 2.4~~{\it S-duality}}}

During the last few years the concept of duality has been the basic 
ingredient for striking results in the analysis of the non-perturbative 
behaviour of supersymmetric gauge theories as well as of superstring 
theories. The fundamental idea of duality is that of a transformation 
relating two different and somehow complementary descriptions of a 
physical system. The r\^ole that duality  plays in the study of the 
non-perturbative properties of a physical model relies on the 
possibility of constructing a transformation that maps the 
non-perturbative regime of the system into the perturbative regime 
of the `dual' system. Examples of this kind of 
duality have been known for a long time; in particular the duality 
shown by Kramers and Wannier \cite{kramerswannier} in the case of 
the 2-D Ising model and that established between the Sine-Gordon and 
the Thirring models \cite{sgthirring1,sgthirring2} appear very 
representative.

The two dimensional Ising model is defined by a set of spin 
variables $\sigma_{i}$ taking values $\pm 1$ and living on a two 
dimensional square lattice. The interaction is restricted to nearest 
neighbours and the partition function reads
\begin{equation}
	Z(K) = \sum_{\{\sigma\}} \exp \left( K \sum_{(i,j)} \sigma_{i}\sigma_{j}
	\right) \, ,
\label{fpartising}
\end{equation}
where $K$ is related to the temperature $T$ and the ferromagnetic 
coupling constant $J$ through $K=J/k_{B}T$ ($k_{B}$ being Boltzmann's 
constant). In (\ref{fpartising}) the sum on $\{\sigma\}$ is over all 
spin configurations and the sum on $(i,j)$ is over all pairs of 
nearest neighbours. The model was explicitly  solved by Onsager and 
exhibits a single phase transition to a ferromagnetic state at a 
critical temperature $T_{c}$. Kramers and Wannier computed the 
critical temperature by using duality arguments, before the exact 
solution of the model was found. The crucial point is that the 
theory can be reformulated in a different (``dual'') way 
by writing the partition function as a sum over the 
elementary plaquettes of a dual lattice whose sites are the centers 
of the plaquettes of the original one, with a new coupling $K^{*}$. 
The two descriptions are then equivalent, namely $Z^{*}(K^{*})=Z(K)$, 
if the couplings are related by 
\begin{displaymath}
	\sinh 2K^{*}=\frac{1}{\sinh 2K} \, .
\end{displaymath}
Notice that this implies that the high temperature (or weak coupling) 
regime, $K\ll 1$, of the original description is mapped by the duality 
transformation to the low temperature (or strong coupling) regime, 
$K^{*}\gg 1$, of the dual system. The computation 
of the critical temperature is based on the hypothesis that the system 
has a single transition point. The critical point must then be the 
self-dual point of the transformation, $K=K^{*}$, which gives 
$\sinh [2J/k_{B}T_{c}]=1$.

A second interesting example of duality occurs in two dimensional 
relativistic field theory and relates the Sine-Gordon and the 
Thirring models. The Sine-Gordon model is defined by the 2-D action
 \begin{equation}
	S_{_{{\rm SG}}}=\int d^{2}x \, \left[\frac{1}{2} 
	\partial_{\mu}\phi\partial^{\mu}\phi+\frac{\alpha}{\beta^{2}}
	(\cos \beta \phi -1) \right] \, ,
	\label{asinegord}
\end{equation}
where $\phi$ is a scalar field and the parameter $\beta$ 
plays the r\^ole of a coupling constant as 
can be established by power expanding the potential term.
The spectrum of the theory contains `meson' excitations of mass 
$M_{m}=\sqrt{\alpha}$ and solitonic states with mass 
$M_{s}=\frac{8\sqrt{\alpha}}{\beta^{2}}$. Therefore the solitonic 
degrees of freedom have large mass in the weak coupling regime. 
It can be shown that the Sine-Gordon model is completely equivalent 
to the Thirring model which describes a system of interacting fermions. 
The action of the latter is 
\begin{equation}
	S_{_{{\rm T}}}=\int d^{2}x \, \left[ {\overline 
	\psi}i\gamma_{\mu}\partial^{\mu}\psi +m{\overline \psi} \psi - 
	\frac{g}{2}{\overline \psi}\gamma_{\mu}\psi {\overline \psi}
	\gamma^{\mu}\psi \right] \, .
	\label{athirring}
\end{equation}
$S_{_{{\rm T}}}$ can be mapped into the action (\ref{asinegord}) through 
bosonization \cite{sgthirring1,sgthirring2}. More precisely it can be 
proved that the duality transformation maps the soliton of the 
Sine-Gordon model into the elementary fermion of the Thirring model 
and the meson states of (\ref{asinegord}) into fermion--anti-fermion 
bound states. The relation between the coupling constants is 
\begin{displaymath}
	\frac{\beta^{2}}{4\pi} = \frac{1}{\begin{displaystyle}1+\frac{g}{\pi}
	\end{displaystyle}} \: ,
\end{displaymath}
so that in this case as well duality establishes a strong/weak 
coupling correspondence. 

The crucial r\^ole that duality plays in the context of supersymmetric 
gauge theories relies on this fundamental feature. The kind of 
duality that is relevant in four dimensional quantum field theories is 
a generalization of the electric-magnetic duality introduced for the 
first time by Dirac in \cite{dirac}. The general features of 
electric-magnetic duality are reviewed in appendix \ref{appb}, 
where the case of the Georgi--Glashow (or Yang--Mills--Higgs) model 
is described. The application to the ${\cal N}$=4 super 
Yang--Mills theory is discussed in this section.
There are various reviews on electric-magnetic duality, see in particular  
\cite{oliverev,harvey,divecchia}.

\subsection{Montonen--Olive and SL(2,$\mathbb{Z}$) duality}
\label{montonenoliveduality}

Most of the analysis of this section will be carried out for a SU(2) 
gauge group, but the results can be generalized to larger groups.
The classical spectrum of non Abelian gauge theories can contain 
non-perturbative states associated with solitonic solutions of the 
equations of motion. The corresponding field configurations are 
characterized by a topological charge that is interpreted as a 
magnetic charge, so that these states describe monopoles and in general 
dyons  (states with both electric and magnetic charge) 
\cite{goddardolive}.
As discussed in appendix \ref{appb} the Georgi--Glashow model with 
Lagrangian
\begin{displaymath}
	{\cal L}=-\frac{1}{4}F^{a}_{\mu\nu}F^{a\mu\nu}+
	\frac{1}{2}D^{\mu}\Phi^{a}D_{\mu}\Phi_{a}-
	\frac{\lambda}{4}\left(\Phi^{a}\Phi_{a}-v^{2}\right)^{2} \, ,
\end{displaymath}
possesses, in the (BPS) limit of vanishing potential, a `classical' 
spectrum, containing one scalar massive Higgs field, one massless 
`photon', massive $W^{\pm}$ bosons and monopoles $M^{\pm}$, that is 
summarized in the following table
\vspace{0.5cm}

\begin{center}
\begin{tabular}{|l|c|c|c|}
	\hline
	 & ~~~{\rm Mass}~~~ & ~~~($Q_{e},Q_{m}$)
	 \rule[-7pt]{0pt}{21pt}~~~ & ~~~{\rm Spin}~~~  \\
	\hline \hline
	~~{\rm Higgs} \rule[-7pt]{0pt}{21pt} & 0 & (0,0) & 0  \\
	\hline
	~~Photon \rule[-7pt]{0pt}{21pt} & 0 & (0,0) & 1  \\
	\hline
	~~$W^{\pm}$ \rule[-7pt]{0pt}{21pt} & $ve$ & ($\pm e$,0) & 1  \\
	\hline
	~~$M^{\pm}$ \rule[-7pt]{0pt}{21pt} & $vg$ & (0,$\pm g$) & 0  \\
	\hline
\end{tabular}
\end{center}
\vspace{0.5cm}
where $Q_{e}$ and $Q_{m}$ are the electric and magnetic charge 
operators defined as
\begin{eqnarray}
	Q_{m} & = & \frac{1}{v} \int d^{3}x \, D_{i}\Phi^{a}B^{i}_{a}
	\nonumber  \\
	Q_{e} & = & \frac{1}{v} \int d^{3}x \, D_{i}\Phi^{a}E^{i}_{a} \, ,
	\label{elmagncharges}
\end{eqnarray}
with $E^{i}_{a}$ and $B^{i}_{a}$ the non Abelian electric and 
magnetic fields respectively.
All of the states saturate the Bogomol'nyi bound 
\cite{bogomolnyi}
\begin{equation}
	m \geq v \sqrt{Q_{e}^{2}+Q_{m}^{2}} \, ,
	\label{bbound}
\end{equation}
where $Q_{m}=\frac{4\pi n_{m}}{e}$ and $Q_{e}=n_{e}e$, with $n_{e},n_{m} 
\in \mathbb{Z}$. 
Of course quantum corrections might in 
principle modify this situation, however the above picture is expected to 
be a good approximation in the weak coupling limit, $e\ll 1$. 
In this limit
\begin{displaymath}
	m_{_{W}}=ev \ll v \:, \hspace{1cm} m_{_{M}} = gv = \frac{4\pi}{e}v 
	\gg v \, .
\end{displaymath}
Montonen and Olive have suggested in \cite{olivemont} that a duality 
transformation exchanging ``electric'' states associated to $W^{\pm}$ 
bosons and ``magnetic'' states associated to monopoles $M^{\pm}$ and 
at the same time electric and magnetic charges could be an exact symmetry 
of the full quantum theory in the BPS limit. The exchange of electric 
and magnetic charges implies, because of the Dirac quantization 
condition (see appendix \ref{appb}), 
\begin{displaymath}
	e \longrightarrow g = \frac{4\pi}{e} \, ,
\end{displaymath}
so that the proposed symmetry is intrinsically non-perturbative.

The conjecture of \cite{olivemont} is based on the classical spectrum 
and on the following arguments. 
First of all the mass formula (\ref{bbound}) is invariant under 
duality as can be easily checked. 
Moreover a semiclassical analysis \cite{manton} based 
on the so called {\em moduli space approximation} (to be discussed in 
the final subsection) shows that the interaction between two 
monopoles vanishes, whereas there is an attractive interaction 
between monopoles and anti-monopoles. The duality symmetry would 
require that the same should be true for the $W^{\pm}$ bosons. It can 
be proved at the semi-classical level that this is indeed the case because 
the repulsive interaction between equal charge $W$ bosons is exactly 
cancelled by an identical contribution due to the exchange of scalars, 
that are massless in the BPS limit. 

There are some basic problems with the Montonen--Olive proposal, namely
 \begin{itemize}
 	\item  Quantum corrections are expected to spoil the results of the 
 	semi-classical analysis. In particular a non-vanishing potential 
 	$V(\Phi)$ could be generated. 
 
 	\item  The elementary $W^{\pm}$ bosons have spin one while the 
 	monopoles that should be related to them by the duality have spin 
 	zero, making an exact matching of the quantum numbers impossible.
 
 	\item  The r\^ole of the dyons has been neglected in the previous 
 	analysis.
 \end{itemize}
 It will be shown that all of these problems can be solved in 
 supersymmetric theories. In particular the ${\cal N}$=4 
 supersymmetric Yang--Mills theory is supposed to realize exactly a 
 generalization of the Montonen--Olive duality called {\em 
 S-duality}. This generalization emerges when the effects of a 
 non-vanishing vacuum angle are taken into account. 
In the presence of a $\theta$-term (see the end of section 
\ref{thooftpolyakov}) the Lagrangian of the Georgi--Glashow model can be 
written
\begin{equation}
	{\cal L}=-\frac{1}{32\pi}{\rm Im}\, \left[ \tau 
	\left(F^{\mu\nu}+i*F^{\mu\nu}\right)
	\left(F_{\mu\nu}+i*F_{\mu\nu}\right) \right] - 
	\frac{1}{2} D^{\mu}\Phi D_{\mu} \Phi \, ,
	\label{georglatau}
\end{equation}
with $\tau=\frac{\theta}{2\pi} + i \frac{4\pi}{e^{2}}$. The electric 
and magnetic charges of the dyon states saturating the Bogomol'nyi 
bound are $Q_{e}=n_{e}e+n_{m}\frac{e \theta}{2\pi}$ and 
$Q_{m}=n_{m}\frac{4\pi}{e}$ respectively. 

Periodicity in $\theta$ requires invariance under 
\begin{equation}
	\tau \longrightarrow \tau+b \, , \qquad b \in \mathbb{Z} \, ,
	\label{t-trans}
\end{equation}
which amounts to a shift in the electric charge $n_{e}\to 
n_{e}-bn_{m}$, $n_{m}\to n_{m}$, or in matrix form
\begin{equation}
	\left( \begin{array}{c} n_{e} \\ n_{m} \end{array} \right) 
	\longrightarrow \left( \begin{array}{cc} 1 & b \\ 0 & 1 \end{array}
	\right) \left( \begin{array}{c} n_{e} \\ n_{m} \end{array} \right) 
	\, .
	\label{t-trans-qn}
\end{equation}
Furthermore at $\theta=0$ the duality of Montonen and Olive becomes
\begin{equation}
	\tau \longrightarrow -\frac{1}{\tau} \, ,
	\label{s-trans}
\end{equation}
that is equivalent to the exchange of electric and magnetic charges, 
$e\to g = \frac{4\pi}{e}$, $g \to -e$. In Montonen--Olive duality 
this is supplemented by the exchange of the corresponding 
quantum numbers, $Q_{e}\leftrightarrow Q_{m}$, implying $n_{e}\to 
n_{m}$, $n_{m}\to -n_{e}$. The latter transformation can be written in 
matrix form as 
\begin{equation}
	\left( \begin{array}{c} n_{e} \\ n_{m} \end{array} \right) 
	\longrightarrow \left( \begin{array}{cc} 0 & 1 \\ -1 & 0 \end{array}
	\right) \left( \begin{array}{c} n_{e} \\ n_{m} \end{array} \right) 
	\, .
	\label{s-trans-qn} 	
\end{equation} 
It is thus natural to generalize the duality symmetry to the set of 
transformations generated by (\ref{t-trans}) and (\ref{s-trans}) with 
arbitrary $\theta$. The resulting group of transformations is 
SL(2,$\mathbb{Z}$) acting projectively on the parameter $\tau$
\begin{equation}
	\tau \longrightarrow \frac{a\tau+b}{c\tau+d} \, ,\hspace{0.5cm} 
	a,b,c,d \in \mathbb{Z} \, , \qquad  ad-bc = 1 \, .
	\label{sdualtrans}
\end{equation}
The corresponding action on the quantum numbers $n_{e}$ and $n_{m}$, 
generated by (\ref{t-trans-qn}) and (\ref{s-trans-qn}), is 
\begin{equation}
	\left( \begin{array}{c} n_{e} \\ n_{m} \end{array} \right) 
	\longrightarrow \left( \begin{array}{cc} -a & b \\ c & -d 
	\end{array} \right)\left( \begin{array}{c} n_{e} \\ n_{m} 
	\end{array} \right) \: .
	\label{quantnumtrans}
\end{equation}
One can then easily check that the mass formula, written in terms 
of $\tau$ as
\begin{equation}
	m^{2} \geq 4\pi v^{2} (n_{e},n_{m})  \, \frac{1}{{\rm Im}\, \tau} 
	\left(\begin{array}{cc} 1 & -{\rm Re}\, \tau \\ -{\rm Re}\, \tau & 
	|\tau|^{2}\end{array} \right) \left(\begin{array}{c} n_{e} \\ n_{m}
	\end{array} \right)
	\label{massformulatau}
\end{equation}
is invariant.
 
\subsection{S-duality in ${\cal N}$=4 supersymmetric Yang--Mills 
theory}

The general properties of the ${\cal N}$=4 supersymmetric Yang--Mills 
theory previously discussed allow to give an intuitive explanation 
of why it is considered the original example of a theory possessing 
exact S-duality. In the Coulomb phase for each scalar with 
non-vanishing vev the same construction leading to the 't 
Hooft--Polyakov monopole (see the appendix) can be carried on. As a 
result the classical spectrum contains monopoles and dyons as well. 
Unlike the case of the Georgi--Glashow model 
the finiteness of the theory makes the 
semiclassical analysis of the spectrum sensible even at the quantum 
level. Moreover the theory has the correct number of fermions in the 
adjoint representation to give spin one dyon/monopole configurations 
through the mechanism sketched in section \ref{monopfermi}. 
More precisely, in the case of a SU(2) gauge group being discussed 
here, creation and annihilation operators $a_{\pm}^{i \dagger}$ and 
$a_{\pm}^{i}$, with $i=1,2$, associated with the fermionic zero modes 
can be defined so that the spectrum of zero-energy states, in the 
charge-one monopole sector, consists of 
the following states \cite{osborn}
\vspace{0.5cm}

\begin{center}
\begin{tabular}{|l|c|}
	\hline
	~~~~~{\rm State}\rule[-7pt]{0pt}{21pt}~~~ & ~~~~
	{\rm Spin}~~($S_{z}$)~~~~ \\
	\hline \hline
	~~~$|\Omega \rangle$~ \rule[-7pt]{0pt}{21pt} & 0  \\
	\hline
	~~~$a_{\pm}^{i\,\dagger}|\Omega \rangle$ \rule[-7pt]{0pt}{21pt} & 
	$\pm \frac{1}{2}$ \\
	\hline
	~~~$a_{-}^{i\,\dagger}a_{+}^{j\,\dagger}|\Omega \rangle$ 
	\rule[-7pt]{0pt}{21pt} & $ 0 $ \\
	\hline
	~~~$a_{+}^{1\,\dagger}a_{+}^{2\,\dagger}|\Omega \rangle$ 
	\rule[-7pt]{0pt}{21pt}~~~ & 1  \\
	\hline	
	~~~$a_{-}^{1\,\dagger}a_{-}^{2\,\dagger}|\Omega \rangle$ 
	\rule[-7pt]{0pt}{21pt}~~~ & -1  \\
	\hline
	~~~$a_{\mp}^{1\,\dagger}a_{\mp}^{2\,\dagger}a_{\pm}^{i\,\dagger}
	|\Omega \rangle$ \rule[-7pt]{0pt}{21pt}~~~ & $\mp\frac{1}{2}$ \\
	\hline
	~~~$a_{+}^{1\,\dagger}a_{+}^{2\,\dagger}a_{-}^{1\,\dagger}
	a_{-}^{2\,\dagger}|\Omega \rangle$ \rule[-7pt]{0pt}{21pt}~~~ & 0 \\
	\hline
\end{tabular}
\end{center}
\vspace{0.5cm}
where $|\Omega \rangle$ is the magnetic charge-one Clifford vacuum. The 
multiplet generated in this way contains sixteen states: eight bosons 
(six spin 0 scalars and two spin $\pm 1$ vectors) and eight spin 
$\pm\frac{1}{2}$ fermions. As a result the spin quantum numbers of the 
monopole multiplet exactly match those of the multiplet of elementary 
fields. 

Furthermore in extended supersymmetric theories the Bogomol'nyi bound 
comes as a consequence of the supersymmetry algebra, so that it holds 
in the full quantum theory. In supersymmetric models a vanishing scalar 
potential in correspondence of non trivial scalar field configurations 
is perfectly natural because of the presence of flat directions. In 
the case of ${\cal N}$=4 super Yang--Mills this scalar field 
configuration corresponds to a generic point of the Coulomb phase.

The relation of the BPS limit with the supersymmetry algebra is better 
understood in the ${\cal N}$=2 case. The ${\cal N}$=2 supersymmetry 
algebra possesses two central charges which are related to the magnetic 
and electric charge operators of the theory \cite{witolive}. Using a 
Majorana notation, the ${\cal N}$=2 algebra contains the anticommutator
 \begin{displaymath}
	\{Q^{i}_{\alpha},Q^{j}_{\beta}\}=\delta^{ij}\gamma^{\mu}_{\alpha\beta}
	P_{\mu}+\delta_{\alpha\beta}U^{ij}+(\gamma_{5})_{\alpha\beta}V^{ij}
	\: , \hspace{0.5cm} i,j=1,2 \, ,
\end{displaymath}
where the central charges have been denoted by $U^{ij}$ and $V^{ij}$. 
By expressing $U^{ij}$ and $V^{ij}$ in terms of the fields of the 
theory Olive and Witten have proved the relations
\begin{displaymath}
	U^{ij}=\varepsilon^{ij}vQ_{e}\:, \hspace{1cm} V^{ij}=\varepsilon^{ij}
	vQ_{m} \: ,
\end{displaymath}
where $Q_{e}$ and $Q_{m}$ are the electric and magnetic charge 
operators respectively, that generalize those in equation 
(\ref{elmagncharges}). In the previous equation $v^{2}= \langle \tr 
\left( S^{2}\right)\rangle + \langle \tr \left( P^{2} \right) \rangle$, 
where $S$ and $P$ denote the complex Higgs fields of the model. 
As a consequence of this relation the Bogomol'nyi bound follows directly 
from the supersymmetry algebra because of the positive definiteness of 
the operator $\{Q,Q\}$ in the rest frame. 

In the ${\cal N}$=4 case the situation is more involved because there are 
twelve central charges. Correspondingly there are six electric and six 
magnetic charge operators associated to the six scalars, $\varphi^{i}$, 
of the ${\cal N}$=4 multiplet 
\begin{eqnarray}
	Q^{i}_{e} & = & \frac{1}{v}\int_{S^{2}_{\infty}} d\vec{\sigma}\cdot 
	\tr \left(\vec{E}\varphi^{i}\right)
	\nonumber  \\
	Q^{i}_{m} & = & \frac{1}{v}\int_{S^{2}_{\infty}} d\vec{\sigma}\cdot 
	\tr \left(\vec{B}\varphi^{i}\right) \, ,
	\label{eq:emchargesn4}
\end{eqnarray}
where $v^{2}=\sum_{i=1}^{6} \langle \tr \left(\varphi^{i}\right)^{2} 
\rangle$. The Bogomol'nyi bound reads
\begin{displaymath}
	m^{2}\geq v^{2}\left[ (Q^{i}_{e})^{2}+(Q^{i}_{m})^{2}\right] 
\end{displaymath}
and can be saturated only by states annihilated by exactly one half 
of the supercharges. The BPS states which saturate the limit correspond 
to short multiplets of the ${\cal N}$=4 superalgebra and for this reason 
are protected in the full quantum theory. 
The mass of the BPS states is completely 
determined  by their electric and magnetic quantum numbers and is 
exact because supersymmetry is preserved at the quantum level.

All the BPS states in the spectrum can be displayed on six 
two-dimensional lattices (one for each couple of charges $Q^{i}_{e}$ 
and $Q^{i}_{m}$) and lie at points of coordinates 
\begin{displaymath}
{\cal Q} = Q_{e}+iQ_{m}=\left(n_{e}e+
n_{m}\frac{e\theta}{2\pi}\right)+in_{m}\frac{4\pi}{e} \, .
\end{displaymath} 
\vspace{0.3cm}
\begin{figure}[!h]
	\centering
	\includegraphics[width=10truecm]{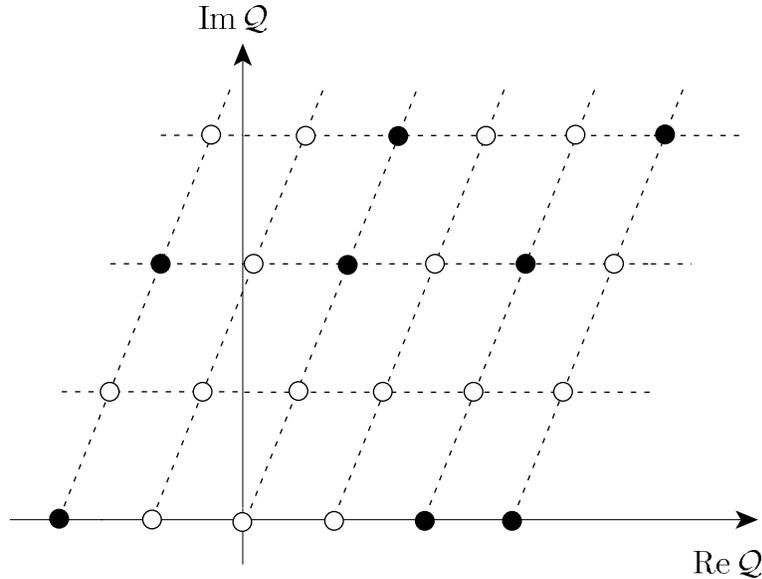}
	\caption{Lattice of BPS states}
\label{bpsstates}
\end{figure}
\vspace{0.3cm}

Supersymmetry implies that the BPS states in the spectrum can only 
decay into a couple of BPS states, so that a generic state of mass 
$m$ is unstable if and only if there exist BPS states of masses 
$m_{1}$ and $m_{2}$ such that
\begin{displaymath}
	m \geq m_{1}+m_{2} \, .
\end{displaymath}
It immediately follows that stable BPS states are all and only those 
which have integer quantum numbers $n_{e}$ and $n_{m}$ that are 
coprime. These are the states marked by empty circles in figure 
\ref{bpsstates}, while the full circles denote unstable states. 

Thanks to the fact that the BPS states are protected by supersymmetry 
in the ${\cal N}$=4 theory a semi-classical analysis of the spectrum 
is sensible at the quantum level as well. Some arguments supporting the 
conjectured S-duality of the ${\cal N}$=4 super Yang--Mills theory based 
on the moduli space approximation are briefly reviewed in the next 
subsection.

\subsection{Semiclassical quantization}

In appendix \ref{appb} it is discussed how to construct a classical 
solution of the equations of motion describing a monopole in the 
Georgi--Glashow model, the aim of this section is to show how the 
semiclassical quantization of the monopole solution can be used to 
study the spectrum of the model \cite{manton,osboreview}.
The notion of collective coordinates will be introduced starting with 
the 't Hooft--Polyakov monopole and then the semiclassical quantization 
will be applied to the ${\cal N}$=4 super Yang--Mills case. A readable 
review of these topics is provided by \cite{gauntlett}.

In the BPS limit the monopole solution satisfies the Bogomol'nyi 
equation $B_{i}^{a}=D_{i}\Phi^{a}$ ($i=1,2,3$, $a=1,\ldots,{\rm 
dim\,G}$). Given a field configuration solving this equation in 
general one can find a multi-parameter family of solutions with the same 
energy. The parameters labeling the degenerate solutions are called 
{\em moduli} or {\em collective coordinates} of the monopole and the 
corresponding space {\em moduli space}. 

For the 't Hooft--Polyakov monopole described in appendix \ref{appb} 
three collective coordinates are associated with the translational 
invariance of the Bogomol'nyi equation. Given a solution 
$(A_{i}^{{\rm cl}\,a}(\vec{x}),\Phi^{{\rm cl}\,a}(\vec{x}))$ the field 
configuration 
\begin{equation}
    A_{i}^{{\rm cl}\,a}(\vec{x}+\vec{X}) \, , \qquad 
    \Phi^{{\rm cl}\,a}(\vec{x}+\vec{X}) \, ,
    \label{eq:translmoduli}
\end{equation}
where $\vec{X}$ is a constant vector, still satisfies the equation 
and has the same energy (and magnetic charge). Thus the three 
coordinates $\vec{X}$ are moduli. 

A fourth collective coordinates is related to gauge transformations. 
Gauge invariance implies that the functional space of physical field 
configurations is 
\begin{displaymath}
    {\cal C}={\cal A}/{\cal G} \, ,
\end{displaymath}
where ${\cal A}=(A_{i}^{a},\Phi^{a})$ and ${\cal G}$ is the group of 
``small'' gauge transformations, \ie gauge transformations that do 
not reduce to the identity at spatial infinity. ``Large'', or 
``global'', gauge transformations 
which do not approach the identity at infinity 
are true symmetries of the theory and so field configurations related 
by large gauge transformations are inequivalent. The BPS solution for 
the 't Hooft--Polyakov monopole is symmetric under the diagonal 
SO(3) subgroup of SO(3)$_{{\rm R}}\times$SU(2)$_{{\rm G}}$, where 
SO(3)$_{{\rm R}}$ is the group of spatial rotations and SU(2)$_{{\rm 
G}}$ the group of large gauge transformations, but it is not invariant 
under SU(2)$_{{\rm G}}$. As a result by acting with the U(1) subgroup 
of SU(2)$_{{\rm G}}$ that is not broken in the Higgs vacuum one 
generates a new solution with the same energy, so that the fourth 
collective coordinate is associated with the U(1) group of global 
`electromagnetic' transformations. By letting $\vec{X}$ depend on time 
one describes a moving monopole, 
($A_{i}^{{\rm cl}\,a}(\vec{x}+\vec{X}(t)),
\Phi^{{\rm cl}\,a}(\vec{x}+\vec{X}(t))$). Since 
the Hamiltonian for the system in the temporal gauge $A_{0}=0$ is 
\begin{displaymath}
    H=T+V=\frac{1}{2}\int d^{3}x \, , \left[{\dot A}_{i}^{a} {\dot 
    A}_{a}^{i}+{\dot \Phi}^{a}{\dot \Phi}_{a} \right]
    + \frac{1}{2}\int d^{3}x \, , 
    \left[B^{a}_{i}B^{i}_{a}+D_{i}\Phi^{a}D^{i}\Phi_{a} \right] 
\end{displaymath}
the time dependence of $\vec{X}$ generates a non vanishing electric 
field so that the kinetic energy increases while the potential energy 
remains constant. This is a general feature of a motion in the moduli 
space. One can then consider a general deformation of a BPS monopole 
in the $A_{0}=0$ gauge
\begin{equation}
    \delta A_{i}^{a}(\vec{x},t) \, , \qquad 
    \delta\Phi^{a}(\vec{x},t) \, .
\label{deform}
\end{equation}
To keep the potential energy fixed (\ref{deform}) must satisfy the 
linearized Bogomol'nyi equation
\begin{equation}
    \varepsilon_{ijk}D^{j}\delta A^{k} - D_{i}\delta\Phi + 
    e [\delta A_{i}, \Phi^{{\rm cl}}] = 0
    \label{linearbogo}
\end{equation}
and the Gauss law constraint
\begin{equation}
    D_{i}\delta A^{i} + e [\Phi^{{\rm cl}},\delta\Phi] = 0 \, .
    \label{gausslaw}
\end{equation}
The general solution can be written in terms of a single function of 
time $\chi(t)$
\begin{eqnarray}
    \delta A_{i} & = & D_{i}\left( \chi(t) \Phi^{{\rm cl}} \right)
    \nonumber  \\
    \delta \Phi & = & 0
    \label{solmoduli}  \\
    \delta A_{0} & = & D_{0}\left(\chi(t)\Phi^{{\rm cl}}\right)-
    {\dot \chi}(t)\Phi^{{\rm cl}}
    \, . \nonumber
\end{eqnarray}
If $\chi$ is taken to be constant and ($\delta A_{i}^{a},\delta\Phi^{a}$) 
does not contain a translation $\delta \vec{x}=\vec{x}+\vec{X}$, 
the deformation reduces to a large gauge transformation, so that $\chi$ 
is identified with the fourth collective coordinate. For $\dot\chi 
\neq 0$ a non-vanishing kinetic energy is generated. Moreover $\chi$ is 
periodic since it parameterizes a compact U(1) (it is a subgroup of 
SU(2)$_{{\rm G}}$), therefore the moduli space with coordinates 
$(\vec{X},\chi)$ is topologically ${\cal M}_{1}=\mathbb{R}^{3}\times S^{1}$.

The previous construction of the moduli space can be generalized to a 
charge $k$ monopole solution. A rigorous derivation of multimonopole 
solutions is quite involved \cite{monopgeo}, the resulting moduli space 
${\cal M}_{k}$ is $4k$-dimensional and roughly speaking the $4k$ 
collective coordinates correspond to the locations of the $k$ monopoles 
and their dyon degrees of freedom. On physical grounds the existence of 
multimonopole BPS configurations is related to the fact that the 
repulsion between monopoles of the same charge can be compensated by 
an attractive interaction mediated by the massless Higgs field, thus 
allowing the construction of static configurations.

For a charge $k$ monopole deformations $(\delta_{\alpha} A^{a}_{i},
\delta_{\alpha}\Phi^{a})$, with $\alpha=1,\ldots,k$, satisfying the 
linearized Bogomol'nyi equations and the Gauss law constraint define 
tangent vectors to ${\cal M}_{k}$. Given the tangent vectors the 
standard metric on ${\cal M}_{k}$ can be constructed as
\begin{equation}
    {\cal G}_{\alpha\beta}=-\int d^{3}x \, \tr \left(
    \delta_{\alpha}A_{i}\delta_{\beta}A^{i}+\delta_{\alpha}\Phi
    \delta_{\beta}\Phi \right) \, .
  \label{metrick}
\end{equation}
This metric is actually induced by the form of the action for the 
gauge theory in ${\cal C}$ as can be seen by writing the monopole 
solution as $A_{i}^{{\rm cl}}(\vec{x},z_{\alpha})$, 
$\Phi^{{\rm cl}}(\vec{x},z_{\alpha})$, 
where $z_{\alpha}$'s denote the moduli. 
The tangent vectors $(\delta_{\alpha}A_{i},\delta_{\alpha}\Phi)$ are 
calculated by differentiating with respect to the $z_{\alpha}$. In 
general one has to include a gauge transformation to ensure that 
$(\delta_{\alpha}A_{i},\delta_{\alpha}\Phi)$ are tangent to the moduli 
space of physical fields, {\em i.e.} that the fields 
$A_{i}^{\prime}=A_{i}^{{\rm cl}}+\delta_{\alpha}A_{i}\delta 
z^{\alpha}$, $\Phi^{\prime}=\Phi^{{\rm cl}}+\delta_{\alpha}\Phi\delta 
z^{\alpha}$ belong to the space ${\cal C}$. The general form of the 
tangent vectors is  
\begin{eqnarray*}
    \delta_{\alpha}A_{i} & = & \frac{\partial A_{i}}{\partial 
    z^{\alpha}}-D_{i}\epsilon_{\alpha}  \\
    \delta_{\alpha}\Phi & = & \frac{\partial \Phi}{\partial z^{\alpha}}
    -e [\Phi^{{\rm cl}},\epsilon_{\alpha}] \, ,
\end{eqnarray*}
where $\epsilon_{\alpha}(\vec{x},z^{\beta})$ is a suitable gauge 
parameter that is fixed by requiring that the linearized Bogomol'nyi 
equation (\ref{linearbogo}) and the Gauss law constraint 
(\ref{gausslaw}) be satisfied.
The action of the theory on the BPS configuration is then obtained 
after substituting 
\begin{eqnarray}
    A_{i}(\vec{x},t) & \longrightarrow & A_{i}^{{\rm cl}}
    (\vec{x},z^{\alpha}(t))  \nonumber \\
    A_{0} & \longrightarrow & {\dot z}^{\alpha}(t) \epsilon_{\alpha}
    \label{ansatzb} \\
    \Phi(\vec{x},t) & \longrightarrow &  \Phi^{{\rm cl}}
    (\vec{x},z^{\alpha}(t)) \nonumber
\end{eqnarray}
and reads
\begin{equation}
    S=-\frac{1}{2} \int d^{3}x dt \, \tr \left(
    F^{{\rm cl}}_{0i}F_{{\rm cl}}^{0i}\right)=
    \int dt \, {\cal G}_{\alpha\beta}{\dot z}^{\alpha}{\dot z}^{\beta}
    -\frac{4\pi v}{e}k \, .
\label{actionmetric}
\end{equation}
The action (\ref{actionmetric}) is the starting point for the 
semiclassical quantization which is achieved by studying the 
corresponding quantum mechanics.

In the $k=1$ case the moduli are $z^{\alpha}=(\vec{X},\chi)$ and the 
action becomes 
\begin{equation}
    S=\frac{1}{2} \int dt \, \left[\frac{4\pi v}{e}
    {\dot {\vec{X}}}^{2} +\frac{4\pi}{ve^{2}}{\dot \chi}^{2}\right]
    -\frac{4\pi v}{e} \, .
    \label{eq:actionk1}
\end{equation}
The resulting wave functions are plane waves 
\begin{displaymath}
	\psi = e^{i\vec{P}\cdot\vec{X}}e^{in_{e}\chi} \: ,
\end{displaymath}
with $n_{e}\in\mathbb{Z}$, and one can compute the electric charge, 
$Q_{e}=-ie\partial_{\chi}$, so that the quantum mechanical system 
possesses an infinite tower of states with $Q_{e}=n_{e}e$. For the 
mass of these states one gets
\begin{displaymath}
	m = \frac{n_{e}^{2}ve^{3}}{8\pi}+\frac{4\pi v}{e} \approx 
	v[Q_{e}^{2}+Q_{m}^{2}] \, ,
\end{displaymath}
showing that the Bogomol'nyi bound is saturated.

The preceding arguments can be extended to the case of the ${\cal N}$=4 
super Yang--Mills theory to find stronger evidence for the exact S-duality 
of the model. In order to be able to use the previous considerations 
the simplest case will be considered, namely the gauge group will be 
taken to be G=SU(2), broken by the vacuum configuration
\begin{eqnarray*}
	&& \langle \varphi^{2} \rangle = \langle \varphi^{3} \rangle = 
	\ldots = \langle \varphi^{6} \rangle = 0 \\
	&& \langle \varphi^{1} \rangle = \Phi \, , \qquad 
	\langle \tr \Phi^{2} \rangle = v^{2} \, .
\end{eqnarray*}
The perturbative states in the bosonic sector of the theory, labeled by 
quantum numbers $(n_{m},n_{e})$ consist of a massless photon multiplet 
with charge $(0,0)$ and massive $W_{\pm}$ boson multiplets with charge 
$(0,\pm 1)$. The tower of stable BPS states predicted by S-duality have 
charges $(l,k)$ (with $l$ and $k$ coprimes) and can be studied within the 
framework of semiclassical quantization since they can be put in 
correspondence with certain geometric structures on ${\cal M}_{k}$. Then 
the non-renormalization properties of the theory make the semiclassical 
result reliable in the full quantum theory.

In the Georgi--Glashow model ${\cal M}_{k}$ is parameterized by $4k$ 
collective coordinates which are bosonic zero modes. In the ${\cal 
N}$=4 supersymmetric Yang--Mills theory there are fermionic zero modes 
as well in the monopole background and the index theorem of 
\cite{index} shows that the number of such zero modes in the charge 
$k$ sector is precisely $4k$. 
As a consequence the moduli space has fermionic coordinates too and 
the low energy ansatz to be substituted into the Bogomol'nyi equation 
is given by (\ref{ansatzb}) supplemented by
\begin{equation}
	\lambda(\vec{x},t) \sim \psi(t) 
	\lambda^{{\rm cl}}(\vec{x},z_{\alpha}(t))
	\label{ansatzf}
\end{equation}
for the Weyl spinors in the monopole background. In equation 
(\ref{ansatzf}) the ansatz for the spinors $\lambda(\vec{x},t)$ has 
been schematically written as the product of a c-number 
$\lambda^{{\rm cl}}(\vec{x},z_{\alpha})$ satisfying a zero mode 
equation in the monopole background and a time dependent fermionic 
collective coordinate $\psi(t)$. The zero modes can be shown to 
constitute a short multiplet preserving half of the supersymmetries 
just like is expected for BPS states. 
The exact ansatz for the fields is rather complicated and was 
discussed in detail in \cite{n4qm}. A technically involved computation 
allows to derive the quantum mechanical action to be employed in the 
semiclassical analysis. The action reads 
\begin{equation}
	S=\frac{1}{2} \int dt \, \left[ {\cal G}_{\alpha\beta} \left(
	{\dot z}^{\alpha}{\dot z}^{\beta}+i{\overline \psi}^{\alpha}
	\gamma^{0}D_{t}\psi^{\beta}\right) + \frac{1}{6} 
	R_{\alpha\beta\gamma\delta}{\overline \psi}^{\alpha}\psi^{\gamma}
	{\overline \psi}^{\beta}\psi^{\delta} \right] -
	\frac{4\pi v}{e^{2}}k \, ,
	\label{n4actionmoduli}
\end{equation}
where a Majorana notation is used. In (\ref{n4actionmoduli}) 
$D_{t}\psi^{\alpha} = {\dot \psi}^{\alpha}+\Gamma^{\alpha}_{\beta\gamma}
{\dot z}^{\beta}\psi^{\gamma}$ is the covariant derivative, with 
$\Gamma^{\alpha}_{\beta\gamma}$ the Christoffel symbols of the metric 
${\cal G}_{\alpha\beta}$, and $R_{\alpha\beta\gamma\delta}$ is the 
corresponding Riemann tensor \cite{n4qm}. It can be shown that thanks 
to the fact that the monopole moduli space is 
a hyper-K\"{a}hler manifold, $S$ in equation (\ref{n4actionmoduli}) 
is invariant under the action of eight real supercharges. 
The quantization of the supersymmetric quantum mechanics 
(\ref{n4actionmoduli}) was performed in \cite{witmoduli}, where it 
was shown that the states can be put in one to one correspondence 
with differential forms on ${\cal M}_{k}$ and the Hamiltonian of the 
system $H$ can be expressed in terms of the exterior derivative acting 
on forms. This construction leads to a basis of sixteen independent 
forms that are eigenstates of $H$ in the $n_{m}=k=1$ sector. These 
states constitute a BPS multiplet as can be verified by a computation 
of the corresponding quantum numbers. The complete wave functions 
contain a plane wave factor from the bosonic sector of the form 
$e^{i\vec{P}\cdot\vec{X}}e^{in_{e}\chi}$ so that again one finds a 
tower of states with $Q_{e}=n_{e}+\frac{e\theta}{2\pi}$. These 
states all saturate the Bogomol'nyi bound and therefore are exactly 
those predicted by S-duality.

For $n_{m}=k > 1$ the problem is much more complicated. The moduli 
space is
\begin{displaymath}
	{\cal M}_{k}=\mathbb{R}^{3}\times(S^{1}\times
	\tilde{{\cal M}}^{0}_{k})/\mathbb{Z}_{k} \, ,
\end{displaymath}
where $\tilde{{\cal M}}^{0}_{k}$ is a $4(k-1)$-dimensional 
hyper-K\"ahler manifold. The states are tensor products of forms on 
$\mathbb{R}^{3}\times S^{1}$ with forms on $\tilde{{\cal 
M}}^{0}_{k}$, $|s\rangle = |\omega,n_{e}\rangle \otimes |\alpha 
\rangle$~\footnote{There are actually further subtleties related to 
the $\mathbb{Z}_{k}$ identification.}. The forms $|\alpha\rangle$ can 
be proved to be unique and much in the same way as in the $k=1$ case 
there are sixteen forms $|\omega,n_{e}\rangle$ for all $k$ that give 
rise to a BPS multiplet. The resulting states carry electric charge 
$Q_{e}=n_{e}e+n_{m}\frac{e\theta}{2\pi}$ and saturate the Bogomol'nyi 
bound. The unique form $|\alpha\rangle$ for the case $k=2$ has been 
explicitly determined by Sen in \cite{sen}. Arguments suggesting a 
possible generalization to $k > 2$ were given in \cite{sdualk>2}.


\chapter{Perturbative analysis of ${\cal N}=4$ supersymmetric 
Yang--Mills theory}
\label{cap3}
\vspace*{2cm}
\fancyhead[RO,LE]{\thepage}
\fancyhead[RE]{{\footnotesize {\rm Chapter 3.}~~{\it Perturbative 
analysis of ${\cal N}=4$ SYM theory}}} 
\fancyhead[LO]{}

\noindent
As discussed in the previous chapter, the ${\cal N}$=4 supersymmetric 
Yang--Mills theory has been proved to be finite at the perturbative 
level up to three loops and strong arguments have been proposed in 
order to extend this result to all orders and even non perturbatively. 
For this reason the theory has long been considered to be rather trivial. 
However recent developments, within the context of the correspondence 
with type IIB superstring theory on anti de Sitter space, that will 
be studied in the final chapter, have shown that it is actually an 
extremely interesting superconformal field theory, possessing very 
peculiar properties.

Even in perturbation theory a careful analysis allows to point out 
various problems mostly related to the choice of gauge. Both in the 
component field formulation and using ${\cal N}$=1 superfields the 
gauge fixing procedure appears very subtle. As will be discussed in 
detail in both cases one finds divergences in off-shell Green 
functions. 

In the component formulation the propagators of the 
elementary fields are ultraviolet divergent in the Wess--Zumino (WZ) 
gauge and these infinities are exactly cancelled when the contributions 
of the gauge-dependent fields, that are put to zero in the WZ gauge, 
are taken into account. The choice of the WZ gauge, that is almost 
unavoidable in explicit computations, introduces divergences that 
require a wave function renormalization. This is a general result that 
applies to theories with less supersymmetry as well, but appears 
particularly dramatic in the ${\cal N}$=4 case because the theory is 
finite, so that the divergences cannot be reabsorbed through a 
redefinition of the bare parameters.

Different problems emerge in the formulation of the theory in terms 
of ${\cal N}$=1 superfields. Almost all the calculations showing 
the vanishing of the quantum corrections to two- and three-points 
functions that are presented in the literature were performed in the 
supersymmetric generalization of the Fermi--Feynman gauge ($\alpha$=1 
in the notation of section \ref{quantization}). With a different 
choice of gauge two and three point functions develop infrared 
singularities of difficult interpretation, leading to the result that 
the choice $\alpha$=1 is somehow privileged. This conclusion was proposed 
for the first time in \cite{piguetrouet} and then it was discussed in 
the case of the ${\cal N}$=4 theory in \cite{storey}; however no possible 
explanation was proposed. Unlike those of the elementary fields 
the correlation functions of gauge invariant composite operators, that 
play a crucial r\^ole in the correspondence with AdS type IIB 
supergravity/superstring theory, should not suffer from problems 
related to the gauge fixing.

Also, it will be discussed, in the superfield formulation, 
the effect of the introduction of a mass term for the chiral 
superfields, which breaks supersymmetry from ${\cal N}$=4 down to 
${\cal N}$=1. It will be shown that this deformation of the model does not 
modify the ultraviolet properties of the original theory. This result 
was first proposed in \cite{parkeswest}, where it was proved that the 
inclusion of  mass terms for the (anti) chiral superfields does not 
generate divergent corrections to the effective action. 
It will be argued that this statement can be reinforced, showing that, at 
least at one loop, no new divergences, not even corresponding to wave  
function renormalizations, appear as a consequence of the addition of 
the mass terms. This result supports the claim put forward in 
\cite{japan,yoshida}, where the `mass deformed' ${\cal N}$=4 theory 
was proposed as a supersymmetry-preserving regularization scheme for 
a class of ${\cal N}$=1 theories. It will also be argued that the 
same approach should work in the case of ${\cal N}$=2 theories. 

All of this problems are examined in detail in this chapter.
Section \ref{wzproblems} deals with the difficulties introduced by the 
choice of the Wess--Zumino gauge when the component formalism is used. 
The subsequent sections report calculations performed in the 
${\cal N}$=1 superfield formalism of two-, three- and four-point 
functions. The material presented in this chapter is an original 
and systematic re-analysis of the perturbative properties of ${\cal 
N}$=4 supersymmetric Yang--Mills theory, the treatment follows closely 
\cite{sk}.

\newpage 

\section{Perturbation theory in components: problems with the 
Wess--Zumino gauge}
\label{wzproblems}
\fancyhead[LO]{{\footnotesize 3.1~~{\it Perturbation theory in components}}}

Throughout this chapter calculations will be carried on in Euclidean 
space-time and unless otherwise stated the gauge group will be taken 
to be SU($N$). To discuss perturbation theory in components in this 
section the formulation of equation (\ref{n1sym4}), 
with a SU(3)$\times$U(1) subgroup 
of the SU(4) R-symmetry group manifest, will be employed. To correctly 
deal with the lower components of the vector superfield $V$, that are 
put to zero in the Wess--Zumino gauge, it is however useful not to 
eliminate the auxiliary fields $F$ and $D$ through their equations of 
motion, but rather keep them in the action: in the computation of 
Green functions the corresponding $x$-space propagators are simply 
$\delta$-functions. The complete expression for the vector 
superfield $V$ was given in equation (\ref{vector}) and contains 
beyond the physical fields, $A_{\mu}$ and $\lambda$, and the 
auxiliary field, $D$, the `gauge-dependent' fields $C$, $\chi$ and $S$. 
Note that the scalar field $C$ and the spinor $\chi$ have ``wrong'' 
physical dimension, resulting in non standard free propagators, as will 
be shown. The non Abelian field strength superfield $W_{\alpha}$ is 
\begin{equation}
	W_{\alpha}=-\frac{1}{4} {\overline D}{\overline D} e^{-V}
	D_{\alpha}e^{V}=\sum_{k=1}^{\infty} W_{\alpha}^{(k)} \, ,
	\label{wexpansion}
\end{equation}
where 
\begin{equation}
\begin{array}{ccl}
	\begin{displaystyle}W_{\alpha}^{(1)} \end{displaystyle}
	&=& \begin{displaystyle} -\frac{1}{4}{\overline D}{\overline D}
	D_{\alpha}V \end{displaystyle} \\
    \begin{displaystyle}W_{\alpha}^{(2)} \end{displaystyle} 
    &=& \begin{displaystyle} \frac{1}{8}{\overline D}{\overline D}
	[V,D_{\alpha}V] \rule{0pt}{24pt}\end{displaystyle}
	\label{w1-2}
\end{array}
\end{equation}
and the terms $W_{\alpha}^{(k)}$, with $k\geq 3$, contain $k$ factors of 
$V$ and vanish in the WZ gauge. 

The Euclidean action in ${\cal N}$=1 superfields can be written 
\begin{eqnarray}
	&& \hspace*{-0.5cm} S^{({\rm E})} 
	= \frac{1}{d_{{\bf r}}} tr \int d^{4}x \, 
	d^{4}\theta \, \left\{ -\Phi^{\dagger}_{I}\Phi^{I}
	-\frac{1}{4g^{2}}\left[ \frac{1}{4} 
	W^{(1)\alpha} W^{(1)}_{\alpha} \delta({\overline \theta})
	+\frac{1}{4} {\overline W}^{(1)}_{\dot \alpha} 
	{\overline W}^{(1)\dot \alpha}\delta(\theta) + 
	\right. \right. \nonumber \\
	&& \hspace*{-0.5cm} \left. - \frac{1}{8\alpha} {\overline D}^{2} 
	V D^{2} V \right] \left. - 
	g [\Phi_{I}^{\dagger} V] \Phi^{I} - \left[ \left( 
	\frac{1}{8g^{2}} W^{(1)\alpha} W^{(2)}_{\alpha} 
	+\frac{1}{16g^{2}} W^{(2)\alpha} W^{(2)}_{\alpha} \right) 
	\delta({\overline \theta}) + \nonumber \right. \right. \\ 
	&& \hspace*{-0.5cm} +  \left. \left. \left( \frac{1}{8g^{2}} 
	{\overline W}^{(1)}_{\dot \alpha} {\overline W}^{(2)\dot \alpha} + 
	\frac{1}{16g^{2}} {\overline W}^{(2)}_{\dot \alpha} 
	{\overline W}^{(2)\dot \alpha} \right) \delta(\theta) \right] - 
	\frac{1}{2}g^{2}[V,[V, \Phi_{I}^{\dagger}]] \Phi^{I} + \ldots 
	\rule{0pt}{18pt}\right\}
	\label{relevantaction} \, ,
\end{eqnarray}
where the gauge fixing term has been included, whereas no ghost term 
is displayed since it will not be relevant for the computations to be 
discussed in this section. 
In equation (\ref{relevantaction}) the dots denote terms 
of higher order in $V$, which do not contribute to the Green functions 
that will be considered and that will be suppressed from now on. 
In the following calculations the 
Fermi--Feynman gauge, $\alpha$=1, will be used. With this choice one 
obtains~\footnote{From now on the superscript E will be suppressed. 
Euclidean signature is understood unless otherwise stated.}
\begin{eqnarray*}
	S \! &=& \! \frac{1}{d_{{\bf r}}} \tr\int d^{4}x \, d^{4}\theta \, 
	\left\{ \left[ V \Box V - \Phi_{I}^{\dagger} \Phi^{I} \right] - 
	\left[ \left( \frac{1}{8g^{2}} W^{(1)\alpha} W^{(2)}_{\alpha} 
	+\frac{1}{16g^{2}} W^{(2)\alpha} W^{(2)}_{\alpha} \right) \delta(
	{\overline \theta}) + \right. \right.  \\ 
	&+& \hspace{-0.2cm} \left. \left. \left( \frac{1}{8g^{2}} 
	{\overline W}^{(1)}_{\dot \alpha} {\overline W}^{(2)\dot \alpha} + 
	\frac{1}{16g^{2}} {\overline W}^{(2)}_{\dot \alpha} 
	{\overline W}^{(2)\dot \alpha} \right) \delta(\theta) \right] - 
	\left[ - g [\Phi_{I}^{\dagger} V] \Phi^{I} + \frac{1}{2} 
	g^{2}[V,[V, \Phi_{I}^{\dagger}]] \Phi^{I} \right] \right\} ,
\end{eqnarray*}
where from the definition (\ref{w1-2}) it follows
\begin{eqnarray}
	W^{(1)}_{\alpha} & = & -i\lambda_{\alpha} + \left[ 
	\delta_{\alpha}^{\beta} D -\frac{1}{2} \delta^{\beta}_{\alpha}\Box C 
	-\frac{i}{2} (\sigma^{\mu}{\overline \sigma}^{\nu})_{\alpha}{}^{\beta}
	\left( \partial_{\mu} A_{\nu}-\partial_{\nu} A_{\mu} \right) \right] 
	\theta_{\beta} + \nonumber \\  
	 && + \theta \theta \sigma^{\mu}_{\alpha{\dot \alpha}} 
	\partial_{\mu}{\overline \lambda}^{\dot \alpha} \, .
	\label{w1}
\end{eqnarray}
The explicit form of $W^{(2)}_{\alpha}$ will not be necessary for 
the moment (see however equation (\ref{w2nonwz})). 
Expansion of the action in components 
using the complete expression of $V$ gives
\begin{displaymath}
	S = S_{0}+S_{{\rm int}} \, .
\end{displaymath}
The free action $S_{0}$ comes from the terms $-V \Box V$ 
and $\Phi_{I}^{\dagger} \Phi^{I}$ and reads
\begin{eqnarray}
	S_{0} & = & \int d^{4}x \, \left\{ \left[ 
	(\partial_{\mu}\varphi^{a\dagger}_{I})(\partial^{\mu}
	\varphi_{a}^{I}) + {\overline \psi}_{I}^{a}{\overline \sigma}^{\mu} 
	(\partial_{\mu}\psi^{I}_{a}) + F^{a\dagger}_{I}F^{I}_{a} \right] + 
	\left[ S^{\dagger}_{a}\Box S^{a} +
	\right. \right. \nonumber \\
	&& \hspace{-1.6cm} \left. \left. -\frac{1}{2} 
	C^{a} \Box D_{a} -\frac{1}{2} D^{a}\Box C_{a} - \frac{1}{2} C^{a} 
	\Box^{2} C_{a} + \frac{1}{2} \chi^{a} \Box \lambda_{a} + \frac{1}{2} 
	{\overline \chi}^{a}\Box {\overline \lambda}_{a} + 
	\frac{1}{2} \lambda^{a}\Box \chi_{a} + 
	\right. \right. \nonumber \\ 
	&& \hspace{-1.6cm} \left. \left. + \frac{1}{2} 
	{\overline \lambda}^{a}\Box {\overline \chi}_{a} + \frac{1}{2} 
	\chi^{a} \Box \sigma^{\mu} (\partial_{\mu}{\overline \chi}_{a}) + 
	\frac{1}{2} {\overline \chi}^{a} \Box {\overline \sigma}^{\mu} 
	(\partial_{\mu}\chi_{a}) +\frac{1}{2} (\partial_{\mu}A^{a}_{\nu})
	(\partial_{\nu}A_{a\mu}) \right] \right\} .
	\label{freeact}
\end{eqnarray}
The interaction part $S_{{\rm int}}$ contains an infinite number of 
terms. The propagators of the fermion $\psi$ and of the scalar 
$\varphi$ belonging to the ${\cal N}$=1 chiral multiplet will now be 
computed at one loop. The terms that are relevant for these 
calculations come from the expansion of $\Phi^{\dagger}V\Phi$ and 
$\Phi^{\dagger}V^{2}\Phi$ in the superfield action. The latter generates 
tadpole type diagrams in the propagator $\langle \varphi^{\dagger}
\varphi \rangle$, but not in $\langle {\overline \psi} \psi \rangle$. 
The interaction part of the action to be considered is 
\begin{eqnarray}
     && \hspace*{-0.6cm}  
	 S_{{\rm int}} = \int d^{4}x \left\{ igf_{abc} \left[ \frac{1}{2}
     \phi_{I}^{a\dagger}D^{b}\phi^{cI} + \frac{1}{2} 
     \phi_{I}^{a\dagger}(\Box C^{b})\phi^{cI} - \frac{1}{2} 
     (\partial_{\mu}\phi_{I}^{a\dagger})C^{b}(\partial^{\mu}\phi^{cI}) 
     + \right. \right. \nonumber \\
     && \hspace*{-0.6cm}  
	 +\frac{i}{2} \left( \phi_{I}^{a\dagger}A_{\mu}^{b}(
     \partial^{\mu} \phi^{cI}) - (\partial^{\mu}\phi_{I}^{a\dagger}) 
     A_{\mu}^{b}\phi^{cI} \right) + \frac{i}{\sqrt{2}}\left( 
     \phi_{I}^{a\dagger}S^{b\dagger}F^{cI}-F_{I}^{a\dagger}S^{b}
     \phi^{cI} \right) + \nonumber \\ 
     && \hspace*{-0.6cm} 
	 - F^{a\dagger}_{I}C^{b}F^{cI} + \frac{i}{\sqrt{2}} \left( 
     {\overline \psi}^{a}_{I}{\overline \lambda}^{b}\phi^{cI} - 
     \phi_{I}^{a\dagger}\lambda^{b}\psi^{cI} \right) + \frac{i}{\sqrt{2}}
     \left( F^{a\dagger}_{I}\chi^{b}\psi^{cI} - {\overline \psi}^{a}_{I} 
     {\overline \chi}^{b}F^{cI} \right) + \nonumber \\
     && \hspace*{-0.6cm} 
	 -\frac{i}{\sqrt{2}} \left( \phi^{a\dagger}(\partial_{\mu}
     {\overline \chi}^{b})\sigma^{\mu}\psi^{cI} + {\overline \psi}^{a}_{I}
     {\overline \sigma}^{\mu}(\partial_{\mu}\chi^{b}) \phi^{cI} \right) +
     \frac{1}{2} \left( C^{b}{\overline \psi}^{a}_{I}{\overline \sigma}^{\mu}
     (\partial_{\mu}\psi^{cI}) - C^{b}(\partial_{\mu}
     {\overline \psi}^{a}_{I}){\overline \sigma}^{\mu} \psi^{cI} 
     \right)  + \nonumber \\ 
     && \hspace*{-0.6cm} 
	 \left. -\frac{i}{2} {\overline \psi}^{a}_{I}
     {\overline \sigma}^{\mu}\psi^{cI}A_{\mu}^{b} \right] 
     - \frac{\sqrt{2}}{3!}gf_{abc} \left[ \rule{0pt}{16pt} 
     \varepsilon^{IJK} \left( 
     \phi^{a\dagger}_{I} \phi^{b\dagger}_{J}F^{c\dagger}_{K} - 
     \phi_{I}^{a\dagger}{\overline\psi}^{b}_{J}{\overline\psi}^{c}_{K} 
     \right) + \varepsilon_{IJK} \left( \phi^{aI}\phi^{bJ} F^{cK} + 
     \right. \right. \nonumber \\
     && \hspace*{-0.6cm} 
	 \left. \left. - \phi^{aI} \psi^{bJ}\psi^{cK} \right) 
     \rule{0pt}{14pt} \right] -
     \frac{g^{2}}{2} f_{abe}f^{e}{}_{cd}\left( \left[ \rule{0pt}
     {18pt} -C^{b}\psi^{dI} \sigma^{\mu}{\overline \psi}^{a}_{I} A^{c}_{\mu}+  
     \right. \right. \nonumber \\
     && \hspace*{-0.6cm} \left. + 
     ({\overline \chi}^{c}{\overline \psi}^{d}_{I})(\chi^{b} \psi^{aI}) +
     \frac{i}{2} \left( \partial_{\mu} \psi^{dI} 
     \sigma^{\mu} {\overline \psi}^{a}_{I} - \psi^{dI} \sigma^{\mu}  
     \partial_{\mu}{\overline \psi}^{a}_{I} \right)
     C^{b}C^{c} \right] + \varphi^{a\dagger}_{I} \left[ C^{b} \left( D^{c}+
     \rule{0pt}{17pt} 
     \right. \right. \nonumber \\ 
     && \hspace*{-0.6cm} \left. +
     \frac{1}{2} \Box C^{c} \right)  - 
     \chi^{b} \left( \lambda^{c} + \frac{i}{2} 
     \sigma^{\mu}\partial_{\mu}{\overline \chi}^{c} \right) - 
     {\overline \chi}^{b} \left( {\overline \lambda}^{c} + \frac{i}{2} 
     \sigma^{\mu} \partial_{\mu} \chi^{c}\right) +  
     S^{b\dagger}S^{c} - \rule{0pt}{17pt} 
     \nonumber \\
     && \hspace*{-0.6cm} \left. 
     -\frac{1}{2} A^{bc}_{\mu}A^{c\mu}\right] 
     \varphi^{dI} + \frac{i}{2} C^{b} A^{c\mu} 
     \left( \varphi^{a\dagger}_{I} (\partial_{\mu} \varphi^{dI}) - 
     (\partial_{\mu}\varphi^{a\dagger}_{I}) \varphi^{dI} \right) +
     \frac{1}{4} C^{b}C^{c}\left(\varphi^{a\dagger}_{I}
     (\Box \varphi^{dI}) + \right. \nonumber \\ 
     && \hspace*{-0.6cm} 
	 \left.\left. \left. + (\Box \varphi^{a\dagger}_{I}) 
     \varphi^{dI} \right) +\frac{i}{2} \chi^{b}\sigma^{\mu}
     {\overline \chi}^{c} \left( \varphi^{a\dagger}_{I}
     (\partial_{\mu} \varphi^{dI}) - 
     (\partial_{\mu}\varphi^{a\dagger}_{I}) \varphi^{dI} \right) 
     \rule{0pt}{18pt}\right) \right\} \, .
     \label{intaction}
\end{eqnarray}
The  quadratic part of the action, $S_{0}$, can be written in a more 
compact form introducing the notation 
\begin{equation}
	{\cal B}^{a}(x) = \left( \begin{array}{c} C^{a}(x) \\ 
	\rule{0pt}{16pt} D^{a}(x) \end{array} \right) \, , 
	\hspace{1cm} {\cal F}^{a}(x) = \left( \begin{array}{c} \chi^{a}(x) \\
	{\overline \chi}^{a}(x) \rule{0pt}{16pt} \\ \lambda^{a}(x) 
	\rule{0pt}{16pt} \\ {\overline \lambda}^{a}(x) \rule{0pt}{16pt} 
	\end{array} \right) \, , 
	\label{fieldvector}
\end{equation}
so that 
\begin{eqnarray}
	S_{0} &=& \int d^{4}x \, \left\{ \left[ 
	(\partial_{\mu}\varphi^{a\dagger}_{I})(\partial^{\mu}
	\varphi_{a}^{I}) + {\overline \psi}_{I}^{a}{\overline \sigma}^{\mu} 
	(\partial_{\mu}\psi^{I}_{a}) + F^{a\dagger}_{I}F^{I}_{a} \right] + 
	\right. \nonumber \\ 
	&& + \left. \left[ S^{\dagger}_{a}\Box S^{a} - 
	\frac{1}{2} A^{a}_{\mu}A_{a}^{\mu} + {\cal B}_{a}^{T} M {\cal B}^{a} + 
	{\cal F}_{a}^{T} N {\cal F}^{a} \right] \right\} \, , 
	\label{compactfree}
\end{eqnarray}
where 
\begin{equation}
	M = \frac{1}{2} \left( \begin{array}{cc} -\Box^{2} & -\Box \\ 
	-\Box & 0 \end{array} \right) \, , \hspace{0.5cm} 
	N = \frac{1}{2} \left( \begin{array}{cccc} 0 & \Box 
	\sigma^{\mu}\partial_{\mu} & -\Box & 0 \\ \Box 
	{\overline \sigma}^{\mu}\partial_{\mu} & 0 & 0 & -\Box \\
	-\Box & 0 & 0 & 0 \\ 0 & -\Box & 0 & 0 \end{array} \right) \, .
	\label{kineticmatrices}
\end{equation}
Inverting the kinetic matrices $M$ and $N$ one gets the free 
propagators. From (\ref{kineticmatrices}) it follows 
\begin{equation}
	M^{-1} = \left( \begin{array}{cc} 0 & \begin{displaystyle}
	\frac{1}{\Box} \end{displaystyle} \\ \begin{displaystyle}
	\frac{1}{\Box} \end{displaystyle} & -1 \end{array} \right) \, ,
	\hspace{0.5cm} N^{-1} = \left( \begin{array}{cccc} 0 & 0 & 
	\begin{displaystyle} -\frac{1}{\Box} \end{displaystyle} & 0 \\
	0 & 0 & 0 & \begin{displaystyle} -\frac{1}{\Box} \end{displaystyle} 
	\\ \begin{displaystyle} -\frac{1}{\Box} \end{displaystyle} & 0 & 0 
	& \begin{displaystyle} \frac{\rule{0pt}{14pt}\sigma^{\mu}
	\partial_{\mu}}{\Box} \end{displaystyle} \\ 
	0 & \begin{displaystyle} -\frac{1}{\Box} 
    \end{displaystyle} & \begin{displaystyle} \frac{\rule{0pt}{14pt}
	{\overline \sigma}^{\mu} \partial_{\mu}}{\Box} \end{displaystyle} 
	& 0 \end{array} \right) \, ,
	\label{inverskinmatr}
\end{equation}
so that the free propagators are \\
\hspace*{0.5cm}
\begin{fmffile}{phi+phi}
\begin{fmfgraph*}(90,50) 
  \fmfleft{i1} \fmfright{o1}
  \fmf{scalar}{i1,o1}
  \fmflabel{\raisebox{22pt}{\hspace{-20pt} $I,a$ \hspace{20pt}} 
  \raisebox{12pt}{$\longrightarrow \hspace{0.2cm} \langle 
  \varphi^{b\dagger}_{J}(x) \varphi^{I}_{a}(y) \rangle_{{\rm free}} = 
  - \begin{displaystyle} \frac{\delta^{b}_{a}\delta^{I}_{J}}{\Box} 
  \end{displaystyle}\delta(x-y)=\Delta^{bI}_{aJ}(x-y)$}}{o1}
  \fmflabel{\raisebox{22pt}{\hspace{18pt} $J,b$ 
  \hspace{-18pt}}}{i1}
\end{fmfgraph*}   
\end{fmffile} \\
\hspace*{0.5cm}
\begin{fmffile}{S+S}
\begin{fmfgraph*}(90,50) 
  \fmfleft{i1} \fmfright{o1}
  \fmf{scalar}{i1,o1}
  \fmflabel{\raisebox{22pt}{\hspace{-20pt} $a$ \hspace{20pt}} 
  \raisebox{12pt}{$\longrightarrow \hspace{0.2cm} 
  \langle S^{b\dagger}(x) S_{a}(y) \rangle_{{\rm free}} = 
  \begin{displaystyle} \frac{\delta^{b}_{a}}{\Box}\end{displaystyle}
  \delta(x-y)=2\Delta^{b}_{a}(x-y)$}}{o1}
  \fmflabel{\raisebox{22pt}{\hspace{18pt} $b$ 
  \hspace{-18pt}}}{i1}
\end{fmfgraph*}   
\end{fmffile} \\ 
\hspace*{0.5cm}
\begin{fmffile}{CD}
\begin{fmfgraph*}(90,50) 
  \fmfleft{i1} \fmfright{o1}
  \fmf{dashes}{i1,v}
  \fmf{dashes}{v,o1}
  \fmfv{d.sh=cross,d.size=3mm}{v}
  \fmflabel{\raisebox{22pt}{\hspace{-20pt} $a$ \hspace{20pt}} 
  \raisebox{12pt}{$\longrightarrow \hspace{0.2cm} 
  \langle C^{b}(x) D_{a}(y) \rangle_{{\rm free}} = 
  \begin{displaystyle} \frac{\delta^{b}_{a}}{\Box}\end{displaystyle}
  \delta(x-y)=\Delta^{b}_{a}(x-y)$}}{o1}
  \fmflabel{\raisebox{22pt}{\hspace{18pt} $b$ 
  \hspace{-18pt}}}{i1}
\end{fmfgraph*}   
\end{fmffile} \\
\hspace*{0.5cm}
\begin{fmffile}{DD}
\begin{fmfgraph*}(90,50) 
  \fmfleft{i1} \fmfright{o1}
  \fmf{dashes}{i1,o1}
  \fmflabel{\raisebox{22pt}{\hspace{-20pt} $a$ \hspace{20pt}} 
  \raisebox{12pt}{$\longrightarrow \hspace{0.2cm} \langle D^{b}(x) 
  D_{a}(y) \rangle_{{\rm free}} = -\delta^{b}_{a}\delta(x-y)$}}{o1}
  \fmflabel{\raisebox{22pt}{\hspace{18pt} $b$ 
  \hspace{-18pt}}}{i1}
\end{fmfgraph*}   
\end{fmffile} \\
\hspace*{0.5cm}
\begin{fmffile}{F+F}
\begin{fmfgraph*}(90,50) 
  \fmfleft{i1} \fmfright{o1}
  \fmf{scalar}{i1,o1}
  \fmflabel{\raisebox{22pt}{\hspace{-20pt} $I,a$ \hspace{20pt}} 
  \raisebox{12pt}{$\longrightarrow \hspace{0.2cm} 
  \langle F^{b\dagger}_{J}(x) F^{I}_{a}(y) \rangle_{{\rm free}} = 
  \delta^{I}_{J}\delta^{b}_{a}\delta(x-y)$}}{o1}
  \fmflabel{\raisebox{22pt}{\hspace{18pt} $J,b$ 
  \hspace{-18pt}}}{i1}
\end{fmfgraph*}   
\end{fmffile} \\
\hspace*{0.5cm}
\begin{fmffile}{AA}
\begin{fmfgraph*}(90,50) 
  \fmfleft{i1} \fmfright{o1}
  \fmf{boson}{i1,o1}
  \fmflabel{\raisebox{22pt}{\hspace{-20pt} $\nu,a$ \hspace{10pt}} 
  \raisebox{12pt}{$\longrightarrow \hspace{0.2cm} 
  \langle A^{b}_{\mu}(x)A_{a\nu}(y)\rangle_{{\rm free}}=\begin{displaystyle} 
  -\frac{\delta_{\mu\nu}\delta^{b}_{a}}{\Box} \end{displaystyle} 
  \delta(x-y)=\Delta^{b}_{a\mu\nu}(x-y)$}}{o1}
  \fmflabel{\raisebox{22pt}{\hspace{18pt} $\mu,b$ 
  \hspace{-18pt}}}{i1}
\end{fmfgraph*}   
\end{fmffile} \\
\hspace*{0.5cm}
\begin{fmffile}{chilam}
\begin{fmfgraph*}(90,50) 
  \fmfleft{i1} \fmfright{o1}
  \fmf{fermion}{v,i1}
  \fmf{fermion}{v,o1}
  \fmfv{d.sh=cross,d.size=3mm}{v}
  \fmflabel{\raisebox{22pt}{\hspace{-20pt} $\beta,a$ \hspace{10pt}} 
  \raisebox{12pt}{$\longrightarrow \hspace{0.2cm} \langle 
  \chi^{b\alpha}(x) \lambda_{a}^{\beta}(y) \rangle_{{\rm free}} = 
  \begin{displaystyle} -\frac{\varepsilon^{\alpha\beta}\delta^{b}_{a}}
  {\Box} \end{displaystyle}
  \delta(x-y)=R_{a}^{b\alpha\beta}(x-y)$}}{o1}  
  \fmflabel{\raisebox{22pt}{\hspace{18pt} $\alpha,b$ 
  \hspace{-18pt}}}{i1}
\end{fmfgraph*}
\end{fmffile} \\
\hspace*{0.5cm}
\begin{fmffile}{bchiblam}
\begin{fmfgraph*}(90,50) 
  \fmfleft{i1} \fmfright{o1}
  \fmf{fermion}{i1,v}
  \fmf{fermion}{o1,v}
  \fmfv{d.sh=cross,d.size=3mm}{v}
  \fmflabel{\raisebox{22pt}{\hspace{-20pt} ${\dot\beta},a$\hspace{10pt}} 
  \raisebox{12pt}{$\longrightarrow \hspace{0.2cm} 
  \langle {\overline \chi}^{b\dot\alpha}
  (x) {\overline \lambda}_{a}^{\dot\beta}(y) \rangle_{{\rm free}} = 
  \begin{displaystyle} - \frac{\varepsilon^{{\dot\alpha}{\dot\beta}}
  \delta^{b}_{a}}{\Box} \end{displaystyle}\delta(x-y)=
  {\overline R}_{a}^{b{\dot\alpha}{\dot\beta}}(x-y)$}}{o1}
  \fmflabel{\raisebox{22pt}{\hspace{18pt} ${\dot\alpha},b$ 
  \hspace{-18pt}}}{i1}  
\end{fmfgraph*} 
\end{fmffile} \\
\hspace*{0.5cm}
\begin{fmffile}{blamlam}
\begin{fmfgraph*}(90,50) 
  \fmfleft{i1} \fmfright{o1}
  \fmf{fermion}{i1,o1}
  \fmflabel{\raisebox{-20pt}{\hspace{-20pt} $\alpha,a$ \hspace{20pt}} 
  \raisebox{-30pt}{$\longrightarrow \hspace{0.2cm} 
  \langle {\overline \lambda}^{b\dot\alpha}(x) \lambda_{a}^{\alpha}(y) 
  \rangle_{{\rm free}} = \begin{displaystyle} 
  \frac{\delta^{b}_{a}{\overline \sigma}_{{\dot\alpha}\alpha}^{\mu}
  \partial_{\mu}}{\Box} 
  \end{displaystyle} \delta(x-y)=$
  \raisebox{-1cm}{\hspace{-4cm}=${\overline S}^{b}_{a{\dot\alpha}\alpha}
  (x-y)$}}}{o1}
  \fmflabel{\raisebox{22pt}{\hspace{18pt} ${\dot\alpha},b$ 
  \hspace{-18pt}}}{i1}  
\end{fmfgraph*}   
\end{fmffile} 

\vspace{0.3cm}
\noindent
\hspace*{0.5cm}
\begin{fmffile}{bpsipsi}
\begin{fmfgraph*}(90,50) 
  \fmfleft{i1} \fmfright{o1}
  \fmf{fermion}{i1,o1}
  \fmflabel{\raisebox{-20pt}{\hspace{-25pt} $\alpha,J,a$ \hspace{20pt}} 
  \raisebox{-30pt}{$\longrightarrow \hspace{0.2cm} 
  \langle {\overline \psi}_{I}^{b\dot\alpha}(x)\psi_{a}^{\alpha J}(y) 
  \rangle_{{\rm free}} = \begin{displaystyle} 
  \frac{\delta^{b}_{a}\delta^{J}_{I}{\overline \sigma}_{{\dot\alpha}
  \alpha}^{\mu}\partial_{\mu}}{\Box} 
  \end{displaystyle} \delta(x-y)=$
  \raisebox{-1cm}{\hspace{-4.35cm}=${\overline S}^{bJ}_{aI{\dot\alpha}
  \alpha}(x-y)$}}}{o1}
  \fmflabel{\raisebox{22pt}{\hspace{23pt} ${\dot\alpha},I,b$ 
  \hspace{-18pt}}}{i1}
\end{fmfgraph*}   
\end{fmffile}

\vspace{0.3cm}
\noindent
In conclusion beyond the ordinary propagators for the physical 
fields, $\varphi$, $\psi$, $\lambda$ and $A_{\mu}$, and those for the 
auxiliary fields, $F$ and $D$, one further obtains the propagators 
$\langle S^{\dagger}S\rangle$, $\langle CD\rangle$, $\langle 
\chi\lambda \rangle$ and $\langle {\overline \chi} {\overline \lambda}
\rangle$. The latter are not present in the Wess--Zumino gauge.

\subsection{One loop corrections to the propagator of the fermions 
belonging to the chiral multiplet}

The one-loop correction to the propagator of the fermions $\psi^{I}$ 
in the chiral multiplet is the simplest to compute. 
In the WZ gauge there are three 
contributions at the one loop level, that will be shown to lead to a 
logarithmically divergent result. 

From the action (\ref{freeact}), (\ref{intaction}), with 
$C$=$\chi$=$S$=0 
one obtains the following three diagrams 

\vspace{0.3cm}
\noindent
\hspace*{2cm}
\begin{fmffile}{psipsiA}
 \begin{fmfgraph*}(150,80) 
  \fmfleft{i1} \fmfright{o1}
  \fmf{fermion,tension=1.2}{i1,v1}
  \fmf{boson,label=$A_{\mu}A_{\nu}$,left=0.6,tension=0.5}{v1,v2}
  \fmf{fermion,label=${\overline\psi}\psi$,right=0.6,tension=0.5}{v1,v2}
  \fmf{fermion,tension=1.2}{v2,o1}
  \fmfv{l=${\overline \psi}^{a}_{J}(x)$,l.a=180}{i1}
  \fmfv{l=$\psi^{I}_{b}(y)$ \hspace{0.5cm} $\longrightarrow ~~~ 
        A^{aI}_{bJ}(x;y)$,l.a=0}{o1}
  \fmfdot{v1,v2}
 \end{fmfgraph*}   
\end{fmffile} \\
\hspace*{2cm}
\begin{fmffile}{psipsiB}
 \begin{fmfgraph*}(150,80) 
  \fmfleft{i1} \fmfright{o1}
  \fmf{fermion,tension=1.2}{i1,v1}
  \fmf{scalar,label=$\varphi\varphi^{\dagger}$,left=0.6,tension=0.5}
      {v2,v1}
  \fmf{fermion,label=$\psi{\overline\psi}$,right=0.6,tension=0.5}{v2,v1}
  \fmf{fermion,tension=1.2}{v2,o1}
  \fmfv{l=${\overline \psi}^{a}_{J}(x)$,l.a=180}{i1}
  \fmfv{l=$\psi^{I}_{b}(y)$ \hspace{0.5cm} $\longrightarrow ~~~ 
        B^{aI}_{bJ}(x;y)$,l.a=0}{o1}
  \fmfdot{v1,v2}
 \end{fmfgraph*}   
\end{fmffile} \\
\hspace*{2cm}
\begin{fmffile}{psipsiC}
 \begin{fmfgraph*}(150,80) 
  \fmfleft{i1} \fmfright{o1}
  \fmf{fermion,tension=1.2}{i1,v1}
  \fmf{scalar,label=$\varphi^{\dagger}\varphi$,right=0.6,tension=0.5}
      {v1,v2}
  \fmf{fermion,label=$\lambda{\overline\lambda}$,right=0.6,tension=0.5}
      {v2,v1}
  \fmf{fermion,tension=1.2}{v2,o1}
  \fmfv{l=${\overline \psi}^{a}_{J}(x)$,l.a=180}{i1}
  \fmfv{l=$\psi^{I}_{b}(y)$ \hspace{0.5cm} $\longrightarrow ~~~ 
        C^{aI}_{bJ}(x;y)$,l.a=0}{o1}
  \fmfdot{v1,v2}
 \end{fmfgraph*}   
\end{fmffile} \\
The three contributions depicted above can be easily evaluated and 
give the results
\begin{eqnarray}
	A^{{\dot\alpha}\alpha aI}_{bJ}(x;y) &=& -\frac{1}{2}g^{2}
	f^{d}{}_{ef}f^{l}{}_{mn}\int d^{4}x_{1}
	d^{4}x_{2} \, \left\{ \Delta^{me}_{\nu\mu}(x_{2}-x_{1}) \left[
	{\overline S}^{{\dot\alpha}\beta 
	aI}_{lL}(x-x_{2})\sigma^{\nu}_{\beta{\dot \gamma}} \cdot
	\right. \right.\nonumber \\
	&& \left. \left. \hspace{2cm}  \cdot 
	{\overline S}^{{\dot\gamma}\gamma nL}_{dK}(x_{2}-x_{1})  
	\sigma^{\mu}_{\gamma{\dot\beta}} {\overline S}^{{\dot\beta}\alpha 
	fK}_{bJ}(x_{1}-y)\right] \right\} \, ,
	\label{bpsipsi-a}
\end{eqnarray}
\begin{eqnarray}
    B^{{\dot\alpha}\alpha aI}_{bJ}(x;y) &=& -\frac{1}{2}g^{2}
    \varepsilon^{LMN}\varepsilon^{PQ}{}_{R}f_{de}{}^{f}f_{lmn}\int 
    d^{4}x_{1}d^{4}x_{2} \, \left\{ \Delta^{ld}_{PL}(x_{2}-x_{1}) 
    \rule{0pt}{13pt} \cdot \right. \nonumber \\ 
    && \left. \cdot
    \left[ {\overline S}^{{\dot\alpha}\beta 
	aI}_{fN}(x-x_{2}) S_{\beta{\dot\beta}MQ}^{em}(x_{2}-x_{1})
	{\overline S}^{{\dot\beta}\alpha 
	nR}_{bJ}(x_{1}-y)\right] \right\} \, ,
\end{eqnarray}
\begin{eqnarray}
    C^{{\dot\alpha}\alpha aI}_{bJ}(x;y) &=& - g^{2}
    f_{de}{}^{f}f_{lm}{}^{n} \int 
    d^{4}x_{1}d^{4}x_{2} \, \left\{ \Delta^{lK}_{fL}(x_{2}-x_{1}) 
    \rule{0pt}{13pt} \cdot \right. \nonumber \\ 
    && \left.  \cdot
    \left[ {\overline S}^{{\dot\alpha}\beta 
	aI}_{nK}(x-x_{2}) S_{\beta{\dot\beta}}^{em}(x_{2}-x_{1})
	{\overline S}^{{\dot\beta}\alpha 
	dL}_{bJ}(x_{1}-y)\right] \right\} \, .
\end{eqnarray}
Taking the Fourier transform one obtains 
\begin{displaymath}
	\tilde{A}^{{\dot\alpha}\alpha aI}_{bJ}(p) = \frac{i}{2} g^{2}
	N(N^{2}-1)\delta^{a}_{b} \delta^{I}_{J} 
	\tilde{\overline{S}}^{{\dot\alpha}\beta}(p) 
	\sigma^{\lambda}_{\beta{\dot\beta}}p_{\lambda} 
	\tilde{\overline{S}}^{{\dot\beta}\alpha}(p) 
	\left[ \int \frac{d^{4}k}{(2\pi)^{4}} 
	\frac{1}{\left(k+\frac{p}{2}\right)^{2}\left(k-\frac{p}{2}
	\right)^{2}} \right] \, ,
\end{displaymath}
\begin{displaymath}
    \tilde{B}^{{\dot\alpha}\alpha aI}_{bJ}(p) = 
    -\tilde{C}^{{\dot\alpha}\alpha aI}_{bJ}(p) = 
    \tilde{A}^{{\dot\alpha}\alpha aI}_{bJ}(p) \, ,
\end{displaymath}
where 
\begin{displaymath}
    \tilde{\overline{S}}^{{\dot\alpha}\alpha}(p) = -i 
    \frac{{\overline \sigma}_{\mu}^{{\dot\alpha}\alpha} 
    p^{\mu}}{p^{2}} \, .
\end{displaymath}
In conclusion the one-loop correction to the fermion propagator 
in the Wess--Zumino gauge turns out to be logarithmically divergent
\begin{eqnarray} 
	&& \hspace*{-0.5cm} 
	\langle \left({\overline \psi}\psi\right)^{{\dot\alpha}
	\alpha aI}_{bJ} \rangle^{{\rm 1-loop,WZ}}_{{\rm FT}} = 
	\tilde{A}^{{\dot\alpha}\alpha aI}_{bJ}(p) + 
    \tilde{B}^{{\dot\alpha}\alpha aI}_{bJ}(p) + 
    \tilde{C}^{{\dot\alpha}\alpha aI}_{bJ}(p) = \nonumber \\ 
    && \hspace*{-0.5cm}  = \frac{i}{2} g^{2} N(N^{2}-1) 
	\delta^{a}_{b} \delta^{I}_{J} 
	\tilde{\overline{S}}^{{\dot\alpha}\beta}(p) 
	\sigma^{\lambda}_{\beta{\dot\beta}}p_{\lambda} 
	\tilde{\overline{S}}^{{\dot\beta}\alpha}(p) 
	\left[ \int \frac{d^{4}k}{(2\pi)^{4}} 
	\frac{1}{\left(k+\frac{p}{2}\right)^{2}\left(k-\frac{p}{2}
	\right)^{2}} \right] \, . ~~~
	\label{onelooppsiwz}
\end{eqnarray}
Equation (\ref{onelooppsiwz}) shows that a logarithmically divergent wave 
function renormalization is required in the WZ gauge.

This divergence will now be proved to be a gauge artifact due to the 
choice of the WZ gauge. In fact if the fields $C$, $\chi$ and $S$ are 
included two further contributions must be taken into account, which 
correspond to the diagrams 

\vspace{0.3cm}
\noindent
\hspace*{2cm}
\begin{fmffile}{psipsiD}
 \begin{fmfgraph*}(150,80) 
  \fmfleft{i1} \fmfright{o1}
  \fmf{fermion,tension=1.5}{i1,v1}
  \fmf{fermion,tension=1.5}{v2,o1}
  \fmf{phantom,right=0.6,tension=0.4,tag=1}{v1,v2}
  \fmf{phantom,right=0.6,tension=0.4,tag=2}{v2,v1}
  \fmf{phantom,left=0.6,tension=0.5,tag=3}{v1,v2} 
  \fmfv{l=${\overline \psi}^{a}_{J}(x)$,l.a=180}{i1}
  \fmfv{l=$\psi^{I}_{b}(y)$ \hspace{0.5cm} $\longrightarrow ~~~ 
        D^{aI}_{bJ}(x;y)$,l.a=0}{o1}
  \fmfdot{v1,v2}
  \fmffreeze
  \fmfipath{p[]}
  \fmfiset{p1}{vpath1(__v1,__v2)}
  \fmfiset{p2}{vpath2(__v2,__v1)}
  \fmfiset{p3}{vpath3(__v1,__v2)}
  \fmfi{scalar,l=$\varphi^{\dagger}\varphi$}{subpath (0,length(p1)) of p1}
  \fmfi{fermion}{subpath (0,length(p3)/2) of p3}
  \fmfi{fermion}{subpath (0,length(p2)/2) of p2}  
  \fmfiv{d.sh=cross,d.si=3mm,l=${\overline\lambda}{\overline\chi}$,
       l.a=90}{point length(p2)/2 of p2}      
 \end{fmfgraph*}   
\end{fmffile} \\
\hspace*{2cm}
\begin{fmffile}{psipsiE}
 \begin{fmfgraph*}(150,80) 
  \fmfleft{i1} \fmfright{o1}
  \fmf{fermion,tension=1.5}{i1,v1}
  \fmf{fermion,tension=1.5}{v2,o1}
  \fmf{phantom,right=0.6,tension=0.4,tag=1}{v1,v2}
  \fmf{phantom,right=0.6,tension=0.4,tag=2}{v2,v1}
  \fmf{phantom,left=0.6,tension=0.5,tag=3}{v1,v2} 
  \fmfv{l=${\overline \psi}^{a}_{J}(x)$,l.a=180}{i1}
  \fmfv{l=$\psi^{I}_{b}(y)$ \hspace{0.5cm} $\longrightarrow ~~~ 
        E^{aI}_{bJ}(x;y)$,l.a=0}{o1}
  \fmfdot{v1,v2}
  \fmffreeze
  \fmfipath{p[]}
  \fmfiset{p1}{vpath1(__v1,__v2)}
  \fmfiset{p2}{vpath2(__v2,__v1)}
  \fmfiset{p3}{vpath3(__v1,__v2)}
  \fmfi{scalar,l=$\varphi^{\dagger}\varphi$}{subpath (0,length(p1)) of p1}
  \fmfi{fermion}{subpath (length(p3)/2,length(p3)) of p3}
  \fmfi{fermion}{subpath (length(p2)/2,length(p2)) of p2}  
  \fmfiv{d.sh=cross,d.si=3mm,l=$\lambda\chi$,
       l.a=90}{point length(p2)/2 of p2}      
 \end{fmfgraph*}   
\end{fmffile} 

The contribution of these two diagrams exactly cancels the divergence 
in equation (\ref{onelooppsiwz}) giving a net vanishing one-loop result 
for the propagator. In fact the computation of $D$ and $E$ gives
\begin{eqnarray}
    && D^{{\dot\alpha}\alpha aI}_{bJ}(x;y) = -\frac{1}{2}g^{2}
    f_{de}{}^{f}f_{lm}{}^{n} \int 
    d^{4}x_{1}d^{4}x_{2} \, \left\{ \Delta^{lK}_{fL}(x_{2}-x_{1}) 
    \rule{0pt}{18pt}\cdot \right. \nonumber \\ 
    && \left.  \cdot
    \left[ {\overline S}^{{\dot\alpha}\gamma aI}_{mK}
	(x-x_{2}) \sigma^{\mu}_{\gamma{\dot\gamma}} 
	\left(\partial_{\mu}^{(2)} R^{{\dot\gamma}ne}_{\dot\beta}
	(x_{2}-x_{1}) \right)
	{\overline S}^{{\dot\beta}\alpha 
	dL}_{bJ}(x_{1}-y)\right] \right\} \, ,
\end{eqnarray}
\begin{displaymath}
	E^{{\dot\alpha}\alpha aI}_{bJ}(x;y) = 
	D^{{\dot\alpha}\alpha aI}_{bJ}(x;y) \, .
\end{displaymath}
In momentum space one has 
\begin{displaymath}
	\tilde{D}^{{\dot\alpha}\alpha aI}_{bJ}(p) = -\frac{i}{4} g^{2}
	N(N^{2}-1)\delta^{a}_{b} \delta^{I}_{J} 
	\tilde{\overline{S}}^{{\dot\alpha}\beta}(p) 
	\sigma^{\lambda}_{\beta{\dot\beta}}p_{\lambda} 
	\tilde{\overline{S}}^{{\dot\beta}\alpha}(p) 
	\left[ \int \frac{d^{4}k}{(2\pi)^{4}} 
	\frac{1}{\left(k+\frac{p}{2}\right)^{2}\left(k-\frac{p}{2}
	\right)^{2}} \right] \, ,
\end{displaymath}
\begin{displaymath}
	\tilde{E}^{{\dot\alpha}\alpha aI}_{bJ}(p) =
	\tilde{D}^{{\dot\alpha}\alpha aI}_{bJ}(p) \, .
\end{displaymath}
In conclusion summing up all the terms gives
\begin{eqnarray}
    && \hspace{-1cm}\langle \left({\overline \psi}\psi
	\right)^{{\dot\alpha}\alpha aI}_{bJ} 
	\rangle^{{\rm 1-loop}}_{{\rm FT}} = 
    \nonumber \\
    && \hspace{-1cm} = \tilde{A}^{{\dot\alpha}\alpha aI}_{bJ}(p) + 
    \tilde{B}^{{\dot\alpha}\alpha aI}_{bJ}(p) + 
    \tilde{C}^{{\dot\alpha}\alpha aI}_{bJ}(p) +
    \tilde{D}^{{\dot\alpha}\alpha aI}_{bJ}(p) +
    \tilde{E}^{{\dot\alpha}\alpha aI}_{bJ}(p) = 0 
    \rule{0pt}{18pt}\, ,
    \label{onelooppsi}
\end{eqnarray}
so that the one-loop correction to the $\langle{\overline 
\psi}\psi\rangle$ propagator, if the Wess--Zumino 
gauge is not exploited to perform the calculation, 
is zero, as expected in ${\cal N}$=4 super Yang--Mills theory.

Notice that the insertion of tadpoles such as 

\vspace*{0.2cm}
\noindent
\hspace*{0.7cm}
\begin{fmffile}{tadpole}
 \begin{fmfgraph*}(100,40) 
  \fmfleft{i1,i2} \fmfright{o1,o2}
  \fmf{phantom,tension=1}{i1,v1}
  \fmf{phantom,tension=1}{i2,v1}
  \fmf{boson,tension=1}{v1,v2}
  \fmf{phantom,tension=1}{v2,o1}
  \fmf{phantom,tension=1}{v2,o2}
  \fmf{fermion,right=30,tension=0.4}{v2,v2}
  \fmfdot{v2}
 \end{fmfgraph*}   
\end{fmffile} \\

\vspace*{-0.8cm}
\noindent 
in a diagram gives a vanishing result because all the propagators are 
diagonal in colour space, so that the tadpole contains a factor 
$\delta_{ab}f^{abc}\equiv 0$. The same is true also for diagrams in 
${\cal N}$=1 superspace that will be discussed in subsequent sections. 

\subsection{One loop corrections to the propagator of the scalars 
belonging to the chiral multiplet}

The calculation of the propagator $\langle\varphi^{\dagger}\varphi
\rangle$ is more complicated because more diagrams are involved. In 
particular tadpole type graphs, coming from the term 
$\Phi^{\dagger}V^{2}\Phi$ in the action, are present. However the 
result is analogous to what was found for the 
$\langle{\overline\psi}\psi\rangle$ propagator: in the WZ gauge the 
one-loop correction is logarithmically divergent, requiring a wave 
function renormalization, but when the contributions neglected in the 
WZ gauge are considered the total one-loop result is vanishing.

The diagrams contributing to $\langle\varphi^{\dagger}\varphi\rangle$ 
at the one-loop level in the Wess--Zumino gauge are the following

\vspace{0.3cm}
\noindent
\hspace*{2cm}
\begin{fmffile}{phiphiA}
 \begin{fmfgraph*}(150,75) 
  \fmfleft{i1} \fmfright{o1}
  \fmf{scalar,tension=1.2}{i1,v1}
  \fmf{boson,label=$A_{\mu}A_{\nu}$,left=0.6,tension=0.5}{v1,v2}
  \fmf{scalar,label=$\varphi^{\dagger}\varphi$,right=0.6,tension=0.5}
      {v1,v2}
  \fmf{scalar,tension=1.2}{v2,o1}
  \fmfv{l=${\varphi}^{a\dagger}_{J}(x)$,l.a=180}{i1}
  \fmfv{l=$\varphi^{I}_{b}(y)$ \hspace{0.5cm} $\longrightarrow ~~~ 
        A^{aI}_{bJ}(x;y)$,l.a=0}{o1}
  \fmfdot{v1,v2}
 \end{fmfgraph*}   
\end{fmffile} 

\vspace*{0.2cm}
\noindent
\hspace*{2cm}
\begin{fmffile}{phiphiB}
 \begin{fmfgraph*}(150,75) 
  \fmfleft{i1} \fmfright{o1}
  \fmf{scalar,tension=1.2}{i1,v1}
  \fmf{fermion,label=${\overline\lambda}\lambda$,left=0.6,tension=0.5}
      {v1,v2}
  \fmf{fermion,label=${\overline\psi}\psi$,right=0.6,tension=0.5}{v1,v2}
  \fmf{scalar,tension=1.2}{v2,o1}
  \fmfv{l=${\varphi}^{a\dagger}_{J}(x)$,l.a=180}{i1}
  \fmfv{l=$\varphi^{I}_{b}(y)$ \hspace{0.5cm} $\longrightarrow ~~~ 
        B^{aI}_{bJ}(x;y)$,l.a=0}{o1}
  \fmfdot{v1,v2}
 \end{fmfgraph*}   
\end{fmffile} \\

\vspace*{-0.3cm}
\noindent
\hspace*{2cm}
\begin{fmffile}{phiphiC}
 \begin{fmfgraph*}(150,75) 
  \fmfleft{i1} \fmfright{o1}
  \fmf{scalar,tension=1.2}{i1,v1}
  \fmf{fermion,label=$\psi{\overline\psi}$,left=0.6,tension=0.5}{v2,v1}
  \fmf{fermion,label=$\psi{\overline\psi}$,right=0.6,tension=0.5}{v2,v1}
  \fmf{scalar,tension=1.2}{v2,o1}
  \fmfv{l=${\varphi}^{a\dagger}_{J}(x)$,l.a=180}{i1}
  \fmfv{l=$\varphi^{I}_{b}(y)$ \hspace{0.5cm} $\longrightarrow ~~~ 
        C^{aI}_{bJ}(x;y)$,l.a=0}{o1}
  \fmfdot{v1,v2}
 \end{fmfgraph*}   
\end{fmffile} \\

\vspace*{0.6cm}
\noindent
\hspace*{2cm}
\begin{fmffile}{phiphiD}
 \begin{fmfgraph*}(150,75) 
  \fmfleft{i1} \fmfright{o1}
  \fmf{scalar,tension=1.4}{i1,v1}
  \fmf{scalar,tension=1.4}{v1,o1}
  \fmf{phantom,tension=1}{v1,v1}
  \fmfv{l=$\varphi^{a\dagger}_{J}(x)$,l.a=180}{i1}
  \fmfv{l=$\varphi^{I}_{b}(y)$ \hspace{0.5cm} $\longrightarrow ~~~ 
        D^{aI}_{bJ}(x;y)$,l.a=0}{o1}
  \fmfdot{v1}
  \fmffreeze
  \fmfipath{p[]}
  \fmfiset{p1}{vpath(__v1,__v1)}
  \fmfi{boson}{subpath (0,length(p1)) of p1}
  \fmfiv{l=$A_{\mu}A_{\nu}$,l.a=90}{point length(p1)/2 of p1}      
 \end{fmfgraph*}   
\end{fmffile} 

\vspace*{0.3cm}
\noindent
\hspace*{2cm}
\begin{fmffile}{phiphiE}
 \begin{fmfgraph*}(150,75) 
  \fmfleft{i1} \fmfright{o1}
  \fmf{scalar,tension=1.2}{i1,v1}
  \fmf{scalar,label=$\varphi^{\dagger}\varphi$,left=0.6,tension=0.5}
      {v1,v2}
  \fmf{dashes,label=$DD$,right=0.6,tension=0.5}{v1,v2}
  \fmf{scalar,tension=1.2}{v2,o1}
  \fmfv{l=${\varphi}^{a\dagger}_{J}(x)$,l.a=180}{i1}
  \fmfv{l=$\varphi^{I}_{b}(y)$ \hspace{0.5cm} $\longrightarrow ~~~ 
        E^{aI}_{bJ}(x;y)$,l.a=0}{o1}
  \fmfdot{v1,v2}
 \end{fmfgraph*}   
\end{fmffile} 

\vspace*{0.3cm}
\noindent
\hspace*{2cm}
\begin{fmffile}{phiphiF}
 \begin{fmfgraph*}(150,75) 
  \fmfleft{i1} \fmfright{o1}
  \fmf{scalar,tension=1.2}{i1,v1}
  \fmf{scalar,label=$\varphi\varphi^{\dagger}$,left=0.6,tension=0.5}
      {v2,v1}
  \fmf{scalar,label=$FF^{\dagger}$,right=0.6,tension=0.5}{v2,v1}
  \fmf{scalar,tension=1.2}{v2,o1}
  \fmfv{l=${\varphi}^{a\dagger}_{J}(x)$,l.a=180}{i1}
  \fmfv{l=$\varphi^{I}_{b}(y)$ \hspace{0.5cm} $\longrightarrow ~~~ 
        F^{aI}_{bJ}(x;y)$,l.a=0}{o1}
  \fmfdot{v1,v2}
 \end{fmfgraph*}   
\end{fmffile} 

\vspace*{0.3cm}
\noindent
The last three diagrams are all tadpole corrections since the free 
propagators for the auxiliary fields $F$ and $D$ are 
$\delta$-functions, so that the corresponding lines shrink to a point. 
Each of these diagrams is quadratically divergent. There is also a 
quadratically divergent contribution coming from the two diagrams 
$B(x,y)$ and $C(x,y)$ and the sum of all these terms exactly vanishes.
The sum of the previous diagrams gives a final result that is 
logarithmically divergent and in momentum space is schematically of 
the form
\begin{equation}
	\langle (\varphi^{\dagger}\varphi)^{aI}_{bJ} 
	\rangle^{{\rm 1-loop,WZ}}_{{\rm FT}} \sim g^{2} \delta^{I}_{J}
	\delta^{a}_{b} p^{2} \int \frac{d^{4}k}{(2\pi)^{4}} 
	\frac{1}{\left(k+\frac{p}{2}\right)^{2}
	\left(k-\frac{p}{2}\right)^{2}} \, .
	\label{phioneloopwz}
\end{equation}
Just like in the case of $\langle {\overline \psi}\psi\rangle$ this 
divergence corresponds to a wave function renormalization for the 
field $\varphi(x)$. 

Taking into account the contribution of the fields $C$, $\chi$ and $S$ 
there are eight more diagrams to be calculated 

\vspace{-0.4cm}
\noindent
\hspace*{2cm}
\begin{fmffile}{phiphiG}
 \begin{fmfgraph*}(150,75) 
  \fmfleft{i1} \fmfright{o1}
  \fmf{scalar,tension=1.5}{i1,v1}
  \fmf{scalar,tension=1.5}{v2,o1}
  \fmf{phantom,right=0.6,tension=0.4,tag=1}{v1,v2}
  \fmf{phantom,right=0.6,tension=0.4,tag=2}{v2,v1}
  \fmf{phantom,left=0.6,tension=0.4,tag=3}{v1,v2} 
  \fmfv{l=$\varphi^{a\dagger}_{J}(x)$,l.a=180}{i1}
  \fmfv{l=$\varphi^{I}_{b}(y)$ \hspace{0.5cm} $\longrightarrow ~~~ 
        G^{aI}_{bJ}(x;y)$,l.a=0}{o1}
  \fmfdot{v1,v2}
  \fmffreeze
  \fmfipath{p[]}
  \fmfiset{p1}{vpath1(__v1,__v2)}
  \fmfiset{p2}{vpath2(__v2,__v1)}
  \fmfiset{p3}{vpath3(__v1,__v2)}
  \fmfi{fermion,l=${\overline\psi}\psi$}{subpath (0,length(p1)) of p1}
  \fmfi{fermion}{subpath (length(p3)/2,length(p3)) of p3}
  \fmfi{fermion}{subpath (length(p2)/2,length(p2)) of p2}  
  \fmfiv{d.sh=cross,d.si=3mm,l=$\chi\lambda$,
       l.a=90}{point length(p2)/2 of p2}      
 \end{fmfgraph*}   
\end{fmffile} 

\vspace*{0.4cm}
\noindent
\hspace*{2cm}
\begin{fmffile}{phiphiH}
 \begin{fmfgraph*}(150,75) 
  \fmfleft{i1} \fmfright{o1}
  \fmf{scalar,tension=1.5}{i1,v1}
  \fmf{scalar,tension=1.5}{v2,o1}
  \fmf{phantom,right=0.6,tension=0.4,tag=1}{v1,v2}
  \fmf{phantom,right=0.6,tension=0.4,tag=2}{v2,v1}
  \fmf{phantom,left=0.6,tension=0.4,tag=3}{v1,v2} 
  \fmfv{l=$\varphi^{a\dagger}_{J}(x)$,l.a=180}{i1}
  \fmfv{l=$\varphi^{I}_{b}(y)$ \hspace{0.5cm} $\longrightarrow ~~~ 
        H^{aI}_{bJ}(x;y)$,l.a=0}{o1}
  \fmfdot{v1,v2}
  \fmffreeze
  \fmfipath{p[]}
  \fmfiset{p1}{vpath1(__v1,__v2)}
  \fmfiset{p2}{vpath2(__v2,__v1)}
  \fmfiset{p3}{vpath3(__v1,__v2)}
  \fmfi{fermion,l=${\overline\psi}\psi$}{subpath (0,length(p1)) of p1}
  \fmfi{fermion}{subpath (0,length(p3)/2) of p3}
  \fmfi{fermion}{subpath (0,length(p2)/2) of p2}  
  \fmfiv{d.sh=cross,d.si=3mm,l=${\overline\lambda}{\overline\chi}$,
       l.a=90}{point length(p2)/2 of p2}      
 \end{fmfgraph*}   
\end{fmffile} 

\vspace*{0.4cm}
\noindent
\hspace*{2cm}
\begin{fmffile}{phiphiL}
 \begin{fmfgraph*}(150,75) 
  \fmfleft{i1} \fmfright{o1}
  \fmf{scalar,tension=1.5}{i1,v1}
  \fmf{scalar,tension=1.5}{v2,o1}
  \fmf{phantom,right=0.6,tension=0.4,tag=1}{v1,v2}
  \fmf{phantom,right=0.6,tension=0.4,tag=2}{v2,v1}
  \fmf{phantom,left=0.6,tension=0.4,tag=3}{v1,v2} 
  \fmfv{l=$\varphi^{a\dagger}_{J}(x)$,l.a=180}{i1}
  \fmfv{l=$\varphi^{I}_{b}(y)$ \hspace{0.5cm} $\longrightarrow ~~~ 
        L^{aI}_{bJ}(x;y)$,l.a=0}{o1}
  \fmfdot{v1,v2}
  \fmffreeze
  \fmfipath{p[]}
  \fmfiset{p1}{vpath1(__v1,__v2)}
  \fmfiset{p2}{vpath2(__v2,__v1)}
  \fmfiset{p3}{vpath3(__v1,__v2)}
  \fmfi{scalar,l=${\overline\psi}\psi$}{subpath (0,length(p1)) of p1}
  \fmfi{dashes}{subpath (0,length(p2)) of p2}
  \fmfiv{d.sh=cross,d.si=3mm,l=$CD$,
       l.a=90}{point length(p2)/2 of p2}      
 \end{fmfgraph*}   
\end{fmffile} 

\vspace*{0.5cm}
\noindent
\hspace*{2cm}
\begin{fmffile}{phiphiM}
 \begin{fmfgraph*}(150,70) 
  \fmfleft{i1} \fmfright{o1}
  \fmf{scalar,tension=1.2}{i1,v1}
  \fmf{scalar,label=$SS^{\dagger}$,left=0.6,tension=0.5}
      {v2,v1}
  \fmf{scalar,label=$F^{\dagger}F$,left=0.6,tension=0.5}{v1,v2}
  \fmf{scalar,tension=1.2}{v2,o1}
  \fmfv{l=${\varphi}^{a\dagger}_{J}(x)$,l.a=180}{i1}
  \fmfv{l=$\varphi^{I}_{b}(y)$ \hspace{0.5cm} $\longrightarrow ~~~ 
        M^{aI}_{bJ}(x;y)$,l.a=0}{o1}
  \fmfdot{v1,v2}
 \end{fmfgraph*}   
\end{fmffile} 

\vspace*{1.7cm}
\noindent
\hspace*{2cm}
\begin{fmffile}{phiphiN}
 \begin{fmfgraph*}(150,75) 
  \fmfleft{i1} \fmfright{o1}
  \fmf{scalar,tension=1.4}{i1,v1}
  \fmf{scalar,tension=1.4}{v1,o1}
  \fmf{phantom,tension=0.9}{v1,v1}
  \fmfv{l=$\varphi^{a\dagger}_{J}(x)$,l.a=180}{i1}
  \fmfv{l=$\varphi^{I}_{b}(y)$ \hspace{0.5cm} $\longrightarrow ~~~ 
        N^{aI}_{bJ}(x;y)$,l.a=0}{o1}
  \fmfdot{v1}
  \fmffreeze
  \fmfipath{p[]}
  \fmfiset{p1}{vpath(__v1,__v1)}
  \fmfi{fermion}{subpath (0,length(p1)/2) of p1}
  \fmfi{fermion}{subpath (length(p1),length(p1)/2) of p1}  
  \fmfiv{d.sh=cross,d.si=3mm,l=${\overline\lambda}{\overline\chi}$,
       l.a=90}{point length(p1)/2 of p1}      
 \end{fmfgraph*}   
\end{fmffile} 

\vspace*{0.8cm}
\noindent
\hspace*{2cm}
\begin{fmffile}{phiphiP}
 \begin{fmfgraph*}(150,75) 
  \fmfleft{i1} \fmfright{o1}
  \fmf{scalar,tension=1.4}{i1,v1}
  \fmf{scalar,tension=1.4}{v1,o1}
  \fmf{phantom,tension=0.9}{v1,v1}
  \fmfv{l=$\varphi^{a\dagger}_{J}(x)$,l.a=180}{i1}
  \fmfv{l=$\varphi^{I}_{b}(y)$ \hspace{0.5cm} $\longrightarrow ~~~ 
        P^{aI}_{bJ}(x;y)$,l.a=0}{o1}
  \fmfdot{v1}
  \fmffreeze
  \fmfipath{p[]}
  \fmfiset{p1}{vpath(__v1,__v1)}
  \fmfi{fermion}{subpath (length(p1)/2,0) of p1}
  \fmfi{fermion}{subpath (length(p1)/2,length(p1)) of p1}  
  \fmfiv{d.sh=cross,d.si=3mm,l=$\lambda\chi$,
       l.a=90}{point length(p1)/2 of p1}      
 \end{fmfgraph*}   
\end{fmffile} 

\vspace*{0.8cm}
\noindent
\hspace*{2cm}
\begin{fmffile}{phiphiQ}
 \begin{fmfgraph*}(150,75) 
  \fmfleft{i1} \fmfright{o1}
  \fmf{scalar,tension=1.4}{i1,v1}
  \fmf{scalar,tension=1.4}{v1,o1}
  \fmf{phantom,tension=0.9}{v1,v1}
  \fmfv{l=$\varphi^{a\dagger}_{J}(x)$,l.a=180}{i1}
  \fmfv{l=$\varphi^{I}_{b}(y)$ \hspace{0.5cm} $\longrightarrow ~~~ 
        Q^{aI}_{bJ}(x;y)$,l.a=0}{o1}
  \fmfdot{v1}
  \fmffreeze
  \fmfipath{p[]}
  \fmfiset{p1}{vpath(__v1,__v1)}
  \fmfi{dashes}{subpath (0,length(p1)) of p1}
  \fmfiv{d.sh=cross,d.si=3mm,l=$CD$,
       l.a=90}{point length(p1)/2 of p1}      
 \end{fmfgraph*}   
\end{fmffile} 

\newpage

\vspace*{0.3cm}
\noindent
\hspace*{2cm}
\begin{fmffile}{phiphiR}
 \begin{fmfgraph*}(150,75) 
  \fmfleft{i1} \fmfright{o1}
  \fmf{scalar,tension=1.4}{i1,v1}
  \fmf{scalar,tension=1.4}{v1,o1}
  \fmf{phantom,tension=0.9}{v1,v1}
  \fmfv{l=$\varphi^{a\dagger}_{J}(x)$,l.a=180}{i1}
  \fmfv{l=$\varphi^{I}_{b}(y)$ \hspace{0.5cm} $\longrightarrow ~~~ 
        R^{aI}_{bJ}(x;y)$,l.a=0}{o1}
  \fmfdot{v1}
  \fmffreeze
  \fmfipath{p[]}
  \fmfiset{p1}{vpath(__v1,__v1)}
  \fmfi{scalar}{subpath (0,length(p1)) of p1}
  \fmfiv{l=$SS^{\dagger}$,l.a=90}{point length(p1)/2 of p1}      
 \end{fmfgraph*}   
\end{fmffile} 

Single diagrams contain quadratically divergent terms that cancel in 
the sum leaving a total contribution that is again logarithmically 
divergent. This non-vanishing correction is exactly what is needed to 
cancel the divergence in (\ref{phioneloopwz}). As a result the complete 
one-loop correction to the $\langle\varphi^{\dagger}\varphi\rangle$ 
propagator is zero, whereas in the Wess--Zumino gauge a wave function 
renormalization is required.

Notice that the situation induced by the choice of the WZ gauge is a 
feature common to every supersymmetric gauge theory, since the 
contribution of the gauge-dependent fields $C$, $\chi$ and $S$ is in 
general logarithmically divergent. However in theories with less 
supersymmetry a logarithmic wave function renormalization is usually 
necessary in any case, so that this effect is completely irrelevant. 
On the contrary it becomes important in the ${\cal N}$=4 Yang--Mills 
theory and in general in finite theories. Of course the divergences 
encountered here are gauge artifacts and disappear in gauge invariant 
correlation functions. Examples of computations of correlators of gauge 
invariant operators, in which the choice of the WZ gauge does not lead 
to this kind of problems, will be considered within the discussion of 
the correspondence with AdS type IIB superstring theory in the final 
chapter. Note also that the divergences encountered in the propagators 
are zero on-shell.

The calculations described here seem to suggest the possibility of 
constructing improved Feynman rules in which the effect of the gauge 
dependent fields, neglected in the WZ gauge, is dealt with by a suitable 
redefinition of the free propagators. However this program proves 
extremely complicated when the propagators of fields in the vector 
multiplet or three- and more-point functions are considered. The 
calculation of these correlation functions requires, already at the 
one-loop level, many other terms to be included in the action 
$S_{{\rm int}}$. In particular terms coming from $W^{(1)}W^{(2)}$ 
and $W^{(2)}W^{(2)}$ must be considered. The calculation of $W^{(2)}$ 
without the simplifications introduced by the Wess--Zumino gauge is 
rather lengthy and gives
\begin{eqnarray}
	W^{a(2)}_{\alpha} &=& -\frac{i}{2} f^{a}{}_{bc} \left\{ 
	-iC^{b}\lambda^{c}_{\alpha}-\frac{1}{2} 
	\sigma^{\mu}_{\alpha{\dot\alpha}}{\overline\chi}^{{\dot\alpha}b} 
	\partial_{\mu}C^{c}-\frac{i}{2}\sigma^{\mu}_{\alpha{\dot\alpha}}
	{\overline\chi}^{{\dot\alpha}b}A^{c}_{\mu} + 
	\frac{1}{\sqrt{2}} S^{b\dagger}\chi^{c}_{\alpha}+\right. \nonumber \\
	&&\left. + 
	\left[ S^{b\dagger}
	S^{c}\delta_{\alpha}{}^{\beta} + \delta_{\alpha}{}^{\beta} 
	C^{b}D^{c}-\frac{i}{2}(\sigma^{\mu}
	{\overline\sigma}^{\nu})_{\alpha}{}^{\beta} C^{b} 
	(\partial_{\mu}A^{c}_{\nu}-\partial_{\nu}
	A^{c}_{\mu}) + \right. \right.\nonumber \\ 
	&& \left. \left. -\frac{i}{2}(\sigma^{\mu}
	{\overline\sigma}^{\nu})_{\alpha}{}^{\beta}
	(A^{b}_{\nu}\partial_{\mu}C^{c}+A_{\mu}^{b}\partial_{\nu}C^{c})-
	\frac{1}{2} \delta_{\alpha}{}^{\beta}C^{b}\Box C^{c} - 
	\delta_{\alpha}{}^{\beta}{\overline\chi}^{b}{\overline\lambda}^{c}+
	\right. \right. \nonumber \\
	&& \left. \left. 
	+\frac{1}{2}(\sigma^{\mu}\sigma^{\nu})_{\alpha}{}^{\beta}
	\partial_{\mu}C^{b}\partial_{\nu}C^{c}-
	\chi^{b\beta}\lambda^{c}_{\alpha} -
	\lambda^{b\beta}\chi^{c}_{\alpha}-i\sigma^{\mu}_{\alpha{\dot\alpha}}
	{\overline\chi}^{b{\dot\alpha}}\partial_{\mu}\chi^{c\beta} + 
	\right. \right. \nonumber \\
	&& \left. \left. +i\varepsilon ^{\gamma\beta}
	\sigma^{\mu}_{\gamma{\dot\beta}}\partial_{\mu}{\overline\chi}^{b} 
	\chi^{c}_{\alpha} + \frac{1}{2} 
	(\sigma^{\mu}\sigma^{\nu})_{\alpha}{}^{\beta}A_{\nu}^{b}A_{\mu}^{c} 
	\right] \theta_{\beta} + \left[ 
	C^{b}\sigma^{\mu}_{\alpha{\dot\alpha}} 
	\partial_{\mu}{\overline\lambda}^{c{\dot\alpha}} + 
	\right. \right. \nonumber \\
	&& \left. \left. -i\chi_{\alpha}^{b}D^{c}-\frac{1}{4} (\sigma^{\mu}
	{\overline\sigma}^{\nu})_{\alpha}{}^{\beta} \chi_{\beta}^{b} 
	(\partial_{\mu}A^{c}_{\nu}-\partial_{\nu} A^{c}_{\mu}) - 
	\frac{1}{2} A_{\nu}^{b} (\sigma^{\mu}
	{\overline\sigma}^{\nu})_{\alpha}{}^{\beta} \partial_{\mu}
	\chi_{\beta}^{c} + \right. \right. \nonumber \\
	&& \left. \left. 
	+ \frac{1}{2}\partial_{\mu}A^{b\mu}\chi_{\alpha}^{c} - 
	\frac{i}{2} \partial_{\nu}C^{b} (\sigma^{\mu}
	{\overline\sigma}^{\nu})_{\alpha}{}^{\beta} \partial_{\mu}
	\chi_{\beta}^{c} - \frac{i}{\sqrt{2}} \sigma^{\mu}_{\alpha{\dot\alpha}}
	\partial_{\mu}{\overline\chi}^{{\dot\alpha}b}S^{c} + 
	\sqrt{2} S^{b}\lambda^{c}_{\alpha}+ \right. \right. \nonumber \\
	&& \left. \left. -i A_{\mu}^{b} \sigma^{\mu}_{\alpha{\dot\alpha}}
	{\overline\lambda}^{c{\dot\alpha}} \right] \theta\theta
	\rule{0pt}{18pt}\right\} \label{w2nonwz}
\end{eqnarray}
Substituting (\ref{w2nonwz}) into $S_{{\rm int}}$ 
results in a number of relevant interaction terms of the order 
of 100, making this formulation totally impractical in explicit 
calculations. In conclusion perturbation theory in components requires 
almost unavoidably the WZ gauge, but the latter 
introduces divergences in gauge dependent quantities. In the following 
sections the superfield formalism, that allows to avoid these problems, 
will be employed. However as will be shown new difficulties related to 
the gauge fixing emerge.

\section{Perturbation theory in ${\cal N}$=1 superspace: propagators}
\label{n1propagators}
\fancyhead[LO]{{\footnotesize 3.2~~{\it Propagators in the ${\cal N}$=1 
formulation}}}

The difficulties encountered in the previous section in the 
calculation of gauge dependent quantities, because of the divergences 
introduced by the Wess--Zumino gauge, can be overcome using the 
${\cal N}$=1 superfield formulation of the model.

In chapter \ref{cap2} it has already been pointed out how the ${\cal N}$=1 
superfield formalism has proved to be a powerful tool in the proof of 
the finiteness of the theory up to three loops. In this approach there 
is no particular difficulty in working without fixing the WZ gauge. If 
one does not choose to work in  the WZ gauge the action is non polynomial, 
however a finite number of terms is relevant at each order in perturbation 
theory, so that only at very high order the Wess--Zumino gauge 
introduces significant simplifications. The aim of this section is to 
show that there are other subtleties related to the gauge fixing, 
even if one does not work in the WZ gauge.

The formulation of the theory in ${\cal N}$=1 superspace that will be 
employed is a `mass deformed' version of that given by equation 
(\ref{actionsuperfields}). In other words the action that will be 
considered is that of equation (\ref{actionsuperfields}) plus mass 
terms for the (anti) chiral superfields
\begin{eqnarray}
	S_{{\rm m}} &=& -\int d^{4}x d^{4}\theta \, \left[ \frac{1}{2} 
	m \delta_{IJ} \Phi_{a}^{I}(x,\theta,{\overline \theta}) 
	\Phi^{Ja}(x,\theta,{\overline \theta}) 
	\delta({\overline \theta}) + \right. \nonumber \\
	&& \left. \hspace{1cm} + \frac{1}{2} m^{*} \delta^{IJ}
    {\Phi^{\dagger}}^{a}_{I}(x,\theta,{\overline \theta}) 
	{\Phi^{\dagger}}_{Ja}(x,\theta,{\overline \theta}) 
	\delta(\theta) \right] \, .
	\label{massterm}
\end{eqnarray}
The addition of this term to the action breaks supersymmetry down 
to ${\cal N}$=1. In \cite{parkeswest} it was argued, by means of 
dimensional arguments, that this term should not affect the ultraviolet 
properties of the ${\cal N}$=4 theory. More precisely the statement of 
\cite{parkeswest} is that no divergences appear in gauge invariant 
quantities, so that no divergent contribution to the quantum 
effective action is generated perturbatively. 
As a result the model obtained in this way would be an 
example of a finite ${\cal N}$=1 theory. The inverse 
construction, in which one deforms a ${\cal N}$=1 model to ${\cal N}$=4 
super Yang--Mills plus a mass term, has been proposed in 
\cite{japan,yoshida} as a regularization procedure preserving 
supersymmetry. The calculations presented in this section will show 
that in the presence of the term (\ref {massterm}) no divergence is 
generated, at the one-loop level, in the two-, three- and four-point 
Green functions that are computed. The results presented suggest 
the possibility of reinforcing the conclusions of \cite{parkeswest}; 
namely the ${\cal N}$=4 theory augmented with (\ref{massterm}) appears 
to be finite in the sense that no divergences appear in the total 
$n$-point irreducible Green functions, at least at one loop. In 
particular no wave function renormalization is required. Notice that 
this result is what is actually necessary for the consistency of 
the approach advocated in \cite{japan,yoshida}. 

The complete action is~\footnote{In the following the mass parameter $m$ 
will be taken to be real for simplicity of notation.} 
\begin{eqnarray}
    && \hspace{-0.65cm} S = \int d^{4}x\,d^{4}\theta \: 
	\left\{  \rule{0pt}{17pt}
	V^{a}(x,\theta,{\overline \theta}) \left[ - \Box P_{T} - \xi 
	(P_{1}+P_{2}) \Box \right] V_{a}(x,\theta,{\overline \theta}) + 
	\right. \nonumber \\ 
	&& \hspace{-0.65cm} + {\Phi^{\dagger}}^{a}_{I}(x,\theta,
	{\overline \theta}) 
	\Phi_{a}^{I}(x,\theta,{\overline \theta}) - \frac{1}{2} 
	m \delta_{IJ} \Phi_{a}^{I}(x,\theta,{\overline \theta}) 
	\Phi^{Ja}(x,\theta,{\overline \theta}) 
	\delta({\overline \theta}) + \nonumber \\
	&& \hspace{-0.65cm} -\frac{1}{2} m \delta^{IJ}
    {\Phi^{\dagger}}^{a}_{I}(x,\theta,{\overline \theta}) 
	{\Phi^{\dagger}}_{Ja}(x,\theta,{\overline \theta}) 
	\delta(\theta) + i g f_{abc}{\Phi^{\dagger}}^{a}_{I}
	(x,\theta,{\overline \theta})
	V^{b}(x,\theta,{\overline \theta}) \cdot 
\end{eqnarray}
\begin{eqnarray}
	&& \hspace{-0.65cm}  \cdot 
	\Phi^{Ic}(x,\theta,{\overline \theta})
	- \frac{1}{2} g^{2} {f_{ab}}^{e}f_{ecd} {\Phi^{\dagger}}^{a}_{I}
	(x,\theta,{\overline \theta})
	V^{b}(x,\theta,{\overline \theta}) V^{c}(x,\theta,{\overline \theta})
	\Phi^{Id}(x,\theta,{\overline \theta}) + \nonumber \\
	&& \hspace{-0.65cm} - \frac{i}{4} g f_{abc} 
	\left[ {\overline D}^{2} 
	\left( D^{\alpha}V^{a}(x,\theta,{\overline \theta}) \right) 
	\right] V^{b}(x,\theta,{\overline \theta}) 
	\left( D_{\alpha}V^{c}(x,\theta,
	{\overline \theta}) \right) + \nonumber \\
	&& \hspace{-0.65cm} -\frac{1}{8} g^{2} f_{ab}{}^{e}f_{ecd} 
    V^{a}(x,\theta,{\overline \theta}) \left( D^{\alpha}
	V^{b}(x,\theta,{\overline 
	\theta}) \right) \left[ \left({\overline D}^{2} 
	V^{c}(x,\theta,{\overline \theta}) 
	\right) \left( D_{\alpha}V^{d}(x,\theta,{\overline \theta}) \right) 
	\right] \hspace{-0.15cm} + \nonumber \\
	&& \hspace{-0.65cm} +\ldots - \frac{\sqrt{2}}{3!} g f^{abc} 
	\left[ \varepsilon_{IJK}
	\Phi_{a}^{I}(x,\theta,{\overline \theta}) 
	\Phi_{b}^{J}(x,\theta,{\overline \theta}) 
	\Phi_{c}^{K}(x,\theta,{\overline \theta}) \delta({\overline 
	\theta}) + \right. \nonumber \\
	&& \hspace{-0.65cm} + \left. \varepsilon^{IJK} 
	{\Phi^{\dagger}}_{Ia}(x,\theta,{\overline \theta}) 
	{\Phi^{\dagger}}_{Jb}(x,\theta,{\overline \theta}) 
	{\Phi^{\dagger}}_{Kc}(x,\theta,{\overline \theta}) 
	\delta(\theta) \right] + 
	\left( {{\overline C}^{\prime}}_{a}(x,\theta,{\overline \theta})
	C^{a}(x,\theta,{\overline \theta}) + \right. \nonumber \\
	&& \hspace{-0.65cm} \left. - {C^{\prime}}_{a}
	(x,\theta,{\overline \theta})
	{\overline C}^{a}(x,\theta,{\overline \theta}) \right)  
    +\frac{i}{2\sqrt{2}} g f_{abc} \left( {C^{\prime}}^{a}
	(x,\theta,{\overline \theta}) +
	{{\overline C}^{\prime}}^{a}(x,\theta,{\overline \theta}) \right) 
	\cdot \nonumber \\ 
	&& \hspace{-0.65cm} \cdot V^{b}(x,\theta,{\overline \theta}) 
	\left( C^{c}(x,\theta,{\overline \theta}) +
	{\overline C}^{c}(x,\theta,{\overline \theta}) \right) - 
	\frac{1}{8} g^{2} {f_{ab}}^{e}f_{ecd}  \left( 
	{C^{\prime}}^{a}(x,\theta,{\overline \theta}) +
	{{\overline C}^{\prime}}^{a}(x,\theta,{\overline \theta}) 
	\right) \cdot \nonumber \\
	&& \hspace{-0.65cm} \left. \cdot  
	V^{b}(x,\theta,{\overline \theta}) 
	V^{c}(x,\theta,{\overline \theta}) 
	\left( C^{d}(x,\theta,{\overline \theta}) +
	{\overline C}^{d}(x,\theta,{\overline \theta}) \right)  
    + \ldots \rule{0pt}{18pt} \right\} \, ,
\label{n1sfaction}	
\end{eqnarray}
where the dots stand for terms that are not relevant for the 
considerations of this chapter.

Notice that in the action (\ref{n1sfaction}) a gauge fixing 
term corresponding to a family of gauges parameterized by 
$\alpha=\frac{1}{\xi}$ has been introduced. It will now be shown, by 
explicitly computing the propagators of both the chiral and the vector 
superfields, that the supersymmetric generalization of the 
Fermi--Feynman gauge, corresponding to $\alpha=1$, is somehow 
privileged (see also \cite{storey,juerstorey2}), since any other 
choice of the parameter $\alpha$ leads to infrared divergences in 
Green functions.

\subsection{Propagator of the chiral superfield} 

\noindent 
The propagator of the chiral superfield is the simplest Green 
function  to compute. The calculation will be reported in detail in 
order to illustrate the superfield technique. 

It follows from the action (\ref{actionsuperfields}) that there are 
three diagrams contributing to the propagator $\langle \Phi
\Phi^{\dagger} \rangle$ at the one loop level. The one-particle 
irreducible parts of these diagrams will be directly evaluated in 
momentum space using the super Feynman rules of section 
\ref{quantization}. The convention employed is that all momenta are 
taken to be incoming. The diagrams are the following

\noindent
\hspace*{2cm}
\begin{fmffile}{sphiphiA}
 \begin{fmfgraph*}(150,80) 
  \fmfleft{i1} \fmfright{o1}
  \fmf{fermion,tension=1.2}{i1,v1}
  \fmf{boson,label=$VV$,left=0.6,tension=0.5}{v1,v2}
  \fmf{fermion,label=$\Phi\Phi^{\dagger}$,right=0.6,tension=0.5}{v1,v2}
  \fmf{fermion,tension=1.2}{v2,o1}
  \fmfv{l=$\Phi^{a\dagger}_{J}(z)$,l.a=180}{i1}
  \fmfv{l=$\Phi^{I}_{b}(z^{\prime})$ \hspace{0.5cm} $\longrightarrow ~~~ 
        A(z;z^{\prime})$,l.a=0}{o1}
  \fmfdot{v1,v2}
 \end{fmfgraph*}   
\end{fmffile} \\
\hspace*{2cm}
\begin{fmffile}{sphiphiB}
 \begin{fmfgraph*}(150,80) 
  \fmfleft{i1} \fmfright{o1}
  \fmf{fermion,tension=1.2}{i1,v1}
  \fmf{fermion,label=$\Phi^{\dagger}\Phi$,left=0.6,tension=0.5}
      {v2,v1}
  \fmf{fermion,label=$\Phi^{\dagger}\Phi$,right=0.6,tension=0.5}{v2,v1}
  \fmf{fermion,tension=1.2}{v2,o1}
  \fmfv{l=$\Phi^{a\dagger}_{J}(z)$,l.a=180}{i1}
  \fmfv{l=$\Phi^{I}_{b}(z^{\prime})$ \hspace{0.5cm} $\longrightarrow ~~~ 
        B(z;z^{\prime})$,l.a=0}{o1}
  \fmfdot{v1,v2}
 \end{fmfgraph*}   
\end{fmffile} 

\vspace*{1.2cm}
\noindent
\hspace*{2cm}
\begin{fmffile}{sphiphiC}
 \begin{fmfgraph*}(150,80) 
  \fmfleft{i1} \fmfright{o1}
  \fmf{fermion,tension=1.4}{i1,v1}
  \fmf{fermion,tension=1.4}{v1,o1}
  \fmf{phantom,tension=0.9}{v1,v1}
  \fmfv{l=$\Phi^{a\dagger}_{J}(z)$,l.a=180}{i1}
  \fmfv{l=$\Phi^{I}_{b}(z^{\prime})$ \hspace{0.5cm} $\longrightarrow ~~~ 
        C(z;z^{\prime})$,l.a=0}{o1}
  \fmfdot{v1}
  \fmffreeze
  \fmfipath{p[]}
  \fmfiset{p1}{vpath(__v1,__v1)}
  \fmfi{boson}{subpath (0,length(p1)) of p1}
  \fmfiv{l=$VV$,l.a=90}{point length(p1)/2 of p1}      
 \end{fmfgraph*}   
\end{fmffile} \\
where $z=(x,\theta,{\overline\theta})$ and the notation for the 
internal propagators is 

\noindent
\begin{fmffile}{sphi+phi}
\begin{fmfgraph*}(90,50) 
  \fmfleft{i1} \fmfright{o1}
  \fmf{fermion}{i1,o1}
  \fmflabel{\raisebox{22pt}{\hspace{-20pt} $J,b$ \hspace{20pt}} 
  \raisebox{12pt}{$\longrightarrow \hspace{0.2cm} 
  \langle \Phi^{aI}(z)\Phi^{b\dagger}_{J}(z^{\prime}) 
  \rangle_{{\rm free}} = \begin{displaystyle}
  \delta^{I}_{J}\delta^{a}_{b}\frac{1}{\Box + m^{2}}
  \end{displaystyle} \delta_{8}(z-z^{\prime})$}}{o1}
  \fmflabel{\raisebox{22pt}{\hspace{18pt} $I,a$ 
  \hspace{-18pt}}}{i1}
\end{fmfgraph*}   
\end{fmffile} \\
\begin{fmffile}{svv}
\begin{fmfgraph*}(90,50) 
  \fmfleft{i1} \fmfright{o1}
  \fmf{boson}{i1,o1}
  \fmflabel{\raisebox{22pt}{\hspace{-20pt} $b$ \hspace{20pt}} 
  \raisebox{12pt}{$\longrightarrow \hspace{0.2cm} 
  \langle V^{a}(z)V_{b}(z^{\prime})\rangle_{{\rm free}}=
  \begin{displaystyle} -\frac{\delta^{a}_{b}}{\Box}
  [1+(\alpha -1)(P_{1}+P_{2})] \end{displaystyle} 
  \delta_{8}(z-z^{\prime})$}}{o1}
  \fmflabel{\raisebox{22pt}{\hspace{18pt} $a$ 
  \hspace{-18pt}}}{i1}
\end{fmfgraph*}   
\end{fmffile} \\
$P_{1}$ and $P_{2}$ in the $\langle VV \rangle$ propagator are the 
projectors (\ref{p1p2project}). Moreover in the presence of mass terms 
for $\Phi$ and $\Phi^{\dagger}$ there are free propagators 
$\langle\Phi\Phi\rangle$ and 
$\langle\Phi^{\dagger}\Phi^{\dagger}\rangle$, that will enter 
the calculation of the vector superfield propagator at one loop

\noindent
\begin{fmffile}{sphiphi}
\begin{fmfgraph*}(90,50) 
  \fmfleft{i1} \fmfright{o1}
  \fmf{fermion}{v1,i1}
  \fmf{fermion}{v1,o1}
  \fmflabel{\raisebox{22pt}{\hspace{-20pt} $J,b$ \hspace{20pt}} 
  \raisebox{12pt}{$\longrightarrow \hspace{0.2cm} 
  \langle \Phi^{aI}(z)\Phi_{b}^{J}(z^{\prime}) 
  \rangle_{{\rm free}} = \begin{displaystyle}
  -\delta^{IJ}\delta^{a}_{b}\,\frac{m}{4}
  \frac{D^{2}}{\Box(\Box + m^{2})} \end{displaystyle}
  \delta_{8}(z-z^{\prime})$}}{o1}
  \fmflabel{\raisebox{22pt}{\hspace{18pt} $I,a$ 
  \hspace{-18pt}}}{i1}
  \fmfv{d.sh=cross,d.si=3mm}{v1}
\end{fmfgraph*}   
\end{fmffile} 

\begin{fmffile}{sphi+phi+}
\begin{fmfgraph*}(90,50) 
  \fmfleft{i1} \fmfright{o1}
  \fmf{fermion}{i1,v1}
  \fmf{fermion}{o1,v1}
  \fmflabel{\raisebox{22pt}{\hspace{-20pt} $J,b$ \hspace{20pt}} 
  \raisebox{12pt}{$\longrightarrow \hspace{0.2cm} 
  \langle \Phi^{a\dagger}_{I}(z)\Phi_{bJ}^{\dagger}(z^{\prime}) 
  \rangle_{{\rm free}} = \begin{displaystyle}
  -\delta_{IJ}\delta^{a}_{b}\,\frac{m}{4}
  \frac{{\overline D}^{2}}{\Box(\Box + m^{2})}
  \end{displaystyle} \delta_{8}(z-z^{\prime})$}}{o1}
  \fmflabel{\raisebox{22pt}{\hspace{18pt} $I,a$ 
  \hspace{-18pt}}}{i1}
  \fmfv{d.sh=cross,d.si=3mm}{v1}
\end{fmfgraph*}   
\end{fmffile} 

Using these free propagators and the rules of section 
\ref{quantization}, with the vertices read from the action 
(\ref{n1sfaction}), the three contributions can be evaluated as 
follows
\begin{eqnarray}
	\tilde{A}(p) & = & \int \frac{d^{4}k}{(2\pi)^{4}} 
	d^{4}\theta_{1}d^{4}\theta_{2} \, \left\{ i g 
	f_{acd}\Phi^{a\dagger}_{I}(p,\theta_{1},{\overline \theta}_{1}) 
	\left[\left(-\frac{1}{4}{\overline D}_{1}^{2}\right) 
	\frac{\delta^{I}_{J}\delta^{cf}\delta(1,2)}{(p-k)^{2}+m^{2}} 
	\right. \cdot \right. \nonumber  \\
	&& \hspace{-1.7cm}\cdot \left. \left.\left(-\frac{1}{4}
	\db{D}_{1}^{2} \right) \right] \left[ \left(1+\gamma(D^{2}_{1}
	{\overline D}^{2}_{1}+ {\overline D}^{2}_{1}D^{2}_{1})\right) 
	\left(-\frac{\delta^{de}\delta(1,2)}{k^{2}}\right)\right]
	i g f_{efb} \Phi^{bJ}(-p,\theta_{2},{\overline \theta}_{2}) 
	\rule{0pt}{20pt}\right\} ,
	\nonumber
\end{eqnarray}
where $\gamma=(\alpha-1)$ and the compact notation 
$\delta(1,2)=\delta_{2}(\theta_{1}-\theta_{2})\delta_{2}({\overline 
\theta_{1}}-{\overline \theta_{2}})$ has been introduced. The 
computation uses the properties of the $\delta$-function, which imply 
for example \cite{grisaruroceksiegel,wessbagger}
\begin{eqnarray*}
	&& D_{1\alpha}\delta(1,2) = - \delta(1,2) \db{D}_{2\alpha} \, , 
	\qquad {\overline D}_{1{\dot \alpha}} \delta(1,2) = - \delta(1,2) 
	\dbo{{\overline D}}_{2{\dot\alpha}} \, , \\ 
	&& {\overline D}_{1{\dot\alpha}} D_{1\alpha}\delta(1,2)=\delta(1,2)
	\dbo{{\overline D}}_{2{\dot\alpha}} \db{D}_{2\alpha} \, , \qquad
	{\overline D}^{2}_{1}D^{2}_{1} \delta(1,2) = \delta(1,2) 
	\dbo{{\overline D}}^{2}_{2} \db{D}^{2}_{2} \, ,
\end{eqnarray*}
and integrations by parts on the Grassmannian variables in order to 
remove the $D$ and ${\overline D}$ derivatives from one $\delta$, so 
that one $\theta$ integration can be performed immediately. In this 
way one obtains an expression that is local in $\theta$ as is expected 
from the ${\cal N}$=1 non-renormalization theorem. 
For $\tilde{A}(p)$ one obtains
\begin{eqnarray*}
	\tilde{A}(p) & = & -g^{2}N(N^{2}-1)\delta^{a}_{b} \delta_{J}^{I}
	\left( \frac{1}{4} \right)^{2} \int \frac{d^{4}k}{(2\pi)^{4}} 
	d^{4}\theta_{1}d^{4}\theta_{2} \, \frac{1}{k^{2}[(p-k)^{2}+m^{2}]}
	\left\{ \Phi^{\dagger}_{aI}(p,1) \cdot \right. \\
	&& \hspace{-1.7cm} \left. \cdot \Phi^{bJ}(-p,2) \left[
	{\overline D}^{2}_{1} D^{2}_{1} \delta(1,2) \right]\left[ 1+\gamma
	(D^{2}_{1}{\overline D}^{2}_{1}+ {\overline D}^{2}_{1}D^{2}_{1}) 
	\delta(1,2) \right] \right\} = \tilde{A}_{1}(p) + 
	\tilde{A}_{2}(p) \rule{0pt}{20pt} \, .
\end{eqnarray*}
The first term is trivially calculated using 
\cite{grisaruroceksiegel,wessbagger}
\begin{eqnarray}
	&& \hspace{-1cm} \int d^{4}\theta \, \left[ 
	{\overline D}^{2} D^{2} \delta(\theta-\theta^{\prime}) \right] 
	\delta(\theta-\theta^{\prime}) = 16 \nonumber \\ 
	&& \hspace{-1cm} \int d^{4}\theta \, \left[ {\overline D}^{m} 
	D^{n} \delta(\theta-\theta^{\prime}) \right] 
	\delta(\theta-\theta^{\prime}) = 0 \quad {\rm if} \quad
	(m,n) \neq (2,2) 
	\label{thetaint}
\end{eqnarray}
and gives
\begin{equation}
    \tilde{A}_{1}(p) = - g^{2}N(N^{2}-1)\delta^{a}_{b} \delta_{J}^{I} 
    \int \frac{d^{4}k}{(2\pi)^{4}} d^{4}\theta \, 
    \frac{1}{k^{2}[(p-k)^{2}+m^{2}]} \left\{ 
	\Phi^{\dagger}_{aI}(p,\theta,{\overline \theta}) 
	\Phi^{bJ}(-p,\theta,{\overline \theta}) \right\} .
	\label{a1tildephiphi} 
\end{equation}
In the computation of the second term one must use the (anti) 
commutators of covariant derivatives, see appendix \ref{appa}, 
which in particular imply \cite{grisaruroceksiegel,wessbagger}
\begin{equation}
	{\overline D}^{2}D^{2}{\overline D}^{2} = 16 \Box {\overline D}^{2} 
	\qquad D^{2}{\overline D}^{2}D^{2} = 16 \Box D^{2} \, .
	\label{dprop1}
\end{equation}
Then integration by parts analogously gives 
\begin{equation}
    \tilde{A}_{2}(p) = \gamma \, g^{2}N(N^{2}-1) \delta^{a}_{b}
	\delta_{J}^{I} \int \frac{d^{4}k}{(2\pi)^{4}} d^{4}\theta \, 
	\frac{[(p-k)^{2}+p^{2}]}{k^{4}[(p-k)^{2}+m^{2}]} 
	\left\{ \Phi^{\dagger}_{aI}(p,\theta,{\overline \theta}) 
	\Phi^{bJ}(-p,\theta,{\overline \theta}) \right\} . 
\end{equation}
In conclusion from the first diagram one obtains two contributions, 
the first proportional to $\gamma$ and the second independent of it. 
The second diagram gives one single $\gamma$-independent contribution. 
The Feynman rules give
\begin{eqnarray*}
	&& \tilde{B}(p) =\int \frac{d^{4}k}{(2 \pi)^{4}} 
	d^{4}{\theta}_{1} d^{4}{\theta}_{2} \left\{ \left( -\frac{\sqrt{2}}{3!} 
	\right) g \varepsilon^{I}{}_{KL} f_{acd} {{\Phi}^{\dagger}}_{I}^{a}
	(p,1) \cdot \right. \\
	&& \cdot \left[ \left( -\frac{1}{4} {D_{1}}^{2} \right)
	\frac{\delta^{ce}\delta^{K}_{M}\delta(1,2)}{(p-k)^{2}+m^{2}} 
	\left( -\frac{1}{4} \dbo{{\overline D}_{2}}^{2} 
	\right) \right] \left[ \frac{\delta^{d}_{f}\delta^{L}_{N}
	\delta(1,2)}{k^{2} + m^{2}} \right] \cdot  \\
	&& \left. \cdot  \left(- 
	\frac{\sqrt{2}}{3!} \right) \varepsilon_{J}{}^{MN} f_{be}{}^{f} 
	{\Phi}^{bJ}(-p,2) \right\} \, .
\end{eqnarray*}
Proceeding exactly as for $\tilde{A}_{1}$ one obtains
\begin{equation}
    \tilde{B}(p) = g^{2}N(N^{2}-1) \delta^{a}_{b} \delta_{J}^{I} 
    \int \frac{d^{4}k}{(2\pi)^{4}} \, d^{4}\theta \, 
    \left\{\frac{ \Phi^{\dagger}_{aI}(p,\theta,{\overline \theta}) 
	\Phi^{bJ}(-p,\theta,{\overline \theta}) }{(k^{2}+m^{2})
	[(p-k)^{2}+m^{2}]} \right\}  \, .
	\label{btildephiphi} 
\end{equation}
From the last diagram one gets
\begin{eqnarray*}
	&& \tilde{C}(p) = \int \frac{d^{4}k}{(2\pi)^{4}} 
	d^{4}\theta_{1} \, \left\{ \left(- \frac{g^{2}}{2} \right) {f_{ac}}^{d} 
	f_{deb} {\Phi}^{a\dagger}_{I}(p,1) \cdot \right. \\
	&& \cdot \left. \left[ - \left(1+ \gamma 
	({\overline D}_{1}^{2}D_{1}^{2}+D_{1}^{2}{\overline D}_{1}^{2}) 
	\right) \frac{\delta^{ce}\delta(1,1)}{k^{2}} \right] 
	\Phi^{bJ}(-p,1) \right\} \, .
\end{eqnarray*}
The only non-vanishing contribution comes from the term in which the 
projection operators act on the $\delta$-function, since 
$\delta_{4}(\theta-\theta)=0$, so that 
\begin{equation}
	\tilde{C}(p) =-\gamma \, g^{2}N(N^{2}-1) \delta_{a}^{b}
	\delta_{J}^{I} \int \frac{d^{4}k}{(2\pi)^{4}} d^{4}\theta
	\frac{1}{k^{4}} \left\{ \Phi^{a\dagger}_{I}(p,\theta,
	{\overline \theta}){\Phi}_{b}^{J}(-p,\theta,{\overline \theta}) 
	\right\} \, ,
	\label{ctildephiphi}
\end{equation}
from which one sees that the $C$ contribution is absent in the gauge 
$\alpha=1$, \ie $\gamma=0$. 

Putting the various corrections together gives the following result. 
The sum of the terms $\tilde{A}_{1}(p)$ and $\tilde{B}(p)$, which does 
not depend on $\gamma$, is 
\begin{eqnarray*}
	{\tilde{A}}_{1}(p) + \tilde{B}(p) &=& g^{2} N(N^{2}-1) \delta_{a}^{b} 
	\delta_{J}^{I} \int \frac{d^{4}k}{(2\pi)^{4}} d^{4}\theta
	\left[ \Phi^{a\dagger}_{I}(p,\theta,{\overline \theta})
	{\Phi}_{b}^{J}(-p,\theta,{\overline \theta}) \right] \cdot \\
	&& \cdot \left\{ \frac{1}{k^{2} \left[ (p-k)^{2} + m^{2} \right]} - 
	\frac{1}{(k^{2} + m^{2}) \left[ (p-k)^{2} + m^{2} \right]} \right\} 
	\, .
\end{eqnarray*}
This is finite and exactly vanishes for $m=0$, \ie in the limit in 
which the ${\cal N}$=4 theory is recovered. 

The sum of the terms $\tilde{A}_{2}(p)$ and $\tilde{C}(p)$ is 
proportional to $\gamma$ and reads
\begin{eqnarray*}
	{\tilde{A}}_{2}(p)+\tilde{C}(p) &=& \gamma \, g^{2}N(N^{2}-1) 
	\delta_{a}^{b} \delta_{J}^{I} \int \frac{d^{4}k}{(2\pi)^{4}} 
	d^{4}\theta \left[ \Phi^{a\dagger}_{I}(p,\theta,{\overline \theta})
	{\Phi}_{b}^{J}(-p,\theta,{\overline \theta}) \right] \cdot \\
	&& \cdot \left\{ \left( \frac{(p-k)^{2}}{k^{4} \left[ (p-k)^{2} + 
	m^{2} \right]} - \frac{1}{k^{4}} \right) + \left( 
	\frac{p^{2}}{k^{4} \left[ (p-k)^{2} + m^{2} \right]} 
	\right) \right\} \, .
\end{eqnarray*}
Both of the terms in the last integral are infrared divergent. The 
first one is zero in the limit $m \to 0$, while the second one gives an 
infrared divergence that is still present in the ${\cal N}$=4 
theory, \ie with $m=0$. More precisely putting 
\begin{displaymath}
	{\tilde{A}}_{2}(p)+\tilde{C}(p)=\gamma \, g^{2}N(N^{2}-1) 
	\delta_{a}^{b} \delta_{J}^{I}\left[ 
	\Phi^{a\dagger}_{I}(p,\theta,{\overline \theta})
	{\Phi}_{b}^{J}(-p,\theta,{\overline \theta}) \right]
	\left[ I_{1}(p) + I_{2}(p) \right] \, ,
\end{displaymath}
one has
\begin{eqnarray*}
	&& I_{1}(p) = \int  \frac{d^{4}k}{(2\pi)^{4}}
	\left\{ \frac{(p-k)^{2}}{k^{4} \left[ (p-k)^{2} + m^{2} \right]} - 
	\frac{1}{k^{4}} \right\} = \\
	&& = \int  \frac{d^{4}k}{(2\pi)^{4}} \left\{ 
	\frac{m^{2}}{k^{4} \left[ (p-k)^{2} + m^{2} \right]} \right\} = \\
	&& =2 m^{2}  \int_{0}^{1}  d\zeta \, \zeta \int 
	\frac{d^{4}k}{(2\pi)^{4}}\frac{1}{{\left[ k^{2} + 
	(p^{2} \zeta + m^{2})(1-\zeta) \right]}^{3}} = \\
	&& = \frac{2 m^{2}}{2(4\pi)^{2}}  \int_{0}^{1}  d\zeta \, \zeta
	\frac{1}{(p^{2}\zeta + m^{2})(1-\zeta)} = \\
	&& = -\frac{m^{2}}{(4\pi)^{2}} \left[ \frac{1}{(p^{2}+m^{2})} \log 
	\epsilon \; + \; \frac{m^{2}}{p^{2}(p^{2}+m^{2})} \log \left(
	\frac{p^{2}+m^{2}}{m^{2}} \right) \right]
\end{eqnarray*}
where a standard Feynman parameterization has been used. Analogously
\begin{eqnarray*}
	&& I_{2}(p) = \int \frac{d^{4}k}{(2\pi)^{4}} 
	\left\{ \frac{p^{2}}{k^{4}\left[(p-k)^{2}+m^{2}\right]}\right\}= \\ 
	&& = 2p^{2}  \int_{0}^{1}  d\zeta \, \zeta 
	\int \frac{d^{4}k}{(2\pi)^{4}}
	\frac{1}{{\left[ k^{2} + (p^{2} \zeta + m^{2})(1-\zeta) 
	\right]}^{3}} = \\ 
	&& = \frac{p^{2}}{(4\pi)^{2}}  \int_{0}^{1}  d\zeta \, \zeta
	\frac{1}{(p^{2}\zeta + m^{2})(1-\zeta)} = \\
	&& = -\frac{1}{(4\pi)^{2}} \left[ \frac{p^{2}}{(p^{2}+m^{2})} \log 
	\epsilon \; + \; \frac{m^{2}}{(p^{2}+m^{2})} \log \left(
	\frac{p^{2}+m^{2}}{m^{2}} \right) \right] \, .
\end{eqnarray*}
In the above expressions an infrared regulator $\epsilon$ has been 
introduced. 

Notice that the total correction is exactly zero on-shell, \ie for 
$p^{2}$=$m^{2}$. 

To summarize the results, the propagator of the chiral superfields of 
the ${\cal N}$=4 super Yang--Mills theory in ${\cal N}$=1 superspace 
is logarithmically infrared-divergent for any choice of the 
gauge parameter $\alpha \neq 1$. This divergence corresponds to a 
wave function renormalization for the superfields $\Phi^{I}$ and 
vanishes on-shell. In the Fermi--Feynman gauge $\alpha$=1 the 
one-loop correction exactly vanishes.

\subsection{Propagator of the vector superfield}

The one-loop calculation of the propagator of the vector superfield 
is much more complicated, because many more diagrams are involved 
producing a large number of contributions. The final result is however 
completely analogous: in the Fermi--Feynman gauge the one-loop 
correction is zero, whereas infrared divergences arise for $\alpha 
\neq 1$. In the presence of a mass term for the (anti) chiral 
superfields no new divergences are generated.

Form a calculational viewpoint the new feature 
with respect to the $\langle \Phi \Phi^{\dagger} \rangle$ 
case is that there are also diagrams involving the 
ghosts. There are two multiplets of ghosts, described by the chiral 
superfields $C$ and $C^{\prime}$. The free propagators for these 
superfields will be denoted by

\vspace*{0.5cm}
\noindent
\hspace*{1cm}
\begin{fmffile}{scprime+c}
\begin{fmfgraph*}(90,50) 
  \fmfleft{i1} \fmfright{o1}
  \fmf{scalar}{i1,o1}
  \fmflabel{\raisebox{22pt}{\hspace{-20pt} $b$ \hspace{20pt}} 
  \raisebox{12pt}{$\longrightarrow \hspace{0.2cm} 
  \langle {\overline C}^{a\prime}(z) C^{b}(z^{\prime}) 
  \rangle_{{\rm free}} = \begin{displaystyle}
  \delta^{a}_{b}\frac{1}{\Box}
  \end{displaystyle} \delta_{8}(z-z^{\prime})$}}{o1}
  \fmflabel{\raisebox{22pt}{\hspace{18pt} $a$ 
  \hspace{-18pt}}}{i1}
\end{fmfgraph*}   
\end{fmffile} 

\noindent
\hspace*{1cm}
\begin{fmffile}{sc+cprime}
\begin{fmfgraph*}(90,50) 
  \fmfleft{i1} \fmfright{o1}
  \fmf{scalar}{i1,o1}
  \fmflabel{\raisebox{22pt}{\hspace{-20pt} $b$ \hspace{20pt}} 
  \raisebox{12pt}{$\longrightarrow \hspace{0.2cm} 
  \langle {\overline C}^{a}(z) C^{b\prime}(z^{\prime}) 
  \rangle_{{\rm free}} = \begin{displaystyle}
  \delta^{a}_{b}\frac{1}{\Box}
  \end{displaystyle} \delta_{8}(z-z^{\prime})$}}{o1}
  \fmflabel{\raisebox{22pt}{\hspace{18pt} $a$ 
  \hspace{-18pt}}}{i1}
\end{fmfgraph*}   
\end{fmffile} 

\vspace*{0.5cm}
\noindent
The ghosts are treated exactly like ordinary chiral superfields 
with the only difference that there is a minus sign associated with the 
loops, because $C$ and $C^{\prime}$ are anticommuting fields 
\cite{grisaruroceksiegel}. 

The corrections to the $\langle VV \rangle$ propagator at the 
one-loop level correspond to the diagrams

\vspace{0.5cm}
\noindent
\hspace*{2cm}
\begin{fmffile}{vvA}
 \begin{fmfgraph*}(150,80) 
  \fmfleft{i1} \fmfright{o1}
  \fmf{boson,tension=1.6}{i1,v1}
  \fmf{boson,tension=1.6}{v2,o1}
  \fmf{phantom,left=0.6,tension=0.3,tag=1}{v1,v2}
  \fmf{phantom,right=0.6,tension=0.3,tag=2}{v2,v1}
  \fmf{phantom,right=0.6,tension=0.3,tag=3}{v1,v2}
  \fmf{phantom,left=0.6,tension=0.3,tag=4}{v2,v1} 
  \fmfv{l=$V^{a}(z)$,l.a=180}{i1}
  \fmfv{l=$V_{b}(z^{\prime})$ \hspace{0.5cm} $\longrightarrow ~~~ 
        A(z;z^{\prime})$,l.a=0}{o1}
  \fmfdot{v1,v2}
  \fmffreeze
  \fmfipath{p[]}
  \fmfiset{p1}{vpath1(__v1,__v2)}
  \fmfiset{p2}{vpath2(__v2,__v1)}
  \fmfiset{p3}{vpath3(__v1,__v2)}
  \fmfiset{p4}{vpath4(__v2,__v1)}
  \fmfi{fermion}{subpath (0,length(p1)/2) of p1}
  \fmfi{fermion}{subpath (0,length(p2)/2) of p2}
  \fmfi{fermion}{subpath (length(p3)/2,length(p3)) of p3}  
  \fmfi{fermion}{subpath (length(p4)/2,length(p3)) of p4}
  \fmfiv{d.sh=cross,d.si=3mm,l=$\Phi\Phi$,
       l.a=90}{point length(p1)/2 of p1}
  \fmfiv{d.sh=cross,d.si=3mm,l=$\Phi^{\dagger}\Phi^{\dagger}$,
       l.a=-90}{point length(p3)/2 of p3}       
 \end{fmfgraph*}   
\end{fmffile} 

\vspace*{1.2cm}
\noindent
\hspace*{2cm}
\begin{fmffile}{vvB}
 \begin{fmfgraph*}(150,80) 
  \fmfleft{i1} \fmfright{o1}
  \fmf{boson,tension=1.6}{i1,v1}
  \fmf{boson,tension=1.6}{v2,o1}
  \fmf{phantom,left=0.6,tension=0.3,tag=1}{v1,v2}
  \fmf{phantom,right=0.6,tension=0.3,tag=2}{v2,v1}
  \fmf{phantom,right=0.6,tension=0.3,tag=3}{v1,v2}
  \fmf{phantom,left=0.6,tension=0.3,tag=4}{v2,v1} 
  \fmfv{l=$V^{a}(z)$,l.a=180}{i1}
  \fmfv{l=$V_{b}(z^{\prime})$ \hspace{0.5cm} $\longrightarrow ~~~ 
        B(z;z^{\prime})$,l.a=0}{o1}
  \fmfdot{v1,v2}
  \fmffreeze
  \fmfipath{p[]}
  \fmfiset{p1}{vpath1(__v1,__v2)}
  \fmfiset{p2}{vpath2(__v2,__v1)}
  \fmfiset{p3}{vpath3(__v1,__v2)}
  \fmfiset{p4}{vpath4(__v2,__v1)}
  \fmfi{fermion}{subpath (0,length(p1)) of p1}
  \fmfi{fermion}{subpath (0,length(p4)) of p4}
  \fmfiv{l=$\Phi\Phi^{\dagger}$,l.a=90}{point length(p1)/2 of p1}
  \fmfiv{l=$\Phi^{\dagger}\Phi$,l.a=-90}{point length(p3)/2 of p3}       
 \end{fmfgraph*}   
\end{fmffile}

\vspace*{2cm}
\noindent
\hspace*{2cm}
\begin{fmffile}{vvC}
 \begin{fmfgraph*}(150,80) 
  \fmfleft{i1} \fmfright{o1}
  \fmf{boson,tension=1.4}{i1,v1}
  \fmf{boson,tension=1.4}{v1,o1}
  \fmf{phantom,tension=0.9}{v1,v1}
  \fmfv{l=$V^{a}(z)$,l.a=180}{i1}
  \fmfv{l=$V_{b}(z^{\prime})$ \hspace{0.5cm} $\longrightarrow ~~~ 
        C(z;z^{\prime})$,l.a=0}{o1}
  \fmfdot{v1}
  \fmffreeze
  \fmfipath{p[]}
  \fmfiset{p1}{vpath(__v1,__v1)}
  \fmfi{fermion}{subpath (0,length(p1)) of p1}
  \fmfiv{l=$\Phi\Phi^{\dagger}$,l.a=90}{point length(p1)/2 of p1}      
 \end{fmfgraph*}   
\end{fmffile} 

\vspace*{0.5cm}
\noindent
\hspace*{2cm}
\begin{fmffile}{vvD1}
 \begin{fmfgraph*}(150,80) 
  \fmfleft{i1} \fmfright{o1}
  \fmf{boson,tension=1.6}{i1,v1}
  \fmf{boson,tension=1.6}{v2,o1}
  \fmf{phantom,left=0.6,tension=0.3,tag=1}{v1,v2}
  \fmf{phantom,right=0.6,tension=0.3,tag=2}{v2,v1}
  \fmf{phantom,right=0.6,tension=0.3,tag=3}{v1,v2}
  \fmf{phantom,left=0.6,tension=0.3,tag=4}{v2,v1} 
  \fmfv{l=$V^{a}(z)$,l.a=180}{i1}
  \fmfv{l=$V_{b}(z^{\prime})$ \hspace{0.5cm} $\longrightarrow ~~~ 
        D_{1}(z;z^{\prime})$,l.a=0}{o1}
  \fmfdot{v1,v2}
  \fmffreeze
  \fmfipath{p[]}
  \fmfiset{p1}{vpath1(__v1,__v2)}
  \fmfiset{p2}{vpath2(__v2,__v1)}
  \fmfiset{p3}{vpath3(__v1,__v2)}
  \fmfiset{p4}{vpath4(__v2,__v1)}
  \fmfi{scalar}{subpath (0,length(p1)) of p1}
  \fmfi{scalar}{subpath (0,length(p4)) of p4}
  \fmfiv{l=${\overline C}C^{\prime}$,l.a=90}{point length(p1)/2 of p1}
  \fmfiv{l=$C^{\prime}{\overline C}$,l.a=-90}{point length(p3)/2 of p3}       
 \end{fmfgraph*}   
\end{fmffile} 

\vspace*{0.5cm}
\noindent
\hspace*{2cm}
\begin{fmffile}{vvD2}
 \begin{fmfgraph*}(150,80) 
  \fmfleft{i1} \fmfright{o1}
  \fmf{boson,tension=1.6}{i1,v1}
  \fmf{boson,tension=1.6}{v2,o1}
  \fmf{phantom,left=0.6,tension=0.3,tag=1}{v1,v2}
  \fmf{phantom,right=0.6,tension=0.3,tag=2}{v2,v1}
  \fmf{phantom,right=0.6,tension=0.3,tag=3}{v1,v2}
  \fmf{phantom,left=0.6,tension=0.3,tag=4}{v2,v1} 
    \fmfv{l=$V^{a}(z)$,l.a=180}{i1}
  \fmfv{l=$V_{b}(z^{\prime})$ \hspace{0.5cm} $\longrightarrow ~~~ 
        D_{2}(z;z^{\prime})$,l.a=0}{o1}
  \fmfdot{v1,v2}
  \fmffreeze
  \fmfipath{p[]}
  \fmfiset{p1}{vpath1(__v1,__v2)}
  \fmfiset{p2}{vpath2(__v2,__v1)}
  \fmfiset{p3}{vpath3(__v1,__v2)}
  \fmfiset{p4}{vpath4(__v2,__v1)}
  \fmfi{scalar}{subpath (0,length(p1)) of p1}
  \fmfi{scalar}{subpath (0,length(p4)) of p4}
  \fmfiv{l=${\overline C}^{\prime}C$,l.a=90}{point length(p1)/2 of p1}
  \fmfiv{l=$C{\overline C}^{\prime}$,l.a=-90}{point length(p3)/2 of p3}       
 \end{fmfgraph*}   
\end{fmffile} 

\vspace*{0.6cm}
\noindent
\hspace*{2cm}
\begin{fmffile}{vvD3}
 \begin{fmfgraph*}(150,80) 
  \fmfleft{i1} \fmfright{o1}
  \fmf{boson,tension=1.4}{i1,v1}
  \fmf{boson,tension=1.4}{v2,o1}
  \fmf{phantom,left=0.6,tension=0.5,tag=1}{v1,v2}
  \fmf{phantom,right=0.6,tension=0.5,tag=2}{v1,v2} 
  \fmfv{l=$V^{a}(z)$,l.a=180}{i1}
  \fmfv{l=$V_{b}(z^{\prime})$ \hspace{0.5cm} $\longrightarrow ~~~ 
        D_{3}(z;z^{\prime})$,l.a=0}{o1}
  \fmfdot{v1,v2}
  \fmffreeze
  \fmfipath{p[]}
  \fmfiset{p1}{vpath1(__v1,__v2)}
  \fmfiset{p2}{vpath2(__v2,__v1)}
  \fmfi{scalar}{subpath (0,length(p1)) of p1}
  \fmfi{scalar}{subpath (0,length(p2)) of p2}
  \fmfiv{l=${\overline C}^{\prime}C$,l.a=90}{point length(p1)/2 of p1}
  \fmfiv{l=${\overline C}C^{\prime}$,l.a=-90}{point length(p2)/2 of p2}       
 \end{fmfgraph*}   
\end{fmffile} 

\newpage
\vspace*{0.5cm}
\noindent
\hspace*{2cm}
\begin{fmffile}{vvE1}
 \begin{fmfgraph*}(150,80) 
  \fmfleft{i1} \fmfright{o1}
  \fmf{boson,tension=1.4}{i1,v1}
  \fmf{boson,tension=1.4}{v1,o1}
  \fmf{phantom,tension=0.9}{v1,v1}
  \fmfv{l=$V^{a}(z)$,l.a=180}{i1}
  \fmfv{l=$V_{b}(z^{\prime})$ \hspace{0.5cm} $\longrightarrow ~~~ 
        E_{1}(z;z^{\prime})$,l.a=0}{o1}
  \fmfdot{v1}
  \fmffreeze
  \fmfipath{p[]}
  \fmfiset{p1}{vpath(__v1,__v1)}
  \fmfi{scalar}{subpath (0,length(p1)) of p1}
  \fmfiv{l=${\overline C}^{\prime}C$,l.a=90}{point length(p1)/2 of p1}      
 \end{fmfgraph*}   
\end{fmffile} 

\vspace*{0.5cm}
\noindent
\hspace*{2cm}
\begin{fmffile}{vvE2}
 \begin{fmfgraph*}(150,80) 
  \fmfleft{i1} \fmfright{o1}
  \fmf{boson,tension=1.4}{i1,v1}
  \fmf{boson,tension=1.4}{v1,o1}
  \fmf{phantom,tension=0.9}{v1,v1}
  \fmfv{l=$V^{a}(z)$,l.a=180}{i1}
  \fmfv{l=$V_{b}(z^{\prime})$ \hspace{0.5cm} $\longrightarrow ~~~ 
        E_{2}(z;z^{\prime})$,l.a=0}{o1}
  \fmfdot{v1}
  \fmffreeze
  \fmfipath{p[]}
  \fmfiset{p1}{vpath(__v1,__v1)}
  \fmfi{scalar}{subpath (0,length(p1)) of p1}
  \fmfiv{l=${\overline C}C^{\prime}$,l.a=90}{point length(p1)/2 of p1}      
 \end{fmfgraph*}   
\end{fmffile} \\
\hspace*{2cm}
\begin{fmffile}{vvF}
 \begin{fmfgraph*}(150,80) 
  \fmfleft{i1} \fmfright{o1}
  \fmf{boson,tension=1.4}{i1,v1}
  \fmf{boson,tension=1.4}{v2,o1}
  \fmf{phantom,left=0.6,tension=0.5,tag=1}{v1,v2}
  \fmf{phantom,right=0.6,tension=0.5,tag=2}{v1,v2} 
  \fmfv{l=$V^{a}(z)$,l.a=180}{i1}
  \fmfv{l=$V_{b}(z^{\prime})$ \hspace{0.5cm} $\longrightarrow ~~~ 
        F(z;z^{\prime})$,l.a=0}{o1}
  \fmfdot{v1,v2}
  \fmffreeze
  \fmfipath{p[]}
  \fmfiset{p1}{vpath1(__v1,__v2)}
  \fmfiset{p2}{vpath2(__v2,__v1)}
  \fmfi{boson}{subpath (0,length(p1)) of p1}
  \fmfi{boson}{subpath (0,length(p2)) of p2}
  \fmfiv{l=$VV$,l.a=90}{point length(p1)/2 of p1}
  \fmfiv{l=$VV$,l.a=-90}{point length(p2)/2 of p2}       
 \end{fmfgraph*}   
\end{fmffile} 

\vspace*{1.3cm}
\noindent
\hspace*{2cm}
\begin{fmffile}{vvG}
 \begin{fmfgraph*}(150,80) 
  \fmfleft{i1} \fmfright{o1}
  \fmf{boson,tension=1.4}{i1,v1}
  \fmf{boson,tension=1.4}{v1,o1}
  \fmf{phantom,tension=0.9}{v1,v1}
  \fmfv{l=$V^{a}(z)$,l.a=180}{i1}
  \fmfv{l=$V_{b}(z^{\prime})$ \hspace{0.5cm} $\longrightarrow ~~~ 
        G(z;z^{\prime})$,l.a=0}{o1}
  \fmfdot{v1}
  \fmffreeze
  \fmfipath{p[]}
  \fmfiset{p1}{vpath(__v1,__v1)}
  \fmfi{boson}{subpath (0,length(p1)) of p1}
  \fmfiv{l=$VV$,l.a=90}{point length(p1)/2 of p1}      
 \end{fmfgraph*}   
\end{fmffile} 

The contributions $A$ to $E$ are rather straightforward to evaluate 
much in the same way as the diagrams entering the $\langle \Phi 
\Phi^{\dagger} \rangle$ propagator. The last two graphs are more 
involved because the free propagator for the $V$ superfield is more 
complicated for generic values of the parameter $\alpha$. Moreover the
cubic and quartic vertices
\begin{eqnarray*}
&& - \frac{i}{16 \sqrt{2}} g f_{abc} \left[ {\overline D}^{2} 
\left( D^{\alpha}V^{a} \right) \right] V^{b} \left( D_{\alpha}V^{c}
 \right) \\ 
 && -\frac{1}{128} g^{2} f_{ab}{}^{e}f_{ecd} V^{a} 
\left( D^{\alpha} V^{b} \right) \left[ \left({\overline D}^{2} 
V^{c} \right) \left( D_{\alpha}V^{d} \right) \right] \, , 
\end{eqnarray*}
lead to various terms corresponding to the ways the covariant 
derivatives can act on the $V$ lines. The $V^{3}$ vertex in particular

\vspace*{0.8cm}
\noindent
\hspace*{2cm}
\begin{fmffile}{vvv}
  \begin{fmfgraph*}(90,80) 
  \fmfleft{i1,i2} \fmfright{o1}
  \fmf{boson}{i1,v1}
  \fmf{boson}{i2,v1}
  \fmf{boson}{v1,o1}
  \fmfv{l=$1\;a$,l.a=180}{i1}
  \fmfv{l=$2\;b$,l.a=180}{i2}
  \fmfv{l=$3\;c$,l.a=0}{o1}
  \end{fmfgraph*}   
\end{fmffile} 

\vspace*{0.6cm}
\noindent
gives rise to six different terms.

Schematically the calculation goes as follows. 
$\tilde{A}(p)$ is a contribution that appears because of the 
addition of the mass terms (\ref{massterm}); it is not present in the 
${\cal N}$=4 theory, \ie when $m$=0. It is useful to discuss 
separately the corrections coming from diagrams $\tilde{A}$, $\tilde{B}$ 
and $\tilde{C}$ and those obtained from $\tilde{D}_{i}$, 
$\tilde{E}_{i}$, $\tilde{F}$ and $\tilde{G}$, since the latter 
correspond to the one-loop contribution to the vector superfield 
propagator in the ${\cal N}$=1 supersymmetric Yang--Mills theory.

The first diagram, $\tilde{A}$, is logarithmically divergent and reads
\begin{equation}
	\tilde{A}(p) = 3 g^{2}N(N^{2}-1)\delta_{ab}  \int 
	\frac{d^{4}k}{(2\pi)^{4}} d^{4}\theta  \, \left\{
	\frac{m^{2}\left[ V^{a}(p,\theta,{\overline\theta}) 
	V^{b}(-p,\theta,{\overline\theta})\right]}
	{(k^{2}+m^{2})[(p-k)^{2}+m^{2}]} \right\} .
    \label{vvmassdiverg}
\end{equation}
For the diagram $\tilde{B}$ application of the Feynman rules gives 
rise to three different contributions, one quadratically divergent 
and two logarithmically divergent
\begin{eqnarray}
	&& \tilde{B}(p)= 
	3 g^{2} N(N^{2}-1) \delta_{ab}
	\int\frac{d^{4}k}{(2\pi)^{4}} d^{4}\theta \, 
	\frac{1}{(k^{2}+m^{2})[(p-k)^{2}+m^{2}]} 
	\cdot \nonumber \\
	&& \cdot \left\{ k^{2}V^{a}
	(p,\theta,{\overline\theta})V^{b}(-p,\theta,{\overline\theta})
	\rule{0pt}{18pt} - 
	\frac{i}{4} p_{\mu}\sigma^{\mu}_{\alpha{\dot\alpha}}V^{a}
	(p,\theta,{\overline\theta}) \left[
	({\overline D}^{\dot\alpha}D^{\alpha}) 
	V^{b}(-p,\theta,{\overline\theta}) \right] + \right. \nonumber \\
	&& \left. + \frac{1}{4}
	V^{a}(p,\theta,{\overline\theta}) \left[({\overline D}^{2}D^{2}) 
	V^{b}(-p,\theta,{\overline\theta}) \right] \right\} \, .
	\label{bcorrvv}
\end{eqnarray}
The tadpole diagram $\tilde{C}$ gives a quadratically divergent 
result of the form
\begin{equation}
	\tilde{C} = -3 g^{2} N(N^{2}-1) \delta^{ab}
	\int \frac{d^{4}k}{(2\pi)^{4}} d^{4}\theta \, 
	\frac{1}{(k^{2}+m^{2})} 
	\left\{ V^{a}(p,\theta,{\overline\theta}) 
	V^{b}(-p,\theta,{\overline\theta}) \right\} \, .
\end{equation}
Summing and subtracting $m^{2}$ in the numerator of the first term 
in (\ref{bcorrvv}) and putting the three corrections $\tilde{A}$, 
$\tilde{B}$ and $\tilde{C}$ together gives a net result that is only 
logarithmically divergent
\begin{eqnarray}
	\tilde{A}(p)+\tilde{B}(p)+\tilde{C}(p) \! &=& \! 
	-3 g^{2} N(N^{2}-1) \delta_{ab}
	\int\frac{d^{4}k}{(2\pi)^{4}} d^{4}\theta \, 
	\frac{1}{(k^{2}+m^{2})[(p-k)^{2}+m^{2}]} \cdot \nonumber \\
	&& \cdot \left\{ \frac{i}{4} 
	p_{\mu}\sigma^{\mu}_{\alpha{\dot\alpha}}V^{a}
	(p,\theta,{\overline\theta}) \left[
	({\overline D}^{\dot\alpha}D^{\alpha}) 
	V^{b}(-p,\theta,{\overline\theta}) \right]+\right.\nonumber \\ 
	&& \left. + \frac{1}{16}
	V^{a}(p,\theta,{\overline\theta}) \left[({\overline D}^{2}D^{2}) 
	V^{b}(-p,\theta,{\overline\theta}) \right] \right\} \, .
	\label{vvlog1}
\end{eqnarray}
Notice that in particular the logarithmically divergent contribution 
proportional to $m^{2}$ exactly cancels out. 
This is crucial because this correction would correspond to a mass 
renormalization for the vector superfield that is known to be 
excluded in any gauge theory as well as in supersymmetric theories in 
general.

The diagrams $\tilde{D}_{i}$ and $\tilde{E}_{i}$ are completely 
analogous to the previous ones, with the only difference that the mass
does not appear in the denominators and there is a minus sign 
associated with the loops. Their sum is logarithmically divergent and 
takes the form 
\begin{eqnarray}
	&& \tilde{D}_{1}(p)+\tilde{D}_{2}(p)+\tilde{D}_{3}(p)+
	\tilde{E}_{1}(p)+\tilde{E}_{2}(p) = \frac{1}{8} g^{2} N(N^{2}-1)
	\delta_{ab} \cdot \nonumber \\ 
	&& \cdot \int\frac{d^{4}k}{(2\pi)^{4}} d^{4}\theta \, 
	\frac{1}{k^{2}(p-k)^{2}} \left\{ i p_{\mu} 
	\sigma^{\mu}_{\alpha{\dot\alpha}}
	V^{a}(p,\theta,{\overline\theta}) \left[({\overline D}^{\dot\alpha}
	D^{\alpha}) V^{b}(-p,\theta,{\overline\theta}) \right] +
	\right. \nonumber \\
	&& \hspace{4.4cm} \left. +
	\frac{1}{8} V^{a}(p,\theta,{\overline\theta}) 
	\left[({\overline D}^{2}D^{2}) 
	V^{b}(-p,\theta,{\overline\theta}) \right] \right\} \, .
	\label{vvlog2} 
\end{eqnarray}

All of these corrections do not depend on the gauge parameter 
$\gamma$=$\alpha -1$ and must be summed to those coming from the last 
two diagrams. $\tilde{F}$ exactly vanishes for any $\alpha$, so that 
only $\tilde{E}$ needs to be considered. This diagram produces in 
principle 72 corrections because the Feynman rules give rise to 18 
terms (distributing the covariant derivatives associated with the two 
vertices), each of which splits into 4, since the free propagator 
itself contains two terms. Many of these contributions can be easily 
shown to vanish using the properties of the covariant derivatives. In 
particular one uses  
\begin{displaymath}
	{\overline D}^{2}D_{\alpha}{\overline D}^{2}D^{2} = 0 \, ,
\end{displaymath}
which follows from ${\overline D}^{3}=0$ and use of the (anti) 
commutation relations for the $D$'s. It is useful to separate in the 
non-vanishing part terms proportional to $\gamma$ and $\gamma^{2}$, 
from the $\gamma$-independent terms. The latter combine to give a 
logarithmically divergent correction that together with that of equation 
(\ref{vvlog2}) cancel the correction (\ref{vvlog1}) coming from the sum 
$\tilde{A}+\tilde{B}+\tilde{C}$. Actually if $m\neq 0$ this 
sum is finite and exactly vanishes at $m$=0. As a result the only 
non-vanishing corrections to the vector superfield propagator at the 
one loop level come from terms proportional to $\gamma$ and to 
$\gamma^{2}$ in $\tilde{E}(p)$. The former contain an infrared divergent 
part of the form
\begin{displaymath}
   J^{(1)}(p) = c_{1}\gamma g^{2}\delta^{a}_{b} \int \frac{d^{4}k}
   {(2\pi)^{4}} d^{4}\theta \, \frac{\sigma^{\mu}_{\alpha{\dot\alpha}}
   \sigma^{\nu}_{\beta{\dot\beta}}p_{\mu}p_{\nu}}{k^{4}(p-k)^{2}} 
   \left\{ V_{a}(p,\theta,{\overline\theta}) \left[D^{\alpha}
   {\overline D}^{\dot\alpha}{\overline D}^{\dot\beta}D^{\beta}
   V^{b}(-p,\theta,{\overline\theta}) \right] \right\} \, ,
\end{displaymath} 
but is ultraviolet finite.
Furthermore there is a correction, finite both in the ultraviolet and 
in the infrared regions, proportional to $\gamma^{2}$, that reads
\begin{eqnarray*}
   J^{(2)}(p) &=& c_{2}\gamma^{2}g^{2}\delta^{a}_{b} \int \frac{d^{4}k}
   {(2\pi)^{4}}d^{4}\theta  \, \frac{(\sigma^{\nu}
   {\overline\sigma}^{\mu}\sigma^{\lambda})_{\alpha{\dot\alpha}}
   (p-k)_{\mu}k_{\nu}p_{\lambda}}{k^{4}(p-k)^{4}} \cdot \\ 
   && \hspace{1.3cm} \cdot 
   \left\{ \left[ ({\overline D}^{2}D^{\alpha})
   V_{a}(p,\theta,{\overline\theta}) \right] 
   \left[ (D^{2}{\overline D}^{\dot\alpha}) 
   V^{b}(-p,\theta,{\overline\theta}) \right] \right\} \, .
\end{eqnarray*}
In the previous expressions for $J^{(1)}$ and $J^{(2)}$ $c_{1}$ and 
$c_{2}$ are non-vanishing numerical constants proportional to 
$N(N^{2}-1)$. 

In conclusion in the ${\cal N}$=4 theory, \ie when $m$=0, the one-loop 
correction to the vector superfield propagator, just like that of the 
chiral superfield, is infrared singular unless the Fermi--Feynman 
gauge, $\alpha$=1, is chosen, in which case it vanishes. Like in the 
case of the chiral superfield propagator the non-vanishing 
$\gamma$-dependent corrections correspond to a wave function 
renormalization for the superfield $V$ and are zero on-shell, 
\ie for $p^{2}$=0.

The proof of the finiteness of the theory in the presence of the mass 
terms (\ref{massterm}) given in \cite{parkeswest} is based on naive power 
counting, which gives the superficial degree of divergence, $d_{s}$, 
of a diagram in ${\cal N}$=1 superspace 
\cite{divdegree}
\begin{displaymath}
	d_{s} = 2 - E -C \, ,
\end{displaymath}
where $E$ is the number of external (anti) chiral lines and $C$ the 
number of $\Phi \Phi$ or $\Phi^{\dagger}\Phi^{\dagger}$ propagators. 
In \cite{parkeswest} it is also argued that for corrections to the 
effective action involving only $V$ superfields the requirement of gauge 
invariance reduces the degree of divergence to
\begin{displaymath}
	d_{s} = -C \, .
\end{displaymath}
However for the purposes of \cite{japan,yoshida} it appears crucial 
that no divergences, not even corresponding to a wave function 
renormalization, are present in total $n$-point functions for any $n$. 
The computation of the two-point functions in this section has shown 
that this actually the case at one loop. An argument for the 
generalization of this result to Green functions with an arbitrary 
number of external $V$ lines will now be briefly discussed. First note 
that diagrams involving only vector and ghost superfields are not 
modified by the inclusion of (\ref{massterm}), so that one must only 
consider graphs containing internal chiral lines. The ultraviolet 
properties of diagrams that are only logarithmically divergent in the 
original ${\cal N}$=4 theory are not modified by the presence of the 
mass in the propagators. Thus the contributions to be analyzed are 
the quadratically divergent ones, that can acquire subleading 
logarithmic singularities, plus eventually new diagrams involving 
$\Phi\Phi$ and $\Phi^{\dagger}\Phi^{\dagger}$ propagators. The 
relevant quadratic divergences come from tadpole diagrams 

\vspace{1cm}
\noindent
\hspace*{2cm}
\begin{fmffile}{vvvvC}
 \begin{fmfgraph*}(160,110) 
  \fmfleft{i1,i2,i3,i4,i5} 
  \fmfright{o1,o2,o3,o4,o5}
  \fmf{boson,tension=0.2,tag=1}{i1,v1}
  \fmf{boson,tension=0.2,tag=2}{i2,v1}
  \fmf{phantom,tension=0.2,tag=3}{i3,v1}
  \fmf{phantom,tension=0.2,tag=4}{i4,v1}
  \fmf{boson,tension=0.2,tag=5}{i5,v1}
  \fmf{boson,tension=0.2,tag=10}{v1,o1}
  \fmf{boson,tension=0.2,tag=11}{v1,o2}
  \fmf{phantom,tension=0.2,tag=12}{v1,o3}
  \fmf{phantom,tension=0.2,tag=13}{v1,o4}
  \fmf{boson,tension=0.2,tag=14}{v1,o5}
  \fmf{fermion,left=-90,tension=0.9}{v1,v1}
  \fmfv{l=$\tilde{V}^{a_{1}}$,l.a=180}{i1}
  \fmfv{l=$\tilde{V}^{a_{2}}$,l.a=180}{i2}
  \fmfv{l=$\tilde{V}^{a_{n}}$,l.a=180}{i5}
  \fmfv{l=$\tilde{V}^{b_{1}}$,l.a=0}{o1}
  \fmfv{l=$\tilde{V}^{b_{2}}$,l.a=0}{o2}
  \fmfv{l=$\tilde{V}^{b_{n}}$,l.a=0}{o5}
  \fmfdot{v1}
  \fmffreeze
  \fmfipath{p[]}
  \fmfiset{p3}{vpath3(__i3,__v1)}
  \fmfiset{p4}{vpath4(__i4,__v1)}
  \fmfiset{p12}{vpath12(__v1,__o3)}
  \fmfiset{p13}{vpath13(__v1,__o4)}
  \fmfi{dots}{point length(p3)/2 of p3
              -- point 14length(p4)/30 of p4}
  \fmfi{dots}{point length(p12)/2 of p12
              -- point 16length(p13)/30 of p13}           
 \end{fmfgraph*}   
\end{fmffile} 

\vspace{0.5cm}
\noindent
The only new diagram containing $\Phi\Phi$ and 
$\Phi^{\dagger}\Phi^{\dagger}$ propagators that must be considered 
is 

\vspace{1cm}
\noindent
\hspace*{2cm}
\begin{fmffile}{vvvvA}
 \begin{fmfgraph*}(160,110) 
  \fmfleft{i1,i2,i3,i4,i5} 
  \fmfright{o1,o2,o3,o4,o5}
  \fmf{boson,tension=0.3,tag=1}{i1,v1}
  \fmf{boson,tension=0.3,tag=2}{i2,v1}
  \fmf{phantom,tension=0.3,tag=3}{i3,v1}
  \fmf{phantom,tension=0.3,tag=4}{i4,v1}
  \fmf{boson,tension=0.3,tag=5}{i5,v1}
  \fmf{phantom,left=0.6,tension=0.2,tag=6}{v1,v2}
  \fmf{phantom,right=0.6,tension=0.2,tag=7}{v1,v2}
  \fmf{phantom,left=0.6,tension=0.2,tag=8}{v2,v1}
  \fmf{phantom,right=0.6,tension=0.2,tag=9}{v2,v1}
  \fmf{boson,tension=0.3,tag=10}{v2,o1}
  \fmf{boson,tension=0.3,tag=11}{v2,o2}
  \fmf{phantom,tension=0.3,tag=12}{v2,o3}
  \fmf{phantom,tension=0.3,tag=13}{v2,o4}
  \fmf{boson,tension=0.3,tag=14}{v2,o5}
  \fmfv{l=$\tilde{V}^{a_{1}}$,l.a=180}{i1}
  \fmfv{l=$\tilde{V}^{a_{2}}$,l.a=180}{i2}
  \fmfv{l=$\tilde{V}^{a_{n}}$,l.a=180}{i5}
  \fmfv{l=$\tilde{V}^{b_{1}}$,l.a=0}{o1}
  \fmfv{l=$\tilde{V}^{b_{2}}$,l.a=0}{o2}
  \fmfv{l=$\tilde{V}^{b_{n}}$,l.a=0}{o5}
  \fmfdot{v1,v2}
  \fmffreeze
  \fmfipath{p[]}
  \fmfiset{p3}{vpath3(__i3,__v1)}
  \fmfiset{p4}{vpath4(__i4,__v1)}
  \fmfiset{p6}{vpath6(__v1,__v2)}
  \fmfiset{p7}{vpath7(__v1,__v2)}
  \fmfiset{p8}{vpath8(__v2,__v1)}
  \fmfiset{p9}{vpath9(__v2,__v1)}
  \fmfiset{p12}{vpath12(__v2,__o3)}
  \fmfiset{p13}{vpath13(__v2,__o4)}
  \fmfi{fermion}{subpath (0,length(p6)/2) of p6}
  \fmfi{fermion}{subpath (0,length(p9)/2) of p9}
  \fmfi{fermion}{subpath (length(p8)/2,length(p8)) of p8}
  \fmfi{fermion}{subpath (length(p7)/2,length(p7)) of p7}
  \fmfiv{d.sh=cross,d.si=3mm}{point length(p6)/2 of p6}
  \fmfiv{d.sh=cross,d.si=3mm}{point length(p7)/2 of p7}
  \fmfi{dots}{point length(p3)/2 of p3
              -- point 9length(p4)/20 of p4}
  \fmfi{dots}{point length(p12)/2 of p12
              -- point 11length(p13)/20 of p13}           
 \end{fmfgraph*}   
\end{fmffile} 

\vspace{0.5cm}
\noindent
Since all the vertices entering these graphs come from the expansion 
of the term $\tr \left[ e^{-V}\Phi^{\dagger}e^{V}\Phi \right]$ 
in the action, the only covariant derivatives that are present come 
from the functional derivatives and must act on the internal (anti) 
chiral lines to give a non-vanishing result. As a consequence the 
loop integral in the above diagrams is exactly the same as for the 
$\tilde{C}$ and $\tilde{A}$ corrections to the $VV$ propagator. 
Hence summing to the previous diagrams the contribution of

\vspace{1cm}
\noindent
\hspace*{2cm}
\begin{fmffile}{vvvvB}
 \begin{fmfgraph*}(160,110) 
  \fmfleft{i1,i2,i3,i4,i5} 
  \fmfright{o1,o2,o3,o4,o5}
  \fmf{boson,tension=0.3,tag=1}{i1,v1}
  \fmf{boson,tension=0.3,tag=2}{i2,v1}
  \fmf{phantom,tension=0.3,tag=3}{i3,v1}
  \fmf{phantom,tension=0.3,tag=4}{i4,v1}
  \fmf{boson,tension=0.3,tag=5}{i5,v1}
  \fmf{phantom,left=0.6,tension=0.2,tag=6}{v1,v2}
  \fmf{phantom,right=0.6,tension=0.2,tag=7}{v1,v2}
  \fmf{phantom,left=0.6,tension=0.2,tag=8}{v2,v1}
  \fmf{phantom,right=0.6,tension=0.2,tag=9}{v2,v1}
  \fmf{boson,tension=0.3,tag=10}{v2,o1}
  \fmf{boson,tension=0.3,tag=11}{v2,o2}
  \fmf{phantom,tension=0.3,tag=12}{v2,o3}
  \fmf{phantom,tension=0.3,tag=13}{v2,o4}
  \fmf{boson,tension=0.3,tag=14}{v2,o5}
  \fmfv{l=$\tilde{V}^{a_{1}}$,l.a=180}{i1}
  \fmfv{l=$\tilde{V}^{a_{2}}$,l.a=180}{i2}
  \fmfv{l=$\tilde{V}^{a_{n}}$,l.a=180}{i5}
  \fmfv{l=$\tilde{V}^{b_{1}}$,l.a=0}{o1}
  \fmfv{l=$\tilde{V}^{b_{2}}$,l.a=0}{o2}
  \fmfv{l=$\tilde{V}^{b_{n}}$,l.a=0}{o5}
  \fmfdot{v1,v2}
  \fmffreeze
  \fmfipath{p[]}
  \fmfiset{p3}{vpath3(__i3,__v1)}
  \fmfiset{p4}{vpath4(__i4,__v1)}
  \fmfiset{p6}{vpath6(__v1,__v2)}
  \fmfiset{p7}{vpath7(__v1,__v2)}
  \fmfiset{p8}{vpath8(__v2,__v1)}
  \fmfiset{p9}{vpath9(__v2,__v1)}
  \fmfiset{p12}{vpath12(__v2,__o3)}
  \fmfiset{p13}{vpath13(__v2,__o4)}
  \fmfi{fermion}{subpath (0,length(p6)) of p6}
  \fmfi{fermion}{subpath (0,length(p8)) of p8}
  \fmfi{dots}{point length(p3)/2 of p3
              -- point 9length(p4)/20 of p4}
  \fmfi{dots}{point length(p12)/2 of p12
              -- point 11length(p13)/20 of p13}           
 \end{fmfgraph*}   
\end{fmffile} 

\vspace{0.5cm}
\noindent
leads to a net logarithmically divergent correction that is exactly 
the same as the one obtained in the original ${\cal N}$=4 theory. 
Like in the latter case this divergence will be cancelled by the 
contributions coming from the other one-loop diagrams involving $V$ 
and ghost internal lines. The argument given here reinforces the results 
of \cite{parkeswest}, at least at the one-loop level, and supports 
the proposal, put forward in \cite{japan,yoshida}, according to which 
the mass deformed ${\cal N}$=4 model can be considered a consistent 
supersymmetry-preserving regularization for a class of ${\cal N}$=1 
theories.

Notice that for the previous discussion it is not necessary to 
consider equal masses for all the (anti) chiral superfields. The same 
results can be proved giving different masses to the three superfields. 
This can be easily understood since in each diagram considered in this 
section only one chiral/anti-chiral pair is involved, because the 
propagators are diagonal in `flavour' space and the vertices containing 
vector and (anti) chiral superfields couple $\Phi^{\dagger}_{I}$ and 
$\Phi^{I}$ with the same index $I$. Basically this means that the 
above discussed cancellations apply separately to the contributions 
of each (anti) chiral superfield. From the viewpoint of the 
dimensional analysis of \cite{parkeswest} having different masses 
$m_{I}$ is irrelevant. 

As a consequence one can in particular give mass to only two of the 
(anti) chiral multiplets. This suggests the possibility of 
generalizing the approach of \cite{japan,yoshida} to the case of 
${\cal N}$=2 super Yang--Mills theories. A discussion of the effect 
of a ${\cal N}$=2 mass term in ${\cal N}$=4 supersymmetric 
Yang--Mills theory can be found in \cite{n2mass}.

\subsection{Discussion}

The infrared divergences found in the calculation of the propagators 
of the chiral and the vector superfields are due to the fact that the 
vector superfield is dimensionless, so 
that it contains in particular, as its lowest component, the scalar 
$C$ that is itself dimensionless and hence has a propagator which 
behaves, in momentum space, like
\begin{displaymath}
	\langle (CC)(k) \rangle \sim \frac{1}{k^{4}} \, .
\end{displaymath}
The contribution of the scalar $C$ to the $\langle VV \rangle$ 
propagator leads to an infrared divergence whenever a diagram contains 
a loop involving a $VV$ line. In the Fermi--Feynman gauge the problem 
is not present because the choice $\alpha=1$ gives a kinetic matrix 
of the form of equation (\ref{kineticmatrices}) in the component 
expansion. The corresponding inverse matrix $M^{-1}$ in  
(\ref{inverskinmatr}) shows that no $\langle CC \rangle$ free 
propagator is present. On the contrary any choice $\alpha \neq 1$ 
produces such a propagator, \ie gives a matrix $M^{-1}$ with a 
non-vanishing $(M^{-1})_{11}$ entry, leading to the previously 
discussed problems. An explicit calculation in components with 
$\alpha \neq 1$ and without fixing the Wess--Zumino gauge should 
show problems with infrared divergences analogous to those encountered 
here. These problems with infrared divergences are not peculiar of 
the ${\cal N}$=4 super Yang--Mills theory, but appear in any  
supersymmetric gauge theory. Analogous infrared divergences in Green 
functions have been observed in \cite{juerstorey2,juerstorey1}. The 
conclusion proposed in these papers is that there exist no way to 
remove the infrared divergences, so that the choice $\alpha$=1 is 
somehow necessary, at least for the computation of gauge-dependent 
quantities.

There are two general theorems concerning infrared divergences in 
quantum field theory. The first is the Kinoshita--Lee--Nauenberg 
theorem \cite{kinoleenauen}, which deals with infrared divergences 
in cross sections. It states that no infrared problem is present 
in physical cross sections of a renormalizable quantum field theory, 
if the appropriate sums over degenerate initial and final states, 
associated with soft and collinear particles, are considered. 
The second theorem was proved by Kinoshita, Poggio and Quinn 
\cite{kinopoggioquinn} and concerns Green functions. The statement of 
the theorem is that the proper (one-particle irreducible) Euclidean 
Green functions with non-exceptional external momenta are free of 
infrared divergences in a renormalizable quantum field theory. A set 
of momenta $p_{i}$, $i=1,2,\ldots,n$ is said to be exceptional if any 
of the partial sums
\begin{displaymath}
	\sum_{i\in I} p_{i}\, , \qquad {\rm with}~~I~~{\rm a~subset~of}~~ 
	\{1,2,\ldots,n\} \, ,
\end{displaymath}
vanishes. The reason for the absence of divergences is that the 
external momenta, if non-exceptional, play the r\^ole of an infrared 
regulator in the Green functions. 

The proof of the theorem is based on dimensional analysis which does 
not work in the presence of a $\frac{1}{k^{4}}$ propagator. 
In particular it fails in the case of 
supersymmetric gauge theories in general gauges, because the 
adimensionality of the $V$ superfield implies that factors of 
$\frac{1}{k^{4}}$ can enter the loop integrals. The apparent violation 
of the Kinoshita--Poggio--Quinn theorem can be understood from the 
viewpoint of the component  formulation: supersymmetric gauge theories, 
being non-polynomial and thus containing an infinite number of 
interaction terms, are not {\em formally} renormalizable. The choice 
of the Wess--Zumino gauge makes the theory polynomial. In fact in 
this case no infrared divergence is found in Green functions and the 
general theorems are satisfied. However in the case of the ${\cal 
N}$=4 theory the choice of the WZ gauge results in a change of the 
ultraviolet properties of the model, at least for what concerns gauge 
dependent quantities, {\em e.g.} the propagators.

In theories with less supersymmetry, in which ultraviolet divergences 
are present the problem might seem less severe, since the subtraction 
of the ultraviolet infinities could cure also the infrared 
singularity. 

The infrared problems discussed here are however gauge artifacts and 
cannot affect gauge invariant quantities. The mechanism leading to 
the cancellation of infrared divergences in gauge independent Green 
functions has not been verified in detail yet, but can be understood 
starting from the super Feynman rules. When the propagator $\langle 
VV \rangle$ is inserted into a Feynman graph it is connected to 
vertices which carry covariant derivatives. These derivatives come 
directly from the form of the action for vertices involving only $V$ 
fields and from the definition of the functional derivatives for 
vertices involving (anti) chiral superfields. 
The situation in diagrams for gauge invariant 
quantities is such that at least one covariant 
derivative can always be brought to act on the $V$ propagator by 
integrations by parts. In this way an additional factor of the momentum 
$k$ is generated in the numerator. More precisely for the infrared 
problematic term $-\frac{1}{k^{4}}({\overline 
D}^{2}D^{2}+D^{2}{\overline D}^{2})\delta_{4}(\theta-\theta^{\prime})$ 
one gets
\begin{displaymath}
	D_{\alpha}\left[-\frac{1}{k^{4}}({\overline D}^{2}
	D^{2}+D^{2}{\overline D}^{2})\delta_{4}(\theta-\theta^{\prime})
	\right] = -4i\frac{k_{\mu}\sigma^{\mu}_{\alpha{\dot\alpha}}}{k^{4}}
	{\overline D}^{\dot\alpha}D^{2}
\end{displaymath}
and
\begin{displaymath}
	{\overline D}^{\dot\alpha}\left[-\frac{1}{k^{4}}({\overline D}^{2}
	D^{2}+D^{2}{\overline D}^{2})\delta_{4}(\theta-\theta^{\prime})
	\right] = 4i\frac{k_{\mu}{\overline \sigma}^{\mu{\dot\alpha}\alpha}}
	{k^{4}} D_{\alpha}{\overline D}^{2} \, .
\end{displaymath}
A rigorous and general check of this mechanism in gauge invariant 
Green functions has not been achieved yet.

\section{Three- and four-point functions of ${\cal N}$=1 superfields}
\label{threefourpointf}

The computation of Green functions with three and four external legs 
will now be considered. From now on the Fermi--Feynman gauge will be 
fixed. The three-point functions should in principle suffer from 
infrared problems of the kind encountered in the preceding section. 
This claim will not be studied here since the calculation with 
$\alpha \neq 1$ is awkward, even at the one-loop level, because of the 
huge number of contributions. Four-point functions on the 
contrary should be infrared finite, because they are directly related 
to physical scattering amplitudes. 
Notice, however, that infrared divergences in the Green function 
with four external $V$ lines were found in \cite{juerstorey1}. In 
that paper beyond the infrared singularity, it was shown that the 
adimensionality of the superfield $V$ implies that it requires a 
non-linear renormalization, \ie the renormalized field $V_{R}$ is a 
non-linear function of the bare field $V$
\begin{displaymath}
	V_{R}=f(V) \, .
\end{displaymath}

An example of three-point function will be briefly discussed here and 
then the more interesting case of a four-point function will be studied 
in greater detail.
 
\subsection{Three-point functions}
\fancyhead[LO]{{\footnotesize 3.3~~{\it Three- and four-point functions}}}

The simplest three point function to consider corresponds to the 
correction to the vertex $\varepsilon_{IJK}\tr\left(\Phi^{I}\Phi^{J}
\Phi^{K}\right)$ (or $\varepsilon^{IJK}\tr\left(\Phi^{\dagger}_{I}
\Phi^{\dagger}_{J}\Phi^{\dagger}_{K}\right)$), which is determined by one 
single diagram at the one loop level. However as a less trivial example of 
computation of three-point functions the one-loop correction to the vertex 

\vspace{0.5cm}
\noindent
\hspace*{1.5cm}
\begin{fmffile}{phi+vphi}
 \begin{fmfgraph*}(150,95) 
  \fmfleft{i1,i2} \fmfright{o1}
  \fmf{phantom,tension=0.7,tag=1}{i1,v1}
  \fmf{phantom,tension=0.7,tag=2}{i2,v1}
  \fmf{phantom,tension=0.7,tag=3}{v1,i1}
  \fmf{phantom,tension=0.7,tag=4}{v1,i2} 
  \fmf{boson,tension=2.5}{v1,o1}
  \fmfv{l=$\Phi^{a\dagger}_{I}(z)$,l.a=180}{i1}
  \fmfv{l=$\Phi^{J}_{b}(z)$,l.a=180}{i2}
  \fmfv{l=$V_{c}(z)$ \hspace{0.5cm} 
          $\longrightarrow ~~~ \left(\Phi^{a\dagger}_{I}
		  V^{c}\Phi_{b}^{I}\right)(z)$,l.a=0}{o1}
  \fmfdot{v1}
  \fmffreeze
  \fmfipath{p[]}
  \fmfiset{p1}{vpath1(__i1,__v1)}
  \fmfiset{p2}{vpath2(__i2,__v1)}
  \fmfiset{p3}{vpath3(__v1,__i1)}
  \fmfiset{p4}{vpath4(__v1,__i2)}
  \fmfi{fermion}{subpath (0,length(p1)) of p1}
  \fmfi{fermion}{subpath (0,length(p4)) of p4}
 \end{fmfgraph*}   
\end{fmffile} 

\vspace{0.5cm}
\noindent
will be considered here. The correction at the one-loop level comes 
from the following diagrams (the notation for the propagators is the 
same as in the previous section)

\vspace{0.5cm}
\noindent
\hspace*{1.5cm}
\begin{fmffile}{phi+vphiA}
 \begin{fmfgraph*}(150,95) 
  \fmfleft{i1,i2} \fmfright{o1}
  \fmf{phantom,tension=0.7,tag=1}{i1,v1}
  \fmf{phantom,tension=0.7,tag=2}{i2,v1}
  \fmf{phantom,tension=0.7,tag=3}{v1,i1}
  \fmf{phantom,tension=0.7,tag=4}{v1,i2} 
  \fmf{boson,tension=2.5}{v1,o1}
  \fmfv{l=$\tilde{\Phi}^{a\dagger}_{I}(p)$,l.a=180}{i1}
  \fmfv{l=$\tilde{\Phi}^{J}_{b}(q)$,l.a=180}{i2}
  \fmfv{l=$\tilde{V}_{c}(-p-q)$ \hspace{0.5cm} 
          $\longrightarrow ~~~ \tilde{A}(p;q)$,l.a=0}{o1}
  \fmfdot{v1}
  \fmffreeze
  \fmfipath{p[]}
  \fmfiset{p1}{vpath1(__i1,__v1)}
  \fmfiset{p2}{vpath2(__i2,__v1)}
  \fmfiset{p3}{vpath3(__v1,__i1)}
  \fmfiset{p4}{vpath4(__v1,__i2)}
  \fmfi{fermion}{subpath (0,length(p1)/2) of p1}
  \fmfi{fermion}{subpath (length(p1)/2,length(p1)) of p1}
  \fmfi{fermion}{subpath (0,length(p4)/2) of p4}
  \fmfi{fermion}{subpath (length(p4)/2,length(p4)) of p4}
  \fmfi{boson}{point length(p1)/2 of p1 -- point
               length(p2)/2 of p2}
  \fmfiv{d.sh=circle,d.fill=full,d.si=1.5mm}{point length(p1)/2 of p1}      
  \fmfiv{d.sh=circle,d.fill=full,d.si=1.5mm}{point length(p2)/2 of p2}
 \end{fmfgraph*}   
\end{fmffile} 

\vspace{0.8cm}
\noindent
\hspace*{1.5cm}
\begin{fmffile}{phi+vphiB}
 \begin{fmfgraph*}(150,95) 
  \fmfleft{i1,i2} \fmfright{o1}
  \fmf{phantom,tension=0.7,tag=1}{i1,v1}
  \fmf{phantom,tension=0.7,tag=2}{i2,v1}
  \fmf{phantom,tension=0.7,tag=3}{v1,i1}
  \fmf{phantom,tension=0.7,tag=4}{v1,i2} 
  \fmf{boson,tension=2.5}{v1,o1}
  \fmfv{l=$\tilde{\Phi}^{a\dagger}_{I}(p)$,l.a=180}{i1}
  \fmfv{l=$\tilde{\Phi}^{J}_{b}(q)$,l.a=180}{i2}
  \fmfv{l=$\tilde{V}_{c}(-p-q)$ \hspace{0.5cm} 
          $\longrightarrow ~~~ \tilde{B}(p;q)$,l.a=0}{o1}
  \fmfdot{v1}
  \fmffreeze
  \fmfipath{p[]}
  \fmfiset{p1}{vpath1(__i1,__v1)}
  \fmfiset{p2}{vpath2(__i2,__v1)}
  \fmfiset{p3}{vpath3(__v1,__i1)}
  \fmfiset{p4}{vpath4(__v1,__i2)}
  \fmfi{fermion}{subpath (0,length(p1)/2) of p1}
  \fmfi{fermion}{subpath (0,length(p3)/2) of p3}
  \fmfi{fermion}{subpath (length(p2)/2,length(p2)) of p2}
  \fmfi{fermion}{subpath (length(p4)/2,length(p4)) of p4}
  \fmfi{fermion}{point length(p2)/2 of p2 -- point length(p1)/2 of p1}
  \fmfiv{d.sh=circle,d.fill=full,d.si=1.5mm}{point length(p1)/2 of p1}      
  \fmfiv{d.sh=circle,d.fill=full,d.si=1.5mm}{point length(p2)/2 of p2}
 \end{fmfgraph*}   
\end{fmffile} 

\vspace{0.8cm}
\noindent
\hspace*{1.5cm}
\begin{fmffile}{phi+vphiC}
 \begin{fmfgraph*}(150,95) 
  \fmfleft{i1,i2} \fmfright{o1}
  \fmf{phantom,tension=0.7,tag=1}{i1,v1}
  \fmf{phantom,tension=0.7,tag=2}{i2,v1}
  \fmf{phantom,tension=0.7,tag=3}{v1,i1}
  \fmf{phantom,tension=0.7,tag=4}{v1,i2} 
  \fmf{boson,tension=2.5}{v1,o1}
  \fmfv{l=$\tilde{\Phi}^{a\dagger}_{I}(p)$,l.a=180}{i1}
  \fmfv{l=$\tilde{\Phi}^{J}_{b}(q)$,l.a=180}{i2}
  \fmfv{l=$\tilde{V}_{c}(-p-q)$ \hspace{0.5cm} 
          $\longrightarrow ~~~ \tilde{C}(p;q)$,l.a=0}{o1}
  \fmfdot{v1}
  \fmffreeze
  \fmfipath{p[]}
  \fmfiset{p1}{vpath1(__i1,__v1)}
  \fmfiset{p2}{vpath2(__i2,__v1)}
  \fmfiset{p3}{vpath3(__v1,__i1)}
  \fmfiset{p4}{vpath4(__v1,__i2)}
  \fmfi{fermion}{subpath (0,length(p1)/2) of p1}
  \fmfi{boson}{subpath (0,length(p3)/2) of p3}
  \fmfi{boson}{subpath (0,length(p4)/2) of p4}
  \fmfi{fermion}{point length(p1)/2 of p1 -- point length(p2)/2 of p2}
  \fmfi{fermion}{subpath (length(p4)/2,length(p4)) of p4}
  \fmfiv{d.sh=circle,d.fill=full,d.si=1.5mm}{point length(p1)/2 of p1}      
  \fmfiv{d.sh=circle,d.fill=full,d.si=1.5mm}{point length(p2)/2 of p2}
 \end{fmfgraph*}   
\end{fmffile} 

\vspace{0.8cm}
\noindent
\hspace*{1.5cm}
\begin{fmffile}{phi+vphiD}
 \begin{fmfgraph*}(150,95) 
  \fmfleft{i1,i2} \fmfright{o1}
  \fmf{fermion}{i1,v1}
  \fmf{fermion}{v1,i2}
  \fmf{boson,left=0.6,tension=0.6}{v1,v2}
  \fmf{boson,right=0.6,tension=0.6}{v1,v2}
  \fmf{boson,tension=1.2}{v2,o1}
  \fmfv{l=$\tilde{\Phi}^{a\dagger}_{I}(p)$,l.a=180}{i1}
  \fmfv{l=$\tilde{\Phi}^{J}_{b}(q)$,l.a=180}{i2}
  \fmfv{l=$\tilde{V}_{c}(-p-q)$ \hspace{0.5cm} 
          $\longrightarrow ~~~ \tilde{D}(p;q)$,l.a=0}{o1}	  
  \fmfdot{v1,v2}
 \end{fmfgraph*}   
\end{fmffile} 

\vspace{0.8cm}
\noindent
\hspace*{1.5cm}
\begin{fmffile}{phi+vphiE}
 \begin{fmfgraph*}(150,95) 
  \fmfleft{i1,i2} \fmfright{o1}
  \fmf{phantom}{i1,v1,v2,v3}
  \fmf{phantom}{v3,v4,v5,i2}
  \fmf{boson,tension=1.5}{v3,o1}
  \fmffreeze
  \fmf{fermion}{i1,v1}
  \fmf{fermion}{v1,v3}
  \fmf{fermion}{v3,i2}
  \fmf{boson,right,tension=0}{v1,v3} 
  \fmfv{l=$\tilde{\Phi}^{a\dagger}_{I}(p)$,l.a=180}{i1}
  \fmfv{l=$\tilde{\Phi}^{J}_{b}(q)$,l.a=180}{i2}
  \fmfv{l=$\tilde{V}_{c}(-p-q)$ \hspace{0.5cm} 
          $\longrightarrow ~~~ \tilde{E}(p;q)$,l.a=0}{o1}
  \fmfdot{v1,v3}
 \end{fmfgraph*}   
\end{fmffile} 

\vspace{1cm}
\noindent
\hspace*{1.5cm}
\begin{fmffile}{phi+vphiF}
 \begin{fmfgraph*}(150,95) 
  \fmfleft{i1,i2} \fmfright{o1}
  \fmf{phantom}{i1,v1,v2,v3}
  \fmf{phantom}{v3,v4,v5,i2}
  \fmf{boson,tension=1.5}{v3,o1}
  \fmffreeze
  \fmf{fermion}{i1,v3}
  \fmf{fermion}{v3,v5}
  \fmf{fermion}{v5,i2}
  \fmf{boson,right,tension=0}{v3,v5} 
  \fmfv{l=$\tilde{\Phi}^{a\dagger}_{I}(p)$,l.a=180}{i1}
  \fmfv{l=$\tilde{\Phi}^{J}_{b}(q)$,l.a=180}{i2}
  \fmfv{l=$\tilde{V}_{c}(-p-q)$ \hspace{0.5cm} 
          $\longrightarrow ~~~ \tilde{F}(p;q)$,l.a=0}{o1}
  \fmfdot{v3,v5}
 \end{fmfgraph*}   
\end{fmffile}

\vspace{0.5cm}
\noindent
In the presence of a mass term for the (anti) chiral superfields the 
expressions given above are modified by the presence of the 
mass in the free propagators and furthermore there are two additional 
sets of contributions corresponding to the diagrams

\vspace*{0.8cm}
\noindent
\hspace*{1.5cm}
\begin{fmffile}{phi+vphiG}
 \begin{fmfgraph*}(150,95) 
  \fmfleft{i1,i2} \fmfright{o1}
  \fmf{phantom,tension=0.7,tag=1}{i1,v1}
  \fmf{phantom,tension=0.7,tag=2}{i2,v1}
  \fmf{phantom,tension=0.7,tag=3}{v1,i1}
  \fmf{phantom,tension=0.7,tag=4}{v1,i2} 
  \fmf{boson,tension=2.5}{v1,o1}
  \fmfv{l=$\tilde{\Phi}^{a\dagger}_{I}(p)$,l.a=180}{i1}
  \fmfv{l=$\tilde{\Phi}^{J}_{b}(q)$,l.a=180}{i2}
  \fmfv{l=$\tilde{V}_{c}(-p-q)$ \hspace{0.5cm} 
          $\longrightarrow ~~~ \tilde{G}(p;q)$,l.a=0}{o1}
  \fmfdot{v1}
  \fmffreeze
  \fmfipath{p[]}
  \fmfiset{p1}{vpath1(__i1,__v1)}
  \fmfiset{p2}{vpath2(__i2,__v1)}
  \fmfiset{p3}{vpath3(__v1,__i1)}
  \fmfiset{p4}{vpath4(__v1,__i2)}
  \fmfi{fermion}{subpath (0,length(p1)/3) of p1}
  \fmfi{fermion}{subpath (length(p1)/3,2length(p1)/3) of p1}
  \fmfi{fermion}{subpath (0,length(p3)/3) of p3}
  \fmfi{fermion}{subpath (2length(p2)/3,length(p2)) of p2}
  \fmfi{fermion}{subpath (length(p4)/3,2length(p4)/3) of p4}
  \fmfi{fermion}{subpath (2length(p4)/3,length(p4)) of p4}
  \fmfi{boson}{point length(p1)/3 of p1 -- point
               length(p2)/3 of p2}
  \fmfiv{d.sh=circle,d.fill=full,d.si=1.5mm}{point length(p1)/3 of p1}      
  \fmfiv{d.sh=circle,d.fill=full,d.si=1.5mm}{point length(p2)/3 of p2}
  \fmfiv{d.sh=cross,d.si=3mm,d.a=42}{point 2length(p1)/3 of p1}      
  \fmfiv{d.sh=cross,d.si=3mm,d.a=42}{point 2length(p2)/3 of p2}
 \end{fmfgraph*}   
\end{fmffile} 

\vspace*{0.8cm}
\noindent
\hspace*{1.5cm}
\begin{fmffile}{phi+vphiH}
 \begin{fmfgraph*}(150,95) 
  \fmfleft{i1,i2} \fmfright{o1}
  \fmf{phantom,tension=0.7,tag=1}{i1,v1}
  \fmf{phantom,tension=0.7,tag=2}{i2,v1}
  \fmf{phantom,tension=0.7,tag=3}{v1,i1}
  \fmf{phantom,tension=0.7,tag=4}{v1,i2} 
  \fmf{boson,tension=2.5}{v1,o1}
  \fmfv{l=$\tilde{\Phi}^{a\dagger}_{I}(p)$,l.a=180}{i1}
  \fmfv{l=$\tilde{\Phi}^{J}_{b}(q)$,l.a=180}{i2}
  \fmfv{l=$\tilde{V}_{c}(-p-q)$ \hspace{0.5cm} 
          $\longrightarrow ~~~ \tilde{H}(p;q)$,l.a=0}{o1}
  \fmfdot{v1}
  \fmffreeze
  \fmfipath{p[]}
  \fmfiset{p1}{vpath1(__i1,__v1)}
  \fmfiset{p2}{vpath2(__i2,__v1)}
  \fmfiset{p3}{vpath3(__v1,__i1)}
  \fmfiset{p4}{vpath4(__v1,__i2)}
  \fmfi{fermion}{subpath (0,length(p1)/3) of p1}
  \fmfi{fermion}{subpath (length(p3)/3,2length(p3)/3) of p3}
  \fmfi{fermion}{subpath (2length(p1)/3,length(p1)) of p1}
  \fmfi{fermion}{subpath (0,length(p4)/3) of p4}
  \fmfi{fermion}{subpath (length(p2)/3,2length(p2)/3) of p2}
  \fmfi{fermion}{subpath (2length(p4)/3,length(p4)) of p4}
  \fmfi{fermion}{point length(p2)/3 of p2 -- point
               length(p1)/3 of p1}
  \fmfiv{d.sh=circle,d.fill=full,d.si=1.5mm}{point length(p1)/3 of p1}      
  \fmfiv{d.sh=circle,d.fill=full,d.si=1.5mm}{point length(p2)/3 of p2}
  \fmfiv{d.sh=cross,d.si=3mm,d.a=42}{point 2length(p1)/3 of p1}      
  \fmfiv{d.sh=cross,d.si=3mm,d.a=42}{point 2length(p2)/3 of p2}
 \end{fmfgraph*}   
\end{fmffile}

\vspace{0.5cm}
\noindent
The calculation of these diagrams is completely analogous to those 
that were presented in the previous section. All of these proper 
contributions can be derived using the Feynman rules. The last two 
diagrams are finite and vanish in the ${\cal N}$=4 theory, \ie at  
$m$=0, as they are proportional to $m^{2}$; they will not be considered 
here. The other contributions will be briefly studied in the limit 
$m\to 0$.

The diagram $\tilde{D}(p,q)$ is zero as a consequence of the 
contraction among the colour indices, so that there are five 
contributions to be calculated. 
The dimensional analysis of section \ref{generalprop} gives a 
vanishing superficial degree of divergence, corresponding to a 
logarithmic divergence, for the Green function under consideration. 
Each diagram actually contains a logarithmically divergent term 
plus finite terms. A straightforward but rather lengthy computation, 
based on elementary $D$-algebra, allows to prove that the divergent 
part of all the diagrams is of the form
\begin{displaymath}
	I_{{\rm log}} = c \, \delta^{I}_{J} f^{abc}
	\int \frac{d^{4}k}{(2\pi)^{4}}d^{4}\theta \, 
	\frac{1}{k^{2}(p-k)^{2}} \left\{ 
	\Phi^{\dagger}_{aI}(p,\theta,{\overline \theta}) V_{b}
	(-p-q,\theta,{\overline \theta}) \Phi^{J}_{c}(q,\theta,
	{\overline \theta}) \right\} \, ,
\end{displaymath}
where $c$ is a constant. Moreover one finds, from each diagram a finite 
contribution of the form 
\begin{eqnarray*}
	I_{{\rm finite}} &=& \delta^{I}_{J} f^{abc}
	\int \frac{d^{4}k}{(2\pi)^{4}}d^{4}\theta \, 
	\frac{1}{k^{2}(q+k)^{2}(p-k)^{2}} \\ 
	&& \hspace{-1cm}\Phi^{\dagger}_{aI}(p,\theta,{\overline \theta}) 
	\left\{ c_{1}\left[ D^{\alpha}{\overline D}^{2}D_{\alpha} 
	V_{b}(-p-q,\theta,{\overline \theta}) \right]
    \Phi^{J}_{c}(q,\theta,{\overline \theta}) 
	\right. + \\
	&& \hspace{-1cm} \left. +
	c_{2}\sigma^{\mu}_{\alpha{\dot\alpha}}(q+k)_{\mu} \left[
	D^{\alpha}{\overline D}^{\dot\alpha}V_{b}
	(-p-q,\theta,{\overline \theta}) \right] 
	\Phi^{J}_{c}(q,\theta,{\overline \theta}) \right\} \, ,
\end{eqnarray*}
where $c_{1}$ and $c_{2}$ are numerical constants.
The sum of all the logarithmically divergent terms contained in the 
diagrams $\tilde{A}$, $\tilde{B}$, $\tilde{C}$, $\tilde{E}$ and $\tilde{F}$ 
vanishes. The residual finite part can be shown to be zero as well. 
In conclusion the one-loop correction to the three-point function 
$\langle \Phi^{\dagger} V \Phi \rangle$ exactly vanishes in the 
Fermi--Feynman gauge. The same result can be shown to hold for the 
other three-point functions.

The superfield formalism does not lead to significant simplifications 
in the calculation of three-point functions with respect to the same 
computation in components (in the Wess--Zumino gauge!); actually for 
the Green function considered here the number of diagrams to be evaluated 
is approximately the same in the component formulation. However the 
power of superspace techniques becomes clear in the computation of 
four-point functions that will be considered in next subsection.

\subsection{Four-point functions}

The computation of four-point functions in components is awkward even 
at the one-loop level and in the Wess--Zumino gauge. In this case the 
choice of the WZ gauge should not lead to divergences because the 
complete four-point function must finally give the physical 
scattering amplitude. In the ${\cal N}$=1 superfield formulation the 
calculation of four-point functions, though rather lengthy, is much 
more simple. 

In this section the computation of the one-particle irreducible 
one-loop correction to the Green function 
$\langle\Phi^{\dagger}\Phi\Phi^{\dagger}\Phi\rangle$ in the 
Fermi--Feynman gauge will be presented. There are several diagrams to 
be considered: each of them is free of ultraviolet divergences, as 
immediately follows from dimensional analysis. Moreover in the 
Fermi--Feynman gauge the single corrections are infrared safe. In 
conclusion one finds a finite and non-vanishing result.

The first subset of diagrams corresponds to 

\vspace{1cm}
\noindent
\hspace*{2cm}
\begin{fmffile}{phi+phiphi+phiA}
 \begin{fmfgraph*}(150,110) 
  \fmfleft{i1,i2} 
  \fmfright{o1,o2}
  \fmf{fermion,tension=0.6}{i1,v1}
  \fmf{fermion,tension=0.6}{v2,i2}
  \fmf{fermion,tension=0.6}{v3,o1}
  \fmf{fermion,tension=0.6}{o2,v4}
  \fmf{phantom,tension=0.1,tag=1}{v1,v2}
  \fmf{phantom,tension=0.1,tag=2}{v2,v1}
  \fmf{phantom,tension=0.1,tag=3}{v2,v4}
  \fmf{phantom,tension=0.1,tag=4}{v4,v2}
  \fmf{phantom,tension=0.1,tag=5}{v4,v3}
  \fmf{phantom,tension=0.1,tag=6}{v3,v4}
  \fmf{phantom,tension=0.1,tag=7}{v3,v1}
  \fmf{phantom,tension=0.1,tag=8}{v1,v3}
  \fmfv{l=$\tilde{\Phi}^{a\dagger}_{I}$,l.a=180}{i1}
  \fmfv{l=$\tilde{\Phi}^{J}_{b}$,l.a=180}{i2}
  \fmfv{l=$\tilde{\Phi}^{L}_{d}$,l.a=0}{o1}
  \fmfv{l=$\tilde{\Phi}^{c\dagger}_{K}$,l.a=0}{o2}
  \fmfdot{v1,v2,v3,v4}
  \fmffreeze
  \fmfipath{p[]}
  \fmfiset{p1}{vpath1(__v1,__v2)}
  \fmfiset{p2}{vpath2(__v2,__v1)}
  \fmfiset{p3}{vpath3(__v2,__v4)}
  \fmfiset{p4}{vpath4(__v4,__v2)}
  \fmfiset{p5}{vpath5(__v4,__v3)}
  \fmfiset{p6}{vpath6(__v3,__v4)}
  \fmfiset{p7}{vpath7(__v3,__v1)}
  \fmfiset{p8}{vpath8(__v1,__v3)}
  \fmfi{fermion}{subpath (0,length(p1)) of p1}
  \fmfi{boson}{subpath (0,length(p7)) of p7}
  \fmfi{boson}{subpath (0,length(p3)) of p3}
  \fmfi{fermion}{subpath (0,length(p5)) of p5}
  \fmfiv{l=\hspace{7cm}$\longrightarrow ~~
            \tilde{A}(p;q;r)$,l.a=90}
            {point 3length(p5)/5 of p5}
 \end{fmfgraph*}   
\end{fmffile} 

\vspace{0.5cm}
\noindent
together with those obtained from this one by crossing. 
There are a total of four diagrams of this kind. Application of the 
Feynman rules and steps completely analogous to those entering the 
evaluation of the propagators allow to find the following results.
\begin{displaymath}
	\tilde{A}_{1}(p,q,r) = \left(\frac{1}{4}\right)^{2}
	g^{4}\,\kappa N^{2}(N^{2}-1)(\delta_{a}^{b}\delta^{d}_{c} + 
	\delta_{ac}\delta^{bd})\delta^{I}_{J}\delta^{K}_{L} 
	I_{1\:bdIK}^{(A)acJL}(p,q,r)
\end{displaymath}
where
\begin{eqnarray}
	I_{1\:bdIK}^{(A)acJL}(p,q,r) & = & \int \frac{d^{4}k}{(2\pi)^{4}}d^{4}
	\theta \frac{1}{k^{2}(p-k)^{2}(q+k)^{2}(q+k-r)^{2}} \cdot  
	\nonumber \\
	&& \hspace{-2.5cm} \cdot \left\{ \left[(D^{2}{\overline D}^{2}) 
	\Phi^{a\dagger}_{I}(p,\theta,{\overline \theta}) \right]
	\Phi^{c\dagger}_{K}(p+q-r,\theta,{\overline \theta})
	\Phi_{d}^{L}(r,\theta,{\overline\theta}) \Phi_{b}^{J}(q,\theta,
	{\overline\theta}) + \Phi^{a\dagger}_{I}(p,\theta,{\overline \theta})
	\cdot \right. \nonumber \\
	&& \hspace{-2.5cm} \left. \cdot\left[(D^{2}{\overline D}^{2}) 
	\Phi^{c\dagger}_{K}(p+q-r,\theta,{\overline \theta})\right] 
	\Phi_{d}^{L}(r,\theta,{\overline\theta}) \Phi_{b}^{J}
	(q,\theta,{\overline\theta}) 
	+ 2\left[(D^{2}{\overline D}_{\dot\alpha})\Phi^{a\dagger}_{I}(p,
	\theta,{\overline \theta})\right] \cdot \right. \nonumber \\ 
	&&\hspace{-2.5cm} \left.\cdot\left[{\overline D}^{\dot\alpha}
	\Phi^{c\dagger}(p+q-r,\theta,{\overline \theta})\right] 
	\Phi_{d}^{L}(r,\theta,{\overline\theta}) \Phi_{b}^{J}(q,\theta,
	{\overline\theta}) + 2\left[{\overline D}_{\dot\alpha}
	\Phi^{a\dagger}_{I}(p,\theta,{\overline \theta})\right]
	\cdot \right. \nonumber \\
	&& \hspace{-2.5cm} \left. \cdot 
	\left[(D^{2}{\overline D}^{\dot\alpha})\Phi^{c\dagger}
	(p+q-r,\theta,{\overline \theta})\right] \Phi_{d}^{L}(r,\theta,
	{\overline\theta}) \Phi_{b}^{J}(q,\theta,{\overline\theta}) + 
	4\left[(D^{\alpha}{\overline D}_{\dot\alpha})\Phi^{a\dagger}_{I}
	(p,\theta,{\overline \theta})\right] \cdot \right. \nonumber \\
	&& \hspace{-2.5cm} \left. \cdot \left[D_{\alpha}
	{\overline D}^{\dot\alpha}\Phi^{c\dagger}(p+q-r,\theta,{\overline 
	\theta})\right] \Phi_{d}^{L}(r,\theta,
	{\overline\theta}) \Phi_{b}^{J}(q,\theta,{\overline\theta}) 
	\right\} \equiv \nonumber \\
	&& \hspace{-2.5cm}\equiv \int \frac{d^{4}k}{(2\pi)^{4}}d^{4}\theta 
	\frac{1}{k^{2}(p-k)^{2}(q+k)^{2}(q+k-r)^{2}} K_{bdIK}^{acJL}
	(p,q,r;\theta;{\overline\theta}) 
	\label{4pointa1}
\end{eqnarray}
and the constant $\kappa$ is a group theory factor defined by
\begin{displaymath}
	f_{aef}f^{bfg}f_{cgh}f^{dhe} = \kappa N^{2}(N^{2}-1)
	(\delta_{a}^{b}\delta_{c}^{d} + \delta_{a}^{d}\delta_{c}^{b}) \, ;
\end{displaymath}
\begin{displaymath}
	\tilde{A}_{2}(p,q,r) = \left(\frac{1}{4}\right)^{2}
	g^{4}\,\kappa N^{2}(N^{2}-1) (\delta_{a}^{d}\delta^{b}_{c} + 
	\delta_{ac}\delta^{bd})\delta^{I}_{L}\delta^{K}_{J} 
	I_{2\:bdIK}^{(A)acJL}(p,q,r) \, ,
\end{displaymath}
where
\begin{eqnarray}
	I_{2\:bdIK}^{(A)acJL}(p,q,r) &=& \int \frac{d^{4}k}{(2\pi)^{4}}d^{4}
	\theta \frac{1}{k^{2}(p-k)^{2}(p-k-r)^{2}(q+k)^{2}(p+q-k-r)^{2}} 
	\cdot \nonumber \\
	&& \hspace{2cm} \rule{0pt}{20pt}
	\cdot K_{bdIK}^{acJL} (p,q,r;\theta;{\overline\theta}) \, ;
	\label{4pointa2}
\end{eqnarray}
\begin{displaymath}
	\tilde{A}_{3}(p,q,r) = \left(\frac{1}{4}\right)^{2}
	g^{4}\,\kappa N^{2}(N^{2}-1)(\delta_{a}^{b}\delta^{d}_{c} + 
	\delta_{a}^{d}\delta^{b}_{c})\delta^{I}_{J}\delta^{K}_{L} 
	I_{3\:bdIK}^{(A)acJL}(p,q,r) \, ,
\end{displaymath}
where
\begin{eqnarray}
	I_{3\:bdIK}^{(A)acJL}(p,q,r) & = & \int \frac{d^{4}k}{(2\pi)^{4}}d^{4}
	\theta \frac{1}{k^{2}(p-k)^{2}(q+k)^{2}(q+k-r)^{2}} \cdot  
	\nonumber \\ 
	&& \hspace{-2.5cm} \cdot \left\{ \left[{\overline 
	D}^{2}\Phi^{a\dagger}_{I}(p,\theta,{\overline \theta})\right] \left[ 
	D^{2} \Phi^{J}_{b}(q,\theta,{\overline \theta})\right] 
	\Phi^{c\dagger}_{K}(p+q-r,\theta,{\overline \theta}) \Phi^{L}_{d}
	(r,\theta,{\overline \theta}) + 
	\right. \nonumber \\ 
	&& \hspace{-2.5cm} \left. + 8i\sigma^{\mu}_{\alpha{\dot\alpha}}
	(p-k-r)_{\mu} \left[ {\overline D}^{\dot\alpha}\Phi^{a\dagger}_{I}
	(p,\theta,{\overline \theta})\right] \left[ 
	D^{\alpha} \Phi^{J}_{b}(q,\theta,{\overline \theta})\right] 
	\Phi^{c\dagger}_{K}(p+q-r,\theta,{\overline \theta}) 
	\cdot \right. \nonumber \\
	&& \hspace{-2.5cm}\cdot\left.\Phi^{L}_{d}(r,\theta,{\overline\theta}) 
	- 16 (p-k-r)^{2} \Phi^{a\dagger}_{I}
	(p,\theta,{\overline \theta})\Phi^{J}_{b}(q,\theta,{\overline\theta})
	\Phi^{c\dagger}_{K}(p+q-r,\theta,{\overline \theta}) \cdot \right.
	\nonumber \\ 
	&& \hspace{-2.5cm}\cdot\left. \Phi^{L}_{d}
	(r,\theta,{\overline \theta}) \right\} \, ;
	\label{4pointa3}
\end{eqnarray}
\begin{displaymath}
	\tilde{A}_{4}(p,q,r) = \left(\frac{1}{4}\right)^{2}
	g^{4}\,\kappa N^{2}(N^{2}-1)(\delta_{a}^{b}\delta^{d}_{c} + 
	\delta_{a}^{d}\delta^{b}_{c})\delta^{I}_{J}\delta^{K}_{L} 
	I_{4\:bdIK}^{(A)acJL}(p,q,r) \, ,
\end{displaymath}
where
\begin{eqnarray}
	I_{4\:bdIK}^{(A)acJL}(p,q,r) & = & \int \frac{d^{4}k}{(2\pi)^{4}}
	d^{4}\theta \frac{1}{k^{2}(p-k)^{2}(p-k-r)^{2}(q+k)^{2}} \cdot  
	\nonumber \\ 
	&& \hspace{-2.5cm} \cdot \left\{ \left[{\overline 
	D}^{2}\Phi^{a\dagger}_{I}(p,\theta,{\overline \theta})\right] 
	\Phi^{J}_{b}(q,\theta,{\overline \theta}) 
	\Phi^{c\dagger}_{K}(p+q-r,\theta,{\overline \theta}) 
	\left[ D^{2} \Phi^{L}_{d}(r,\theta,{\overline \theta}) \right]+ 
	\right. \nonumber \\ 
	&& \hspace{-2.5cm} \left. + 8i\sigma^{\mu}_{\alpha{\dot\alpha}}
	(q+k)_{\mu} \left[ {\overline D}^{\dot\alpha}\Phi^{a\dagger}_{I}
	(p,\theta,{\overline \theta})\right] \Phi^{J}_{b}(q,\theta,
	{\overline \theta}) \Phi^{c\dagger}_{K}
	(p+q-r,\theta,{\overline \theta}) 
	\cdot \right. \nonumber \\
	&& \hspace{-2.5cm}\cdot\left.\left[ 
	D^{\alpha} \Phi^{L}_{d}(r,\theta,{\overline\theta}) \right]
	- 16 (q+k)^{2} \Phi^{a\dagger}_{I}
	(p,\theta,{\overline \theta})\Phi^{J}_{b}(q,\theta,{\overline\theta})
	\Phi^{c\dagger}_{K}(p+q-r,\theta,{\overline \theta}) \cdot \right.
	\nonumber \\ 
	&& \hspace{-2.5cm}\cdot\left. \Phi^{L}_{d}
	(r,\theta,{\overline \theta}) \right\} \, .
	\label{4pointa4}
\end{eqnarray}

The second kind of contribution corresponds to the diagram

\vspace{1cm}
\noindent
\hspace*{2cm}
\begin{fmffile}{phi+phiphi+phiB}
 \begin{fmfgraph*}(150,110) 
  \fmfleft{i1,i2} 
  \fmfright{o1,o2}
  \fmf{fermion,tension=0.6}{i1,v1}
  \fmf{fermion,tension=0.6}{v2,i2}
  \fmf{fermion,tension=0.6}{v3,o1}
  \fmf{fermion,tension=0.6}{o2,v4}
  \fmf{phantom,tension=0.1,tag=1}{v1,v2}
  \fmf{phantom,tension=0.1,tag=2}{v2,v1}
  \fmf{phantom,tension=0.1,tag=3}{v2,v4}
  \fmf{phantom,tension=0.1,tag=4}{v4,v2}
  \fmf{phantom,tension=0.1,tag=5}{v4,v3}
  \fmf{phantom,tension=0.1,tag=6}{v3,v4}
  \fmf{phantom,tension=0.1,tag=7}{v3,v1}
  \fmf{phantom,tension=0.1,tag=8}{v1,v3}
  \fmfv{l=$\tilde{\Phi}^{a\dagger}_{I}$,l.a=180}{i1}
  \fmfv{l=$\tilde{\Phi}^{J}_{b}$,l.a=180}{i2}
  \fmfv{l=$\tilde{\Phi}^{L}_{d}$,l.a=0}{o1}
  \fmfv{l=$\tilde{\Phi}^{c\dagger}_{K}$,l.a=0}{o2}
  \fmfdot{v1,v2,v3,v4}
  \fmffreeze
  \fmfipath{p[]}
  \fmfiset{p1}{vpath1(__v1,__v2)}
  \fmfiset{p2}{vpath2(__v2,__v1)}
  \fmfiset{p3}{vpath3(__v2,__v4)}
  \fmfiset{p4}{vpath4(__v4,__v2)}
  \fmfiset{p5}{vpath5(__v4,__v3)}
  \fmfiset{p6}{vpath6(__v3,__v4)}
  \fmfiset{p7}{vpath7(__v3,__v1)}
  \fmfiset{p8}{vpath8(__v1,__v3)}
  \fmfi{fermion}{subpath (0,length(p2)) of p2}
  \fmfi{fermion}{subpath (0,length(p7)) of p7}
  \fmfi{fermion}{subpath (0,length(p3)) of p3}
  \fmfi{fermion}{subpath (0,length(p6)) of p6}
  \fmfiv{l=\hspace{7cm}$\longrightarrow ~~
            \tilde{B}(p;q;r)$,l.a=90}
            {point 3length(p5)/5 of p5}
 \end{fmfgraph*}   
\end{fmffile} 

\vspace{0.5cm}
\noindent 
In this case the diagrams obtained by crossing are identical to the 
one depicted, so that they are accounted for by giving the correct 
weight to $\tilde{B}(p,q,r)$. For this contribution one finds
\begin{displaymath}
	\tilde{B}(p,q,r) = \frac{g^{4}}{6}
	\,\kappa N^{2}(N^{2}-1) (\delta_{a}^{b}\delta^{d}_{c} + 
	\delta_{a}^{d}\delta^{b}_{c})(\delta^{I}_{J}\delta^{K}_{L}
	+\delta^{I}_{L}\delta^{K}_{J}) I_{bdIK}^{(B)acJL}(p,q,r) \, ,
\end{displaymath}
where
\begin{eqnarray}
	I_{bdIK}^{(B)acJL}(p,q,r) & = & \int \frac{d^{4}k}{(2\pi)^{4}}d^{4}
	\theta \frac{1}{k^{2}(p-k)^{2}(p+q-r)^{2}(q+k)^{2}} \cdot  
	\nonumber \\ 
	&& \hspace{-2.5cm} \cdot \left\{ \left[{\overline 
	D}^{2}\Phi^{a\dagger}_{I}(p,\theta,{\overline \theta})\right] 
	\Phi^{J}_{b}(q,\theta,{\overline \theta}) 
	\Phi^{c\dagger}_{K}(p+q-r,\theta,{\overline \theta}) 
	\left[ D^{2} \Phi^{L}_{d}(r,\theta,{\overline \theta}) \right]+ 
	\right. \nonumber \\ 
	&& \hspace{-2.5cm} \left. + 8i\sigma^{\mu}_{\alpha{\dot\alpha}}
	(p-k)_{\mu} \left[ {\overline D}^{\dot\alpha}\Phi^{a\dagger}_{I}
	(p,\theta,{\overline \theta})\right] \Phi^{J}_{b}(q,\theta,
	{\overline \theta}) \Phi^{c\dagger}_{K}
	(p+q-r,\theta,{\overline \theta}) 
	\cdot \right. \nonumber \\
	&& \hspace{-2.5cm}\cdot\left.\left[ 
	D^{\alpha} \Phi^{L}_{d}(r,\theta,{\overline\theta}) \right]
	- 16 (p-k)^{2} \Phi^{a\dagger}_{I}
	(p,\theta,{\overline \theta})\Phi^{J}_{b}(q,\theta,{\overline\theta})
	\Phi^{c\dagger}_{K}(p+q-r,\theta,{\overline \theta}) \cdot \right.
	\nonumber \\ 
	&& \hspace{-2.5cm}\cdot\left. \Phi^{L}_{d}
	(r,\theta,{\overline \theta}) \right\} \, .
	\label{4pointb}
\end{eqnarray}

The next one-loop correction is 

\vspace{1cm}
\noindent
\hspace*{2cm}
\begin{fmffile}{phi+phiphi+phiC}
 \begin{fmfgraph*}(160,90) 
  \fmfleft{i1,i2} 
  \fmfright{o1,o2}
  \fmf{fermion,tension=1}{i1,v1}
  \fmf{fermion,tension=1}{v1,i2}
  \fmf{boson,left=0.6,tension=0.3}{v1,v2}
  \fmf{boson,right=0.6,tension=0.3}{v1,v2}
  \fmf{fermion,tension=1}{v2,o1}
  \fmf{fermion,tension=1}{o2,v2}
  \fmfv{l=$\tilde{\Phi}^{a\dagger}_{I}$,l.a=180}{i1}
  \fmfv{l=$\tilde{\Phi}^{J}_{b}$,l.a=180}{i2}
  \fmfv{l=$\tilde{\Phi}^{L}_{d}$,l.a=0}{o1}
  \fmfv{l=$\tilde{\Phi}^{c\dagger}_{K}$,l.a=0}{o2}
  \fmfv{l=\hspace{2cm}$\longrightarrow ~~
            \tilde{C}(p;q;r)$,l.a=0}{v2}
  \fmfdot{v1,v2}
 \end{fmfgraph*}   
\end{fmffile} 

\vspace{0.5cm}
\noindent
This contribution is trivially zero because it contains the product 
\begin{displaymath}
	\delta_{4}(\theta_{1}-\theta_{2})
	\delta_{4}(\theta_{1}-\theta_{2}) \equiv 0 \, .
\end{displaymath}
This would not be true with a choice of gauge different from the 
Fermi--Feynman one, \ie with $\alpha \neq 1$, because in that case 
there would be projectors acting on the $\delta$'s. The vanishing 
of $\tilde{C}(p,q,r)$, that is completely trivial in the superfield 
formulation, corresponds, in the component formulation, to a 
complicated cancellation among various terms coming from graphs with 
the same topology. 

Another subset of diagrams contains

\vspace{1cm}
\noindent
\hspace*{2cm}
\begin{fmffile}{phi+phiphi+phiD}
 \begin{fmfgraph*}(150,110) 
  \fmfleft{i1,i2} 
  \fmfright{o1,o2}
  \fmf{fermion,tension=0.6}{i1,v1}
  \fmf{fermion,tension=0.6}{i2,v2}
  \fmf{fermion,tension=0.6}{v3,o1}
  \fmf{fermion,tension=0.6}{v4,o2}
  \fmf{phantom,tension=0.1,tag=1}{v1,v2}
  \fmf{phantom,tension=0.1,tag=2}{v2,v1}
  \fmf{phantom,tension=0.1,tag=3}{v2,v4}
  \fmf{phantom,tension=0.1,tag=4}{v4,v2}
  \fmf{phantom,tension=0.1,tag=5}{v4,v3}
  \fmf{phantom,tension=0.1,tag=6}{v3,v4}
  \fmf{phantom,tension=0.1,tag=7}{v3,v1}
  \fmf{phantom,tension=0.1,tag=8}{v1,v3}
  \fmfv{l=$\tilde{\Phi}^{a\dagger}_{I}$,l.a=180}{i1}
  \fmfv{l=$\tilde{\Phi}^{J}_{b}$,l.a=0}{o2}
  \fmfv{l=$\tilde{\Phi}^{L}_{d}$,l.a=0}{o1}
  \fmfv{l=$\tilde{\Phi}^{c\dagger}_{K}$,l.a=180}{i2}
  \fmfdot{v1,v2,v3,v4}
  \fmffreeze
  \fmfipath{p[]}
  \fmfiset{p1}{vpath1(__v1,__v2)}
  \fmfiset{p2}{vpath2(__v2,__v1)}
  \fmfiset{p3}{vpath3(__v2,__v4)}
  \fmfiset{p4}{vpath4(__v4,__v2)}
  \fmfiset{p5}{vpath5(__v4,__v3)}
  \fmfiset{p6}{vpath6(__v3,__v4)}
  \fmfiset{p7}{vpath7(__v3,__v1)}
  \fmfiset{p8}{vpath8(__v1,__v3)}
  \fmfi{fermion}{subpath (0,length(p1)) of p1}
  \fmfi{fermion}{subpath (0,length(p4)) of p4}
  \fmfi{fermion}{subpath (0,length(p5)) of p5}
  \fmfi{boson}{subpath (0,length(p7)) of p7}
  \fmfiv{l=\hspace{7cm}$\longrightarrow ~~
            \tilde{D}(p;q;r)$,l.a=90}
            {point 3length(p5)/5 of p5}
 \end{fmfgraph*}   
\end{fmffile} 

\vspace{0.5cm}
\noindent
plus the crossed versions of this one. There are three inequivalent 
crossed diagrams. The result of the calculation is the following.
\begin{displaymath}
	\tilde{D}_{1}(p,q,r) = \left(\frac{1}{4}\right)^{2}
	g^{4}\,\kappa N^{2}(N^{2}-1) (\delta_{ac}\delta^{bd} + 
	\delta_{a}^{d}\delta^{b}_{c})(\delta^{I}_{J}\delta^{K}_{L}
	-\delta^{I}_{L}\delta^{K}_{J}) I_{1\:bdIK}^{(D)acJL}(p,q,r) \, ,
\end{displaymath}
where
\begin{eqnarray}
	I_{1\:bdIK}^{(D)acJL}(p,q,r) & = & \int \frac{d^{4}k}{(2\pi)^{4}}
	d^{4}\theta \frac{1}{k^{2}(p-k)^{2}(k+r-p-q)^{2}(p-k-r)^{2}} \cdot  
	\nonumber \\ 
	&& \hspace{-2.5cm} \cdot \left\{ \left[{\overline 
	D}^{2}\Phi^{a\dagger}_{I}(p,\theta,{\overline \theta})\right] 
	\left[ D^{2} \Phi^{J}_{b}(q,\theta,{\overline \theta})\right] 
	\Phi^{c\dagger}_{K}(p+q-r,\theta,{\overline \theta}) 
	\Phi^{L}_{d}(r,\theta,{\overline \theta}) + 
	\right. \nonumber \\ 
	&& \hspace{-2.5cm} \left. + 8i\sigma^{\mu}_{\alpha{\dot\alpha}}
	(k+r-p)_{\mu} \left[ {\overline D}^{\dot\alpha}\Phi^{a\dagger}_{I}
	(p,\theta,{\overline \theta})\right] \left[ D^{\alpha} 
	\Phi^{J}_{b}(q,\theta,{\overline \theta}) \right] 
	\Phi^{c\dagger}_{K}(p+q-r,\theta,{\overline \theta}) 
	\cdot \right. \nonumber \\
	&& \hspace{-2.5cm}\cdot\left.
	\Phi^{L}_{d}(r,\theta,{\overline\theta}) 
	- 16 (k+r-p)^{2} \Phi^{a\dagger}_{I}
	(p,\theta,{\overline \theta})\Phi^{J}_{b}(q,\theta,{\overline\theta})
	\Phi^{c\dagger}_{K}(p+q-r,\theta,{\overline \theta}) \cdot \right.
	\nonumber \\ 
	&& \hspace{-2.5cm}\cdot\left. \Phi^{L}_{d}
	(r,\theta,{\overline \theta}) \right\} \, ;
\end{eqnarray}
\begin{displaymath}
	\tilde{D}_{2}(p,q,r) = \left(\frac{1}{4}\right)^{2}
	g^{4}\,\kappa N^{2}(N^{2}-1) (\delta_{ac}\delta^{bd} + 
	\delta_{a}^{b}\delta^{d}_{c})(\delta^{I}_{L}\delta^{K}_{J}
	-\delta^{I}_{J}\delta^{K}_{L}) I_{2\:bdIK}^{(D)acJL}(p,q,r) \, ,
\end{displaymath}
where
\begin{eqnarray}
	I_{2\:bdIK}^{(D)acJL}(p,q,r) & = & \int \frac{d^{4}k}{(2\pi)^{4}}
	d^{4}\theta \frac{1}{k^{2}(p-k)^{2}(k+r-p-q)^{2}(p+q-k)^{2}} \cdot  
	\nonumber \\ 
	&& \hspace{-2.5cm} \cdot \left\{ \left[{\overline 
	D}^{2}\Phi^{a\dagger}_{I}(p,\theta,{\overline \theta})\right] 
	\Phi^{J}_{b}(q,\theta,{\overline \theta}) 
	\Phi^{c\dagger}_{K}(p+q-r,\theta,{\overline \theta}) 
	\left[ D^{2} \Phi^{L}_{d}(r,\theta,{\overline \theta}) \right] + 
	\right. \nonumber \\ 
	&& \hspace{-2.5cm} \left. + 8i\sigma^{\mu}_{\alpha{\dot\alpha}}
	(k-p-q)_{\mu} \left[ {\overline D}^{\dot\alpha}\Phi^{a\dagger}_{I}
	(p,\theta,{\overline \theta})\right] 
	\Phi^{J}_{b}(q,\theta,{\overline \theta}) 
	\Phi^{c\dagger}_{K}(p+q-r,\theta,{\overline \theta}) 
	\cdot \right. \nonumber \\
	&& \hspace{-2.5cm}\cdot\left.
	\left[ D^{\alpha}\Phi^{L}_{d}(r,\theta,{\overline\theta})\right] 
	- 16 (k-p-q)^{2} \Phi^{a\dagger}_{I}
	(p,\theta,{\overline \theta})\Phi^{J}_{b}(q,\theta,{\overline\theta})
	\Phi^{c\dagger}_{K}(p+q-r,\theta,{\overline \theta}) \cdot \right.
	\nonumber \\ 
	&& \hspace{-2.5cm}\cdot\left. \Phi^{L}_{d}
	(r,\theta,{\overline \theta}) \right\} \, ;
\end{eqnarray}
\begin{displaymath}
	\tilde{D}_{3}(p,q,r) = \left(\frac{1}{4}\right)^{2}
	g^{4}\,\kappa N^{2}(N^{2}-1) ( \delta_{a}^{d}\delta^{b}_{c}
	+ \delta_{ac}\delta^{bd})(\delta^{I}_{J}\delta^{K}_{L}
	-\delta^{I}_{L}\delta^{K}_{J}) I_{3\:bdIK}^{(D)acJL}(p,q,r) \, ,
\end{displaymath}
where
\begin{eqnarray}
	I_{3\:bdIK}^{(D)acJL}(p,q,r) & = & \int \frac{d^{4}k}{(2\pi)^{4}}
	d^{4}\theta \frac{1}{k^{2}(p-k)^{2}(k+r-p-q)^{2}(p+q-k)^{2}} \cdot  
	\nonumber \\ 
	&& \hspace{-2.5cm} \cdot \left\{ \Phi^{a\dagger}_{I}
	(p,\theta,{\overline \theta})
	\Phi^{J}_{b}(q,\theta,{\overline \theta}) 
	\left[{\overline D}^{2}\Phi^{c\dagger}_{K}
	(p+q-r,\theta,{\overline \theta}) \right] 
	\left[ D^{2} \Phi^{L}_{d}(r,\theta,{\overline \theta}) \right] + 
	\right. \nonumber \\ 
	&& \hspace{-2.5cm} \left. - 8i\sigma^{\mu}_{\alpha{\dot\alpha}}
	(p-k-r)_{\mu} \Phi^{a\dagger}_{I}(p,\theta,{\overline \theta}) 
	\Phi^{J}_{b}(q,\theta,{\overline \theta}) 
	\left[ {\overline D}^{\dot\alpha}\Phi^{c\dagger}_{K}
	(p+q-r,\theta,{\overline \theta}) \right]
	\cdot \right. \nonumber \\
	&& \hspace{-2.5cm}\cdot\left.
	\left[ D^{\alpha}\Phi^{L}_{d}(r,\theta,{\overline\theta})\right] 
	- 16 (p-k-r)^{2} \Phi^{a\dagger}_{I}
	(p,\theta,{\overline \theta})\Phi^{J}_{b}(q,\theta,{\overline\theta})
	\Phi^{c\dagger}_{K}(p+q-r,\theta,{\overline \theta}) \cdot \right.
	\nonumber \\ 
	&& \hspace{-2.5cm}\cdot\left. \Phi^{L}_{d}
	(r,\theta,{\overline \theta}) \right\} \, ;
\end{eqnarray}
\begin{displaymath}
	\tilde{D}_{4}(p,q,r) = \left(\frac{1}{4}\right)^{2}
	g^{4}\,\kappa N^{2}(N^{2}-1) ( \delta_{a}^{b}\delta^{d}_{c}
	+ \delta_{ac}\delta^{bd})(\delta^{I}_{L}\delta^{K}_{J}
	-\delta^{I}_{J}\delta^{K}_{L}) I_{4\:bdIK}^{(D)acJL}(p,q,r) \, ,
\end{displaymath}
where
\begin{eqnarray}
	I_{4\:bdIK}^{(D)acJL}(p,q,r) & = & \int \frac{d^{4}k}{(2\pi)^{4}}
	d^{4}\theta \frac{1}{k^{2}(p-k)^{2}(k+r-p-q)^{2}(p+q-k)^{2}} \cdot  
	\nonumber \\ 
	&& \hspace{-2.5cm} \cdot \left\{ \Phi^{a\dagger}_{I}
	(p,\theta,{\overline \theta})
	\left[ D^{2} \Phi^{J}_{b}(q,\theta,{\overline \theta})\right] 
	\left[{\overline D}^{2}\Phi^{c\dagger}_{K}
	(p+q-r,\theta,{\overline \theta}) \right] 
	\Phi^{L}_{d}(r,\theta,{\overline \theta}) + 
	\right. \nonumber \\ 
	&& \hspace{-2.5cm} \left. - 8i\sigma^{\mu}_{\alpha{\dot\alpha}}
	(p+q-k)_{\mu} \Phi^{a\dagger}_{I}(p,\theta,{\overline \theta}) 
	\left[ D^{\alpha}\Phi^{J}_{b}(q,\theta,{\overline \theta})\right] 
	\left[ {\overline D}^{\dot\alpha}\Phi^{c\dagger}_{K}
	(p+q-r,\theta,{\overline \theta}) \right]
	\cdot \right. \nonumber \\
	&& \hspace{-2.5cm}\cdot\left.
	\Phi^{L}_{d}(r,\theta,{\overline\theta}) 
	- 16 (p+q-k)^{2} \Phi^{a\dagger}_{I}
	(p,\theta,{\overline \theta})\Phi^{J}_{b}(q,\theta,{\overline\theta})
	\Phi^{c\dagger}_{K}(p+q-r,\theta,{\overline \theta}) \cdot \right.
	\nonumber \\ 
	&& \hspace{-2.5cm}\cdot\left. \Phi^{L}_{d}
	(r,\theta,{\overline \theta}) \right\} \, .
\end{eqnarray}

The last one-loop family of diagrams is formed by the one below plus 
those obtained by crossing.

\vspace{1cm}
\noindent
\hspace*{2cm}
\begin{fmffile}{phi+phiphi+phiE}
 \begin{fmfgraph*}(150,110) 
  \fmfleft{i1,i2} 
  \fmfright{o1,o2}
  \fmf{fermion,tension=1}{i1,v1}
  \fmf{fermion,tension=1}{v2,i2}
  \fmf{fermion,tension=0.2}{v1,v2}
  \fmf{boson,tension=0.6}{v1,v3}
  \fmf{boson,tension=0.6}{v2,v3}
  \fmf{fermion,tension=1}{v3,o1}
  \fmf{fermion,tension=1}{o2,v3}
  \fmfv{l=$\tilde{\Phi}^{a\dagger}_{I}$,l.a=180}{i1}
  \fmfv{l=$\tilde{\Phi}^{J}_{b}$,l.a=180}{i2}
  \fmfv{l=$\tilde{\Phi}^{L}_{d}$,l.a=0}{o1}
  \fmfv{l=$\tilde{\Phi}^{c\dagger}_{K}$,l.a=0}{o2}
  \fmfv{l=\hspace{2cm}$\longrightarrow ~~
            \tilde{E}(p;q;r)$,l.a=0}{v3}
  \fmfdot{v1,v2,v3}
 \end{fmfgraph*}   
\end{fmffile} 

\vspace{0.5cm}
\noindent
The resulting contributions to the Green function are
\begin{displaymath}
	\tilde{E}_{1}(p,q,r) = \left(\frac{1}{4}\right)^{2}
	\frac{g^{4}}{2}\,\kappa N^{2}(N^{2}-1) \delta_{a}^{b}\delta^{d}_{c}
	+ \delta_{ac}\delta^{bd})\delta^{I}_{J}\delta^{K}_{L}
	I_{1\:bdIK}^{(E)acJL}(p,q,r) \, ,
\end{displaymath}
where
\begin{eqnarray}
	I_{1\:bdIK}^{(E)acJL}(p,q,r) & = & \int \frac{d^{4}k}{(2\pi)^{4}}
	d^{4}\theta \frac{1}{k^{2}(p-k)^{2}(p+q-k)^{2}} \cdot  
	\nonumber \\ 
	&& \hspace{-2.5cm} \cdot \left\{ \Phi^{a\dagger}_{I}
	(p,\theta,{\overline \theta})\Phi^{J}_{b}(q,\theta,{\overline\theta})
	\Phi^{c\dagger}_{K}(p+q-r,\theta,{\overline \theta}) 
	\Phi^{L}_{d}(r,\theta,{\overline \theta}) \right\} \, ;
\end{eqnarray}
\begin{displaymath}
	\tilde{E}_{2}(p,q,r) = \left(\frac{1}{4}\right)^{2}
	\frac{g^{4}}{2}\,\kappa N^{2}(N^{2}-1)( \delta_{a}^{d}\delta^{b}_{c}
	+ \delta_{ac}\delta^{bd})\delta^{I}_{L}\delta^{K}_{J}
	I_{2\:bdIK}^{(E)acJL}(p,q,r) \, ,
\end{displaymath}
where
\begin{eqnarray}
	I_{2\:bdIK}^{(E)acJL}(p,q,r) & = & \int \frac{d^{4}k}{(2\pi)^{4}}
	d^{4}\theta \frac{1}{k^{2}(p-k)^{2}(r-k)^{2}} \cdot  
	\nonumber \\ 
	&& \hspace{-2.5cm} \cdot \left\{ \Phi^{a\dagger}_{I}
	(p,\theta,{\overline \theta})\Phi^{J}_{b}(q,\theta,{\overline\theta})
	\Phi^{c\dagger}_{K}(p+q-r,\theta,{\overline \theta}) 
	\Phi^{L}_{d}(r,\theta,{\overline \theta}) \right\} \, ;
\end{eqnarray}
\begin{displaymath}
	\tilde{E}_{3}(p,q,r) = \left(\frac{1}{4}\right)^{2}
	\frac{g^{4}}{2}\,\kappa N^{2}(N^{2}-1)( \delta_{a}^{b}\delta^{d}_{c}
	+ \delta_{ac}\delta^{bd})\delta^{I}_{L}\delta^{K}_{J}
	I_{3\:bdIK}^{(E)acJL}(p,q,r) \, ,
\end{displaymath}
where
\begin{eqnarray}
	I_{3\:bdIK}^{(E)acJL}(p,q,r) & = & \int \frac{d^{4}k}{(2\pi)^{4}}
	d^{4}\theta \frac{1}{k^{2}(p+q-k)^{2}(k-r)^{2}} \cdot  
	\nonumber \\ 
	&& \hspace{-2.5cm} \cdot \left\{ \Phi^{a\dagger}_{I}
	(p,\theta,{\overline \theta})\Phi^{J}_{b}(q,\theta,{\overline\theta})
	\Phi^{c\dagger}_{K}(p+q-r,\theta,{\overline \theta}) 
	\Phi^{L}_{d}(r,\theta,{\overline \theta}) \right\} \, ;
\end{eqnarray}
\begin{displaymath}
	\tilde{E}_{4}(p,q,r) = \left(\frac{1}{4}\right)^{2}
	\frac{g^{4}}{2}\,\kappa N^{2}(N^{2}-1)( \delta_{a}^{d}\delta^{b}_{c}
	+ \delta_{ac}\delta^{bd})\delta^{I}_{L}\delta^{K}_{J}
	I_{4\:bdIK}^{(E)acJL}(p,q,r) \, ,
\end{displaymath}
where
\begin{eqnarray}
	I_{4\:bdIK}^{(E)acJL}(p,q,r) & = & \int \frac{d^{4}k}{(2\pi)^{4}}
	d^{4}\theta \frac{1}{k^{2}(q+k)^{2}(k+r-p)^{2}} \cdot  
	\nonumber \\ 
	&& \hspace{-2.5cm} \cdot \left\{ \Phi^{a\dagger}_{I}
	(p,\theta,{\overline \theta})\Phi^{J}_{b}(q,\theta,{\overline\theta})
	\Phi^{c\dagger}_{K}(p+q-r,\theta,{\overline \theta}) 
	\Phi^{L}_{d}(r,\theta,{\overline \theta}) \right\} \, .
\end{eqnarray}

The sum of all the preceding terms results in a finite and non 
vanishing total one-loop correction to the four point function.
The final expression contains terms with six different tensorial 
structures
\begin{displaymath}
	\langle \Phi^{\dagger}\Phi\Phi^{\dagger}\Phi \rangle = 
	\kappa \left( \frac{1}{4}\right)^{2} g^{4}N^{2}(N^{2}-1) 
	\sum_{i=1}^{6} G^{(i)} \, ,
\end{displaymath}
where
\begin{eqnarray*}
	G^{(1)} &=&  
	\delta_{a}^{b}\delta_{c}^{d}\delta_{J}^{I}\delta_{L}^{K}\left(
	I_{1\:bdIK}^{(A)acJL} + I_{3\:bdIK}^{(A)acJL}
	+\frac{2}{3}I_{bdIK}^{(B)acJL}-I_{2\:bdIK}^{(D)acJL}+ 
	\right. \\
	&-& \left.
	I_{4\:bdIK}^{(D)acJL}+I_{1\:bdIK}^{(E)acJL}+
	I_{3\:bdIK}^{(E)acJL} \right) \\
	G^{(2)} &=&  
	\delta_{ac}\delta^{bd}\delta_{J}^{I}\delta_{L}^{K}\left(
	I_{1\:bdIK}^{(A)acJL} + I_{1\:bdIK}^{(D)acJL}-
	I_{2\:bdIK}^{(D)acJL}+I_{3\:bdIK}^{(D)acJL}+ 
	\right. \\
	&-& \left.
	I_{4\:bdIK}^{(D)acJL}+I_{1\:bdIK}^{(E)acJL}+
	I_{3\:bdIK}^{(E)acJL} \right) \\
	G^{(3)} &=&  
	\delta_{a}^{b}\delta_{c}^{d}\delta_{L}^{I}\delta_{J}^{K}\left(
	I_{2\:bdIK}^{(A)acJL} + \frac{2}{3}I_{bdIK}^{(B)acJL}+
	I_{2\:bdIK}^{(D)acJL}+I_{4\:bdIK}^{(D)acJL}
	\right) 
\end{eqnarray*}
\begin{eqnarray*}
	G^{(4)} &=& 
	\delta_{a}^{d}\delta_{c}^{b}\delta_{L}^{I}\delta_{J}^{K}\left(
	I_{2\:bdIK}^{(A)acJL} + I_{4\:bdIK}^{(A)acJL}
	+\frac{2}{3}I_{bdIK}^{(B)acJL}-I_{1\:bdIK}^{(D)acJL}- 
	\right. \\
	&-& \left.
	I_{3\:bdIK}^{(D)acJL}+I_{2\:bdIK}^{(E)acJL}+
	I_{4\:bdIK}^{(E)acJL} \right) \\
	G^{(5)} &=& 
	\delta_{a}^{d}\delta_{c}^{b}\delta_{J}^{I}\delta_{L}^{K}\left(
	I_{3\:bdIK}^{(A)acJL}+\frac{2}{3}I_{bdIK}^{(B)acJL}
	+I_{1\:bdIK}^{(D)acJL}+I_{3\:bdIK}^{(E)acJL}
	\right) \\
	G^{(6)} &=& 
	\delta_{ac}\delta^{bd}\delta_{L}^{I}\delta_{J}^{K}\left(
	I_{4\:bdIK}^{(A)acJL} - I_{1\:bdIK}^{(D)acJL}+
	I_{2\:bdIK}^{(D)acJL}-I_{3\:bdIK}^{(D)acJL}+ 
	\right. \\ 
	&+& \left. 
	I_{4\:bdIK}^{(D)acJL}+I_{2\:bdIK}^{(E)acJL}+
	I_{4\:bdIK}^{(E)acJL} \right)
\end{eqnarray*}

In the presence of a mass term for the (anti) chiral superfields the 
expressions given above are slightly modified by the presence of the 
mass in the free propagators and furthermore there are two additional 
sets of contributions corresponding to the diagrams

\vspace{1cm}
\noindent
\hspace*{2cm}
\begin{fmffile}{phi+phiphi+phiF}
 \begin{fmfgraph*}(150,110) 
  \fmfleft{i1,i2} 
  \fmfright{o1,o2}
  \fmf{fermion,tension=0.6}{i1,v1}
  \fmf{fermion,tension=0.6}{i2,v2}
  \fmf{fermion,tension=0.6}{v3,o1}
  \fmf{fermion,tension=0.6}{v4,o2}
  \fmf{phantom,tension=0.1,tag=1}{v1,v2}
  \fmf{phantom,tension=0.1,tag=2}{v2,v1}
  \fmf{phantom,tension=0.1,tag=3}{v2,v4}
  \fmf{phantom,tension=0.1,tag=4}{v4,v2}
  \fmf{phantom,tension=0.1,tag=5}{v4,v3}
  \fmf{phantom,tension=0.1,tag=6}{v3,v4}
  \fmf{phantom,tension=0.1,tag=7}{v3,v1}
  \fmf{phantom,tension=0.1,tag=8}{v1,v3}
  \fmfv{l=$\tilde{\Phi}^{a\dagger}_{I}$,l.a=180}{i1}
  \fmfv{l=$\tilde{\Phi}^{J}_{b}$,l.a=0}{o1}
  \fmfv{l=$\tilde{\Phi}^{L}_{d}$,l.a=0}{o2}
  \fmfv{l=$\tilde{\Phi}^{c\dagger}_{K}$,l.a=180}{i2}
  \fmfdot{v1,v2,v3,v4}
  \fmffreeze
  \fmfipath{p[]}
  \fmfiset{p1}{vpath1(__v1,__v2)}
  \fmfiset{p2}{vpath2(__v2,__v1)}
  \fmfiset{p3}{vpath3(__v2,__v4)}
  \fmfiset{p4}{vpath4(__v4,__v2)}
  \fmfiset{p5}{vpath5(__v4,__v3)}
  \fmfiset{p6}{vpath6(__v3,__v4)}
  \fmfiset{p7}{vpath7(__v3,__v1)}
  \fmfiset{p8}{vpath8(__v1,__v3)}
  \fmfi{fermion}{subpath (0,length(p1)/2) of p1}
  \fmfi{fermion}{subpath (0,length(p2)/2) of p2}
  \fmfi{fermion}{subpath (length(p6)/2,length(p6)) of p6}
  \fmfi{fermion}{subpath (length(p5)/2,length(p5)) of p5}
  \fmfi{boson}{subpath (0,length(p3)) of p3}
  \fmfi{boson}{subpath (0,length(p7)) of p7}
  \fmfiv{d.sh=cross,d.si=3mm}{point length(p1)/2 of p1}
  \fmfiv{d.sh=cross,d.si=3mm,l=\raisebox{-20pt}{\hspace{7cm}
        $\longrightarrow ~~\tilde{F}(p;q;r)$},l.a=90}
        {point length(p5)/2 of p5}
 \end{fmfgraph*}   
\end{fmffile} 

\vspace{1cm}
\noindent
\hspace*{2cm}
\begin{fmffile}{phi+phiphi+phiG}
 \begin{fmfgraph*}(150,110) 
  \fmfleft{i1,i2} 
  \fmfright{o1,o2}
  \fmf{fermion,tension=0.6}{i1,v1}
  \fmf{fermion,tension=0.6}{v2,i2}
  \fmf{fermion,tension=0.6}{o1,v3}
  \fmf{fermion,tension=0.6}{v4,o2}
  \fmf{phantom,tension=0.1,tag=1}{v1,v2}
  \fmf{phantom,tension=0.1,tag=2}{v2,v1}
  \fmf{phantom,tension=0.1,tag=3}{v2,v4}
  \fmf{phantom,tension=0.1,tag=4}{v4,v2}
  \fmf{phantom,tension=0.1,tag=5}{v4,v3}
  \fmf{phantom,tension=0.1,tag=6}{v3,v4}
  \fmf{phantom,tension=0.1,tag=7}{v3,v1}
  \fmf{phantom,tension=0.1,tag=8}{v1,v3}
  \fmfv{l=$\tilde{\Phi}^{a\dagger}_{I}$,l.a=180}{i1}
  \fmfv{l=$\tilde{\Phi}^{J}_{b}$,l.a=180}{i2}
  \fmfv{l=$\tilde{\Phi}^{L}_{d}$,l.a=0}{o2}
  \fmfv{l=$\tilde{\Phi}^{c\dagger}_{K}$,l.a=0}{o1}
  \fmfdot{v1,v2,v3,v4}
  \fmffreeze
  \fmfipath{p[]}
  \fmfiset{p1}{vpath1(__v1,__v2)}
  \fmfiset{p2}{vpath2(__v2,__v1)}
  \fmfiset{p3}{vpath3(__v2,__v4)}
  \fmfiset{p4}{vpath4(__v4,__v2)}
  \fmfiset{p5}{vpath5(__v4,__v3)}
  \fmfiset{p6}{vpath6(__v3,__v4)}
  \fmfiset{p7}{vpath7(__v3,__v1)}
  \fmfiset{p8}{vpath8(__v1,__v3)}
  \fmfi{fermion}{subpath (0,length(p2)) of p2}
  \fmfi{fermion}{subpath (0,length(p3)/2) of p3}
  \fmfi{fermion}{subpath (0,length(p4)/2) of p4}
  \fmfi{fermion}{subpath (0,length(p5)) of p5}
  \fmfi{fermion}{subpath (length(p8)/2,length(p8)) of p8}
  \fmfi{fermion}{subpath (length(p7)/2,length(p7)) of p7}
  \fmfiv{d.sh=cross,d.si=3mm}{point length(p3)/2 of p3}
  \fmfiv{d.sh=cross,d.si=3mm}{point length(p7)/2 of p7}
  \fmfiv{l=\raisebox{-20pt}{\hspace{7cm}
        $\longrightarrow ~~\tilde{G}(p;q;r)$},l.a=90}
        {point length(p5)/2 of p5}
 \end{fmfgraph*}   
\end{fmffile}

\vspace{0.5cm}
\noindent
Both of these graphs give corrections proportional to $m^{2}$, that can 
be calculated much in the same way as the previous ones.

With a different choice of gauge the computation of four-point Green 
functions is more complicated. Single diagrams involving vector 
superfield propagators contain new contributions, some of which are 
infrared divergent. The correction $\tilde{C}$ is not zero anymore, 
because there are projection operators acting on the $\delta$-functions. 
Moreover further diagrams must be included in the calculation at the 
same order as a consequence of the non-vanishing of the one-loop 
correction to the vertices. For example one must consider diagrams 
such as

\vspace{1cm}
\noindent
\hspace*{2cm}
\begin{fmffile}{phi+phiphi+phiH}
 \begin{fmfgraph*}(150,100) 
  \fmfleft{i1,i2} 
  \fmfright{o1,o2}
  \fmf{fermion,tension=1}{i1,v1}
  \fmf{fermion,tension=1}{v1,i2}
  \fmf{boson,tension=1.2}{v1,v2}
  \fmf{fermion,tension=0.8}{v2,v3}
  \fmf{fermion,tension=0.8}{v4,v2}
  \fmf{fermion}{o1,v4}
  \fmf{fermion}{v3,o2}
  \fmffreeze
  \fmf{boson}{v3,v4}
  \fmfv{l=$\tilde{\Phi}^{J}_{b}$,l.a=180}{i2}
  \fmfv{l=$\tilde{\Phi}^{L}_{d}$,l.a=0}{o2}
  \fmfv{l=$\tilde{\Phi}^{c\dagger}_{K}$,l.a=0}{o1}
  \fmfv{l=$\tilde{\Phi}^{a\dagger}_{I}$,l.a=180}{i1}
  \fmfdot{v1,v2,v3,v4}
 \end{fmfgraph*}   
\end{fmffile} 

\vspace{1cm}
The techniques illustrated in this chapter for the calculation of 
Green functions in the ${\cal N}$=1 formalism can be applied to 
correlation functions of composite operators as well. Green functions 
of gauge invariant composite operators such as those that form 
the multiplet of currents (equations (\ref{currentdef1}) and 
(\ref{currentdef2})) play  a crucial r\^ole in the correspondence 
with type IIB superstring theory on AdS space to be discussed in 
chapter 5. The application of ${\cal N}$=1 superspace 
to this problem will be sketched in the final chapter. It can be 
shown that the extension of the formalism to the case of composite 
operators is rather straightforward, the fundamental difference being 
that the complete Green functions and not the proper parts must be 
considered.


\chapter{Nonperturbative effects in ${\cal N}=4$ super Yang--Mills 
theory}
\label{cap4}
\vspace*{2cm}
\fancyhead[RO,LE]{\thepage}
\fancyhead[RE]{{\footnotesize {\rm Chapter 4.}~~{\it Non-perturbative 
effects in ${\cal N}=4$ SYM theory}}} 
\fancyhead[LO]{}

\noindent
In the last chapter the quantum properties of ${\cal N}$=4 
supersymmetric Yang--Mills theory have been analyzed at the 
perturbative level, this chapter is devoted to the study of some 
non-perturbative effects. Instanton calculus will be employed 
to compute various Green functions of gauge invariant composite 
operators in the semiclassical approximation.

It is well known that in gauge field theories non-perturbative effects 
can modify in a dramatic way the perturbative structure of the 
vacuum. Instanton calculus has proved to be an extremely powerful tool 
in the analysis of non-perturbative phenomena in quantum field 
theories, particularly in the case of supersymmetric gauge theories. 
The r\^ole of instantons in Yang--Mills theories was 
pointed out in \cite{belpolyschtyup} at the classical level and then 
at the quantum level in a fundamental work by `t Hooft \cite{thooftb}. 
Instantons are non trivial finite action solutions of the classical 
Euclidean equations of motion of the model. The computation of Green 
functions in the instanton background allows to study 
non-perturbative properties of the quantum theory. In 
non-supersymmetric non Abelian gauge theories the contribution of 
instantons to Green functions turns out to be divergent in the 
semiclassical approximation \cite{thooftb}. However in the 
supersymmetric case explicit instanton calculations can be performed 
leading to finite and well defined results, thanks to the cancellation 
between fermionic and bosonic quantum fluctuations. Compensations 
of quantum effects due bosons and fermions take place in the 
instanton background and allow to explicitly compute vacuum 
expectation values (vev's) of composite operators (condensates). 
These cancellations are reminiscent of boson and fermion 
compensations in internal loops at the perturbative level.
The condensates play the r\^ole of order parameters in 
non-perturbative phenomena such as for instance the spontaneous 
breaking of chiral symmetries or even of supersymmetry in certain 
models \cite{akmrv}. Furthermore instanton calculus together with 
holomorphy properties, has allowed to derive some exact results 
in supersymmetric theories \cite{nsvz} and, more recently, to 
obtain non trivial checks of the exact result proposed in \cite{sw}. 

As was discussed in previous chapters supersymmetric gauge theories 
typically possess a moduli space of vacua parameterized by the vev's 
of scalar fields. The literature on instanton effects in asymptotic 
free theories includes examples in which the computations are 
performed with zero vev's for all the scalars \cite{akmrv} as well as 
examples in which non vanishing vev's for the scalar fields 
parameterizing the moduli space are considered \cite{shifman2}. 
The ${\cal N}$=4 super Yang--Mills theory has a (exactly) 
vanishing $\beta$-function and a moduli space of vacua, 
${\cal M}$=$\mathbb{R}^{6k}\times {\cal S}_{k}$, 
where $k$ is the rank of the gauge group and ${\cal S}_{k}$ the group 
of permutations of $k$ elements, parameterized by the vev's 
of six real scalar fields, $\langle\varphi^{i}\rangle$. The 
superconformal phase, corresponding to the origin of ${\cal M}$, \ie 
$\langle\varphi^{i}\rangle$=0, $\forall i$, cannot be obtained 
naively as the limit of the theory in the Coulomb phase. In this 
chapter the approach of \cite{akmrv} will be applied to the 
computation of various condensates corresponding to operators in 
the supermultiplet of currents (\ref{currentdef1}), (\ref{currentdef2}). 
Four-, eight- and sixteen-point functions of bilinear operators 
belonging to the current supermultiplet will be computed in the 
semiclassical approximation in the case of a SU(2) gauge group. 
In particular the exact complete spatial dependence will be determined 
for a four-point function at lowest order.  

Due to the vanishing of the $\beta$-function and of the associated 
chiral anomalies, the correlation functions that will be studied, 
unlike the cases encountered in theories with less supersymmetry, 
receive a non-vanishing perturbative contribution as well. 
These Green functions will reconsidered in the final chapter in the 
context of the correspondence with type IIB superstring theory 
compactified on AdS$_{5}\times S^{5}$. It will be shown that the 
difficulties encountered in perturbative calculations in chapter 
\ref{cap3} do not affect the computations discussed here.

The chapter is organized as follows. The general features of instanton 
calculus in supersymmetric gauge theories are briefly discussed in 
sections 1, 2 and 3. Sections \ref{instantons-n=4}, \ref{inst-4-point} 
and \ref{inst-8-16-point} report on original results in the 
case of ${\cal N}$=4 supersymmetric Yang--Mills theory, following 
\cite{bgkr}. The final section describes the generalizations of 
the results of \cite{bgkr} to the case of a SU($N$) gauge group and 
to the $K$-instanton sector in the large $N$ limit, that were given 
respectively in \cite{mattis1} and \cite{mattis2,mattislong}. 

\section{Instanton calculus in non Abelian gauge theories}
\label{yminstantons}
\fancyhead[LO]{{\footnotesize 4.1~~{\it Instantons in Yang--Mills 
theories}}}

In this section the semiclassical quantization in the background of 
an instanton configuration will be reviewed for the case of a pure 
Yang--Mills theory. The extension to fermions will be discussed in 
the next section.

In quantum field theory one is interested in computing vacuum 
expectation values of operators ${\cal O}(\phi)$, that in general are 
polynomials in the elementary fields $\phi$. In Euclidean space one 
must calculate
\begin{equation}
	\langle {\cal O}(x_{1},\ldots,x_{n})\rangle = 
	\int [{\cal D}\phi] e^{-\frac{S[\phi]}{\hslash}}{\cal O}(\phi(x_{1}),
	\ldots,\phi(x_{n})) \, ,
	\label{expectop}
\end{equation}
where the dependence on $\hslash$ has been explicitly indicated in 
order to discuss the semiclassical approximation. 

The classical vacuum of the theory corresponds to a constant 
expectation value $\langle \phi \rangle$ for the elementary fields 
and a perturbative series is obtained expanding the 
interaction part of the action $S[\phi]$ for small fluctuations, 
$\delta \phi$, around the classical vacuum. In instanton calculations 
one considers a situation in which there exists a non-trivial solution 
of the classical equations of motion with finite Euclidean action. 
Denoting by ${\overline \phi}$ such a field configuration one can then 
study quantum fluctuation around $\phi={\overline \phi}$ in the 
limit $\hslash \to 0$. Putting 
\begin{displaymath}
	\phi(x)  = {\overline\phi}(x) + \delta\phi(x) = 
	{\overline \phi}(x) + \hslash^{1/2} \eta(x) 
\end{displaymath}
and expanding $S[\phi]$ in powers of $\hslash$ in the form 
\begin{displaymath}
	S[\phi] = S[{\overline \phi}] + \frac{\hslash}{2}\int d^{4}xd^{4}y 
	\left. \left( \frac{\delta^{2}S}{\delta\phi(x)\delta\phi(y)} 
	\right) \right|_{\phi={\overline\phi}} \eta(x) \eta(y) + 
	O(\hslash^{3/2}) \, ,
\end{displaymath}
the vacuum expectation value of an operator ${\cal O}[\phi]$ can be 
evaluated in the semiclassical limit by a saddle point 
approximation~\footnote{This relation holds for bosonic fields 
$\phi$, see the next section for the case of fermions.}
\begin{eqnarray}
	\langle{\cal O}(x_{1},\ldots,x_{n})\rangle &=&\int [{\cal D}\eta]
	e^{-\frac{S[{\overline\phi}+\eta]}{\hslash}} {\cal 
	O}[({\overline\phi}+\eta)(x_{i})] = \nonumber \\
	&& \hspace{-3cm} = {\cal O} ({\overline\phi}) \, 
	e^{-\frac{{\overline S}}{\hslash}}
	\left[ \det \left.\left(\frac{\delta^{2}S}{\delta\phi(x)
	\delta\phi(y)}\right)\right|_{\phi={\overline\phi}} 
	\right]^{-\frac{1}{2}} \left( 1 + O(\hslash) \right) \, .
	\label{osemiclass}
\end{eqnarray}
Notice that in principle one could consider a classical solution 
${\overline\phi}$ with infinite Euclidean action, but this would give 
a vanishing contribution in the semiclassical approximation because of 
the factor $e^{-S[{\overline\phi}]/\hslash}$ in (\ref{osemiclass}).

\subsection{Yang--Mills instantons}

The classification of non-trivial solutions of the classical 
Euclidean equations of motion for the Yang--Mills system has been 
given in \cite{belpolyschtyup}, where an explicit solution has also 
been constructed. Another (multi-center) solution has been proposed 
by 't Hooft, who also studied the semiclassical quantization 
\cite{thooftb}. The case of a SU(2) gauge group will be 
discussed first and then the extension to SU($N$) will be 
sketched. 

The Euclidean Yang--Mills action is 
\begin{equation}
	S_{{\rm YM}}[A] = -\frac{1}{4 d_{{\bf  r}} g^{2}} \int d^{4}x \, 
	\tr \left( F_{\mu\nu} F^{\mu\nu} \right) \, ,
	\label{actionym}
\end{equation}
where the gauge field $A_{\mu}$ and the field strength $F_{\mu\nu}$ are
respectively 
\begin{eqnarray*}
	A_{\mu} &=& A_{\mu}^{a}T_{a} \\
	F^{\mu\nu} &=& \partial^{\mu}A^{\nu} -
	\partial^{\nu}A^{\mu} + [A^{\mu},A^{\nu}] = \left(
	\partial_{\mu}A_{\mu}^{a} - \partial_{\nu}A_{\mu}^{a} + i f^{a}{}_{bc}
	A_{\mu}^{b}A_{\nu}^{c} \right) T_{a} 
\end{eqnarray*}
and $T^{a}$ are generators of the gauge group in the representation 
${\bf r}$, satisfying $[T^{a},T^{b}] = i f^{ab}{}_{c}T^{c}$ and 
$\tr \left(T^{a}T^{b}\right)=\delta^{ab} d_{{\bf r}}$, $d_{{\bf r}}$ 
being the Dynkin index of the representation. Gauge transformations take 
the form 
\begin{displaymath}
	A_{\mu}(x) \longrightarrow A^{\Omega}_{\mu}(x) = \Omega(x) \left[
	A_{\mu}(x)+i\partial_{\mu} \right] \Omega^{\dagger}(x) \, ,
\end{displaymath}
with $\Omega(x)=e^{i\alpha_{a}(x) T^{a}}$. 

The study of finite Euclidean action configurations is performed much 
in the same way as for the monopoles of the Georgi--Glashow model in 
appendix \ref{appb}. The requirement of finite action implies that 
$A_{\mu}$ must approach asymptotically, \ie for $|x|\to\infty$, a pure 
gauge configuration. This leads to a topological classification of 
the field configurations in homotopy classes. The quantity which 
characterizes the finite energy solutions is the Pontryagin index
\begin{equation}
	K[A] = \frac{1}{32\pi^{2}} \int d^{4}x 
	F_{\mu\nu}^{a}\tilde{F}_{a}^{\mu\nu} \, ,
	\label{pontryagin}
\end{equation}
where $\tilde{F}=\frac{1}{2} \varepsilon_{\mu\nu\rho\sigma} 
F^{\rho\sigma}$. The Yang--Mills action satisfies a ``Bogomol'nyi 
bound''  
\begin{equation}
	S_{{\rm YM}}[A] = \pm \frac{8\pi^{2}}{g^{2}} K[A]+\frac{1}{8\pi^{2}}
	\int d^{4}x \left(F_{\mu\nu}^{a}\mp \tilde{F}_{\mu\nu}^{a}\right) 
	\left(F^{\mu\nu}_{a}\mp \tilde{F}^{\mu\nu}_{a}\right) \geq 
	\frac{8\pi^{2}}{g^{2}} \left| K[A] \right| \, . 
	\label{sbound}
\end{equation}
Solutions of the classical equations of motion correspond to local 
minima of the action functional, so that in each homotopy class they 
saturate the bound in (\ref{sbound}). As a consequence they are 
determined by the condition 
\begin{displaymath}
	F_{\mu\nu}^{a} = \pm \tilde{F}_{\mu\nu}^{a} \, .
\end{displaymath}
In the $K$=0 sector the minimum is reached on the trivial vacuum, 
$A_{\mu}$=0, while the configurations minimizing $S[A]$ for $|K|\neq 0$ 
are (anti) instantons. Self-dual and anti self-dual 
configurations are referred to as respectively instantons and 
anti-instantons. 

In the $K$=1 sector and for a gauge group SU(2) the explicit solution 
was found in \cite{belpolyschtyup,thooftb} through the ansatz
\begin{equation}
	{\overline A}_{\mu} = i \frac{{\overline\sigma}_{\mu\nu}}{2} 
	\partial^{\nu} \log \Phi \, ,
	\label{thooftansatz}
\end{equation}
where $\Phi$ is real scalar field. Substituting into the equations of 
motion one obtains the equation for $\Phi$
\begin{displaymath}
	\frac{\Box\Phi}{\Phi} = 0 \, .
\end{displaymath}
A solution can be written in the form 
\begin{equation}
	{\overline\Phi} = 1 + \frac{\rho^{2}}{(x-x_{0})^{2}} \, .
	\label{K1ansatz}
\end{equation}
${\overline\Phi}$ depends on 5 free parameters, $\rho$ associated with 
the ``size'' of the instanton and $x_{0}$ with its position. The 
corresponding instanton solution is 
\begin{equation}
	{\overline A}_{\mu} = -2i \rho^{2} {\overline\eta}^{a}_{\mu\nu} 
	\frac{y^{\nu}}{y^{2}(y^{2}+\rho^{2})} \, ,
	\label{singinstanton}
\end{equation}
where $y=x-x_{0}$ and ${\overline\eta}^{a}_{\mu\nu}$ are the anti 
self-dual 't Hooft symbols
\begin{displaymath}
	{\overline\eta}^{a}_{\mu\nu} = -{\overline\eta}^{a}_{\nu\mu} = 
	\left\{ \begin{array}{l}
	\varepsilon^{a\mu\nu} \qquad {\rm for} \quad \mu,\nu=1,2,3 \\
	-\delta^{a\mu} \qquad {\rm for} \quad \nu=4 \, . 
    \end{array} \right.
\end{displaymath}
The general solution depends on 3 additional parameters corresponding 
to the freedom of performing global gauge rotations. Actually the 
Yang--Mills system is classically invariant under the 
four-dimensional conformal group, which partially overlaps
with global gauge transformations. In \cite{jackiwrebbi} it has been 
proved that the most general one-instanton solution depends on 8 
parameters, referred to as collective coordinates or moduli, that in 
the following will be denoted by $\beta_{i}$. The moduli space, \ie the 
space parameterized by the coordinates $\beta_{i}$, will be indicated 
by ${\cal M}$. The field (\ref{singinstanton}) has a singularity at 
$y$=0, which can be removed by a correspondingly singular gauge 
transformation, so that one can construct a solution with no 
singularities at any $x$
\begin{equation}
	{\overline A}_{\mu} = -2i e^{i\theta_{a}T^{a}} 
	\eta^{a}_{\mu\nu} \frac{y^{\nu}}{(y^{2}+\rho^{2})}
	e^{-i\theta_{a}T^{a}}\, ,
	\label{instanton}
\end{equation}
where $\eta^{a}_{\mu\nu}$ are the self-dual 't Hooft symbols
\begin{displaymath}
	\eta^{a}_{\mu\nu} = -\eta^{a}_{\nu\mu} = \left\{ 
	\begin{array}{l}
	\varepsilon^{a\mu\nu} \qquad {\rm for} \quad \mu,\nu=1,2,3 \\
	\delta^{a\mu} \qquad {\rm for} \quad \nu=4 \, , 
    \end{array} \right.
\end{displaymath}
which satisfy
\begin{displaymath}
	\sigma_{\mu\nu} = \eta^{a}_{\mu\nu} \sigma_{a} \, ,
\end{displaymath}
with $\sigma^{a}$, $a=1,2,3$ the standard Pauli matrices.
The associated field strength is 
\begin{equation}
	{\overline F}_{\mu\nu} = F_{\mu\nu}[{\overline A}] = -2i
	e^{i\theta_{a}T^{a}} \sigma_{\mu\nu} 
	\frac{\rho^{2}}{(y^{2}+\rho^{2})^{2}} e^{-i\theta_{a}T^{a}} \, .
	\label{instfieldstrength}
\end{equation}
The ansatz of (\ref{K1ansatz}) can be generalized to a $n$-instanton 
solution considering the scalar field
\begin{displaymath}
	\Phi(x) = 1+ \sum_{i=1}^{n}\frac{\rho^{2}_{i}}{(x-x_{0}^{(i)})^{2}} 
	\, .
\end{displaymath}
Substituting into (\ref{thooftansatz}) yields a solution, 
${\overline A}^{(n)}_{\mu}$, with singularities at 
$x_{\mu}=x_{0\,\mu}^{(i)}$, which can be made 
regular by a suitable singular gauge transformation. For such a 
configuration one can prove that  
\begin{displaymath}
	S[{\overline A}^{(n)}] = n S[{\overline A}^{(1)}] \, ,
\end{displaymath}
so that ${\overline A}^{(n)}_{\mu}(x)$ is a solution in the homotopy 
class $K[A]$=$n$. The number of collective coordinates for the 
$n$-instanton solution is $8n-3$ (with $n>1$). There are 8 moduli for 
each single instanton, but one must subtract 3 global rotations that 
are already accounted for by the relative rotations. This result can 
be proved to hold in general, even if the $n$ instantons are not widely 
separated, \cite{atiyahward}. 

\subsection{Semiclassical quantization}

To perform a semiclassical quantization around an instanton 
configuration according to (\ref{osemiclass}) one needs the expansion 
of $S[A]$ around $A_{\mu}={\overline A}_{\mu}$, 
obtained putting $A_{\mu}={\overline A}_{\mu}+Q_{\mu}$, 
where $Q_{\mu}$ denotes the quantum fluctuation. 
A straightforward calculation gives the gauge kinetic operator 
in the instanton background 
\begin{displaymath}
	M_{\mu\nu}^{ab} = \left.\left(\frac{\delta^{2}S[A]}{\delta 
	A^{a}_{\mu}\delta A^{b}_{\nu}} \right) \right|_{{\overline A}} = 
	\left[-D^{2}({\overline A})\delta_{\mu\nu} + D_{\mu}({\overline A})
	D_{\nu}({\overline A}) - 2 {\overline F}_{\mu\nu}\right]
	\delta^{ab} \, ,
\end{displaymath}
where $D_{\mu}({\overline A}) = \partial_{\mu} + i [{\overline 
A}_{\mu},\,.\, ]$ is the covariant derivative.

In the instanton background the Faddeev--Popov quantization must be 
suitably modified, because the kinetic operator $M_{\mu\nu}$ possesses 
8 new zero modes (in the case of a SU(2) gauge group under 
consideration) beyond those associated with gauge symmetry. These new 
zero-modes correspond to the symmetries of the action $S[A]$ that are 
broken by the instanton configuration and are related to the 8 collective 
coordinates $\beta_{i}$ in the general solution (\ref{instanton}). 
In fact since ${\overline A}_{\mu}$ is a solution for any $\beta_{i}$ 
\begin{displaymath}
	\int d^{4}y \left.\left(\frac{\delta^{2}S[A]}{\delta 
	A_{\mu}\delta A_{\nu}} \right) \right|_{{\overline A}_{\beta}} 
	\frac{\partial{\overline A}_{\nu}}{\partial\beta_{i}} (y,\beta)
	= 0 \, ,
\end{displaymath}
so that (\ref{osemiclass}) is formally divergent. This divergence is 
completely analogous to those introduced in the partition function of 
gauge theories by the functional integration along the gauge orbits, 
that is dealt with by the Faddeev--Popov procedure. The correct way of 
performing the quantization in the instanton background was discussed 
in \cite{yaffe}. The general instanton configuration will be 
parameterized as
\begin{displaymath}
	{\overline A}^{\Omega}_{\mu}(x,\beta)=\Omega^{\dagger}{\overline A}_{\mu}
	\Omega + i \Omega^{\dagger} \partial_{\mu} \Omega \, ,
\end{displaymath}
where $\Omega$ denotes an arbitrary gauge transformation and $\beta$ 
the set collective coordinates, 8 in the case of SU(2) and $K[A]$=1 
and in general $4NK[A]$ in the $K$-instanton sector for a SU($N$) gauge 
group. The basic idea is then to constrain {\em \`a la} Faddeev--Popov 
the quantum fluctuations $Q_{\mu}$ to be orthogonal to 
${\overline A}^{\Omega}_{\mu}$ as $\Omega$ varies in G and $\beta_{i}$
in the moduli space ${\cal M}$. In the partition function of the 
system~\footnote{From now on $\hslash$ will be set equal 1.} 
\begin{displaymath}
	Z = \int [{\cal D}A] e^{-S[A]}
\end{displaymath}
one inserts 
\begin{equation}
	1 = \Delta_{{\rm FP}} \int_{G} \prod_{a,x} [{\cal D}h^{a}(x) ]
	\int_{{\cal M}} \prod_{i}d\beta_{i} \delta\left( (A_{\mu}-
	{\overline A}^{\Omega}_{\mu},\frac{\delta
	{\overline A}^{\Omega}_{\mu}}{\delta h^{a}}) \right) 
	\delta\left( (A_{\mu}-{\overline A}^{\Omega}_{\mu},\frac{
	\partial{\overline A}^{\Omega}_{\mu}}{\partial \beta_{i}}) 
	\right) \, ,
	\label{deltafp}
\end{equation}
where $\Omega(x)$ is parameterized as $\Omega(x) =e^{ih^{a}(x)T_{a}}$
and the scalar product in the arguments of the $\delta$-functions is 
\begin{equation}
	(f_{\mu},g^{\mu}) = \frac{1}{2} \int d^{4}x \, \tr \left( 
	f_{\mu}(x) g^{\mu}(x) \right) \, .
	\label{scalarprod}
\end{equation}
Equation (\ref{deltafp}) defines the Faddeev--Popov determinant 
$\Delta_{{\rm FP}}$. Some algebraic steps allow write $Z$ in the form
\begin{eqnarray*}
    && Z = \int [{\cal D} A_{\mu}] \int \prod_{i} d \beta_{i} \, 
    e^{-S[A]} \Delta_{{\rm FP}}(A,\beta) \cdot \\
    && \cdot \delta \left( \tr \left[ T^{a} D_{\mu}({\overline A}) 
    \left( A^{\mu}-{\overline A} \right) (x) \right] \right) 
    \delta\left( (A_{\mu}-{\overline A}_{\mu},\frac{
    \partial{\overline A}_{\mu}}{\partial \beta_{i}}) 
    \right) \, .
\end{eqnarray*}
In this relation an infinite constant, coming from the integration 
over the group algebra, has been dropped, since in the following 
the above construction will be employed in the computation of normalized 
vev's of operators. The Faddeev--Popov determinant in the above 
expression can be computed much in the same way as in the case of the 
ordinary of gauge theories. 

To compute expectation values in the semiclassical approximation one 
only needs $\Delta_{{\rm FP}}(A,\beta)$ on the classical solution 
${\overline A}$. It can be shown that in this case $\Delta_{{\rm FP}}$ 
can be written as
\begin{equation}
	\left. \Delta_{{\rm FP}}(A,\beta)\rule[-10pt]{0pt}{10pt}
	\right|_{A={\overline A}} = 
	\det_{\begin{array}{c} \scriptstyle{x,y} \\ 
	\raisebox{5pt}{$\scriptstyle{a,b}$} \end{array}}
	\left( D_{ab}^{2}({\overline A}) \delta(x-y) \right) 
	\, \det_{i,j} \left( d_{ij}(A,\beta) \right) \, ,
	\label{detfp}
\end{equation}
where 
\begin{displaymath}
	d_{ij}(A,\beta) = (\hat{a}_{\mu}^{(i)}(\beta),
	\hat{a}^{\mu(j)}(\beta) )
\end{displaymath}
The vectors $\hat{a}$ in the above equation are given by
\begin{displaymath}
	\hat{a}^{(i)\,a}_{\mu}(x,\beta) = \frac{\partial 
	{\overline A}^{a}_{\mu}}{\partial\beta_{i}} + 
	\xi^{(i)\,a}_{\mu}(x) 
\end{displaymath}
and $\xi^{(i)\,a}_{\mu})$ are determined by the transversality 
condition 
\begin{displaymath}
	D^{ab}_{\mu}({\overline A}) \hat{a}^{(i)}_{\mu\,b}
	(x,\beta) = 0 \, . 
\end{displaymath}

In conclusion introducing ghost fields, $c$ and ${\overline c}$ to 
rewrite the determinants as Gaussian integrals the expectation value 
of a gauge invariant operator ${\cal O}[A]$ in the semiclassical 
approximation takes the form
\begin{eqnarray}
	&&\langle {\cal O}(x_{1},\ldots,x_{n})\rangle_{{\overline A}} = 
	e^{-\frac{8\pi^{2}}{g^{2}}K[{\overline A}]} \cdot 
	\nonumber \\
	&& \cdot \frac{\begin{displaystyle}
	\int [{\cal D}Q_{\mu}{\cal D}c{\cal D}{\overline c}] \prod_{i}
	d\beta_{i} \frac{\|\hat{a}\|}{\sqrt{2\pi}} \, 
	e^{-\frac{1}{2}\int Q^{\mu}M^{{\rm g.f.}}_{\mu\nu}
	Q^{\nu}-\int{\overline c}D^{2}({\overline A})c}
	{\cal O}[{\overline A}]\end{displaystyle}}{\begin{displaystyle}
	\int [{\cal D}Q_{\mu}{\cal D}c{\cal D}{\overline c}] 
	e^{-\frac{1}{2}\int Q^{\mu}M^{(0){\rm g.f.}}_{\mu\nu}
	Q^{\nu}-\int{\overline c}\Box c}
	\end{displaystyle}} = \nonumber \\
	&& = e^{-\frac{8\pi^{2}}{g^{2}}K[{\overline A}]} 
	\int \prod_{i}d\beta_{i}\frac{\|\hat{a}\|}{\sqrt{2\pi}} \,
	{\cal O}[{\overline A}] \frac{\left(\det^{\prime}
	M^{{\rm g.f.}}_{\mu\nu}\right)^{-\frac{1}{2}}\,\det D^{2}({\overline 
	A})}{\left(\det^{\prime}M^{(0){\rm 
	g.f.}}_{\mu\nu}\right)^{-\frac{1}{2}}\det \Box} 
	\, , \label{finalsemiclass} 
\end{eqnarray}
where the `gauge-fixed' kinetic operators $M^{{\rm g.f.}}_{\mu\nu}$ and 
$M^{(0){\rm g.f.}}_{\mu\nu}$ are
\begin{displaymath}
	M^{{\rm g.f.}}_{\mu\nu} = -D^{2}({\overline A}) \delta_{\mu\nu}
	-2F_{\mu\nu}[{\overline A}] \, ,\qquad M^{(0){\rm g.f.}}_{\mu\nu}
	=-\Box \delta_{\mu\nu} \, .
\end{displaymath}

\subsection{Bosonic zero-modes}

To explicitly compute expectation values through equation 
(\ref{finalsemiclass}) one needs to know the zero-mode vectors 
$\hat{a}^{(i)}_{\mu}$, whose norms enter the integration measure. Since 
the kinetic operator is modified by the gauge-fixing term the vectors 
$\hat{a}^{(i)}_{\mu}$ are given by
\begin{displaymath}
	\hat{a}^{(i)}_{\mu}(x) = \frac{\partial {\overline A}_{\mu}}
	{\partial \beta_{i}}(x,\beta) + D_{\mu}({\overline A}) 
	\Lambda^{(i)}(x) \, ,
\end{displaymath}
for a suitable function $\Lambda^{(i)}(x)$, such that 
\begin{displaymath}
	\hat{a}^{(i)}_{\mu}(x) \in {\rm Ker}\,\left[M_{\mu\nu}^{{\rm g.f.}}
	\right] \, .
\end{displaymath}
To compute the zero-mode vectors it is convenient to refer to the 
singular instanton solution. For the case of a SU(2) gauge 
group, after rescaling $A_{\mu} \to \frac{1}{g} A_{\mu}$, it reads
\begin{equation}
	{\overline A}_{\mu}(x,\beta) = -\frac{i}{g} e^{i\theta_{a}T^{a}} 
	\left[ {\overline\sigma}_{\mu\nu}\frac{\rho^{2}(x-x_{0})_{\mu}}
	{(x-x_{0})^{2}[(x-x_{0})^{2}+\rho^{2}]} \right]
	e^{-i\theta_{a}T^{a}} \, ,
	\label{explicitsinginst}
\end{equation}
where the 8 moduli are $\beta_{i}=(x_{0}^{\mu}, \rho,\theta_{a})$.

\vspace{0.7cm}
{\sl Zero-modes associated with translations.} 
\vspace{0.3cm}

\noindent
In this case $\beta_{i}\to x_{0}^{\mu}$ and there are four zero-modes.
The transverse vectors are
\begin{displaymath}
	\hat{a}_{\mu(\nu)} = \frac{\partial}{\partial x_{0}^{\nu}} 
	{\overline A}_{\mu} + D_{\mu}({\overline A}){\overline A}_{\nu} = 
	F_{\mu\nu}[{\overline A}] \, .
\end{displaymath}
The computation of the norm is straightforward using 
(\ref{explicitsinginst}), the result is 
\begin{displaymath}
	\|\hat{a}_{\mu(\nu)}\| = \frac{2\sqrt{2} \,\pi}{g} \, .
\end{displaymath}

\vspace{0.7cm}
{\sl Zero-modes associated with dilatations.} 
\vspace{0.3cm}

\noindent
The corresponding collective coordinate is the size of the 
instanton $\rho$. $\hat{a}_{\mu}^{({\rm dil})}$ is simply given by 
\begin{displaymath}
	\hat{a}_{\mu}^{({\rm dil})} = \frac{\partial {\overline A}_{\mu}}
	{\partial \rho} = -\frac{2i}{g} {\overline \sigma}_{\mu\nu} 
	\frac{\rho y^{\nu}}{(y^{2}+\rho^{2})^{2}} \, ,
\end{displaymath}
so that for the norm one obtains
\begin{displaymath}
		\|\hat{a}_{\mu}^{({\rm dil})}\| = \frac{4\pi}{g} \, .
\end{displaymath}

\vspace{0.7cm}
{\sl Zero-modes associated with gauge rotations.} 
\vspace{0.3cm}

\noindent
The relevant collective coordinates are the angles $\theta_{a}$ in 
(\ref{explicitsinginst}) and one can put 
\begin{displaymath}
	\hat{a}_{\mu}^{(a)} = \frac{\partial {\overline A}_{\mu}}
	{\partial\theta_{a}} + D_{\mu}({\overline A}) \Lambda^{(a)} = 
	D_{\mu}({\overline A}) \left[ -\frac{i}{g}T^{a}+\Lambda^{(a)} 
	\right] \equiv D_{\mu}({\overline A}) \phi^{(a)} \, ,	
\end{displaymath}
where the scalar field $\phi^{(a)}$, defined by the last equality in 
the above equation, must satisfy
\begin{displaymath}
	D^{2}({\overline A})\phi^{(a)}=0 \, , \hspace{1cm} \lim_{|x|\to 
	\infty} \phi^{(a)} = -\frac{i}{g}T^{(a)} \, .
\end{displaymath}
The solution is 
\begin{displaymath}
	\phi^{(a)}= \left(-\frac{i}{g}T^{(a)}\right) 
	\frac{r^{2}}{r^{2}+\rho^{2}} \, \hspace{1cm} {\rm with} 
	\hspace{0.5cm} r=|x| \, .
\end{displaymath}
The norms are then
\begin{displaymath}
	\|\hat{a}_{\mu}^{(a)} \| = \frac{2\pi\rho}{g} \, .
\end{displaymath}
The above eight vectors can be shown to be orthogonal with respect 
to the scalar product (\ref{scalarprod}). 

Having determined the vectors $\hat{a}_{\mu}$ one can thus replace the 
integration over the zero-modes in (\ref{finalsemiclass}) with an 
integration over the instanton collective coordinates and write 
the integration measure as
\begin{eqnarray}
	d\mu_{{\rm B}}&=&\prod_{i}d\beta_{i}\frac{\|\hat{a}\|}{\sqrt{2\pi}}= 
	\frac{1}{(2\pi)^{4}} \left(\frac{2\sqrt{2}\pi}{g}\right)^{4} 
	\frac{4\pi}{g} \left(\frac{2\pi\rho}{g}\right)^{3}d^{4}x_{0}d\rho
	d^{3}\theta = \nonumber \\
	&& = \frac{2^{7}\pi^{4}}{g^{8}}\rho^{3}d^{4}x_{0}d\rho
	d^{3}\theta \, .
	\label{bosonicmeasure}
\end{eqnarray}

The generalization to larger gauge groups, in particular to SU($N$), is 
rather straightforward \cite{bernard2,bernard1}. 
First it can be shown \cite{coleman}, that 
the mapping of $S^{3}$ into a generic group G, which determines the 
topological classification of the instanton solutions, can be 
continuously deformed to a mapping into SU(2). Hence the 
classification of classical solutions into homotopy classes holds in 
general. Furthermore one can construct a solution in the case of a 
gauge group G by the embedding of a SU(2) instanton. In particular 
for G=SU($N$) one can consider a configuration
\begin{displaymath}
	{\overline A}_{\mu} = \frac{2i}{g} \frac{{\overline 
	\eta}^{a}_{\mu\nu} x^{\nu}}{x^{2}(x^{2}+\rho^{2})} 
	\frac{\lambda_{k}}{2} \, ,
\end{displaymath}
where $\lambda_{k}$, $k=1,2,3$, are the first three generators of the 
fundamental representation of SU($N$). Under the SU(2) subgroup of 
SU($N$) 
($\lambda_{1},\lambda_{2},\lambda_{3}$) transforms as a triplet, whereas 
the other generators are organized into $2(N-2)$ doublets and the 
remaining are singlets. In conclusion, in addition to eight zero-modes 
of the SU(2) instanton, there are $4(N-2)$ additional zero-modes. The 
corresponding vectors take the form \cite{bernard1}
\begin{displaymath}
	\hat{a}_{\mu}^{r} = D_{\mu}({\overline A}) \left[ 
	\frac{\lambda^{r}}{g} \left( \frac{x^{2}}{x^{2}+\rho^{2}}
	\right)^{\frac{1}{2}}\right] \, . 
\end{displaymath}

\section{Instanton calculus in the presence of \\ 
fermionic fields}
\label{instantons-fermions}
\fancyhead[LO]{{\footnotesize 4.2~~{\it Instantons and fermions}}}

In this section the semiclassical quantization in the background of 
an instanton configuration will be generalized to the case of a non 
Abelian gauge theory with interacting fermions. 

The Euclidean action of the model is 
\begin{equation}
	S[\psi,{\overline \psi},A] = S_{{\rm YM}}[A] + 
	S_{A}[\psi,{\overline\psi}] = S_{{\rm YM}}[A] +
	\int d^{4}x \, {\overline\psi}\left[ iD_{\mu}({\bf r}) \right]
	{\overline\sigma}^{\mu} \psi \, ,
	\label{actionfermions}
\end{equation}
where $S_{{\rm YM}}$ is the Yang--Mills action (\ref{actionym}) and 
$D_{\mu}({\bf r})$ denotes the covariant derivative in the 
representation ${\bf r}$ of the gauge group G (see appendix 
\ref{appa} for the notations). In (\ref{actionfermions}) $\psi$ is 
a Weyl fermion; an equivalent description can be given in terms of a 
Dirac spinor $\Psi$, introducing a fictitious right chirality Weyl 
spinor and writing the action as
\begin{displaymath}
	S_{A}[\Psi,{\overline\Psi}] = \int d^{4}x \, {\overline \Psi}
	\left[ i\Dsm_{L}({\bf r}) \right] \Psi \, ,
\end{displaymath}
where 
\begin{displaymath}
	\Dsm_{L}({\bf r}) = \left( \begin{array}{cc} 0 & 
	\sigma^{\mu}\partial_{\mu} \\ 
	{\overline\sigma}^{\mu}(\partial_{\mu} + A_{\mu})\rule{0pt}{16pt} 
	& 0 \end{array} \right) \, .
\end{displaymath}

\subsection{Semiclassical approximation}

In the full interacting theory one is interested in computing 
expectation values of fields ${\cal O}(x_{1},\ldots,x_{n})$ that are 
composite operators made up from the elementary fields $\psi$, 
${\overline\psi}$ and $A_{\mu}$. In the semiclassical approximation 
one considers fluctuations, $A_{\mu}={\overline A}_{\mu}+Q_{\mu}$, 
around a classical solution ${\overline A}_{\mu}$; for the fermionic 
part of the action it is sufficient to restrict oneself to 
$S_{{\overline A}}[\psi,{\overline\psi}]$. In the spirit of the 
semiclassical approximation, in the computation of expectation values 
of operators depending on $A_{\mu}$ as well as on fermionic fields, 
the functional integration over the latter is performed first and then 
the result is substituted into (\ref{finalsemiclass}). Thus as a 
starting point one considers 
\begin{displaymath}
	\langle{\cal O}[\Psi,{\overline\Psi}]\rangle_{A} = 
	\int[{\cal D}\Psi{\cal D}{\overline\Psi}] 
	e^{-S_{A}[\Psi,{\overline\Psi}]} {\cal O}[\Psi,{\overline\Psi}]  
\end{displaymath}
and evaluates this integral through a saddle point approximation.

According to the properties of Berezin integration (see appendix 
\ref{appa}) the last integral yields a non-vanishing result only if 
all the integrations over the fermionic variables are correctly 
saturated. Expanding $\Psi$ and ${\overline \Psi}$ into eigenfunctions 
of the Dirac operator
\begin{displaymath}
	i\Dsm_{L}({\bf r}) f^{(n)} = \lambda_{n} f^{(n)} \, ,
	\hspace{2cm} i\Dsm_{R}({\bf r}) g^{(n)} = \lambda^{*}_{n} g^{(n)} 
	\, ,
\end{displaymath}
gives
\begin{eqnarray}
	\Psi &=& \sum_{i=1}^{m} b_{i}^{(0)}f_{i}^{(0)} +\sum_{n}b_{n}f^{(n)} 
	\nonumber \\
	{\overline\Psi} &=& \sum_{j=1}^{{\overline m}}{\overline 
	c}_{j}^{(0)}g_{j}^{(0)\dagger} + \sum_{n}{\overline c}_{n}
	g^{(n)\dagger} \, ,
	\label{expansionpsibpsi}
\end{eqnarray}
where $b$ and ${\overline c}$ are Grassmannian c-numbers. 
Substituting into the action one gets
\begin{displaymath}
	S_{A}[\Psi,{\overline\Psi}] = \sum_{n\neq 0}\lambda_{n}{\overline 
	c}_{n}b_{n} \, ,
\end{displaymath}
so that one obtains 
\begin{equation}
	\langle{\cal O} \rangle_{A} = \int \prod_{i=1}^{m}[d b_{i}^{(0)}]
	\prod_{j=1}^{{\overline m}} [d {\overline c}_{j}^{(0)}] 
	\prod_{n\neq 0}[d b_{n} \, d {\overline c}_{n}] \,
	e^{-\sum_{n\neq 0}\lambda_{n}{\overline c}_{n}b_{n}} 
	{\cal O}(b,{\overline c}) \, .
	\label{expofermi}
\end{equation}
This expression shows that the operator ${\cal O}$ must contain 
exactly the number of zero-modes of the Dirac operator that are 
present in the instanton background in order to have a non-vanishing 
vev. 

Hence one must study the eigenvalue problem for the Dirac operator in 
the instanton background. The number, $n_{L}$, of left-chirality 
zero-modes of the Dirac operator (\ie the number of zero-modes of 
$i\Dsm_{L}$) and the number of right-chirality zero-modes (\ie the 
number of zero-modes of $i\Dsm_{R}$), $n_{R}$, are related by an index 
theorem \cite{atiyahsinger}
\begin{equation}
	n_{L}-n_{R} = 2 d_{{\bf r}} K[{\overline A}] \, .
	\label{indexth}
\end{equation}
Moreover for configurations with self-dual (anti self-dual) 
field strength one has $n_{R}$=0 ($n_{L}$=0).  Thus from (\ref{expofermi}) 
in the $K$-instanton sector it immediately follows that the operator 
${\cal O}$ must contain an excess of exactly $2Kd_{{\bf r}}$ factors 
of $\Psi$ with respect to ${\overline \Psi}$ to yield a non-zero result. 

For instance in the case of a theory containing two Weyl fermions, 
$\psi$ and $\chi$, in the fundamental representation of the gauge 
group SU(2) the relation 
\begin{equation}
	n_{L} = 2 d_{{\bf r}} K[{\overline A}] \, , \qquad n_{R}=0 \, 
	\hspace{1cm} {\rm when} \hspace{0.5cm} F_{\mu\nu}[A]=
	\tilde{F}_{\mu\nu}[A]
	\label{leftnumber} 
\end{equation}
gives one single zero-mode, $f^{(0)}$ for $\psi$ and $\chi$ in the 
$K$=1 sector, so that
\begin{eqnarray*}
	&& \psi = a_{0}f^{(0)} + \sum_{n\neq 0} a_{n}f^{(n)} \\
	&& \chi = b_{0}f^{(0)} + \sum_{n\neq 0} b_{n}f^{(n)} \, ,
\end{eqnarray*}
while ${\overline\psi}=\sum_{n\neq 0} {\overline c}_{n}g^{(n)\dagger}$ 
and ${\overline\chi}=\sum_{n\neq 0} {\overline d}_{n}g^{(n)\dagger}$ 
do not possess zero-modes.
An operator with non-vanishing vev is ${\cal O}(x,y) = 
\psi^{\alpha}(x)\chi_{\alpha}(y)$. The saddle point semiclassical 
approximation to the expectation value of ${\cal O}$ is 
\begin{eqnarray*}
	\langle {\cal O} \rangle_{A} &=& \int [da_{0}\,db_{0}] 
	\prod_{n\neq 0} [da_{n}\,db_{n}\,d{\overline c}_{n} \, 
	d{\overline d}_{n}] e^{-\sum_{n\neq 0}\lambda_{n}({\overline c}_{n} 
	a_{n} +{\overline d}_{n}b_{n})} \cdot \\ 
	&& \cdot \left(\sum_{n=0}^{\infty}a_{n}f^{(n)}\right)
	\left(\sum_{n=0}^{\infty}b_{n}f^{(n)}\right) = 
	\left[\det{}^{\prime}(i\Dsm_{L}({\bf 2})) \right]^{2} f^{(0)}(x)
	f^{(0)}(y) \, ,
\end{eqnarray*}
where $\det^{\prime}$ denotes the product of the non-zero eigenvalues.
The generalization of this elementary example to more complicated 
cases is rather straightforward. In the following sections computations 
of vev's of composite operators in ${\cal N}$=4 supersymmetric 
Yang--Mills theory will be described. The ${\cal N}$=4 theory 
contains four Weyl spinors in the adjoint representation. In the case 
of a SU(2) gauge group equation (\ref{leftnumber}) with $d_{\bf 3}$=2 
yields sixteen zero-modes in the one-instanton sector. 

\subsection{Fermionic zero-modes}

The discussion of the previous subsection shows that in order to 
compute the instanton contribution to expectation values of fermionic 
fields one needs to know the zero-value eigenstates of the Dirac 
operator. In particular their norms are necessary to convert the 
integrations over the zero-modes into integrations over (Grassmannian) 
collective coordinates as in the case of the bosonic zero-modes. 
For a Weyl spinor in the fundamental representation of SU(2) there 
is one single zero mode which satisfies the equation 
\begin{displaymath}
	iD_{\mu}^{({\rm f})}({\overline A}) 
	{\overline\sigma}^{\mu\,{\dot\alpha}\alpha} 
	\psi_{\alpha}^{(0)}=0 \, ,
\end{displaymath}
where $D_{\mu}^{({\rm f})}({\overline A})$ denotes the Dirac operator in 
the fundamental of SU(2) in the $K$=1 instanton sector. The solution 
is 
\begin{displaymath}
	\psi^{(0)}_{\alpha,s}(x) = \frac{\varepsilon_{\alpha s}}
	{[(x-x_{0})^{2}\rho^{2}]^{3/2}} \, ,
\end{displaymath}
where $s$=1,2 is an index of the fundamental of SU(2). The norm of 
this zero eigenfunction is
\begin{displaymath}
	\|\psi^{(0)}\| = \frac{\pi}{\rho} \, .
\end{displaymath}

In the case of the adjoint representation the equation for the 
zero-modes becomes
\begin{equation}
	\left[iD_{\mu}^{{\rm (adj)}}({\overline A})\right]_{a}{}^{b} 
	\, {\overline\sigma}^{\mu\,{\dot\alpha}\alpha} 
	\lambda_{\alpha b}^{(0)}=0 \, .
	\label{zeromodeadj}
\end{equation}
This equation is more complicated and according to the Atiyah--Singer 
theorem \cite{atiyahsinger} possesses four independent solutions in 
the case $K$=1 and G=SU(2). Instead of explicitly solving equation 
(\ref{zeromodeadj}) the zero eigenfunctions can be obtained acting 
with the symmetries of the model~\footnote{Note that this construction 
applies to non-supersymmetric cases as well.} on the configuration 
\begin{displaymath}
	\lambda={\overline\lambda}=0 \, , \qquad A_{\mu} = 
	{\overline A}_{\mu} \, ,
\end{displaymath}
which trivially satisfies (\ref{zeromodeadj}). More precisely the 
zero-modes are obtained acting by the symmetries broken by the 
instanton background. Two of the four zero-modes are generated by a 
supersymmetry transformation
\begin{displaymath}
	\delta_{1}\lambda = \frac{\sqrt{\rho}}{2} {\overline 
	F}_{\mu\nu}\sigma^{\mu\nu}\eta 
\end{displaymath}
and are associated with the supersymmetry parameter $\eta$. The 
remaining two eigenfunctions are obtained by a superconformal 
transformation with parameter ${\overline \xi}$
\begin{displaymath}
	\delta_{2}\lambda = \frac{1}{2\sqrt{\rho}} {\overline 
	F}_{\mu\nu}\sigma^{\mu\nu} \left[ (x-x_{0})_{\kappa}\sigma^{\kappa} 
	{\overline \xi} \right] \, .
\end{displaymath}
The above four zero-modes can be written in the compact form
\begin{equation}
	\lambda^{(0)}_{i} = \frac{1}{2} {\overline 
	F}_{\mu\nu}\sigma^{\mu\nu} \zeta_{i}(x) \, ,
	\label{compzeromodes}
\end{equation}
where 
\begin{equation}
	\zeta_{i}^{(0)}(x) = \frac{1}{\sqrt{\rho}} \left[ \rho \eta_{i} + 
	(x-x_{0})^{\mu}\sigma_{\mu} \xi_{i-2} \right] 
	\, , \hspace{1cm} i=1,2,3,4
	\label{zetazerodimsusy}
\end{equation}
and
\begin{displaymath}
	\eta_{1}={\overline\xi}_{1}=\left( \begin{array}{c} 1 \\ 0 
	\end{array} \right) \, , \hspace{1cm} 
	\eta_{2}={\overline\xi}_{2}= \left( \begin{array}{c} 0 \\ 1 
	\end{array} \right) \, , \hspace{1cm} 
	\eta_{3,4}={\overline\xi}_{-1,-2} = 0 \, .
\end{displaymath}
The powers of $\rho$ in (\ref{zetazerodimsusy}) are chosen in such a 
way as to give to $\zeta_{i}^{(0)}$ the correct dimension, 
$[{\rm mass}]^{-\frac{1}{2}}$, for a supersymmetry parameter. 
The zero-modes $\lambda_{i}^{(0)}$ are orthogonal and their norms are
\begin{displaymath}
	\|\lambda^{(0)}_{j}\| = \frac{4\sqrt{2}\pi\sqrt{\rho}}{g} 
	\, , \qquad j=1,2 \:,
	\qquad \|\lambda^{(0)}_{k}\| = \frac{8 \pi \sqrt{\rho}}{g}
	\, , \qquad k=3,4 \, .
\end{displaymath}
The extension of the analysis described here to the case of a SU($N$) 
gauge group, in the $K$=1 sector, can be obtained without too much 
effort and is achieved by the embedding of a SU(2) instanton in the 
SU($N$) field configuration, see \cite{arv}. The generalization to 
multi-instanton sectors on the contrary is quite involved and has 
been developed in \cite{adhm,corrfairgoddtamp,corrfairtamp,
criweinstan,osb}.

\section{Instanton calculus in supersymmetric \\ 
gauge theories}
\label{geninstcalculus}
\fancyhead[LO]{{\footnotesize 4.3~~{\it Instanton calculus in SYM 
theories}}}

The machinery of the previous sections allows to compute the 
instanton contribution to the vacuum expectation value of composite 
operators in supersymmetric gauge theories. The problem is reduced to 
the computation of the ``primed'' determinants of the kinetic operators, 
\ie the product of non-zero eigenvalues, and then to the integration 
over the collective coordinates with the measure constructed in  
sections \ref{yminstantons} and \ref{instantons-fermions}.

In non-supersymmetric theories the integration over the bosonic 
collective coordinates is IR-divergent because of the singular behaviour 
at $\rho\to 0$. In supersymmetric gauge theories, on the contrary, in 
all the known cases the integration turns out to be finite thanks to the 
balance of fermionic and bosonic degrees of freedom that controls 
the $\rho$-dependence. This exact balance produces another significant 
simplification: the product of non-zero eigenvalues of bosonic and 
fermionic kinetic operators exactly cancels out. As a result in 
various models instanton calculus allows to obtain finite and exact 
results (in the semiclassical approximation) for the vev's of 
composite operators, see \cite{akmrv} for a review.

\vspace{0.7cm}
{\sl Cancellation of determinants in supersymmetric theories.} 
\vspace{0.3cm}

\noindent
Supersymmetry implies relations among the eigenvalues of the kinetic 
operators for scalars, vectors and spinors, that lead to an exact 
compensation in the ratio of determinants that enters the expression 
of the expectation values in the semiclassical approximation. 

First notice that the chiral Dirac operator $i\Dsm_{L}({\bf r})$ 
satisfies 
\begin{displaymath}
	\left| \det \left( i\Dsm_{L}({\bf r}) \right) \right|^{2} = 
	\det\left[\left(i\Dsm_{L}({\bf r}) i\Dsm_{R}({\bf r}) \right) 
	\right] = \det\left(i\Dsm({\bf r})\right) \, \det(-\Box) \, ,
\end{displaymath} 
where $i\Dsm ({\bf r}) = i\gamma^{\mu}D_{\mu}({\bf r})$ is the kinetic 
operator for a Dirac spinor, so it is sufficient to solve the eigenvalue 
problem for $i\Dsm({\bf r})$. Let $\lambda_{n}$ be the non zero 
solutions of the eigenvalue problem for a scalar field in the instanton 
background 
\begin{displaymath}
	D^{2}({\overline A}) \varphi^{(n)} = -\lambda_{n}^{2} \varphi^{(n)} 
	\, \hspace{1.5cm} \lambda_{n}\neq 0 \, .
\end{displaymath}
The two eigenvalue problems to be solved are 
\begin{eqnarray*}
	&& i\Dsm({\overline A}) \Psi^{(n)} = \mu_{n}\Psi^{(n)} \\
	&& \left(-D^{2}({\overline A}) \delta_{\mu\nu} - 2F_{\mu\nu}
	[{\overline A}]\right)Q_{\nu}^{(n)}=\rho^{2}_{n} Q_{\mu}^{(n)} 
\end{eqnarray*}
and one can prove the following relations
\begin{displaymath}
	\mu_{n} = \lambda_{n} \qquad \longleftrightarrow \qquad 
	\Psi^{(n)}(x) = \left( \begin{array}{c} \xi^{(n)} \\ 
	{\overline\chi}^{(n)} \end{array} \right) \, ,
\end{displaymath}
with $\xi^{(n)}=\frac{i}{\lambda_{n}} D_{\mu}({\overline A}) 
\varphi^{(n)} \sigma^{\mu} {\overline \epsilon}$,~~${\overline
\chi}^{(n)} = {\overline \epsilon}\varphi^{(n)}$ and 
\begin{displaymath}
	\rho_{n}=\lambda_{n} \qquad \longleftrightarrow \qquad 
	Q^{(n)}_{\mu} = {\overline \eta} {\overline\sigma}_{\mu}\xi^{(n)} 
	\, ,
\end{displaymath}
where ${\overline\eta}$ and ${\overline\epsilon}$ are two constant 
Weyl spinors.

The ratio of determinants that enters the expression for the vev of an 
operator in the semiclassical approximation is 
\begin{displaymath}
	\left[ \frac{\det(M_{\mu\nu}^{{\rm 
	g.f.}}+\mu^{2})}{\det^{\prime}M_{\mu\nu}^{{\rm 
	g.f.}}}\right]^{\frac{1}{2}} \frac{\det D^{2}({\overline 
	A})}{\det(D^{2}({\overline A})+\mu^{2})} 
	\frac{\det^{\prime}\left(i\Dsm_{L}^{({\rm adj})}\right)}
	{\det \left(i\Dsm_{L}^{({\rm adj})}+i\mu \right)} \, ,
\end{displaymath}
where a regularization {\em \`a la} Pauli--Villars has been introduced.
Now substituting the eigenvalues found in the previous sections one 
can show that there is a an exact cancellation between bosons and 
fermions, so that the contribution of the non-zero modes is 1! 
The total contribution of the zero-modes is 
\begin{equation}
	\mu^{n_{B}-\frac{1}{2}n_{F}} \, ,
	\label{zeromodedetcontrib} 
\end{equation}
where $n_{B}$ and $n_{F}$ are respectively the number of bosonic and 
fermionic zero-modes. Notice that in the case of ${\cal N}$=4 super 
Yang--Mills theory (\ref{zeromodedetcontrib}) reduces to 1, as is 
expected, since the theory, being finite, cannot lead to a result 
depending on the renormalization scale. 

For the calculation of vev's of operators ${\cal O}$ there are two 
different situations to be considered. If the operator ${\cal O}$ 
contains precisely the number of zero-modes to saturate the fermionic 
integrations in the instanton background, one must simply substitute 
the fermionic fields in ${\cal O}$ with the zero-modes and perform the 
integration over the collective coordinates. This is the case that 
will be discussed in section \ref{inst-8-16-point}. If instead the 
operator ${\cal O}$ does not contain a sufficient number of fermionic 
insertions to saturate the Grassmannian integrals, then the correlator 
is zero at the lowest order and one must consider in $\langle {\cal 
O}\rangle$ the insertion of terms lowered from the action, until the 
correct number of spinors is obtained. This case will be described in 
section \ref{inst-4-point} for the correlation function of four 
scalar fields.

\section{Instanton calculations in ${\cal N}$=4 supersymmetric 
Yang--Mills theory}
\label{instantons-n=4}
\fancyhead[LO]{{\footnotesize 4.4~~{\it Instanton calculations in 
${\cal N}$=4 SYM theory}}}

The formalism described in the preceding sections will now be applied 
to the computation of Green functions of gauge invariant composite 
operators in ${\cal N}$=4 supersymmetric Yang--Mills theory. In 
particular correlation functions of operators belonging to the 
multiplet of currents, see equations (\ref{currentdef1}) and 
(\ref{currentdef2}), will be calculated to lowest order in the 
one-instanton sector and with gauge group SU(2) \cite{bgkr}. 
The extension of the results presented here to the case of SU($N$) and 
to the $K$-instanton sector in the large $N$ limit was given in 
\cite{mattis1} and \cite{mattis2,mattislong} and will be briefly 
reviewed in the concluding section. The operators in the current 
multiplet that will be considered play a central r\^ole in the 
correspondence with type IIB superstring theory compactified on 
AdS$_{5}\times S^{5}$. This issue will be discussed in chapter 5, 
where the instanton calculations will be compared with D-instanton 
effects in the type IIB superstring on AdS$_{5}\times S^{5}$. 

In  \cite{bgkr} a closed form for the instanton contribution to a 
four-point function of composite scalar operators, ${\cal Q}^{ij}$ in the 
{\bf 20} of the SU(4) R-symmetry group has been given. In the next section 
the complete space-time dependence of this Green function will be computed.
As will be discussed it will prove useful to employ for this 
calculation a description of the model in terms of ${\cal N}$=2 
multiplets. Also, it will be shown that when a limit of coincident 
points is considered, in order to study the operator product expansion 
(OPE), the function develops a logarithmic singularity. Analogous 
results have been found by various groups 
\cite{freed4point1,brodiegut,bk,liu}. In these papers a similar 
behaviour was extracted without explicitly performing all the 
integrations, on the contrary in the computation presented here 
the singular behaviour is shown with no ambiguities after a complete 
evaluation of the integrals which yield the Green function under 
consideration in the semiclassical approximation. A more detailed 
analysis of this subject and related questions will be presented 
elsewhere \cite{bkrs}. 

In section \ref{inst-8-16-point} the computation of correlation 
functions of sixteen fermionic currents $\hat{\Lambda}^{A}_{\alpha}$ 
and of eight gaugino bilinears ${\cal E}^{AB}$ will be reported. For 
these Green functions the integration over the collective coordinates 
will not be performed, nevertheless the unintegrated form presented 
here will be sufficient for the comparison with the results from type 
IIB superstring theory discussed in the final chapter.

\section{A correlation function of four scalar supercurrents}
\label{inst-4-point}
\fancyhead[LO]{{\footnotesize 4.5~~{\it Instanton contribution to 
a four-point function}}}

The multiplet of superconformal currents of ${\cal N}$=4 super 
Yang--Mills theory has been given in section \ref{symmandcurr} for 
the Abelian case. The natural four-point function of superconformal 
currents to consider would be a correlator of four stress energy 
tensors. However, due to its complicated 
tensorial structure even the free-field expression for this correlator
is awkward to express compactly and it is much simpler to consider 
 correlations of four gauge-invariant  composite scalar operators  
\begin{equation}
{\cal Q}^{ij}=\varphi^i\varphi^j-\frac{1}{6} \delta^{ij} 
\varphi_k\varphi^k \, , \hspace{1cm} i,j,k=1,2,\dots ,6 \, ,
\label{qdef}
\end{equation}
belonging to the representation {\bf 20} of the SU(4) R-symmetry group. 
As discussed in chapter \ref{cap2} ${\cal Q}^{ij}$ 
is the  lowest  component of the composite  twisted chiral current 
superfield \cite{howest} ${\cal W}_{(2)}^{ij}$ 
\begin{displaymath}
{\cal W}_{(2)}^{ij} = \tr\left( W^i W^j - 
\frac{\delta^{ij}}{6} W_k W^k \right) \, . 
\end{displaymath}
After calculating correlation functions of these components  
one can derive those of any other component in the ${\cal N}=4$ 
supercurrent multiplet by making use of the superconformal 
symmetry \cite{howest}. A way to explicitly do it may be to resort 
to analytic superspace, recalling that 
\begin{displaymath} 
{\cal Q}^{ij}(x)  \sim \left. 
{\cal W}_{(2)}^{ij}(\Upsilon) \right|_{\theta=\bar\theta=0}  \, , 
\label{npoint} 
\end{displaymath}
where $\Upsilon$ are the supercoordinates of analytic superspace  
\cite{extsuperspace}. Therefore, by computing correlators of ${\cal 
Q}^{ij}(x)$ and  substituting $x\rightarrow \Upsilon$ any of the 
other correlation functions can in principle be obtained by expanding 
in the fermionic as well as in the auxiliary bosonic coordinates of 
analytic superspace. 

The correlation function that will be considered is therefore 
\begin{equation}
	\langle {\cal Q}^{i_1j_1}(x_1)  {\cal Q}^{i_2j_2}(x_2)
	{\cal Q}^{i_3j_3}(x_3) {\cal Q}^{i_4j_4}(x_4)\rangle \, .
	\label{qfourpoint} 
\end{equation}
The value of this  correlation function in the  free field theory is
determined from the expression for the  free two-point scalar 
Green function which is
\begin{equation} 
\langle \varphi^{ia}(x) \varphi^{jb}(y)\rangle_{_{{\rm free}}} =  
\frac{1}{(2 \pi)^2} \frac{\delta^{ij}\delta^{ab}}{(x-y)^2} \, . 
\label{twopoint} 
\end{equation}   
Hence, the free-field expression for the correlation function  
that follows by  Wick contractions is   
\begin{eqnarray}
	&& \langle {\cal Q}^{i_1j_1}(x_1)  {\cal Q}^{i_2j_2}(x_2) 
	{\cal Q}^{i_3j_3}(x_3) {\cal Q}^{i_4j_4}(x_4)
	\rangle_{_{{\rm free}}}  \\  \rule{0pt}{26pt}
	&& \hspace{-0.1cm} = \frac{1}{(4 \pi^2)^{4}} \left[ {N}^{4} 
	\frac{\delta^{i_1 i_3}\delta^{j_1 j_3 }\delta^{i_2 i_4}
	\delta^{j_2 j_4}}{x_{13}^4 x_{24}^4} + N^{2} 
	\frac{\delta^{j_4 i_1}\delta^{j_1 i_3}\delta^{j_3 i_2}
	\delta^{j_2 i_4}}{x_{41}^2 x_{13}^2 x_{32}^2 x_{24}^2} + 
	{\rm permutations} \right] \nonumber \, , 
	\label{wick} 
\end{eqnarray}   
where, 
\begin{displaymath}
	x_{ij}=x_i-x_j \: .
\end{displaymath}

The first term in this expression is simply the product of two 
two-point functions and is known to be exact. The second term, 
which is the free-field contribution to the connected four-point 
function, certainly gets corrections from interactions. 

As has been discussed in detail in the preceding sections, in ${\cal
N}$=4 Yang--Mills theory there are sixteen zero-modes in the 
one-instanton sector in the case at hand of a SU(2) gauge group.
In the background of the one-instanton solution, (\ref{instanton}), 
(\ref{instfieldstrength}), the correlation function 
\begin{equation}
	\langle {\cal Q}^{i_1j_1}(x_1)  {\cal Q}^{i_2j_2}(x_2)
	{\cal Q}^{i_3j_3}(x_3) {\cal Q}^{i_4j_4}(x_4)\rangle_{_{K=1}} \, 
	\label{instq4point}
\end{equation}
where the subscript $K=1$ denotes the winding number of the 
background, is zero at the lowest order, because it does not contain any 
fermionic insertion to saturate the integrations over the Grassmannian 
collective coordinates of the sixteen zero-modes. 
However it is easy to check 
that the correlation function (\ref{instq4point}) soaks up these 
sixteen gaugino zero-modes when the perturbative corrections to the 
standard instanton configuration are considered. This can be seen 
either from the form of the supersymmetry transformations, see equation 
(\ref{transs}), or from the equation of motion for the scalar fields 
$\varphi^{i}$. The expression (\ref{instfieldstrength}) is annihilated 
by the conserved supersymmetry transformations (those associated with 
the parameter ${\overline \eta}^{\dot \alpha}_{\ A}$ in (\ref{transs})), 
while the transformations corresponding to supersymmetries 
associated with $\eta_{\alpha}^{A}$ are broken and generate 
eight of the sixteen fermionic zero modes. The other eight zero-modes 
are generated by superconformal transformations. As a consequence 
acting with the broken supersymmetry and superconformal 
transformations on the scalar fields in (\ref{instq4point}) produces 
a configuration which possesses the required sixteen zero-modes. 
To obtain the first non-vanishing contribution to (\ref{instq4point}) 
one must consider insertions of terms lowered from the action until 
the correct number of fermionic fields is obtained. The first non 
vanishing correction comes from the insertion of a Yukawa term for 
each field $\varphi^{i}$ (see the form (\ref{action}) of the action)
\begin{eqnarray}
	&& \hspace{-0.6cm} \langle {\cal Q}^{i_1j_1}(x_1)  
	{\cal Q}^{i_2j_2}(x_2){\cal Q}^{i_3j_3}(x_3){\cal Q}^{i_4j_4}(x_4) 
	\left( \int d^{4}z_{1} \left[\frac{i}{4} g f^{abc}
	{\overline t}^{i}_{AB}\left(\lambda^{\alpha A}_{a}
	\lambda^{B}_{\alpha b}\varphi_{i c}\right)(z_{1}) \right] \right)
	\cdot \nonumber \\
	&& \hspace{-0.6cm} \ldots \cdot \left( \int d^{4}z_{8} 
	\left[\frac{i}{4} g f^{abc}{\overline t}^{i}_{AB}
	\left(\lambda^{\alpha A}_{a}\lambda^{B}_{\alpha b}\varphi_{i c}
	\right)(z_{8}) \right] \right) \rangle_{K} \, .
	\label{q4insert}
\end{eqnarray}
Wick contractions among the scalar fields produce propagators, so that 
(\ref{q4insert}) becomes schematically 
\begin{equation}
	\langle \left( \int d^{4}z_{1} \left[ g\Delta(x_{1}-z_{1})
	\left(\lambda\lambda\right)(z_{1}) \right] \right) \ldots 
	\left( \int d^{4}z_{8} \left[ g\Delta(x_{4}-z_{8})
	\left(\lambda\lambda\right)(z_{8}) \right] \right) 
	\rangle ,
	\label{q4insertcontracted}
\end{equation}
where for simplicity of notation not all the indices have been indicated 
explicitly. Computing the integrations in (\ref{q4insertcontracted}) 
amounts to substitute each factor with the solution, 
$\varphi^{(0)}_{i}$, of an equation of the form
\begin{equation}
	\left[D^{2}({\overline A}) \varphi^{i}\right](x) = J^{i}(x)  \, ,
	\label{lowestphi} 
\end{equation}
with `source' $J^{i}(x) = {\overline t}^{i}_{AB}
\left(\lambda^{A}\lambda^{B}\right)(x)$. 

In order to evaluate this  instanton contribution  
in the most convenient fashion it will be convenient to use the 
${\cal N}$=2 supersymmetric description in which, as explained 
in section (\ref{formulations}), $\varphi^i$ decomposes into 
the complex ${\cal N}$=2  singlet $\varphi$ with U(1) charge $+2$ 
and a neutral ${\cal N}$=2 `quaternion' $q_{_T}$ in the  $({\bf 2,2})_0$
representation of SU(2)$_{\cal V}\times$ SU(2)$_{\cal H}$ which resides 
in the ${\cal N}$=2 hypermultiplet. The scalar component 
${{\cal Q}^{AB}}_{CD}$ of the ${\cal N}$=4 current decomposes in the 
following way in terms of ${\cal N}$=2 fields, 
\begin{eqnarray} 
	{\cal Q}^{S 0} &=& q^{S}\varphi\qquad {{\cal Q}^{S 0}}^{\dagger} 
	= q^{S} {\overline \varphi}  \qquad {\cal Q}^{S T} = 
	q^{S}q^{T} - {\rm trace} \nonumber \\ 
	{\cal Q}_{(+)} &=& \varphi^{2} \qquad\quad
	{\cal Q}_{(-)} = {\overline \varphi}^{2}\qquad\quad  
	{\cal Q}_{(0)} = {\overline \varphi} \varphi - \mbox{ trace}~~,  
\label{dqfields} 
\end{eqnarray} 
where the notations are those of section (\ref{formulations}) and 
$q_{u {\dot u}} = q_{_S} {\sigma^{^{S}}}_{u {\dot u}}$ 
and $\sigma^{^{S}}=(\I,\vec\sigma)$, with $\vec\sigma$ the standard Pauli 
matrices (and $S,T = 1,\dots,4$ are SU(2)$_{\cal V}\times$ 
SU(2)$_{\cal H}$ vector indices). 
The correlation function of two $\varphi^2$ and two 
${\overline{\varphi}}^2$ will be discussed here. From (\ref{wick}) 
the free-field expression for this particular correlator is 
\begin{eqnarray}
   && \langle\varphi^2(x_1){\varphi}^2(x_2){\overline {\varphi}}^2(x_3) 
   {\overline{\varphi}}^2(x_4)
   \rangle_{_{{\rm free}}} = \nonumber \\
   && = \frac{\rule{0pt}{16pt}1}{(4 \pi^2)^{4}} 
   \left(\frac{4 N^{4}}{x_{13}^4 x_{24}^4} 
   + \frac{4 N^{4}}{x_{14}^4 x_{23}^4} +  
   \frac{16 N^{2}}{x_{41}^2 x_{13}^2 x_{32}^2  x_{24}^2} \right) \, ,
   \label{corrpert}
\end{eqnarray}
where in the previous notation $\varphi^{2} = {\cal Q}_{(+)} \equiv 
{\cal Q}^{55} - {\cal Q}^{66} + 2i {\cal Q}^{56}$.
 
The ${\cal N}$=2 formalism is particularly suitable to perform the 
integrations over the fermionic zero-modes because  
it follows from the structure of the  Yukawa couplings (\ref{yukawa}) 
that $\varphi$ only absorbs those zero modes of the ${\cal N}$=4
gauginos that belong to the ${\cal N}$=2 vector multiplet 
($\lambda^u_\alpha$), while $\overline{\varphi}$ absorbs the zero modes 
belonging to the ${\cal N}$=2 hypermultiplet ($\psi^{\dot u}_\alpha$).  
The expressions for the zero-modes that will be used are those suggested 
by the supersymmetry transformations of the instanton 
\begin{equation}
\lambda_{(0) \alpha}^u =  \frac{1}{2} 
{\overline F}_{\mu\nu} \sigma_\alpha^{\mu\nu\, \beta} 
\frac{1}{\sqrt{\rho_0}} \left( \rho_0\eta_\beta^u + 
(x-x_0)_\kappa \sigma^\kappa_{\beta\dot \beta }
\bar\xi^{\dot \beta u} \right) \, .
\label{susyzero}
\end{equation}
and similarly for $\psi^{\dot u}_{(0)\alpha}$.
Here and in the following the `size' of the instanton will be denoted 
by $\rho_{0}$. The reason for this choice of notation will become 
clear when the results of this chapter will be reanalyzed in the 
context of the AdS/CFT correspondence in chapter {\ref{cap5}. 
In this decomposition $\eta_{\beta}$ and 
$x_{\beta{\dot \beta}}{\overline \xi}^{{\dot \beta}}$
are the parameters of the broken supersymmetry and special 
supersymmetry transformations respectively. One can assemble the 
fermionic collective coordinates into the spinors 
$\zeta^{u}_{\pm}(\rho_0,x)$ (where $\pm$ refers to the U(1)  
R-symmetry charge) to  parameterize the fermionic zero modes 
\begin{equation}
	\zeta^u_{\pm \alpha} (\rho_0,x-x_0) = \frac{1}{\sqrt{\rho_0}} 
	\left( \rho_0 \eta^u_{\pm \alpha} +  
	{\sigma^{\mu}}_{\alpha{\dot \alpha}} (x_{\mu} - x_{0\, \mu})  
	{\overline \xi}^{{\dot \alpha\, u}}_{\pm} \right) \, ,
	\label{zetpmdef} 
\end{equation}
Using the bosonic measure for SU(2) instantons derived in section 
\ref{yminstantons} and the norms of the fermionic zero-modes given 
in section \ref{instantons-fermions} and proceeding as described 
after equation (\ref{instq4point}) one finds that in the 
semi-classical approximation the one-instanton contribution to 
the four bosonic current Green function is~\footnote{Here and in the 
following computations of correlators of current bilinears in the
fundamental Yang--Mills fields a factor of $g^2$ is included 
for each external insertion. It would be equivalent to rescale the 
fundamental Yang--Mills fields according to $A \to A^{\prime}=g A$ 
and compute the Green functions of primed fields. This 
leads to a common overall dependence on $g^2$ for the three 
correlation functions that will be considered.}
\begin{eqnarray}
&& G_{{\cal Q}^4}(x_{p}) = \langle  g^2 \varphi^2(x_1) 
 g^2  \varphi^2(x_2) g^2 \overline{\varphi}^2(x_3) 
 g^2 \overline{\varphi}^2(x_4) \rangle_{_{K =1}} = 
 \frac{g^8}{2^{32}\pi^{10}}e^{-\frac{8\pi^2}{g^2} + i\theta} 
 \nonumber \\ \rule{0pt}{20pt}
 && \int \frac{d\rho_0 d^4x_0}{\rho_0^5} d^4\eta_+  
 d^4\overline{\xi}_+d^4\eta_-d^4\overline{\xi}_- \:  
 \varphi_{(0)}^2(x_1)\varphi_{(0)}^2(x_2) 
 \overline{\varphi}_{(0)}^2(x_3)\overline{\varphi}_{(0)}^2(x_4) \, , 
 \label{corrinst} 
\end {eqnarray} 
where $(+)$ (or $(-)$) refers to the U(1) charges of the gauginos 
in the vector (or hyper) multiplet (see below). 
In equation (\ref{corrinst}) the explicit dependence on the 
vacuum angle, $\theta$, has been indicated. The Green function
$G_{{\cal Q}^4}$ receives a contribution also from the
$K=-1$ sector that is the complex conjugate of (\ref{corrinst}).

In (\ref{corrinst})  the fields $\varphi$ and 
$\overline{\varphi}$ have been  replaced  by the expressions 
\begin{eqnarray} 
&& \varphi(x) \rightarrow \varphi_{(0)}(x) =  \frac{1}{2\sqrt{2}}
\varepsilon_{uv} \zeta^{u}_{+} \sigma^{\mu\nu} \zeta^{v}_{+}   
{\overline F}_{\mu\nu}  \nonumber \\ 
&& \overline{\varphi}(x) \rightarrow 
\overline{\varphi}_{(0)}(x) =  
\frac{1}{2\sqrt{2}} \varepsilon_{\dot{u}\dot{v}} \zeta^{\dot u}_{-} 
\sigma^{\mu\nu} \zeta^{\dot v}_{-}  
{\overline F}_{\mu\nu} \, , 
\label{scalarsol} 
\end{eqnarray}  
which satisfy an equation of the form of (\ref{lowestphi}) and are the 
leading nonvanishing terms that  result from  Wick contractions in 
which Yukawa couplings are lowered from the exponential of the action 
until a sufficient number of fermion fields are present to saturate 
the fermionic integrals. Of course, these expressions can also be 
obtained directly from the supersymmetry transformations (\ref{transs}) 
by acting twice on ${\overline F}_{\mu\nu}$ with the broken supersymmetry 
generators. After some elementary Fierz transformations on the fermionic  
collective coordinates the fermionic integrations can be performed in a 
standard manner and the result is 
\begin{equation}  
G_{{\cal Q}^4}(x_{p}) = \frac{3^{4}}{4\pi^{10}}\, g^8
\,e^{-\frac{8\pi^2}{g^2} + i\theta} \,   
\int \frac{d\rho_0 d^4 x_{0}}{\rho_0^5} \: 
x_{12}^4 x_{34}^4 \, \prod_{p=1}^4 {\left[ 
\frac{\rho_0}{{\rho_0^2 + (x_p - x_{0})^2}} \right]}^4 . 
\label{collectint} 
\end{equation}
The integration  in (\ref{collectint}) resembles that of a standard 
Feynman diagram with momenta replaced by position differences and can 
be performed by introducing the Feynman parameterization, 
\begin{eqnarray}  
G_{{\cal Q}^4}(x_p) &=& \frac{3^{4}\Gamma(16)}{4\pi^{10}\left(
\Gamma(4)\right)^{4}} \, g^8   \, 
e^{-\frac{8\pi^2}{g^2} + i\theta} \,   
\int \prod_{p} {\alpha}_p^3 d {\alpha}_p \: \delta 
\left( 1-\sum_q \alpha_q \right) \cdot \nonumber \\ 
&& \hspace{-0.5cm} \cdot \int \frac{d\rho_0 d^4 x_{0}}{\rho_0^5}   
\frac{x_{12}^4 x_{34}^4 \: \rho_0^{16}}{\left(\rho_0^2 + 
x_{0}^2 - 2 x_{0} \cdot \sum_p \alpha_p x_p + 
\sum_p x_p^2\right)^{16}} \, . 
\label{feynman} 
\end{eqnarray}
The five-dimensional integral yields 
\begin{eqnarray}  
G_{{\cal Q}^4}(x_p) &=& \frac{3^{3} \, \Gamma(11)}
{2^{7} \left( \pi^3 \Gamma(4) \right)^{4}} \:
 g^{8} e^{-\frac{8\pi^2}{g^2} + i\theta} \cdot  
\nonumber \\
&& \hspace{-0.7cm} \cdot \int \prod_p \alpha_p^3 d\alpha_p \, 
\delta \left(1-{\sum}_q  \alpha_q \right) 
\frac{x_{12}^4 x_{34}^4}{\left(\sum_p \alpha_p \alpha_q 
x_{pq}^2 \right)^{8}} \; . 
\label{integral} 
\end{eqnarray} 
This integral can be simplified by observing that it is essentially  
obtained by acting with derivatives on the box-integral with four  
massless external particles, 
\begin{equation}  
G_{{\cal Q}^4}(x_p) =\frac{3^{3} \, \Gamma(11)}
{2^{7}\left(\pi^3 \Gamma(4) \right)^{4}} \:
 g^{8} e^{-\frac{8\pi^2}{g^2} + i\theta}  \,
 x_{12}^4 x_{34}^4 \prod_{p<q} 
 \frac{\partial}{\partial x_{pq}^2} B(x_{pq}) \, , 
 \label{intbox} 
\end{equation}
where the box integral is 
\begin{equation} 
  B(x_{pq}) = \int \prod_{p} d\alpha_p \: 
  \delta \left( 1-\sum_q \alpha_q \right) \frac{1}{\left(\sum_p 
  \alpha_p \alpha_q x_{pq}^2 \right)^{2}} \: . 
  \label{boxint} 
\end{equation} 
 
The result, correcting a sign error in \cite{bgkr}, may be expressed 
as a combination of logarithms and dilogarithms \cite{bern} 
\begin{eqnarray}  
 B(x_{pq}) &=& \frac{4}{\sqrt{\Delta(x_{pq})}} \left[ 
 \frac{1}{2} \log\left (\frac{u_+u_-}{(1-u_+)^2(1-u_-)^2}\right) 
 \log\left(\frac{u_+}{u_-} \right) \right. + \nonumber  \\  
 && \hspace{-2.3cm} - \left. {\rm Li}_2(1-u_+) + {\rm Li}_2 (1-u_-)  
 -{\rm Li}_2\left( 1-\frac{1}{u_-} \right) + 
 {\rm Li}_2\left( 1-\frac{1}{u_+} \right) \right] \, ,
 \label{boxeplicit} 
\end{eqnarray}   
where 
\begin{equation} 
\Delta = \det_{4\times 4}((x^2_{pq})) =    
X^2 + Y^2 + Z^2 - 2 XY - 2 YZ - 2 ZX 
\label{det} 
\end{equation}   
and 
\begin{equation}
u_{\pm} = \frac{Y + X - Z \pm \sqrt{\Delta}}{2 Y} \: , 
\label{ratio} 
\end{equation}  
with $X = x_{12}^2 x_{34}^2$, $Y = x_{13}^2 x_{24}^2$ and  
$Z=x_{14}^2 x_{23}^2$. 
 
Notice that up to an overall dimensional factor needed for the correct
scaling, $G_{{\cal Q}^4}$ turns out to be a function of the two
independent superconformally invariant cross ratios  $X/Z$ and $Y/Z$.
 Although not immediately apparent, the expression $B(x_{pq})$
is   symmetric under any permutation of the external legs, 
as can be seen by making use of the properties  of the dilogarithms, 
\begin{eqnarray}
{\rm Li}_2(z) + {\rm Li}_2 (1-z) &=& \frac{\pi^2}{6} - \log(z) 
\log(1-z) \nonumber \\   
{\rm Li}_2(z) + {\rm Li}_2 \left(\frac{1}{z} \right) &=& 
-\frac{\pi^2}{6} - \frac{1}{2}[\log(-z)]^2 \: , 
\label{dilog} 
\end{eqnarray}
and observing that the relevant permutations correspond to  
permutations of $X= Y u_+u_-$, $Y$   and $Z= Y(1-u_+)(1-u_-)$, 
that are generated by the two 
transformations: a) $u_+\rightarrow 1/u_-$, $u_-\rightarrow 1/u_+$,  
$Y\rightarrow Y u_- u_+$, which is equivalent to the exchange of $X$ 
and $Y$, leaving $Z$ fixed  (\ie to the exchange of $x_1$ and $x_4$ 
or, equivalently, of $x_2$ and $x_3$) and  b) 
$u_+\rightarrow 1-u_-$, $u_-\rightarrow 1-u_+$ 
at fixed $Y$, which is  equivalent to the exchange of $X$ and 
$Z$ (or the exchange of $x_2$ and $x_4$ or, equivalently, of 
$x_1$ and $x_3$). 

To compute the explicit dependence of $G_{{\cal Q}^{4}}$ on $x_{p}$ one 
must perform the six derivatives with respect to $x_{pq}$. These can be 
more conveniently calculated rewriting the differential operator in 
(\ref{intbox}) as 
\begin{equation}
	\prod_{p<q} \frac{\partial}{\partial x_{pq}^2} = {\cal D}_{Z}
	{\cal D}_{Y} {\cal D}_{X} \label{diffop} \, ,
\end{equation}
where
\begin{displaymath}
	{\cal D}_{T} = \left(\frac{\partial}{\partial T}+
	T \frac{\partial^{2}}{\partial T^{2}} \right) \, , 
	\qquad T = X,Y,Z \, .
\end{displaymath}
Defining 
\begin{displaymath}
	Q(x_{pq}) = \prod_{p<q} \frac{\partial}{\partial x_{pq}^2}
	B(x_{pq}) \, 
\end{displaymath}
after a lengthy computation one obtains 
\begin{eqnarray*}
    && \hspace*{-0.75cm} Q(x_{pq}) =
    \frac{2}{{{\Delta }^5}} \bigg(193 {X^6}-114 {X^5} Y-1281 {X^4} {Y^2}+
    2404 {X^3} {Y^3}- 1281 {X^2} {Y^4} + \\
    && \hspace*{-0.75cm} 
    -114 X {Y^5}+193 {Y^6}-114 {X^5} Z + 4734 {X^4} Y Z-
    4620 {X^3} {Y^2} Z-4620 {X^2} {Y^3} Z + \\
    && \hspace*{-0.75cm} 
    +4734 X {Y^4} Z-114 {Y^5} Z-1281 {X^4} {Z^2}- 4620 {X^3} Y {Z^2}+
    15402 {X^2} {Y^2} {Z^2}+ \\
    && \hspace*{-0.75cm} 
    -4620 X {Y^3} {Z^2}-1281 {Y^4} {Z^2}+ 2404 {X^3} {Z^3}-
    4620 {X^2} Y {Z^3}-4620 X {Y^2} {Z^3} + \\
    && \hspace*{-0.75cm} 
    +2404 {Y^3} {Z^3}-1281 {X^2} {Z^4}+4734 X Y {Z^4}-1281 {Y^2} {Z^4}-
    114 X {Z^5}-114 Y {Z^5} + \\
    && \hspace*{-0.75cm} 
    +193 {Z^6}\bigg) + \frac{6}{{{\Delta }^6}} \bigg\{
    \bigg( -33 {X^8}-96 {X^7} Y+714 {X^6} {Y^2}- 
    1008 {X^5} {Y^3}+1008 {X^3} {Y^5} + \\
    && \hspace*{-0.75cm}
    -714 {X^2} {Y^6}+96 X {Y^7}+33 {Y^8}-6 {X^7} Z- 1770 {X^6} Y Z+
    198 {X^5} {Y^2} Z+8298 {X^4} {Y^3} Z + \\
    && \hspace*{-0.75cm} 
    - 8298 {X^3} {Y^4} Z - 198 {X^2} {Y^5} Z +
    1770 X {Y^6} Z+6 {Y^7} Z+435 {X^6} {Z^2}+
    3312 {X^5} Y {Z^2} + \\
    && \hspace*{-0.75cm} 
    - 13473 {X^4} {Y^2} {Z^2} +13473 {X^2} {Y^4} {Z^2}
    -3312 X {Y^5} {Z^2}-435 {Y^6} {Z^2}- 
    930 {X^5} {Z^3} + \\
    && \hspace*{-0.75cm} 
    + 1914 {X^4} Y {Z^3}+15396 {X^3} {Y^2} {Z^3}-
    15396 {X^2} {Y^3} {Z^3} - 1914 X {Y^4} {Z^3} + 930 {Y^5} {Z^3} + \\
    && \hspace*{-0.75cm} 
    + 645 {X^4} {Z^4}-
    6384 {X^3} Y {Z^4}+6384 X {Y^3} {Z^4}- 645{Y^4}{Z^4}+
    78{X^3}{Z^5} +3114 {X^2} Y {Z^5} + \\
    && \hspace*{-0.75cm} 
    - 3114 X {Y^2} {Z^5}-78 {Y^3} {Z^5}-  
    279 {X^2} {Z^6}+279 {Y^2} {Z^6}+90 X {Z^7}-90 Y {Z^7}\bigg) 
    \log 
    \bigg[\begin{displaystyle}\frac{X}{Y}\end{displaystyle}\bigg] + 
\end{eqnarray*}
\begin{eqnarray*}
    && \hspace*{-0.75cm} 
    -\bigg(11 {X^8}-28 {X^7} Y-52 {X^6} {Y^2}+284 {X^5} {Y^3}-
    430 {X^4} {Y^4}+284 {X^3} {Y^5} - 52 {X^2} {Y^6} + \\
    && \hspace*{-0.75cm} 
    -28 X {Y^7} +11 {Y^8}+62 {X^7} Z+ 590 {X^6} Y Z-2142 {X^5} {Y^2} Z+ 
    1490 {X^4} {Y^3} Z + \\
    && \hspace*{-0.75cm} 
    +1490 {X^3} {Y^4} Z-2142 {X^2} {Y^5} Z+ 590 X {Y^6} Z+
    62 {Y^7} Z- 331 {X^6} {Z^2}+972 {X^5} Y {Z^2} +  \\
    && \hspace*{-0.75cm} 
    + 4491 {X^4} {Y^2} {Z^2}-10264 {X^3} {Y^3} {Z^2}+ 
    4491 {X^2} {Y^4} {Z^2}+972 X {Y^5} {Z^2}-331 {Y^6} {Z^2} + \\
    && \hspace*{-0.75cm} 
    +362 {X^5} {Z^3}-4894 {X^4} Y {Z^3}+ 5132 {X^3} {Y^2} {Z^3}+ 
    5132 {X^2} {Y^3} {Z^3} -  4894 X {Y^4} {Z^3} + \\ 
    && \hspace*{-0.75cm} 
    +362 {Y^5} {Z^3}+  215 {X^4} {Z^4}+3404 {X^3} Y {Z^4}
    - 8982 {X^2} {Y^2} {Z^4}+ 3404 X {Y^3} {Z^4} + \\
    && \hspace*{-0.75cm} 
    + 215 {Y^4} {Z^4}-646 {X^3} {Z^5}+1170 {X^2} Y {Z^5}+
    1170 X {Y^2} {Z^5}-646 {Y^3} {Z^5}+ 383 {X^2} {Z^6} + \\
    && \hspace*{-0.75cm} 
    -1180 X Y {Z^6}+383 {Y^2} {Z^6}-
    34 X {Z^7}-34 Y {Z^7}-22 {Z^8}\bigg) \log \bigg[\begin{displaystyle}
    \frac{X Y}{Z^{2}} \end{displaystyle}\bigg] + \\
    && \hspace*{-0.75cm} 
    + \frac{36}{{{\Delta }^{13/2}}}\bigg\{ \bigg( {X^9}+
    3 {X^8} (Y+Z) - 6 {X^7} \big(5 {Y^2}-19 Y Z+5 {Z^2}\big)+{{(Y-Z)}^6} 
    \big({Y^3}+9 {Y^2} Z +  \\
    && \hspace*{-0.75cm} 
    +9 Y {Z^2}+{Z^3}\big)+{X^6} \big(62 {Y^3}-144 {Y^2} Z-
    144 Y {Z^2}+62 {Z^3}\big) + 3 X {{(Y-Z)}^4} \cdot  \\
    && \hspace*{-0.75cm} 
    \cdot \big({Y^4}+42 {Y^3} Z+
    114 {Y^2} {Z^2}+ 42 Y {Z^3}+{Z^4}\big) 
    - 6 {X^5} \big(6 {Y^4}+83 {Y^3} Z-252 {Y^2} {Z^2}+ \\
    && \hspace*{-0.75cm} 
    + 83 Y {Z^3}+6 {Z^4}\big)-6 {X^2} {{(Y-Z)}^2} \big(5 {Y^5}+34 {Y^4} Z-
    189 {Y^3} {Z^2}-189 {Y^2} {Z^3}+34 Y {Z^4}+  \\
    && \hspace*{-0.75cm} 
    +5 {Z^5}\big) - 6 {X^4} \big(6 {Y^5}-175 {Y^4} Z+
    223 {Y^3} {Z^2}+ 223 {Y^2} {Z^3}-175 Y {Z^4}+6 {Z^5}\big)+ \\ 
    && \hspace*{-0.75cm} 
    + {X^3} \big(62 {Y^6}-498 {Y^5} Z-1338 {Y^4} {Z^2}+  
    3948 {Y^3} {Z^3}-1338 {Y^2} {Z^4}-498 Y {Z^5}+ 62 {Z^6}\big)\bigg) 
    \cdot  \\
    && \hspace*{-0.75cm} 
    \cdot \bigg[ \frac{1}{2} \log \bigg(\frac{X Y}{{Z^2}}\bigg) 
    \log \bigg(1+\frac{(X+Y-Z) {\sqrt{\Delta }}+\Delta }{2 X Y}\bigg) -
    \rm{Li}_{2}\bigg(\frac{X-Y+Z-{\sqrt{\Delta }}}{2 X}\bigg) + \\
    && \hspace*{-0.75cm} 
    -\rm{Li}_{2}\bigg(\frac{-X+Y+Z-{\sqrt{\Delta }}}{2 Y}\bigg) +  
    \rm{Li}_{2}\bigg(\frac{X-Y+Z+{\sqrt{\Delta }}}
    {2 X}\bigg) + \rm{Li}_{2}\bigg(\frac{-X+Y+Z+{\sqrt{\Delta }}}
    {2 Y}\bigg)\bigg]\bigg\} \, .
\end{eqnarray*}
In conclusion the one-instanton contribution to the correlation 
function $G_{{\cal Q}^{4}}(x_{p})$ is 
\begin{equation}
	G_{{\cal Q}^4}(x_p) =\frac{3^{3} \, \Gamma(11)}
    {2^{5}\left(\pi^3 \Gamma(4) \right)^{4}} \:
    g^{8} e^{-\frac{8\pi^2}{g^2} + i\theta}  \,
    x_{12}^4 x_{34}^4 \, Q(x_{pq}) \, .
	\label{final-4-point}
\end{equation}

Unlike  correlation functions of elementary 
fields that are  infra-red problematic and gauge-dependent, the above 
correlator is well defined at non-coincident points. Moreover it 
possesses the correct symmetry properties. This can be derived from 
the symmetry of $B(x_{pq})$ discussed above and the observation that 
the differential operator (\ref{diffop}) is completely symmetric. 
Of course the symmetry properties can be checked directly on the final 
expression for $Q(x_{pq})$. Notice that the 
above form for $Q(x_{pq})$ is valid only for $X,Y,Z$ in the `physical 
domain', \ie in the region obtainable for allowed choices of 
$x_{pq}^{2}$, that is defined by the condition 
\begin{displaymath}
	\Delta = \Delta(X,Y,Z) = X^{2}+Y^{2}+Z^{2}-2XY-2XZ-2YZ \leq 0 \, . 
\end{displaymath} 
In this region single terms in the above expression are complex, but 
the complete function is real. To describe the physical region it is 
convenient to rewrite $Q(x_{pq})$ in terms of two independent cross 
ratios, 
\begin{displaymath}
	r = \frac{X}{Y} = \frac{x_{12}^{2}x_{34}^{2}}{x_{13}^{2}x_{24}^{2}}
	\, , \hspace{1.5cm} 
	s = \frac{Z}{Y} = \frac{x_{14}^{2}x_{23}^{2}}{x_{13}^{2}x_{24}^{2}}
	\, ,
\end{displaymath}
extracting a factor $\frac{1}{Y^{4}}$ in $Q(x_{pq})$. Then the 
physical domain corresponds to the region 
\begin{displaymath}
	\delta(r,s) = \frac{\Delta}{Y^{2}} = 1+r^{2}+s^{2}-2r-2s-2rs \leq 0
\end{displaymath}
in the $(r,s)$ plane. This is the region inside the parabola in 
figure \ref{physicaldom}
\begin{figure}[!h]
	\centering
	\includegraphics[width=10truecm]{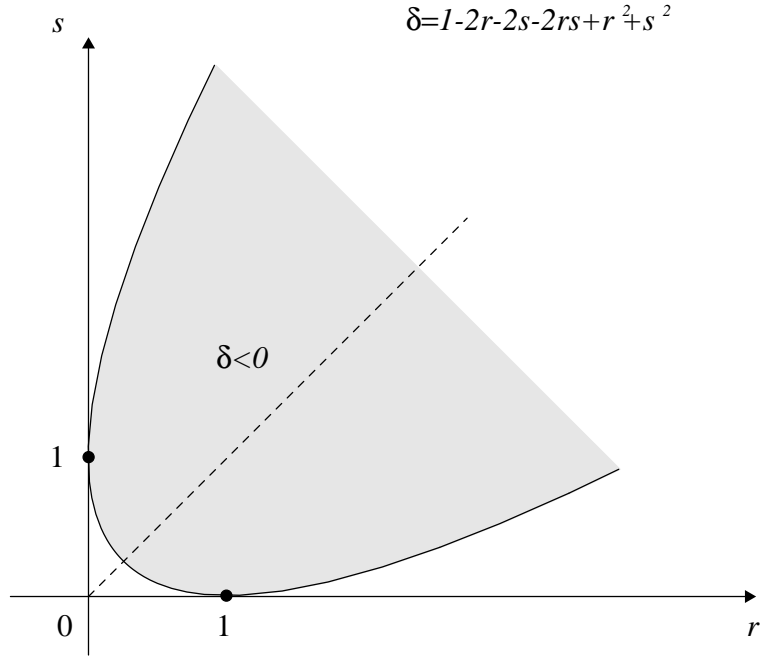}
	\caption{Physical domain for $B(x_{pq})$}
	\label{physicaldom}
\end{figure}

Having obtained a closed form for $G_{{\cal Q}^{4}}(x_{p})$ 
one can study the behaviour when any two points are taken close 
to one another, $x_{pq}\to 0$. For instance 
$x_{12}\to 0$ corresponds to $r\to 0$, $s\to 1$ simultaneously. 
In this limit $B(x_{pq})$ develops a logarithmic singularity. 

The logarithmic divergence in the Green function ${\cal G}_{{\cal 
Q}^{4}}$ can be shown more clearly considering the particular 
configuration in which the points $x_{p}$, $p=1,2,3,4$, are taken on a 
line. In this case the cross ratios $r$ and $s$ are not independent. 
Assuming $x_{1}<x_{2}<x_{3}<x_{4}$ $r$ and $s$ are related by 
\begin{displaymath}
	\sqrt{r}+\sqrt{s} = 1 \, .
\end{displaymath}
Thus one can express the result in terms of a single variable $\eta$ 
defined by 
\begin{equation}
	r = \eta^{2} \, , \hspace{1cm} s = \left(1-\eta\right)^{2} \, .
	\label{etavar}
\end{equation}
In this limit the above computed function $Q(x_{pq})$ reduces 
to~\footnote{This expression has been obtained by Yassen Stanev 
starting from a different but equivalent formula for $Q(x_{pq})$.}
\begin{eqnarray}
    && \tilde{Q}=\tilde{Q}(\eta) = 3\,{\displaystyle 
    \frac{\log(\eta ^{2})}{( - 1 + \eta )^{7}}} (100\,
    \eta ^{6} + 429 + 2431\,\eta ^{2} - 2717\,\eta ^{3} + 1794\,
    \eta ^{4} +  \nonumber \\ 
    && - 650\,\eta ^{5} - 1287\,\eta )\, - 
    3\,{\displaystyle \frac{\log[( - 1 + 
    \eta )^{2}]}{\eta ^{7}}} \,(100\,\eta ^{6} + 50\,\eta ^{5} 
    + 44\,\eta ^{4} + 41\,\eta^{3} + \nonumber \\ 
    &&  + 44\,\eta ^{2} + 50\,\eta  + 
    100)  - 2\,{\displaystyle \frac{(\eta ^{2} - \eta  + 1)^{2}}
    {\eta ^{6}\,( - 1 + \eta )^{6}}}(300 - 900\,\eta  + 307\,
	\eta ^{2} + 886\,\eta ^{3} + \nonumber \\ 
    && + 307\,\eta ^{4} - 900\,\eta ^{5} + 300\,\eta ^{6}) \, .
    \label{linelim}
\end{eqnarray}
and it can be shown that $\tilde{Q}(\eta)$ has a logarithmic 
singularity as $\eta \to 0$ or $\eta \to 1$.

This peculiar behaviour has been observed by various 
authors in four-point functions computed in the context of the 
AdS/CFT correspondence \cite{freed4point1,brodiegut,bk,liu}, 
but here it is derived from an exact field theoretical computation 
and thus shown with no ambiguities. Notice that single terms in 
$Q(x_{pq})$ have pole-type singularities in the limit $x_{12}\to 0$, 
but in the sum the singularity is only logarithmic.
Note in particular that the poles associated with operators of 
dimension lower or equal to ${\cal Q}^{2}$ in the OPE are absent. 
Moreover with non-vanishing $Y$ the only singularities in $B(x_{pq})$ 
correspond to the points $(r=0,s=1)$ and $(r=1,s=0)$ marked in 
figure \ref{physicaldom}. Taking into account the pre-factor, 
$x_{12}^{4}x_{34}^{4}$, in the complete expression of $G_{{\cal 
Q}^{4}}$ one finds that the limit in which two $\varphi^{2}$ 
operators are taken to coincide gives a vanishing result, 
while a logarithmic singularity is  actually present when 
$\varphi^{2}$ approaches $\overline{\varphi}^{2}$.

The logarithms by themselves do not 
violate the superconformal symmetry of the theory, that is manifest in 
the final expression of the four-point function. In some 
respect, logarithmic behaviours in four-point functions should not sound 
unexpected in a superconformal theory, such as ${\cal N}$=4 
supersymmetric Yang--Mills theory, 
that contains a large number of primary fields all of whose 
(protected) dimensions are integer. The singularity may be due to the 
presence of an infinite tower of stable BPS dyons that collapse to 
vanishing mass and size in the superconformal phase or to a gas of 
instantons of small size from the Euclidean viewpoint. This problem 
is under active investigation \cite{bkrs}.

\section{Eight- and sixteen-point correlation functions of current 
bilinears}
\label{inst-8-16-point}

In this section the one-instanton contribution to correlation 
functions of sixteen fermionic bilinears $\hat{\Lambda}^{A}_{\alpha}$ 
and of eight gaugino bilinears ${\cal E}^{AB}$ in the current 
supermultiplet will be calculated in the semiclassical approximation.
The expressions obtained will be reconsidered in the next chapter and a 
comparison will be made with D-instanton effects in type IIB 
superstring theory compactified on AdS$_{5}\times S^{5}$.

\subsection{The correlation function of sixteen fermionic currents}
\fancyhead[LO]{{\footnotesize 4.6~~{\it Instanton contribution to 
eight- and sixteen-point functions}}}

As could have been anticipated, it is particularly simple to analyze the 
contribution of the Yang--Mills instanton to the correlation function 
of sixteen of the fermionic superconformal current bilinears, 
${\hat \Lambda}_\alpha^{A} = \tr\left(\sigma^{\mu \nu}
{}_{\alpha}{}^{\beta} F^{-}_{\mu \nu}  {\lambda_{\beta}}^{A}\right)$, 
\begin{equation} 
	G_{\hat \Lambda^{16}}(x_p) = 	\langle \prod_{p=1}^{16} g^2 
	\hat \Lambda^{A_{p}}_{\alpha_{p}}(x_{p})  
	\rangle_{_{K=1}} \; , 
	\label{l16}
\end{equation} 
Since each factor of $\hat\Lambda$ in the product 
provides a single fermion zero mode it is necessary to consider the 
product of sixteen currents  in order  to saturate the sixteen  
Grassmannian integrals. To leading order in $g$, 
$G_{\hat \Lambda^{16}}$ does not receive contribution 
from anti-instantons.
The leading term in the one-instanton sector is simply obtained 
by replacing  each $F^{-}_{\mu\nu}$ with the instanton 
profile  ${\overline F}_{\mu \nu}$ (equation 
(\ref{instfieldstrength})) and each  $\lambda_\alpha^{\ A}$ 
with the corresponding zero mode, $\lambda_{(0)\alpha}^{A}$
\begin{equation}
	\lambda_{(0) \alpha}^A =  \frac{1}{2} 
	{\overline F}_{\mu\nu} \sigma_\alpha^{\mu\nu\,\beta} 
	\frac{1}{\sqrt{\rho_0}} \left( \rho_0\eta_\beta^A + 
	(x-x_0)_\mu \sigma^\mu_{\beta{\dot\beta}}
	{\overline \xi}^{\dot \beta A} \right) \, .
	\label{lamzero}
\end{equation}
The resulting correlation function thus has the form 
\begin{eqnarray} 
	&& \hspace*{-1cm} G_{\hat \Lambda^{16}}(x_p) 
    = \frac{2^{62}3^{16}}{\pi^{10}}\, 
    g^{8} e^{-\frac{8\pi^2}{g^2} + i\theta} 
    \int \frac{d^{4}x_{0} \, d\rho_0}{\rho_0^{5}}  
	\int d^{8}\eta d^{8}{\overline \xi} \cdot \nonumber \\  
	&& \hspace*{-1cm} \cdot \prod_{p=1}^{16} \left[ 
	\frac{\rho_{0}^{4}}{[\rho_0^{2}+(x_{p}-x_{0})^{2}]^{4}} \,
	\frac{1}{\sqrt{\rho_{0}}}
	\left( \rho_0 \eta^{A_{p}}_{\alpha_{p}}+
	{(x_{p}-x_{0})}_{\mu}  \sigma^{\mu}_{\alpha_{p}{\dot \alpha}_{p}}  
	{\overline \xi}^{{\dot \alpha}_{p}A_{p}} \right)  \right] \, . 
	\label{l16-sym} 
\end{eqnarray} 
The integration over the fermion zero modes leads to a 
sixteen-index invariant  tensor, $t_{16}$, of the product of the  
SU(4) and Lorentz groups. Assembling the 16 fermionic collective
coordinates into a sixteen-dimensional spinor $t_{16}$ would simply read
$t_{16}^{a_1a_2\dots a_{16}} = \varepsilon^{a_1a_2\dots a_{16}}$, with
$a_i = 1,2 \dots 16$. 
Further integration over the instanton moduli space would determine 
the dependence on the coordinates $x_{p}$. However, for the purpose 
of the comparison with the corresponding expression obtained in the 
type IIB string theory in AdS$_{5}\times S^{5}$, that will be 
discussed in the next chapter, it is sufficient to leave the expression 
in the unintegrated form (\ref{l16-sym}).

\subsection{The correlation function of eight gaugino bilinears}

In a similar fashion it is easy to deduce the one-instanton 
contribution to other related processes, such as the eight-point 
correlation function, 
\begin{equation}
	G_{{\cal E}^8}(x_p) = \langle g^{2} {\cal E}^{A_1B_1}(x_1) \dots 
	g^{2}{\cal E}^{A_8B_8}(x_8) \rangle_{_{K=1}} \, ,
	\label{eeightpoint} 
\end{equation}   
which also saturates the sixteen fermionic zero-modes present 
in the SU(2) one-instanton background. To leading order in 
$g$, $G_{{\cal E}^8}$ does not receive contribution from anti-instantons.
The complete non-abelian expression for ${\cal E}^{AB}$ was given in 
equation (\ref{enonabelian}) and reads 
\begin{equation}
	{\cal E}^{AB} = \lambda^{\alpha a A} {\lambda_{\alpha a}}^{B} + 
	g f_{abc} \, t_{ijk}^{(AB)_+} \phi^{ia}
	\phi^{jb}\phi^{kc} \, , 
	\label{exacte} 
\end{equation}
but  at leading order in the gauge coupling constant only the term  
proportional to the gaugino bilinear is relevant.  
In the instanton background the gaugino bilinear is given by 
\begin{equation}
	\lambda_{(0)}^{\alpha a A} {\lambda_{(0)\alpha a}}^{B} =  
	\frac{3\cdot 2^6}{g^{2}} 
	\frac{\rho_0^{4}}{\left( \rho_0^{2}+(x-x_{0})^{2}\right)^{4}}
	\: \zeta^{\alpha A} {\zeta_{\alpha}}^{B} \, . 
	\label{bilin} 
\end{equation}
Then it follows 
\begin{eqnarray}
	\hspace{-0.5cm} && G_{{\cal E}^8}(x_p)
	= \frac{3^8 2^{14}}{\pi^{10}} \, 
	g^{8} \, e^{-\frac{8\pi^2}{g^2} + i\theta} 
	\int \frac{d^{4}x_{0} \, d\rho_0}{\rho_0^{5}}\int d^{8}\eta
	d^{8}{\overline \xi} \cdot \nonumber \\ 
	&& \cdot \prod_{p=1}^{8} \left[ 
	\frac{\rho_0^{4}}{\left( \rho_0^{2}+(x_{p}-x_{0})^{2}\right)^{4}}
	\frac{1}{\sqrt{\rho_0}}
	\left( \rho_0 \eta^{A_{p}}_{\alpha_{p}}+{(x_{p}-x_{0})}_{\mu}  
	\sigma^{\mu}_{\alpha_{p}{\dot \alpha}_{p}}  
	{\overline \xi}^{A_p {\dot \alpha}_p}\right) 
	\right. \cdot \nonumber \\ 
	\hspace{-0.5cm} && \left. \cdot  
	\varepsilon^{\alpha_p \beta_p} \frac{1}{\sqrt{\rho_0}}
	\left( \rho_0 \eta^{B_{p}}_{\beta_{p}}+{(x_{p}-x_{0})}_{\nu}  
	\sigma^{\nu}_{\beta_{p}{\dot \beta}_{p}}  
	{\overline \xi}^{ B_p {\dot \beta}_p}\right)\right] \, .  
	\label{e8-sym} 
\end{eqnarray}
The integration over the fermion zero modes leads to an SU(4) invariant 
contraction of the sixteen-index tensor $t_{16}$ defined after 
(\ref{l16-sym}) and further integration over the instanton moduli space 
would determine the exact dependence on the coordinates $x_{p}$.  
Again the unintegrated expression (\ref{e8-sym}) is sufficient for 
comparison with the D-instanton contribution to the corresponding 
AdS$_{5}\times S^{5}$ amplitude that will be considered in the next 
chapter. 

\section{Generalization to arbitrary $N$ and to any $K$ for large $N$}
\label{anynandk}

In the previous sections instanton calculus in ${\cal N}$=4 
supersymmetric Yang--Mills theory with gauge group SU(2) in the $K$=1 
sector have been presented following \cite{bgkr}. These results have 
been extended to the case of SU($N$) for generic $N$ in 
\cite{mattis1} and to arbitrary instanton number in the large $N$ 
limit in \cite{mattis2,mattislong}. 
These extensions are of fundamental importance for the 
re-interpretation of the results of this chapter in the context of 
the AdS/CFT correspondence that will be discussed in chapter 
\ref{cap5}. In this section the main steps of the extensions of 
\cite{mattis1,mattis2,mattislong} will be briefly reviewed. 
The generalization to $N>2$ is rather straightforward, on the contrary 
instanton calculus in sectors $K>1$ presents an enormous increase of 
computational complexity. The formalism for multi-instanton 
calculations (ADHM formalism) has been developed in 
\cite{adhm,corrfairgoddtamp,corrfairtamp,criweinstan,osb} and a review of 
the application of these techniques to SU($N$) supersymmetric gauge 
theories can be found in \cite{insthunter}. As has been explained in 
\cite{mattis2,mattislong} there are remarkable simplifications in the 
ADHM formalism in the large $N$ limit. 

Since for a gauge group SU($N$) there are $2N\cdot{\cal N}$ gaugino 
zero modes in the one-instanton background it might appear from a 
superficial analysis that the $\hat \Lambda^{16}$-correlation 
function should vanish for $N>2$. However it has been argued in 
\cite{bgkr} and then clearly shown in \cite{mattis1} that only the 
sixteen `geometric' (8 supersymmetric + 8 superconformal)
zero-modes are actually relevant in the general case of SU($N$), 
since all the remaining ones, $\mu_{i}^{A}$, ${\overline\mu}_{i}^{A}$, 
are `lifted' by a four-fermion term that is generated in the one-instanton 
action. The instanton action computed in \cite{mattis1} has the form
\begin{equation}
	S_{_{{\rm inst}}}=\frac{8 \pi^2 K}{g^{2}} + 
	S_{_{{\rm 4F}}} \, ,
	\label{liftaction}
\end{equation}
where $S_{_{{\rm 4F}}}$ contains a quartic fermionic interaction 
built up with the $8(N-2)$ additional fermionic collective 
coordinates. The exact form of $S_{_{{\rm 4F}}}$ in the $K$=1 
sector is
\begin{equation}
	S_{_{{\rm 4F}}}=\frac{\pi^2}{16\rho^2 g^{2}} \, 
	\varepsilon_{ABCD} \left[
	\sum_{i=1}^{N-2}{\overline \mu}^A_i \mu^B_i\right] \left[
	\sum_{j=1}^{N-2}{\overline \mu}^C_j \mu^D_j\right] \, .
	\label{4faction}
\end{equation}
$S_{_{{\rm 4F}}}$ is supersymmetric and lifts all but the 16 
`geometric' gaugino zero modes. 
\fancyhead[LO]{{\footnotesize 4.7~~{\it Generalization to any $N$ and 
to $K>1$ for large $N$}}}

For definiteness in the following the attention will be focused mainly 
on the $\hat{\Lambda}^{16}$ correlator, that is studied in 
\cite{mattis1}. In the case of SU($N$) there are additional zero-modes 
contributing to ${\lambda_{(0)\alpha}}^{A}$ in (\ref{lamzero}). 
However these are lifted by (\ref{4faction}) and the lowest order 
contribution in the one-instanton background is still obtained by 
substituting (\ref{lamzero}) in each factor of $\hat{\Lambda}$. 
As a result the expression for the $\hat{\Lambda}^{16}$-correlator 
reviewed above for the case $N$=2 is actually true for all $N$, apart 
from an overall $N$-dependent factor
\begin{eqnarray} 
	&& \hspace*{-1.2cm} G_{\hat \Lambda^{16}}^{N}(x_p) = C_{N} \, g^{8} 
	e^{-\frac{8\pi^2}{g^2} + i\theta} 
    \int \frac{d^{4}x_{0} \, d\rho_0}{\rho_0^{5}}  
	\int d^{8}\eta d^{8}{\overline \xi} 
	\int \prod_{A=1}^{4}\prod_{i=1}^{N-2} d\mu_{i}^{A}
	d{\overline\mu}_{i}^{A} \, e^{-S_{_{{\rm 4F}}}} \cdot \nonumber \\  
	&& \hspace*{-1.2cm} \cdot \prod_{p=1}^{16} \left[ 
	\frac{\rho_{0}^{4}}{[\rho_0^{2}+(x_{p}-x_{0})^{2}]^{4}} \,
	\frac{1}{\sqrt{\rho_{0}}}
	\left( \rho_0 \eta^{A_{p}}_{\alpha_{p}}+
	{(x_{p}-x_{0})}_{\mu}  \sigma^{\mu}_{\alpha_{p}{\dot \alpha}_{p}}  
	{\overline \xi}^{{\dot \alpha}_{p}A_{p}} \right)  \right] \, , 
	\label{l16-sym-n} 
\end{eqnarray} 
where $C_{N}$ is a $N$-dependent constant determined by the norms of the 
bosonic and fermionic zero-modes. To completely determine the 
dependence on $N$ one must still compute the integrals over $\mu_{i}^{A}$ 
and ${\overline\mu}_{i}^{A}$
\begin{equation}
	I_{N} = \int \prod_{A=1}^{4}\prod_{i=1}^{N-2} d\mu_{i}^{A}
	d{\overline\mu}_{i}^{A} e^{-S_{_{{\rm 4F}}}} \, . 
	\label{nongeoint}
\end{equation}
In computing the last integral it is convenient to perform a 
Hubbard-Stratonovich transformation of the fermion bilinears and 
represent $S_{_{{\rm 4F}}}$ as a Gaussian integral over auxiliary bosonic 
`collective coordinates' $\chi^i_{AB}$ \cite{mattis1}. 
In conclusion one finds the following overall $N$ dependence 
\cite{mattis1} 
\begin{equation}
	G_{\hat \Lambda^{16}}^{N}(x_p) \sim \sqrt{N} \, 
	G_{\hat \Lambda^{16}}(x_p) \, ,
	\label{lam16n}
\end{equation}
where $G_{\hat \Lambda^{16}}(x_p)$ is the expression valid in the 
$N$=2 case. This factor of $\sqrt{N}$ will be of relevance in the next 
chapter for the comparison with D-instanton effects in type IIB 
superstring theory.

The generalization to multi-instanton sectors in the large $N$ limit, 
as derived in \cite{mattis2}, is much more involved and would require 
an extended discussion, so only the basic ideas will be summarized 
here. A detailed report on the construction can be found 
in \cite{mattislong}. The moduli space of a SU($N$) $K$-instanton 
configuration, known as ADHM moduli space, has a very complicated 
structure for generic $N$. In \cite{mattis2,mattislong} it has been 
proved that in the large $N$ limit it is dominated by $K$ instantons 
living in $K$ mutually orthogonal SU(2) subgroups of SU($N$). Then a 
saddle point approximation is used to show that the geometry is 
actually described by (AdS$_{5}\times S^{5}$)$^{K}$, where in each 
factor AdS$_{5}$ is parameterized by the positions and sizes of the 
$K$ instantons, ($x_{0\,\mu}^{i},\rho^{i}$), $i=1,\ldots,K$, and 
$S^{5}$ by the auxiliary bosonic coordinates, $\chi^i_{AB}$, 
introduced in the Hubbard--Stratonovich transformation. Moreover the 
integrations in the vicinity of the large $N$ saddle point generate 
an attractive potential which actually forces all the $K$ instantons 
to a single point reducing the moduli space to a single copy of 
AdS$_{5}\times S^{5}$ \cite{mattislong}. Finally the small 
fluctuations about this reduced moduli space describe the dimensional 
reduction of SU($K$) ${\cal N}$=1 super Yang--Mills from $d$=10 to 
$d$=0+0, so that the $K$-instanton measure in the large $N$ limit 
factorizes into the product of the measure on AdS$_{5}\times S^{5}$ 
times the partition function ${\cal Z}_{K}$ for the SU($K$) 
(0+0)-dimensional theory. In conclusion, after evaluating ${\cal 
Z}_{K}$, see \cite{greengut,greengut2,zkpartfunction1,zkpartfunction2}, 
one obtains for a $n$-point correlation function like 
those considered in this chapter an expression of the form 
\cite{mattislong}
\begin{equation}
	G^{(K)}_{n}(x_{p}) = g^{8} \sqrt{N} K^{n-\frac{7}{2}} 
	e^{-\frac{8\pi^{2}K}{g^{2}}} \sum_{d|K}\frac{1}{d^{2}} 
	F_{n}(x_{p}) \, ,
	\label{greenkinst}
\end{equation}
where the sum is over the divisors of $K$ and the function 
$F_{n}(x_{p})$ is independent of $K$.


\chapter{AdS/SCFT correspondence}
\label{cap5}
\vspace*{2cm}
\fancyhead[RO,LE]{\thepage}
\fancyhead[RE]{{\footnotesize {\rm Chapter 5.}~~{\it AdS/SCFT 
correspondence}}} 
\fancyhead[LO]{}

\noindent
The remarkable proposal suggested by Maldacena in \cite{maldacena} of 
a correspondence relating type IIB superstring theory in 
AdS$_{5}\times S^{5}$ to ${\cal N}$=4 super Yang--Mills theory has led 
to a great renewal of interest in the latter. According to the 
formalization of Maldacena's original idea given by Gubser, Klebanov and 
Polyakov \cite{gkp} and by Witten \cite{witten} ${\cal N}$=4 
supersymmetric Yang--Mills theory in the superconformal phase is dual 
to type IIB superstring theory in AdS$_{5}\times S^{5}$ in the sense 
that correlation functions of Yang--Mills gauge invariant composite 
operators can be derived from the computation of amplitudes in the 
compactification of the type IIB superstring on AdS$_{5}\times S^{5}$. 
The precise interpretation of this duality will be discussed later. 

More generally the main idea of \cite{maldacena} is that type IIB 
superstring theory in AdS$_{d+1}\times M$, where $M$ is a compact 
manifold with positive curvature, should be equivalent to a (super) 
conformal field theory living on the $d$-dimensional boundary of 
AdS$_{d+1}$. In this respect the proposed correspondence is 
holographic in the sense of \cite{holography1,holography2}. 

Moreover in the case of superconformal field theories with gauge 
group SU($N$) Maldacena has argued that the large $N$ limit should be 
equivalent to the supergravity approximation to the AdS 
compactification of the type IIB superstring. This consideration makes 
the proposed duality particularly intriguing because it has long been 
believed that the large $N$ behaviour should be crucial in the 
understanding of the non-perturbative dynamics of non Abelian gauge 
theories \cite{thooft}. 

Non trivial checks of the suggested correspondence have been obtained 
in the large $N$ limit, \ie employing the supergravity approximation, 
both at the perturbative and at the non-perturbative level. However 
very little has been said about the extension to finite $N$ and the 
r\^ole of string corrections, although arguments have been developed 
supporting the validity of the correspondence for finite $N$ as well. 

The computation of four-point functions in the context of the 
AdS/SCFT correspondence has also allowed to point out peculiar 
behaviours in Green functions of the (super) conformal field theories 
under consideration at short distances. Four-point functions develop 
logarithmic singularities in the limit of coincident points. This 
kind of behaviour has already been shown in the instanton contribution 
to a four-point function in ${\cal N}$=4 super Yang--Mills theory 
in section \ref{inst-4-point}. Analogous results have been observed by 
various authors in Green functions computed from supergravity 
amplitudes in AdS$_{5}\times S^{5}$.

Most of the work on the AdS/CFT correspondence has concentrated on the 
case of AdS$_{5}\times S^{5}$ which is related to ${\cal N}$=4 super 
Yang--Mills. However considering the compactification of the type IIB 
superstring on AdS$_{d+1}\times M_{9-d}$, with $M_{9-d}$ a different 
compact manifold, one can study superconformal field theories with 
less supersymmetry. This is a very interesting subject for future 
developments. In particular the possibility of generalizing 
the results to non-supersymmetric theories would be of great 
relevance. 

Some recent reviews on the subject can be found in \cite{revads}.

The chapter is organized as follows. In section \ref{typeiib} a very 
brief introduction to type IIB superstring theory and to its low 
energy supergravity limit is given. The concept of D-branes, that has 
played a crucial r\^ole in Maldacena's construction, is recalled in 
section \ref{dbranes}. The original conjecture of \cite{maldacena} 
together with the successive formalizations are reviewed in section 
\ref{theconjecture}. The results by various authors in the 
computation of two- and three-point functions are reported in 
section \ref{2-3-point}, while section \ref{4-point} contains a 
discussion on four-point functions. The remaining sections report on 
original results in the comparison between instanton effects in ${\cal 
N}$=4 super Yang--Mills theory and D-instanton contributions to 
amplitudes in type IIB superstring theory, following \cite{bgkr}. 

\section{Type IIB superstring theory: a bird's eye view}
\label{typeiib}
\fancyhead[LO]{{\footnotesize 5.1~~{\it Type IIB superstring theory}}}

After their introduction in the late sixties in the context of ``dual 
models'' for the description of hadronic processes, string theories 
have emerged as a prime (and up to now unique) consistent candidate 
for a unified theory of fundamental interactions including gravity.

A form of the action describing the motion of a string in $d$-dimensional 
Minkowski space-time that is suitable for quantization has been 
proposed in \cite{bosonicstring,polyakovstring}
\begin{equation}
	S_{{\rm P}} = \frac{1}{4\pi \alpr} \int_{\Sigma} d\sigma d\tau \, 
	\sqrt{\gamma} \gamma^{ab}\partial_{a}X^{\mu}\partial_{b}X_{\mu}\, ,
	\label{polyaction}
\end{equation}
where $X^{\mu}=X^{\mu}(\sigma,\tau)$ are the coordinates of the string 
which give the embedding of the two-dimensional surface spanned by 
the string (worldsheet) into the $d$-dimensional space-time (target 
space), $\alpr$ is a parameter with dimension $({\rm mass})^{-2}$, 
related to the string tension $T$ by $T=\frac{1}{2\pi\alpr}$,  
and $\gamma^{ab}$ is the two-dimensional metric.  

Roughly speaking the quantization of the bosonic string is achieved by 
expanding in normal modes the solution of the classical equations of 
motion which follow from (\ref{polyaction}) and interpreting the 
coefficients of the expansion as creation and annihilation operators.
For open strings this gives rise to a spectrum made of a tachyon with 
mass $m^{2}=-\frac{1}{\alpr}$ and a massless vector plus an infinite 
tower of massive states, with masses proportional to 
$\frac{1}{\sqrt{\alpr}}$. For closed strings one obtains, beyond the 
tachyon and a tower of massive states, a rank two tensor, which 
decomposes into a symmetric traceless tensor, an antisymmetric tensor 
and a scalar. 

The presence of tachyons in the spectrum of bosonic string theories 
is very problematic. An interesting extension is 
represented by superstring theories, in which worldsheet 
supersymmetry is obtained by supplementing the bosonic fields 
$X^{\mu}(\sigma,\tau)$ with their superpartners, $\Psi^{\mu}
(\sigma,\tau)$, that are two dimensional Majorana spinors and 
space-time vectors. Superstring theories are consistently formulated 
in ten space-time dimensions and, as will be briefly discussed, 
possess a spectrum of states including a finite set of massless states 
plus an infinite tower of massive states, but no tachyons. 

The observation that two-dimensional gravity is conformally invariant 
at the classical level 
is the key point that allows to exploit the powerful tools of 
conformal field theory to describe the dynamics of the string. This is 
achieved by mapping the string worldsheet to the complex plane through 
$(\sigma,\tau)\to z=e^{\tau+i\sigma}$ and introducing bosonic and 
fermionic coordinates for the string as functions of $z$ and 
${\overline z}$. The quantization is then performed employing a BRS 
formalism to identify the physical states in the Hilbert space 
generated by ``vertex operators''. However since the aim of this section 
is to briefly review the basic ingredients of superstring theories 
(particularly the type IIB superstring) the more intuitive formalism of 
oscillators will be used.

The supersymmetric generalization of (\ref{polyaction}) is 
\begin{eqnarray*}
	S &=& \frac{1}{4\pi\alpr} \int_{\Sigma}d\sigma d\tau \, \left\{ 
	\sqrt{\gamma} \gamma^{ab} \left(\partial_{a}X^{\mu}\partial_{b}
	X_{\mu} + i {\overline \Psi}^{\mu}\rho_{a}\partial_{b}\Psi_{\mu}
	\right) + \rule{0pt}{18pt} \right. \\
	&& \hspace{1cm} \left. + {\overline \chi}_{a}\rho^{b}\rho^{a} 
	\left( X_{\mu}\partial_{b}X^{\mu} +\frac{1}{2} {\overline 
	\Psi}^{\mu}\Psi_{\mu} \chi_{b}\right) \right\} \, ,
\end{eqnarray*}
where $\rho^{a}$ are worldsheet Dirac matrices. $\chi^{a}$ is a 
Rarita--Schwinger field, the worldsheet gravitino (superpartner 
of $\gamma_{ab}$), which is a space-time scalar. Exploiting 
worldsheet superconformal symmetry one can put the action in the form 
\begin{equation}
	S=\frac{1}{4\pi\alpr} \int d^{2}\sigma \left[ 
	\partial_{+}X^{\mu}\partial_{-}X_{\mu} - i\psi_{L}\cdot\partial_{+} 
	\psi_{L} - i \psi_{R}\cdot\partial_{-} \psi_{R} \right] \, ,
	\label{superpolyact}
\end{equation}
where $+$ and $-$ indices refer to light-cone coordinates, 
$\sigma_{\pm} = \tau \pm \sigma$. The equations of motion coming 
from (\ref{superpolyact}) lead to an expansion of $X^{\mu}$ and 
$\psi^{\mu}$ in terms of independent left- and right-moving 
oscillation modes for closed strings. For open strings imposing 
Neumann (free) boundary conditions makes left- and right-moving 
degrees equivalent, so that roughly speaking the open string has half 
the degrees of freedom of a closed string. 

The coordinate $\sigma$ is taken in the range $[0,2\pi]$ and is periodic; 
for the fields $\psi_{L,R}$ there are two consistent boundary conditions, 
either periodic, giving rise to the so called {\em Ramond sector} ($R$), or 
antiperiodic, corresponding to the {\em Neveu--Schwarz sector} ($NS$). 

The solution of the equations of motion can be written schematically in the 
form 
\begin{eqnarray}
	\partial X^{\mu}_{L} &=& \sum_{n} \alpha_{-n}^{\mu} 
	e^{in(\tau+\sigma)} \label{xsolution} \\
	\psi^{\mu}_{L} &=& \sum_{n} \psi_{-n}^{\mu} e^{in(\tau+\sigma)}
	\label{psisolution}
\end{eqnarray}
and analogously for the right-moving part. The sum in 
(\ref{xsolution}) is over integer $n$, while for the fermions in 
(\ref{psisolution}) $n$ takes integer or semi-integer values in the 
$R$ and $NS$ sectors respectively. The zero-modes in the expansion of 
$\partial X^{\mu}_{L,R}$ correspond to the momentum of the center of 
mass of the string, $P_{L}=P_{R}=P$. For the fermions there are only 
zero-modes in the $R$ sector that satisfy a Clifford algebra. The 
ground state of the quantum theory is constructed acting with the 
creation operators associated with the zero-modes on the trivial vacuum. 
It is a SO(9,1) spinor in the $R$ sector and a (tachyonic) scalar in the 
$NS$ sector. A generic state in the spectrum of the string is of the form
\begin{displaymath}
	\alpha_{-n_{1}}^{L\mu_{1}}\ldots\psi_{-n_{k}}^{L\mu_{k}}\ldots 
	|P_{L},a \rangle \otimes \alpha_{-m_{1}}^{R\nu_{1}}\ldots
	\psi_{-m_{h}}^{R\nu_{h}}\ldots |P_{R},b \rangle \, ,
\end{displaymath}
where conventionally the negative modes are taken as creation 
operators and $a$ and $b$ refer to the choice of the spin ground 
state in the $R$ sector. The states 
are built considering separately the $NS$ and $R$ sectors for left- and 
right-moving degrees of freedom and then combining them. For the 
left-movers for instance this construction leads to a tachyonic ground 
state in the $NS$ sector, while the first higher mass state is a massless 
vector. In the $R$ sector the ground state corresponds to massless 
fermions. The correct way of dealing with tachyons was worked out in 
\cite{gso}, where a prescription, known as {\em GSO projection}, was 
given which leads to space-time supersymmetric string theories, 
containing no tachyonic state. In the $NS$ sector the GSO projection 
eliminates the tachyon and leaves with a massless vector, while in the 
$R$ sector it produces a spinor with definite chirality. Denoting, 
after the GSO projection, by {\bf v} the $NS$ ground state and with 
{\bf s} and {\bf s}$^{\prime}$ the spinors with opposite chirality of 
the $R$ sector one has the following possibilities. If the GSO projections 
on $L$ and $R$ moving degrees of freedom are different one gets the type 
IIA superstring theory, which has space-time supersymmetry 
${\cal N}$=(1,1) and is non chiral, whereas using the same projection 
on the $L$ and $R$ parts gives the type IIB string theory, which has 
${\cal N}$=(2,0) supersymmetry and is chiral. In particular for the 
type IIB theory, on which the attention will be focused in the following, 
the massless bosonic degrees of freedom are obtained combining 
\begin{displaymath}
	NS \otimes NS \quad \longleftrightarrow \quad 
	{\rm {\bf v}} \otimes {\rm {\bf v}}
\end{displaymath}
or 
\begin{displaymath}
	R \otimes R \quad \longleftrightarrow \quad 
	{\rm {\bf s}} \otimes {\rm {\bf s}} \, ,
\end{displaymath}
and correspond to the following field content 
\begin{eqnarray*}
	NS \otimes NS \quad & \longleftrightarrow & \quad \left( 
	g_{\mu\nu},B_{\mu\nu}, \phi \right) \\
	R\otimes R \quad & \longleftrightarrow & \quad \left(
	\chi=C^{(0)}, B^{\prime}_{\mu\nu}, C^{(4)} \right) \, ,
\end{eqnarray*}
where $g_{\mu\nu}$ is a rank two symmetric traceless tensor (the 
graviton), $B_{\mu\nu}$ and $B^{\prime}_{\mu\nu}$ antisymmetric tensors, 
$\phi$ a scalar (the dilaton), $\chi$ a pseudoscalar (the axion), and 
$C^{(4)}$ a four-form with self-dual field strength, 
$F^{(5)}=dC^{(4)}=(*F)^{(5)}$. From the mixed sectors ($R \otimes NS$ 
and $NS \otimes R$) one obtains the fermionic superpartners of the 
previous fields. 

The low energy effective description of type IIB superstring theory,  
constructed in terms of the above fields, is type IIB supergravity 
in ten dimensions, which has chiral ${\cal N}$=(2,0) supersymmetry. 
There are subtleties in writing down a covariant action for the model 
because of the self duality constraint on the four form field strength. 
The covariant equations of motion of the theory have been derived in 
\cite{schw}. Combining the \NSNS scalar and the \RR pseudoscalar into 
a complex field $\tau$, 
\begin{equation}
	\tau = \tau_{1}+i\tau_{2} = \chi + ie^{-\phi} \, ,
	\label{complexscalar}
\end{equation}
and denoting by $H_{1}$ and $H_{2}$ the field strengths of the \NSNS 
and \RR two forms, the low energy action of type IIB supergravity can 
be written, in the Einstein frame, in the form
\begin{eqnarray}
	S_{{\rm IIB}} &=& \frac{1}{2\kappa_{0}^{2}} \int d^{10}X \, 
	\sqrt{g} \left\{ R-\frac{1}{2\tau_{2}^{2}} \partial_{\Lambda}\tau 
	\partial^{\Lambda}{\overline\tau} + \left( F^{(5)} \right)^{2} - 
	\right. \nonumber \\
	&& \left. - \frac{1}{12\tau_{2}} \left[
	\tau H_{1}+H_{2} \right]_{\Lambda\Gamma\Pi} \left[ {\overline\tau} 
	H_{1} + H_{2} \right]^{\Lambda\Gamma\Pi} \right\} + \ldots \, ,  
	\label{iibsugra}
\end{eqnarray}
where $\kappa_{0}$ is the ten-dimensional Newton constant 
($\kappa_{0}\sim (\alpr)^{2}$). In (\ref{iibsugra}) the dots stand for 
higher derivative terms and the fermionic fields have not been displayed. 
Here and in the following capital Greek letters refer to flat 
ten-dimensional space-time. In (\ref{iibsugra}) the notation 
$\left(F^{(5)}\right)^{2}$ is schematic, since a covariant kinetic 
term for the \RR 4-form cannot be written down naively because of the 
self-duality constraint. The correct way to deal with this problem 
was described in \cite{pst}. The classical action is invariant under 
SL(2,$\mathbb{R}$) transformations that act projectively on the 
complex scalar $\tau$. Under the SL(2,$\mathbb{R}$) group the \NSNS 
and \RR antisymmetric tensors form a doublet. The scalar $\tau$ 
parameterizes the coset space SL(2,$\mathbb{R}$)/U(1)$_{B}$, where 
the U(1)$_{B}$ is an anomalous R-symmetry acting on the fermions. In 
particular the left chirality gravitino carries charge $+\frac{1}{2}$ 
under this symmetry and the right-chirality dilatino present in the 
type IIB spectrum has charge $+\frac{3}{2}$. At the quantum level the 
continuous SL(2,$\mathbb{R}$) symmetry is expected to be broken to 
SL(2,$\mathbb{Z}$) \cite{schwarz}. 

For the calculations presented in the following the only 
relevant terms in the effective action are those involving only 
derivatives of the metric apart from some overall function of the complex 
scalar. For future purposes it is useful to write these terms in the 
string frame that is related to the Einstein frame by a Weyl rescaling
\begin{displaymath}
	g_{\mu\nu}^{(s)} = e^{\phi/2}\,g_{\mu\nu}^{(E)} \, .
\end{displaymath}
The leading terms in the derivative expansion that will be 
considered take the form \cite{greengut}
\begin{equation}
    (\alpha')^{-4} \int d^{10}X \sqrt{g} \left[ e^{-2\phi} R + 
    \kappa (\alpha^\prime)^3 f_4(\tau,\bar\tau) 
    e^{-\phi/2} {\cal R}^4 \right] \, .
    \label{r4term}
\end{equation}
where $\kappa$ is a numerical constant. The Riemann curvature enters 
the ${\cal R}^4$ factor in a manner that may be most compactly 
described by writing it as an integral over a sixteen-component 
Grassmann spinor, 
\begin{displaymath} 
{\cal R}^4 \equiv \int d^{16}\Theta (R_{\Theta^4})^4 \, , 
\end{displaymath}
where  
\begin{displaymath}
R_{\Theta^4} = \overline{\Theta} \Gamma^{\Lambda_1 \Lambda_2  
\Lambda} \Theta \, \overline{\Theta} 
\Gamma^{\Lambda_3\Lambda_4}{}_{\Lambda} \Theta \, 
R_{\Lambda_1\Lambda_2\Lambda_3\Lambda_4} \, , 
\end{displaymath}
which only includes the Weyl tensor piece of the Riemann 
tensor. Here, $\Gamma^{\Lambda_1 \Lambda_2 \Lambda_3}$ 
are the totally antisymmetric products of three ten-dimensional
$\Gamma$-matrices and the Grassmann parameter $\Theta^a$ 
($a=1,\dots,16$) is a  chiral spinor of the ten-dimensional theory.   
This expresses ${\cal R}^4$ as an integral over half of the  
on-shell type IIB superspace. 

Beyond these terms there is a sixteen-dilatino term in the effective 
action that is related to the ${\cal R}^{4}$ term by supersymmetry and 
will play a central r\^ole in the comparison with the results of 
instanton calculus presented in the previous chapter. The sixteen 
fermion interaction reads 
\cite{greenvanhove,greengutkwon}
\begin{equation}
(\alpha')^{-1} \int d^{10}X \sqrt g \, e^{-\phi/2} 
f_{16}(\tau,\overline{\tau})  \Lambda^{16}   + \mbox{c.c.} \, ,
\label{lambint} 
\end{equation}
where the complex chiral dilatino has been denoted by $\Lambda$ and 
the interaction is (totally) antisymmetric in the sixteen spinor indices.
The properties of the modular functions $f_{4}(\tau,{\overline\tau})$ 
and $f_{16}(\tau,{\overline\tau})$ in equations (\ref{r4term}) and 
(\ref{lambint}) will be discussed later. They receive both 
perturbative corrections and non-perturbative contributions from 
D-instantons \cite{greengut}. The latter will be related to instanton 
effects in ${\cal N}$=4 super Yang--Mills theory.

\section{D-branes}
\label{dbranes}
\fancyhead[LO]{{\footnotesize 5.2~~{\it D-branes}}}

The perturbative spectrum of type II (A and B) superstring theories 
does not contain states charged with respect to the fields in the \RR 
sector \cite{norrcoupling}. However the flurry of work on string dualities 
during the last years has lead to a much deeper understanding of 
non-perturbative aspects of string theories, changing dramatically the 
scenario which was established at the perturbative level. 

From a low energy viewpoint the existence of solitonic solutions of 
the supergravity equations of motion has been known for a long time 
\cite{oldpbrane}. These solitonic configurations describe extended 
objects, {\em p-branes}, that are naturally coupled to the \RR forms 
and their duals. The r\^ole of these non perturbative states in the 
fundamental string theory has been clarified by Polchinski 
\cite{polchinski} through the concept of {\em D-branes}. 

A $p$-brane is $p$-dimensional object, whose world-volume is 
$p+1$-dimensional. It naturally couples to a $(p+1)$-form, $C^{(p+1)}$, 
through an action of the form 
\begin{displaymath}
	\int_{V_{p+1}} C^{(p+1)} \, ,
\end{displaymath}
where $V_{p+1}$ is the world-volume of the brane. In $d$ space-time 
dimensions the corresponding ``electric'' charge would be
\begin{displaymath}
	Q_{e} = \int_{S^{d-p-2}_{\infty}} *dC^{(p+1)} \, .
\end{displaymath}
The magnetic dual of a $p$-brane is a ($d-p-4$)-brane, which 
couples to a ($d-p-3$)-form and carries ``magnetic'' charge
\begin{displaymath}
	Q_{m} = \int_{S^{p+2}_{\infty}} dC^{(p+1)} \, .
\end{displaymath}

In particular in the case of the type IIB theory in the \RR sector 
there is a scalar, the axion $\chi$, \ie a zero-form $C^{(0)}\equiv 
\chi$, a rank two anti-symmetric tensor, $B_{\mu\nu}^{\prime}$, \ie a 
two-form $C^{(2)}$ and a four form $C^{(4)}$. Coupled to $C^{(0)}$ 
one has a (-1)-brane which is an object localized in space-time, \ie an 
instanton in Euclidean space-time. Effects of D-instantons in type 
IIB superstring theory will be discussed later. The two-form couples to a 
one-dimensional object, a D-string that should not be confused with the 
fundamental string which couples to the \NSNS two-form. Finally there 
is a 3-brane charged with respect to the four-form $C^{(4)}$. 
Moreover one can consider the corresponding magnetically dual 
objects. 

Starting from the type II supergravity action, extremal (\ie saturating 
a BPS bound) solitonic configurations with the previously discussed 
properties can be constructed assuming that only the metric $g_{\mu\nu}$, 
the dilaton $\phi$ and a $p+1$ form are non-vanishing. The ansatz that one 
considers for these fields is
\begin{eqnarray*}
	d s^{2} &=& e^{2A(r)}d\vec{x}^{2}+e^{2B(r)}d\vec{y}^{2} \\
	\phi &=& \phi(r) \\
	C^{(p+1)}_{01\ldots p} &=& -e^{C(r)} \, ,
\end{eqnarray*}
where the ten-dimensional coordinates $X^{\Lambda}$ are split into 
longitudinal ($x^{\mu}$, $\mu=0,1,\ldots,p$) and transverse ($y^{i}$, 
$i=p+1,\ldots,9$) coordinates and $r^{2}=y_{i}y^{i}$. The solution of 
the supergravity equations can be written in terms of a single 
harmonic function $H_{p}(r)$, \ie a solution of the Laplace equation 
in the $9-p$ transverse dimensions, 
\begin{displaymath}
	H_{p}(r) = 1+\frac{a_{p}}{r^{7-p}} \, ,
\end{displaymath}
where $a_{p}$ is a constant with dimension $L^{7-p}$, related to the 
charge carried by the $p$-brane. The $p$-brane solution is (in the 
Einstein frame) 
\begin{eqnarray}
	ds^{2} &=& \left[ H_{p}(r) \right]^{\frac{p-7}{8}} d\vec{x}^{2} +
	\left[ H_{p}(r) \right]^{\frac{p+1}{8}} d\vec{y}^{2} \\
	C^{(p+1)}_{01\ldots p}(r) &=& \left[ H_{p}(r) \right]^{-1}-1 \\
	e^{2\phi(r)} &=& \left[ H_{p}(r) \right]^{\frac{3-p}{2}} \, .
	\label{pbranesolution}
\end{eqnarray}
Notice in particular that in the case of a 3-brane the dilaton $\phi$ 
reduces to a constant. The solution simplifies in the string frame 
where the metric becomes 
\begin{displaymath}
	ds^{2} = \left[ H_{p}(r) \right]^{-\frac{1}{2}} d\vec{x}^{2} +
	\left[ H_{p}(r) \right]^{\frac{1}{2}} d\vec{y}^{2} \, .
\end{displaymath}

It can be proved that these solitonic solutions 
are BPS saturated and preserve one half of the supersymmetries of the 
original flat background, \ie 16 supercharges out of 32. This 
implies for instance that there exist a relation between the mass 
and charge of the solitons. 

Many of the recent developments in string theory have been made 
possible by the suggestion \cite{polchinski} of a 
way to incorporate the above solitonic objects into type II string 
theories, see \cite{polchitasi} for a review. 

The solitonic $p$-branes are introduced in the microscopic string 
theory in relation with a (Dirichlet) open string subsector. More 
precisely one introduces {\em Dirichlet $p$-branes} (or D$p$-branes) 
as $p$-dimensional hyperplanes where open strings can end, 
satisfying Neumann (standard) boundary conditions on the $p$ 
longitudinal coordinates $X^{\mu}$, $\mu=0,\ldots,p-1$ and Dirichlet 
(fixed) conditions on the remaining transverse coordinates $X^{m}$, 
$m=p,\ldots,9$. D-branes defined in this way can be seen as 
`topological defects' where open strings can end. These hyperplanes 
are actually dynamical objects, whose elementary excitations are 
described by the open strings attached to them. These open strings can 
be consistently introduced in type II superstring theories, without 
considering `standard' open strings propagating independently in the 
target space. There are various arguments that lead to identify  
D-branes with the solitonic $p$-branes of type II supergravities. 
For instance considering two parallel D-branes their interaction is 
mediated by the exchange of closed strings and one can prove that this 
results in a vanishing total force just like for example is found for 
a monopole pair. Furthermore it can be shown that the boundary 
conditions imposed on the open strings ending on the D-brane break 
half the supersymmetries, so that the D-branes are BPS saturated 
objects. 

The dynamics of D-branes, in a microscopic string theory perspective, 
is described by the open strings attached to the brane. 
Comparing the effective world-volume dynamics of the D-brane with the 
dynamics of solitonic $p$-brane solutions in the moduli space 
approximation one can find further evidence for the identification of 
the two. 

The massless spectrum of the subsector of open strings attached to a 
$p$ D-brane is a maximally supersymmetric U(1) gauge theory in 
$p+1$ dimensions. The effective world-volume degrees of freedom of a 
$p$ D-brane correspond to the dimensional reduction of a ${\cal N}$=1 
vector multiplet from $d$=10 to $d=p+1$. This gives rise to a 
supersymmetric Maxwell theory (with sixteen supercharges) 
including $9-p$ scalars, one vector and their superpartners. 

Considering $N$ D-branes one can have also open strings stretched 
between them. They describe the exchange of massive vector states. 
There are $N^{2}$ ways of stretching the strings among the branes. 
In the limit in which the $N$ D-branes are taken to coincide these 
$N^{2}$ different states become massless. The symmetry is enhanced to 
U($N$) and one is left with an effective U($N$) Yang--Mills theory in 
which the $N^{2}$ vectors form the adjoint. In general the relative 
separations among the branes determine the expectation values of 
the $9-p$ scalars in the effective theory. 

In particular in the case of a collection of $N$ coincident D3-branes 
the world-volume effective action is a U($N$) supersymmetric 
Yang--Mills theory with sixteen supercharges in $d$=4, \ie the 
${\cal N}$=4 super Yang--Mills theory with gauge group 
U($N$)~\footnote{Actually there is a U(1) factor that describes the 
center of mass motion and decouples, so that the gauge group 
is SU($N$) \cite{wittdb}.}. 

\newpage

\section{The AdS/SCFT correspondence conjecture}
\label{theconjecture}
\fancyhead[LO]{{\footnotesize 5.3~~{\it AdS/SCFT correspondence}}}

In the case of D3-branes the solution for the metric yields (in the 
string frame) 
\begin{equation}
	ds^{2} = \left[ 1+\frac{L^{4}}{r^{4}} \right]^{-\frac{1}{2}} 
	d\vec{x}^{2} + \left[ 1+\frac{L^{4}}{r^{4}} \right]^{\frac{1}{2}} 
	d\vec{y}^{2} \, ,
	\label{d3metric}
\end{equation}
where $(x^{\mu},y^{i})$ are the previously introduced Cartesian 
coordinates and $L$ is a length scale. As already noticed in the case 
of a D3-brane the dilaton is constant, $\phi=\phi_{0}$, and determines 
the string coupling constant
\begin{displaymath}
	g_{s}=e^{\phi_{0}} \, .
\end{displaymath}
Considering a D3-brane configuration with $N$ units of \RR 5-form 
field strength, \ie a configuration of $N$ coincident D3-branes, 
the length $L$ is 
\begin{equation}
	L^{4} = 4 \pi g_{s}N \alpr^{2} 
	\label{L-alpr} \, .
\end{equation}
The key observation which led Maldacena to the formulation of the 
AdS/CFT correspondence conjecture \cite{maldacena} is that the metric 
(\ref{d3metric}), in the ``near horizon'' limit $r\to 0$, reduces to
\begin{equation}
	ds^{2}= \frac{L^{2}}{\rho^{2}} \left( dx\cdot dx + 
	d\rho^{2} \right) +   d\omega_{5}^{2} \, ,
	\label{d3nearhorizon}
\end{equation}
where $\rho^{2}=\frac{L^{4}}{r^{2}}$ and $d\omega_{5}^{2}$ is the 
spherically symmetric constant curvature metric on a 5-sphere, which 
is the metric of the AdS$_{5}\times S^{5}$ space. In (\ref{d3nearhorizon}) 
$L$ is the radius of curvature of both the 5-sphere and the AdS factor. 
The D3-brane metric can be viewed as a solitonic configuration 
interpolating between two maximally supersymmetric backgrounds, the 
near horizon AdS$_{5}\times S^{5}$ and flat ten dimensional Minkowski 
space-time at infinity. 

Maldacena has argued, on the basis of previous results in the context 
of the description of black holes by D-branes \cite{maldastrominger}, 
that the region that should be identified with the Yang--Mills low 
energy description is the ``throat'', \ie the region $r\ll L$, in the 
D3-brane metric (\ref{d3metric}). See \cite{gubskleb} for related work. 
This leads to the conjecture that ${\cal N}$=4 supersymmetric Yang--Mills 
theory should be dual to type IIB superstring theory compactified on 
AdS$_{5}\times S^{5}$. The original conjecture relates type IIB 
supergravity to SU($N$) ${\cal N}$=4 super Yang--Mills in the large 
$N$ limit. This is related to the fact that the limit $g_{s}\ll 1$ at 
large length scale with respect to the string scale, 
$\frac{\alpr}{L^{2}}\ll 1$, in which one can trust the supergravity 
approximation, requires $N\to \infty$ with $4\pi g_{s}N$ fixed and large. 
The correspondence relates the complexified Yang--Mills coupling constant 
to the one of the type IIB superstring in AdS$_{5}\times S^{5}$ by 
\begin{equation}
	g_{s} = \frac{\gym^{2}}{4\pi} \, ,\qquad 
	\chi_{0} = \frac{\theta_{_{{\rm YM}}}}{2\pi} \, ,
	\label{dict}
\end{equation}
where $\chi_{0}$ is the constant \RR scalar and from now on the 
Yang--Mills coupling constant and vacuum-angle will be denoted 
respectively by $\gym$ and $\tym$. 

This implies that the large $N$ limit of relevance here coincides with 
the large $N$ limit considered by 't Hooft in \cite{thooft}, with the 
effective coupling $\hat{g}=\gym N ^{2}$ fixed at a large value.

The AdS$_{5}$ space has constant negative curvature and possesses a 
boundary that is the four dimensional Minkowski 
space~\footnote{Actually this is better understood considering the 
Euclidean version of the AdS$_{5}$ space, see \cite{witten} for a 
detailed description.}. According to Maldacena's proposal, the boundary 
of AdS$_{5}$, which corresponds to $\rho \to 0$, \ie $r\to \infty$ in 
the notation of (\ref{d3nearhorizon}), is exactly the location of the 
${\cal N}$=4 theory and the boundary values of the bulk supergravity 
fields act as sources that couple to gauge-invariant composite 
operators in ${\cal N}$=4 super Yang--Mills theory. The correct 
interpretation of this statement has been explained in \cite{gkp,witten} 
and will be discussed soon. 

The AdS$_{5}\times S^{5}$ background is characterized by the 
non-vanishing fields, 
\begin{eqnarray*}
    F_{_{MNPQR}} &=& \frac{1}{L} \varepsilon_{_{MNPQR}}   
    \qquad R_{_{MNPQ}} = - \frac{1}{L^2} (g_{_{MP}}g_{_{NQ}} - 
    g_{_{MQ}}g_{_{NP}}) \\ 
    F_{mnpqr} &=& \frac{1}{L} \varepsilon_{mnpqr} 
    \qquad\quad R_{mnpq} = + \frac{1}{L^2} 
    (g_{mp}g_{nq} - g_{mq}g_{np}) \, ,
\end{eqnarray*}
where upper case Latin indices, $M,N,\dots=0,1,2,3,5$, span the
AdS$_5$ coordinates and lower case Latin indices, $m,n,\dots 
=1,2,3,4,5$ span the $S^5$ coordinates.  
The only non-vanishing components of the Ricci tensor are 
\begin{equation}  
    R_{MN} = - \frac{4}{L^2} g_{MN} \qquad 
    R_{mn} = + \frac{4}{L^2} g_{mn} \, .
	\label{ricci}
\end{equation}
Upon contracting (\ref{ricci}) with the  metric tensor it follows 
that the total (5+5)-dimensional scalar curvature vanishes. The Weyl 
tensor, defined in $d$-dimensional space-time as
\begin{eqnarray*}
	 {C}_{\mu \nu\rho \sigma} &=& {R}_{\mu \nu\rho \sigma} - 
	 \frac{1}{d-2} \left[ R_{\mu\nu} g_{\nu \sigma} + (3~{\rm terms}) 
	 \right] + \nonumber \\ 
	 &+& \frac{1}{(d-1)(d-2)} R \left( g_{\mu\rho} g_{\nu\sigma}-
	 g_{\mu\sigma}g_{\nu\rho} \right) \, ,
\end{eqnarray*}
vanishes as well in AdS$_{5}\times S^{5}$ because of the conformal 
flatness of the metric. 

This background is maximally supersymmetric (just like the Minkowski 
vacuum) so there are 32 conserved supercharges, that transform as a
complex chiral spinor of the tangent-space group, SO(4,1)$\times$SO(5). 
In the basis where the ten dimensional $\Gamma_\Lambda$ matrices 
are given by $\Gamma_M = \sigma_1\otimes\gamma_M\otimes \I$ and
$\Gamma_m=\sigma_2\otimes \I\otimes\gamma_m$, the supersymmetries 
are generated by the Killing spinors that satisfy  
\begin{equation}
    D_\Lambda \epsilon - {1 \over 2L}(\sigma_1\otimes \I\otimes\I) 
    \Gamma_\Lambda \epsilon = 0 \, ,
    \label{kilspina} 
\end{equation}
which follows from the requirement that the gravitino supersymmetry 
transformation should vanish. In this basis the complex chiral 
supersymmetry parameters read
\begin{eqnarray*}
    \epsilon_{\pm}=\left( \begin{array}{c} 1 \\ 0 \end{array} \right)
    \otimes \zeta_{\pm} \otimes \kappa_{\pm} \, , 
\end{eqnarray*}
where $\zeta_{\pm}$ are complex four-component SO(4,1) spinors and 
$\kappa_{\pm}$ complex four-component SO(5) spinors, satisfying 
\begin{eqnarray*}
    D_M \zeta_{\pm} \mp {1 \over 2L} \gamma_M \zeta_{\pm} &=& 0 \\ 
    D_m \kappa_{\pm} \mp i {1 \over 2L} \gamma_m \kappa_{\pm} &=&0 \, .
\end{eqnarray*}

Strong support to the conjecture comes from symmetry considerations. 
AdS$_{5}$ has a SO(2,4) isometry group which is to be identified with the 
four-dimensional conformal group in the boundary theory. The isometry 
group of the $S^{5}$ factor is SO(6)$\sim$SU(4) and is related to the 
SU(4) R-symmetry group of the ${\cal N}$=4 Yang--Mills theory. 
Including the associated fermionic generators the super-isometry group of 
the type IIB background (\ref{d3nearhorizon}) is exactly the superconformal 
group SU(2,2$|$4) of ${\cal N}$=4 super Yang--Mills in four dimensions. 
Moreover the SL(2,$\mathbb{Z}$) S-duality group that is conjectured to 
be preserved in the type IIB superstring at the quantum level is 
identified with the (conjectured) exact S-duality symmetry group of 
four dimensional SU($N$) ${\cal N}$=4 Yang--Mills theory. The former 
is connected to the existence of an infinite set of stable dyonic BPS 
string solitons \cite{schwarz}, the latter to the existence an infinite 
set of stable dyonic BPS states \cite{sen}, as discussed in chapter 
\ref{cap2}. 

The spectrum of excitations of the maximally supersymmetric 
compactification of the type IIB superstring on the AdS$_{5}\times S^{5}$ 
background, first considered in \cite{schw}, was studied in great detail 
in \cite{gunmar,kiromvan}. The map that associates fields in the multiplet of 
type IIB supergravity to the corresponding operators in ${\cal N}$=4 
super Yang--Mills has been worked out in \cite{ff,adf,ferrzaff}. 
States in the spectrum of $d$=5 ${\cal N}$=8 gauged supergravity 
admit a natural classification in terms of spherical harmonics on the 
5-sphere. In the conjectured correspondence the lowest lying states, 
assembled into the ${\cal N}$=8 gauged supergravity multiplet, couple to the 
supermultiplet of super Yang--Mills currents, equations (\ref{currentdef1}) 
and (\ref{currentdef2}), that are ``bilinear'' in the elementary Yang--Mills 
fields. Higher $\ell$ spherical harmonics, \ie Kaluza--Klein (KK) descendants
of the supergravity fields, couple to other ``chiral primary fields'',
that correspond to ``$\ell$th-order'' gauge-invariant polynomials in 
the elementary super Yang--Mills fields. Moreover, stable string 
excitations that couple to non-chiral primary fields are needed for the 
closure of the operator algebra \cite{konishi}. Despite some interesting 
progress \cite{gsads}, a useful worldsheet description for the type IIB 
superstring in AdS$_5\times S^5$ is still lacking. Nevertheless the 
stringy truncation of the KK spectrum to $\ell \leq N$ seems to be a 
robust result that matches with the naive expectations on the super 
Yang--Mills side. 

The explicit connection between the bulk theory and the boundary 
theory has been given a precise formulation in \cite{gkp,witten}. 
The partition function of the type IIB superstring, computed with suitable 
boundary conditions on the four-dimensional boundary of AdS$_{5}$, 
plays the r\^ole of generating functional for connected Green functions 
of gauge-invariant composite operators in ${\cal N}$=4 Yang--Mills 
theory. More precisely one identifies
\begin{equation} 
    Z_{_{{\rm IIB}}}[J] = \int [DA] \, 
    \exp(-S_{_{{\rm YM}}}[A] + 
{\cal O}[A] J) \,  , 
\label{main} 
\end{equation}
where $Z_{_{{\rm IIB}}}[J]$ is the partition function of the type IIB 
superstring, evaluated in terms of the bulk string fields, which is 
computed with prescribed boundary values $J$ for the fields. 
Elementary fields of the ${\cal N}$=4 super Yang--Mills theory on the 
boundary are denoted by $A$ and ${\cal O}(A)$ are gauge-invariant  
composite operators to which $J(x)$'s couple. 
 
The left hand side of (\ref{main}) is computed performing the 
functional integration over the bulk fields $\Phi(z;\omega)$, where 
$\omega$ are the coordinates on $S^5$ and $z^M\equiv (x^\mu,\rho)$ 
($M=0,1,2,3,5$ and $\mu=0,1,2,3$) the coordinates on AdS$_5$ ($\rho 
\equiv z_5$ is the coordinate transverse to the boundary)
\begin{equation}
	Z_{_{{\rm IIB}}}[J] = \int [D\Phi]_J\exp(-S_{_{{\rm IIB}}} 
	[\Phi]) \, .
	\label{intfunz}
\end{equation}
In (\ref{intfunz}) the notation $\Phi(z;\omega)$ refers collectively  to 
the `massless' supergravity fields, their Kaluza--Klein descendents, and 
possibly the string excitations. The functional integral 
depends on the boundary values, $J(x)$, of the bulk fields.  

The exact map which associates fields in the type IIB compactification 
with operators in ${\cal N}$=4 super Yang--Mills theory  is dictated by 
the matching of quantum numbers corresponding to the previously discussed 
global symmetries. Furthermore the conformal dimension, $\Delta$, of a 
Yang--Mills operator ${\cal O}$ is related to the AdS `mass' $m$ of 
the corresponding supergravity field. For instance in the case of 
scalar operators the relation is 
\begin{equation}
	(mL)^{2} = \Delta(\Delta-4) \, ,
	\label{masstodimension}
\end{equation}
where only the positive branch of $\Delta_{\pm}=2\pm\sqrt{4+(mL)^{2}}$ 
is relevant for the lowest-`mass' supergravity multiplet. 

As already remarked in the Kaluza--Klein reduction of type IIB 
supergravity on $S^{5}$ the states are classified expanding the 
ten-dimensional fields in spherical harmonics on the 5-sphere. 
This procedure generates an infinite tower of states with 
spins up to 2. The masses of the states associated with single modes and 
their behaviour under the SO(2,4) isometry group of AdS$_{5}$ are 
determined by solving the corresponding linearized equations of motion 
\cite{kiromvan}. 

Alternatively the Kaluza--Klein excitations of all the fields in the 
type IIB gauged supergravity can be constructed, using the Killing spinors 
defined in equation (\ref{kilspina}), from the modes of the massless singlet 
dilaton $e^{\phi}$. The construction is rather involved and will be 
briefly sketched only for fields that will be relevant in the following.

One obtains a classification of the states according to the 
representations of the isometry group of AdS$_{5}\times S^{5}$, which 
adding the associated fermionic generators becomes SU(2,2$|$4). In 
particular the mass of the states is identified with the eigenvalue of 
one of the generators of SO(2,4). This allows to construct a map which 
relates each field in the supergravity multiplet to the corresponding 
composite operator in ${\cal N}$=4 super Yang--Mills theory by matching 
the SU(2,2$|$4) quantum numbers (recall that on the super Yang--Mills side 
SU(2,2$|$4) is the ${\cal N}$=4 superconformal group). Notice that since the 
states in the supergravity multiplet as well as their KK descendants have 
spins not larger than 2, they belong to short multiplets of SU(2,2$|$4).

The massless dilaton is associated with the constant mode on $S^5$, \ie 
with the scalar spherical harmonic $Y_{\ell}(\omega)$ with $\ell=0$. Also
at the $\ell=0$ level there is the five-dimensional graviton, describing 
the excitations of the AdS$_{5}$ metric. The other scalars in the 
supermultiplet are associated with excitations on the 5-sphere, \ie 
they have $\ell>0$. In particular there are real scalars $Q^{ij}$ with 
mass $m^2 = -\frac{4}{L^2}$ in the {\bf 20}$_{\mathbb{R}}$ of the $SO(6)$ 
isometry group of $S^5$. They result from a combination of the trace of 
the internal metric and the self-dual \RR\ five-form field, $F^{(5)}$, 
with $\ell=2$ ($Q^{ij}$ are quadrupole moments of $S^5$). The trace 
part of the metric fluctuations can be written in the form
\begin{displaymath}
	\delta g_{MN} = \frac{16}{15} f \delta_{MN} \hspace{1cm} 
	\delta g_{mn} = f \delta_{mn} \, , 
\end{displaymath}
where $f=f(z,\omega)$. Analogously for the \RR 4-form one gets
\begin{displaymath}
	A_{MNPQ} = -\frac{L}{4} \varepsilon_{MNPQR} {\nabla}^{R} f \hspace{1cm}
	A_{mnpq} = -\frac{L}{4} \varepsilon_{mnpqr} \nabla^r f \, .
\end{displaymath}
The relation with the scalar fields $Q_{ij}(z)$ is of the form
\begin{displaymath}
	f(z,\omega) = Q_{ij}(z) Y^{ij}_{\ell=2}(\omega) \, .
\end{displaymath}
Similarly the complex scalars $E^{AB}$ with mass $m^2 = -\frac{3}{L^2}$ 
and their conjugates are associated with the pure two-form fluctuations 
with $\ell=1$ of the complexified antisymmetric tensor, constructed from 
the \RR\ and \NSNS\ two-forms, in the internal directions. The massless 
vectors $V_M^{[ij]}$ in the {\bf 15}, that gauge the SO(6) isometry group, 
are in one-to-one correspondence with the Killing vectors of $S^5$ and 
result from a linear combination with $\ell=1$ of the mixed components of 
the metric and the internal three-form components of the \RR four-form 
potential, $C^{(4)}$. The {\bf 6} complex antisymmetric tensors 
$B_{MN}^{[AB]}$ with $m^2=\frac{1}{L^2}$, that have peculiar first order 
equations of motion, result from scalar spherical harmonics with $\ell=1$. 
The analysis of the fermions is similar. The dilatini $\Lambda^A$ in the 
{\bf 4} of SO(6) are proportional to the internal Killing spinors 
$\kappa_+$ and have mass $m = -\frac{3}{2L}$. They play a central r\^ole 
in the comparison between the effects of instantons in ${\cal N}$=4 
Yang--Mills and D-instantons in type IIB string theory that has been 
carried out in \cite{bgkr} and will be reviewed in section 
\ref{dinstantons}. The {\bf 20}$_{\mathbb{C}}$ spinors $\chi_{BC}^A$ with 
mass $m=-\frac{1}{2L}$ correspond to internal components of the gravitino 
with $\ell=1$. Finally the supergravity multiplet is completed by the 
massless $\overline{{\bf 4}}$ gravitinos $\Psi_{MA}$ which  are 
proportional to the internal Killing spinors $\kappa_-$.  

The above fields are the ones that act as sources for the superconformal
currents (\ref{currentdef1}), (\ref{currentdef2}). More precisely the 
graviton and dilaton are associated with the SU(4) singlets in the 
${\cal N}$=4 current multiplet, the stress-energy tensor 
${\cal T}_{\mu\nu}$ and the scalar ${\cal C}$ respectively. The other 
scalars, $E^{AB}$ and $Q^{ij}$, correspond to ${\cal E}^{AB}$ and 
${\cal Q}^{ij}$ in (\ref{currentdef2}) respectively. The massless 
vectors $V_{M}^{[ij]}$ are dual to the SU(4) currents ${\cal 
J}^{A}_{\mu B}$ and the anti-symmetric tensors $B_{MN}^{[AB]}$ to 
${\cal B}^{AB}_{\mu\nu}$. The identification can be carried out 
analogously for the fermions. The dilatini $\Lambda^{A}$ and the 
spinors $\chi^{A}_{BC}$ are associated respectively to the 
fermions $\hat{\Lambda}^{A}$ and $\hat{\chi}^{A}_{BC}$ in 
(\ref{currentdef2}). The gravitinos $\Psi_{M\,A}$ correspond to the 
supersymmetry currents $\Sigma_{\mu A}$.

Higher Kaluza--Klein modes correspond to higher values of $\ell$. For 
example, there are other scalar modes with $\ell>2$. Each of these can be 
put in one-to-one correspondence with a gauge singlet composite operator 
$W_{(\ell)}$ that starts with 
\begin{equation}
    \left. W^{(i_1...i_\ell)}\right|_{\theta=0} = 
    \tr \left( \phi^{(i_1} \ldots \phi^{i_\ell)} \right) 
    - {\rm traces} \, ,
    \label{kkwell} 
\end{equation}
which has dimension 
$\Delta=\ell$ and belongs to the $\ell$-fold symmetric traceless 
tensor representation of $SO(6)$ with Dynkin labels $(0,\ell,0)$. 
The multiplet contains $256\ell^2(\ell^2-1)/12$ states with different 
$\Delta$ and SU(4) quantum numbers. 

The correct recipe for calculating correlations follows from (\ref{main}) 
and amounts to computing the truncated Green functions in the bulk and 
attaching {\em bulk-to-boundary propagators} to the external legs 
\cite{gkp,witten,freedman}. The precise forms of the latter
depend on the spin and `mass' of the field. For instance, 
the bulk-to-boundary Green function for a bulk scalar field 
$\Phi_{m}$ associated to a dimension $\Delta$ operator 
${\cal O}_{\Delta}$ on the boundary reads
\begin{equation}
	K_{\Delta} (x^\mu, \rho; x^{\prime\mu},0) = c_{_{\Delta}}  
	{\rho^{\Delta} \over  (\rho^2 
	+ (x-x^{\prime})^2)^{\Delta}} \, , 
	\label{greenfun} 
\end{equation}
where $c_{_{\Delta}}$ is a constant given by 
$c_{_{\Delta}}=\Gamma(\Delta)/[\pi^{2} \Gamma(\Delta - 2)]$.
For simplicity the $\omega$ dependence has been suppressed, so that 
the expression (\ref{greenfun}) is appropriate for an `S-wave' process 
in which there are no excitations in the directions of the five-sphere, 
$S^5$. In terms of $K_{\Delta}$ the bulk field 
\begin{equation}
    \Phi_m (z;J) = \int d^4 x^{\prime} K_{\Delta}(x,\rho; x^{\prime},0) 
    J_{\Delta}(x^{\prime}) 
    \label{btob}
\end{equation}
satisfies the boundary condition $\Phi_m (x,\rho;J) \approx 
\rho^{4-\Delta} J_{\Delta}(x)$ as $\rho \to 0$,
since $\rho^{\Delta-4}K_{\Delta}$ reduces to a $\delta$-function on the 
boundary. As observed in \cite{freedman} the bulk-to-boundary 
propagator must be modified for $\Delta$=2. In this case the previous 
expression vanishes since $c_{_{\Delta}}$=0 for $\Delta$=2. In the 
special case $\Delta$=2 one must require $\Phi_{m}(x,\rho;J) \to 
\rho^{2} \log\rho \, J(x)$, which yields 
$c_{_{\Delta=2}} = \frac{\Gamma(2)}{2\pi^{2}}$. 
Before discussing specific calculations, that will be related 
to instanton effects in ${\cal N}$=4 supersymmetric Yang--Mills theory, 
this prescription for evaluating Green functions will 
be illustrated in more detail in the case of two- and three-point 
functions in the next section. 

The generalization of this construction to fermionic fields has been 
considered in \cite{henningsfet} and will be relevant later when 
the calculation of a correlation function of sixteen fermionic 
dilatini will be presented. 

The original conjecture involves the supergravity approximation and 
holds in the large $N$ limit. However in \cite{maldacena} it has been 
suggested that the correspondence can be generalized to finite $N$ 
by taking in account string theory corrections. This claim has been 
made more explicit in \cite{witten}. For finite $N$ the 
r\^ole of generating functional is played by the full 
partition function of the string in AdS$_{5}\times S^{5}$, computed 
with suitable boundary conditions. Moreover for finite $N$, 
\ie for finite radius, one expects a truncation to $\ell \leq N$ 
of the Kaluza--Klein spectrum, that should result by taking 
into account full-fledged stringy geometry 
as in any string compactifications. 

Perturbative checks of the conjecture at finite $N$ for two- and 
three-point functions have been given in \cite{freedhokskib,sei} and 
will be discussed in the next section. The non-perturbative computations 
of \cite{bgkr} suggest that the correspondence remain valid at finite $N$ 
also for four- or higher-point functions of some protected operators at 
least. 

Witten has also suggested an intuitive way to represent diagrammatically 
the above described construction through graphs like those depicted 
in figure \ref{witdiagr}.

\begin{figure}[!h]
	\centering
	\includegraphics[width=12truecm]{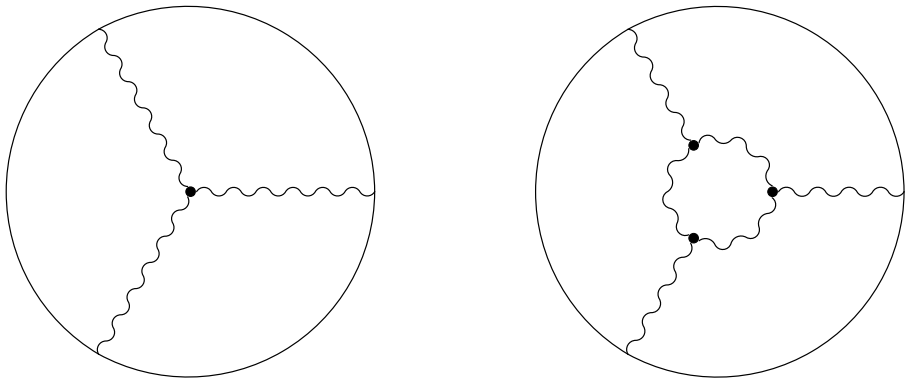}
	\caption{Witten's representation of three point functions in the 
	AdS/CFT correspondence.}
	\label{witdiagr}
\end{figure}

\noindent
The circles in figure \ref{witdiagr} represents the four-dimensional 
boundary of AdS$_{5}$ and the interior is the bulk. Lines connecting an 
internal point with a point on the boundary are the above discussed 
bulk-to-boundary propagators (see \eg equation (\ref{greenfun})). 
In the supergravity approximation, \ie in the limit of large $N$ and 
large and fixed $\gym^{2}N$, only tree diagrams like the first one in 
figure \ref{witdiagr} contribute, while the loop-diagrams become relevant 
for finite $N$, when string effects must be taken into account. This can 
be understood as the loops correspond to closed string exchange which 
gives a contribution that is suppressed in the large $N$ limit. 

Attempts have also been made to generalize the correspondence to 
conformal theories with less supersymmetry and hopefully to 
non-supersymmetric theories. This is achieved by studying the 
compactification of the type IIB superstring on AdS$_{5}\times X_{5}$, 
where $X_{5}$ is a compact Einstein manifold, \ie a positively curved 
compact manifold with metric satisfying $R_{\mu\nu} = \lambda g_{\mu\nu}$, 
with $\lambda>0$. The simplest examples are obtained considering 
$X_{5}$=$S^{5}/\Gamma$, where $\Gamma$ is a discrete subgroup of 
SO(6). In this case $X_{5}$ presents orbifold type singularities, but 
it has locally the geometry of $S^{5}$. In 
\cite{kachrusilv,nekrasov} this models have been discussed in detail 
and it has been shown that taking $N$ coincident D3-branes placed at 
an orbifold singularity one can construct ${\cal N}$=2 and ${\cal 
N}$=1 theories. Such theories have product gauge groups SU($N$)$^{k}$ 
and contain matter fields in the fundamental and adjoint representations.  

A possible approach to the study of non conformal field theories 
through the AdS correspondence has been suggested in 
\cite{wittenterm}. The model proposed in \cite{wittenterm} is based on 
the observation that field theories at finite temperature can be 
described by near-extremal $p$-brane solutions. Non supersymmetric 
theories are then obtained by imposing anti-periodic boundary 
conditions for the fermions on the compactified direction, corresponding 
to the raising of temperature, which explicitly break supersymmetry. 
Qualitative agreement with the 
expectations for QCD at strong coupling has been found following this 
approach. In particular the results from supergravity calculations 
include the area low behaviour of Wilson loops, the presence of a 
mass gap for glueball states, the relation between confinement and 
dual superconductivity and the existence of heavy quark baryonic 
states \cite{wittenterm,wilsbaryons}. Numerical calculations of 
glueball masses have been performed in \cite{glueball} applying the 
same method. The results appear to be in agreement with those 
obtained from lattice simulations. The limit of this approach is that 
for the supergravity approximation to be reliable one must take the 
effective 't Hooft coupling to be large, while the weak coupling 
region is the one relevant for the continuum limit on the lattice. As 
a consequence to actually compare the results from AdS with lattice 
extrapolations to the continuum, it seems to be necessary to extend the 
correspondence to finite $N$, which requires a full string theory 
description in AdS. 

More recently the near-horizon geometry of a stack of D3-branes in 
type 0 theories has been proposed as a candidate for a non 
supersymmetric version of the above discussed correspondence. In these 
models the {\it ``tachyon should\ldots peacefully condense in the 
bulk''} \cite{newpolyakov}. The basic idea suggested by Polyakov in 
\cite{newpolyakov} is that non-supersymmetric theories should have as 
their duals string theories displaying worldsheet supersymmetry, but 
no space-time supersymmetry. This is the property characterizing type 
0A and 0B string theories, which have been introduced in 
\cite{type0}. This theories are constructed by a non-chiral GSO 
projection, which explicitly breaks space-time supersymmetry and leads 
to a spectrum containing no space-time fermions. The bosonic spectrum 
is the same as in the corresponding type II theories in the \NSNS 
sector and contains a doubled set of \RR fields. The interest in 
these theories has been renewed in the systematic construction of 
their open string descendants \cite{biasa}. The inclusion of 
D-branes in such theories has been studied in \cite{biasa,dbratype0} 
and allows to construct non-supersymmetric gauge theories as 
world-volume theories, which do not contain open string tachyons 
\cite{biasa,newpolyakov,klebtse1type0}. In particular in 
\cite{klebtse1type0} a SU($N$) Yang--Mills theory coupled to six real 
massless scalars has been constructed as world-volume theory of $N$ 
coincident D3-branes in type 0B theory. In this model the 5-form field 
strength is not constrained to be self dual and so one can consider 
both electrically and magnetically charged D3-branes. Moreover in the 
D3-brane solution the dilaton is not constant, but acquires a 
spatial (radial) dependence as a consequence of the presence of the 
\NSNS tachyon. In \cite{klebtse1type0} it is argued that this radial 
dependence may be related to a dependence of the coupling constant  
on the energy scale in the gauge theory description. In 
\cite{minahan,klebtse2type0} a field theory with asymptotically free 
behaviour is constructed from a D3-brane type 0 background with 
AdS$_{5}\times S^{5}$ geometry. 

This very interesting issues will not be further addressed here, in the 
following the attention will be focused on the correspondence between 
type IIB string theory in AdS$_{5}\times S^{5}$ and ${\cal N}$=4 super 
Yang--Mills theory.  

\section{Calculations of two- and three-point functions}
\label{2-3-point}
\fancyhead[LO]{{\footnotesize 5.4~~{\it Two- and three-point functions}}}

In this section the prescription for the calculation of correlation 
functions will be discussed in more detail. First the general idea in 
the simplest case of scalar fields and for spinors will be reviewed 
considering two-point functions, then some general considerations on 
the properties of correlators of R-current operators and some 
non-renormalization theorems will be reported. Finally the computation 
of two- and three-point functions of generic chiral primary operators 
will be discussed. 

The computation of two-point functions according to the prescription 
of the previous section goes as follows. The Euclidean action for a  
real massive scalar field in AdS$_{5}$ is 
\begin{equation}
	S = \frac{1}{2} \int d^{5}z \, \sqrt{g} \left[ g^{MN}(\partial_{M} 
	\Phi) (\partial_{N}\Phi) + m^{2} \Phi^{2} \right] \, .
	\label{actionscalarads5}
\end{equation}
As has been explained in \cite{gkp,witten} in order to compute the 
two-point Green function of an operator ${\cal O}[A]$, related by the 
correspondence to $\Phi$ and coupled to a source $J$ in the boundary 
theory, in the supergravity approximation valid for large $N$, one must 
solve the equation of motion obtained from (\ref{actionscalarads5}) with 
the condition that $\Phi(x,\rho)$ approach $J(x)$ as $\rho \to 0$. 
The equation of motion is 
\begin{displaymath}
	\frac{1}{\sqrt{g}} \partial_{M}\left(\sqrt{g}g^{MN}\partial_{N}
	\Phi \right) - m^{2}\Phi = 0 \,.
\end{displaymath}
Then, after substituting into (\ref{actionscalarads5}) the solution 
with the prescribed behaviour in the limit $\rho\to 0$, taking the 
functional derivatives with respect to $J$ one obtains the Green 
function. The relevant solution is of the form of equation (\ref{btob}). 
As a result one gets for the two-point functions
\begin{eqnarray}
	\langle{\cal O}(x){\cal O}(y)\rangle &=& -\int 
	\frac{d^{4}x^{\prime}d\rho}{\rho^{5}} \, \left[ \partial_{M}
	K_{\Delta}(x^{\prime},\rho;x,0) \rho^{2} \partial^{M}
	K_{\Delta}(x^{\prime},\rho;y,0) + \right. \nonumber \\ 
	&& \left. \hspace{1.5cm} + m^{2}K_{\Delta}(x^{\prime},\rho;x,0)
	K_{\Delta}(x^{\prime},\rho;y,0) \right] \, .
	\label{twosubst}
\end{eqnarray}
Then integration by parts and use of the equation of motion 
yields~\footnote{Again this expression must be modified in the case 
$\Delta$=2, see the previous section.}
\begin{equation}
	\langle{\cal O}(x){\cal O}(y)\rangle = \frac{\Gamma(\Delta+1)}
	{\pi^{2}\Gamma(\Delta-2)} \frac{1}{(x-y)^{2\Delta}} \, .
	\label{twopointdimdelta}
\end{equation}
This computation of two point functions has led Witten to 
establish the relation (\ref{masstodimension}) between the AdS `mass' 
$m$ of the field $\Phi$ and the conformal dimension $\Delta$ of the 
corresponding  operator ${\cal O}$. In \cite{witten} the same analysis 
has been carried out for massive vector fields. In general for a 
$p$-form field the relation between the AdS `mass' and the scaling 
dimension of the dual composite operator becomes 
$(\Delta-4+p)(\Delta-p)=(mL)^{2}$, which yields
\begin{displaymath}
	\Delta = 2 \pm \sqrt{4+(mL)^{2}+p(p-4)} \, .
\end{displaymath}
The case of the graviton $g_{MN}$, related to the stress-energy tensor, has 
been considered in \cite{gkp,liutse3graviton}. 

Repeating the above construction for a massive spinor in 
AdS$_{5}$ provides the bulk to boundary propagator for fermionic 
fields \cite{henningsfet}. One starts from the action for `massive' 
spinors in AdS$_{5}$ 
\begin{equation}
	S = \int d^{5}z \, \sqrt{g} \left[ {\overline\Psi}  
	(\Dsm - m) \Psi \right] \, ,
	\label{actionfermionads}
\end{equation}
which yields the Dirac operator acting on spin-$\frac{1}{2}$ fields in 
AdS$_{5}$ 
\begin{equation} 
	\ms{D}\Psi  = {e_{{\hat L}}}^{M} \gamma^{\hat{L}}  
	\left( \partial_{M} +\frac{1}{4} 
	\omega^{\hat{M}\hat{N}}_{M} \gamma_{\hat{M}\hat{N}}
	\right) \Psi = (\rho  \gamma^{\hat 5}  \partial_{5}  
	+ \rho \gamma^{\hat\mu} \partial_{\mu} - 2 \gamma^{\hat 5})
	\Psi \, , 
	\label{dirac-op} 
\end{equation} 
where ${e_{\hat{L}}}^{M}$ is the vielbein, $\omega^{\hat{M}\hat{N}}_{M}$  
the spin connection (hatted indices refer to the tangent
space) and $\gamma^{ \hat\mu}$ are the four-dimensional Dirac matrices.  
Just like in the case of scalar fields in order to compute correlation 
functions one must solve the equations of motion with assigned boundary 
conditions. However the action (\ref{actionfermionads}) vanishes for 
any field configuration which satisfies the equations of motion. 
This implies that the partition function is actually independent of 
the boundary condition on the field $\Psi$, so that it cannot be used 
as a generating functional. To solve this problem a suitable 
modification of the action (\ref{actionfermionads}) by the addition 
of a boundary term has been proposed in \cite{henningsfet}. The 
uniqueness of such a boundary term has been recently discussed in 
\cite{henneaux}.

The suggestion of \cite{henningsfet} is to supplement the action 
(\ref{actionfermionads}) by a term of the form 
\begin{equation}
	S_{{\rm b}} = \lim_{\epsilon\to 0}\int_{M^{\epsilon}} d^{4}x 
	\sqrt{g^{\epsilon}} {\overline\Psi}\Psi \, ,
	\label{boundactferm}
\end{equation}
where $M^{\epsilon}$ is a closed four-dimensional submanifold of 
AdS$_{5}$, which approaches the boundary of AdS$_{5}$ as $\epsilon\to 
0$. In (\ref{boundactferm}) $g^{\epsilon}$ is the metric on 
$M^{\epsilon}$ induced by the one of AdS$_{5}$. This boundary term 
does not modify the equations of motion in the bulk and furthermore is 
invariant under the isometry group of AdS$_{5}$. As a consequence of 
the vanishing of the bulk action on field configurations satisfying 
the equations of motion the entire contribution to the partition 
function, to be employed as generating functional in the AdS/CFT 
correspondence, comes from the boundary term (\ref{boundactferm}). 

This allows to construct the bulk to boundary propagator 
$K^{F}_{\Delta}(x^{\prime},\rho;y,\rho_{0})$ for spinors of mass $m$. 
For fermionic fields, as discussed in \cite{henningsfet}, the relation 
between the AdS `mass' of a supergravity field and the scaling 
dimension, $\Delta$, becomes
\begin{displaymath}
	\Delta = 2 - mL \, .
\end{displaymath}
The case of the dilatino, $\Lambda$, with AdS mass $m=-\frac{3}{2L}$, 
corresponding to an operator of dimension $\Delta=\frac{7}{2}$ will 
be studied in section \ref{dinstantons}. 

In the construction of \cite{henningsfet} there is a subtlety related 
to the apparent arbitrariness of the coefficient of the boundary 
term (\ref{boundactferm}) to be added to to the bulk action. This 
point has been recently clarified in \cite{henneaux}. 
In the derivation of the equations of motion from a variational 
principle only the negative chirality part of the boundary limit of 
the spinor $\Psi$ is fixed. The variation of the positive chirality 
component of $\Psi$ on the boundary generates the boundary term in 
the action, which is uniquely fixed by the requirement that the 
total variation of the action must be zero on the solution of the 
equations of motion in the class of histories defined by the boundary 
conditions. 

To compute three-point functions one must consider interaction terms 
in the supergravity action and solve the corresponding equations of 
motion, with a suitable prescription for the behaviour on the 
boundary. For instance, in the case of scalar fields, 
interaction terms of the form
\begin{eqnarray*}
	{\cal L}^{(A)} &=& \Phi_{1}\Phi_{2}\Phi_{3} \\
	{\cal L}^{(B)} &=& g^{MN}\Phi_{1}(\partial_{M}\Phi_{2})
	(\partial_{N}\Phi_{3}) 
\end{eqnarray*}
give the following contributions to three point functions 
\begin{eqnarray}
	\langle{\cal O}_{1}(x){\cal O}_{2}(y){\cal O}_{3}(z)\rangle^{(A)}
	&=& - \int \frac{d^{4}x^{\prime}d\rho}{\rho^{5}} \, 
	K_{\Delta_{1}}(x^{\prime},\rho;x,0) K_{\Delta_{2}}
	(x^{\prime},\rho;y,0) \cdot \nonumber \\ 
	&& \hspace{2cm} \cdot K_{\Delta_{3}}(x^{\prime},\rho;z,0)
	\nonumber \\
	\langle{\cal O}_{1}(x){\cal O}_{2}(y){\cal O}_{3}(z)\rangle^{(B)} 
	&=& - \int \frac{d^{4}x^{\prime}d\rho}{\rho^{5}} \, 
	K_{\Delta_{1}}(x^{\prime},\rho;x,0) \partial_{M} K_{\Delta_{2}}
	(x^{\prime},\rho;y,0) \cdot \nonumber \\
	&& \hspace{2cm} \cdot \rho^{2} \partial^{M} K_{\Delta_{3}}
	(x^{\prime},\rho;z,0) \, ,
	\label{scalar3pointcorr} 
\end{eqnarray}
where $\Delta_{1}$, $\Delta_{2}$ and $\Delta_{3}$ are the dimensions 
of the operators coupled to $\Phi_{1}$, $\Phi_{2}$ and $\Phi_{3}$ 
respectively. Equations (\ref{scalar3pointcorr}) are obtained, 
analogously to the expressions for two-point functions, by solving the 
equations of motion in the presence of the above interaction terms and 
then substituting back into the action, whose exponential is used as 
generating functional. To derive (\ref{scalar3pointcorr}) one must 
solve the equations recursively as a series in the coupling constant 
$g_{s}$. To generate three-point functions it is sufficient to stop to 
the lowest order, so that the bulk to boundary propagators, 
$K_{\Delta_{i}}$ in (\ref{scalar3pointcorr}) are the same entering 
two-point functions.

The contributions to three-point functions can be represented through 
a diagram like the one depicted in figure \ref{3pointwitdiagr}, where 
the derivatives, read from the interaction, eventually act on the bulk 
to boundary propagators. 
\begin{figure}[!h]
	\centering
	\includegraphics[width=4.5truecm]{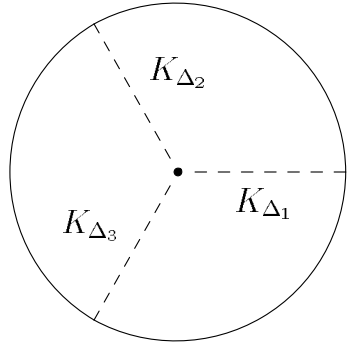}
	\caption{AdS diagram for the correlator of three scalar operators}
	\label{3pointwitdiagr}
\end{figure}

By writing explicitly (\ref{scalar3pointcorr}) one immediately sees 
that these correlation functions have the correct form required by 
conformal symmetry. Actually, as will be discussed in the case of 
R-currents, two-point functions in the boundary conformal field theory 
are determined by superconformal symmetry up to a constant, so that 
the comparison with the results obtained from AdS$_{5}$ only allows to 
establish the map between the fields on the two sides of the 
correspondence and to fix the normalizations. Analogously three-point 
functions are strongly constrained. Their supergravity calculation  
in AdS$_{5}$ and comparison with field theory only provides a check of 
the normalization constants (OPE coefficients). A completely 
non-trivial check of the proposed correspondence requires the 
computation of four- or higher-point functions, that will be examined 
in subsequent sections.

The R-symmetry conserved currents of ${\cal N}$=4 super Yang--Mills 
theory were given in equation (\ref{currentdef1}). The two-point 
function $\langle{\cal J}^{a}_{\mu}{\cal J}^{b}_{\nu}\rangle$ is 
completely fixed by superconformal invariance. The dimension of the 
operators ${\cal J}^{a}_{\mu}$ is $\Delta$=3 and the form of the 
correlator of two currents is \cite{freedman}
\begin{displaymath}
	\langle{\cal J}^{a}_{\mu}(x){\cal J}^{b}_{\nu}(y)\rangle =
	c \frac{\delta^{ab}}{(2\pi)^{4}} \left(\delta_{\mu\nu} \Box -
	\partial_{\mu}\partial_{\nu} \right) \frac{1}{(x-y)^{4            }} \, ,
\end{displaymath}
where $c$ is a positive constant, which is only determined by one-loop 
corrections thanks to a non-renormalization theorem. $c$ is the 
central charge of the ${\cal J}{\cal J}$ OPE and equals the trace 
anomaly. Since the latter is related by ${\cal N}$=1 supersymmetry to 
the R-current anomaly which is one-loop exact $c$ is determined by 
the one-loop result as well. In the case of ${\cal N}$=4 super 
Yang--Mills with gauge group SU($N$) one finds \cite{freedman}
 \begin{displaymath}
 	c = \frac{1}{2} (N^{2}-1) \, .
 \end{displaymath}
The three point function of currents is also strongly constrained 
by superconformal symmetry; it contains a normal parity part and an 
abnormal parity part \cite{freedman}
\begin{eqnarray*}
	&& \langle{\cal J}^{a}_{\mu}(x){\cal J}^{b}_{\nu}(y){\cal 
	J}^{c}_{\lambda}(z)\rangle = \langle{\cal J}^{a}_{\mu}(x)
	{\cal J}^{b}_{\nu}(y){\cal J}^{c}_{\lambda}(z)\rangle_{+} + 
	\langle{\cal J}^{a}_{\mu}(x){\cal J}^{b}_{\nu}(y){\cal 
	J}^{c}_{\lambda}(z)\rangle_{-} = \\
	&& = f^{abc} \left[k_{1}^{(+)} D_{\mu\nu\lambda}(x,y,z) +
	k_{2}^{(+)} C_{\mu\nu\lambda}(x,y,z) \right] + k^{(-)}
	d^{abc} M_{\mu\nu\lambda}(x,y,z) \, ,
\end{eqnarray*}
where $D_{\mu\nu\lambda}$, $C_{\mu\nu\lambda}$ and 
$M_{\mu\nu\lambda}$ are known tensor functions and the $f^{abc}$ and 
$d^{abc}$ SU($N$) symbols are defined by 
$\tr\left(T^{a}T^{b}T^{c}\right) = \frac{1}{4} (if^{abc}+d^{abc})$. 
$k_{1}^{(+)}$, $k_{2}^{(+)}$ and $k^{(-)}$ are numerical constants; 
$k_{1}^{(+)}$ is fixed by the Ward identity connecting two- and 
three-point functions, which relates it to $c$ by 
$k_{1}^{(+)}$=$\frac{c}{16\pi^{2}}$, $k_{2}^{(+)}$ is independent and 
it is not protected by general non-renormalization theorems. The third 
constant, $k^{(-)}$, is controlled by the Adler--Badeen theorem which 
implies that it is one-loop exact. The same non-renormalization 
properties of the R-currents should of course hold true for the other 
composite operators in the current multiplet. 

The computation of two- and three-point functions of chiral primary 
operators (CPO's) will now be reviewed following \cite{sei,freedman}. 
The chiral primary operators that will be considered are of the form
\begin{equation}
	{\cal O}^{I}_{\ell} = \gym^{\ell}t^{I}_{i_{1}\ldots i_{\ell}}
	\tr\left(\varphi^{i_{1}}(x)\ldots \varphi^{i_{\ell}}(x)\right) \, ,
	\label{cpo}
\end{equation}
where $\varphi^{i}$ are the six real matrix-valued scalar fields of 
${\cal N}$=4 supersymmetric Yang--Mills theory and $t^{I}_{i_{1}\ldots
i_{\ell}}$ is a totally symmetric (traceless) rank $\ell$ tensor of 
SU(4)$\sim$SO(6), whose normalization is chosen such that 
$t^{I_{1}}_{i_{1}\ldots i_{\ell}}t^{I_{2}\,i_{1}\ldots i_{\ell}}=
\delta^{I_{1}I_{2}}$. The free propagator for the fields $\varphi^{i}$ 
is 
\begin{displaymath}
	\langle\varphi^{i}_{a}\varphi^{j}_{b}\rangle = 
	\frac{\delta^{ij}\delta_{ab}}{(2\pi)^{2}} \frac{1}{(x-y)^{2}}\, .
\end{displaymath}

Chiral primary operators (\ref{cpo}) belong to representations of the 
SU(4) R-symmetry group characterized by Dynkin labels $(0,\ell,0)$, they 
are Lorentz scalars, \ie they belong to the representation 
$j_{1}$=$j_{2}$=0 of the Euclidean Lorentz group 
SO(4)=SU(2)$\times$ SU(2), and their conformal dimension is 
$\Delta$=$\ell$. The class of CPO's comprises the operators ${\cal Q}^{ij}$ 
and in particular ${\cal Q}$ and $\overline{{\cal Q}}$ considered in 
section \ref{inst-4-point}. 

In general operators involved in the AdS/CFT correspondence have a
conformal dimension $\Delta$ that is not an integer. Only operators 
belonging to short supermultiplets of the SU(2,2$|$4) superconformal 
group, which are dual to the massless AdS supergravity fields and 
their KK descendants, have a dimension that is protected 
under renormalization \cite{ferstr98}. In supersymmetric 
field theories there are chiral primary operators, like hose 
considered here, whose dimensions are integer and protected, even if 
they are not conserved currents. In the case  of ${\cal N}$=4 super 
Yang--Mills theory under consideration the conformal dimension of a 
generic operator satisfies the bound
\begin{equation}
	\Delta\geq q = \ell \, ,
	\label{protopbound}
\end{equation}
where $q$ is the charge of the field under a U(1) subgroup of the SU(4) 
R-symmetry group. (\ref{protopbound}) is a ``BPS-type'' bound, which 
follows from the superconformal algebra (\ref{supconfalg}). 

As already discussed in the previous section the supergravity fields 
and the Kaluza--Klein states in $d$=5 ${\cal N}$=8 gauged 
supergravity belong to short multiplets of the super-isometry group 
SU(2,2$|$4). The same is true for their counterparts in ${\cal N}$=4 
Super Yang--Mills theory (where SU(2,2$|$4) is the superconformal 
group). The CPO's introduced above play a central r\^ole in the 
construction of these multiplets. As explained in \cite{ferstr98} all 
these multiplets can be obtained starting from scalar chiral primary 
operators like those in (\ref{cpo}) and resorting to analytic 
superspace. The generic short multiplet is described by a composite 
twisted chiral superfield \cite{howest} of the form 
\begin{displaymath}
    {\cal W}_{(\ell)} = \tr \left(W^{(i_{1}}\ldots 
    W^{i_{\ell})}\right) - {\rm ~traces} \, , 
\end{displaymath}
where the superfields $W^{i}=\frac{1}{2} t^{i}_{AB} W^{[AB]}$ are the 
same introduced in chapter \ref{cap2}, equations (\ref{wrealcond}) and 
(\ref{wchiralcond}).

Massive string states on the contrary are in long multiplets as well 
as their duals, which are non chiral primary operators. An example of 
operators of this kind is represented by the ${\cal N}$=4 embedding of 
the Konishi multiplet \cite{konimult,ferstr98}, which, in terms of 
${\cal N}$=1 superfields, is 
\begin{equation}
	\Sigma = \Phi^{I}e^{V}\Phi_{I}^{\dagger} \, ,
	\label{konishimult}
\end{equation}
where $\Phi^{I}$, $I=1,2,3$ are the ${\cal N}$=1 chiral multiplets 
and $V$ the ${\cal N}$=1 vector multiplet of ${\cal N}$=4 super 
Yang--Mills theory. 

In \cite{sei} two- and three-point Green functions of CPO's have been 
computed using type IIB supergravity in AdS$_{5}$ and Witten's 
prescription. The result is 
\begin{eqnarray}
	\langle{\cal O}^{I_{1}}_{\ell}(x){\cal O}^{I_{2}}_{\ell}(y)
	\rangle &=& \hat{g}^{2\ell} \frac{\ell}{(2\pi)^{2\ell}}
	\delta^{I_{1}I_{2}}\frac{1}{(x-y)^{2\ell}} 
	\nonumber \\
	\langle{\cal O}^{I_{1}}_{\ell_{1}}(x){\cal O}^{I_{2}}_{\ell_{2}}(y)
	{\cal O}^{I_{3}}_{\ell_{3}}(z)\rangle &=& 
	\frac{\hat{g}^{\Sigma}}{N}
	\frac{\ell_{1}\ell_{2}\ell_{3}}{(2\pi)^{\Sigma}}t^{I_{1}I_{2}I_{3}}
	\frac{1}{(x-y)^{2\alpha_{3}}(y-z)^{2\alpha_{1}}(x-z)^{2\alpha_{2}}}
	\nonumber \, , 
\end{eqnarray}
where $\Sigma$=$\ell_{1}+\ell_{2}+\ell_{3}$ and the tensor 
$t^{I_{1}I_{2}I_{3}}$ is the unique invariant that can be constructed 
contracting $\alpha_{1}=\frac{\ell_{2}+\ell_{3}-\ell_{1}}{2}$ indices 
between $t^{I_{2}}_{i_{1}\ldots i_{\ell_{2}}}$ and $t^{I_{3}}_{i_{1}\ldots 
i_{\ell_{3}}}$ and $\alpha_{2}=\frac{\ell_{1}+\ell_{3}-\ell_{2}}{2}$ and 
$\alpha_{3}=\frac{\ell_{1}+\ell_{2}-\ell_{3}}{2}$ respectively between 
the other two pairs. After rescaling the CPO's according to 
\begin{displaymath}
	{\cal O}^{I}_{\ell} \longrightarrow \tilde{{\cal O}}^{I}_{\ell} = 
	\frac{(2\pi)^{\ell}}{\hat{g}^{\ell}\sqrt{\ell}}{\cal O}^{I}_{\ell} \, ,
\end{displaymath}
the above correlators are found to agree perfectly with the free field 
calculation in ${\cal N}$=4 super Yang--Mills. 

This result is expected to be valid only in the large $N$ limit at fixed 
and large $\hat{g}$, when the Yang--Mills coupling $\gym$ goes to zero. 
However in \cite{freedhokskib} it was proved that the one-loop 
corrections to the above correlators are zero. 
As a consequence the AdS calculation agrees with the {\em exact} 
perturbative contribution to two- and three-point functions of CPO's 
for arbitrary $\gym$. This strongly suggests that the conjectured 
correspondence should be valid at finite $N$ as well. In 
\cite{freedhokskib} a component formulation in terms of ${\cal N}$=1 
multiplets is employed. The chiral primary operators that are 
considered are of the form 
\begin{eqnarray}
	\tr\big(X^{\ell}\big) &=& \tr\left(\varphi^{I_{1}}\ldots
	\varphi^{I_{\ell}}\right)  \\
	\tr\big({X^{\dagger}}^{\ell}\big) &=& \tr\left(
	\varphi^{\dagger}_{I_{1}}\ldots\varphi^{\dagger}_{I_{\ell}}
	\right)  \, ,
	\label{n1cpos}
\end{eqnarray}
where $\varphi^{I}$ and $\varphi^{\dagger}_{I}$, with $I=1,2,3$, are the 
complex scalar fields belonging to the chiral multiplets. The action 
of the model in this formulation was given in equation (\ref{n1sym4}).

The proof of the vanishing of the one-loop corrections to correlators 
of CPO's given in \cite{freedhokskib} is based on a 
non-renormalization `theorem' \cite{ansfreed} for the two-point function 
\begin{displaymath}
	\langle\tr\big(X^{2}\big)\tr\big({X^{\dagger}}^{2}\big)\rangle =
	\hspace{2cm} \raisebox{-0.92cm}{
  \begin{fmffile}{X2X2}
   \begin{fmfgraph*}(70,60) 
    \fmfleft{i1} 
    \fmfright{o1}
    \fmf{phantom,left=0.6,tension=0.5,tag=1}{i1,o1}
    \fmf{phantom,right=0.6,tension=0.5,tag=2}{i1,o1}
    \fmfv{l=$\tr \big(X^{2}\big)$,l.a=180}{i1}
    \fmfv{l=$\tr \big({X^{\dagger}}^{2}\big)$,l.a=0}{o1}
    \fmfdot{i1,o1}
    \fmffreeze
    \fmfipath{p[]}
    \fmfiset{p1}{vpath1(__i1,__o1)}
    \fmfiset{p2}{vpath2(__i1,__o1)}
    \fmfi{fermion}{subpath (0,length(p1)) of p1}
    \fmfi{fermion}{subpath (0,length(p2)) of p2}
   \end{fmfgraph*}   
  \end{fmffile}}
\end{displaymath}
and on clever colour combinatorics. The non-renormalization theorem 
says that the one loop correction to the above correlator vanishes.  
This is a consequence of the following relations. The one-loop 
corrections to $\langle\tr\big(X^{2}\big)\tr\big({X^{\dagger}}^{2}
\big)\rangle$ come from the diagrams 
\begin{equation}
  \raisebox{-0.92cm}{
  \begin{fmffile}{X2X2a}
   \begin{fmfgraph*}(70,60) 
    \fmfleft{i1} 
    \fmfright{o1}
    \fmf{phantom,left=0.5,tension=0.5,tag=1}{i1,o1}
    \fmf{phantom,right=0.5,tension=0.5,tag=2}{i1,o1}
    \fmfdot{i1,o1}
    \fmffreeze
    \fmfipath{p[]}
    \fmfiset{p1}{vpath1(__i1,__o1)}
    \fmfiset{p2}{vpath2(__i1,__o1)}
    \fmfi{fermion}{subpath (0,length(p1)/2) of p1}
    \fmfi{fermion}{subpath (length(p1)/2,length(p1)) of p1}
    \fmfi{fermion}{subpath (0,length(p2)/2) of p2}
    \fmfi{fermion}{subpath (length(p2)/2,length(p2)) of p2}
    \fmfi{boson}{point length(p1)/2 of p1 --
                   point length(p2)/2 of p2}
    \fmfiv{d.sh=circle,d.fill=full,d.si=1.5mm}{point 
           length(p1)/2 of p1}
    \fmfiv{d.sh=circle,d.fill=full,d.si=1.5mm}{point 
           length(p2)/2 of p2}
   \end{fmfgraph*}   
  \end{fmffile} \hspace{0.3cm}} + \hspace{0.3cm}
  \raisebox{-0.92cm}{
  \begin{fmffile}{X2X2b}
   \begin{fmfgraph*}(70,60) 
    \fmfleft{i1} 
    \fmfright{o1}
    \fmf{phantom,left=0.6,tension=0.4,tag=1}{i1,v1}
    \fmf{phantom,left=0.6,tension=0.4,tag=2}{v1,o1}
    \fmf{phantom,right=0.6,tension=0.4,tag=3}{i1,v1}
    \fmf{phantom,right=0.6,tension=0.4,tag=4}{v1,o1}
    \fmfdot{i1,v1,o1}
    \fmffreeze
    \fmfipath{p[]}
    \fmfiset{p1}{vpath1(__i1,__v1)}
    \fmfiset{p2}{vpath2(__v1,__o1)}
    \fmfiset{p3}{vpath3(__i1,__v1)}
    \fmfiset{p4}{vpath4(__v1,__o1)}
    \fmfi{fermion}{subpath (0,length(p1)) of p1}
    \fmfi{fermion}{subpath (0,length(p2)) of p2}
    \fmfi{fermion}{subpath (0,length(p3)) of p3}
    \fmfi{fermion}{subpath (0,length(p4)) of p4}    
   \end{fmfgraph*}   
  \end{fmffile}} \hspace{0.5cm} = ~ c \frac{B(x,y)}{(x-y)^{4}}
  \label{x2x2bcorr}
\end{equation}
\begin{equation}
  \raisebox{-0.92cm}{
  \begin{fmffile}{X2X2c}
   \begin{fmfgraph*}(70,60) 
    \fmfleft{i1} 
    \fmfright{o1}
    \fmf{phantom,left=0.5,tension=0.5,tag=1}{i1,o1}
    \fmf{phantom,right=0.5,tension=0.5,tag=2}{i1,o1}
    \fmfdot{i1,o1}
    \fmffreeze
    \fmfipath{p[]}
    \fmfiset{p1}{vpath1(__i1,__o1)}
    \fmfiset{p2}{vpath2(__i1,__o1)}
    \fmfi{fermion}{subpath (0,3length(p1)/7) of p1}
    \fmfi{fermion}{subpath (4length(p1)/7,length(p1)) of p1}
    \fmfi{fermion}{subpath (0,length(p2)) of p2}
    \fmfiv{d.sh=circle,d.si=5mm,d.fi=shaded}{point length(p1)/2 of p1}
   \end{fmfgraph*}      
  \end{fmffile} \hspace{0.3cm}} + \hspace{0.3cm}
  \raisebox{-0.92cm}{
  \begin{fmffile}{X2X2d}
   \begin{fmfgraph*}(70,60) 
    \fmfleft{i1} 
    \fmfright{o1}
    \fmf{phantom,left=0.5,tension=0.5,tag=1}{i1,o1}
    \fmf{phantom,right=0.5,tension=0.5,tag=2}{i1,o1}
    \fmfdot{i1,o1}
    \fmffreeze
    \fmfipath{p[]}
    \fmfiset{p1}{vpath1(__i1,__o1)}
    \fmfiset{p2}{vpath2(__i1,__o1)}
    \fmfi{fermion}{subpath (0,length(p1)) of p1}
    \fmfi{fermion}{subpath (0,3length(p2)/7) of p2}
    \fmfi{fermion}{subpath (4length(p2)/7,length(p2)) of p2}
    \fmfiv{d.sh=circle,d.si=5mm,d.fi=shaded}{point length(p2)/2 of p2}
   \end{fmfgraph*}   
  \end{fmffile}} \hspace{0.5cm} = ~ 2c \frac{A(x,y)}{(x-y)^{4}}
  \label{x2x2acorr} 
\end{equation}

\vspace*{0.5cm}
\noindent
Notice that in the formulation of (\ref{n1sym4}), that is used in 
\cite{freedhokskib}, the Wess--Zumino gauge is exploited, so that the 
one loop corrections to the propagator, represented by the bubbles in 
(\ref{x2x2acorr}), are not zero. Moreover the second term in 
(\ref{x2x2bcorr}) is the contribution of the `$D$-term', while the 
`$F$-term' does not enter. Both $B(x,y)$ and $A(x,y)$ are logarithmically 
divergent (see section \ref{wzproblems}), but one can prove the 
non-renormalization theorem
\begin{equation}
	B(x,y) + 2 A(x,y) = 0 \, .
	\label{2phi2nonrenorm}
\end{equation}

The one-loop correction to the two-point function $\langle\tr
\big(X^{\ell}\big) \tr\big({X^{\dagger}}^{\ell}\big)\rangle$ 

\vspace*{1cm}
\hspace*{2cm}
\begin{fmffile}{XkXk}
 \begin{fmfgraph*}(90,80) 
  \fmfleft{i1} 
  \fmfright{o1}
  \fmf{phantom,left=0.7,tension=0.5,tag=1}{i1,o1}
  \fmf{phantom,left=0.25,tension=0.5,tag=2}{i1,o1}
  \fmf{phantom,right=0.3,tension=0.5,tag=3}{i1,o1}
  \fmf{phantom,right=0.5,tension=0.5,tag=4}{i1,o1}
  \fmf{phantom,right=0.7,tension=0.5,tag=5}{i1,o1}
  \fmfv{l=$\tr \big(X^{\ell}\big)$,l.a=180}{i1}
  \fmfv{l=$\tr \big({X^{\dagger}}^{\ell}\big)$,l.a=0}{o1}
  \fmfdot{i1,o1}
  \fmffreeze
  \fmfipath{p[]}
  \fmfiset{p1}{vpath1(__i1,__o1)}
  \fmfiset{p2}{vpath2(__i1,__o1)}
  \fmfiset{p3}{vpath3(__i1,__o1)}
  \fmfiset{p4}{vpath4(__i1,__o1)}
  \fmfiset{p5}{vpath5(__i1,__o1)}
  \fmfi{fermion}{subpath (0,length(p1)) of p1}
  \fmfi{fermion}{subpath (0,length(p2)) of p2}
  \fmfi{fermion}{subpath (0,length(p3)) of p3}
  \fmfi{fermion}{subpath (0,length(p4)) of p4}
  \fmfi{fermion}{subpath (0,length(p5)) of p5}
 \end{fmfgraph*}   
\end{fmffile} \\
\vspace*{0.5cm}

\noindent
comes from the sum of diagrams of the following types

\vspace*{0.7cm}
\hspace*{2cm}
\begin{fmffile}{XkXka}
 \begin{fmfgraph*}(90,80) 
  \fmfleft{i1} 
  \fmfright{o1}
  \fmf{phantom,left=0.7,tension=0.5,tag=1}{i1,o1}
  \fmf{phantom,left=0.25,tension=0.5,tag=2}{i1,o1}
  \fmf{phantom,right=0.3,tension=0.5,tag=3}{i1,o1}
  \fmf{phantom,right=0.5,tension=0.5,tag=4}{i1,o1}
  \fmf{phantom,right=0.7,tension=0.5,tag=5}{i1,o1}
  \fmfv{l=$\tr \big(X^{\ell}\big)$,l.a=180}{i1}
  \fmfv{l=$\tr \big({X^{\dagger}}^{\ell}\big)$,l.a=0}{o1}
  \fmfdot{i1,o1}
  \fmffreeze
  \fmfipath{p[]}
  \fmfiset{p1}{vpath1(__i1,__o1)}
  \fmfiset{p2}{vpath2(__i1,__o1)}
  \fmfiset{p3}{vpath3(__i1,__o1)}
  \fmfiset{p4}{vpath4(__i1,__o1)}
  \fmfiset{p5}{vpath5(__i1,__o1)}
  \fmfi{fermion}{subpath (0,length(p1)) of p1}
  \fmfi{fermion}{subpath (0,3length(p2)/7) of p2}
  \fmfi{fermion}{subpath (4length(p2)/7,length(p2)) of p2}
  \fmfi{fermion}{subpath (0,length(p3)) of p3}
  \fmfi{fermion}{subpath (0,length(p4)) of p4}
  \fmfi{fermion}{subpath (0,length(p5)) of p5}
  \fmfiv{d.sh=circle,d.si=5mm,d.fi=shaded}{point length(p2)/2 of p2}
 \end{fmfgraph*}   
\end{fmffile} 

\vspace*{0.8cm}
\hspace*{2cm}
\begin{fmffile}{XkXkb}
 \begin{fmfgraph*}(90,80) 
  \fmfleft{i1} 
  \fmfright{o1}
  \fmf{phantom,left=0.7,tension=0.5,tag=1}{i1,o1}
  \fmf{phantom,left=0.25,tension=0.5,tag=2}{i1,o1}
  \fmf{phantom,right=0.3,tension=0.5,tag=3}{i1,o1}
  \fmf{phantom,right=0.5,tension=0.5,tag=4}{i1,o1}
  \fmf{phantom,right=0.7,tension=0.5,tag=5}{i1,o1}
  \fmfv{l=$\tr \big(X^{\ell}\big)$,l.a=180}{i1}
  \fmfv{l=$\tr \big({X^{\dagger}}^{\ell}\big)$,l.a=0}{o1}
  \fmfdot{i1,o1}
  \fmffreeze
  \fmfipath{p[]}
  \fmfiset{p1}{vpath1(__i1,__o1)}
  \fmfiset{p2}{vpath2(__i1,__o1)}
  \fmfiset{p3}{vpath3(__i1,__o1)}
  \fmfiset{p4}{vpath4(__i1,__o1)}
  \fmfiset{p5}{vpath5(__i1,__o1)}
  \fmfi{fermion}{subpath (0,length(p1)) of p1}
  \fmfi{fermion}{subpath (0,length(p2)/2) of p2}
  \fmfi{fermion}{subpath (length(p2)/2,length(p2)) of p2}
  \fmfi{fermion}{subpath (0,length(p3)/2) of p3}
  \fmfi{fermion}{subpath (length(p3)/2,length(p3)) of p3}
  \fmfi{fermion}{subpath (0,length(p1)) of p4}
  \fmfi{fermion}{subpath (0,length(p1)) of p5}
  \fmfi{boson}{point length(p2)/2 of p2 --
               point length(p3)/2 of p3}
 \fmfiv{d.sh=circle,d.fill=full,d.si=1.5mm}{point length(p2)/2 of p2}
 \fmfiv{d.sh=circle,d.fill=full,d.si=1.5mm}{point length(p3)/2 of p3}
 \end{fmfgraph*}   
\end{fmffile} 

\newpage
\hspace*{2cm}
\begin{fmffile}{XkXkc}
 \begin{fmfgraph*}(90,80) 
  \fmfleft{i1} 
  \fmfright{o1}
  \fmf{phantom,left=0.7,tension=0.5,tag=1}{i1,o1}
  \fmf{phantom,left=0.4,tension=0.5,tag=2}{i1,v1}
  \fmf{phantom,right=0.4,tension=0.5,tag=3}{v1,o1}
  \fmf{phantom,right=0.4,tension=0.5,tag=4}{i1,v1}
  \fmf{phantom,left=0.4,tension=0.5,tag=5}{v1,o1}
  \fmf{phantom,right=0.5,tension=0.5,tag=6}{i1,o1}
  \fmf{phantom,right=0.7,tension=0.5,tag=7}{i1,o1}
  \fmfv{l=$\tr \big(X^{\ell}\big)$,l.a=180}{i1}
  \fmfv{l=$\tr \big({X^{\dagger}}^{\ell}\big)$,l.a=0}{o1}
  \fmfdot{i1,o1,v1}
  \fmffreeze
  \fmfipath{p[]}
  \fmfiset{p1}{vpath1(__i1,__o1)}
  \fmfiset{p2}{vpath2(__i1,__v1)}
  \fmfiset{p3}{vpath3(__o1,__v1)}
  \fmfiset{p4}{vpath4(__i1,__v1)}
  \fmfiset{p5}{vpath5(__o1,__v1)}
  \fmfiset{p6}{vpath6(__i1,__o1)}
  \fmfiset{p7}{vpath7(__i1,__o1)}
  \fmfi{fermion}{subpath (0,length(p1)) of p1}
  \fmfi{fermion}{subpath (0,length(p2)) of p2}
  \fmfi{fermion}{subpath (0,length(p3)) of p3}
  \fmfi{fermion}{subpath (0,length(p4)) of p4}
  \fmfi{fermion}{subpath (0,length(p5)) of p5}
  \fmfi{fermion}{subpath (0,length(p6)) of p6}
  \fmfi{fermion}{subpath (0,length(p7)) of p7}
 \end{fmfgraph*}   
\end{fmffile} 

\vspace*{0.5cm}
\noindent 
One must consider the sum of contributions in which the `two-particle' 
interaction structures (\ref{x2x2bcorr}) and (\ref{x2x2acorr}) are 
inserted between each pair of lines. In \cite{freedhokskib} it is shown, 
through a combinatoric analysis, that this sum leads to a total one-loop 
correction of the form
\begin{displaymath}
	\langle \tr\big(X^{k}{X^{\dagger}}^{k}\big) \sim 
	kN\frac{[B(x,y)+2A(x,y)]}{(x-y)^{2k}} \, ,
\end{displaymath}
which vanishes thanks to the non-renormalization theorem for the $k$=2 
case.

Analogously one can prove the vanishing of the one-loop correction 
to the three point function $\langle \tr\big(X^{\ell_{1}}\big) 
\tr\big(X^{\ell_{2}}\big) \tr\big({X^{\dagger}}^{\ell_{3}}\big) 
\rangle$

\vspace*{0.5cm}
\hspace*{2cm}
\begin{fmffile}{XkXkXk}
 \begin{fmfgraph*}(90,80) 
  \fmfleft{i1,i2} 
  \fmfright{o1}
  \fmf{phantom,right=0.35,tension=0.5,tag=1}{i1,o1}
  \fmf{phantom,right=0.15,tension=0.5,tag=2}{i1,o1}
  \fmf{phantom,left=0.2,tension=0.5,tag=3}{i1,o1}
  \fmf{phantom,right=0.23,tension=0.5,tag=4}{i2,o1}
  \fmf{phantom,tension=0.5,tag=5}{i2,o1}
  \fmf{phantom,left=0.2,tension=0.5,tag=6}{i2,o1}
  \fmf{phantom,left=0.35,tension=0.5,tag=7}{i2,o1}
  \fmfv{l=$\tr \big(X^{\ell_{1}}\big)$,l.a=180}{i1}
  \fmfv{l=$\tr \big(X^{\ell_{2}}\big)$,l.a=180}{i2}
  \fmfv{l=$\tr \big({X^{\dagger}}^{\ell_{3}}\big)$,l.a=0}{o1}
  \fmfdot{i1,i2,o1}
  \fmffreeze
  \fmfipath{p[]}
  \fmfiset{p1}{vpath1(__i1,__o1)}
  \fmfiset{p2}{vpath2(__i1,__o1)}
  \fmfiset{p3}{vpath3(__i1,__o1)}
  \fmfiset{p4}{vpath4(__i2,__o1)}
  \fmfiset{p5}{vpath5(__i2,__o1)}
  \fmfiset{p6}{vpath6(__i2,__o1)}
  \fmfiset{p7}{vpath7(__i2,__o1)}
  \fmfi{fermion}{subpath (0,length(p1)) of p1}
  \fmfi{fermion}{subpath (0,length(p2)) of p2}
  \fmfi{fermion}{subpath (0,length(p3)) of p3}
  \fmfi{fermion}{subpath (0,length(p4)) of p4}
  \fmfi{fermion}{subpath (0,length(p5)) of p5}
  \fmfi{fermion}{subpath (0,length(p6)) of p6}
  \fmfi{fermion}{subpath (0,length(p7)) of p7}
 \end{fmfgraph*}   
\end{fmffile} 

\vspace*{0.5cm}
To summarize the expressions of two- and three-point functions of 
CPO's in perturbation theory are given by the free-field result and, 
as shown in \cite{sei}, this coincides with what is found in the 
supergravity calculation in AdS$_{5}$. As already remarked 
supersymmetry allows to generalize this result to all the 
composite operators of the current multiplet. In \cite{freedhokskib} 
this has been explicitly verified for a three-point function involving 
the operator ${\cal E}^{AB}$ of equation (\ref{currentdef2}), which is a 
descendant~\footnote{${\cal E}^{AB}$ is a descendant under superconformal 
transformations, but a primary field from the viewpoint of conformal 
symmetry.} of CPO's. Notice that in this computation it is crucial to 
take into account the modification (\ref{enonabelian}) of the 
operator ${\cal E}^{AB}$ in the non Abelian case \cite{bgkr}.

\section{Four-point functions}
\label{4-point}

Unlike two- and three-point functions which, as discussed in the 
previous section, are completely determined by superconformal symmetry 
up to constants, four-point functions are only determined up to an 
arbitrary function of two independent cross-ratios, for instance 
those introduced in section \ref{inst-4-point}
\begin{equation}
	r = \frac{x_{12}^{2}x_{34}^{2}}{x_{13}^{2}x_{24}^{2}}
	\, , \hspace{1.5cm} 
	s = \frac{x_{14}^{2}x_{23}^{2}}{x_{13}^{2}x_{24}^{2}}
	\, .
	\label{crossratios}
\end{equation}
As a consequence the comparison of four-point amplitudes computed in  
supergravity/superstring theory in AdS$_{5}$ with super Yang--Mills 
correlation functions would represent a truly `dynamical' and 
non-trivial check of the correspondence. Such a check has been carried 
out, in an indirect way, for the non-perturbative correction to a 
four-point function in \cite{bgkr} using the exact match of a correlation 
function of sixteen fermionic operators in the same supersymmetry 
multiplet and will be reviewed in the next section. 

The calculation of four-point Green functions of composite operators 
in ${\cal N}$=4 supersymmetric Yang--Mills is rather complicated. 
The application of ${\cal N}$=1 superspace techniques to this problem 
will be examined later. 
\fancyhead[LO]{{\footnotesize 5.5~~{\it Four-point functions}}}

On the supergravity/superstring side the computation of four- and 
higher-point amplitudes with the approach of section \ref{2-3-point} 
presents a new difficulty, with respect to the case of two- and 
three--point functions, related to the existence of two kinds of 
contributions, associated with ``contact'' and ``exchange'' diagrams. 
\begin{figure}[!h]
	\centering
    \begin{minipage}[t]{0.4\linewidth}
    \centering
    \includegraphics[width=4cm]{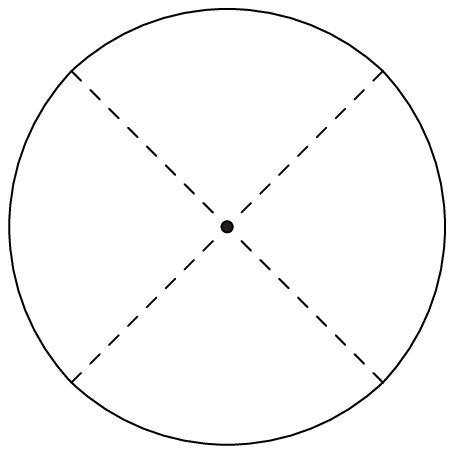}
    \caption{Contact 4-point amplitude}
    \label{4contact}
    \end{minipage}
	\begin{minipage}[t]{0.1\linewidth} ~
	\end{minipage}
    \begin{minipage}[t]{0.4\linewidth}
	\centering
    \includegraphics[width=4cm]{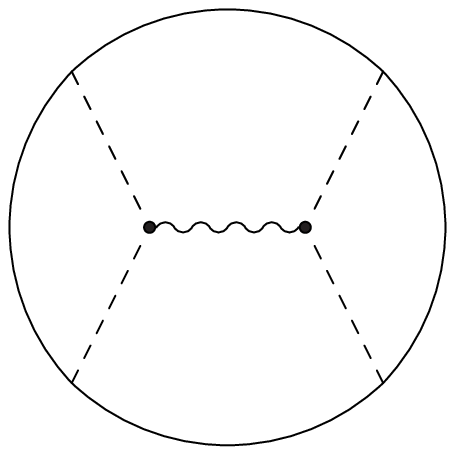}
    \caption{Exchange 4-point amplitude}
    \label{4exchange}
    \end{minipage}
\end{figure}

\noindent
The former, see figure \ref{4contact}, are completely analogous to the 
above discussed contributions to two- and three-point functions, while 
the latter correspond to diagrams like the one in figure \ref{4exchange} 
and, as a new feature, involve also {\em bulk to bulk propagators} in 
AdS$_{5}$. The diagram of figure \ref{4exchange} represents an 
$s$-channel amplitude, analogously one must take into account $t$ 
and $u$ channels. Moreover one must consider the possible exchange of 
all the states, for finite $N$ also string states, coupled to the 
external vertices in the supergravity/superstring action. The 
corresponding amplitude has the form
\begin{eqnarray*}
	&& A(x_{1},x_{2},x_{3},x_{4}) = \int 
	\frac{d^{4}x^{\prime}d\rho^{\prime}}{{\rho^{\prime}}^{5}}
	\frac{d^{4}x^{\prime\prime}d\rho^{\prime\prime}}
	{{\rho^{\prime\prime}}^{5}} \, \left[ K(x^{\prime},
	\rho^{\prime};x_{1},0) \cdot \right. \\
	&& \left. \cdot K(x^{\prime},\rho^{\prime};x_{2},0) G(x^{\prime},
	\rho^{\prime};x^{\prime\prime},\rho^{\prime\prime}) K(x^{\prime
	\prime},\rho^{\prime\prime};x_{3},0) K(x^{\prime\prime},
	\rho^{\prime\prime};x_{4},0) \right] \, ,
\end{eqnarray*}
where $G(z^{\prime};z^{\prime\prime})$ denotes the bulk to bulk 
propagator and, for compactness of notation, indices and possible 
derivatives acting on the various lines, depending the explicit form 
of the trilinear couplings, have been suppressed. 

The first computations of this kind have been presented in 
\cite{liutse4point1,freed4point1} for scalar four-point functions. 
The amplitudes considered are 
\begin{eqnarray}
	&& A_{\phi\phi\phi\phi}(x_{1},x_{2},x_{3},x_{4}) \nonumber \\
	&& A_{\phi\chi\phi\chi}(x_{1},x_{2},x_{3},x_{4}) \label{scalaramplit} \\
	&& A_{\chi\chi\chi\chi}(x_{1},x_{2},x_{3},x_{4}) \, , \nonumber
\end{eqnarray}
where $\phi$ and $\chi$=$C^{(0)}$ are the 
the two massless scalars of the type IIB superstring, the dilaton and the 
axion respectively. The operator corresponding to $\phi$ is the 
``glueball field'' ${\cal G}=\tr\left(F^{2}\right)$, whereas $\chi$ 
corresponds to ${\cal C} = \tr\left(F\tilde{F}\right)$, so that the 
amplitudes (\ref{scalaramplit}) should give the Green functions
\begin{eqnarray}
	&& \langle{\cal G}(x_{1}){\cal G}(x_{2}){\cal G}(x_{3})
	{\cal G}(x_{4})\rangle \nonumber \\
	&& \langle{\cal G}(x_{1}){\cal C}(x_{2}){\cal G}(x_{3})
	{\cal C}(x_{4})\rangle \label{scalargreen4} \\
	&& \langle{\cal C}(x_{1}){\cal C}(x_{2}){\cal C}(x_{3})
	{\cal C}(x_{4})\rangle  \nonumber
\end{eqnarray} 
in ${\cal N}$=4 super Yang--Mills. 

The relevant part of the type IIB supergravity action in AdS$_{5}$ is
\begin{equation}
	S = \frac{1}{2\kappa_{0}^{2}} \int d^{5}z \sqrt{g} \left[ R -\frac{1}{2} 
	\left(\partial\phi\right)^{2}-\frac{1}{2}e^{2\phi}\left(\partial 
	\chi\right)^{2} \right] \, .
	\label{adsscalaraction}
\end{equation}
Notice that since the scalar fields $\phi$ and $\chi$ correspond 
to zeroth-order spherical harmonics, one can forget about the $S^{5}$ factor 
and consider $\phi=\phi(z)$ and $\chi=\chi(z)$ as depending only on 
the AdS$_{5}$ coordinate.

In the case of the $A_{\phi\phi\phi\phi}$ amplitude the only exchange 
diagrams that contribute involve the graviton propagator and at this 
order there are no contact terms, in the $A_{\phi\chi\phi\chi}$ amplitude 
both the graviton and the axion propagators enter, while 
$A_{\chi\chi\chi\chi}$ receives contribution from the exchange of gravitons 
and dilatons. Furthermore contact diagrams contribute to 
$A_{\phi\chi\phi\chi}$ and $A_{\chi\chi\chi\chi}$. 

The fundamental difficulty in the computation of the above amplitudes 
is the construction of the propagators in the AdS$_{5}$ space. An 
interesting result has been proved in \cite{freed4point1}. It has 
been shown that the graphs corresponding to the $\chi$-exchange in the 
amplitude $A_{\phi\chi\phi\chi}$ actually coincide with the contact 
graph. This is a consequence of the fact that the scalar bulk to bulk 
propagator $G(z;z^{\prime})$ satisfies an equation of the form
\begin{displaymath}
	\Delta_{z} G(z;z^{\prime}) = \delta(z-z^{\prime}) 
\end{displaymath}
in AdS$_{5}$. Integrating by parts in the exchange amplitude, 
one obtains exactly a factor $\Delta_{z} G(z;z^{\prime})$ and hence 
one of the two integrations over AdS$_{5}$ can be performed using 
the $\delta$-function. However this result does not hold in general 
and so the knowledge of the propagators for all the relevant fields 
in AdS$_{5}$ is necessary. 

The AdS$_{5}$ propagator for scalar fields has been computed in 
\cite{freddhokscalar}, while the propagator for gauge fields was given 
in \cite{freddhokvector}. The graviton propagator in AdS$_{5}$ has 
been computed only very recently \cite{freedmangraviton}. 

As already remarked in section \ref{typeiib} the type IIB supergravity 
action contains terms that receive also non-perturbative corrections 
from D-instantons. In particular D-instantons contribute to the ${\cal 
R}^{4}$ term (\ref{r4term}) and to the $\Lambda^{16}$ term 
(\ref{lambint}). As argued in \cite{banksgreen,bgkr} the 
corresponding contributions to supergravity/superstring amplitudes in 
AdS$_{5}$ should be compared with instanton contributions to the 
correlation functions of the associated composite operators in ${\cal 
N}$=4 supersymmetric Yang--Mills theory. This comparison has been 
achieved in \cite{bgkr} for the correlation functions presented in the 
previous chapter (see sections \ref{inst-4-point} and 
\ref{inst-8-16-point}). The results of \cite{bgkr} will be reviewed in 
the following sections. Notice that for the non-perturbative contribution 
to scattering amplitudes, with the ``minimal'' number of insertions 
leading to a non-vanishing result, {\em only contact terms are 
relevant}~\footnote{Two- and three-point functions do not receive 
non-perturbative corrections at lowest order.}. As a consequence it is 
easy to single out the contributions to be compared with the 
Yang--Mills computations. In particular notice that there is no 
ambiguity related to possible cancellations among contact and exchange 
integrals. Moreover in general the contact non-perturbative contribution 
to an amplitude has schematically the form 
\begin{displaymath}
	\int \frac{d^{4}xd{\rho}}{\rho^{5}} K_{\Delta_{1}}K_{\Delta_{2}}
	\ldots K_{\Delta_{n}} \, ,
\end{displaymath}
which exactly coincides with the form expected for an instanton 
contribution in ${\cal N}$=4 Yang--Mills theory after the integrations 
over the fermionic collective coordinates have been performed, since 
the one-instanton moduli space is exactly AdS$_{5}$ 
\cite{bgkr,mattis1,mattis2,mattislong} and the instanton 
`profile' $\tr\left({\overline F}^{2}_{\mu\nu}\right)$ is 
identical to the bulk to boundary propagator $K_{\Delta}$ for $\Delta=4$. 
This implies that the comparison with the Yang--Mills calculation can be 
performed {\em before} computing the integrals. These considerations will 
be exploited in the next section, where specific examples will be 
considered. 

Some remarks on the computation of perturbative contributions to 
four-point Green functions will now be made. 
The calculation of correlation functions of composite operators in 
${\cal N}$=4 super Yang--Mills theory can be simplified using 
superspace techniques. Although the explicit calculations have not yet 
been completed the application of this method to chiral primary 
operators will be briefly sketched. 

The approach followed in \cite{freedhokskib} and reviewed in the 
previous section for the computation of three-point functions of 
CPO's can be generalized to correlation functions of ${\cal N}$=1 
superfields whose lowest components are chiral primary operators of 
the form of equation (\ref{n1cpos}). The superfields to be considered 
are of the form
\begin{equation}
	\tr \left({\Phi^{I}}^{\,\ell}\right) \, , \hspace{1cm}
	\tr \left({\Phi^{\dagger}_{I}}^{\,\ell}\right)  \, ,
	\hspace{1.5cm} I=1,2,3 \, . 
	\label{n1supercpo}
\end{equation}
The simplest case, $\ell$=2, will be discussed here. 
Exploiting the SU(3) symmetry, that is manifest in the ${\cal N}$=1 
formulation, one can restrict to Green functions with fixed values of 
the indices $I$, which allows to simplify the calculation. 
The chiral primary operators with $\ell = 2$ are the scalars ${\cal 
Q}^{ij}$ in (\ref{currentdef2}). The operators ${\cal Q}^{ij}$ belong to 
the {\bf 20}$_{\mathbb{R}}$ of the SU(4) R-symmetry group, which 
decomposes with respect to the SU(3)$\times$U(1) subgroup that is 
manifest in the ${\cal N}$=1 formulation according to 
\begin{equation}
	{\bf 20}_{\mathbb{R}} \supset {\overline{\bf 6}}_{-\frac{4}{3}} + 
	{\bf 6}_{\frac{4}{3}} + {\bf 8}_{0} \, ,
	\label{20decomp}
\end{equation}
where the subscript on the right hand side refers to the U(1) charge. 
The corresponding decomposition of ${\cal Q}^{ij}$ reads 
\begin{eqnarray}
	&& {\cal O}^{IJ}_{{\bf 6}} = \tr \left(\Phi^{I}\Phi^{J}\right) \, 
	\in \, {\bf 6} \, , \hspace{1cm} {\cal O}_{{\overline {\bf 6}}\, 
	IJ} = \tr \left(\Phi_{I}^{\dagger}\Phi_{J}^{\dagger} \right) \, 
	\in \, {\overline {\bf 6}} \, , \nonumber \\
	&& {\cal O}^{I}_{{\bf 8}\, J} = \tr\left( \Phi^{I}e^{V}
	\Phi^{\dagger}_{J} - \frac{\delta^{I}{}_{J}}{3} \Phi^{K} e^{V} 
	\Phi^{\dagger}_{K} \right) \, \in \, {\bf 8} \, .
	\label{superfielddecomp}
\end{eqnarray}
For the sake of simplicity the attention will be restricted to Green 
functions of operators ${\cal O}^{IJ}_{{\bf 6}}$ and ${\cal O}_{{\overline 
{\bf 6}}\,IJ}$. 

Hence one type of four-point functions that one must consider is of the 
form 
\begin{displaymath}
	G^{(a)}_{I} = 
	\langle \tr\left({\Phi_{I}^{\dagger}}^{\,2}\right) 
	\tr\left({\Phi^{I}}^{\,2}\right)\tr\left({\Phi_{I}^{\dagger}}^{\,2}
	\right) \tr\left({\Phi^{I}}^{\,2}\right) \rangle \, , 
\end{displaymath}
with fixed $I$. There are two different tree-level contributions, 
corresponding to the diagrams (the notations are the same as in 
chapter \ref{cap3})

\vspace*{1cm}
\hspace*{0.3cm}
\begin{fmffile}{P2P+2P2P+2a}
 \begin{fmfgraph*}(100,80) 
  \fmfleft{i1,i2} 
  \fmfright{o1,o2}
  \fmf{phantom,tension=0.5,tag=1}{i1,i2}
  \fmf{phantom,tension=0.5,tag=2}{i1,o1}
  \fmf{phantom,tension=0.5,tag=3}{o2,o1}
  \fmf{phantom,tension=0.5,tag=4}{o2,i2}
  \fmfv{l=$\tr \left({\Phi^{I}}^{2}\right)$,l.a=180}{i1}
  \fmfv{l=$\tr \left({\Phi_{I}^{\dagger}}^{2}\right)$,l.a=180}{i2}
  \fmfv{l=$\tr \left({\Phi_{I}^{\dagger}}^{2}\right)$,l.a=0}{o1}
  \fmfv{l=$\tr \left({\Phi^{I}}^{2}\right)$,l.a=0}{o2}
  \fmfdot{i1,i2,o1,o2}
  \fmffreeze
  \fmfipath{p[]}
  \fmfiset{p1}{vpath1(__i1,__i2)}
  \fmfiset{p2}{vpath2(__i1,__o1)}
  \fmfiset{p3}{vpath3(__o2,__o1)}
  \fmfiset{p4}{vpath4(__o2,__i2)}
  \fmfi{fermion}{subpath (0,length(p1)) of p1}
  \fmfi{fermion}{subpath (0,length(p2)) of p2}
  \fmfi{fermion}{subpath (0,length(p3)) of p3}
  \fmfi{fermion}{subpath (0,length(p4)) of p4}
 \end{fmfgraph*}   
\end{fmffile} 
\hspace*{3cm}
\begin{fmffile}{P2P+2P2P+2b}
 \begin{fmfgraph*}(90,80) 
  \fmfleft{i1,i2} 
  \fmfright{o1,o2}
  \fmf{phantom,left=0.4,tension=0.5,tag=1}{i1,i2}
  \fmf{phantom,right=0.4,tension=0.5,tag=2}{i1,i2}
  \fmf{phantom,left=0.4,tension=0.5,tag=3}{o2,o1}
  \fmf{phantom,right=0.4,tension=0.5,tag=4}{o2,o1}
  \fmfv{l=$\tr \left({\Phi^{I}}^{2}\right)~~$,l.a=180}{i1}
  \fmfv{l=$\tr \left({\Phi_{I}^{\dagger}}^{2}\right)~~$,l.a=180}{i2}
  \fmfv{l=$~~\tr \left({\Phi_{I}^{\dagger}}^{2}\right)$,l.a=0}{o1}
  \fmfv{l=$~~\tr \left({\Phi^{I}}^{2}\right)$,l.a=0}{o2}
  \fmfdot{i1,i2,o1,o2}
  \fmffreeze
  \fmfipath{p[]}
  \fmfiset{p1}{vpath1(__i1,__i2)}
  \fmfiset{p2}{vpath2(__i1,__i2)}
  \fmfiset{p3}{vpath3(__o1,__o2)}
  \fmfiset{p4}{vpath4(__o1,__o2)}
  \fmfi{fermion}{subpath (0,length(p1)) of p1}
  \fmfi{fermion}{subpath (0,length(p2)) of p2}
  \fmfi{fermion}{subpath (0,length(p3)) of p3}
  \fmfi{fermion}{subpath (0,length(p4)) of p4}
 \end{fmfgraph*}   
\end{fmffile}

\vspace*{1cm}
\noindent
The connected diagram is proportional to $N^{2}$ while the 
disconnected one goes like $N^{4}$. The only non-vanishing corrections 
to this Green function come from the diagrams 

\vspace*{1cm}
\hspace*{0.3cm}
\begin{fmffile}{aP2P+2P2P+2}
 \begin{fmfgraph*}(100,80) 
  \fmfleft{i1,i2} 
  \fmfright{o1,o2}
  \fmf{phantom,tension=0.5,tag=1}{i1,i2}
  \fmf{phantom,tension=0.5,tag=2}{i1,o1}
  \fmf{phantom,tension=0.5,tag=3}{o2,o1}
  \fmf{phantom,tension=0.5,tag=4}{o2,i2}
  \fmfv{l=$\tr \left({\Phi^{I}}^{2}\right)$,l.a=180}{i1}
  \fmfv{l=$\tr \left({\Phi_{I}^{\dagger}}^{2}\right)$,l.a=180}{i2}
  \fmfv{l=$\tr \left({\Phi_{I}^{\dagger}}^{2}\right)$,l.a=0}{o1}
  \fmfv{l=$\tr \left({\Phi^{I}}^{2}\right)$,l.a=0}{o2}
  \fmfdot{i1,i2,o1,o2}
  \fmffreeze
  \fmfipath{p[]}
  \fmfiset{p1}{vpath1(__i1,__i2)}
  \fmfiset{p2}{vpath2(__i1,__o1)}
  \fmfiset{p3}{vpath3(__o2,__o1)}
  \fmfiset{p4}{vpath4(__o2,__i2)}
  \fmfi{fermion}{subpath (0,length(p1)) of p1}
  \fmfi{fermion}{subpath (0,length(p2)/2) of p2}
  \fmfi{fermion}{subpath (length(p2)/2,length(p2)) of p2}
  \fmfi{fermion}{subpath (0,length(p3)) of p3}
  \fmfi{fermion}{subpath (0,length(p4)/2) of p4}
  \fmfi{fermion}{subpath (length(p4)/2,length(p4)) of p4}
  \fmfi{boson}{point length(p2)/2 of p2 -- 
               point length(p4)/2 of p4}
  \fmfiv{d.sh=circle,d.fill=full,d.si=1.5mm}{point length(p2)/2 of p2}
  \fmfiv{d.sh=circle,d.fill=full,d.si=1.5mm}{point length(p4)/2 of p4}
 \end{fmfgraph*}   
\end{fmffile} 
\hspace*{3cm}
\begin{fmffile}{bP2P+2P2P+2}
 \begin{fmfgraph*}(100,80) 
  \fmfleft{i1,i2} 
  \fmfright{o1,o2}
  \fmf{phantom,tension=0.5,tag=1}{i1,i2}
  \fmf{phantom,tension=0.5,tag=2}{i1,o1}
  \fmf{phantom,tension=0.5,tag=3}{o2,o1}
  \fmf{phantom,tension=0.5,tag=4}{o2,i2}
  \fmfv{l=$\tr \left({\Phi^{I}}^{2}\right)$,l.a=180}{i1}
  \fmfv{l=$\tr \left({\Phi_{I}^{\dagger}}^{2}\right)$,l.a=180}{i2}
  \fmfv{l=$\tr \left({\Phi_{I}^{\dagger}}^{2}\right)$,l.a=0}{o1}
  \fmfv{l=$\tr \left({\Phi^{I}}^{2}\right)$,l.a=0}{o2}
  \fmfdot{i1,i2,o1,o2}
  \fmffreeze
  \fmfipath{p[]}
  \fmfiset{p1}{vpath1(__i1,__i2)}
  \fmfiset{p2}{vpath2(__i1,__o1)}
  \fmfiset{p3}{vpath3(__o2,__o1)}
  \fmfiset{p4}{vpath4(__o2,__i2)}
  \fmfi{fermion}{subpath (0,length(p2)) of p2}
  \fmfi{fermion}{subpath (0,length(p1)/2) of p1}
  \fmfi{fermion}{subpath (length(p1)/2,length(p1)) of p1}
  \fmfi{fermion}{subpath (0,length(p4)/2) of p4}
  \fmfi{fermion}{subpath (length(p4)/2,length(p4)) of p4}
  \fmfi{fermion}{subpath (0,length(p3)) of p3}
  \fmfi{boson}{point length(p1)/2 of p1 -- 
               point length(p4)/2 of p4}
  \fmfiv{d.sh=circle,d.fill=full,d.si=1.5mm}{point length(p1)/2 of p1}
  \fmfiv{d.sh=circle,d.fill=full,d.si=1.5mm}{point length(p4)/2 of p4}
 \end{fmfgraph*}   
\end{fmffile} 

\vspace*{1cm}
\noindent
plus the analogous diagrams in which the $VV$ internal propagator is 
inserted between different pairs of external lines.
Notice in particular that the following correction, obtained from the 
disconnected diagram, 

\vspace*{1cm}
\hspace*{2cm}
\begin{fmffile}{cP2P+2P2P+2b}
 \begin{fmfgraph*}(90,80) 
  \fmfleft{i1,i2} 
  \fmfright{o1,o2}
  \fmf{phantom,left=0.4,tension=0.5,tag=1}{i1,i2}
  \fmf{phantom,right=0.4,tension=0.5,tag=2}{i1,i2}
  \fmf{phantom,left=0.4,tension=0.5,tag=3}{o2,o1}
  \fmf{phantom,right=0.4,tension=0.5,tag=4}{o2,o1}
  \fmfv{l=$\tr \left({\Phi^{I}}^{2}\right)~~$,l.a=180}{i1}
  \fmfv{l=$\tr \left({\Phi_{I}^{\dagger}}^{2}\right)~~$,l.a=180}{i2}
  \fmfv{l=$~~\tr \left({\Phi_{I}^{\dagger}}^{2}\right)$,l.a=0}{o1}
  \fmfv{l=$~~\tr \left({\Phi^{I}}^{2}\right)$,l.a=0}{o2}
  \fmfdot{i1,i2,o1,o2}
  \fmffreeze
  \fmfipath{p[]}
  \fmfiset{p1}{vpath1(__i1,__i2)}
  \fmfiset{p2}{vpath2(__i1,__i2)}
  \fmfiset{p3}{vpath3(__o1,__o2)}
  \fmfiset{p4}{vpath4(__o1,__o2)}
  \fmfi{fermion}{subpath (0,length(p1)) of p1}
  \fmfi{fermion}{subpath (0,length(p2)/2) of p2}
  \fmfi{fermion}{subpath (length(p2)/2,length(p2)) of p2}
  \fmfi{fermion}{subpath (0,length(p3)) of p3}
  \fmfi{fermion}{subpath (0,length(p4)/2) of p4}
  \fmfi{fermion}{subpath (length(p4)/2,length(p4)) of p4}
  \fmfi{boson}{point length(p2)/2 of p2 --
                 point length(p4)/2 of p4}
  \fmfiv{d.sh=circle,d.fill=full,d.si=1.5mm}{point length(p2)/2 of p2}
  \fmfiv{d.sh=circle,d.fill=full,d.si=1.5mm}{point length(p4)/2 of p4}
 \end{fmfgraph*}   
\end{fmffile}

\vspace*{1cm}
\noindent
is trivially zero because of colour contractions. Moreover working in 
the superfield formalism and choosing the Fermi--Feynman gauge, there 
are no corrections to the propagators to be included. Also, because of 
the non-renormalization properties of two-point functions discussed in 
the previous section, there is no non-vanishing disconnected diagram 
contributing to the four-point function at first or higher order. 

A second kind of four-point functions is of the form 
\begin{displaymath}
	G_{IJ}^{(b)} = 
	\langle \tr\left({\Phi_{I}^{\dagger}}^{\,2}\right) 
	\tr\left({\Phi^{I}}^{\,2}\right)\tr\left({\Phi_{J}^{\dagger}}^{\,2}
	\right) \tr\left({\Phi^{J}}^{\,2}\right) \rangle \, ,
\end{displaymath}
with $I\neq J$. In this case the only tree-level diagram is disconnected

\vspace*{1cm}
\hspace*{2cm}
\begin{fmffile}{P2P+2PP2PP+2}
 \begin{fmfgraph*}(90,80) 
  \fmfleft{i1,i2} 
  \fmfright{o1,o2}
  \fmf{phantom,left=0.4,tension=0.5,tag=1}{i1,i2}
  \fmf{phantom,right=0.4,tension=0.5,tag=2}{i1,i2}
  \fmf{phantom,left=0.4,tension=0.5,tag=3}{o1,o2}
  \fmf{phantom,right=0.4,tension=0.5,tag=4}{o1,o2}
  \fmfv{l=$\tr \left({\Phi^{I}}^{2}\right)~~$,l.a=180}{i1}
  \fmfv{l=$\tr \left({\Phi_{I}^{\dagger}}^{2}\right)~~$,l.a=180}{i2}
  \fmfv{l=$~~\tr \left({\Phi^{J}}^{2}\right)$,l.a=0}{o1}
  \fmfv{l=$~~\tr \left({\Phi_{J}^{\dagger}}^{2}\right)$,l.a=0}{o2}
  \fmfdot{i1,i2,o1,o2}
  \fmffreeze
  \fmfipath{p[]}
  \fmfiset{p1}{vpath1(__i1,__i2)}
  \fmfiset{p2}{vpath2(__i1,__i2)}
  \fmfiset{p3}{vpath3(__o1,__o2)}
  \fmfiset{p4}{vpath4(__o1,__o2)}
  \fmfi{fermion}{subpath (0,length(p1)) of p1}
  \fmfi{fermion}{subpath (0,length(p2)) of p2}
  \fmfi{fermion}{subpath (0,length(p3)) of p3}
  \fmfi{fermion}{subpath (0,length(p4)) of p4}
 \end{fmfgraph*}   
\end{fmffile} 

\vspace*{1cm}
\noindent
and goes as $N^{4}$. There is a unique non vanishing first order 
correction corresponding to the connected diagram 

\vspace*{1cm}
\hspace*{2cm}
\begin{fmffile}{aP2P+2PP2PP+2}
 \begin{fmfgraph*}(100,80) 
  \fmfleft{i1,i2} 
  \fmfright{o1,o2}
  \fmf{phantom,tension=0.5,tag=1}{i1,i2}
  \fmf{phantom,tension=0.5,tag=2}{i1,o1}
  \fmf{phantom,tension=0.5,tag=3}{o1,i1}
  \fmf{phantom,tension=0.5,tag=4}{o1,o2}
  \fmf{phantom,tension=0.5,tag=5}{i2,o2}
  \fmf{phantom,tension=0.5,tag=6}{o2,i2}
  \fmfv{l=$\tr \left({\Phi^{I}}^{2}\right)~~$,l.a=180}{i1}
  \fmfv{l=$\tr \left({\Phi_{I}^{\dagger}}^{2}\right)~~$,l.a=180}{i2}
  \fmfv{l=$~~\tr \left({\Phi^{J}}^{2}\right)$,l.a=0}{o1}
  \fmfv{l=$~~\tr \left({\Phi_{J}^{\dagger}}^{2}\right)$,l.a=0}{o2}
  \fmfdot{i1,i2,o1,o2}
  \fmffreeze
  \fmfipath{p[]}
  \fmfiset{p1}{vpath1(__i1,__i2)}
  \fmfiset{p2}{vpath2(__i1,__o1)}
  \fmfiset{p3}{vpath3(__o1,__i1)}
  \fmfiset{p4}{vpath4(__o1,__o2)}
  \fmfiset{p5}{vpath5(__i2,__o2)}
  \fmfiset{p6}{vpath6(__o2,__i2)}
  \fmfi{fermion}{subpath (0,length(p1)) of p1}
  \fmfi{fermion}{subpath (0,length(p2)/2) of p2}
  \fmfi{fermion}{subpath (0,length(p3)/2) of p3}
  \fmfi{fermion}{subpath (0,length(p4)) of p4}
  \fmfi{fermion}{subpath (length(p5)/2,length(p5)) of p5}
  \fmfi{fermion}{subpath (length(p6)/2,length(p6)) of p6}
  \fmfi{fermion}{point length(p5)/2 of p5 --
                 point length(p2)/2 of p2}
  \fmfiv{d.sh=circle,d.fill=full,d.si=1.5mm}{point length(p2)/2 of p2}
  \fmfiv{d.sh=circle,d.fill=full,d.si=1.5mm}{point length(p5)/2 of p5}
 \end{fmfgraph*}   
\end{fmffile} 

\vspace*{1cm}
\noindent
where the  internal propagator corresponds to the exchange of a chiral 
superfield with index $K\neq I\neq J$. Just like in the previous 
case there is no disconnected first order correction because 
of the non-renormalization properties of two-point functions. 

The previous diagrams can be evaluated using the techniques of 
chapter \ref{cap3}. Since the points where the composite 
operators are inserted are external and not interaction points the 
result will not be local in the fermionic variables of ${\cal N}$=1 
superspace. This is not in agreement with the results of a calculation 
presented in \cite{gonzrparkschalm}. 

An analysis similar to the one described here has been carried out 
in \cite{westsokatchev} employing a formulation of the model in ${\cal 
N}$=2 harmonic superspace. In \cite{westsokatchev} four-point functions 
are explicitly computed in terms of polylogarithm functions. The 
expressions obtained resemble the one that was obtained in section 
\ref{inst-4-point} from the instanton contribution to the correlator 
$\langle{\cal Q}{\cal Q}\overline{{\cal Q}}\overline{{\cal 
Q}}\rangle$. 

It has been observed by various authors \cite{liutse4point1,freed4point1,
liu,brodiegut,bk,chalmschalm1,chalmschalm2} that the four-point 
functions computed in the AdS/CFT correspondence develop logarithmic 
singularities, when the limit of two coincident points is considered. 
This kind of behaviour has already been pointed out in section 
\ref{inst-4-point}. In all the papers 
\cite{liutse4point1,freed4point1,liu,brodiegut,chalmschalm1,chalmschalm2} 
the short-distance behaviour has been shown taking the limit in the 
unintegrated expression obtained from the supergravity amplitude in 
AdS$_{5}$. This is rather subtle because the exchange of limit and 
integration might be problematic, moreover, as observed in 
\cite{freed4point1}, there could be compensations among various 
AdS diagrams. However this kind of cancellations cannot take 
place in the cases considered in \cite{brodiegut,bk}. In particular 
in \cite{brodiegut} some non-perturbative contributions from D-instantons 
to the four-point functions of scalar operators corresponding the 
dilaton and axion fields has been considered. 
The results of section \ref{inst-4-point}, where the instanton 
contribution to the four-point function $G_{{\cal Q}^{4}}$ has been 
given explicitly, unconfutably show the logarithmic behaviour. 
A satisfactory explanation of the origin of these singularities as 
well as of the reason for the lack of the expected pole-type singularities 
has not been proposed yet. In \cite{liu} an interesting analysis relating 
the contribution of contact and exchange amplitudes in AdS to the OPE 
of the corresponding operators in ${\cal N}$=4 super Yang--Mills has 
been given, using the technique of `conformal partial wave expansion', 
but the possible cancellation of logarithms between the two types of 
contributions has not been excluded. 

From the point of view of the boundary conformal field theory the 
central point appears to be the presence of anomalous dimensions for 
some operators that does not violate the conformal symmetry. This 
issue is under active investigation \cite{bkrs} and will be briefly 
addressed in the concluding remarks. Notice however that the behaviour of 
the correlation function $G_{{\cal Q}^{4}}$ for coincident points does not 
violate the non-renormalization theorems for three-point functions of 
the previous section, so that such non-renormalization properties could 
be generalizable to the non-perturbative level, at least for 
protected operators. In fact the limit in which two ${\cal Q}$ 
or two $\overline{{\cal Q}}$ operators are taken to coincide in 
$G_{{\cal Q}^{4}}$, that would correspond to one of the three-point 
functions that are expected to be tree-level exact, gives a vanishing 
result thanks to the pre-factor in equation (\ref{intbox}). 

\section{D-instanton effects in $d$=10 type IIB superstring theory}
\label{dinstantons}

As already discussed the low energy type IIB supergravity action 
contains a $f_{4}(\tau,{\overline\tau}){\cal R}^{4}$ term as well 
as a supersymmetry related $f_{16}(\tau,{\overline\tau})\Lambda^{16}$ 
term. Both of these terms receive contributions from D-instantons of 
the type IIB theory. 
 
The AdS/CFT correspondence described in the previous section allows to 
associate to these terms well defined contributions to correlation 
functions of composite operators in the boundary Yang--Mills theory.
It is natural to expect that the D-instanton generated terms 
should be related to instanton effects in ${\cal N}$=4 super 
Yang--Mills.
\fancyhead[LO]{{\footnotesize 5.6~~{\it D-instanton effects}}}

The type IIB D-instantons are Dirichlet $p=-1$ branes, \ie they are 
localized objects in ten-dimensional space-time, which means that they 
have a zero-dimensional world-volume. The charge-$K$ D-instanton in flat 
ten-dimensional space-time is a finite-action solution of the classical 
Euclidean equations of motion in which the metric (in the Einstein frame) 
is flat and the complex scalar, $\tau=\chi+ie^{-\phi}$, has a non-trivial 
profile with a singularity at the position of the D-instanton, whereas 
the other fields vanish \cite{ggp}. Moreover the D-instanton is a BPS 
saturated configuration preserving half of the supersymmetries. In the 
following the constant expectation values of the dilaton and the axion 
in the zero instanton sector will be denoted by $\tilde{\phi}$ and 
$\tilde{\chi}$ respectively, while hatted symbols will refer to the 
classical D-instanton solution. 

To compute the scalar part of the type IIB action in the D-instanton 
background it is useful to dualize the \RR scalar, $\chi=C^{(0)}$, to 
an eight-form potential, $C^{(8)}$. Since the metric is trivial, the 
action (up to constants) can be put in the form 
\begin{eqnarray}
	S &=& \frac{1}{4} \int \left(d\phi\pm e^{-\phi}*dC^{(8)}\right) 
	\wedge * \left(d\phi\pm e^{-\phi}*dC^{(8)}\right) \pm \frac{1}{2}
	\int e^{-\phi}d\phi\wedge dC^{(8)} = \nonumber \\
	&=& \frac{1}{4} \int \left(d\phi\pm e^{-\phi}*dC^{(8)}
	\right) \wedge * \left(d\phi\pm e^{-\phi}*dC^{(8)}\right) \mp 
	\frac{1}{2} \int_{\partial M} e^{-\phi}dC^{(8)} 
	\label{dinstantonact} \, ,
\end{eqnarray}
where $\partial M$ denotes the boundary of space-time, that in the 
presence of the D-instanton is understood as made of the sphere 
at infinity and an infinitesimal sphere around the location of the 
instanton. The first term in (\ref{dinstantonact}) is 
semi-positive definite, so that it follows that the action satisfies a 
Bogomol'nyi bound 
\begin{equation}
	S \geq \frac{1}{2} \left| \int_{\partial M} e^{-\phi}dC^{(8)} 
	\right| \, . 
	\label{dinstbogo}
\end{equation}
Minimizing the action yields a BPS condition relating $\phi$ and 
$C^{(8)}$, which, expressed in terms of $\chi$, becomes \cite{ggp}
\begin{equation}
	\partial_\Sigma \chi = \pm i \partial_\Sigma e^{-\phi} \, .
	\label{bogochi}
\end{equation}
The D-instanton solution is constructed through the ansatz
\begin{displaymath}
	\chi = \tilde{\chi}+if(r) \, ,
\end{displaymath}
where $r=|X-X_{0}|$ and $f(r)\to 0$, as $r\to\infty$. After imposing 
(\ref{bogochi}) the equation for the dilaton outside a sphere centered at 
the D-instanton location becomes 
\begin{equation}
	e^{-\phi}g^{\Lambda\Sigma} \nabla_\Lambda 
	\nabla_\Sigma e^\phi = 0 \, .
	\label{eqdinstdilaton}
\end{equation}
The solution is the harmonic function \cite{ggp}
\begin{equation}
	e^{\hat\phi^{(10)}} = g_s+{3 K{\alpr}^4 \over \pi^4 |X-X_0|^8}\, . 
	\label{tendim} 
\end{equation}
where $X^\Lambda$ is the ten-dimensional coordinate and $X_0^\Lambda$ 
is the location of the D-instanton. In (\ref{tendim}) $g_s$ is the 
asymptotic value of the string coupling and the normalization of the 
second term has a quantized value \cite{ggp} by virtue of a condition 
analogous to the Dirac--Nepomechie--Teitelboim condition that quantizes 
the charge of an electrically charged $p$-brane and of its magnetically 
charged $p'$-brane dual \cite{nepoteit}. It is notable that the solution 
in (\ref{tendim}) is simply the Green function for a scalar field to 
propagate from $X_0$ to $X$ subject to the boundary condition that $e^\phi 
= g_s$ at $|X| \to \infty$ or $|X_0| \to \infty$. The solution for $\chi$ 
follows from (\ref{bogochi}) and is of the form 
\begin{equation}
	\chi = \tilde{\chi} +i\left( A - \frac{1}{g_{s}} + 
	e^{-\hat{\phi}} \right) \, ,
	\label{chisolu}
\end{equation}
where $A$ is an arbitrary constant. 

The action for a single D-instanton receives contribution only from 
the boundary term in (\ref{dinstantonact}), \ie from the integration 
over the sphere at infinity and over an infinitesimal sphere around 
$X_{0}$. The arbitrariness in the constant $A$ can be used to simplify 
the computation. In particular with the choice $A$=0 in (\ref{chisolu}) 
the entire D-instanton action comes from the boundary of the infinitesimal 
sphere. Substituting for $f$ from (\ref{chisolu}) gives 
\begin{equation}
	S_K = {2\pi |K|\over g_s} \, .
	\label{flatdinstact} 
\end{equation}
On the other hand, with the choice $A=\frac{1}{g_s}$ in (\ref{chisolu}) the  
expression (\ref{dinstantonact}) reduces to an integral over the  boundary 
at infinity, but the total action remains the same as $S_K$ in 
(\ref{flatdinstact}).

The bosonic zero-modes associated with the D-instanton solution are 
parameterized by the coordinates $X_{0}^{\Lambda}$ of the center of 
the D-instanton. As already remarked the D-instanton background 
breaks half of the supersymmetries. The fermionic zero-modes can be 
generated by acting with the broken supersymmetries on the scalar 
solution. 

The D-instanton contributions to the effective action of type IIB 
supergravity can be deduced by studying scattering amplitudes in the 
instanton background. In this way one obtains in particular the 
corrections to the ${\cal R}^{4}$ and $\Lambda^{16}$ terms in 
equations (\ref{r4term}) and (\ref{lambint}) \cite{greengut,greengutkwon}. 

In \cite{banksgreen} it was pointed out that the ${\cal R}^4$ term 
vanishes when computed in the AdS$_{5}\times S^{5}$ background, 
because it involves a fourth power of the (vanishing) Weyl tensor.
The first, second and third functional derivatives of ${\cal R}^4$
vanish as well, and as a result one finds no corrections to zero, one,
two and three-point amplitudes when the ${\cal R}^{4}$ term acts as 
source for Yang--Mills operators on the boundary.
However, there is a non-zero four-graviton amplitude  
arising from this term. The boundary values of these gravitons are
sources for various bosonic components of the Yang--Mills current 
supermultiplet. For example, the components of the metric in the
AdS$_5$ directions couple to the stress tensor, ${\cal T}^{\mu\nu}$,
whereas the traceless components polarized in the $S^5$ directions couple to  
massive Kaluza-Klein states. A linear combination of the trace of the
metric on $S^5$ and the fluctuation of the \RR four-form potential
couples to the scalar components, ${\cal Q}^{ij}$. As a consequence 
the instanton contribution to the four-point correlation function 
$G_{{\cal Q}^{4}}$ computed in the previous chapter should be related 
to the appropriate D-instanton correction extracted from 
$f_{4}(\tau,{\overline\tau}){\cal R}^{4}$.
However it is extremely laborious to single out the exact terms to 
be compared with the Yang--Mills computation, so that it will be 
convenient to show the matching of the D-instanton correction to the 
$\Lambda^{16}$ term with the $G_{{\hat{\Lambda}}^{16}}$ Green function 
of section \ref{inst-8-16-point}. 

All the other terms of the same dimension, that are related to ${\cal R}^4$ 
by supersymmetry can be picked out, for example, associating the physical 
fields with the components of an on-shell type IIB superfield \cite{howesttwo}. 
As already remarked included among these is the sixteen-fermion 
interaction (\ref{lambint}). Supersymmetry arguments \cite{pk,grse} imply 
that the functions $f_{16}$ and $f_4$ are related by the action of 
SL(2,$\mathbb{Z}$) modular covariant derivatives
\begin{equation}
	f_{16}(\tau,{\overline\tau}) = (\tau_{2} {\cal D})^{12} 
	f_{4}(\tau,{\overline\tau}) \, , 
	\label{relf}
\end{equation}
where ${\cal D}$=$\frac{\partial}{\partial\tau} + 2i\frac{q}{\tau_{2}}$  
is the covariant  derivative that maps a $(q,p)$ modular form into a 
$(q+2,p)$ form (where the notation $(q,p)$ labels the holomorphic and 
anti-holomorphic SL(2,$\mathbb{Z}$) weights of the form). 
Whereas $f_4$ transforms 
with modular weight $(0,0)$, the function $f_{16}$ has 
weight $(12,-12)$  and therefore transforms with a non-trivial
phase under SL(2,$\mathbb{Z}$). This is precisely the phase required to
compensate for the anomalous U(1)$_B$ transformation of the 16 
$\Lambda$'s, so that the full expression (\ref{lambint}) is invariant 
under the SL(2,$\mathbb{Z}$) S-duality group.

In the small coupling limit, all of these terms can be expanded in the form 
\begin{eqnarray} 
	&& e^{-\phi/2}  f_n \; = \; a_n \zeta(3) e^{-2\phi} + b_n + \nonumber \\
	&& + \sum_{K =1}^\infty \mu_{K} (K e^{-\phi})^{n -7/2} 
	e^{2\pi i K\tau} \left(1 + \sum_{k=1}^\infty 
	c_{k,n}^K (K e^{-\phi})^{-k} \right) \, . 
\label{fnexp} 
\end{eqnarray}  
The first two terms have the form of string tree-level and one-loop 
contributions, while the last one has the appropriate weight of 
$e^{2\pi i K\tau}$ to be associated with a charge-$K$ D-instanton effect. 
Anti-D-instanton contributions have not been displayed.
The coefficients $c_{k,n}^K$ as well as the `D-instanton partition functions'
$\mu_{K}$ are explicitly given in \cite{greengutkwon}. 
The coefficient of the leading term in the series of perturbative  
fluctuations around a charge-$K$ D-instanton is independent of which 
particular interaction term is being discussed and reads 
\begin{displaymath}
{\cal Z}_K = \mu_{K} (K e^{-\phi})^{-7/2} e^{2\pi i K\tau} \: .
\label{znew} 
\end{displaymath}
It should be identified with the contribution of a charge-$K$  
D-instanton to the measure in string frame which, up to an overall 
numerical factor, reads
\begin{equation}
d\Omega_K^{(s)} = (\alpha')^{-1} d^{10} X \, d^{16}\,  
\Theta \,{\cal Z}_K \: .
\label{dmeasdef}
\end{equation}
Actually there are subtleties in doing this, since the full series is 
not convergent, it is rather an asymptotic approximation to a Bessel 
function.

\subsection{Testing the AdS/CFT correspondence: type IIB D-instantons 
vs. Yang--Mills instantons}

As observed in \cite{banksgreen}\ the charge-$K$  
D-instanton action that appears in the exponent in (\ref{znew}) 
coincides with the action of a charge-$K$ Yang--Mills instanton in the 
boundary theory, which indicates a correspondence between these sources 
of non-perturbative effects. This idea  is reinforced by the correspondence 
between other factors. For example, after substituting 
$e^\phi = \frac{\gym^2}{4\pi}$ and 
$\alpha^{\prime} = L^2 N^{-1/2} \gym^{-1}$, the measure 
(\ref{dmeasdef}) contains an overall factor of the coupling constant 
in the form $\gym^8$.
Indeed, this is exactly the power expected on the basis of the  
AdS/CFT correspondence, since 
the one-instanton contribution to the Green functions
in the ${\cal N}=4$ Yang--Mills theory, considered in chapter \ref{cap4},
also has a factor of $\gym^8$ arising from the combination  
of the bosonic and fermionic zero-mode norms. In this section 
the leading instanton contributions to type IIB superstring amplitudes 
will be compared with the corresponding  ${\cal N}$=4 current correlators 
considered in chapter \ref{cap4}.  
In the next section the same analysis will be carried out studying the 
semiclassical fluctuations around the AdS$_{5}\times S^{5}$ 
D-instanton solution. 

To compare the Yang--Mills instanton calculation with an amplitude 
obtained according to Maldacena's prescription one must expand 
the function $f_{16}(\tau,{\overline\tau})$ in (\ref{relf}) to extract the
one-instanton term. In order to compare with the Yang--Mills
sixteen-point correlator one needs to consider the situation in which
all sixteen fermions propagate to well defined configurations
on the boundary.  

Equation (\ref{dirac-op}) leads to the normalized 
bulk-to-boundary propagator of the fermionic field $\Lambda$ 
of AdS `mass' $m=-\frac{3}{2L}$, associated to the composite operator  
${\hat \Lambda}$ of dimension $\Delta = \frac{7}{2}$, see equation 
(\ref{currentdef2}), 
\begin{equation}
	K^{{\rm F}}_{7/2}(\rho_0,x_0;x) = K_{4}(\rho_0,x_{0};x) {1 \over 
	\sqrt{\rho_0}} \left( \rho_0 \gamma_{\hat5} + {(x_{0}-x)}^{\mu}
	\gamma_{\hat\mu} \right) \, .
	\label{prop-7/2} 
\end{equation}   
Suppressing all spinor indices, one obtains  
\begin{equation}
 \Lambda_{J}(x_0,\rho_0) = \int d^4 x K^{\rm F}_{7/2}(\rho_0,x_{0};x) 
J_{\Lambda}(x) \, ,
\label{boundarylambda}
\end{equation} 
where $J_{\Lambda}(x)$ is a left-handed  boundary value of $\Lambda$ and 
acts as the source for the composite operator ${\hat \Lambda}$ in the 
boundary ${\cal N}$=4 Yang--Mills theory. As a result, the classical action 
for the operator ${(\Lambda)}^{16}$ in the type IIB supergravity on 
AdS$_{5}\times S^{5}$ is
\begin{eqnarray}  
  && S_{\Lambda} [J] = e^{-2\pi ({1\over g_s} + i\chi)} g_s^{-12} \, 
  V_{S^5} \, \int \frac{d^{4}x_{0}d\rho_0}{\rho_0^{5}} \nonumber\\
  && t_{16} \prod_{p=1}^{16}  
  \left[ K_{4}(\rho_0,x_{0};x_{p}) \frac{1}{\sqrt{\rho_0}}
  \left( \rho_0 \gamma^{\hat 5} +  
  {(x_{0}-x_{p})}^{\mu} \gamma_{\hat\mu} \right) \,
  J_{\Lambda}(x_{p}) \right] \, ,
\label{source-l16} 
\end{eqnarray} 
where $e^\phi = g_s$ and $\chi = \tilde{\chi}$ 
(since the scalar fields are taken to be constant in the 
AdS$_{5}\times S^{5}$ background) and $V_{S^5}=\pi^3$ 
is the $S^5$ volume. The 16-index invariant tensor $t_{16}$ is the same
as the one defined after (\ref{l16-sym}). 
The overall power of the coupling constant comes from the factor of 
$(e^{-\phi})^{-\frac{25}{2}}=\left(\frac{1}{g_{s}}\right)^{\frac{25}{2}}$ 
in the expansion (\ref{fnexp}) and the factor of $\frac{1}{\alpr}$ in 
front of the $\Lambda^{16}$ term in the type IIB action, which using 
the relation (\ref{L-alpr}) yields 
\begin{displaymath}
	\frac{1}{\alpr} \left(\frac{1}{g_{s}}\right)^{\frac{25}{2}} \sim 
	\left(\frac{1}{g_{s}}\right)^{12} \, ,
\end{displaymath}
where an overall numerical constant has been dropped. Using the dictionary 
(\ref{dict}) and differentiating with respect to the chiral sources this 
result agrees with the expression (\ref{l16-sym}) of section 
\ref{inst-8-16-point} obtained in the Yang Mills calculation. In 
particular the overall power of $\gym$ in (\ref{l16-sym}) is 
reproduced after introducing the same rescaling of each $\Lambda$ in 
(\ref{source-l16}) by a factor of $4\pi g_{s}$ as in equation 
(\ref{l16}). Actually the overall numerical constant has not been 
checked, see however \cite{mattislong}. Notice that to match the result 
of the Yang--Mills calculation it is crucial that in the product of sixteen 
matrices $K^{{\rm F}}_{7/2}$ exactly eight factors of 
$\sqrt{\rho_{0}}\gamma^{5}$ and eight factors of $\frac{(x-x_{0})_{\mu}}
{\sqrt{\rho_{0}}}\gamma^{\mu}$ are picked out. This 
is ensured by the contraction with $t_{16}$, which selects the 
correct number of factors of each kind. In the next section this 
expression will also be motivated directly by semi-classical quantization 
around a D-instanton field configuration that is a Euclidean solution of 
the type IIB supergravity in AdS$_{5}\times S^{5}$. 
 
The agreement of the $\Lambda^{16}$ amplitude  with the corresponding 
sixteen-current correlation function of chapter \ref{cap4} is sufficient 
to guarantee that the instanton contributions to all the other 
Yang--Mills  correlation functions, that are related to this term  
by ${\cal N}$=4 supersymmetry, will also agree with their type IIB 
superstring counterparts.  For example, the correlation function that 
has been considered in most detail in chapter \ref{cap4} was the one 
with four superconformal scalar currents 
${\cal Q}^{ij}$ which are in the {\bf 20}$_{\mathbb{R}}$ of SU(4), 
and have  dimension $\Delta$=2 and AdS `mass' $m^2=-\frac{4}{L^2}$. 
As already discussed the supergravity field, $Q^{ij}$, that couples 
to ${\cal Q}^{ij}$ is a linear combination of the fluctuation of the 
trace of the metric on $S^5$, $h_{m}^{\ m}$,  and of the four-form 
field potential, $C^{(4)}_{MNPQ}= \epsilon_{MNPQR} \nabla^Rf$.
Therefore, contributions to the correlation of four of these 
composite scalars in the one D-instanton background should correspond to 
the leading parts of the  $K$=1 terms in the expansion of the ${\cal R}^4$  
interaction (\ref{r4term}) as well as terms  of the form 
${\cal R}^2 \left(\nabla F^{(5)}\right)^2$ and $\left(\nabla 
F^{(5)}\right)^4$.  These last two terms involve
$F^{(5)}$, the self-dual  field strength of the antisymmetric four-form 
potential, and are related by supersymmetry to the ${\cal R}^4$ term.

It follows from the structure of (\ref{collectint}) that the Yang--Mills  
instanton contribution to each factor of ${\cal Q}^{ij}$ is of the form   
${\nabla\nabla} K_2(x_0^\mu,\rho_0 ;x^\mu,0)$, where the two 
derivatives are not necessarily contracted. But this is the expected form
for a propagator from the AdS$_{5}\times S^{5}$ bulk to the boundary for a 
scalar field of dimension 2. Therefore, at least the general form of the   
expression obtained from the $K$=1 terms in the expansion of ${\cal R}^4$ 
and the related $F^{(5)}$ interactions agrees with the four-${\cal Q}^{ij}$ 
correlation function in a Yang--Mills instanton background. 
In order to see this agreement in more detail it would first be necessary 
to determine the precise form of the $\left(\nabla F^{(5)}\right)^2$ and 
$\left(\nabla F^{(5)}\right)^4$  
interactions that contain the fluctuations of $F^{(5)}$.   
The $K$=1 contribution  to the amplitude of four fluctuations of the 
appropriate combination of $h_{m}^{\ m}$ and $f$ can then be 
extracted from these interaction terms. Although this computation has 
not been performed explicitly, the result is guaranteed to reproduce the 
$K$=1 expression obtained from the Yang--Mills theory in chapter 
\ref{cap4} since it is related by supersymmetry to the $\Lambda^{16}$ 
amplitude.  

The analogous comparison of the  correlation  function of eight  
${\cal E}^{AB}$'s with the amplitude for eight $E^{AB}$'s in the type IIB
superstring theory proceeds in much the same way. The supergravity fields,
$E^{AB}$, that couple to ${\cal E}^{AB}$ arise from the internal 
components of the complex type IIB antisymmetric tensor. Supersymmetry 
relates an $H^8$ term, where $H$ is the complex combination of the type IIB 
three-form field-strengths, $H_{1}$ and $H_{2}$, introduced in 
(\ref{iibsugra}), to the ${\cal R}^4$ term in the type IIB effective action. 
One thus expects interactions schematically of the form $(\nabla K_3)^8$, 
with various contractions of the derivatives. Using the explicit form of 
the spinor collective coordinates $\zeta^A=(\rho_0\eta^A +
(x-x_0)_\mu \sigma^{\mu}{\bar\xi}^A)/\sqrt{\rho_0}$ one may check that the 
gaugino bilinears ${\cal E}^{AB}$ in the one-instanton background exactly 
give rise to $K_3$ and derivatives thereof, as expected for the scalar 
propagator of a scalar field of dimension $\Delta$=3 and AdS mass 
$m^2=-\frac{1}{L^2}$. Although a precise matching of the resulting 
amplitudes has not been performed, one may appeal to supersymmetry 
arguments to determine the complete structure of these terms.

Notice that the above considered non-perturbative terms in the type IIB 
supergravity effective action when expanded around the 
AdS$_5\times S^5$ background give rise to both derivative and 
non-derivative interaction terms. The matching of the non-derivative 
``mass-related'' terms considered here with corresponding terms in the 
Yang--Mills Green functions is rather straightforward, but clearly it 
is only a hint to the conjectured correspondence. 

\section{The D-instanton solution in AdS$_{5}\times S^{5}$}
\label{ads5dinstanton}

The results of the previous section will now be reinforced by studying 
how the $(\alpha')^{-1}$ terms in the type IIB effective action can be 
determined by semi-classical type IIB supergravity field theory 
calculation in a D-instanton background.

The solution of the equations of motion of the type IIB theory in 
Euclidean  AdS$_{5}\times S^{5}$ will be now discussed. 
The BPS condition for a D-instanton in this background is identical 
to (\ref{bogochi}) and together with the equation of motion for 
the dilaton leads to
\begin{equation}
	g^{\Lambda\Sigma} \nabla_\Lambda \nabla_\Sigma e^\phi = 0 
	\label{disteq} 
\end{equation}
and determines the complex scalar $\tau$.
In the Einstein frame the Einstein equations are unaltered by
the presence of the D-instanton, because the associated Euclidean stress 
energy tensor vanishes. As a consequence AdS$_{5}\times S^{5}$ remains a 
solution. Equation (\ref{disteq}) is identical to the equation for the 
Green function of a massless scalar propagating between the location of the 
D-instanton ($x_0^\mu, y_0^i$) and the point ($x^\mu,y^i$), which is the
bulk-to-bulk propagator (subject to the  boundary condition that it is
constant in the limits $\rho\to 0$ and  $\rho \to \infty$).
This is easy to solve using the conformal flatness of 
AdS$_{5}\times S^{5}$. In Cartesian coordinates 
$X^\Lambda=(x^\mu,y^i)$,  (\ref{d3nearhorizon}) reads 
$ds^2 = L^2\rho^{-2} dX\cdot dX$, where $\rho^2 = y \cdot y$ and the
solution of (\ref{disteq}) with a constant asymptotic behavior is of 
the form
\begin{equation}
	e^{\hat \phi}  = g_s +  { \rho_0^4 \rho^4\over L^8} 
	\left(e^{\hat \phi^{(10)}} - g_s \right) \, ,
	\label{conflat}
\end{equation}
where  $\rho_0 = |y_0|$ and $e^{\hat \phi^{(10)}}$ 
is the harmonic function that appeared in the flat ten-dimensional 
case, (\ref{tendim}). In evaluating D-instanton dominated  amplitudes 
one is only interested in the case in which the point
$(x^\mu,y^i$) approaches the boundary ($\rho \equiv |y| 
\to 0$), in which case it is necessary to rescale the 
dilaton profile (just as it is necessary to rescale the scalar
bulk-to-bulk propagator, \cite{gkp,witten}), so that the combination
\begin{equation}
	\rho^{-4} \left(e^{\hat \phi} - g_s \right) =   
	{3 K(\alpha')^4 \over L^8 \pi^4}{  \rho_0^4\over  ((x-x_0)^2 + 
	\rho_0^2)^4 }  \, , 
	\label{classd}
\end{equation}
will be of relevance in the $\rho \to 0$ limit.     
\fancyhead[LO]{{\footnotesize 5.7~~{\it AdS$_{5}\times S^{5}$ 
D-instanton solution}}}

As mentioned earlier, the correspondence with the Yang--Mills instanton 
follows from the fact that $  \rho_0^4/ ((x-x_0)^2 + \rho_0^2)^4 = 
K_4$ is proportional to the instanton number density, $({\overline 
F}_{\mu\nu})^2$, in the ${\cal N}$=4 Yang--Mills theory. Strikingly, 
the scale size of the Yang--Mills instanton is replaced by the distance 
$\rho_0$ of the D-instanton from the boundary. This is another indication 
of how the geometry of the Yang--Mills theory is encoded in the type IIB 
superstring. Note, in particular, that as the D-instanton approaches the 
boundary $\rho_0 \to 0$, the expression for  $\rho^{-4} e^{\hat \phi}$ 
reduces to a $\delta$ function that corresponds to a zero-size Yang--Mills 
instanton.  

The BPS condition implies that one can write the solution for the \RR\
scalar as
\begin{equation}
	\hat{\chi} = \tilde{\chi} + i f(x,y) \, ,
	\label{rrhat}
\end{equation}
where $\tilde{\chi}$ is the constant real part of the field (which
corresponds to $\frac{\tym}{2\pi}$, see equation (\ref{dict})) and 
just like in the flat ten-dimensional case
\begin{equation}
	f = A -  {1\over g_s} + e^{-\hat \phi} \, .
	\label{fsoln}
\end{equation}
Since the action is independent of constant shifts of $\chi$, 
actually it does not depend on the arbitrary constant, $A$. 
In a manner that follows closely  the flat ten-dimensional case 
considered in the appendix of \cite{greengut} and reviewed in the 
preceding section, the action for a charge $K$ D-instanton can be 
written as 
\begin{equation}
	S_K = - {L^{10}\over (\alpha')^4} \int  
	{d\rho d^4x d^5 \omega \over \rho^5}  g^{\Lambda\Sigma} 
	\nabla_\Lambda (e^{2\hat \phi} f \partial_\Sigma f) \, ,
	\label{actint}
\end{equation}
which reduces to an integral over the boundaries of 
AdS$_{5}\times S^{5}$ and the surface of an infinitesimal sphere centered 
on the D-instanton at $x=x_0$, $y=y_0$. The result is exactly the same 
as in the flat case
\begin{equation}
	S_K = {2\pi |K|\over g_s} \, .
	\label{dinstact} 
\end{equation}
By choosing $A=\frac{1}{g_s}$ in (\ref{fsoln}) the  
expression (\ref{actint}) reduces to an integral over the  boundary 
at $\rho=0$. Remarkably, in this case the boundary integrand is 
{\em identical} to the action density of the standard four-dimensional 
Yang--Mills instanton. 

Whereas the AdS$_{5}\times S^{5}$ metric remains unchanged by the 
presence of the D-instanton in the Einstein frame, it is radically altered 
in the string frame where the instanton is manifested as a space-time
wormhole (as in the flat ten-dimensional case \cite{ggp}). For finite
values of $K$ the dilaton becomes large in the Planck-scale neck and the 
classical solution is not reliable in that region. However, for very large
instanton number, the neck region becomes much larger than the Planck
scale so, by analogy with the D-brane examples studied in
\cite{maldacena}, it should be very interesting to study the
implications of the modified AdS$_{5}\times S^{5}$ geometry in the 
large-$K$ limit of the large-$N$ theory.

The D-instanton contribution to the amplitude with
sixteen external dilatinos, $\Lambda_{\alpha}^A$, may now be 
obtained directly by semi-classical quantization around the classical
D-instanton solution in AdS$_{5}\times S^{5}$. 
The leading instanton contribution can be determined by applying
supersymmetry transformations to the scalar field  which has an 
instanton profile given by  (\ref{classd}). Since 
the D-instanton background breaks half the supersymmetries the
relevant transformations are those in which the supersymmetry
parameter corresponds to the  Killing spinors for the sixteen broken 
supersymmetries. These Killing spinors have  U(1)$_{B}$ charge 
$\frac{1}{2}$ and are defined by a modified version of (\ref{kilspina})
that includes the non-trivial composite U(1)$_{B}$ connection, $Q_M$
\cite{schw}, that is made from the type IIB scalar field \cite{ggp} 
\begin{equation}
	{\cal D}_M \zeta \equiv \left( D_M - {i\over 2} Q_M \right) 
	\zeta = {1 \over 2L} \gamma_M \zeta \, .
	\label{modsixty}
\end{equation}  
Substituting the Euclidean D-instanton solution into the expression
for the composite connection gives 
\begin{equation}
	Q_M =  {i\over 2} e^{-\hat \phi} \partial_M e^{ \hat \phi} \, ,
	\label{Qdef}
\end{equation}  
with $\hat \phi$ defined by (\ref{classd}). The solution of 
(\ref{modsixty}) is
\begin{equation}
	\zeta_{\pm}=e^{-\frac{\hat{\phi}}{4}}\frac{z_{M}\gamma^{\hat{M}}}
	{\sqrt{\rho_0}} \zeta^{(0)}_{\pm} \, , 
\label{spinkil} 
\end{equation}  
where $\zeta^{(0)}_{\pm}$ is a constant spinor satisfying 
$\gamma_5 \zeta^{(0)}_{\pm} = \pm \zeta^{(0)}_{\pm}$.  
 
The sixteen broken supersymmetry transformations associated
with $\zeta^{(0)}_{-}$  give rise to the  dilatino
zero-modes,
\begin{equation}
	\Lambda_{(0)} = \delta \Lambda = (\gamma^M \hat P_M) \zeta_{-} \, , 
	\label{dil5d} 
\end{equation} 
where $\hat P_M$ is the expression for $P_M \equiv \frac{i}{2\tau_2}
\partial_M \tau^*$ in the D-instanton background \cite{greengut}, \ie 
\begin{equation}
	\hat P_M =   e^{-\hat \phi} \partial_M  e^{\hat \phi} \, .
	\label{pmdef} 
\end{equation} 
Using the Killing spinor equation and the D-instanton equation
${\cal D}^M \hat P_M=0$ it is easy to check (recalling that $P_M$ has 
U(1)$_{B}$ charge 2) that 
\begin{equation}
	\gamma^M {\cal D}_M \Lambda_{(0)} = - {3 \over 2L}  
	\Lambda_{(0)} \, ,
	\label{lambdeq}
\end{equation} 
so that $\Lambda_{(0)}$ is a solution of the appropriate massive Dirac 
equation. The relevant amplitudes are those with external states 
located on the boundary, in which case one may use the fact that for 
$\rho \sim 0$
\begin{equation}
	P_M \sim{1\over  g_s} \partial_M e^{\hat \phi}
	\label{farapp}
\end{equation} 
in (\ref{dil5d}), which leads to 
\begin{equation}
	\Lambda_{(0)} \sim  {4\over g_s} (e^{\hat \phi} -  
	g_s) \zeta_{-} \: .  
	\label{lamzer}
\end{equation} 
This means that near $\rho=0$ the dilatino profile in the D-instanton 
background is proportional to $\rho^{4} K_4(x_0,\rho_0;x, 0)$.  

As a result the leading contribution to the sixteen-dilatino  
amplitude again reproduces the corresponding sixteen-current 
correlator in ${\cal N}=4$  supersymmetric Yang--Mills theory. 
Explicitly, the D-instanton approximation to the amplitude with  
sixteen external dilatinos, $\Lambda_{\alpha}^{A}$, at points on the 
$\rho=0$ boundary is (up to an overall constant factor)
\begin{eqnarray} 
	&& \hspace*{-1.5cm} \langle \prod_{p=1}^{16} 
	\Lambda^{A_{p}}_{\alpha_{p}}(x_{p},0) \rangle_J 
    = g_s^{-12} e^{-2\pi K ({1\over g_s} +  i  C^{(0)}) } 
    V_{S^5} \,  \int \frac{d^4x_0 d\rho_0}{\rho_0^{5}}  
	\int d^{16}\zeta^{(0)}_{-} \nonumber \\  
	&& \hspace*{-1.5cm} \cdot \prod_{p=1}^{16} \left[ K_{4} 
	\left(x_0,\rho_{0};x_{p}\right) {1\over\sqrt{\rho_0}}   
	\left(\rho_0 \eta^{A_{p}}_{\alpha_{p}}+{(x_{p}-x_{0})}_{\mu}  
	\sigma^{\mu}_{\alpha_{p}{\dot \alpha}_{p}}  
	{\overline \xi}^{{\dot \alpha}_{p}A_{p}} \right)
    J_{\Lambda}(x_p) \right],  
	\label{l16dil5d} 
\end{eqnarray} 
where $J_{\Lambda}(x_p)$ is the wave-function of the dilatino
evaluated at the boundary  point $(x_p,0)$ and the Grassmann spinor 
$\zeta_-^{(0)}$ is defined in terms of $\eta$ and $\bar \xi$ by
\begin{displaymath} 
\label{zetdef}
 \zeta (\rho_0,x-x_0) = \left( \frac{1 - \gamma^{5}}{2} \right) 
  { z^{M}  \gamma_{M} \zeta^{(0)} \over \sqrt \rho_0} \: , 
\end{displaymath}
where
\begin{displaymath}
\label{zetzdef} 
	\zeta^{(0)} = \left( 
	\begin{array}{c} 
	    {\eta}_{\alpha} \\  \\ {\overline \xi}^{{\dot \alpha}}  
	\end{array} 
	\right) \, , \hspace{0,5cm} 
	z^{M} = (x^{\mu} - x_0^\mu,\rho_0) \, . 
\end{displaymath}
The power of $g_s^{-12}$ has been inserted in 
(\ref{l16dil5d}) from the expression given in section (\ref{typeiib}) 
although this power should also follow directly by 
considering the normalization of the bosonic and fermionic zero modes.
Up to the overall constant factor, the  amplitude (\ref{l16dil5d})  
agrees with (\ref{source-l16}) and therefore with
(\ref{l16-sym}).   
 
In similar manner the instanton profiles of all the fields in the
supergravity multiplet follow by applying the broken supersymmetries
to $P_M$ the appropriate number of times, just as in the flat
ten-dimensional case \cite{greengut}. The single D-instanton
contributions to any correlation function can then be determined.
This is guaranteed to agree with the corresponding term in the
expansion of the effective type IIB action as well as with
the corresponding ${\cal N}$=4 Yang--Mills correlation function.


\chapter*{Conclusions}
\label{concl}
\addcontentsline{toc}{chapter}{Conclusions}
\vspace*{2cm}
\fancyhead[RO,LE]{\thepage}
\fancyhead[RE,LO]{{\footnotesize {\rm Conclusions}}} 

\noindent
This dissertation reports the study of various aspects of ${\cal 
N}$=4 supersymmetric Yang--Mills theory. 

The perturbative finiteness properties of the theory have been known 
for a long time. However the analysis presented here allows to point 
out several problems in the perturbative computation of Green 
functions of elementary (super) fields. The calculations reviewed in 
chapter \ref{cap3} show that the gauge-fixing procedure is quite 
problematic and presents subtleties both in the description in terms 
of component fields and using the ${\cal N}$=1 superfield formalism. 
The appearance of UV divergences in off-shell propagators, when the 
Wess--Zumino gauge in components is exploited, has been discussed and it 
has been shown that relaxing the choice of the WZ gauge such divergences
are cancelled. Different difficulties have been encountered in superspace 
computations: the supersymmetric generalization of the Fermi--Feynman 
gauge appears to be privileged since any different choice of gauge 
leads to IR divergences in off-shell Green functions. 

These issues have been studied in a systematic way in chapter 
\ref{cap3}. The divergences found in one-loop calculations are 
expected to be gauge artifacts, and it would be interesting to obtain 
an explicit check of the cancellation of such divergences in gauge 
invariant Green functions. As briefly discussed in the final chapter 
in the context of the AdS/SCFT correspondence, superspace techniques 
can be applied to the calculation of correlation functions of 
gauge-invariant composite operators as well. ${\cal N}$=1 superfields 
should prove a powerful tool in this computations like in the case of 
elementary fields. 

At the perturbative level the effect of the inclusion of mass terms 
for the (anti) chiral superfields has also been discussed. The 
considered mass terms break supersymmetry down to ${\cal N}$=1, but 
it is explicitly proved, at the one loop-level, that the ultraviolet 
properties of the model are not affected by this modification. The 
same result had been obtained long ago by means of dimensional 
arguments and is reinforced by the analysis presented here. 

The possibility of employing this mass deformed ${\cal N}$=4 model as 
a supersymmetry preserving regularization scheme for ${\cal N}$=1 
theories has been recently proposed. As observed in chapter 
\ref{cap3} this approach should be generalizable to ${\cal N}$=2 
theories. 

Non perturbative effects in ${\cal N}$=4 supersymmetric Yang--Mills 
theory are studied in chapter \ref{cap4} with the aid of instanton 
calculus. The literature on ${\cal N}$=4 super Yang--Mills contains 
examples of instanton computations of corrections to the effective 
action in the Abelian Coulomb phase.  The calculations presented in 
chapter \ref{cap4} have been performed in the superconformal phase, 
corresponding to the origin of the moduli space, which is related to 
type IIB superstring theory in AdS$_{5}\times S^{5}$ by the duality 
recently proposed by Maldacena. In particular an exact field 
theoretical computation of the one-instanton contribution (in the 
semiclassical approximation) to a four-point function of scalar composite 
operators belonging to the current supermultiplet has been reported. 

The non-perturbative corrections to Green functions in ${\cal N}$=4 
supersymmetric Yang--Mills theory have been subsequently reconsidered 
in the context of the AdS/CFT correspondence. Previous checks of the 
conjectured correspondence had been obtained for two- and three-point 
functions at the perturbative level. The results reviewed here 
strongly support the possibility of extending the correspondence at 
the non-perturbative level as well. The agreement found between the 
contributions to four- and higher-point functions due to two 
different sources of non-perturbative effects, instantons in ${\cal 
N}$=4 super Yang--Mills and D-instantons in type IIB superstring 
theory, represents a highly non trivial check of the conjecture. 
Furthermore these results have been obtained in the case of a SU(2) 
gauge group and therefore suggest that the correspondence, originally 
formulated for a SU($N$) gauge group in the large $N$ limit, may 
actually hold at finite $N$ as well. Perturbative checks of this claim 
have already been proposed, but the non-perturbative results discussed 
here bring stronger evidence in this direction, since they  concern 
four- and higher-point functions that are not fixed by superconformal 
symmetry. Despite these interesting results on the field theory side, 
there is still very little understanding of the correct way of dealing 
with string corrections to the supergravity approximation for finite 
$N$. This is a very interesting subject for future developments. In 
particular generalizations of the original conjecture to theories 
with less supersymmetry and even to non-supersymmetric theories have 
been proposed. In this context computations of quantities of 
phenomenological interest, such as glueball spectra, have been 
performed. However the possibility of comparing this results with, for 
instance, the continuum extrapolations from lattice simulations 
requires the extension of the correspondence to weak 't Hooft 
coupling, \ie to finite $N$ including string effects.

The four-point correlation function computed in chapter \ref{cap4} 
displays logarithmic singularities when any two insertion points are 
taken to coincide. This peculiar behaviour in the operator product 
expansion (OPE) has also been observed by other authors in supergravity 
calculations in AdS$_{5}$ and is a very interesting feature of the 
superconformal field theory at hand. 

In understanding this phenomenon a crucial r\^ole is played by the 
fact that the chiral primary operators, ${\cal O}^{\Delta}_{q}$, 
that enter the Green functions under consideration are protected, 
\ie their dimension $\Delta$ and charge $q$ are related by $\Delta=q$. 
The first point to observe is the absence of pole-type singularities 
in the OPE. This is a direct consequence of the condition $\Delta=q$. 
If one considers the OPE for chiral primary operators with charges 
$q_{1}$ and $q_{2}$, 
\begin{displaymath}
	{\cal O}_{(q_{1})}(x) {\cal O}_{(q_{2})}(y) \hspace{1cm} 
	{\rm for} \hspace{0.6cm} |x-y|\to 0 \, ,
\end{displaymath}
there are two possibilities. Either the OPE contains only chiral 
primary operators and one has
\begin{displaymath}
	{\cal O}_{(q_{1})}(x) {\cal O}_{(q_{2})}(y) \sim c \, {\cal 
	O}^{\Delta}_{(q_{1}+q_{2})}\left(\frac{x+y}{2}\right)  + \, 
	{\rm terms~vanishing~as} \quad |x-y|\to 0\, ,
\end{displaymath}
with $\Delta=q_{1}+q_{2}$ and $c$ a numerical constant, or there is a 
mixing with a non chiral primary operator, whose dimension is strictly 
larger than the charge, and then
\begin{displaymath}
	{\cal O}_{(q_{1})}(x) {\cal O}_{(q_{2})}(y) \sim 
	{\cal V}_{q}^{\Delta}\left(\frac{x+y}{2}\right) + \ldots \: ,
\end{displaymath}
where $q=q_{1}+q_{2}$ and $\Delta = q_{1}+q_{2}+\gamma > 
q_{1}+q_{2}$. As a consequence in any case there cannot be a pole-type 
singularity. The second situation described here could be responsible 
for the logarithmic behaviour. The anomalous dimension $\gamma$ is 
determined by the calculation of the three point function $\langle 
{\cal V}{\cal O}{\cal O}\rangle$, that is obtained by taking the limit 
of coincident points in a four-point function of chiral primary 
operators like the one that is computed in chapter \ref{cap4}. Notice 
that one cannot study the limit in a three-point function, since 
three-point functions of chiral primaries are not renormalized. The 
logarithmically divergent coefficient in the OPE of ${\cal 
O}_{(q_{1})}$ and ${\cal O}_{(q_{2})}$ could originate from a 
perturbative expansion of $\gamma=\gamma(g)$. More precisely if the 
operator ${\cal V}$ has an anomalous dimension, $\gamma$, in perturbation 
theory the latter is expected to be small and to admit an expansion in the
coupling constant $g$. Expanding ${\cal O}_{(q_{1})}(x) 
{\cal O}_{(q_{2})}(y)$ in powers of $\gamma$ yields  
\begin{eqnarray*}
	\langle {\cal O}_{(q_{1})}(x) {\cal O}_{(q_{2})}(y) \rangle
	&\sim& \frac{1}{(x-y)^{2\Delta}} \left[ \rule{0pt}{16pt} 
	1 - \gamma \log(\mu^2 (x-y)^2) + \right. \\
	&& \left.+\frac{1}{2} \gamma^2 (\log (\mu^2 (x-y)^2))^2 + \ldots
	\right] \, ,
\end{eqnarray*}
where an arbitrary mass scale $\mu$ has been introduced for dimensional 
reasons. At each non-trivial order in $g$, the last equation seems
to violate conformal invariance. The exact expression is however 
conformally invariant by construction. One can reconcile the
two results by observing that because of the presence of an 
anomalous dimension, the operator has to be renormalized at each 
order in $g$ and as a consequence it does not provide a linear 
representation of the conformal group.



\appendix

\chapter{Notations and spinor algebra}
\label{appa}
\vspace*{2cm}
\fancyhead[RO,LE]{\thepage}
\fancyhead[RE]{{\footnotesize {\rm Appendix A.}~~{\it 
Notations and Algebra}}} 
\fancyhead[LO]{{\footnotesize {\it Notations and Algebra}}}

\noindent
Minkowskian metric
\begin{equation}
	\eta_{\mu\nu}=\left(\begin{array}{cccc} -1 & 0 & 0 & 0 \\ 
	0 & 1 & 0 & 0 \\ 0 & 0 & 1 & 0 \\ 0 & 0 & 0 & 1 \end{array}
	\right) \, .
	\label{minkowskmetr}
\end{equation}
$\sigma$-matrices 
\begin{equation}
	\sigma^{\mu}=(-\I_{2},\sigma^{i}) \, , \qquad \quad 
	{\overline\sigma}^{\mu}=(-\I_{2},-\sigma^{i}) \, , \qquad i=1,2,3 \, ,
	\label{sigmamatr}
\end{equation}
where
\begin{equation}
    \sigma^{1} = \left( \begin{array}{cc} 0 & 1 \\ 1 & 0 \end{array}
    \right) \, , \quad \sigma^{2} = \left( \begin{array}{cc} 0 & -i \\ 
    i & 0 \end{array}\right) \, , \quad \sigma^{3} = \left( 
    \begin{array}{cc} 1 & 0 \\ 0 & -1 \end{array}\right) \, .
    \label{pauli}
\end{equation}
Weyl spinors are denoted as follows
\begin{equation}
	\psi_{\alpha} \, , \quad {\overline \psi}^{\dot \alpha} \, , \quad 
	{\rm with} \quad \alpha,{\dot\alpha}=1,2 \, .
	\label{weylspin}
\end{equation}
Spinor indices are raised and lowered by the antisymmetric tensors 
$\varepsilon^{\alpha\beta}$ ($\varepsilon^{12}=- \varepsilon^{21}=-
\varepsilon_{12}=\varepsilon_{21}=1$) and $\varepsilon^{{\dot \alpha}
{\dot \beta}}$ ($\varepsilon^{{\dot 1}{\dot 2}}=- \varepsilon^{{\dot 2}
{\dot 1}}=-\varepsilon_{{\dot 1}{\dot 2}}=
\varepsilon_{{\dot 2}{\dot 1}}=1$) 
\begin{eqnarray}
	&& \psi^{\alpha}=\varepsilon^{\alpha\beta}\psi_{\beta} \, , \quad 
	\psi_{\alpha}=\varepsilon_{\alpha\beta}\psi^{\beta} \, , \nonumber \\ 
	&& {\overline \psi}^{\dot\alpha}=\varepsilon^{{\dot \alpha}{\dot \beta}}
	{\overline \psi}_{\dot \beta} \, , \quad 
	{\overline \psi}_{\dot\alpha}=\varepsilon_{{\dot \alpha}{\dot \beta}}
	{\overline \psi}^{\dot \beta} \, .
	\label{raiselower}
\end{eqnarray}
The matrices $\sigma^{\mu}$ have the following index structure
\begin{equation}
	\sigma^{\mu}=\sigma^{\mu}{}_{\alpha{\dot \alpha}} \, .
	\label{sigmaindex}
\end{equation}
Indices of the $\sigma^{\mu}$ matrices are raised and lowered by the 
$\varepsilon$ tensor as well
\begin{equation}
	{\overline \sigma}^{\mu\,{\dot \alpha}\alpha} = 
	\varepsilon^{{\dot \alpha}{\dot \beta}}\varepsilon^{\alpha \beta} 
	\sigma^{\mu}{}_{\beta{\dot \beta}} \, .
	\label{sigmabar}
\end{equation}

\vspace{0.7cm}
{\sl Properties of the $\sigma$-matrices.} 
\vspace{0.3cm}

The following properties of the $\sigma$-matrices can be easily 
derived from their definition
\begin{eqnarray}
	(\sigma^{\mu}{\overline \sigma}^{\nu} + \sigma^{\nu}
	{\overline \sigma}^{\mu})_{\alpha}{}^{\beta} & = & 
	-2\eta^{\mu\nu}\delta_{\alpha}{}^{\beta} \nonumber \\
	({\overline \sigma}^{\mu}\sigma^{\nu} + {\overline \sigma}^{\nu}
	\sigma^{\mu})^{\dot\alpha}{}_{\dot\beta} & = & -2\eta^{\mu\nu}
	\delta^{\dot\alpha}{}_{\dot\beta} \, ,
	\label{sigmasigmabar}
\end{eqnarray} 
as well as the completeness relations
\begin{eqnarray}
	\tr \left( \sigma^{\mu}{\overline \sigma}^{\nu} \right) & = & 
	-2\eta^{\mu\nu}
	\nonumber  \\
	\sigma^{\mu}_{\alpha{\dot\alpha}}
	{\overline \sigma}_{\mu}^{{\dot\beta}\beta} & = & 
	-2\delta_{\alpha}{}^{\beta}\delta_{\dot\alpha}{}^{\dot\beta} \, .
	\label{completeness}
\end{eqnarray}
The generators of Lorentz transformations read
\begin{eqnarray}
	\sigma^{\mu\nu}{}_{\alpha}{}^{\beta} & = & \frac{1}{4}
	\left( \sigma_{\alpha{\dot\alpha}}^{\mu}{\overline 
	\sigma}^{\nu\,{\dot\alpha}\beta}-\sigma^{\nu}_{\alpha{\dot\alpha}}
	{\overline \sigma}^{\mu\,{\dot\alpha}\beta} \right) \nonumber \\
	{\overline \sigma}^{\mu\nu}{}^{\dot\alpha}{}_{\dot\beta} & = & 
	\frac{1}{4}\left( {\overline \sigma}^{\mu\,{\dot\alpha}\alpha}
	\sigma_{\alpha{\dot\beta}}^{\nu}-{\overline \sigma}^{\nu\,{\dot\alpha}
	\alpha}\sigma_{\alpha{\dot\beta}}^{\mu} \right) \, .
	\label{lorentzgen}
\end{eqnarray}

\vspace{0.7cm}
{\sl Euclidean $\sigma$-matrices.} 
\vspace{0.3cm}

They are defined as~\footnote{The subscript E is suppressed in the text.}
\begin{equation}
	\sigma^{\mu}_{{\rm E}}=(\I_{2},i\sigma^{i}) \, , \qquad \quad 
	{\overline\sigma}^{\mu}_{{\rm E}}=(\I_{2},-i\sigma^{i}) \, , 
	\qquad i=1,2,3 \, .
	\label{euclidsigmamatr}
\end{equation}
The Euclidean $\sigma$-matrices satisfy properties analogous to their 
Minkowskian couterparts. In particular
\begin{eqnarray}
	(\sigma^{\mu}_{{\rm E}}{\overline \sigma}^{\nu}_{{\rm E}} + 
	\sigma^{\nu}_{{\rm E}}
	{\overline \sigma}^{\mu}_{{\rm E}})_{\alpha}{}^{\beta} & = & 
	2\delta^{\mu\nu}\delta_{\alpha}{}^{\beta} \nonumber \\
	({\overline \sigma}^{\mu}_{{\rm E}}\sigma^{\nu}_{{\rm E}} + 
	{\overline \sigma}^{\nu}_{{\rm E}}
	\sigma^{\mu}_{{\rm E}})^{\dot\alpha}{}_{\dot\beta} & = & 
	2\delta^{\mu\nu} \delta^{\dot\alpha}{}_{\dot\beta} \, ,
	\label{euclsigmasigmabar}
\end{eqnarray} 
\begin{eqnarray}
	\tr \left( \sigma^{\mu}_{{\rm E}}{\overline \sigma}^{\nu}_{{\rm E}} 
	\right) & = & 2\delta^{\mu\nu}
	\nonumber  \\
	\sigma^{\mu}_{{\rm E}\,\alpha{\dot\alpha}}
	{\overline \sigma}_{{\rm E}\,\mu}^{{\dot\beta}\beta} & = & 
	2\delta_{\alpha}{}^{\beta}\delta_{\dot\alpha}{}^{\dot\beta} \, .
	\label{euclcompleteness}
\end{eqnarray}
The Euclidean Lorentz generators are
\begin{eqnarray}
	\sigma^{\mu\nu}_{{\rm E}}{}_{\alpha}{}^{\beta} & = & \frac{1}{4}
	\left( \sigma_{{\rm E}\,\alpha{\dot\alpha}}^{\mu}{\overline 
	\sigma}^{\nu\,{\dot\alpha}\beta}_{{\rm E}}
	-\sigma^{\nu}_{{\rm E}\,\alpha{\dot\alpha}}
	{\overline \sigma}^{\mu\,{\dot\alpha}\beta}_{{\rm E}} 
	\right) \nonumber \\
	{\overline \sigma}_{{\rm E}}^{\mu\nu}{}^{\dot\alpha}
	{}_{\dot\beta} & = & \frac{1}{4}\left( 
	{\overline \sigma}^{\mu\,{\dot\alpha}\alpha}_{{\rm E}}
	\sigma_{{\rm E}\,\alpha{\dot\beta}}^{\nu}-
	{\overline \sigma}^{\nu\,{\dot\alpha} \alpha}_{{\rm E}} 
	\sigma_{{\rm E}\,\alpha{\dot\beta}}^{\mu} \right) \, .
	\label{eucllorentzgen}
\end{eqnarray}
The following realization of the Dirac $\gamma$-matrices is used
\begin{equation}
	\gamma^{\mu} = \left( \begin{array}{cc} 0 & \sigma^{\mu} \\
	{\overline\sigma}^{\mu} & 0 \end{array} \right) \, , 
	\qquad \gamma^{5} = i\gamma^{0}\gamma^{1}\gamma^{2}\gamma^{3} = 
	\left( \begin{array}{cc} 1 & 0 \\ 0 & -1 \end{array} \right) \, .
	\label{gammamatr}
\end{equation}

\vspace{0.7cm}
{\sl Spinor algebra.} 
\vspace{0.3cm}

Contractions between spinor indices are taken according to the 
conventions
\begin{eqnarray}
	&& \psi\chi = \psi^{\alpha}\chi_{\alpha} = - \psi_{\alpha}
	\chi^{\alpha} = \chi\psi \nonumber \\
	&& {\overline\psi}{\overline\chi} = {\overline\psi}_{\dot\alpha}
	{\overline\chi}^{\dot\alpha} = -{\overline\psi}^{\dot\alpha}
	{\overline\chi}_{\dot\alpha} = {\overline\chi}{\overline\psi} \, .
	\label{spincontraction}
\end{eqnarray}
Fierz rearrangements
\begin{eqnarray}
	&& \psi^{\alpha}\psi^{\beta} = -\frac{1}{2}\varepsilon^{\alpha\beta}
	\psi\psi \: , \hspace{1cm} \psi_{\alpha}\psi_{\beta} = \frac{1}{2}
	\varepsilon_{\alpha\beta}\psi\psi \nonumber \\
	&& {\overline\psi}^{\dot\alpha}{\overline\psi}^{\dot\beta} = 
	\frac{1}{2} \varepsilon^{{\dot\alpha}{\dot\beta}}
	{\overline\psi}{\overline\psi} \: , \hspace{1cm} 
	{\overline\psi}_{\dot\alpha}{\overline\psi}_{\dot\beta} = 
	-\frac{1}{2} \varepsilon_{{\dot\alpha}{\dot\beta}}
	{\overline\psi}{\overline\psi} \, .
	\label{psipsifierz}
\end{eqnarray}
Miscellaneous relations
\begin{eqnarray}
	\theta\sigma^{\mu}{\overline\theta}\theta\sigma^{\nu}
	{\overline\theta} = -\frac{1}{2} \theta\theta{\overline\theta}
	{\overline\theta} \eta^{\mu\nu} && \nonumber \\
	(\theta\psi)(\theta\chi) = -\frac{1}{2}(\theta\theta)(\psi\chi) 
	\, &,& \hspace{0.5cm} 
	({\overline\theta}{\overline\psi})({\overline\theta}
	{\overline\chi}) = -\frac{1}{2}({\overline\theta}{\overline\theta})
	({\overline\psi}{\overline\chi}) \nonumber \\
	\chi\sigma^{\mu}{\overline\psi} = -\psi\sigma^{\mu}{\overline\chi}
	\, &,& \hspace{0.5cm} (\chi\sigma^{\mu}{\overline\psi})^{\dagger} = 
	\psi\sigma^{\mu}{\overline\chi} \nonumber \\
	\chi\sigma^{\mu}{\overline\sigma}^{\nu}{\overline\psi} = \psi
	\sigma^{\nu}{\overline\sigma}^{\mu}{\overline\chi} \,&,&\hspace{0.5cm} 
	(\chi\sigma^{\mu}{\overline\sigma}^{\nu}{\overline\psi})^{\dagger} = 
	{\overline\psi}{\overline\sigma}^{\nu}\sigma^{\mu}{\overline\chi}
	\nonumber \\
	(\psi\chi){\overline\lambda}_{\dot\alpha} = -\frac{1}{2} 
	(\chi\sigma^{\mu}{\overline\lambda})
	(\psi\sigma_{\mu})_{\dot\alpha} \, . &&
	\label{variousformulae}
\end{eqnarray}

\vspace{0.7cm}
{\sl Superspace covariant derivatives.} 
\vspace{0.3cm}

Covariant derivatives in Minkowskian superspace are defined as
\begin{equation}
	D_{\alpha}=\frac{\partial}{\partial\theta^{\alpha}}+
	i\sigma^{\mu}_{\alpha{\dot\alpha}}{\overline\theta}^{\dot\alpha}
	\partial_{\mu}\, , \hspace{1cm} 
	{\overline D}_{\dot\alpha}= -\frac{\partial}{\partial
	{\overline\theta}^{\dot\alpha}}-
	i\theta^{\alpha}\sigma^{\mu}_{\alpha{\dot\alpha}}
	\partial_{\mu}
	\label{supercovderiv}
\end{equation}
and satisfy
\begin{eqnarray}
	&& \{D_{\alpha},{\overline D}_{\dot\alpha}\} = 
	-2i\sigma^{\mu}_{\alpha{\dot\alpha}}\partial_{\mu}
	\\ \nonumber
	&& \{D_{\alpha},D_{\beta}\} = \{{\overline D}_{\dot\alpha},
	{\overline D}_{\dot\beta}\} = 0 \, .
	\label{anticommderiv}
\end{eqnarray}
Moreover 
\begin{equation}
	\varepsilon^{\alpha\beta}\frac{\partial}{\partial\theta^{\alpha}} = 
	-\frac{\partial}{\partial\theta_{\alpha}} \, , \hspace{1cm} 
	\varepsilon^{\alpha\beta}\frac{\partial}{\partial\theta^{\alpha}}
	\frac{\partial}{\partial\theta^{\beta}} \theta\theta = 
	\varepsilon^{{\dot\alpha}{\dot\beta}}\frac{\partial}
	{\partial{\overline\theta}^{\dot\alpha}}
	\frac{\partial}{\partial{\overline\theta}^{\dot\beta}} 
	{\overline\theta}{\overline\theta} = 4 \, .
	\label{raiselowderiv}
\end{equation}
Other properties
\begin{eqnarray}
	&& [{\overline D}_{\dot\alpha},\{{\overline 
	D}_{\dot\beta},D_{\gamma}\}] = 0 \, , \hspace{1cm} 
	[D_{\alpha},{\overline D}_{\dot\beta}{\overline D}^{\dot\beta}] =
	-4i\sigma^{\mu}_{\alpha{\dot\alpha}}\partial_{\mu} {\overline 
	D}^{\dot\alpha}
	\nonumber \\
	&& [D^{2},{\overline D}^{2}] = 
	-8iD^{\alpha}\sigma^{\mu}_{\alpha{\dot\alpha}}{\overline D}^{\dot\alpha}
	\partial_{\mu} -16 \Box = 8i{\overline D}^{\dot\alpha} 
	\sigma^{\mu}_{\alpha{\dot\alpha}} D^{\alpha}\partial_{\mu}+16\Box \, . 
	\label{commvarideriv}
\end{eqnarray}

The corresponding Euclidean superspace derivatives read
\begin{equation}
	D^{{\rm E}}_{\alpha}=\frac{\partial}{\partial\theta^{\alpha}}+
	\sigma^{\mu}_{\alpha{\dot\alpha}}{\overline\theta}^{\dot\alpha}
	\partial_{\mu}\, , \hspace{1cm} 
	{\overline D}^{{\rm E}}_{\dot\alpha}= -\frac{\partial}{\partial
	{\overline\theta}^{\dot\alpha}} -
	\theta^{\alpha}\sigma^{\mu}_{\alpha{\dot\alpha}}
	\partial_{\mu}
	\label{euclsupercovderiv}
\end{equation}
and possess analogous properties
\begin{eqnarray}
	&& \{D^{{\rm E}}_{\alpha},{\overline D}^{{\rm E}}_{\dot\alpha}\} = 
	-2\sigma^{\mu}_{\alpha{\dot\alpha}}\partial_{\mu}
	\\ \nonumber
	&& \{D^{{\rm E}}_{\alpha},D^{{\rm E}}_{\beta}\} = 
	\{{\overline D}^{{\rm E}}_{\dot\alpha},
	{\overline D}^{{\rm E}}_{\dot\beta}\} = 0 \, .
	\label{euclanticommderiv}
\end{eqnarray}
\begin{eqnarray}
	&& [{\overline D}^{{\rm E}}_{\dot\alpha},\{{\overline 
	D}^{{\rm E}}_{\dot\beta},D^{{\rm E}}_{\gamma}\}] = 0 \nonumber \\ 
	&& [D^{{\rm E}}_{\alpha},{\overline D}^{{\rm E}}_{\dot\beta}
	{\overline D}^{{\rm E}\,\dot\beta}] =
	-4\sigma^{\mu}_{{\rm E}\,\alpha{\dot\alpha}}\partial_{\mu} {\overline 
	D}^{{\rm E}\,\dot\alpha} \, , \hspace{1cm} 
	[D^{\alpha}_{{\rm E}}D_{\alpha}^{{\rm E}}, {\overline 
	D}_{\dot\alpha}^{{\rm E}}] = -4\sigma^{\mu}_{{\rm E}\,\alpha{\dot\alpha}}
	\partial_{\mu}D^{\alpha}_{{\rm E}} 
	\nonumber \\
	&& [D_{{\rm E}}^{2},{\overline D}^{2}_{{\rm E}}] = 
	-8D^{\alpha}_{{\rm E}}\sigma^{\mu}_{{\rm E}\,\alpha{\dot\alpha}}
	{\overline D}^{\dot\alpha}_{{\rm E}}
	\partial_{\mu} -16 \Box = 8{\overline D}^{\dot\alpha}_{{\rm E}} 
	\sigma^{\mu}_{{\rm E}\, \alpha{\dot\alpha}} 
	D^{\alpha}_{{\rm E}}\partial_{\mu}+16\Box \, . 
	\label{euclcommvarideriv}
\end{eqnarray}
The following relations are useful in superspace calculations
\begin{eqnarray}
	&& D^{2}_{{\rm E}}{\overline D}^{2}_{{\rm E}}D^{2}_{{\rm E}} = 
	16\Box D^{2}_{{\rm E}} \, , \hspace{1cm} {\overline D}^{2}_{{\rm E}}
	D^{2}_{{\rm E}}{\overline D}^{2}_{{\rm E}} = 16 \Box 
	{\overline D}^{2}_{{\rm E}} \nonumber \\
	&& {\overline D}^{2}_{{\rm E}}D_{\alpha}^{{\rm E}} 
	D_{\beta}^{{\rm E}}{\overline D}^{2}_{{\rm E}} = 
	-16 \sigma_{{\rm E}\, \alpha{\dot\alpha}}^{\mu}
	\sigma_{{\rm E}\,\beta{\dot\beta}}^{\nu}\partial_{\mu}\partial_{\nu}
	{\overline D}_{{\rm E}}^{\dot\alpha}{\overline D}_{{\rm 
	E}}^{\dot\beta} \nonumber \\
	&& D^{2}_{{\rm E}} {\overline D}_{\dot\alpha}^{{\rm E}}
	{\overline D}_{\dot\beta}^{{\rm E}} D^{2}_{{\rm E}} = 
	-16 \sigma_{{\rm E}\, \alpha{\dot\alpha}}^{\mu}
	\sigma_{{\rm E}\,\beta{\dot\beta}}^{\nu}\partial_{\mu}\partial_{\nu}
	D_{{\rm E}}^{\alpha}D_{{\rm E}}^{\beta} \, .
	\label{usefulcovrel}
\end{eqnarray}

\vspace{0.7cm}
{\sl Berezin integration.} 
\vspace{0.3cm}

Integration over a Grassmannian variable $\eta$ is defined as
\begin{equation}
	\int d\eta = 0 \, , \hspace{1cm} \int d\eta \, \eta = 1 \, .
	\label{eq:berezinint}
\end{equation}
The power expansion of a function $f(\eta)$ is simply
\begin{equation}
	f(\eta) = a + b \eta \, ,
	\label{powergrass}
\end{equation}
so that
\begin{equation}
	\int d\eta \, f(\eta) = b \, , \hspace{1cm} 
	\int d\eta \, \eta f(\eta) = a \, .
	\label{intfunction}
\end{equation}
Integration by parts in Berezin integrals is allowed since 
\begin{equation}
	\int d \eta \, \frac{\partial}{\partial\eta} f(\eta) = 0 \, .
	\label{graspartialint}
\end{equation}
The $\delta$-function is defined by 
\begin{equation}
	\int d\eta \, f(\eta) \delta(\eta) = f(0) \, ,
	\label{grasdeltafunct}
\end{equation}
which implies $\delta(\eta)=\eta$.
The integration measure over the Grassmannian variables of ${\cal 
N}$=1 superspace is defined as follows
\begin{eqnarray}
	&& d^{4}\theta = d^{2}\theta d^{2}{\overline\theta}  \nonumber \\
	&& d^{2}\theta = \frac{1}{4} \varepsilon_{\alpha\beta} d\theta^{\alpha}
	d\theta^{\beta} \, , \hspace{1cm} 
	d^{2}{\overline\theta} = \frac{1}{4} 
	\varepsilon^{{\dot\alpha}{\dot\beta}} d{\overline\theta}_{\dot\alpha}
	d{\overline\theta}_{\dot\beta} \, ,
	\label{n1supspmeasure}
\end{eqnarray}
so that 
\begin{equation}
	\int d^{2}\theta \, \theta\theta = \int d^{2}{\overline\theta} \, 
	{\overline\theta}{\overline\theta} = 1 \, .
	\label{eq:normalizintssp}
\end{equation}

Other useful formulas can be found in the appendices of 
\cite{wessbagger,west}


\chapter{Electric-magnetic duality}
\label{appb}
\vspace*{2cm}
\fancyhead[RO,LE]{\thepage}
\fancyhead[RE]{{\footnotesize {\rm Appendix B.}~~{\it 
Electric-magnetic duality}}} 

\fancyhead[LO]{{\footnotesize B.1~~{\it Electric-magnetic duality}}}
\section{Electric-magnetic duality and the Dirac monopole}
\label{emduality}

The free Maxwell equations 
\begin{equation}
\begin{array}{ll}
  \begin{displaystyle}\vec{\nabla} \cdot \vec{E} = 0 \end{displaystyle}
  & \hspace{1cm} 
  \begin{displaystyle}\vec{\nabla} \cdot \vec{B} = 0 
  \end{displaystyle} \\ \begin{displaystyle}\vec{\nabla} \times 
  \vec{B} - \frac{\partial E}{\partial t} = 0 \end{displaystyle}  &
  \hspace{1cm} \begin{displaystyle}\vec{\nabla} \times 
  \vec{E} + \frac{\partial B}{\partial t} = 0 \end{displaystyle} \, ,
\end{array}
\end{equation}
are invariant under a duality transformation $D$, exchanging electric 
and magnetic fields according to 
\begin{displaymath}
	D: \qquad \vec{E} \longrightarrow \vec{B} \, , \qquad 
	\vec{B} \longrightarrow -\vec{E} \: .
\end{displaymath}
More generally the equations are invariant under a continuous 
transformation 
\begin{equation}
	\left( \begin{array}{c} \vec{E} \\ \vec{B} \end{array} \right) 
	\longrightarrow \left(\begin{array}{cc} \cos \theta & \sin \theta \\
	-\sin \theta & \cos \theta \end{array} \right) \left( \begin{array}{c}
	\vec{E} \\ \vec{B} \end{array} \right) \: .
	\label{continuousdual}
\end{equation}
A covariant formulation can be given using the language of 
differential forms. The free equations can be written 
\begin{displaymath}
	d\,F=0 \; , \qquad *^{-1}d*F=0 \:,
\end{displaymath}
where $(*F)_{\mu\nu}=\frac{1}{2}\varepsilon_{\mu\nu\rho\sigma} 
F^{\rho\sigma}$ and $d$ denotes the external derivative. The duality 
transformation simply becomes $F \, \rightarrow \, *F$. In the presence 
of electric sources the equations become 
\begin{displaymath}
	dF=0 \; , \qquad *^{-1}d*F=j \: ,
\end{displaymath}
where $j$ is a one-form of components $j_{e}^{\mu} = 
(\rho_{e},\vec{j}_{e})$, so that the symmetry is broken. To 
recover the duality symmetry magnetic source terms must be included 
symmetrizing the equations as follows
\begin{displaymath}
  dF=j_{m} \; , \qquad *^{-1}d*F=j_{e} \: .
\end{displaymath}
The system is then symmetric under the simultaneous exchange of $F$ 
with $*F$ and of $j_{e}$ with $j_{m}$.

Even if the inclusion of magnetic point-like particles in classical 
electromagnetism seems to be ruled out by the experimental results, it 
allows to derive a consequence that is of fundamental importance in 
the more interesting case of non Abelian theories, the {\em 
Dirac quantization condition}. A consistent 
quantization~\footnote{Strikingly the Dirac quantization condition can 
be already derived in the context of classical electromagnetism.} of a 
theory with both electric and magnetic charges $q_{i}$ and $g_{j}$ 
requires that they satisfy the condition 
\begin{equation}
	q_{i}\cdot g_{j} = 2\pi \hslash n_{ij} \, ,
\label{diracquant}
\end{equation}
where $n_{ij}$ are arbitrary integers.

The quantum mechanical description of a charged particle in an 
electromagnetic field requires the introduction of the vector 
potential $A_{\mu}$, related to the field strength by 
$F_{\mu\nu}=\partial_{\mu}A_{\nu}-\partial_{\nu}A_{\mu}$, through the 
substitution $\vec{\nabla}\rightarrow(\vec{\nabla}-
\frac{iq}{\hslash}\vec{A})$. This leads to the minimal coupling in 
the Schroedinger equation
\begin{displaymath}
	i\hslash\frac{\partial \psi}{\partial t} = 
	-\frac{\hslash^{2}}{2m}(\vec{\nabla}-\frac{iq}{\hslash}\vec{A})^{2}
	\psi +V\psi \, .
\end{displaymath}
Gauge transformations for this system read
\begin{equation}
	\psi\longrightarrow e^{-i\frac{q}{\hslash}\chi(\theta)} \psi
	\, , \hspace{1cm} \vec{A} \longrightarrow 
	\vec{A}-\vec{\nabla}\chi(\theta) \, .
	\label{equatgauge}
\end{equation}
In the presence of a point like magnetic charge it is not possible to 
associate a globally defined vector potential to the magnetic field 
through $\vec{B}=\vec{\nabla}\times\vec{A}$. However it is possible 
to construct a well defined vector potential in the northern 
hemisfere ($\vec{A}_{_{N}}$) and in the southern hemisfere 
($\vec{A}_{_{S}}$). This is physically consistent if the fields 
$\vec{A}_{_{N}}$ and $\vec{A}_{_{S}}$ differ by a gauge 
transformation on the overlapping region (the plane of the equator)
\begin{displaymath}
	\left.\vec{A}_{_{N}}\right|_{{\rm eq}} = 
	\left.\vec{A}_{_{S}}\right|_{{\rm eq}}+\vec{\nabla}\chi
\end{displaymath}
and analogously for the wave function
\begin{displaymath}
	\left.\psi_{_{N}}\right|_{{\rm eq}} = 
	e^{-i\frac{q}{\hslash}\chi}\left.\psi_{_{S}}\right|_{{\rm eq}} 
	\, .
\end{displaymath}
By means of Stokes' theorem one obtains 
\begin{displaymath}
	g=\int_{S^{3}} d\vec{\sigma}\cdot\vec{B} = \int_{{\rm eq}}
	d\vec{\ell} \cdot \vec{A}_{_{N}} - \int_{{\rm eq}}
	d\vec{\ell} \cdot \vec{A}_{_{S}} = \chi(2\pi)-\chi(0) \, .
\end{displaymath}
But the wave function $\psi$ must be single valued along the equator 
requiring from (\ref{equatgauge}) 
$\chi(2\pi)-\chi(0)=\frac{2\pi\hslash}{q}n$ which leads to the Dirac 
quantization condition, $qg=2\pi\hslash n$, with $n\in \mathbb{Z}$.
The relation (\ref{diracquant}) is not invariant under the continuous 
duality transformation (\ref{continuousdual}), but it can be generalized 
to the so called Dirac-Schwinger-Zwanziger condition \cite{schwizwanz} 
that is invariant. The latter applies to a system containing two dyonic 
states (\ie states possessing both electric and magnetic charge) of 
charges $(q_{i},g_{i})$ and $(q_{j},g_{j})$ and reads
\begin{displaymath}
	q_{i}g_{j}-q_{j}g_{i}=2\pi\hslash n_{ij} \, .
\end{displaymath}

\section{Non Abelian case: the 't Hooft--Polyakov \\ monopole}
\label{thooftpolyakov}
\fancyhead[LO]{{\footnotesize B.2~~{\it 't Hooft--Polyakov monopole}}}

As discussed in the preceding section magnetic monopoles must be 
included at hand in a purely electromagnetic model and there exist no 
experimental evidence justifying this construction. In non Abelian 
gauge theories on the contrary monopoles and dyons emerge naturally 
in association with particular field configurations. Indeed non 
Abelian gauge theories admit solitonic solutions of the classical 
equations of motion that can be interpreted as describing monopole 
(and in general dyon) states in the quantum theory. The case of the 
{\em 't Hooft--Polyakov monopole} \cite{thooftmonop,polymonop} in the 
Georgi--Glashow model with gauge group SU(2) will be considered 
here. The results can be generalized to a larger gauge group G by, 
for example, embedding a SU(2) factor in G.

The Georgi--Glashow model is defined by the Lagrangian
\begin{equation}
	{\cal L}=-\frac{1}{4}F^{a}_{\mu\nu}F^{a\mu\nu}+
	\frac{1}{2}D^{\mu}\Phi^{a}D_{\mu}\Phi_{a}-V(\Phi) \, ,
\label{lgeorgiglash}
\end{equation}
where 
$F_{\mu\nu}^{a}=\partial_{\mu}A^{a}_{\nu}-\partial_{\nu}A^{a}_{\mu}+
e\varepsilon^{abc}A_{b\mu}A_{c\nu}$ is the non Abelian field strength 
and the scalar fields $\Phi^{a}$ are in the adjoint representation of 
the gauge group SU(2) with covariant derivative~\footnote{From now on 
natural units will be used so that $\hslash = c = 1$.} 
$D_{\mu}\Phi^{a}=\partial_{\mu}\Phi^{a}+e\varepsilon^{abc}A_{b\mu}\Phi_{c}$. 
The potential $V(\Phi)$ is 
\begin{displaymath}
	V(\Phi)=\frac{\lambda}{4}\left(\Phi^{a}\Phi_{a}-v^{2}\right)^{2} \, ,
\end{displaymath}
so that the classical equations of motion are 
\begin{eqnarray}
	D_{\mu}F^{a\mu\nu} & = & \varepsilon^{abc}\Phi_{b}D^{\nu}\Phi_{c}
	\nonumber  \\
	(D^{\mu}D_{\mu}\Phi)^{a} & = & 
	-\lambda\Phi^{a}(\Phi^{b}\Phi_{b}-v^{2}) \, .
	\label{eq:eqmotgg}
\end{eqnarray}
Furthermore $F_{\mu\nu}$ satisfies the Bianchi identity 
$D_{\mu}(*F)^{\mu\nu}=0$.

The total energy for a given field configuration is 
\begin{equation}
	E=\int d^{3}x \, \Theta_{00}=\int d^{3}x\,\left\{\left[\frac{1}{2}
	(B^{a}_{i})^{2}+(E^{a}_{i})^{2}+(\Pi^{a})^{2}+[(D_{i}\Phi)^{a}]^{2}
	\right] + V(\Phi) \right\} \, ,
	\label{energy}
\end{equation}
where $\Pi^{a}$ is the momentum conjugate to $\Phi^{a}$ and 
$E^{a}_{i}$ and $B^{a}_{i}$ are the `non Abelian' electric and 
magnetic fields
\begin{displaymath}
	F^{a}_{0i} = E^{a}_{i} \, , \hspace{1cm} 
	F^{a}_{ij}=-\varepsilon_{ijk}B^{ak} \, .
\end{displaymath}
The energy density $\Theta_{00}$ is semi-positive definite, 
$\Theta_{00}\geq 0$, with equality if and only if 
$F^{a}_{\mu\nu}=D_{\mu}\Phi^{a}=V(\Phi)=0$, implying that the vacuum 
of the theory corresponds to configurations with vanishing 
$A_{\mu}^{a}$ and constant $\Phi^{a}$ satisfying 
$\Phi^{a}\Phi_{a}=v^{2}$. The set of configurations corresponding to 
the Higgs vacuum (moduli space) will be denoted by ${\cal M}_{H}$
\begin{displaymath}
	{\cal M}_{H}= \left\{ \Phi\,: D_{\mu}\Phi^{a}=0\, , \; V(\Phi)=0
	\right\}
\end{displaymath}
In the case at hand ${\cal M}_{H}$ is the sphere $S^{2}$ parametrized by 
the `coordinates' $\Phi^{a}$ constrained by $\Phi^{a}\Phi_{a}=v^{2}$. 

For $v\neq 0$ the gauge symmetry is broken to U(1) and the Higgs 
mechanism produces   a perturbative spectrum which consists of a 
massless `photon', massive bosons $W^{\pm}$ with mass $m_{_{W}}=ev$ 
and a  massive Higgs scalar with mass $m_{_{H}}=\sqrt{2\lambda}v$.

In addition to the constant vacuum solution the classical equations 
of motion admit solitonic solutions corresponding to static field 
configurations with finite (and non-vanishing) energy. In order to 
have finite energy the fields $\Phi^{a}$ must approach (sufficiently 
rapidly) values in ${\cal M}_{H}$ for large values of the radial 
coordinate $r$, so that they define a map $\Phi\, : 
S^{2}_{\infty}\rightarrow {\cal M}_{H}$. For the Georgi--Glashow model 
one has ${\cal M}_{H}=S^{2}$ and the corresponding solitonic 
configurations are characterized by an integer winding number $n\in 
\Pi_{2}(S^{2})\equiv \mathbb{Z}$. 
It can be shown \cite{corrigan} that the most general solution for the 
gauge field is 
\begin{displaymath}
	A_{\mu}^{a}=\frac{1}{v^{2}e} 
	\varepsilon^{abc}\Phi_{b}\partial_{\mu}\Phi_{c}+
	\frac{1}{v}\Phi^{a}K_{\mu} \, , 
\end{displaymath}
with an arbitrary $K_{\mu}$. The resulting field strength is
\begin{displaymath}
	F^{a}_{\mu\nu} = \frac{1}{v} \Phi^{a}F_{\mu\nu} \, ,
\end{displaymath}
where $F_{\mu\nu}$ is an `Abelian' field strength 
\begin{displaymath}
  F_{\mu\nu}=\frac{1}{v^{3}e}\varepsilon_{abc}\Phi^{a}
  \partial_{\mu}\Phi^{b}\partial_{\nu}\Phi^{c}+\partial_{\mu}
  K_{\nu}-\partial_{\nu}K_{\mu} \, ,
\end{displaymath}
satisfying the Maxwell equations. The corresponding magnetic field 
$B_{i}=\frac{1}{2} \varepsilon_{ijk}F^{jk}$ allows to define a 
magnetic charge obeying the Dirac quantization condition
\begin{equation}
	g = \int_{S^{2}_{\infty}} d \vec{\sigma}\cdot \vec{B} = 
	\frac{1}{2ev^{3}} \int_{S^{2}_{\infty}} d^{2}\sigma^{i} \, 
	\varepsilon_{ijk}\varepsilon^{abc}
	\Phi_{a}\partial^{j}\Phi_{b}\partial^{k}\Phi_{c} = 
	\frac{4\pi n}{e} \, ,
	\label{abmagncharge}
\end{equation}
where $n$ is the winding number. 

The configuration $(A_{\mu}^{a},\Phi^{a})$ is interpreted as 
describing a magnetic monopole. In fact the energy density decreases 
to zero outside a finite region as is expected for a `localized' object. 
Furthermore computing the non Abelian magnetic field 
\begin{displaymath}
	B^{a}_{i}=\frac{1}{2}\varepsilon_{ijk}F^{ajk}=
	\frac{1}{2v^{4}e}\Phi^{a}\varepsilon_{ijk}\varepsilon^{bcd}
	\Phi_{b}\partial^{j}\Phi_{c}\partial^{k}\Phi_{d} = 
	\frac{1}{v} \Phi^{a}B_{i}
\end{displaymath}
allows to associate a magnetic charge to the configuration using the 
fact that for large $r$ $\Phi^{a}$ must approach values in 
${\cal M}_{H}$, {\em i.e.} $\Phi^{a}\Phi_{a} = v^{2}$ on 
$S_{\infty}^{2}$, so that 
\begin{equation}
	g = \int_{S^{2}_{\infty}} d\vec{\sigma} \cdot \frac{1}{v} \Phi^{a} 
	\vec{B}_{a} = \frac{1}{v^{2}} \int_{S^{2}_{\infty}} d\vec{\sigma} 
	\cdot \Phi_{a}\Phi^{a} \vec{B} = \int_{S^{2}_{\infty}} 
	d\vec{\sigma}\cdot \vec{B} = \frac{4\pi n}{e} \, .
	\label{magncharge}
\end{equation}
Equation (\ref{magncharge}) shows that the ``magnetic particle'' 
satisfies the Dirac quantization condition, $eg=4\pi n$. Notice that 
the non Abelian field strength is aligned along the Higgs field 
outside the core of the monopole. The magnetic charge $g$ can be 
written in the form
\begin{displaymath}
	g = \frac{1}{v}\int_{S^{2}_{\infty}} d\vec{\sigma} \cdot 
	\vec{B}^{a} \Phi_{a} = \frac{1}{v} \int d^{3}x \, \vec{B}^{a}
	(\vec{D}\Phi)^{a} \, ,
\end{displaymath}
which follows from Stokes' theorem and the Bianchi identity, which 
implies $D_{i}B^{i}=0$.

For a static configuration with vanishing electric field one then 
finds that the `mass' given by (\ref{energy}) satisfies the {\em 
Bogomol'nyi bound}
\begin{equation}
	m_{_{M}} \geq v |g| \, .
	\label{bound}
\end{equation}
In (\ref{bound}) the equality holds if and only if $V(\Phi) = 0$ 
identically (\ie $\lambda=0$ in the Georgi--Glashow model) and the 
first order Bogomol'nyi equation
\begin{equation}
	B^{a}_{i} = D_{i} \Phi^{a}
	\label{bogomeq}
\end{equation}
is satisfied. 

An explicit solution for the monopole field configuration can be found 
in the special case of spherical symmetry and if the Bogomol'nyi 
bound is saturated. 
A field configuration corresponding to a non trivial vacuum cannot be 
invariant either under the group of spatial rotations 
SO(3)$_{{\rm R}}$ or under the SU(2)$_{{\rm G}}$ group of global 
gauge transformations~\footnote{Gauge transformations that do not reduce 
to the identity at spatial infinity are referred to as ``global'' or 
``large'' gauge transformations.}. 
However a solution invariant under the diagonal 
SO(3) subgroup of SO(3)$_{{\rm R}}\times$SU(2)$_{{\rm G}}$ can 
be obtained through the ansatz
\begin{displaymath}
	\Phi^{a}=\frac{x^{a}}{e r^{2}} H(\xi) \, , \hspace{0.5cm}
	A_{0}^{a}=0 \, ,\hspace{0.5cm} A_{i}^{a}= 
	-\varepsilon^{a}{}_{ij}\frac{x^{j}}{er^{2}}[1-K(\xi)] \, ,
\end{displaymath}
where $r=|x|$, $\xi=ver$ and the functions $H$ and $K$ satisfy 
$K(\xi)\to 1$, $H(\xi)\to 0$ for $r\to 0$ and $K(\xi)\to 0$, 
$H(\xi)/\xi \to 1$ for $r\to \infty$. In the limit in which the 
Bogomol'nyi bound is saturated an explicit solution was derived by 
Prasad and Sommerfield \cite{prasadsommer}. The exact form of the 
solution will not be discussed here since it will not be relevant for 
future considerations.

In general a non trivial vacuum in the limit $V(\Phi)=0$ (BPS 
limit~\footnote{Field configurations saturating the 
Bogomol'nyi bound are called ``BPS saturated''.}) 
is rather unnatural and moreover the condition of vanishing potential 
is expected to be broken by quantum effects. However in 
supersymmetric theories this situation is perfectly consistent because 
of the presence of flat directions that are protected 
by supersymmetry. The construction 
leading to the monopoles considered so far can be generalized to the 
case of non vanishing electric field giving rise to configurations 
with both electric and magnetic charge (dyons) 
\cite{julia}~\footnote{Static dyon solutions corresponding to finite 
energy configurations cannot be constructed in the temporal gauge 
$A_{0}=0$}; for such states the Bogomol'nyi bound becomes 
\begin{displaymath}
	m_{_{D}}\geq v \sqrt{e^{2}+g^{2}} \, .
\end{displaymath}

The Lagrangian (\ref{lgeorgiglash}) can be modified by the introduction 
of a $\theta$-term
\begin{equation}
	{\cal L}_{\theta} = -\frac{\theta e^{2}}{32 \pi^{2}}
	F^{a}_{\mu\nu}*F_{a}^{\mu\nu} 
	\label{tetaterm}
\end{equation}
without spoiling renormalizability. The addition of ${\cal 
L}_{\theta}$ amounts to consider a complexified coupling constant
\begin{equation}
	\tau = \frac{\theta}{2\pi}+\frac{4i\pi}{e^{2}} 
	\label{tau}
\end{equation}
and write the Lagrangian as 
\begin{equation}
	{\cal L}=-\frac{1}{32\pi}{\rm Im}\, \tau 
 	\left(F^{\mu\nu}+i*F^{\mu\nu}\right)
 	\left(F_{\mu\nu}+i*F_{\mu\nu}\right) - \frac{1}{2} D^{\mu}\Phi 
 	D_{\mu} \Phi \, .
	\label{taulagrang}
\end{equation}
In the presence of a $\theta$ term the dyons acquire an additional 
electric charge \cite{witteneffect} ({\em Witten effect}). More precisely 
the electric and magnetic charge operators are
\begin{eqnarray}
	Q_{m} & = & \frac{1}{v} \int d^{3}x \, D_{i}\Phi^{a}B_{a}^{i} 
	\nonumber  \\
	Q_{e} & = & \frac{1}{v} \int d^{3}x \, D_{i}\Phi^{a}E_{a}^{i}
	\label{emcharges}
\end{eqnarray}
and one finds for the electric charge of a dyon
\begin{displaymath}
	q=n_{e}e-\frac{e\theta n_{m}}{2\pi} \, ,
\end{displaymath}
with $n_{e}\in \mathbb{Z}$ and $n_{m}=\begin{displaystyle}\frac{eg}{4\pi}
\end{displaystyle}\in \mathbb{Z}$.


\section{Monopoles and fermions}
\label{monopfermi}
\fancyhead[LO]{{\footnotesize B.3~~{\it Monopoles and fermions}}}

In realistic models the gauge fields are coupled to fermions and one 
is then led to consider monopole configurations in the presence of 
fermions. The coupling to spinors in a monopole background has 
important consequences that will be briefly discussed here in  the 
simplest case of gauge group SU(2).

Dirac fermions can be coupled to the Georgi--Glashow model by adding 
to (\ref{lgeorgiglash}) the term 
\begin{equation}
	{\cal L}_{_{{\rm F}}} = \sum_{k=1}^{N_{f}} i {\overline \psi}^{k}
	\gamma^{\mu}D_{\mu}\psi^{k}-if{\overline \psi}^{k}\Phi\psi^{k} \, .
	\label{lferm}
\end{equation}
The Dirac equation reads
\begin{displaymath}
	(i\gamma^{\mu}D_{\mu}-\Phi)\psi^{k}=0 
\end{displaymath}
and a suitable choice for the $\gamma$-matrices is 
\begin{displaymath}
	\gamma^{0}=\left( \begin{array}{cc} 0 & -i \\ i & 0 \end{array} 
	\right) \; , \hspace{1cm} \gamma^{i}=\left( \begin{array}{cc} 
	-i\sigma^{i} & 0 \\ 0 & i\sigma^{i} \end{array} \right) \: .
\end{displaymath}
In the background of a monopole one can look for a solution for 
$\psi^{k}$ in the form 
\begin{displaymath}
	\psi^{k}(x,t)=e^{iEt}\psi^{k}(x) \, ,
\end{displaymath}
with $\psi^{k}(x) = \left( \begin{array}{c} \chi^{k}_{+}(x) \\ 
\chi^{k}_{-}(x) \rule{0pt}{20pt} \end{array} \right)$. 
The Weyl spinors $\chi^{k}_{\pm}$ then satisfy the equations
\begin{eqnarray*}
	\Dsm \chi^{k}_{+} & = & (i\sigma^{i}D_{i}+\Phi)\chi^{k}_{-}=
	E\chi^{k}_{+} \\
	\Dsm \chi^{k}_{-} & = & (i\sigma^{i}D_{i}-\Phi)\chi^{k}_{+}=
	E\chi^{k}_{-} \, ,
\end{eqnarray*}
which in general have both positive energy ($E > $0) and zero energy 
($E$=0) solutions. The monopole ground state is then constructed as a 
combination of the zero-energy eigenfunctions that are interpreted as 
`fermionic collective coordinates' describing a Grassmannian 
deformation of the original monopole. The generic solution of the 
Dirac equation can be written in the form 
\begin{equation}
	\psi^{k}=\sum_{i=1}^{n} a_{_{0}}^{i} \psi^{k}_{_{0}}{}_{i} +
	\sum_{j=1}^{\infty} a^{j}\psi_{j}^{k} \, ,
	\label{diracsolution}
\end{equation}
where $\psi^{k}_{_{0}}{}_{i}$ are the zero-energy eigenfunctions. The 
coefficients of the zero-energy eigenfunctions in the expansion 
(\ref{diracsolution}) become creation 
and annihilation operators and zero-energy solitonic states are 
obtained by acting with such operators on the ``bosonic'' monopole 
ground state. The number of fermionic zero-modes is determined by an 
index theorem \cite{index}. 

In the case of fermions in the fundamental 
representation it can be proved that there is one zero-mode for each 
flavor for a charge one monopole. Rewriting the $N_{f}$ Dirac fermions 
in terms of $2N_{f}$ Weyl fermions one finds that the creation and 
annihilation operators generate a Clifford algebra of dimension 
$2^{2N_{f}/2}=2^{N_{f}}$, that splits into two irreducible 
representations of dimension $2^{N_{f}-1}$. As a result the monopole 
ground state is a spinor of SO(2$N_{f}$). The angular momentum 
generator for a symmetric monopole is 
$\vec{K}=\vec{L}+\vec{S}+\vec{T}$, where $\vec{T}$ are the 
SU(2)$_{{\rm G}}$ 
generators of global gauge transformations. Fermions in the 
fundamental can have `spin' zero since they transform under 
SU(2)$_{{\rm G}}\times$SO(3)$_{{\rm R}}$ in the representation  
${\bf 2}\times{\bf 2}={\bf 1}+{\bf 3}$ which contains the singlet. 
Hence the fermionic zero modes carry zero total angular momentum and 
the resulting monopole configurations have spin zero.

If the fermions are in the adjoint representation there are two 
zero-modes for each flavor in the one-monopole background \cite{index}. 
Furthermore these zero modes necessarily have $\vec{K}\neq 0$ since the 
adjoint fermions transform in the ${\bf 3}\times {\bf 2} =
{\bf 2}+{\bf 4}$ of SU(2)$_{{\rm G}}\times$SO(3)$_{{\rm R}}$. 
Since the zero modes are two-fold degenerate they must carry spin 
$\pm \frac{1}{2}$. Consequently in the simplest case of $N_{f}$=1 the 
corresponding creation and annihilation operators (that will be denoted 
by $a_{\pm}^{\dagger}$ and $a_{\pm}$) can be used to generate the 
following multiplet of zero-energy states acting on the charge-one 
Clifford vacuum $|\Omega \rangle$
\vspace{0.5cm}

\begin{center}
\begin{tabular}{|l|c|}
	\hline
	~~~~~{\rm State}\rule[-7pt]{0pt}{21pt}~~~ & ~~~~Spin~~~~  \\
	\hline \hline
	~~~$|\Omega \rangle$~ \rule[-7pt]{0pt}{21pt} & 0  \\
	\hline
	~~~$a_{+}^{\dagger}|\Omega \rangle$ \rule[-7pt]{0pt}{21pt} & 
	$\frac{1}{2}$ \\
	\hline
	~~~$a_{-}^{\dagger}|\Omega \rangle$ \rule[-7pt]{0pt}{21pt} & 
	$ -\frac{1}{2} $ \\
	\hline
	~~~$a_{+}^{\dagger}a_{-}^{\dagger}|\Omega \rangle$ 
	\rule[-7pt]{0pt}{21pt}~~~ & 0  \\
	\hline
\end{tabular}
\end{center}
\vspace{0.5cm}

In conclusion the coupling to fermions in the adjoint representation 
gives a non-vanishing spin to the monopole.




\end{document}